\documentclass[12pt,twoside]{report}
\usepackage{epsfig}
\usepackage{amssymb}
\usepackage{amsfonts}
\begin{document}
%\Large
\begin{titlepage}
\title{Introduction to Chiral Perturbation Theory}
\author{Stefan Scherer\thanks{scherer@kph.uni-mainz.de, 
http://www.kph.uni-mainz.de/T/}
\\Institut f\"ur Kernphysik\\
Johannes Gutenberg-Universit\"at Mainz\\
J.~J.~Becher Weg 45\\
D-55099 Mainz\\
Germany}
\date{MKPH-T-02-09\\July 23, 2002}
\end{titlepage}
\maketitle
\begin{abstract}
   This article provides a pedagogical introduction to the basic 
concepts of chiral perturbation theory and is designed as a text for a 
two-semester course on that topic.
   Chapter 1 serves as a general introduction to the empirical and 
theoretical foundations which led to the development of chiral perturbation 
theory. 
   Chapter 2 deals with QCD and its global symmetries in the chiral limit;
the concept of Green functions and Ward identities reflecting the 
underlying chiral symmetry is elaborated.
   In Chap.\ 3 the idea of a spontaneous breakdown of a global symmetry
is discussed and its consequences in terms of the Goldstone theorem
are demonstrated.  
   Chapter 4 deals with mesonic chiral perturbation theory and  
the principles entering the construction of the chiral Lagrangian
are outlined.
   Various examples with increasing chiral orders and complexity are given.
   Finally, in Chap.\ 5 the methods are extended to include the interaction
between Goldstone bosons and baryons in the single-baryon sector, with
the main emphasis put on the heavy-baryon formulation.
   At the end, the method of infrared regularization in the relativistic 
framework is discussed.
\end{abstract}
\setcounter{page}{1}
\tableofcontents
\chapter{Introduction} 
\label{chap_i}

\section{Scope and Aim of the Review}
\label{sec_sar}
   The present review has evolved from two courses I have taught 
several times at the Johannes Gutenberg-Universit\"at, Mainz. 
   The first course was an introduction to chiral perturbation theory (ChPT)
which only covered the purely mesonic sector of the theory.  
   In the second course the methods were extended to also include baryons. 
   I have tried to preserve the spirit of those lectures in this article 
in the sense that it is meant to be a {\em pedagogical introduction} to the 
basic concepts of chiral perturbation theory. 
   By this I do not mean that the material covered is trivial, but that 
rather I have deliberately also worked out those pieces which by 
the ``experts'' are considered as well known. 
   In particular, I have often included intermediate steps in derivations  
in order to facilitate the understanding of the origin of the final results.  
   My intention was to keep a balance between mathematical rigor and 
illustrations by means of (numerous) simple examples.

   This article addresses both experimentalists and theorists! 
   Ideally, it would help a graduate student interested in theoretical
physics getting started in the field of chiral perturbation theory.
   However, it is also written for an experimental graduate student with
the purpose of conveying some ideas why the experiment she/he is performing 
is important for our theoretical understanding of the strong interactions.
   My precedent in this context is the review by A.~W.~Thomas
\cite{Thomas} which appeared in this series many years ago and 
served for me as an introduction to the cloudy bag model. 
    
   Finally, this article clearly is not intended to be a comprehensive  
overview of the numerous results which have been obtained over the past two 
decades. 
   For obvious reasons, I would, right at the beginning, like to apologize 
to all the researchers who have made important contributions to the field
that have not been mentioned in this work.
   Readers interested in the present status of applications are
referred to lecture notes and review articles 
\cite{Leutwyler:1991mz,Bijnens:xi,Meissner:1993ah,Leutwyler:1994fi,%
Bernard:1995dp,deRafael:1995zv,Pich:1995bw,Ecker:1995gg,Manohar:1996cq,%
Pich:1998xt,Burgess:1998ku} 
as well as conference proceedings 
\cite{Bernstein:zq,Bernstein:pm,Bernstein:2002}.

   The present article is organized as follows. 
   Chapter 1 contains a general introduction to the empirical and theoretical  
foundations which led to the development of chiral perturbation theory. 
   Many of the technical aspects mentioned in the introduction will be
treated in great detail later on.
   Chapter 2 deals with QCD and its global symmetries in the chiral limit
and the concept of Green functions and Ward identities reflecting the 
underlying chiral symmetry is elaborated.
   In Chap.\ 3 the idea of a spontaneous breakdown of a global symmetry
is discussed and its consequences in terms of the Goldstone theorem
are demonstrated.  
   Chapter 4 deals with mesonic chiral perturbation theory and  
the principles entering the construction of the chiral Lagrangian
are outlined.
   Various examples with increasing chiral orders and complexity are given.
   Finally, in Chap.\ 5 the methods are extended to include the interaction
between Goldstone bosons and baryons in the single-baryon sector with
the main emphasis put on the heavy-baryon formulation.
   At the end, the method of infrared regularization in the relativistic 
framework is discussed.
   Some technical details and simple illustrations are relegated to the
Appendices.

\section{Introduction to Chiral Symmetry and Its Application to Mesons and 
Single Baryons}   
\label{sec_icsamsb}
   In the 1950's a description of the strong interactions in the framework 
of quantum field theory seemed to fail due to an ever increasing number of 
observed hadrons as well as a coupling constant which was too large to allow 
for a sensible application of perturbation theory 
\cite{Gross:1998jx}.
   The rich spectrum of hadrons together with their finite sizes (i.e.,
non-point-like behavior showing 
up, e.g., in elastic electron-proton scattering through the existence 
of form factors) were the first hints pointing to a substructure in terms  
of more fundamental constituents. 
   A calculation of the anomalous magnetic moments of 
protons and neutrons in the framework of a pseudoscalar pion-nucleon  
interaction gave rise to values which were far off the empirical 
ones (see, e.g., \cite{Bethe:1956}). 
   On the other hand, a simple quark model analysis 
\cite{Beg:1964nm,Morpurgo:1965}
gave a prediction $-3/2$ for the ratio $\mu_p/\mu_n$ which is very
close to the empirical value of $-1.46$. 
  Nevertheless, the existence of quarks was hotly debated for a long time,  
since these elementary building blocks, in contrast to the constituents of
atomic or nuclear physics, could not be isolated as free particles, no matter 
what amount of energy was supplied to, say, the proton.  
   Until the early 1970's it was common to talk about ``fictitious''  
constituents allowing for a simplified group-theoretical classification of 
the hadron spectrum \cite{Gell-Mann:nj,Zweig:1964jf}, which, however, could 
not be interpreted as dynamical degrees of freedom in the context of 
quantum field theory. 
 
    In our present understanding, hadrons are highly complex  
objects built from more fundamental degrees of freedom. 
   These are on the one hand matter fields with spin 1/2 (quarks) 
and on the other hand massless spin-1 fields (gluons)
mediating the strong interactions.  
   Many empirical results of medium- and high-energy physics 
\cite{Altarelli:1989ue}
such as, e.g., deep-inelastic lepton-hadron scattering, 
hadron production in electron-positron annihilation, and 
lepton-pair production in Drell-Yan processes 
may successfully be described using 
perturbative methods in the framework of an 
SU(3) gauge theory, which is referred to as quantum chromodynamics 
(QCD) \cite{Gross:1973id,Weinberg:un,Fritzsch:pi}.
   Of particular importance in this context is the concept of 
asymptotic freedom \cite{Gross:1973id,Gross:ju,Politzer:fx}, 
referring to the fact that the coupling strength decreases for increasing 
momentum transfer $Q^2$, providing an explanation of 
(approximate) Bjorken scaling in deep inelastic scattering and allowing 
more generally for a perturbative approach at high energies. 
   Sometimes perturbative QCD is used as a synonym for asymptotic freedom.
   In Refs.\ \cite{Coleman:1973sx,Zee:gn} it was shown that Yang-Mills 
theories, i.e., gauge theories based on non-Abelian Lie groups, provide the 
only possibility for asymptotically free theories in four dimensions.
   At present, QCD is compatible with all empirical phenomena of 
the strong interactions in the asymptotic region. 
   However, one should also keep in mind that many phenomena cannot be
treated by perturbation theory.
   For example, simple (static) properties of hadrons cannot yet be described 
by {\em ab initio} calculations from QCD and this remains one of the largest 
challenges in theoretical particle physics \cite{Gross:1998jx}.
   In this context it is interesting to note that, of the three open
problems of QCD at the quantum level, namely, the ``gap problem,''
``quark confinement,'' and (spontaneous)
``chiral symmetry breaking'' \cite{Jaffe:2000},
the Yang-Mills existence of a mass gap has been chosen 
as one of the Millennium Prize Problems \cite{prizeproblems}
of the Clay Mathematics Institute.
   From a physical point of view this problem relates to the fact that
nuclear forces are strong and short-ranged. 
 
   One distinguishes among six quark flavors $u$ (up), $d$ (down),  
$s$ (strange), $c$ (charm), $b$ (bottom),  and $t$ (top), 
each of which coming in three different 
color degrees of freedom and transforming 
as a triplet under the fundamental representation of color SU(3).  
   The interaction between the quarks and the eight gauge bosons does
not depend on flavor, i.e., gluons themselves are flavor neutral.
   On the other hand, due to the non-Abelian character of the group SU(3), 
also gluons  carry ``color charges'' such that the QCD Lagrangian contains 
gluon self interactions involving vertices with three and four gluon fields. 
   As a result, the structure of QCD is much more complicated than that of
Quantum Electrodynamics (QED) which is based on a local, Abelian U(1) 
invariance. 
   However, it is exactly the non-Abelian nature of the theory which 
provides an anti-screening due to gluons that prevails over the screening
due to $q\bar{q}$ pairs, 
leading to an asymptotically free theory \cite{Nielsen:1981}.
  Since neither quarks nor gluons have been observed as free, asymptotic 
states, one assumes that any observable hadron  must be in a so-called 
color singlet state, i.e., a physically observable state is invariant under 
SU(3) color transformations. 
   The strong increase of the running coupling for large distances possibly 
provides a mechanism for color confinement
\cite{Gross:ju}. 
   In the framework of lattice QCD this can be shown in the so-called strong 
coupling limit \cite{Wilson:1974sk}.
   However, one has to keep in mind that the continuum limit of lattice
gauge theory is approached for a weak coupling
and a mathematical proof for color confinement is still missing
\cite{Jaffe:2000}.
 
   There still exists no {\em analytical} method for the  
description of QCD at large distances, i.e., at low energies. 
   For example, how the asymptotically observed hadrons, including their
rich resonance spectrum, are created from
QCD dynamics is still insufficiently understood.\footnote{
   For a prediction of hadron masses in the framework of lattice QCD see,
e.g., Refs.\ \cite{Butler:1994em,AliKhan:2001tx}.}
   This is one of the reasons why, for many practical purposes, one
makes use of phenomenological, more-or-less QCD-inspired, models of 
hadrons (see, e.g., \cite{Thomas,Holzwarth:1985rb,Meissner:1985qs,%
Zahed:1986qz,Gottfried,Bhaduri:gc,Mosel,Giannini:pc,Vogl:1991qt,%
Donoghue:dd,Thomas:kw}). 
  
   Besides the {\em local} SU(3) color symmetry, QCD exhibits further 
{\em global} symmetries. 
   For example, in a strong interaction process, a given quark cannot change 
its flavor, and if quark-antiquark pairs are created or annihilated during
the interaction, these pairs must be flavor neutral.
   In other words, for each flavor the difference in the number of quarks and 
antiquarks (flavor number) is a constant of the motion.
   This symmetry originates in a global invariance under a direct
product of U(1) transformations for each quark flavor and is
an exact symmetry of QCD independent of the value of the quark masses.
   Other symmetries are more or less satisfied. 
   It is well known that the hadron spectrum may be organized in terms of 
(approximately) degenerate basis states carrying irreducible representations 
of isospin SU(2).
   Neglecting electromagnetic effects, such a symmetry in QCD
results from equal $u$- and $d$-quark (current) masses.
   The extension including the  $s$ quark leads to the famous
flavor SU(3) symmetry \cite{EightfoldWay} which, however, is already
significantly broken due to the larger $s$-quark mass.

   The masses of the three light quarks $u$, $d$, and $s$ are small 
in comparison with the masses of ``typical'' light hadrons such as, e.g., 
the $\rho$ meson (770 MeV) or the proton (938 MeV). 
   On the other hand, the eight lightest pseudoscalar mesons are
distinguished by their comparatively small masses.\footnote{They
are not considered as ``typical'' hadrons due to their special
role as the (approximate) Goldstone bosons of spontaneous chiral
symmetry breaking.}
   Within the pseudoscalar octet, the isospin triplet of pions has a
significantly smaller mass (135 MeV) than the mesons containing 
strange quarks. 
   One finds a relatively large mass gap of about 630 MeV between the 
isospin triplets of the pseudoscalar and the vector
mesons, with the gap between the corresponding multiplets involving strange
mesons being somewhat smaller.

   In the limit in which the masses of the light quarks go to zero,
the left-handed and right-handed quark fields are decoupled from each other
in the QCD Lagrangian.
   At the ``classical'' level QCD then exhibits a global 
$\mbox{U(3)}_L\times\mbox{U(3)}_R$ symmetry.
   However, at the quantum level (including loops) the singlet axial-vector 
current develops an anomaly 
\cite{Adler:1969gk,Adler:1969er,Bardeen:1969md,Bell:1969ts,Adler:1970} 
such that the difference in left-handed and right-handed quark numbers 
is not a constant of the motion.
   In other words, in the so-called chiral limit, the QCD-Hamiltonian
has a $\mbox{SU(3)}_L\times\mbox{SU(3)}_R\times\mbox{U(1)}_V$ symmetry.
    
   Naturally the question arises, whether the hadron spectrum is,
at least approximately, in accordance with such a symmetry of the 
Hamiltonian.
   The $\mbox{U(1)}_V$ symmetry is connected to baryon number conservation, 
where quarks and antiquarks are assigned the baryon numbers
$B=1/3$ and $B=-1/3$, respectively.
   Mesons and baryons differ by their respective baryon numbers
$B=0$ and $B=1$. 
   Since baryon number is additive, a nucleus containing $A$ nucleons
has baryon number $B=A$.

   On the other hand, the $\mbox{SU(3)}_L\times\mbox{SU(3)}_R$ symmetry 
is not even approximately realized by the low-energy spectrum.
   If one constructs from the 16 generators of the group 
$G=\mbox{SU(3)}_L\times\mbox{SU(3)}_R$ the linear 
combinations $Q^a_V=Q^a_R+Q^a_L$ and $Q^a_A=Q^a_R-Q^a_L$,  
$a=1,\cdots,8$, the generators $Q^a_V$ form a Lie algebra
corresponding to a SU(3)$_V$ subgroup $H$ of $G$.
   It was shown in Ref.\ \cite{Vafa:tf} that, in the chiral limit,
the ground state is necessarily invariant
under the group $H$, i.e., the eight generators $Q^a_V$ annihilate the
ground state.
   The symmetry with respect to $H$ is said to be realized in 
the so-called Wigner-Weyl mode.
   As a consequence of Coleman's theorem \cite{Coleman:1966},   
the symmetry pattern of the spectrum follows the symmetry of
the ground state.
   Applying one of the axial generators $Q^a_A$ to an arbitrary state 
of a given multiplet of well-defined parity, one would obtain
a degenerate state of opposite parity, since $Q^a_A$ has negative
parity and, by definition, commutes with the Hamiltonian in the
chiral limit.
   However, due to Coleman's theorem such a conclusion tacitly
assumes that the ground state is annihilated by the $Q^a_A$.
   Since such a parity doubling is not observed in the spectrum
one reaches the conclusion that the $Q^a_A$ do {\em not}
annihilate the ground state.
   In other words, the ground state is not invariant under the full symmetry
group of the Hamiltonian, a situation which is referred to as spontaneous
symmetry breaking or the Nambu-Goldstone realization of a symmetry
\cite{Nambu:xd,Nambu:tp,Nambu:fr}.
   As a consequence of the Goldstone theorem \cite{Goldstone:eq,Goldstone:es},
each generator which commutes with the Hamiltonian 
but does not annihilate the ground state is associated with 
a massless Goldstone boson, whose properties are tightly connected with
the generator in question.

   The eight generators $Q_A^a$ have negative parity, baryon number zero,
and transform as an octet under the subgroup SU(3)$_V$ leaving the vacuum 
invariant.
   Thus one expects the same properties of the Goldstone bosons, 
and the light pseudoscalar octet ($\pi,K,\eta$) qualifies as candidates
for these Goldstone bosons.
   The finite masses of the physical multiplet are interpreted as
a consequence of the explicit symmetry breaking due to the finite
$u$-, $d$-, and $s$-quark masses in the QCD Lagrangian \cite{Gasser:1982ap}.

   Of course, the above (global) symmetry considerations were long known
before the formulation of QCD.
   In the 1960's they were the cornerstones of the description of 
low-energy interactions of hadrons in the framework of various techniques, 
such as the current-algebra approach in combination with the hypothesis of
a partially conserved axial-vector current (PCAC) 
\cite{Gell-Mann:1964tf,Adler:1968,Treiman:1972,DeAlfaro:1973},
the application of phenomenological Lagrangians
\cite{Weinberg:1966fm,Schwinger:1967tc,Weinberg:de,Coleman:sm,Callan:sn,%
Gasiorowicz:kn},
and perturbation theory about a symmetry realized in the Nambu-Goldstone
mode \cite{Dashen:eg,Dashen:ez,Li:1971vr,Pagels:se}.
   All these methods were equivalent in the sense that they produced the
same results for ``soft'' pions \cite{Dashen:ez} .
    
   Although QCD is widely accepted as the fundamental gauge
theory underlying the strong interactions, we still 
lack the analytical tools for {\em ab initio} descriptions of
low-energy properties and processes.
   However, new techniques have been developed to extend the results of
the current-algebra days and {\em systematically} explore corrections
to the soft-pion predictions based on symmetry properties of QCD Green 
functions.
   The method is called chiral perturbation theory (ChPT) 
\cite{Weinberg:1978kz,Gasser:1983yg,Gasser:1984gg} and describes the 
dynamics of Goldstone bosons in the framework of an effective field theory.
   Although one returns to a field theory in terms of non-elementary
hadrons, there is an important distinction between the early
quantum field theories of the strong interactions and the new 
approach in the sense that, now, one is dealing with a so-called 
{\em effective} field theory.
   Such a theory allows for a {\em perturbative} treatment in terms of a
momentum---as opposed to a coupling-constant---expansion.
 
   The starting point is a theorem of Weinberg stating that a perturbative
description in terms of the most general effective Lagrangian containing all
possible terms compatible with assumed symmetry principles yields the most
general $S$ matrix consistent with the fundamental principles of 
quantum field theory and the assumed symmetry principles
\cite{Weinberg:1978kz}.
   The proof of the theorem relies on Lorentz invariance and the 
absence of anomalies \cite{Leutwyler:1993iq,D'Hoker:1994ti} 
and starts from the observation
that the Ward identities satisfied by the Green functions of the symmetry 
currents are equivalent to an invariance of the generating functional
under {\em local} transformations \cite{Leutwyler:1993iq}.
   For that reason, one considers a {\em locally} invariant, effective
Lagrangian although the symmetries of the underlying theory may originate
in a global symmetry.
   If the Ward identities contain anomalies, they show up as a modification
of the generating functional, which can explicitly be incorporated through
the Wess-Zumino-Witten construction \cite{Wess:yu,Witten:tw}.
   All other terms of the effective Lagrangian remain locally invariant.  

   In the present case, the assumed symmetry is the  
$\mbox{SU(3)}_L\times\mbox{SU(3)}_R\times\mbox{U(1)}_V$ symmetry of
the QCD Hamilton operator in the chiral limit,
in combination with a restricted $\mbox{SU(3)}_V\times\mbox{U(1)}_V$
symmetry of the ground state.
   For center-of-mass energies below the $\rho$-meson mass, the only
asymptotic states which can explicitly be produced are the Goldstone
bosons.
   For the description of processes in this energy regime one organizes
the most general, chirally invariant Lagrangian for the pseudoscalar
meson octet in an expansion in terms of momenta and quark masses.
   Such an ansatz is naturally suggested by the fact that the interactions 
of Goldstone bosons are known to vanish in the zero-energy limit.
   Since the effective Lagrangian by construction contains an infinite
number of interaction terms, one needs for any practical purpose an
organization scheme allowing one to compare the importance of, say,
two given diagrams. 
    To that end, for a given diagram, one analyzes its behavior under
a linear rescaling of the {\em external} momenta,
$p_i\to t p_i$, and a quadratic rescaling of the light quark masses,
$m_i\to t^2 m_i$.
   Applying Weinberg's power counting scheme \cite{Weinberg:1978kz},
one finds that any given diagram behaves as $t^D$, where $D\geq 2$ is
determined by the structure of the vertices and the topology of the diagram 
in question.
   For a given $D$, Weinberg's formula unambiguously determines
to which order in the momentum and quark mass expansion the Lagrangian 
needs to be known. 
   Furthermore, the number of loops is restricted to be smaller than
or equal to $D/2\!-\!1$, i.e., Weinberg's power counting establishes
a relation between the momentum expansion and the loop 
expansion.\footnote{The counting refers to ordinary chiral 
perturbation theory in the mesonic sector, where $D$ is an even number.}

   Effective field theories are not renormalizable in the usual sense.
   However, this is no longer regarded as a serious problem, 
since by means of Weinberg's counting scheme the infinities arising from
loops may be identified order by order in the momentum expansion
and then absorbed in a renormalization of the coefficients of the most
general Lagrangian.
   Thus, in any arbitrary order the results are finite.
   Of course, there is a price to pay: the rapid increase
in the number of possible terms as the order increases. 
   Practical applications will hence be restricted to low orders.

   The lowest-order mesonic Lagrangian, ${\cal L}_2$, is given
by the nonlinear $\sigma$ model coupled to external fields
\cite{Gasser:1983yg,Gasser:1984gg}.
   It contains two free parameters:  the pion-decay constant
and the scalar quark condensate, both in the chiral limit.
   The specific values are not determined by chiral symmetry and 
must, ultimately, be explained from QCD dynamics.
   When calculating processes in the phenomenological approximation 
to ${\cal L}_2$, i.e., considering only tree-level diagrams,
one reproduces the results of current algebra \cite{Weinberg:1978kz}.
   Since tree-level diagrams involving vertices derived from a Hermitian
Lagrangian are always real, one has to go beyond the tree level in
order not to violate the unitarity of the $S$ matrix.
   A calculation of one-loop diagrams with ${\cal L}_2$, on the one hand,
leads to infinities which are not of the original type, but also contributes
to a perturbative restoration of unitarity.
   Due to Weinberg's power counting, the divergent terms are of order
${\cal O}(p^4)$ and can thus be 
compensated by means of a renormalization of the 
most general Lagrangian at ${\cal O}(p^4)$.
    
   The most general, effective Lagrangian at ${\cal O}(p^4)$ was first
constructed by Gasser and Leutwyler \cite{Gasser:1984gg} and
contains 10 physical low-energy constants as well as two additional
terms containing only external fields.
   Out of the ten physically relevant structures, eight are required for
the renormalization of the infinities due to the one-loop diagrams
involving ${\cal L}_2$.
   The finite parts of the constants represent free parameters, reflecting
our ignorance regarding the underlying theory, namely QCD, 
in this order of the momentum expansion.
   These parameters may be fixed phenomenologically by comparison with
experimental data \cite{Gasser:1983yg,Gasser:1984gg,Bijnens:1994qh}. 
   There are also theoretical approaches for estimating the low-energy
constants in the framework of QCD-inspired models 
\cite{Ebert:1985kz,Espriu:1989ff,Ebert:1991xd,Bijnens:1992uz},
meson-resonance saturation 
\cite{Ecker:yg,Ecker:1988te,Donoghue:ed,Knecht:2001xc,Leupold:2001vs} and
lattice QCD  \cite{Myint:yw,Golterman:2000mg}.
  
   Without external fields (i.e., pure QCD) or including electromagnetic
processes only, the effective Lagrangians ${\cal L}_2$ and ${\cal L}_4$
have an additional symmetry: they contain
interaction terms involving exclusively an even number of Goldstone
bosons.
   This property is often referred to as normal or even {\em intrinsic} parity,
but is obviously not a symmetry of QCD, because it would exclude reactions
of the type   $\pi^0\to\gamma\gamma$ or $K^+K^-\to\pi^+\pi^-\pi^0$.
   In Ref.\ \cite{Witten:tw}, Witten discussed how to remove this symmetry
from the effective Lagrangian and essentially re-derived the Wess-Zumino 
anomalous effective action which describes the chiral anomaly \cite{Wess:yu}.
   The corresponding Lagrangian, which is of ${\cal O}(p^4)$,
cannot be written as a standard local effective Lagrangian in terms of
the usual chiral matrix $U$ but can be expressed directly in terms of
the Goldstone boson fields.
   In particular, for the above case, by construction it contains interaction 
terms with an odd number of Goldstone bosons (odd intrinsic parity).
   In contrast to the Lagrangian of Gasser and Leutwyler, the 
Wess-Zumino-Witten (WZW) effective action does not contain any free parameter
apart from the number of colors.
   The excellent description of the neutral pion decay $\pi^0\to\gamma\gamma$ 
for $N_c=3$ is regarded as a key evidence for the existence of 
{\em three} color degrees of freedom.

   Chiral perturbation theory to ${\cal O}(p^4)$ has become a well-established
method for describing the low-energy interactions of the pseudoscalar
octet.
   For an overview of its many successful applications the interested 
reader is referred to Refs.\ 
\cite{Leutwyler:1991mz,Bijnens:xi,Meissner:1993ah,Leutwyler:1994fi,%
Bernstein:zq,Bernard:1995dp,deRafael:1995zv,Pich:1995bw,Ecker:1995gg,%
Manohar:1996cq,Bernstein:pm,Pich:1998xt,Burgess:1998ku,Bernstein:2002}. 
   In general, due to the relatively large mass of
the $s$ quark, the convergence in the SU(3) sector is somewhat slower
as compared with the SU(2) version.
   Nevertheless, ChPT in the SU(3) sector has significantly contributed
to our understanding of previously open questions. 
   A prime example is the decay rate of $\eta\to\pi\pi\pi$ which
current algebra predicts to be much too small.
   In Ref.\ \cite{Gasser:1984pr} it was shown that one-loop corrections 
substantially increase the theoretical value and remove the previous
discrepancy between theory and experiment.

   For obvious reasons, the question of convergence of the method is of 
utmost importance.
   The so-called chiral symmetry breaking scale $\Lambda_{\rm CSB}$ is
the dimensional parameter which characterizes the convergence of
the momentum expansion \cite{Manohar:1983md,Georgi}.
   A ``naive'' dimensional analysis of loop diagrams suggests that this 
scale is given by $\Lambda_{\rm CSB}\approx 4\pi F_0$, 
where $F_0\approx 93\,\mbox{MeV}$ denotes the pion-decay 
constant in the chiral limit and the factor $4\pi$ originates from
a geometric factor in the calculation of loop diagrams in four dimensions.
   A second dimensional scale is provided by the masses of the lightest
excitations which have been ``integrated out'' as explicit dynamical
degrees of freedom of the theory---in the present case, typically the 
lightest vector mesons.   
   In a phenomenological approach the exchange of such particles
leads to a propagator of the type 
$(q^2-M^2)^{-1}\approx-M^{-2}(1+q^2/M^2+\cdots)$, where
the expansion only converges for $|q^2|< M^2$.
   The corresponding scale is approximately of the same size as $4\pi F_0$.
   Assuming a reasonable behavior of the coefficients of the momentum
expansion leads to the expectation that ChPT converges for
center-of-mass energies sufficiently below the $\rho$-meson
mass.
   Of course, the validity of such a statement depends on the 
specific process under consideration and the quantum numbers of
the intermediate states.

   Clearly, for a given process, it would be desirable to have an idea about 
the size of the next-to-leading-order corrections.
   In the odd-intrinsic-parity sector such a calculation is at least of
order ${\cal O}(p^6)$, because the WZW action itself is already of order 
${\cal O}(p^4)$.
   Thus, according to Weinberg's power counting, one-loop diagrams
involving exactly one WZW vertex and an arbitrary number of 
${\cal L}_2$ vertices result in corrections of ${\cal O}(p^6)$.
   Several authors have shown that quantum corrections to the 
Wess-Zumino-Witten classical action do not renormalize the coefficient of 
the  Wess-Zumino-Witten term 
\cite{Donoghue:ct,Issler:1990nj,Bijnens:1989jb,Akhoury:1990px,%
Ebertshaeuser:2001,Bijnens:2001bb}.
   Furthermore, the one-loop counter terms lead to conventional
chirally invariant structures at ${\cal O}(p^6)$
\cite{Donoghue:ct,Issler:1990nj,Bijnens:1989jb,Akhoury:1990px,%
Ebertshaeuser:2001,Bijnens:2001bb}.
   There have been several attempts to construct the most general 
odd-intrinsic-parity Lagrangian at ${\cal O}(p^6)$ and only recently
two independent calculations have found the same number of 23 independent 
structures in the SU(3) sector \cite{Ebertshauser:2001nj,Bijnens:2001bb}.  
   For an overview of the application of ChPT to anomalous processes, the 
interested reader is referred to Ref.\ \cite{Bijnens:xi}.
   
   In general, next-to-leading-order corrections to processes in the
even-intrinsic-parity sector are of ${\cal O}(p^4)$.
   However, there are also processes which receive their leading-order
contributions at ${\cal O}(p^4)$. 
   In particular, the reactions $\gamma\gamma\to\pi^0\pi^0$ 
\cite{Morgan:1991zx,Donoghue:1993kw,Bellucci:1994eb,Knecht:1994ug,%
Bellucci:1995ay,Bel'kov:1995fj} and 
$\eta\to\pi^0\gamma\gamma$ 
\cite{Ametller:1991dp,Bellucci:1995ay,Ko:rg,Bel'kov:1995fj,Jetter:1995js}
have received considerable attention, because the predictions at
${\cal O}(p^4)$ \cite{Bijnens:1987dc,Donoghue:ee} and 
\cite{Ametller:1991dp}, 
respectively, were in disagreement with experimental results
(\cite{Marsiske:1990hx} and \cite{Groom:in}, respectively). 
   In the case of $\gamma\gamma\to\pi^0\pi^0$ loop corrections at 
${\cal O}(p^6)$
lead to a considerably improved description, with the result only
little sensitive to the tree-level diagrams at 
${\cal O}(p^6)$ \cite{Bellucci:1994eb}.
   The opposite picture emerges for the decay $\eta\to\pi^0\gamma\gamma$,
where the tree-level diagrams at ${\cal O}(p^6)$ play an important
role.

   A second class of ${\cal O}(p^6)$ calculations includes processes which
already receive contributions at ${\cal O}(p^2)$ such as $\pi\pi$ scattering 
\cite{Bijnens:1995yn} or $\gamma\gamma\to\pi^+\pi^-$
\cite{Burgi:1996mm}.
   Here, ${\cal O}(p^6)$ calculations may be viewed as precision tests of
ChPT. 
   The first process is of fundamental importance because it provides 
information on the mechanism of spontaneous symmetry breaking in QCD
\cite{Bijnens:1995yn}.
   The second reaction is of particular interest because an old 
current-algebra low-energy theorem \cite{Terentev:ix} relates the 
electromagnetic polarizabilities $\bar{\alpha}$ and $\bar{\beta}$
of the charged pion at ${\cal O}(p^4)$
to radiative pion decay $\pi^+\to e^+\nu_e\gamma$. 
   Corrections at ${\cal O}(p^6)$ were 
shown to be 12\% and 24\% of the ${\cal O}(p^4)$ values for  
$\bar{\alpha}$ and $\bar{\beta}$, respectively \cite{Burgi:1996mm}.
   On the other hand, experimental results for the polarizabilities
scatter substantially and still have large uncertainties (see, e.g.,
Ref.\ \cite{Unkmeir:2001gw}) and new experimental data are clearly
needed to test the accuracy of the chiral predictions.

   In the SU(3) sector, the first construction of the most general
even-intrinsic parity Lagrangian at ${\cal O}(p^6)$
was performed in Ref.\ \cite{Fearing:1994ga}.
   Although it was later shown that the original list of terms
contained redundant
structures \cite{Bijnens:1999sh}, even the final number of 90 + 4 
free parameters is very large, such that, in contrast to the Lagrangian 
${\cal L}_4$ of Gasser and Leutwyler, it seems unlikely that all parameters 
can be fixed through comparison with experimental data.
   However, chiral symmetry relates different processes to each 
other, such that groups of interaction terms may be connected with each
other and through comparison with experiment the consistency conditions
of chiral symmetry may be tested.
   Furthermore, the same theoretical methods which have been applied
to predict the coefficients of ${\cal O}(p^4)$ may be extended to the next
order \cite{Belkov:1994qg} which, however, involves much more work.

   Chiral perturbation theory has proven to be highly successful in the
mesonic sector and, for obvious reasons, one would like to have a 
generalization including the interaction of Goldstone bosons with baryons.
   The group-theoretical foundations for a nonlinear realization of
chiral symmetry were developed in 
Refs.\ \cite{Weinberg:de,Coleman:sm,Callan:sn},
which also included the coupling of Goldstone bosons to other 
isospin or, for the more general case, SU($N$)-flavor multiplets.
   Numerous low-energy theorems involving the pion-nucleon interaction
and its SU(3) extension were derived in the 1960's by use of current-algebra 
methods and PCAC.      
   However, a {\em systematic} study of chiral corrections to the 
low-energy theorems has 
only become possible when the methods of mesonic ChPT were extended
to processes with one external nucleon line \cite{Gasser:1987rb}.
   The situation turned out to be more involved than in the pure
mesonic sector because the loops have a more complicated structure due
to the nucleon mass which, in contrast to the Goldstone boson masses, 
does not vanish in the chiral limit.
   This introduces a third scale into the
problem beyond the pion decay constant and the scalar quark condensate. 
   In particular, it was shown that the relativistic formulation, at
first sight, does not provide such a simple connection between the 
chiral expansion and the loop expansion as in the mesonic sector
\cite{Gasser:1987rb}, i.e., higher-order loop diagrams also contribute to 
lower orders in the chiral expansion of a physical quantity.
   This observation was taken as evidence for a breakdown of power counting
in the relativistic formulation.   
   Subsequently, techniques borrowed from heavy-quark physics were applied
to the baryon sector \cite{Jenkins:1990jv,Bernard:1992qa}, providing a 
heavy-baryon formulation of ChPT
(HBChPT), where the Lagrangian is not only expanded in the number of 
derivatives and quark masses but also in powers of inverse nucleon
masses.
   The technique is very similar to the Foldy-Wouthuysen method
\cite{Foldy:1949wa}. 
   
   There have been many successful applications of HBChPT to 
``traditional'' current-algebra processes such as pion photoproduction 
\cite{Bernard:1992nc,Bernard:1996ti} and radiative pion capture
\cite{Fearing:2000uy}, pion electroproduction \linebreak
\cite{Bernard:1992ys,Bernard:1993bq,Bernard:1994dt,Bernard:2000qz},
pion-nucleon scattering 
\cite{Bernard:1995pa,Mojzis:1997tu,Fettes:1998ud,Fettes:2001cr},
to name just a few 
(for an extensive overview, see Ref.\ \cite{Bernard:1995dp}).
   In all these cases, ChPT has allowed one to either systematically 
calculate corrections to the old current-algebra results or to obtain
new predictions which are beyond the scope of the old techniques. 
   Other applications include the calculation of static properties 
such as masses 
\cite{Jenkins:1991ts,Bernard:1993nj,Lebed:1994gt,Borasoy:1996bx,%
McGovern:1998tm}
and various form factors of baryons \linebreak
\cite{Bernard:1992ys,Bernard:1996cc,Fearing:1997dp,Kubis:1999xb}.
   The role of the pionic degrees of freedom has been extensively discussed
for real Compton scattering off the nucleon in terms of the electromagnetic 
polarizabilities  
\cite{Bernard:xi,Bernard:1993bg,Gellas:2000mx,McGovern:1999ar,Kumar:2000pv,%
Griesshammer:2001uw}.
   The new frontier of virtual \mbox{Compton} scattering off the nucleon
\cite{Guichon:1995pu,Drechsel:1996ag,Roche:2000ng}
has also been addressed in the framework of ChPT
\cite{Hemmert:1996gr,Hemmert:1997at,Hemmert:1999pz,L'vov:2001fz}.
   As in the mesonic sector, the most general chiral Lagrangian in
the single-baryon sector is needed which, due to the spin degree
of freedom, is more complicated 
\cite{Gasser:1987rb,Krause:xc,Ecker:1995rk,Fettes:2000gb}.
 
   In the baryonic sector, the $\Delta(1232)$ resonance plays a prominent role
because its excitation energy is only about two times the pion mass 
and its (almost) 100 \% branching ratio to the decay mode $N \pi$.
   In Ref.\ \cite{Hemmert:1996xg}, the formalism of the so-called small 
scale expansion was developed, which also treats the nucleon-delta mass 
splitting as a ``small'' quantity like the pion mass. 
   Subsequently, the formalism was applied to Compton scattering
\cite{Hemmert:1997tj}, baryon form factors \cite{Bernard:1998gv},
the $N\Delta$ transition \cite{Gellas:1998wx} and virtual Compton
scattering \cite{Hemmert:1999pz}.

   While the heavy-baryon formulation provided a useful low-energy
expansion scheme, it was realized in the context of the isovector
spectral function entering the calculation of the nucleon electromagnetic
form factor that the corresponding perturbation 
series fails to converge in part of the low-energy region 
\cite{Bernard:1996cc}.
   Various methods have been suggested to generate a power counting
which is also valid for the relativistic approach and which respects
the singularity structure of Green functions 
\cite{Tang:1996ca,Ellis:1997kc,Becher:1999he,Gegelia:1999gf,Gegelia:1999qt,%
Lutz:1999yr,Lutz:2001yb}.
   The so-called ``infrared regularization'' of Ref.\ \cite{Becher:1999he}
decomposes one-loop diagrams into singular and regular parts. 
   The singular parts satisfy power counting, whereas the regular parts can be 
absorbed into local counter terms of the Lagrangian.      
   This technique solves the power counting 
problem of relativistic baryon chiral perturbation theory at 
the one-loop level and has already been applied to the calculation of 
baryon masses in SU(3) ChPT \cite{Ellis:1999jt}, of form factors
\cite{Kubis:2000zd,Zhu:2000zf,Kubis:2000aa}, pion-nucleon scattering 
\cite{Becher:2001hv} as well as the generalized Gerasimov-Drell-Hearn
sum rule \cite{Bernard:2002bs}.
   At present, the procedure has not yet been generalized to
higher-order loop diagrams.
   In Ref.\ \cite{Gegelia:1999qt} another approach, based on choosing
appropriate renormalization conditions, was proposed, leading
to the correct analyticity structure and a consistent power counting,
which can also be extended to higher loops.

   Finally, the techniques of effective field theory have also been
applied to the nucleon-nucleon interaction (see, e.g., Refs.\ 
\cite{Weinberg:1991um,Ordonez:1996rz,Kaiser:1997mw,vanKolck:1999mw,%
Epelbaum:2000dj,Beane:2001bc,Finelli:2002na}).
   Clearly this is a very important topic in its own right but is
beyond the scope of the present work.

\chapter{QCD and Chiral Symmetry}
\label{chap_qcdcs}

   Chiral perturbation theory (ChPT) provides a systematic framework for 
investigating strong-interaction processes at {\em low} energies,
as opposed to a perturbative treatment of quantum chromodynamics
(QCD) at high momentum transfers
in terms of the ``running coupling constant.''
   The basis of ChPT is the global 
$\mbox{SU(3)}_L\times \mbox{SU(3)}_R\times\mbox{U}(1)_V$
symmetry of the QCD Lagrangian in the limit of massless 
$u$, $d$, and $s$ quarks.
    This symmetry is assumed to be spontaneously broken down to 
$\mbox{SU(3)}_V\times\mbox{U(1)}_V$ giving rise to eight massless
Goldstone bosons.
   In this chapter we will describe in detail one of the foundations
of ChPT, namely the symmetries of QCD and their consequences in terms of 
QCD Green functions.

\section{Some Remarks on SU(3)}
\label{sec_srsu3}
   The group SU(3) plays an important role in the context of strong
interactions, because on the one hand it is the gauge group of QCD
and, on the other hand, 
flavor SU(3) is approximately realized
as a global symmetry of the hadron spectrum (Eightfold Way
\cite{Ne'eman:1961cd,Gell-Mann:xb,EightfoldWay}),
so that the observed (low-mass) hadrons can be 
organized in approximately degenerate multiplets fitting the dimensionalities
of irreducible representations of SU(3).
   Finally, as will be discussed later in this chapter, the direct product 
$\mbox{SU(3)}_L\times\mbox{SU(3)}_R$ is the chiral-symmetry group of 
QCD for vanishing $u$-, $d$-, and $s$-quark masses.
   Thus, it is appropriate to first recall a few basic properties of SU(3)
and its Lie algebra su(3) \cite{Balachandran:ab,O'Raifeartaigh:vq,Jones:ti}.

   The group SU(3) is defined as the set of all 
unitary, unimodular, $3 \times 3$ matrices $U$, i.e.\
$U^\dagger U=1$,\footnote{In this report we often adopt the convention
that 1 stands for the unit matrix in $n$ dimensions. It should be clear
from the respective context which dimensionality actually applies.}
and $\mbox{det}(U)=1$.
   In mathematical terms, SU(3) is an eight-parameter,
simply connected, compact Lie group.
   This implies that any group element can be parameterized by a set of 
eight independent real parameters $\Theta=(\Theta_1,\cdots, \Theta_8)$ 
varying over a continuous range.
   The Lie-group property refers to the fact that the group 
multiplication of two elements $U(\Theta)$ and $U(\Psi)$ 
is expressed in terms of eight {\em analytic} functions 
$\Phi_i(\Theta;\Psi)$, i.e.\ $U(\Theta)U(\Psi)=U(\Phi)$, where 
$\Phi=\Phi(\Theta;\Psi)$.
   It is simply connected because every element can be connected to the
identity by a continuous path in the parameter space and compactness requires 
the parameters to be confined in a finite volume.
   Finally, for compact Lie groups, every finite-dimensional representation
is equivalent to a unitary one and can be decomposed into a direct
sum of irreducible representations (Clebsch-Gordan series).

   Elements of SU(3) are conveniently written in terms
of the exponential representation\footnote{In our notation, the 
indices denoting group parameters and generators will appear as subscripts or 
superscripts depending on what is notationally convenient.
   We do not distinguish between upper and lower indices, i.e., we 
abandon the methods of tensor analysis.}
\begin{equation}
\label{2:1:uexp}
U(\Theta)=\exp\left(-i\sum_{a=1}^8 \Theta_a \frac{\lambda_a}{2}\right),
\end{equation}
with $\Theta_a$ real numbers, and where the eight linearly independent 
matrices $\lambda_a$ are the 
so-called Gell-Mann matrices, satisfying
\begin{eqnarray}
\label{2:1:gmme1}
\frac{\lambda_a}{2}&=&i\frac{\partial U}{\partial \Theta_a}(0,\cdots,0),\\
\label{2:1:gmme2}
\lambda_a&=&\lambda_a^\dagger,\\
\label{2:1:gmme3}
\mbox{Tr}(\lambda_a \lambda_b)&=&2\delta_{ab},\\
\label{2:1:gmme4}
\mbox{Tr}(\lambda_a)&=&0.
\end{eqnarray}
   An explicit representation of the Gell-Mann matrices is given by
\begin{eqnarray}
\label{2:1:gmm} 
&&\lambda_1=\left(\begin{array}{rrr}
0&1&0\\1&0&0\\0&0&0
\end{array}
\right),\quad
\lambda_2=\left(\begin{array}{rrr}
0&-i&0\\i&0&0\\0&0&0
\end{array}
\right),\quad
\lambda_3=\left(\begin{array}{rrr}
1&0&0\\0&-1&0\\0&0&0
\end{array}
\right),\nonumber\\
&&\lambda_4=\left(\begin{array}{rrr}
0&0&1\\0&0&0\\1&0&0
\end{array}
\right),\quad
\lambda_5=\left(\begin{array}{rrr}
0&0&-i\\0&0&0\\i&0&0
\end{array}
\right),\quad
\lambda_6=\left(\begin{array}{rrr}
0&0&0\\0&0&1\\0&1&0
\end{array}
\right),\nonumber\\
&&
\lambda_7=\left(\begin{array}{rrr}
0&0&0\\0&0&-i\\0&i&0
\end{array}
\right),\quad
\lambda_8=\sqrt{\frac{1}{3}}\left(\begin{array}{rrr}
1&0&0\\0&1&0\\0&0&-2
\end{array}
\right).
\end{eqnarray}  
   The set $\{i\lambda_a\}$ constitutes a basis of the 
Lie algebra su(3) of SU(3), i.e., the set of all complex traceless 
skew Hermitian $3\times 3$ matrices.
   The Lie product is then defined in terms of ordinary matrix multiplication
as the commutator of two elements of su(3).
   Such a definition naturally satisfies the Lie properties of 
anti-commutativity 
\begin{equation}
\label{2:1:anticom}
[A,B]=-[B,A]
\end{equation}
as well as the Jacobi identity
\begin{equation}
\label{2:1:jacobi}
[A,[B,C]]+[B,[C,A]]+[C,[A,B]]=0.
\end{equation}
   In accordance with Eqs.\ (\ref{2:1:uexp}) and (\ref{2:1:gmme1}),
elements of su(3) can be interpreted as tangent vectors in the identity of 
SU(3).

   The structure of the Lie group is encoded in the commutation relations
of the Gell-Mann matrices, 
\begin{equation}
\label{2:1:crgmm}
\left[\frac{\lambda_a}{2},\frac{\lambda_b}{2}\right]
=i f_{abc}\frac{\lambda_c}{2},
\end{equation}
where the totally antisymmetric real structure constants $f_{abc}$ are 
obtained from Eq.\ (\ref{2:1:gmme3}) as
\begin{equation}
\label{2:1:fabc}
f_{abc}=\frac{1}{4i}\mbox{Tr}([\lambda_a,\lambda_b]\lambda_c).
\end{equation}
  The independent non-vanishing values are explicitly summarized in the
scheme of Table \ref{table:2:1:su3structurconstants}.
   Roughly speaking, these structure constants are a measure of
the non-commutativity of the group SU(3).

\begin{table}
\begin{center}
\begin{tabular}{|r|r|r|r|r|r|r|r|r|r|}
\hline
$abc$&123&147&156&246&257&345&367&458&678\\ 
\hline
$f_{abc}$&1&$\frac{1}{2}$&$-\frac{1}{2}$&$\frac{1}{2}$&
$\frac{1}{2}$&$\frac{1}{2}$&$-\frac{1}{2}$&$\frac{1}{2}\sqrt{3}$&
$\frac{1}{2}\sqrt{3}$\\
\hline
\end{tabular}
\caption{\label{table:2:1:su3structurconstants}
Totally antisymmetric non-vanishing structure constants of SU(3).} 
\end{center}
\end{table}
   The anticommutation relations read
 \begin{equation}
\label{2:1:acrgmm}
\{\lambda_a,\lambda_b\}
= \frac{4}{3}\delta_{ab} +2 d_{abc} \lambda_c,
\end{equation}  
   where the totally symmetric $d_{abc}$ are given by
\begin{equation}
\label{2:1:dabc}
d_{abc}=\frac{1}{4}\mbox{Tr}(\{\lambda_a,\lambda_b\}\lambda_c),
\end{equation}
and are summarized in Table 
\ref{table:2:1:su3dsymbols}.
   Clearly, the anticommutator of two Gell-Mann matrices is 
not necessarily a Gell-Mann matrix.
   For example, the square of a (nontrivial) skew-Hermitian matrix 
is not skew Hermitian.

\begin{table}
\begin{center}
\begin{tabular}{|r|r|r|r|r|r|r|r|r|}
\hline
$abc$&
118&
146&
157&
228&
247&
256&
338&
344\\
\hline
$d_{abc}$&
$\frac{1}{\sqrt{3}}$&
$\frac{1}{2}$&
$\frac{1}{2}$&
$\frac{1}{\sqrt{3}}$&
$-\frac{1}{2}$&
$\frac{1}{2}$&
$\frac{1}{\sqrt{3}}$&
$\frac{1}{2}$\\
\hline
$abc$&
355&
366&
377&
448&
558&
668&
778&
888\\ 
\hline
$d_{abc}$&
$\frac{1}{2}$&
$-\frac{1}{2}$&
$-\frac{1}{2}$&
$-\frac{1}{2\sqrt{3}}$&
$-\frac{1}{2\sqrt{3}}$&
$-\frac{1}{2\sqrt{3}}$&
$-\frac{1}{2\sqrt{3}}$&
$-\frac{1}{\sqrt{3}}$\\
\hline
\end{tabular}
\caption{\label{table:2:1:su3dsymbols}
Totally symmetric non-vanishing $d$ symbols of SU(3).}
\end{center}
\end{table}
 
  Moreover, it is convenient to introduce as a ninth matrix 
$$\lambda_0 =\sqrt{2/3}\,\mbox{diag}(1,1,1),$$ 
such that Eqs.\ (\ref{2:1:gmme2}) and (\ref{2:1:gmme3}) are still
satisfied by the nine matrices $\lambda_a$.
   In particular, the set $\{i\lambda_a|a=0,\cdots, 8\}$ constitutes a 
basis of the 
Lie algebra u(3) of U(3), i.e., the set of all complex 
skew Hermitian $3\times 3$ matrices.
   Finally, an {\em arbitrary} $3\times 3$ matrix $M$ can then be written as
\begin{equation}
\label{2:1:matrixa}
M=\sum_{a=0}^8 \lambda_a M_a,
\end{equation}
where $M_a$ are complex numbers given by
$$
M_a=\frac{1}{2}\mbox{Tr}(\lambda_a M).
$$

\section{The QCD Lagrangian}
\label{sec_qcdl}
   The gauge principle has proven to be a tremendously successful method
in elementary particle physics to generate interactions between matter fields 
through the exchange of massless gauge bosons 
(for a detailed account see, e.g., \cite{Abers:qs,O'Raifeartaigh:vq}).
   The best-known example is, of course, quantum electrodynamics (QED) which
is obtained from promoting the global U(1) symmetry 
of the Lagrangian describing a free electron,\footnote{We use the
standard representation for the Dirac matrices (see, e.g., 
Ref.\ \cite{Bjorken_1964}).}
\begin{equation}
\label{2:2:freel}
\Psi\mapsto \mbox{exp}(-i\Theta)\Psi:
{\cal L}_{\rm free}=\bar{\Psi}(i\gamma^\mu\partial_\mu -m)\Psi
\mapsto
{\cal L}_{\rm free},
\end{equation}
to a local symmetry.
   In this process the parameter $0\leq \Theta\leq 2\pi$ describing an
element of U(1) is allowed to vary smoothly in space-time,
$\Theta\to\Theta(x)$, which is referred to as gauging the U(1) group. 
   To keep the invariance of the Lagrangian under local transformations
one introduces a four-potential ${\cal A}_\mu$ into the 
theory which transforms under the gauge transformation 
${\cal A}_\mu\mapsto {\cal A}_\mu-\partial_\mu\Theta/e$.
   The method is referred to as gauging the Lagrangian with respect to U(1): 
\begin{equation}
\label{2:2:lqed}
{\cal L}_{\rm QED}=\bar{\Psi}[i\gamma^\mu(\partial_\mu-i e {\cal A}_\mu)-m]\Psi
-\frac{1}{4} {\cal F}_{\mu\nu} {\cal F}^{\mu\nu},
\end{equation}
   where ${\cal F}_{\mu\nu}=\partial_\mu {\cal A}_\nu
-\partial_\nu {\cal A}_\mu$.\footnote{
We use natural units, i.e., $\hbar=c=1, e>0$, and $\alpha=e^2/4\pi
\approx 1/137$.}
   The covariant derivative of $\Psi$,
$$D_\mu \Psi\equiv(\partial_\mu -ie {\cal  A}_\mu)\Psi,$$ 
is defined such that under a so-called gauge transformation of the second
kind
\begin{equation}
\label{2:2:gtsk}
\Psi(x)\mapsto\exp[-i\Theta(x)]\Psi(x),\quad
{\cal A}_\mu(x)\mapsto {\cal A}_\mu(x)-\partial_\mu\Theta(x)/e,
\end{equation}
it transforms in the same way as $\Psi$ itself:
\begin{equation}
\label{2:2:cdt}
D_\mu\Psi(x)\mapsto\exp[-i\Theta(x)] D_\mu\Psi(x).
\end{equation}
   In Eq.\ (\ref{2:2:lqed}), the term containing the squared field strength 
makes the gauge potential a dynamical degree of freedom as opposed to a pure 
external field.
   A mass term $M^2 {\cal A}^2/2$ is not included since it would violate gauge 
invariance and thus the gauge principle requires massless gauge 
bosons.\footnote{Masses of gauge fields can
be induced through a spontaneous breakdown of the gauge symmetry.}
   In the present case we identify the ${\cal A}_\mu$ with the electromagnetic
four-potential and ${\cal F}_{\mu\nu}$ with the field strength tensor 
containing the electric and magnetic fields.
   The gauge principle has (naturally) generated the interaction of the 
electromagnetic field with matter.
   If the underlying gauge group is non-Abelian, the gauge principle
associates an independent gauge field with each independent continuous 
parameter of the gauge group.

   QCD is the gauge theory of the strong interactions 
\cite{Gross:1973id,Weinberg:un,Fritzsch:pi}
with color SU(3) as the underlying gauge 
group.\footnote{Historically, the color degree of freedom was introduced into
the quark model to account for the Pauli principle in the description
of baryons as three-quark states \cite{Greenberg:pe,Han:pf}.}
   The matter fields of QCD are the so-called quarks which are
spin-1/2 fermions, with six different flavors in addition to their
three possible colors (see Table \ref{2:2:table:quarks}).
   Since quarks have not been observed as asymptotically free
states, the meaning of quark masses and their numerical
values are tightly connected with the method by which they are extracted 
from hadronic properties 
(see Ref.\ \cite{Manohar_PDG} for a thorough discussion).
   Regarding the so-called current-quark-mass values of the light quarks,
one should view the quark mass terms merely as symmetry breaking
parameters with their magnitude providing a measure for the extent
to which chiral symmetry is broken \cite{Scheck:ur}.
   For example, {\em ratios} of the light quark masses can be inferred
from the masses of the light pseudoscalar octet 
(see Ref.\  \cite{Leutwyler:1996qg}).
   Comparing the proton mass, $m_p$ = 938 MeV, with the sum of
two up and one down current-quark masses (see Table \ref{2:2:table:quarks}),
\begin{equation}
\label{2:2:mp}
m_p\gg 2m_u+m_d,
\end{equation}
shows that an interpretation of the proton mass in terms of current-quark 
mass parameters must be very different from, say, the situation in the 
hydrogen atom, where the mass is essentially given by the sum of the electron 
and proton masses, corrected by a small amount of binding energy.
\begin{table}
\begin{center}
\begin{tabular}{|l|c|c|c|}
\hline
flavor&u&d&s\\ 
\hline
charge [e] &$2/3$&$-1/3$&$-1/3$\\
\hline
mass [MeV]&$5.1\pm 0.9$ & $9.3\pm 1.4$ & $175\pm 25$\\
&\cite{Leutwyler:1996qg}&\cite{Leutwyler:1996qg}&\cite{Bijnens:1994ci}\\
\hline
\hline
flavor&c&b&t\\ 
\hline
charge [e] &$2/3$&$-1/3$&$2/3$\\
\hline
mass [GeV] & $1.15-1.35$ & $4.0 - 4.4$ &$174.3\pm 3.2\pm 4.0$\\
& \cite{Manohar_PDG}
&\cite{Manohar_PDG}
&\cite{Manohar_PDG}
\\
\hline
\end{tabular}
\caption{\label{2:2:table:quarks} Quark flavors and their charges and masses.
   The absolute magnitude of $m_s$ is determined using QCD sum rules.
   The result is given for the $\overline{\mbox{MS}}$ running mass at scale
$\mu = 1$\, GeV.
   The light quark masses are obtained from the mass ratios found
using chiral perturbation theory, using the strange quark mass as input.
   The heavy-quark masses $m_c$ and $m_b$ are determined by the charmonium 
and D masses, and the bottomium and B masses, respectively.
   The top quark mass $m_t$ results from the measurement of
lepton + jets and dilepton + jets channels in the D$\emptyset$ and CDF 
experiments at Fermilab.}
\end{center}
\end{table}

  The QCD Lagrangian obtained from the gauge principle reads
\cite{Marciano:su,Altarelli:1981ax}
\begin{equation}
\label{2:2:lqcd}
{\cal L}_{\rm QCD}=\sum_{f={u,d,s, \atop c,b,t}}
\bar{q}_f(i D\hspace{-.6em}/ -m_f)q_f
-\frac{1}{4}{\cal G}_{\mu\nu,a}{\cal G}^{\mu\nu}_a.
\end{equation}
   For each quark flavor $f$ the quark field $q_f$ consists of a color triplet
(subscripts $r$, $g$, and $b$ standing for ``red,'' ``green,'' and 
``blue''),
\begin{equation}
\label{2:2:qf}
q_f=\left(\begin{array}{c}q_{f,r}
\\q_{f,g}\\q_{f,b}\end{array}\right),
\end{equation}
   which transforms under a gauge transformation $g(x)$ described
by the set of parameters $\Theta(x)=[\Theta_1(x),\cdots,\Theta_8(x)]$ 
according to\footnote{For the sake of clarity, the Gell-Mann matrices
contain a superscript $C$, indicating the action in color space.}
\begin{equation}
\label{2:2:qft}
q_f\mapsto q_f'=\exp\left[-i\sum_{a=1}^8 \Theta_a(x)
\frac{\lambda_a^C}{2}\right]q_f=U[g(x)]q_f.
\end{equation}
   Technically speaking, each quark field $q_f$ transforms according to the
fundamental representation of color SU(3).
   Because SU(3) is an eight-parameter group, the covariant derivative
of Eq.\ (\ref{2:2:lqcd}) contains eight independent gauge potentials 
${\cal A}_{\mu,a}$, 
\begin{equation}
\label{2:2:ka}
D_\mu\left(\begin{array}{l}
q_{f,r}\\q_{f,g}\\q_{f,b}\end{array}
\right)
=\partial_\mu
\left(\begin{array}{l}
q_{f,r}\\q_{f,g}\\q_{f,b}\end{array}
\right)
-ig\sum_{a=1}^8
\frac{\lambda_a^C}{2}{\cal A}_{\mu,a} \left(\begin{array}{l}
q_{f,r}\\q_{f,g}\\q_{f,b}\end{array}
\right).
\end{equation}
   We note that the interaction between quarks and gluons is independent
of the quark flavors.
   Demanding gauge invariance of ${\cal L}_{\rm QCD}$
imposes the following transformation property of the gauge fields
\begin{equation}
\label{2:2:atraf}
\frac{\lambda_a^C}{2} {\cal A}_{\mu,a}(x)\mapsto
U[g(x)]\frac{\lambda_a^C}{2} {\cal A}_{\mu,a}(x)U^\dagger[g(x)]
-\frac{i}{g}\partial_\mu U[g(x)]U^\dagger[g(x)].
\end{equation}
   Again, with this requirement the covariant derivative $D_\mu q_f$  
transforms as $q_f$, i.e. $D_\mu q\mapsto D'_\mu q'=U(g)D_\mu q$.      
   Under a gauge transformation of the first kind, i.e., a global
SU(3) transformation, the second term on the right-hand side
of Eq.\ (\ref{2:2:atraf}) would vanish and the gauge fields would
transform according to the adjoint representation.

   So far we have only considered the matter-field part of ${\cal L}_{\rm
QCD}$ including its interaction with the gauge fields.
   Equation (\ref{2:2:lqcd}) also contains the generalization of 
the field strength tensor to the non-Abelian case,
\begin{equation}
\label{2:2:gmunu}
{\cal G}_{\mu\nu,a}=\partial_\mu {\cal A}_{\nu,a}-\partial_\nu {\cal A}_{\mu,a}
+g f_{abc}{\cal A}_{\mu,b} {\cal A}_{\nu,c},
\end{equation}
   with the SU(3) structure constants given in Table
\ref{table:2:1:su3structurconstants} and a summation over repeated indices
implied.
   Given Eq.\ (\ref{2:2:atraf}) the field strength tensor transforms under 
SU(3) as 
\begin{equation}
\label{2:2:gtrafo}
{\cal G}_{\mu\nu}\equiv
\frac{\lambda_a^C}{2} {\cal G}_{\mu\nu,a}
\mapsto
U[g(x)]{\cal G}_{\mu\nu} U^\dagger[g(x)].
\end{equation}
   Using Eq.\ (\ref{2:1:gmme3}) the purely gluonic part 
of ${\cal L}_{\rm QCD}$ can be written as
$$-\frac{1}{2}\mbox{Tr}_C({\cal G}_{\mu\nu} {\cal G}^{\mu\nu}),
$$ 
which, using the cyclic property of traces, 
$\mbox{Tr}(AB)=\mbox{Tr}(BA)$, together
with $UU^\dagger=1$,
is easily seen to be invariant under the transformation of 
Eq.\ (\ref{2:2:gtrafo}).

   In contradistinction to the Abelian case of QED, the squared field 
strength tensor gives rise to gauge-field self interactions involving 
vertices with three and four gauge fields of strength $g$ and $g^2$, 
respectively.
   Such interaction terms are characteristic of non-Abelian gauge
theories and make them much more complicated than Abelian theories.

   From the point of view of gauge invariance the strong-interaction 
Lagrangian could also involve a term of the type
\begin{equation}
\label{2:2:ltheta}
{\cal L}_\theta=\frac{g^2\bar{\theta}}{64\pi^2}\epsilon^{\mu\nu\rho\sigma}
\sum_{a=1}^8{\cal G}^a_{\mu\nu}{\cal G}^a_{\rho\sigma},
\end{equation}
where $\epsilon_{\mu\nu\rho\sigma}$ denotes the totally antisymmetric
Levi-Civita tensor.\footnote{
\begin{displaymath}
\epsilon_{\mu\nu\rho\sigma}=\left\{
\begin{array}{rl}
+1& \mbox{if $\{\mu,\nu,\rho,\sigma\}$ is an even permutation of $\{0,1,2,3\}$}
\\
-1& \mbox{if $\{\mu,\nu,\rho,\sigma\}$ is an odd permutation of $\{0,1,2,3\}$}
\\
0& \mbox{otherwise}
\end{array}
\right.
\end{displaymath}}
   The so-called $\theta$ term of Eq.\ (\ref{2:2:ltheta}) implies an explicit 
$P$ and $CP$ violation of the strong interactions which, for example, would 
give rise to an electric dipole moment of the neutron (for an upper limit, see
Ref.\ \cite{Harris:jx}).
   The present empirical information indicates that the $\theta$ term is 
small and, in the following, we will omit Eq.\ (\ref{2:2:ltheta}) from our 
discussion and refer the interested reader to Refs.\ 
\cite{Pich:1991fq,Borasoy:2000pq,Kaiser:2000gs}.

\section{Accidental, Global Symmetries of ${\cal L}_{\rm QCD}$}
\label{sec_agsl}
\subsection{Light and Heavy Quarks}
\label{subsec_lhq}
   The six quark flavors are commonly divided into the three light quarks
$u$, $d$, and $s$ and the three heavy flavors $c$, $b$, and $t$,
\begin{equation}
\label{2:3:mq}
\left(\begin{array}{r}m_u=0.005\,\mbox{GeV}\\
m_d=0.009\,\mbox{GeV}\\
m_s=0.175\,\mbox{GeV}\end{array}\right)
\ll 1\, \mbox{GeV}\le
\left(\begin{array}{r}
m_c= (1.15 - 1.35)\, \mbox{GeV}\\
m_b= (4.0 - 4.4)\, \mbox{GeV}\\
m_t=174\,\mbox{GeV}\end{array}\right),
\end{equation}
   where the scale of 1 GeV is associated with the masses of the lightest
hadrons containing light quarks, e.g., \ $m_\rho$= 770 MeV,
which are not Goldstone bosons resulting from spontaneous 
symmetry breaking. 
   The scale associated with spontaneous symmetry breaking,
$4\pi F_\pi\approx$ 1170 MeV, is of the same order of magnitude
\cite{Pagels:se,Manohar:1983md,Georgi}.
   
   The masses of the lightest meson and baryon containing a charmed quark, 
$D^+=c\bar{d}$ and $\Lambda^+_c=udc$, are $(1869.4\pm 0.5)\, \mbox{MeV}$
and $(2284.9\pm 0.6)\,\mbox{MeV}$, respectively \cite{Groom:in}.
   The threshold center-of-mass energy to produce, say, a $D^+ D^-$ pair
in $e^+ e^-$ collisions is approximately 3.74 GeV, and thus way beyond the 
low-energy regime which we are interested in.
   In the following, we will approximate the full QCD Lagrangian by its
light-flavor version, i.e., we will ignore effects due to (virtual) 
heavy quark-antiquark pairs $h\bar{h}$.
  In particular, Eq.\ (\ref{2:2:mp}) suggests that the Lagrangian 
${\cal L}_{\rm QCD}^0$, containing only the light-flavor quarks in
the so-called chiral limit $m_u,m_d,m_s\to 0$,
might be a good starting point in the discussion
of low-energy QCD: 
\begin{equation}
\label{2:3:lqcd0}
{\cal L}^0_{\rm QCD}=
\sum_{l=u,d,s}\bar{q}_l i D\hspace{-.6em}/\hspace{.3em} q_l
%+\sum_{h=c,b,t}\bar{q}_h(i D\hspace{-.6em}/\hspace{.3em} -m_h) q_h
-\frac{1}{4}{\cal G}_{\mu\nu,a} {\cal G}^{\mu\nu}_a.
\end{equation}
   We repeat that the covariant derivative $D\hspace{-.6em}/\hspace{.3em} 
q_{l}$ acts on color and Dirac indices only, but is independent of flavor.

\subsection{Left-Handed and Right-Handed Quark Fields}
\label{subsec_lhrhqf}

   In order to fully exhibit the global symmetries of Eq.\ (\ref{2:3:lqcd0}),
we consider the chirality matrix $\gamma_5=\gamma^5=i\gamma^0\gamma^1
\gamma^2\gamma^3$, $\{\gamma^\mu,\gamma_5\}=0$, 
$\gamma_5^2=1$,\footnote{Unless stated otherwise, we use the convention of
Ref.\ \cite{Bjorken_1964}.}
and introduce projection operators
\begin{equation}
\label{2:3:prpl}
P_R=\frac{1}{2}(1+\gamma_5)=P_R^\dagger,\quad
P_L=\frac{1}{2}(1-\gamma_5)=P_L^\dagger,
\end{equation}
  where the indices $R$ and $L$ refer to right-handed and 
left-handed, respectively, as will become more clear below.
   Obviously, the 
$4\times 4$ matrices $P_R$ and $P_L$ satisfy a completeness relation,
\begin{equation}
\label{2:3:prplcompleteness}
P_R+P_L=1,
\end{equation}
are idempotent, i.e.,
\begin{equation}
\label{2:3:prplidempotent}
P_R^2=P_R,\quad P_L^2=P_L,
\end{equation}
and respect the orthogonality relations
\begin{equation}
\label{2:3:prplorthogonality}
P_R P_L=P_L P_R=0.
\end{equation}
   The combined properties of Eqs.\ (\ref{2:3:prplcompleteness}) --
(\ref{2:3:prplorthogonality}) guarantee that $P_R$ and $P_L$ are
indeed projection operators which project from the Dirac field variable $q$ to
its chiral components $q_R$ and $q_L$,
\begin{equation}
\label{2:3:qlr}
q_R=P_R q,\quad 
q_L=P_L q.
\end{equation}
   We recall in this context that a chiral (field) variable is one
which under parity is transformed into neither the original variable
nor its negative \cite{Doughty_1990}.\footnote{In case of fields,
a transformation of the argument $\vec{x}\to -\vec{x}$ is implied.}  
   Under parity, the quark field is transformed into its parity conjugate,
$$
P:q(\vec{x},t)\mapsto \gamma_0 q(-\vec{x},t),
$$
   and hence 
$$q_R(\vec{x},t)=P_R \,q(\vec{x},t)
\mapsto P_R \gamma_0 q(-\vec{x},t)
=\gamma_0 P_L q(-\vec{x},t) \neq \pm q_R(-\vec{x},t),
$$
   and similarly for $q_L$.\footnote{Note that in the above sense,
also $q$ is a chiral variable. 
However, the assignment of handedness
does not have such an intuitive meaning as in the case of $q_L$ and
$q_R$.}
 
   The terminology right-handed and left-handed fields can easily be 
visualized in terms of the solution to the free Dirac equation.
   For that purpose, let us consider an extreme relativistic 
positive-energy solution 
with three-momentum $\vec{p}$,\footnote{Here we adopt
a covariant normalization of the spinors, 
$u^{(\alpha)\dagger}(\vec{p}\,)u^{(\beta)}(\vec{p}\,)
= 2 E\delta_{\alpha\beta}$, etc.} 
$$ u(\vec{p},\pm)=\sqrt{E+M}\left(\begin{array}{c}
\chi_\pm\\
\frac{\vec{\sigma}\cdot\vec{p}}{E+M}\chi_\pm\end{array}\right)
\stackrel{\mbox{$E\gg M$}}{\to}
\sqrt{E} 
\left(\begin{array}{r}\chi_\pm\\ \pm\chi_\pm\end{array}
\right)=u_\pm(\vec{p}\,),
$$
   where we assume that the spin in the rest frame is either parallel
or antiparallel to the direction of momentum
$$
\vec{\sigma}\cdot \hat{p} \chi_{\pm}=\pm \chi_\pm.
$$
   In the standard representation of Dirac matrices we find
$$ P_R=\frac{1}{2}\left(\begin{array}{rr}1_{2\times2}&
1_{2\times 2}\\1_{2\times 2}&1_{2\times 2}\end{array}\right),\quad 
P_L=\frac{1}{2}\left(\begin{array}{rr}1_{2\times 2}&-1_{2\times 2}\\
-1_{2\times 2}&1_{2\times 2}\end{array}\right),
$$
such that
$$ 
P_R u_+=\sqrt{E}\frac{1}{2}
\left(\begin{array}{rr}1_{2\times 2}&1_{2\times 2}\\
1_{2\times 2}&1_{2\times 2}\end{array}\right)
\left(\begin{array}{r}\chi_+\\ \chi_+\end{array}\right)
=\sqrt{E}\left(\begin{array}{r}\chi_+\\ \chi_+\end{array}\right)=u_+,
$$
and similarly
$$
P_L u_+=0,\quad
P_R u_-=0,\quad
P_L u_-=u_-.
$$
    In the extreme relativistic limit (or better, in
the zero-mass limit), the operators $P_R$ and $P_R$ project to the 
positive and negative helicity eigenstates, 
i.e., in this limit chirality equals helicity.

   Our goal is to analyze the symmetry of the QCD Lagrangian with
respect to independent global transformations of the left- and right-handed
fields.
   In order to decompose the 16 quadratic forms into their respective
projections to right- and left-handed fields, we make use of 
\cite{Gasser_1989}
\begin{equation}
\label{2:3:qgq}
\bar{q}\Gamma_i q=\left \{\begin{array}{lcl}
\bar{q}_R\Gamma_1 q_R+\bar{q}_L\Gamma_1 q_L&\mbox{for}&
\Gamma_1\in\{\gamma^\mu,\gamma^\mu\gamma_5\}\\
\bar{q}_R\Gamma_2 q_L +\bar{q}_L\Gamma_2 q_R&\mbox{for}& \Gamma_2
\in\{1,\gamma_5,\sigma^{\mu\nu}\}
\end{array}
\right.,
\end{equation}
where $\bar{q}_R=\bar{q}P_L$ and $\bar{q}_L=\bar{q}P_R.$
   Equation (\ref{2:3:qgq}) is easily proven by inserting the completeness
relation of Eq.\ (\ref{2:3:prplcompleteness}) both to the left and the 
right of $\Gamma_i$,
$$\bar{q}\Gamma_i q=\bar{q}(P_R+P_L)\Gamma_i(P_R+P_L)q,$$
and by noting $\{\Gamma_1,\gamma_5\}=0$ and $[\Gamma_2,\gamma_5]=0$.
   Together with the orthogonality relations of Eq.\ 
(\ref{2:3:prplorthogonality}) we then obtain
$$
P_R\Gamma_1 P_R=\Gamma_1P_L P_R=0,
$$
and similarly
$$P_L \Gamma_1 P_L=0,\quad 
P_R \Gamma_2 P_L=0,\quad
P_L \Gamma_2 P_R=0.
$$
   We stress that the validity of Eq.\ (\ref{2:3:qgq}) is general
and does not refer to ``massless'' quark fields.

   We now apply Eq.\ (\ref{2:3:qgq}) to the term containing the contraction
of the covariant derivative with $\gamma^\mu$.
   This quadratic quark form decouples into the sum of two
terms which connect only left-handed with left-handed and right-handed
with right-handed quark fields.
   The QCD Lagrangian in the chiral limit can then be written as
\begin{equation}
\label{2:3:lqcd0lr}
{\cal L}^0_{\rm QCD}=\sum_{l=u,d,s}
(\bar{q}_{R,l}iD\hspace{-.6em}/\hspace{.3em}q_{R,l}+\bar{q}_{L,l}iD
\hspace{-.6em}/\hspace{.3em}
q_{L,l})-\frac{1}{4}{\cal G}_{\mu\nu,a} {\cal G}^{\mu\nu}_a.
\end{equation}
    Due to the flavor independence of the covariant derivative
${\cal L}^0_{\rm QCD}$ is invariant under 
\begin{eqnarray}
\label{2:3:u3lu3r}
\left(\begin{array}{c}u_L\\d_L\\s_L\end{array}\right)
\mapsto U_L\left(\begin{array}{c}u_L\\d_L\\s_L\end{array}\right)
=\exp\left(-i\sum_{a=1}^8 \Theta^L_a \frac{\lambda_a}{2}\right)
e^{-i\Theta^L}\left(\begin{array}{c}u_L\\d_L\\s_L\end{array}\right),
\nonumber\\
\left(\begin{array}{c}u_R\\d_R\\s_R\end{array}\right)
\mapsto U_R\left(\begin{array}{c}u_R\\d_R\\s_R\end{array}\right)
=\exp\left(-i\sum_{a=1}^8 \Theta^R_a \frac{\lambda_a}{2}\right)
e^{-i\Theta^R}\left(\begin{array}{c}u_R\\d_R\\s_R\end{array}\right),
\end{eqnarray}
   where $U_L$ and $U_R$ are independent unitary $3\times 3$ matrices.
   Note that the Gell-Mann matrices act in flavor space.  

   ${\cal L}^0_{\rm QCD}$ is said to have a classical 
{\em global} $\mbox{U(3)}_L\times\mbox{U(3)}_R$ symmetry.
   Applying Noether's theorem (see, for example, 
\cite{Hill_1951,DeAlfaro:1973}) from such an invariance
one would expect a total of $2\times(8+1)=18$ conserved currents.

\subsection{Noether's Theorem}
\label{subsec_nt}

   In order to identify the conserved currents associated with this
invariance, we briefly recall the method of Ref.\ \cite{Gell-Mann:np}
and consider the variation of Eq.\ (\ref{2:3:lqcd0lr}) under a 
{\em local} infinitesimal 
transformation.\footnote{By exponentiating elements of the Lie algebra 
u($N$) any
element of U($N$) can be obtained.}
   For simplicity we consider only internal symmetries.
   To that end we start with a Lagrangian ${\cal L}$ depending on
$n$ independent fields $\Phi_i$ and their first partial 
derivatives,
\begin{equation}
\label{2:3:l}
{\cal L}={\cal L}(\Phi_i,\partial_\mu\Phi_i),
\end{equation}
from which one obtains $n$ equations of motion: 
\begin{equation}
\label{2:3:eom}
\frac{\partial \cal L}{\partial\Phi_i}-\partial_\mu
\frac{\partial\cal L}{\partial\partial_\mu\Phi_i}=0,\quad i=1,\cdots,n.
\end{equation}
   For each of the $r$ generators of infinitesimal transformations representing
the underlying symmetry group, 
we consider a {\em local} infinitesimal transformation of the fields
\cite{Gell-Mann:np},\footnote{Note that the transformation need not
be realized linearly on the fields.}
\begin{equation}
\label{2:3:ltraf}
\Phi_i(x)\mapsto\Phi'_i(x)=\Phi_i(x)+\delta\Phi_i(x)
=\Phi_i(x)-i\epsilon_a(x) F^a_i[\Phi_j(x)], 
\end{equation}
and obtain, neglecting terms of order $\epsilon^2$,
as the variation of the Lagrangian,
\begin{eqnarray}
\label{2:3:dl}
\delta{\cal L}&=& {\cal L}(\Phi'_i,\partial_\mu\Phi'_i)
-{\cal L}(\Phi_i,\partial_\mu\Phi_i)\nonumber\\
&=&\frac{\partial\cal L}{\partial\Phi_i}\delta\Phi_i
+\frac{\partial\cal L}{\partial\partial_\mu\Phi_i}
\partial_\mu\delta\Phi_i\nonumber\\
&=&\epsilon_a(x)\left(-i\frac{\partial\cal L}{\partial\Phi_i}
F^a_i-i\frac{\partial\cal L}{\partial\partial_\mu\Phi_i}
\partial_\mu F^a_i\right)
+\partial_\mu\epsilon_a(x)\left(-i\frac{\partial\cal L}{
\partial\partial_\mu\Phi_i}F^a_i\right)\nonumber\\
&\equiv&\epsilon_a(x)\partial_\mu J^{\mu,a}
+\partial_\mu\epsilon_a(x)J^{\mu,a}.
\end{eqnarray}
   According to this equation we define for each infinitesimal
transformation a four-current density as
\begin{equation}
\label{2:3:strom}
   J^{\mu,a}=-i\frac{\partial\cal L}{\partial\partial_\mu\Phi_i}
F^a_i.
\end{equation}
     By calculating the divergence $\partial_\mu J^{\mu,a}$ of 
Eq.\ (\ref{2:3:strom})
\begin{eqnarray*}
\partial_\mu J^{\mu,a}&=&-i\left(\partial_\mu\frac{\partial\cal L}{\partial
\partial_\mu\Phi_i}\right)F^a_i-i\frac{\partial\cal L}{\partial
\partial_\mu\Phi_i}\partial_\mu F^a_i\\
&=&
-i\frac{\partial\cal L}{\partial \Phi_i}
F^a_i-i\frac{\partial\cal L}{\partial\partial_\mu\Phi_i}
\partial_\mu F^a_i,
\end{eqnarray*}
where we made use of the equations of motion, Eq.\ (\ref{2:3:eom}),
we explicitly verify the consistency with the definition of 
$\partial_\mu J^{\mu,a}$
according to Eq.\ (\ref{2:3:dl}).
   From Eq.\ (\ref{2:3:dl}) it is straightforward to obtain the four-currents
as well as their divergences as
\begin{eqnarray}
\label{2:3:strom2}
J^{\mu,a}&=&\frac{\partial \delta\cal L}{\partial \partial_\mu
\epsilon_a},\\
\label{2:3:divergenz}
\partial_\mu J^{\mu,a}&=&\frac{\partial \delta\cal L}{\partial
\epsilon_a}.
\end{eqnarray}
   For a conserved current, $\partial_\mu J^{\mu,a}=0$, the charge
\begin{equation}
\label{2:3:charge}
Q^a(t)=\int d^3 x J^a_0(\vec{x},t)
\end{equation}
is time independent, i.e., a constant of the motion, which is shown in
the standard fashion by applying the divergence theorem for an infinite
volume with appropriate boundary conditions for $R\to \infty$.

   So far we have discussed Noether's theorem on the classical level, implying
that the charges $Q^a(t)$ can have any continuous real value.
   However, we also need to discuss the implications of a transition to
a quantum theory.
   After canonical quantization, the fields $\Phi_i$ and their conjugate 
momenta $\Pi_i=\partial{\cal L}/\partial (\partial_0 \Phi_i)$ 
are considered as linear operators acting 
on a Hilbert space which, in the Heisenberg picture, are subject to the 
equal-time commutation relations
\begin{eqnarray}
\label{2:3:gzvr}
[\Phi_i(\vec{x},t),\Pi_j(\vec{y},t)]&=&i\delta^3(\vec{x}-\vec{y})
\delta_{ij},\nonumber\\
{[}\Phi_i(\vec{x},t),\Phi_j(\vec{y},t)]&=&0,\nonumber\\
{[}\Pi_i(\vec{x},t),\Pi_j(\vec{y},t)]&=&0.
\end{eqnarray}
   As a special case of Eq.\ (\ref{2:3:ltraf}) let us consider infinitesimal
transformations which are {\em linear} in the fields,
\begin{equation}
\label{2:3:lt}
\Phi_i(x)\mapsto \Phi'_i(x)=\Phi_i(x)-i\epsilon_a(x)t^a_{ij}\Phi_j(x),
\end{equation}
where the $t^a_{ij}$ are constants generating a mixing of the fields.
   From Eq.\ (\ref{2:3:strom}) we then obtain\footnote{Normal ordering
symbols are suppressed.}
\begin{eqnarray}
\label{2:3:j}
J^{\mu,a}(x)&=&-it^a_{ij}\frac{\partial {\cal L}}{\partial
\partial_\mu\Phi_i}\Phi_j,\\
\label{2:3:q}
Q^{a}(t)&=&-i\int d^3x\, \Pi_i(x) t^a_{ij}\Phi_j(x),
\end{eqnarray}
where $J^{\mu,a}(x)$ and $Q^{a}(t)$ are now operators. 
   In order to interpret the charge operators $Q^a(t)$, let us make use of
the equal-time commutation relations, Eqs.\ (\ref{2:3:gzvr}), 
and calculate their commutators with the field operators,
\begin{eqnarray}
\label{2:3:qphi}
[Q^a(t),\Phi_k(\vec{y},t)]&=&-it^a_{ij}\int d^3x\,
[\Pi_i(\vec{x},t)\Phi_j(\vec{x},t),\Phi_k(\vec{y},t)]\nonumber\\
&=&-t^a_{kj}\Phi_j(\vec{y},t).
\end{eqnarray}
   Note that we did not require the charge operators to be time independent.
   On the other hand, for the transformation behavior of the Hilbert space 
associated with a global infinitesimal transformation, we make an
ansatz in terms of an infinitesimal unitary transformation\footnote{
We have chosen to have the fields (field operators) rotate actively
and thus must transform the states of Hilbert space in the opposite
direction.}
\begin{equation}
\label{1:5:tz}
|\alpha'\rangle=[1+i\epsilon_a G^a(t)]|\alpha\rangle,
\end{equation}
with Hermitian operators $G^a$.
   Demanding
\begin{equation}
\label{2:3:at}
\langle\beta|A|\alpha\rangle=\langle\beta'|A'|\alpha'\rangle\quad
\forall\, |\alpha\rangle, |\beta\rangle, \epsilon_a,
\end{equation}
in combination with Eq.\ (\ref{2:3:lt}) yields the condition
\begin{eqnarray*}
\langle\beta|\Phi_i(x)|\alpha\rangle
&=&\langle\beta'|\Phi_i'(x)|\alpha'\rangle\nonumber\\
&=&\langle\beta|[1-i\epsilon_a G^a(t)][\Phi_i(x)-i\epsilon_b
t^b_{ij}\Phi_j(x)][1+i\epsilon_c G^c(t)]|\alpha\rangle.
\end{eqnarray*}
   By comparing the terms linear in $\epsilon_a$ on both sides,
\begin{equation}
0=-i\epsilon_a[G^a(t),\Phi_i(x)]
\underbrace{-i\epsilon_a t^a_{ij}\Phi_j(x)}_{\mbox{$
i\epsilon_a[Q^a(t),\Phi_i(x)]$}},
\end{equation}
   we see that the infinitesimal generators acting on the states of Hilbert 
space which are associated with the transformation of the
fields are identical with the charge operators $Q^a(t)$ of 
Eq.\ (\ref{2:3:q}).

   Finally, evaluating the commutation relations for the case of several
generators,
\begin{eqnarray}
\label{2:3:qaqbkom}
[Q^a(t),Q^b(t)]
&=&-i(t^a_{ij}t^b_{jk}-t^b_{ij}t^a_{jk})
\int d^3x\,\Pi_i(\vec{x},t)\Phi_k(\vec{x},t),
\end{eqnarray}
we find the right-hand side of Eq.\ (\ref{2:3:qaqbkom}) to be
again proportional to a charge operator, if
\begin{equation}
\label{2:3:lrel}
t^a_{ij}t^b_{jk}-t^b_{ij}t^a_{jk}=iC_{abc}t^c_{ik},
\end{equation}
i.e., in that case the charge operators $Q^a(t)$ form a Lie algebra
\begin{equation}
\label{2:3:liealgebra}
[Q^a(t),Q^b(t)]=iC_{abc}Q^c(t)
\end{equation}
with structure constants $C_{abc}$.
   The quantization of the charges (as opposed to continuous values in
the classical case) can be understood in analogy to the algebraic
construction of the angular momentum eigenvalues in quantum
mechanics starting from the su(2) algebra.
   Of course, for conserved currents, the charge operators are time 
independent, i.e., they commute with the Hamilton operator of the system.

   From now on we assume the validity of Eq.\ (\ref{2:3:lrel}) and interpret 
the constants $t^a_{ij}$ as the entries in the $i$th row and $j$th column
of an $n\times n$ matrix $T^a$, 
\begin{displaymath}
T^a=\left(\begin{array}{ccc}
t^a_{11}&\cdots&t^a_{1n}\\
\vdots & &\vdots\\
t^a_{n1}&\cdots &t^a_{nn}
\end{array}
\right).
\end{displaymath}
   Because of Eq.\ (\ref{2:3:lrel}), these matrices form an $n$-dimensional
representation of a Lie algebra, 
$$[T^a,T^b]=iC_{abc}T^c.$$
   The infinitesimal, linear transformations of the fields $\Phi_i$ may
then be written in a compact form,
\begin{equation}
\label{2:3:ltrafo}
\left(\begin{array}{c}\Phi_1(x)\\\vdots\\ \Phi_n(x)\end{array}
\right)=\Phi(x)\mapsto\Phi'(x)=(1-i\epsilon_a T^a)\Phi(x).
\end{equation}
   In general, through an appropriate unitary transformation,
the matrices $T_a$ may be decomposed into their irreducible components, i.e., 
brought into block-diagonal form, such that only fields belonging
to the same multiplet transform into each other under the symmetry
group.

\subsection{Global Symmetry Currents of the Light Quark Sector}
\label{subsec_gsclqs}
   The method of Ref.\ \cite{Gell-Mann:np} can now easily be applied to the 
QCD Lagrangian by calculating the variation under the infinitesimal, local 
form of Eqs.\ (\ref{2:3:u3lu3r}), 
\begin{equation}
\label{2:3:dlqcd0}
\delta{\cal L}^0_{\rm QCD}=\bar{q}_R\left(\sum_{a=1}^8 \partial_\mu \Theta^R_a 
\frac{\lambda_a}{2}+\partial_\mu \Theta^R\right)\gamma^\mu q_R
+\bar{q}_L\left(\sum_{a=1}^8 \partial_\mu \Theta^L_a 
\frac{\lambda_a}{2}+\partial_\mu \Theta^L\right)\gamma^\mu q_L,
\end{equation}
from which, by virtue of Eqs.\ (\ref{2:3:strom2}) and (\ref{2:3:divergenz}),
one obtains the currents associated with the transformations of the
left-handed or right-handed quarks
\begin{eqnarray}
\label{2:3:str}
L^{\mu,a}&=&\bar{q}_L\gamma^\mu \frac{\lambda^a}{2}q_L,\quad
\partial_\mu L^{\mu,a}=0,\nonumber\\
R^{\mu,a}&=&\bar{q}_R\gamma^\mu \frac{\lambda^a}{2}q_R,\quad
\partial_\mu R^{\mu,a}=0.
\end{eqnarray}
   The eight currents $L^{\mu,a}$ transform under 
$\mbox{SU(3)}_L\times\mbox{SU(3)}_R$ 
as an $(8,1)$ multiplet, i.e., as octet and singlet under transformations
of the left- and right-handed fields, respectively. 
   Similarly, the right-handed currents transform as a $(1,8)$ multiplet
under $\mbox{SU(3)}_L\times\mbox{SU(3)}_R$.
   Instead of these chiral currents one often uses linear combinations,
\begin{eqnarray}
\label{2:3:v}
V^{\mu,a}&=& R^{\mu,a}+L^{\mu,a}=\bar{q}\gamma^\mu\frac{\lambda^a}{2}q,\\
\label{2:3:a}
A^{\mu,a}&=&R^{\mu,a}-L^{\mu,a}=\bar{q}\gamma^\mu\gamma_5 \frac{\lambda^a}{2}q,
\end{eqnarray}
   transforming under parity as vector and axial-vector current
densities, respectively,
\begin{eqnarray}
\label{2:3:pv}
P:V^{\mu,a}(\vec{x},t)\mapsto V^a_\mu(-\vec{x},t),\\
\label{2:3pa}
P: A^{\mu,a}(\vec{x},t)\mapsto -A_\mu^a(-\vec{x},t).
\end{eqnarray}

   From Eqs.\ (\ref{2:3:strom2}) and (\ref{2:3:divergenz}) one also obtains a 
conserved singlet vector current resulting from a transformation of
all left-handed and right-handed quark fields by the {\em same} phase, 
\begin{equation}
\label{2:3:sv}
V^\mu=\bar{q}_R\gamma^\mu q_R+\bar{q}_L\gamma^\mu q_L=
\bar{q}\gamma^\mu q, \quad \partial_\mu V^\mu=0.
\end{equation}
   The singlet axial-vector current,
\begin{eqnarray}
\label{2:3:sa}
A^\mu&=&\bar{q}_R \gamma^\mu q_R -\bar{q}_L\gamma^\mu q_L
=\bar{q}P_L\gamma^\mu P_R q -\bar{q}P_R \gamma^\mu P_Lq \nonumber\\
&=& \bar{q}\gamma^\mu P_R q -\bar{q}\gamma^\mu P_L q
=\bar{q}\gamma^\mu\gamma_5q,
\end{eqnarray}
originates from a transformation of all left-handed quark fields
with one phase and all right-handed with the {\em opposite}
phase.
   However, such a singlet axial-vector current is only conserved on
the {\em classical} level. 
    This symmetry is not preserved by quantization and there will be
extra terms, referred to as anomalies 
\cite{Adler:1969gk,Adler:1969er,Bardeen:1969md,Bell:1969ts,Adler:1970}, 
resulting 
in\footnote{In the large $N_c$ (number of colors) 
limit of Ref.\ \cite{'tHooft:1973jz} 
the singlet axial-vector current is conserved, because the strong coupling 
constant behaves as $g^2\sim N_c^{-1}$.}
\begin{equation}
\label{2:3:divsa}
\partial_\mu A^\mu=\frac{3 g^2}{32\pi^2}\epsilon_{\mu\nu\rho\sigma}
{\cal G}^{\mu\nu}_a {\cal G}^{\rho\sigma}_a,\quad \epsilon_{0123}=1,
\end{equation}
where the factor of 3 originates from the number of flavors.

\subsection{The Chiral Algebra}
\label{subsec_ca}
   The invariance of ${\cal L}^0_{\rm QCD}$ under global 
$\mbox{SU(3)}_L\times\mbox{SU(3)}_R\times\mbox{U(1)}_V$ transformations
implies that also the QCD Hamilton operator in the chiral limit,
$H^0_{\rm QCD}$, exhibits a global 
$\mbox{SU(3)}_L\times\mbox{SU(3)}_R\times\mbox{U(1)}_V$ symmetry. 
   As usual, the ``charge operators'' are defined as the space integrals
of the charge densities,
\begin{eqnarray}
\label{2:3:ql}
Q^a_L(t)&=&\int d^3x\, q^\dagger_L(\vec{x},t)\frac{\lambda^a}{2}
q_L(\vec{x},t),\quad a=1,\cdots, 8,\\
\label{2:3:qr}
Q^a_R(t)&=&\int d^3x\, q^\dagger_R(\vec{x},t)\frac{\lambda^a}{2}
q_R(\vec{x},t),\quad a=1,\cdots, 8,\\
\label{2:3:qv}
Q_V(t)&=&\int d^3x\, \left[q^\dagger_L(\vec{x},t)q_L(\vec{x},t)+
q^\dagger_R(\vec{x},t)q_R(\vec{x},t)\right].
\end{eqnarray}
   For conserved symmetry currents, these operators are time independent,
i.e., they commute with the Hamiltonian,
\begin{equation}
\label{2:3:vrhq}
[Q_L^a,H^0_{\rm QCD}]=[Q_R^a,H^0_{\rm QCD}]=[Q_V,H^0_{\rm QCD}]=0.
\end{equation}
   The commutation relations of the charge operators with each other
are obtained by using the equal-time commutation relations of
the quark fields in the Heisenberg picture,
\begin{eqnarray}
\label{2:3:comrelf1}
\{q_{\alpha,r}(\vec{x},t),q^\dagger_{\beta,s}(\vec{y},t)\}&=&
\delta^3(\vec{x}-\vec{y})\delta_{\alpha\beta}\delta_{rs},\\
\label{2:3:comrelf2}
\{q_{\alpha,r}(\vec{x},t),q_{\beta,s}(\vec{y},t)\}&=&0,\\
\label{2:3:comrelf3}
\{q_{\alpha,r}^\dagger(\vec{x},t),q^\dagger_{\beta,s}(\vec{y},t)\}&=&0,
\end{eqnarray}
where $\alpha$ and $\beta$ are Dirac indices and $r$ and $s$ 
flavor indices, respectively.\footnote{Strictly speaking, we should
also include the color indices. However, since we are only discussing
color-neutral quadratic forms a summation over such indices is
always implied, with the net effect that one can completely omit them from
the discussion.}
   The equal-time commutator of two quadratic quark forms is of the type
\begin{eqnarray}
\label{2:3:fk}
\lefteqn{[q^\dagger(\vec{x},t) \Gamma_1 F_1 q(\vec{x},t),
q^\dagger(\vec{y},t)\Gamma_2 F_2 q(\vec{y},t)]=}\nonumber\\
&&\Gamma_{1,\alpha\beta}\Gamma_{2,\gamma\delta} F_{1,rs}F_{2,tu}
[q^\dagger_{\alpha,r}(\vec{x},t)
q_{\beta,s}(\vec{x},t),q^\dagger_{\gamma,t}(\vec{y},t)
q_{\delta,u}(\vec{y},t)],
\end{eqnarray}
   where $\Gamma_i$ and $F_i$ are $4\times 4$ Dirac matrices and
$3\times 3$ flavor matrices, respectively.
   Using    
\begin{equation}
\label{2:3:abcdfk}
[ab,cd]=a\{b,c\}d-ac\{b,d\}+\{a,c\}db-c\{a,d\}b,
\end{equation}
we express the commutator of Fermi fields in terms of anticommutators and make
use of the equal-time commutation relations of Eqs.\ (\ref{2:3:comrelf1})
-- (\ref{2:3:comrelf3}) to obtain
\begin{eqnarray*}
\lefteqn{[q^\dagger_{\alpha,r}(\vec{x},t)
q_{\beta,s}(\vec{x},t),q^\dagger_{\gamma,t}(\vec{y},t)
q_{\delta,u}(\vec{y},t)]=}\\
&&
q^\dagger_{\alpha,r}(\vec{x},t)q_{\delta,u}(\vec{y},t)\delta^3(
\vec{x}-\vec{y})\delta_{\beta\gamma}\delta_{st}
-q^\dagger_{\gamma,t}(\vec{y},t)q_{\beta,s}(\vec{x},t)
\delta^3(\vec{x}-\vec{y})\delta_{\alpha\delta}\delta_{ru}.
\end{eqnarray*}
   With this result Eq.\ (\ref{2:3:fk}) reads
\begin{eqnarray}
\label{2:3:fkf}
\lefteqn{[q^\dagger(\vec{x},t) \Gamma_1 F_1 q(\vec{x},t),
q^\dagger(\vec{y},t)\Gamma_2 F_2 q(\vec{y},t)]=}\nonumber\\
&&\delta^3(\vec{x}-\vec{y})\left[
q^\dagger(\vec{x},t)\Gamma_1\Gamma_2 F_1 F_2 q(\vec{y},t)
-q^\dagger(\vec{y},t)\Gamma_2 \Gamma_1 F_2 F_1 q(\vec{x},t)\right].
\end{eqnarray}

   After inserting appropriate projectors $P_{L/R}$, Eq.\ (\ref{2:3:fkf}) 
is easily applied to the charge operators of Eqs.\ (\ref{2:3:ql}), 
(\ref{2:3:qr}), 
and (\ref{2:3:qv}),  showing that these operators
indeed satisfy the commutation relations corresponding to the Lie algebra
of $\mbox{SU(3)}_L\times\mbox{SU(3)}_R\times\mbox{U(1)}_V$,
\begin{eqnarray}
\label{2:2:crqll}
[Q_L^a,Q_L^b]&=&if_{abc}Q_L^c,\\
\label{2:3:crqrr}
{[Q_R^a,Q_R^b]}&=&if_{abc}Q_R^c,\\
\label{2:3:crqlr}
{[Q_L^a,Q_R^b]}&=&0,\\
\label{2:3:crqlvrv}
{[Q_L^a,Q_V]}&=&[Q_R^a,Q_V]=0.
\end{eqnarray}
    It should be stressed that, even without being able to explicitly solve 
the equation of motion of the quark fields entering the charge operators of 
Eqs.\ (\ref{2:3:ql}) - (\ref{2:3:qv}), we know from the equal-time 
commutation relations
and the symmetry of the Lagrangian that these charge operators are the 
generators of infinitesimal transformations of the Hilbert space associated 
with $H^0_{\rm QCD}$.
   Furthermore, their commutation relations with a given operator
specify the transformation behavior of the operator in question
under the group $\mbox{SU(3)}_L\times\mbox{SU(3)}_R\times\mbox{U(1)}_V$.

\subsection{Chiral Symmetry Breaking Due to Quark Masses}
\label{subsec_csbdqm}
   The finite $u$-, $d$-, and $s$-quark masses in the QCD Lagrangian
result in explicit divergences of the symmetry currents.
   As a consequence, the charge operators are, in general, no longer
time independent. 
   However, as first pointed out by Gell-Mann, the equal-time-commutation
relations still play an important role
even if the symmetry is explicitly broken
\cite{Gell-Mann:xb}.
   As will be discussed later on in more detail, the symmetry currents will 
give rise to chiral Ward identities relating various QCD Green functions 
to each other.    
   Equation (\ref {2:3:divergenz}) allows one to discuss the
divergences in the presence of quark masses.
   To that end, let us consider the quark-mass matrix of the three
light quarks and project it on the nine $\lambda$ matrices of
Eq.\ (\ref{2:1:matrixa}), 
\begin{eqnarray}
\label{2:3:qmm}
M&=&\left(\begin{array}{ccc}
m_u&0&0\\
0&m_d&0\\
0&0&m_s
\end{array}
\right)\nonumber\\
&=&\frac{m_u+m_d+m_s}{\sqrt{6}}\lambda_0
+\frac{(m_u+m_d)/2-m_s}{\sqrt{3}}\lambda_8
+\frac{m_u-m_d}{2}\lambda_3.
\end{eqnarray}
   In particular, applying Eq.\ (\ref{2:3:qgq}) we see that the quark mass
term mixes left- and right-handed fields,
\begin{equation}
\label{2:3:lm}
{\cal L}_M= -\bar{q}Mq=
-(\bar{q}_R M q_L +\bar{q}_L M q_R).
\end{equation}
   The symmetry-breaking term transforms under $\mbox{SU(3)}_L\times
\mbox{SU(3)}_R$ as a member of a $(3,3^\ast)+(3^\ast,3)$ representation,
i.e.,
$$
\bar{q}_{R,i} M_{ij} q_{L,j} +\bar{q}_{L,i} M_{ij} q_{R,j}
\mapsto 
U_{L,jk}U^\ast_{R,il} \bar{q}_{R,l}M_{ij} q_{L,k}+ (L\leftrightarrow R),
$$
where $(U_L,U_R)\in\mbox{SU(3)}_L\times
\mbox{SU(3)}_R$.
   Such symmetry-breaking {\em patterns} were already discussed in
the pre-QCD era in Refs.\
\cite{Glashow:1967rx,Gell-Mann:rz}.

   From ${\cal L}_M$ one obtains as the variation $\delta{\cal L}_M$
under the transformations of Eqs.\ (\ref{2:3:u3lu3r}),
\begin{eqnarray}
\label{2:3:dlm}
\delta {\cal L}_M&=&
-i\left[ \sum_{a=1}^8 \Theta_a^R \left(
\bar{q}_R \frac{\lambda_a}{2}M q_L -\bar{q}_L M \frac{\lambda_a}{2} q_R
\right)
+\Theta^R  \left(\bar{q}_RMq_L-\bar{q}_LM q_R\right)\right.\nonumber\\
&&+\left.\sum_{a=1}^8 \Theta_a^L \left(
\bar{q}_L \frac{\lambda_a}{2}M q_R -\bar{q}_R M \frac{\lambda_a}{2} q_L
\right)
+\Theta^L  \left(\bar{q}_LMq_R-\bar{q}_RM q_L\right)\right],\nonumber\\
\end{eqnarray}
which results in the following divergences,
\begin{eqnarray}
\label{2:3:dslr}
\partial_\mu L^{\mu,a}&=&\frac{\partial \delta {\cal L}_M}{\partial \Theta^L_a}
=-i\left(\bar{q}_L\frac{\lambda_a}{2}M q_R -\bar{q}_R M \frac{\lambda_a}{2}
q_L\right),\nonumber\\
\partial_\mu R^{\mu,a}&=&\frac{\partial \delta {\cal L}_M}{\partial \Theta^R_a}
=-i\left(\bar{q}_R\frac{\lambda_a}{2}M q_L -\bar{q}_L M \frac{\lambda_a}{2}
q_R\right),\nonumber\\
\partial_\mu L^{\mu}&=&\frac{\partial \delta {\cal L}_M}{\partial \Theta^L}
=-i\left(\bar{q}_L M q_R -\bar{q}_R M q_L\right),\nonumber\\
\partial_\mu R^{\mu}&=&\frac{\partial \delta {\cal L}_M}{\partial \Theta^R}
=-i\left(\bar{q}_R M q_L -\bar{q}_L M q_R\right).
\end{eqnarray}
   The anomaly has not yet been considered.
   Applying Eq.\ (\ref{2:3:qgq}) to the case of the vector currents
and inserting projection operators as in the derivation of Eq.\ (\ref{2:3:sa})
for the axial-vector current, the corresponding divergences read
\begin{eqnarray}
\label{2:3:dsva}
\partial_\mu V^{\mu,a}&=&
i\bar{q}[M,\frac{\lambda_a}{2}]q,\nonumber\\
\partial_\mu A^{\mu,a}&=&
i\left(\bar{q}_L\{\frac{\lambda_a}{2},M\}q_R
-\bar{q}_R\{\frac{\lambda_a}{2},M\}q_L\right)
=i\bar{q}\{\frac{\lambda_a}{2},M\}\gamma_5q,\nonumber\\
\partial_\mu V^\mu&=&0,\nonumber\\
\partial_\mu A^\mu&=&2i\bar{q}M\gamma_5 q+
\frac{3 g^2}{32\pi^2}\epsilon_{\mu\nu\rho\sigma}
{\cal G}^{\mu\nu}_a {\cal G}^{\rho\sigma}_a,\quad \epsilon_{0123}=1, 
\end{eqnarray}
   where the axial anomaly has also been taken into account.
   We are now in the position to summarize the various (approximate) 
symmetries of the strong interactions in combination with the corresponding 
currents and their divergences.

\begin{itemize}
\item In the limit of massless quarks, the sixteen currents $L^{\mu,a}$
and $R^{\mu,a}$ or, alternatively, $V^{\mu,a}$ and $A^{\mu,a}$ are
conserved.
   The same is true for the singlet vector current $V^\mu$, whereas the
singlet axial-vector current $A^\mu$ has an anomaly.
\item For any value of quark masses, the individual flavor currents
$\bar{u}\gamma^\mu u$, $\bar{d}\gamma^\mu d$, and 
$\bar{s}\gamma^\mu s$ are always conserved in strong interactions
reflecting the flavor independence of the strong coupling and the
diagonality of the quark mass matrix.
   Of course, the singlet vector current $V^\mu$, being the sum of
the three flavor currents, is always conserved.
\item In addition to the anomaly, the singlet axial-vector current 
has an explicit divergence due to the quark masses.
\item For equal quark masses, $m_u=m_d=m_s$, the eight vector currents
$V^{\mu,a}$ are conserved, because $[\lambda_a,1]=0$.
   Such a scenario is the origin of the SU(3) symmetry
originally proposed by Gell-Mann and Ne'eman \cite{EightfoldWay}.
   The eight axial currents $A^{\mu,a}$ are not conserved.
   The divergences of the octet axial-vector currents of Eq.\ (\ref{2:3:dsva})
are proportional to pseudoscalar quadratic forms.
   This can be interpreted as the microscopic origin of the 
PCAC relation (partially conserved axial-vector current) which
states that the divergences of the axial-vector currents are proportional to 
renormalized field operators representing the lowest lying pseudoscalar
octet (for a comprehensive discussion of the meaning of PCAC see 
Refs.\ \cite{Gell-Mann:1964tf,Adler:1968,Treiman:1972,DeAlfaro:1973}).
\item Various  symmetry-breaking patterns are discussed in great detail in
Ref.\ \cite{Pagels:se}.
\end{itemize}

\section{Green Functions and Chiral Ward Identities}
\label{sec_gfcwi}
\subsection{Chiral Green Functions}
\label{subsec_cgf}
   For conserved currents, the spatial integrals of the
charge densities are time independent, i.e., in a quantized theory
the corresponding charge operators commute with the Hamilton operator.
   These operators are generators of infinitesimal transformations 
on the Hilbert space of the theory.
   The mass eigenstates 
should organize themselves in
degenerate multiplets with dimensionalities corresponding to 
irreducible representations of the Lie group in 
question.\footnote{Here we assume that the dynamical system described
by the Hamiltonian does not lead to a spontaneous symmetry breakdown.
   We will come back to this point later.}
   Which irreducible representations ultimately appear, and what
the actual energy eigenvalues are, is determined by the dynamics of
the Hamiltonian.
   For example, SU(2) isospin symmetry of the strong interactions reflects
itself in degenerate SU(2) multiplets such as the nucleon doublet,
the pion triplet and so on.
   Ultimately, the actual masses of the nucleon and the pion should
follow from QCD (for a prediction of hadron masses in lattice QCD see,
e.g., Refs.\ \cite{Butler:1994em,AliKhan:2001tx}).

   It is also well-known that symmetries imply relations between $S$-matrix
elements. For example, applying the Wigner-Eckart theorem to pion-nucleon
scattering, assuming the strong-interaction Hamiltonian to be an isoscalar,
it is sufficient to consider two isospin amplitudes describing transitions
between states of total isospin $I=1/2$ or $I=3/2$ (see, for example,
\cite{Ericson:gk}). 
   All the dynamical information is contained in these isospin amplitudes and 
the results for physical processes can be expressed in terms of
these amplitudes together with geometrical coefficients, namely, 
the Clebsch-Gordan coefficients.

   In quantum field theory, the objects of interest are the
Green functions which are vacuum expectation values of time-ordered
products.\footnote{Later on, we will also refer to matrix elements
of time-ordered products between states other than the vacuum as
Green functions.} 
   Pictorially, these Green functions can be understood as vertices
and are related to physical scattering amplitudes through the
Lehmann-Symanzik-Zimmermann (LSZ) reduction formalism \cite{Lehmann:1955rq}.
   Symmetries provide strong constraints not only for scattering amplitudes,
i.e.~their transformation behavior, but, more generally speaking, also for
 Green functions and, in particular, 
{\em among} Green functions.
   The famous example in this context is, of course, the Ward identity
of QED associated with U(1) gauge invariance
\cite{Ward:1950xp}, 
\begin{equation}
\label{2:4:qedwardidentity}
\Gamma^\mu(p,p)=-\frac{\partial}{\partial p_\mu}\Sigma(p),
\end{equation}
which relates the electromagnetic vertex of an electron at zero
momentum transfer, $\gamma^\mu+\Gamma^\mu(p,p)$,
to the electron self energy, $\Sigma(p)$.

   Such symmetry relations can be extended to non-vanishing momentum transfer
and also to more complicated groups and are referred to as 
Ward-Fradkin-Takahashi identities 
\cite{Ward:1950xp,Fradkin:1955jr,Takahashi:xn}
(or Ward identities for short).
   Furthermore, even if a symmetry is broken, i.e., the infinitesimal
generators are time dependent, conditions related to
the symmetry breaking terms can still be obtained using equal-time
commutation relations \cite{Gell-Mann:xb}.

   At first, we are interested in 
time-ordered products of color-neutral, Hermitian quadratic forms involving 
the light quark fields evaluated between the vacuum of QCD.
   Using the LSZ reduction formalism 
\cite{Lehmann:1955rq,Itzykson:rh} such Green
functions can be related to physical processes
involving mesons as well as their interactions with the electroweak
gauge fields of the Standard Model.
   The interpretation depends on the transformation properties
and quantum numbers of the quadratic forms, determining for which mesons
they may serve as an interpolating field.
  In addition to the vector and axial-vector currents of 
Eqs.\ (\ref{2:3:v}), (\ref{2:3:a}), and (\ref{2:3:sv})
we want to investigate scalar and pseudoscalar 
densities,\footnote{The singlet axial-vector current involves an
anomaly such that the Green functions involving this current operator
are related to Green functions containing the contraction of the
gluon field-strength tensor with its dual.}
\begin{equation}
\label{2:4:quadraticforms}
S_a(x)=\bar{q}(x)\lambda_a q(x),\quad
P_a(x)=i\bar{q}(x)\gamma_5 \lambda_a  q(x),\quad
a=0,\cdots, 8,
\end{equation}
   which enter, for example, in Eqs.\ (\ref{2:3:dsva}) as the divergences of 
the vector and axial-vector currents for nonzero quark masses.
   Whenever it is more convenient, we will also use 
\begin{equation}
\label{2:4:SP}
S(x)=\bar{q}(x) q(x),\quad
P(x)=i\bar{q}(x)\gamma_5 q(x),
\end{equation}
instead of $S_0$ and $P_0$.

   Later on, we will also consider similar time-ordered products
evaluated between a single nucleon in the initial and final states 
in addition to the vacuum Green functions.
   This will allow us to discuss properties of the nucleon as well as
dynamical processes involving a single nucleon.

   Generally speaking, a chiral Ward identity relates the divergence
of a Green function containing at least one factor of $V^{\mu,a}$ or
$A^{\mu,a}$ [see Eqs.\ (\ref{2:3:v}) and (\ref{2:3:a})] to some linear 
combination of other Green functions. 
   The terminology {\em chiral} refers to the underlying $\mbox{SU(3)}_L\times
\mbox{SU(3)}_R$ group.
   To make this statement more precise, let us consider as a simple example
the two-point Green function involving an axial-vector current and a 
pseudoscalar density,\footnote{The time ordering of $n$ points 
$x_1,\cdots,x_n$ gives rise to $n$! distinct orderings, each
involving products of $n-1$ theta functions.}
\begin{eqnarray}
\label{2:4:gfaav}
G^{\mu,ab}_{AP}(x,y)&=&\langle 0| T[A^\mu_a(x) P_b(y)]|0\rangle\nonumber\\
&=&\Theta(x_0-y_0)\langle 0|A^\mu_a(x) P_b(y)|0\rangle
+\Theta(y_0-x_0)\langle 0|P_b(y) A^\mu_a(x)|0\rangle,\nonumber\\
\end{eqnarray}
   and evaluate the divergence
\begin{eqnarray*}
\lefteqn{\partial_\mu^x G^{\mu,ab}_{AP}(x,y)}\\
&=&
\delta(x_0-y_0)\langle 0| A_0^a(x) P_b(y)|0\rangle
-\delta(x_0-y_0)\langle 0|  P_b(y)A_0^a (x)|0\rangle\\
&&
+\Theta(x_0-y_0)\langle 0|\partial_\mu^x A^\mu_a(x) P_b(y)|0\rangle
+\Theta(y_0-x_0)\langle 0| P_b(y)\partial_\mu^x A^\mu_a(x)|0\rangle\\
&=&\delta(x_0-y_0)\langle 0|[A^a_0(x),P_b(y)]|0\rangle
+\langle 0|T[\partial_\mu^x A^\mu_a(x) P_b(y)]|0\rangle,
\end{eqnarray*}
   where we made use of $\partial_\mu^x \Theta(x_0-y_0)=\delta(x_0-y_0) 
g_{0\mu}=-\partial_\mu^x \Theta(y_0-x_0)$.
   This simple example already shows the main features of (chiral) Ward 
identities.
   From the differentiation of the theta functions 
one obtains equal-time commutators between a charge density and the
remaining quadratic forms.
   The results of such commutators are a reflection of the underlying symmetry,
as will be shown below.
   As a second term, one obtains the divergence of the current operator in 
question.
   If the symmetry is perfect, such terms vanish identically.
   For example, this is always true for the electromagnetic case with its
U(1) symmetry.
   If the symmetry is only approximate, an additional term involving the
symmetry breaking appears.
   For a soft breaking such a divergence can be treated as a perturbation.

   Via induction, the generalization of the above simple example to an
\mbox{$(n+1)$}-point Green function is symbolically of the form
\begin{eqnarray}
\label{2:4:gendmug}
\lefteqn{\partial_\mu^x \langle 0|T\{J^\mu(x) A_1(x_1)\cdots A_n(x_n)
\}|0\rangle=}\nonumber\\
&&\langle 0|T\{[\partial_\mu^x J^\mu(x)] 
A_1(x_1)\cdots A_n(x_n)\}|0\rangle\nonumber\\
&&+\delta(x^0-x_1^0)\langle 0|T\{[J_0(x),A_1(x_1)] A_2(x_2)\cdots A_n(x_n)\}|
0\rangle\nonumber\\
&&+\delta(x^0-x_2^0)\langle 0|T\{A_1(x_1)[J_0(x),A_2(x_2)]
\cdots A_n(x_n)\}|0\rangle\nonumber\\
&&+\cdots+\delta(x^0-x_n^0)
\langle 0|T\{A_1(x_1)\cdots [J_0(x),A_n(x_n)]\}|0\rangle,
\end{eqnarray}
where $J^\mu$ stands generically for any of the Noether currents.

\subsection{The Algebra of Currents}
\label{subsec_ac}
   In the above example, we have seen that chiral Ward identities 
depend on the equal-time commutation relations of the {\em charge densities} 
of the symmetry currents with the relevant quadratic quark forms.
   Unfortunately, a naive application of Eq.\ (\ref{2:3:fkf}) may lead
to erroneous results.
   Let us illustrate this by means of a simplified example, 
the equal-time commutator of the time and space components
of the ordinary electromagnetic current in QED.
   A naive use of the canonical commutation relations leads to
\begin{eqnarray}
\label{2:4:schwinger}
[J_0(\vec{x},t), J_i(\vec{y},t)]&=&
[\Psi^\dagger(\vec{x},t)\Psi(\vec{x},t),\Psi^\dagger(\vec{y},t)\gamma_0\gamma_i
\Psi(\vec{y},t)]\nonumber\\
&=&\delta^3(\vec{x}-\vec{y})\Psi^\dagger(\vec{x},t)[1,\gamma_0\gamma_i]
\Psi(\vec{x},t)=0,
\end{eqnarray}
   where we made use of the delta function to evaluate the fields at
$\vec{x}=\vec{y}$. 
   It was noticed a long time ago by Schwinger that this result 
cannot be true \cite{Schwinger:xd}. 
   In order to see this, consider the commutator
\begin{displaymath}
[J_0(\vec{x},t),\vec{\nabla}_y\cdot \vec{J}(\vec{y},t)]=
-[J_0(\vec{x},t),\partial_t J_0(\vec{y},t)],
\end{displaymath}
   where we made use of current conservation, $\partial_\mu J^\mu=0$.
   If Eq.\ (\ref{2:4:schwinger}) were true, one would necessarily also have
\begin{displaymath}
0=[J_0(\vec{x},t),\partial_t J_0(\vec{y},t)],
\end{displaymath}
which we evaluate for $\vec{x}=\vec{y}$ between the ground state,
\begin{eqnarray*}
0&=&\langle 0|[J_0(\vec{x},t),\partial_t J_0(\vec{x},t)]|0\rangle\\
&=&\sum_n \Big(\langle 0|J_0(\vec{x},t)|n\rangle\langle n| 
\partial_t J_0(\vec{x},t)|0\rangle
-\langle 0|\partial_t J_0(\vec{x},t)|n\rangle\langle n |
J_0(\vec{x},t)|0\rangle\Big)\\
&=&2i\sum_n(E_n-E_0)|\langle 0|J_0(\vec{x},t)|n\rangle|^2.
\end{eqnarray*}
   Here, we inserted a complete set of states and made use of 
\begin{displaymath}
\partial_t J_0(\vec{x},t)=i[H,J_0(\vec{x},t)].
\end{displaymath}
   Since every individual term in the sum is non-negative, one would
need $\langle 0|J_0(\vec{x},t)|n\rangle=0$ for any intermediate 
state which is obviously unphysical. 
   The solution is that the starting point,  Eq.\ (\ref{2:4:schwinger}), 
is not true.
   The corrected version of Eq.\ (\ref{2:4:schwinger}) picks up an
additional, so-called Schwinger term containing a derivative of the
delta function. 

   Quite generally, by evaluating commutation relations with the component 
$\Theta^{00}$ of the energy-momentum tensor one can show that the equal-time 
commutation relation between a charge density and a current density can be
determined up to one derivative of the $\delta$ function \cite{Jackiw:1972},
\begin{equation}
\label{2:4:j0jigen}
[J_0^a(\vec{x},0),J_i^b(\vec{y},0)]=iC_{abc} J_i^c(\vec{x},0)\delta^3
(\vec{x}-\vec{y})+S_{ij}^{ab}(\vec{y},0)\partial^j\delta^3(\vec{x}-\vec{y}),
\end{equation}
where the Schwinger term possesses the symmetry
\begin{displaymath}
S_{ij}^{ab}(\vec{y},0)=S_{ji}^{ba}(\vec{y},0),
\end{displaymath}
and $C_{abc}$ denote the structure constants of the group in question.

    However, in our above derivation of the chiral Ward identity,
we also made use of the {\em naive} time-ordered product ($T$) as opposed to 
the {\em covariant} one ($T^\ast$) which, typically, differ by another 
non-covariant term which is called a seagull.
   Feynman's conjecture \cite{Jackiw:1972}
states that there is a cancelation between Schwinger terms 
and seagull terms such that a Ward identity obtained by 
using the naive T product and by simultaneously omitting Schwinger terms 
ultimately yields the correct result to be satisfied by the Green
function (involving the covariant $T^\ast$ product).
   Although this will not be true in general, a sufficient condition for it 
to happen is that the time component algebra of the full theory remains
the same as the one derived canonically and does not possess a Schwinger
term.   
   For a detailed discussion, the interested reader is referred to
Ref.\ \cite{Jackiw:1972}.
   
   Keeping the above discussion in mind, the complete list of equal-time 
commutation relations, omitting Schwinger terms, reads
\begin{eqnarray}
\label{2:4:letcr}
[V^a_0(\vec{x},t),V^\mu_b(\vec{y},t)]
&=&\delta^3(\vec{x}-\vec{y})if_{abc} V^\mu_c(\vec{x},t),\nonumber\\
{[}V^a_0(\vec{x},t),V^\mu(\vec{y},t)]&=&0,\nonumber\\
{[}V^a_0(\vec{x},t),A^\mu_b(\vec{y},t)]
&=&\delta^3(\vec{x}-\vec{y})if_{abc} A^\mu_c(\vec{x},t),\nonumber\\
{[}V^a_0(\vec{x},t),S_b(\vec{y},t)]
&=&\delta^3(\vec{x}-\vec{y})if_{abc} S_c(\vec{x},t),\quad b=1,\cdots,8,
\nonumber\\
{[}V^a_0(\vec{x},t),S_0(\vec{y},t)]&=&0,\nonumber\\
{[}V^a_0(\vec{x},t),P_b(\vec{y},t)]
&=&\delta^3(\vec{x}-\vec{y})if_{abc} P_c(\vec{x},t),\quad b=1,\cdots,8,
\nonumber\\
{[}V^a_0(\vec{x},t),P_0(\vec{y},t)]&=&0,\nonumber\\
{[}A^a_0(\vec{x},t),V^\mu_b(\vec{y},t)]
&=&\delta^3(\vec{x}-\vec{y})if_{abc} A^\mu_c(\vec{x},t),\nonumber\\
{[}A^a_0(\vec{x},t),V^\mu(\vec{y},t)]&=&0,\nonumber\\
{[}A^a_0(\vec{x},t),A^\mu_b(\vec{y},t)]
&=&\delta^3(\vec{x}-\vec{y})if_{abc} V^\mu_c(\vec{x},t),\nonumber\\
{[}A^a_0(\vec{x},t),S_b(\vec{y},t)]
&=&\delta^3(\vec{x}-\vec{y})if_{abc} P_c(\vec{x},t),\quad b=1,\cdots,8,
\nonumber\\
{[}A^a_0(\vec{x},t),S_0(\vec{y},t)]&=&0,\nonumber\\
{[}A^a_0(\vec{x},t),P_b(\vec{y},t)]
&=&\delta^3(\vec{x}-\vec{y})if_{abc} S_c(\vec{x},t),\quad b=1,\cdots,8,
\nonumber\\
{[}A^a_0(\vec{x},t),P_0(\vec{y},t)]&=&0.
\end{eqnarray}

\subsection{Two Simple Examples}
\label{subsec_tse}
   We now return to our specific example, namely, the divergence of 
Eq.\ (\ref{2:4:gfaav}).
   Inserting the results of Eqs.\ (\ref{2:3:dsva})
and  (\ref{2:4:letcr}) one obtains
\begin{eqnarray}
\label{2:4:dgfaav}
\partial_\mu^x G^{\mu,ab}_{AP}(x,y)
&=&\delta^4(x-y) i f_{abc}\langle 0|S_c(x)|0\rangle\nonumber\\
&&+i\langle 0|T[\bar{q}(x)\{\frac{\lambda_a}{2},M\}\gamma_5
q(x) P_b(y)]|0\rangle.
\end{eqnarray}
   The second term on the right-hand side of Eq.\ (\ref{2:4:dgfaav}) 
can be re-expressed using Eq.\ (\ref{2:3:qmm}) and the anti-commutation 
relations of Eq.\ (\ref{2:1:acrgmm}) in combination
with the $d$ coefficients of Table \ref{table:2:1:su3dsymbols}
(no summation over $a$ implied),
\begin{eqnarray*}
\lefteqn{i\bar{q}(x)\{\frac{\lambda_a}{2},M\}\gamma_5q(x)=}\\
&&\left[\frac{1}{3}(m_u+m_d+m_s)
+\frac{1}{\sqrt{3}}\left(\frac{m_u+m_d}{2}- m_s\right)d_{aa8}\right]P_a(x)\\
&&+\left[\sqrt{\frac{1}{6}}(m_u-m_d)\delta_{a3}
+\frac{\sqrt{2}}{3}\left(\frac{m_u+m_d}{2}-m_s\right)\delta_{a8}\right]P_0(x)\\
&&+\frac{m_u-m_d}{2} \sum_{c=1}^8d_{a3c} P_c(x).
\end{eqnarray*}
   Equation (\ref{2:4:dgfaav}) serves to illustrate two distinct features of 
chiral Ward identities. 
   The first term of Eq.\ (\ref{2:4:dgfaav}) originates in the algebra
of currents and thus represents a consequence of the {\em transformation 
properties} of the quadratic quark forms entering the Green function.  
   In general, depending on whether the appropriate equal-time commutation 
relation of Eq.\ (\ref{2:4:letcr}) vanishes or not, the resulting term
in the divergence of an $n$-point Green function vanishes or is proportional 
to an $(n-1)$-point Green function.  
   In our specific example, the divergence of the Green function involving
the axial-vector current and the pseudoscalar density is related to the 
so-called scalar quark condensate which will be discussed in more detail
in Sec.\ \ref{subsec_sqc}.
   The second term of Eq.\ (\ref{2:4:dgfaav}) is due to an explicit symmetry 
breaking resulting from the quark masses.
   This shows the second property
of chiral Ward identities, namely, symmetry breaking terms give rise to 
another $n$-point Green function.
   To summarize, chiral Ward identities incorporate both transformation
properties of quadratic quark forms as well as symmetry breaking patterns. 

   As another well-known and simple example, let us briefly consider, 
for the two-flavor case, the nucleon matrix element of the axial-vector 
current operator\footnote{This matrix element will be dealt with 
in Sec.\ \ref{subsec_gtravcme}.}
\begin{equation}
\label{2:4:avcn}
\langle N(p_f)|A^i_\mu(x)|N(p_i)\rangle=
\langle N(p_f)|\bar{q}(x)\gamma_\mu\gamma_5 \frac{\tau_i}{2}q(x)|N(p_i)\rangle.
\end{equation}
   This matrix element serves as an illustration of chiral Ward identities
which are taken between one-nucleon states instead of the vacuum.
   According to Eq.\ (\ref{2:3:dsva}), the divergence of Eq.\ (\ref{2:4:avcn})
is related to the pseudoscalar density evaluated between one-nucleon
states.
   Of course, in the chiral limit $M=0$ and the axial-vector current is 
conserved.

\subsection{QCD in the Presence of External Fields and the Generating
Functional}
\label{subsec_qcdpefgf}
   Here, we want to consider the consequences of Eqs.\ (\ref{2:4:letcr}) 
for the Green functions of QCD (in particular, at low energies).
   In principle, using the techniques of the last section, for each
Green function one can {\em explicitly} work out the chiral Ward identity
which, however, becomes more and more tedious as the number $n$ of quark 
quadratic forms increases.
   However, there exists an elegant way of formally combining all Green
functions in a generating functional.
   The (infinite) set of {\em all} chiral Ward identities is encoded
as an invariance property of that functional.
    To see this, one has to consider a coupling to external c-number fields 
such that through functional methods one can, in principle, obtain all
Green functions from a generating functional.
   The rationale behind this approach is that, in the absence of anomalies,
the Ward identities obeyed by the Green functions are equivalent to 
an invariance of the generating functional under a {\em local} transformation 
of the external fields \cite{Leutwyler:1993iq}.
   The use of local transformations allows one to also consider divergences
of Green functions.
   For an illustration of this statement, the reader is referred to Appendix A.

   Following the procedure of Gasser and Leutwyler 
\cite{Gasser:1983yg,Gasser:1984gg}, we introduce into the Lagrangian of QCD 
the couplings of the nine vector currents and the eight axial-vector currents 
as well as the scalar and pseudoscalar quark densities 
to external c-number fields 
$v^\mu (x)$, $v^\mu_{(s)}$, $a^\mu (x)$, $s(x)$, and $p(x)$,
\begin{equation}
\label{2:4:lqcds}
{\cal L}={\cal L}^0_{\rm QCD}+{\cal L}_{\rm ext}
={\cal L}^0_{\rm QCD}+\bar{q}\gamma_\mu (v^\mu +\frac{1}{3}v^\mu_{(s)}
+\gamma_5 a^\mu )q
-\bar{q}(s-i\gamma_5 p)q.
\end{equation}
     The external fields are color-neutral, Hermitian $3\times 3$ matrices,
where the matrix character, with respect to the (suppressed) flavor indices
$u$, $d$, and $s$ of the quark fields, is\footnote{As in Refs.\
\cite{Gasser:1983yg,Gasser:1984gg}, we omit the coupling to the 
singlet axial-vector current which has an anomaly, 
but include a singlet vector current $v^\mu_{(s)}$ which is of some physical
relevance in the two-flavor sector.}
\begin{equation}
\label{2:4:mch}
v^\mu=\sum_{a=1}^8\frac{\lambda_a}{2}v_a^\mu,\quad 
a^\mu=\sum_{a=1}^8\frac{\lambda_a}{2}a_a^\mu,\quad 
s=\sum_{a=0}^8 \lambda_a s_a,\quad 
p=\sum_{a=0}^8\lambda_a p_a.
\end{equation} 
The ordinary three flavor QCD Lagrangian is recovered by setting 
$v^\mu=v^\mu_{(s)}=a^\mu=p=0$ and $s=\mbox{diag}(m_u,m_d,m_s)$ in 
Eq.\ (\ref{2:4:lqcds}).  
   
   If one defines the generating functional\footnote{Many
books on quantum field theory such as Refs.\ 
\cite{Itzykson:rh,Collins:xc,Ryder:wq,Rivers:hi}
reserve the symbol $Z[v,a,s,p]$ for the generating functional of all
Green functions as opposed to the argument of the exponential which denotes
the generating functional of connected Green functions.}
\begin{equation}
\label{2:4:genfun}
\exp(i Z[v,a,s,p])=\langle 0|T\exp\left[i\int d^4 x 
{\cal L}_{\rm ext}(x)\right]|0\rangle,
\end{equation}
then any Green function consisting of the time-ordered product of 
color-neutral, Hermitian quadratic forms can be obtained from 
Eq.\ (\ref{2:4:genfun}) through a functional derivative with respect 
to the external fields.
   The quark fields are operators in the Heisenberg picture and 
have to satisfy the equation of motion and the canonical
anti-commutation relations.
   The actual value of the generating functional for a given configuration
of external fields $v$, $a$, $s$, and $p$ reflects the dynamics generated
by the QCD Lagrangian.
   The generating functional is related to the vacuum-to-vacuum transition
amplitude in the presence of external fields 
\cite{Gasser:1983yg,Gasser:1984gg},
\begin{equation}
\label{2:4:genfunvv}
\exp[i Z(v,a,s,p)]=
\langle 0_{\rm out}|0_{\rm in}\rangle_{v,a,s,p},
\end{equation}
   where the dynamics is determined by the Lagrangian of 
Eq.\ (\ref{2:4:lqcds}).

   For example,\footnote{In order to obtain Green functions from the
generating functional the simple rule 
$$\frac{\delta f(x)}{\delta f(y)}=\delta(x-y)$$
is extremely useful. Furthermore, the functional derivative satisfies
properties similar to the ordinary differentiation, namely linearity,
the product and chain rules.}
the $\bar{u} u$ component of the scalar quark condensate in the chiral limit, 
$\langle 0| \bar{u}u|0\rangle_0$, is given by
\begin{eqnarray}
\label{2:4:sqc}
\lefteqn{\langle 0|\bar{u}(x) u(x)|0\rangle_0 =}\nonumber\\
&&\left.\frac{i}{2}\left[\sqrt{\frac{2}{3}}\frac{\delta}{\delta s_0(x)}
+\frac{\delta}{\delta s_3(x)}
+\frac{1}{\sqrt{3}}
\frac{\delta}{\delta s_8(x)}\right]
\exp(iZ [v,a,s,p])\right|_{v=a=s=p=0},\nonumber\\
\end{eqnarray}
where we made use of Eq.\ (\ref{2:1:matrixa}).
   Note that both the quark field operators and the ground state are considered
in the chiral limit, which is denoted by the subscript 0.
 
    As another example, let us consider the two-point function of the
axial-vector currents of Eq.\ (\ref{2:3:a}) of the ``real world,''
i.e., for $s=\mbox{diag}(m_u,m_d,m_s)$, and the ``true vacuum'' $|0\rangle$,
\begin{eqnarray}
\label{2:4:tpfavc}
\lefteqn{\langle 0|T[A_\mu^a(x) A_\nu^b(0)]|0\rangle =}\nonumber\\
&&\left.
(-i)^2 \frac{\delta^2}{\delta a^\mu_a(x)\delta a^\nu_b(0)}
\exp(iZ[v,a,s,p])\right|_{v=a=p=0,s=\mbox{diag}(m_u,m_d,m_s)}.\nonumber\\
\end{eqnarray}  

   Requiring the total Lagrangian of Eq.\ (\ref{2:4:lqcds}) to be Hermitian
and invariant under $P$, $C$, and $T$ leads to constraints on the 
transformation behavior of the external fields.
   In fact, it is sufficient to consider $P$ and $C$, only, because
$T$ is then automatically incorporated owing to the $CPT$ theorem.

   Under parity, the quark fields transform as 
\begin{equation}
\label{2:4:qtrafop}
q_f(\vec{x},t)\stackrel{\mbox{$P$}}{\mapsto}\gamma^0 q_f(-\vec{x},t),
\end{equation}
   and the requirement of parity conservation,  
\begin{equation}
\label{2:4:parinv}
{\cal L}(\vec{x},t) \stackrel{\mbox{$P$}}{\mapsto} {\cal L}(-\vec{x},t),
\end{equation}
leads, using the results of Table \ref{2:4:parity}, to the following 
constraints for the external fields,
\begin{equation}
\label{2:4:eftrafop}
v^\mu\stackrel{\mbox{$P$}}{\mapsto}v_\mu,\quad
v^\mu_{(s)}\stackrel{\mbox{$P$}}{\mapsto}v_\mu^{(s)},\quad
a^\mu\stackrel{\mbox{$P$}}{\mapsto}-a_\mu,\quad
s\stackrel{\mbox{$P$}}{\mapsto}s,\quad
p\stackrel{\mbox{$P$}}{\mapsto}-p.
\end{equation}
   In Eq.\ (\ref{2:4:eftrafop}) it is understood that the arguments change 
from $(\vec{x},t)$ to $(-\vec{x},t)$.

\begin{table}
\begin{center}
\begin{tabular}{|c|c|c|c|c|c|}
\hline
$\Gamma$&
$1$&
$\gamma^\mu$&
$\sigma^{\mu\nu}$&
$\gamma_5$&
$\gamma^\mu\gamma_5$\\
\hline
$\gamma_0 \Gamma \gamma_0$&
$1$&
$\gamma_\mu$&
$\sigma_{\mu\nu}$&
$-\gamma_5$&
$-\gamma_\mu\gamma_5$
\\
\hline
\end{tabular}
\end{center}
\caption{\label{2:4:parity} Transformation properties of the Dirac
matrices $\Gamma$ under parity.}
\end{table}
   Similarly, under charge conjugation the quark fields transform as 
\begin{equation}
\label{2:4:qtrafc}
q_{\alpha,f}\stackrel{\mbox{$C$}}{\mapsto}C_{\alpha\beta}\bar{q}_{\beta,f},
\quad
\bar{q}_{\alpha,f}\stackrel{\mbox{$C$}}{\mapsto}
-q_{\beta,f}C^{-1}_{\beta\alpha},
\end{equation}
   where the subscripts $\alpha$ and $\beta$ are Dirac
spinor indices, $C=i\gamma^2\gamma^0=-C^{-1}=-C^\dagger=-C^T$ is 
the usual charge conjugation
matrix in the convention of Ref.\ \cite{Bjorken_1964} and $f$ refers to flavor.
   Using Eq.\ (\ref{2:4:qtrafc}) in combination with Table 
\ref{2:4:chargeconjugation} it is straightforward to show that invariance 
of ${\cal L}_{\rm ext}$ under charge conjugation requires the transformation
properties\footnote{In deriving these results we need to make use of
$q_{\gamma,f}\bar{q}_{\delta,f'}=-\bar{q}_{\delta,f'}q_{\gamma,f}$,
since the quark fields are anti-commuting field operators.}
\begin{equation}
\label{2:4:eftrafoc}
v_\mu\stackrel{C}{\rightarrow}-v_\mu^T,\quad
v_\mu^{(s)}\stackrel{C}{\rightarrow}-v_\mu^{(s)T},\quad
a_\mu\stackrel{C}{\rightarrow}a_\mu^T,\quad
s,p\stackrel{C}{\rightarrow}s^T,p^T,
\end{equation}
where the transposition refers to the flavor space.

\begin{table}
\begin{center}
\begin{tabular}{|c|c|c|c|c|c|}
\hline
$\Gamma$&$1$&$\gamma^\mu$&$\sigma^{\mu\nu}$&$\gamma_5$&$\gamma^\mu\gamma_5$\\
\hline
$-C\Gamma^TC
$&$1$&$-\gamma^\mu$&$-\sigma^{\mu\nu}$&$\gamma_5$&$\gamma^\mu\gamma_5$
\\
\hline
\end{tabular}
\end{center}
\caption{\label{2:4:chargeconjugation} 
Transformation properties of the Dirac matrices $\Gamma$
under charge conjugation.}
\end{table}

   Finally, we need to discuss the requirements to be met by the external
fields under local $\mbox{SU(3)}_L\times\mbox{SU(3)}_R\times\mbox{U}(1)_V$ 
transformations.   
   In a first step, we write Eq.\ (\ref{2:4:lqcds}) in terms of the
left- and right-handed quark fields. 
   Besides the properties of Eqs.\ (\ref{2:3:prplcompleteness}) -
(\ref{2:3:prplorthogonality})
we make use of the auxiliary formulae
$$
\gamma_5 P_R=P_R\gamma_5=P_R, \quad \gamma_5 P_L=P_L\gamma_5=-P_L,
$$
and 
$$
\gamma^\mu P_R=P_L\gamma^\mu,\quad \gamma^\mu P_L=P_R\gamma^\mu, 
$$
to obtain
\begin{eqnarray*}
\bar{q}\gamma^\mu(v_\mu+\frac{1}{3}v_\mu^{(s)}
+\gamma_5 a_\mu)q&=&\frac{1}{2}\bar{q}
\gamma^\mu[r_\mu+l_\mu+\frac{2}{3}v_\mu^{(s)}+\gamma_5(r_\mu-l_\mu)]
q\\
&=&\bar{q}_R\gamma^\mu \left(r_\mu +\frac{1}{3}v_\mu^{(s)}\right)q_R
+\bar{q}_L\gamma^\mu \left(l_\mu+\frac{1}{3}v_\mu^{(s)}\right) q_L,
\end{eqnarray*}
where 
\begin{equation}
\label{2:4:vrlarl}
v_\mu=\frac{1}{2}(r_\mu+l_\mu),\quad
a_\mu=\frac{1}{2}(r_\mu-l_\mu).
\end{equation}
   Similarly, we rewrite the second part containing the external scalar
and pseudoscalar fields,
\begin{eqnarray*}
\bar{q}(s-i\gamma_5 p)q&=&
\bar{q}_L(s-ip)q_R+\bar{q}_R(s+ip)q_L,
\end{eqnarray*}
   yielding for the Lagrangian of Eq.\ (\ref{2:4:lqcds})
\begin{eqnarray}
\label{2:4:lqcdsn}
{\cal L}&=&{\cal L}_{\rm QCD}^0
+\bar{q}_L\gamma^\mu\left(l_\mu+\frac{1}{3}v^{(s)}_\mu\right)q_L
+\bar{q}_R\gamma^\mu\left(r_\mu+\frac{1}{3}v^{(s)}_\mu\right)q_R\nonumber\\
&&-\bar{q}_R(s+ip)q_L-\bar{q}_L(s-ip)q_R.
\end{eqnarray}
   Equation (\ref{2:4:lqcdsn}) remains invariant under {\em local} 
transformations 
\begin{eqnarray}
\label{2:4:qrl}
q_R&\mapsto&\exp\left(-i\frac{\Theta(x)}{3}\right) V_R(x) q_R,\nonumber\\
q_L&\mapsto&\exp\left(-i\frac{\Theta(x)}{3}\right) V_L(x) q_L,
\end{eqnarray}
where $V_R(x)$ and $V_L(x)$ are independent space-time-dependent SU(3)
matrices, provided the external fields are subject
to the transformations
\begin{eqnarray}
\label{2:4:sg}
r_\mu&\mapsto& V_R r_\mu V_R^{\dagger}
+iV_R\partial_\mu V_R^{\dagger},\nonumber\\
l_\mu&\mapsto& V_L l_\mu V_L^{\dagger}
+iV_L\partial_\mu V_L^{\dagger},
\nonumber\\
v_\mu^{(s)}&\mapsto&v_\mu^{(s)}-\partial_\mu\Theta,\nonumber\\
s+ip&\mapsto& V_R(s+ip)V_L^{\dagger},\nonumber\\
s-ip&\mapsto& V_L(s-ip)V_R^{\dagger}.
\end{eqnarray}
   The derivative terms in Eq.\ (\ref{2:4:sg}) serve the same purpose as
in the construction of gauge theories, i.e., they cancel analogous
terms originating from the kinetic part of the quark Lagrangian.   

   There is another, yet, more practical aspect of the local invariance,
namely: such a procedure allows one to also discuss a coupling to external 
gauge fields in the transition to the effective theory to be discussed later.
   For example, we have seen in Sec.\ 2.2 that a coupling of the 
electromagnetic field to point-like fundamental particles results from
gauging a U(1) symmetry.
  Here, the corresponding U(1) group is to be understood
as a subgroup of a local $\mbox{SU(3)}_L\times\mbox{SU(3)}_R$.
   Another example deals with the interaction of the light quarks
with the charged and neutral gauge bosons of the weak interactions.

   Let us consider both examples explicitly. The coupling of quarks
to an external electromagnetic field ${\cal A}_\mu$ is given by
\begin{equation}
\label{2:4:rla}
r_\mu=l_\mu=-e Q {\cal A}_\mu,
\end{equation}
where $Q=\mbox{diag}(2/3,-1/3,-1/3)$ is the quark charge matrix:
\begin{eqnarray*}
{\cal L}_{\rm ext}&=&-e {\cal A}_\mu(\bar{q}_L Q\gamma^\mu q_L
+\bar{q}_R Q \gamma^\mu q_R)\\
&=&-e {\cal A}_\mu \bar{q}Q\gamma^\mu q\\
&=&-e {\cal A}_\mu\left(\frac{2}{3}\bar{u}\gamma^\mu u
-\frac{1}{3} \bar{d}\gamma^\mu d -\frac{1}{3}\bar{s}\gamma^\mu s\right)\\
&=&-e {\cal A}_\mu J^\mu.
\end{eqnarray*}
   On the other hand, if one considers only the two-flavor version of ChPT one
has to insert for the external fields
\begin{equation}
\label{2:4:rlasu2}
r_\mu=l_\mu=-e\frac{\tau_3}{2}{\cal A}_\mu,\quad
v_\mu^{(s)}=-\frac{e}{2}{\cal A}_\mu.
\end{equation}

   In the description of semileptonic interactions such as 
$\pi^-\to \mu^-\bar{\nu}_\mu$,  $\pi^-\to\pi^0e^-\bar{\nu}_e$, or
neutron decay $n\to p e^-\bar{\nu}_e$ one needs the interaction of quarks with 
the massive charged weak bosons 
${\cal W}^\pm_\mu=({\cal W}_{1\mu}\mp i {\cal W}_{2\mu})/\sqrt{2}$,
\begin{equation}
\label{2:4:rlw}
r_\mu=0,\quad l_\mu=-\frac{g}{\sqrt{2}}
({\cal W}^+_\mu T_+ + h.c.),
\end{equation}
where $h.c.$ refers to the Hermitian conjugate
and
$$
T_+=\left(\begin{array}{rrr}0&V_{ud}&V_{us}\\0&0&0\\0&0&0\end{array}\right).
$$
   Here, $V_{ij}$ denote the elements of the 
Cabibbo-Kobayashi-Maskawa quark-mixing matrix 
describing the transformation between the mass eigenstates of QCD and the
weak eigenstates \cite{Groom:in},
$$|V_{ud}|=0.9735\pm 0.0008,\quad
|V_{us}|=0.2196\pm 0.0023.
$$
   At lowest order in perturbation theory, the Fermi constant is related
to the gauge coupling $g$ and the $W$ mass as
$$ 
G_F=\sqrt{2} \frac{g^2}{8 M^2_W}=1.16639(1)\times 10^{-5}\,\mbox{GeV}^{-2}.
$$
   Making use of 
\begin{eqnarray*}
\bar{q}_L\gamma^\mu {\cal W}_\mu^+ T_+ q_L&=&
{\cal W}_\mu^+ (\bar{u}\,\,\bar{d}\,\, \bar{s}) P_R\gamma^\mu
\left(\begin{array}{rrr}0&V_{ud}&V_{us}\\0&0&0\\0&0&0\end{array}
\right)
P_L
\left(\begin{array}{c}u\\ d\\ s\end{array}\right)\\
&=&{\cal W}_\mu^
+(\bar{u}\,\,\bar{d}\,\,\bar{s})\gamma^\mu \frac{1}{2}(1-\gamma_5)
\left(\begin{array}{c}V_{ud} d+ V_{us} s\\0\\0\end{array}
\right)\\
&=&\frac{1}{2}{\cal W}_\mu^+[V_{ud}\bar{u}\gamma^\mu(1-\gamma_5)d
+V_{us}\bar{u}\gamma^\mu(1-\gamma_5)s],
\end{eqnarray*}
   we see that inserting Eq.\ (\ref{2:4:rlw}) into Eq.\ (\ref{2:4:lqcdsn})
leads to the standard charged-current weak interaction in the light
quark sector, 
\begin{eqnarray*}
{\cal L}_{\rm ext}&=&-\frac{g}{2\sqrt{2}}\left\{{\cal W}^+_\mu[
V_{ud}\bar{u}\gamma^\mu(1-\gamma_5)d+V_{us}\bar{u}\gamma^\mu(1-\gamma_5)s]
+h.c.\right\}.
\end{eqnarray*}

   The situation is slightly different for the neutral weak interaction.
Here, the SU(3) version requires a coupling to the singlet axial-vector
current which, because of the anomaly of Eq.\ (\ref{2:3:divsa}), we have 
dropped from our discussion.
   On the other hand, in the SU(2) version the axial-vector current part
is traceless and we have  
\begin{eqnarray}
\label{2:4:rlz}
r_\mu&=&e \tan(\theta_W) \frac{\tau_3}{2} {\cal Z}_\mu,\nonumber\\
l_\mu&=&-\frac{g}{\cos(\theta_W)}\frac{\tau_3}{2} {\cal Z}_\mu+
e \tan(\theta_W) \frac{\tau_3}{2} {\cal Z}_\mu,
\nonumber\\
v_\mu^{(s)}&=&\frac{e\tan(\theta_W)}{2}{\cal Z}_\mu,
\end{eqnarray}
where  $\theta_W$ is the weak angle. 
   With these external fields, we obtain the standard weak neutral-current 
interaction \cite{Groom:in}
\begin{eqnarray*}
{\cal L}_{\rm ext}&=&-\frac{g}{2\cos(\theta_W)}{\cal Z}_\mu\left(
\bar{u}\gamma^\mu\left\{\left[\frac{1}{2}-\frac{4}{3}\sin^2(\theta_W)\right]
-\frac{1}{2}\gamma_5\right\}u\right.\nonumber\\
&&\left.+\bar{d}\gamma^\mu\left\{\left[-\frac{1}{2}
+\frac{2}{3}\sin^2(\theta_W)\right]
+\frac{1}{2}\gamma_5\right\}d\right),
\end{eqnarray*}
where we made use of $e=g\sin(\theta_W)$.

\subsection{PCAC in the Presence of an External Electromagnetic Field}
\label{subsec_pcacpeef}
   Finally, the technique of coupling the QCD Lagrangian to external fields 
also allows us to determine the current divergences for rigid external fields,
i.e., which are {\em not} simultaneously transformed.
   For the sake of simplicity we restrict ourselves to the SU(2) sector. 
   (The generalization to the SU(3) case is straightforward.)
   If the external fields are not simultaneously transformed and one considers
a {\em global} chiral transformation only, the divergences of the currents
read [see Eq.\ (\ref{2:3:divergenz})]
\begin{eqnarray}
\label{2:4:divv}
\partial_\mu V^\mu_i&=&i\bar{q}\gamma^\mu[\frac{\tau_i}{2},v_\mu]q
+i\bar{q}\gamma^\mu\gamma_5[\frac{\tau_i}{2},a_\mu]q
-i\bar{q}[\frac{\tau_i}{2},s]q-\bar{q}\gamma_5[\frac{\tau_i}{2},p]q,
\nonumber\\
\label{2:4:diva}\\
\partial_\mu A^\mu_i&=&i\bar{q}\gamma^\mu\gamma_5[\frac{\tau_i}{2},v_\mu]q
+i\bar{q}\gamma^\mu[\frac{\tau_i}{2},a_\mu]q
+i\bar{q}\gamma_5\{\frac{\tau_i}{2},s\}q
+\bar{q}\{\frac{\tau_i}{2},p\}q.\nonumber\\
\end{eqnarray}
   As an example, let us consider the QCD Lagrangian for a finite light quark 
mass $m_q$ in combination with a coupling to an external 
electromagnetic field ${\cal A}_\mu$ [see Eq.\ (\ref{2:4:rlasu2}),
$a_\mu=0=p$].
   In this case the expressions for the divergence of the vector and
axial-vector currents, respectively, read
\begin{eqnarray}
\label{2:4:divvsc}
\partial_\mu V^\mu_i&=&-\epsilon_{3ij}e{\cal A}_\mu \bar{q}\gamma^\mu
\frac{\tau_j}{2}q=-\epsilon_{3ij}e{\cal A}_\mu V^\mu_j,\\
\label{2:4:divasc}
\partial_\mu A^\mu_i
&=&m_q P_i-e {\cal A}_\mu \epsilon_{3ij} A^\mu_j
+\delta_{i3}
\frac{e^2 N_c}{96\pi^2}\epsilon_{\mu\nu\rho\sigma}{\cal F}^{\mu\nu} 
{\cal F}^{\rho\sigma},
\end{eqnarray}
where we have introduced the isovector pseudoscalar density
\begin{equation}
\label{2:4:psd}
P_i=i\bar{q}\gamma_5 \tau_i q,
\end{equation}
and ${\cal F}_{\mu\nu}=\partial_\mu{\cal A}_\nu-\partial_\nu{\cal A}_\mu$ is
the electromagnetic field strength tensor.
   The third component of the axial-vector current, $A^\mu_3$, has an anomaly
\cite{Adler:1969gk,Adler:1969er,Bardeen:1969md,Bell:1969ts,Adler:1970} 
which is related to
the decay $\pi^0\to\gamma\gamma$.
   We emphasize the formal similarity of Eq.\ (\ref{2:4:divasc}) to the 
(pre-QCD) PCAC relation obtained by Adler through the inclusion of the 
electromagnetic interactions
with minimal electromagnetic coupling 
(see the Appendix of Ref.\ \cite{Adler:1965}).\footnote{
   In Adler's version, the right-hand side of Eq.\ (\ref{2:4:divasc}) contains
a renormalized field operator creating and destroying pions instead of
$m_q P_i$.  From a modern point of view, the combination 
$m_q P_i/(M_\pi^2 F_\pi)$  serves as an interpolating
pion field (see Sec.\ \ref{subsec_pps}).
    Furthermore, the anomaly term is not yet present in Ref.\
\cite{Adler:1965}.}
   Since in QCD the quarks are taken as truly elementary, their interaction
with an (external) electromagnetic field is of such a minimal type.

\chapter{Spontaneous Symmetry Breaking and the Goldstone Theorem}
\label{chap_ssbgt}

   So far we have concentrated on the chiral symmetry of the QCD Hamiltonian 
and the {\em explicit} symmetry breaking through the quark masses.
   We have discussed the importance of chiral symmetry for the properties
of Green functions with particular emphasis on the relations
{\em among} different Green functions as expressed through the chiral Ward 
identities. 
   Now it is time to address a second aspect which, for the low-energy 
structure of QCD, is equally important, namely, the concept of 
{\em spontaneous} symmetry breaking.
   A (continuous) symmetry is said to be spontaneously broken or hidden,
if the ground state of the system is no longer invariant under the 
full symmetry group of the Hamiltonian.
   In this chapter we will first illustrate this by means of a discrete 
symmetry and then turn to the case of a spontaneously broken
continuous global symmetry.

\section{Degenerate Ground States}
\label{sec_dgs}
   Before discussing the case of a {\em continuous} symmetry, we will first
have a look at a field theory with a {\em discrete} internal symmetry.
   This will allow us to distinguish between two possibilities: a 
dynamical system with a unique ground state or a system with a 
finite number of distinct degenerate ground states.
   In particular, we will see how, for the second case, an infinitesimal 
perturbation selects a particular vacuum state.
   
\begin{figure}
\begin{center}
\epsfig{file=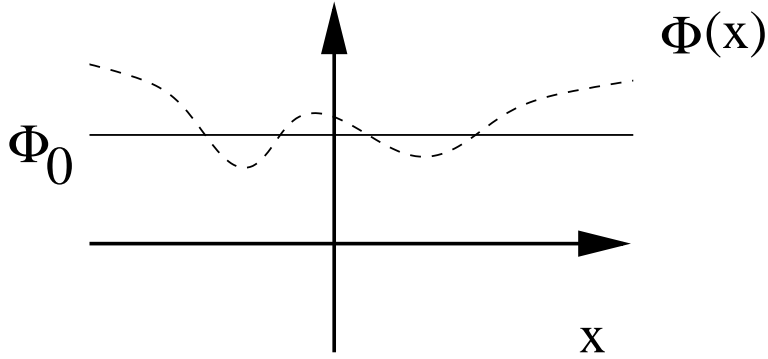,width=10cm}
\caption{\label{3:1:potphi02} Solid line: constant field configuration
$\Phi_0$ minimizing the potential; dashed line: arbitrary
field configuration $\Phi(x)$.}
\end{center}
\end{figure}

\begin{figure}
\begin{center}
\epsfig{file=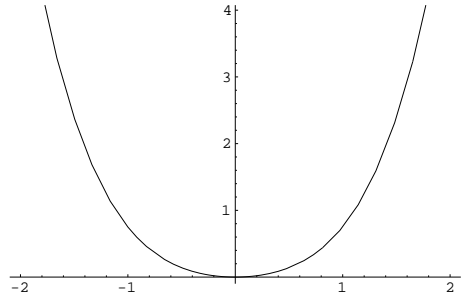,width=6cm}
\caption{\label{3:1:potww} ${\cal V}(x)=x^2/2+x^4/4$ (Wigner-Weyl mode).}
\end{center}
\end{figure}

\begin{figure}
\begin{center}
\epsfig{file=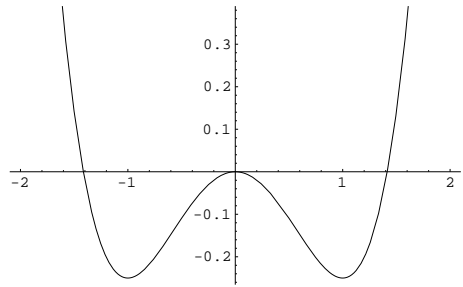,width=6cm}
\caption{\label{3:1:potng} ${\cal V}(x)=-x^2/2+x^4/4$ (Nambu-Goldstone mode).}
\end{center}
\end{figure}

   To that end we consider the Lagrangian of a real scalar field $\Phi(x)$ 
\cite{Georgi}
\begin{equation}
\label{3:1:lphi}
{\cal L}(\Phi,\partial_\mu\Phi)=\frac{1}{2}\partial_\mu \Phi \partial^\mu \Phi
-\frac{m^2}{2}\Phi^2-\frac{\lambda}{4}\Phi^4,
\end{equation}
   which is invariant under the discrete transformation $R: \Phi\to -\Phi$. 
The corresponding classical energy density reads 
\begin{equation}
\label{3:1:ked}
{\cal H}=\Pi\dot{\Phi}-{\cal L}=\frac{1}{2}\dot{\Phi}^2
+\frac{1}{2}(\vec{\nabla}\Phi)^2+
\underbrace{\frac{m^2}{2}\Phi^2+\frac{\lambda}{4}\Phi^4}_{
\mbox{${\cal V}(\Phi)$}},
\end{equation}
   where one chooses $\lambda > 0$ so that $\cal H$ is bounded from below.
   The field $\Phi_0$ which minimizes the Hamilton density ${\cal H}$ 
must be constant and uniform since in that case the first two terms
take everywhere their minimum values of zero.
   It must also minimize the potential since 
${\cal V}(\Phi(x))\geq {\cal V}(\Phi_0)$ (see Fig.~\ref{3:1:potphi02}),
from which we obtain the condition
$$
{\cal V}'(\Phi)=\Phi(m^2+\lambda \Phi^2)=0.
$$
We now distinguish two different cases:
\begin{itemize}
\item $m^2>0$ (see Fig.\ \ref{3:1:potww}): In this case the potential
$\cal V$ has its minimum for $\Phi=0$.
   In the quantized theory we associate a unique ground state $|0\rangle$
with this minimum.
   Later on, in the case of a continuous symmetry, this situation will
be referred to as the Wigner-Weyl realization of the symmetry.
\item $m^2<0$ (see Fig.\ \ref{3:1:potng}): Now the potential exhibits
two distinct minima.
   (In  the continuous symmetry case this will be referred to as the 
Nambu-Goldstone realization of the symmetry.)
\end{itemize}
   We will concentrate on the second situation, because this is the 
one which we would like to generalize to a continuous symmetry and which
ultimately leads to the appearance of Goldstone bosons.
   In the present case, ${\cal V}(\Phi)$ has a maximum for $\Phi=0$ and
{\em two} minima for 
\begin{equation}
\label{3:1:vmin}
\Phi_\pm=\pm \sqrt{\frac{-m^2}{\lambda}}\equiv \pm \Phi_0.
\end{equation}
   As will be explained below, the quantized theory develops two degenerate 
vacua $|0,+\rangle$ and $|0,-\rangle$ which are distinguished 
through their vacuum expectation values of the field $\Phi(x)$:\footnote{
The case of a quantum field theory with an infinite volume $V$ has
to be distinguished from, say, a nonrelativistic particle in a one-dimensional
potential of a shape similar to the function of Fig.\ \ref{3:1:potng}.
   For example, in the case of a symmetric double-well potential, the solutions
with positive parity have always lower energy eigenvalues than those with
negative parity (see, e.g., Ref.\ \cite{Greiner:mm}).}
\begin{eqnarray}
\label{3:1:vewphi}
\langle0,+|\Phi(x)|0,+\rangle&=&
\langle0,+|e^{iP\cdot x}\Phi(0)e^{-iP\cdot x}|0,+\rangle
=\langle0,+|\Phi(0)|0,+\rangle\equiv \Phi_0,\nonumber\\
\langle0,-|\Phi(x)|0,-\rangle&=&-\Phi_0.
\end{eqnarray}
   We made use of translational invariance, 
$\Phi(x)=e^{iP\cdot x}\Phi(0)e^{-iP\cdot x}$, 
and the fact that the ground state is an eigenstate of energy and momentum.
   We associate with the transformation $R:\Phi\mapsto\Phi'=-\Phi$
a unitary operator $\cal R$ acting on the Hilbert space of our model,
with the properties
$${\cal R}^2=1, \quad {\cal R}={\cal R}^{-1}={\cal R}^\dagger.$$
   In accord with Eq.\ (\ref{3:1:vewphi}) the action of the operator 
$\cal R$ on the ground states is given by
\begin{equation}
\label{3:1:rvpm}
{\cal R}|0,\pm\rangle=|0,\mp\rangle.
\end{equation}

   For the moment we select one of the two expectation values and expand
the field with respect to $\pm \Phi_0$:\footnote{The field $\Phi'$ instead
of $\Phi$ is assumed to vanish at infinity.}
\begin{eqnarray}
\label{3:1:entw}
\Phi&=&\pm \Phi_0+\Phi',\nonumber\\
\partial_\mu \Phi&=&\partial_\mu \Phi'.
\end{eqnarray}
   A short calculation yields
\begin{eqnarray*}
{\cal V}(\Phi)&=&
\tilde{\cal V}(\Phi')=
-\frac{\lambda}{4}\Phi_0^4+\frac{1}{2}(-2m^2)\Phi'^2
\pm\lambda\Phi_0\Phi'^3+\frac{\lambda}{4}\Phi'^4,
\end{eqnarray*}
such that the Lagrangian in terms of the shifted dynamical variable
reads 
\begin{equation}
\label{3:1:lphip}
{\cal L}'(\Phi',\partial_\mu \Phi')
=\frac{1}{2}\partial_\mu\Phi'\partial^\mu \Phi'
-\frac{1}{2}(-2m^2)\Phi'^2\mp\lambda\Phi_0\Phi'^3-\frac{\lambda}{4}\Phi'^4
+\frac{\lambda}{4}\Phi_0^4.
\end{equation}
   In terms of the new dynamical variable $\Phi'$, 
the symmetry $R$ is no longer
manifest, i.e., it is hidden. 
   Selecting one of the ground states has led to a spontaneous symmetry
breaking which is always related to the existence of several degenerate
vacua.   

\begin{figure}
\begin{center}
\epsfig{file=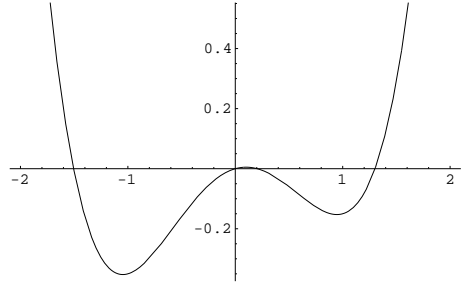,width=6cm}
\caption{\label{3:1:potngsb} Potential with a small odd component:
${\cal V}(x)=x/10-x^2/2+x^4/4$.}
\end{center}
\end{figure}

   At this stage it is not clear why the quantum mechanical 
ground state should be one or the other of $|0,\pm\rangle$ and not a 
superposition of both.
   For example, the linear combination 
$$\frac{1}{\sqrt{2}}\left(|0,+\rangle+|0,-\rangle\right)$$
is invariant under ${\cal R}$ as is the original Lagrangian
of Eq.\ (\ref{3:1:lphi}).
   However, this superposition is not stable against any infinitesimal
external perturbation which is odd in $\Phi$ (see Fig.\ \ref{3:1:potngsb}),
$${\cal R}(\epsilon H'){\cal R}^\dagger =-\epsilon H'.$$
   Any such perturbation will drive the ground state into the vicinity of 
either $|0,+\rangle$ or $|0,-\rangle$ rather than  
$\frac{1}{\sqrt{2}}(|0,+\rangle\pm|0,-\rangle)$.
   This can easily be seen in the framework of perturbation theory for
degenerate states.
   Consider 
$$ |1\rangle=\frac{1}{\sqrt{2}}(|0,+\rangle+|0,-\rangle),\quad
|2\rangle=\frac{1}{\sqrt{2}}(|0,+\rangle-|0,-\rangle),$$ 
such that
$${\cal R}|1\rangle=|1\rangle\quad {\cal R}|2\rangle=-|2\rangle.$$
   The condition for the energy eigenvalues of the ground state,
$E=E^{(0)}+\epsilon E^{(1)}+\cdots$, to first order in $\epsilon$
results from
$$
\mbox{det}\left(\begin{array}{cc}
\langle1|H'|1\rangle-E^{(1)}&\langle1|H'|2\rangle\\\
\langle2|H'|1\rangle&\langle2|H'|2\rangle-E^{(1)}\end{array}\right)
=0.$$      
   Due to the symmetry properties of Eq.\ (\ref{3:1:rvpm}), we obtain
$$\langle 1|H'|1\rangle
=\langle 1|{\cal R}^{-1}{\cal R}H'{\cal R}^{-1}{\cal R}|1\rangle=
\langle 1|-H'|1\rangle=0$$
and similarly $\langle2|H'|2\rangle=0$.
   Setting $\langle1|H'|2\rangle=a>0$, which can always be achieved by 
multiplication of one of the two states by an appropriate phase, one finds
$$\langle2|H'|1\rangle
\stackrel{H'=H'^\dagger}{=}
\langle1|H'|2\rangle^\ast=a^\ast=a=\langle1|H'|2\rangle,$$ 
resulting in
$$\mbox{det}\left(\begin{array}{cc}-E^{(1)}&a\\a&-E^{(1)}\end{array}\right)
={E^{(1)}}^2-a^2\stackrel{!}{=}0,\quad \Rightarrow\quad
E^{(1)}_{1/2}=\pm a.$$
   In other words, the degeneracy has been lifted and we get for the
energy eigenvalues
\begin{equation}
\label{3:1:eew}
E_{1/2}=E^{(0)}\pm\epsilon a+\cdots.
\end{equation}
   The corresponding eigenstates of zeroth order in $\epsilon$ are
$|0,+\rangle$ and $|0,-\rangle$, respectively.
   We thus conclude that an arbitrarily small external perturbation
which is odd with respect to $R$ will push the ground state
to either $|0,+\rangle$ or $|0,-\rangle$.

   In the above discussion, we have tacitly assumed that the Hamiltonian
and the field $\Phi(x)$ can simultaneously be diagonalized in the vacuum
sector, i.e.\ $\langle 0,+|0,-\rangle =0$.
   Following Ref.\ \cite{Weinberg:kr}, we will justify
this assumption which will also be crucial for the continuous case to be
discussed later. 
\begin{figure}
\begin{center}
\epsfig{file=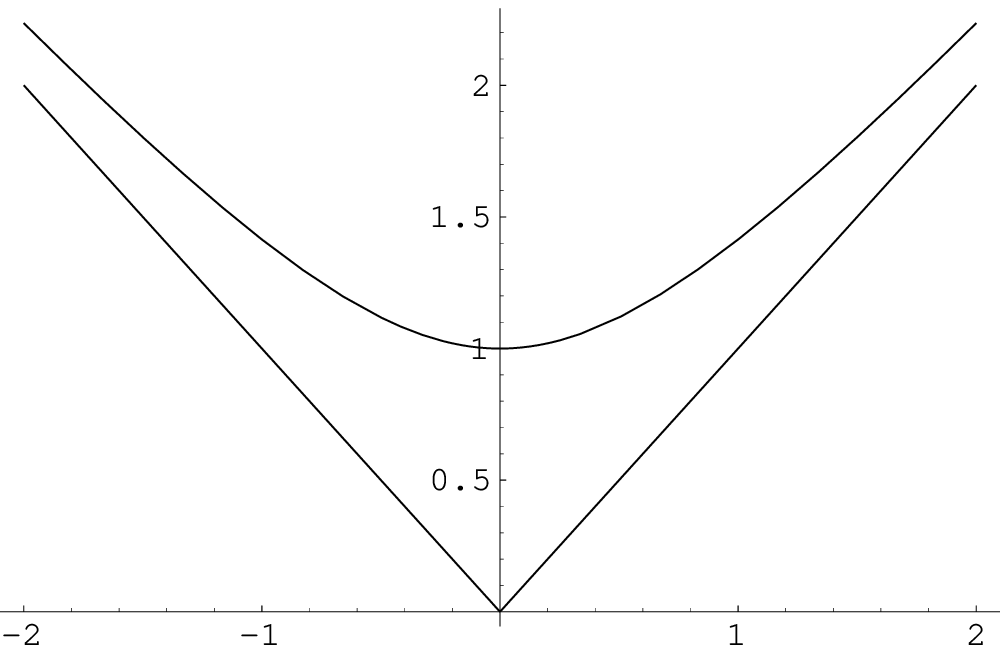,width=6cm}
\caption{\label{3:1:energie}}
Dispersion relation $E=\sqrt{1+\vec{p}\,^2}$ and asymptote
$E=|\vec{p}\,|$. 
\end{center}
\end{figure}
 
 For an infinite volume, a general vacuum state $|v\rangle$ is defined
as a state with momentum eigenvalue $\vec{0}$,
   $$\vec{P}|v\rangle=\vec{0},$$
where $\vec{0}$ is a {\em discrete} eigenvalue as opposed
to an eigenvalue of single- or many-particle states for which
$\vec{p}=0$ is an element of a continuous spectrum 
(see Fig.\ \ref{3:1:energie}).
   We deal with the situation of several degenerate ground states 
which will be denoted by $|u\rangle$, $|v\rangle$, 
{\em etc}\footnote{For continuous 
symmetry groups one may have a non-countably infinite number of ground states.}
and start from the identity
\begin{equation}
\label{3:1:0ucv}
0=\langle u|[H,\Phi(x)]|v\rangle\,\,\forall\,\,x,
\end{equation}
from which we obtain for $t=0$
\begin{eqnarray}
\label{3:1:0uvh}
\int d^3 y \langle u|{\cal H}(\vec{y},0) \Phi(\vec{x},0)|v\rangle&=&
\int d^3 y \langle u|\Phi(\vec{x},0) {\cal H}(\vec{y},0)|v\rangle.
\end{eqnarray}
   Let us consider the left-hand side,
\begin{eqnarray*}
\lefteqn{\int d^3 y \langle u|{\cal H}(\vec{y},0) \Phi(\vec{x},0)|v\rangle
=\sum_w \langle u|H|w\rangle\langle w|\Phi(0)|v\rangle}\\
&&
+\int d^3 y\int d^3p\sum_n\langle u|{\cal H}(\vec{y},0)|n,\vec{p}\,\rangle
\langle n,\vec{p}\,|\Phi(0)|v\rangle e^{-i\vec{p}\cdot \vec{x}},
\end{eqnarray*}
where we inserted a complete set of states which we split 
into the vacuum contribution and the rest, and made use of translational
invariance.
       We now define
$$f_n(\vec{y},\vec{p}\,)=\langle u|{\cal H}(\vec{y},0)|n,\vec{p}\,\rangle
\langle n,\vec{p}\,|\Phi(0)|v\rangle$$
and assume $f_n$ to be reasonably behaved
such that one can apply the lemma of Riemann and Lebesgue,
$$\lim_{|\vec{x}|\to\infty}\int d^3p f(\vec{p}\,)e^{-i\vec{p}\cdot\vec{x}}=0.$$
   At this point the assumption of an infinite volume,
$|\vec{x}|\to\infty$, is crucial.
   Repeating the argument for the right-hand side and taking the limit
$|\vec{x}|\to\infty$, only the vacuum contributions survive in Eq.\
(\ref{3:1:0uvh}) and we obtain
$$\sum_w\langle u|H|w\rangle\langle w|\Phi(0)|v\rangle
=\sum_w\langle u|\Phi(0)|w\rangle\langle w|H|v\rangle
$$ 
for arbitrary ground states $|u\rangle$ and $|v\rangle$.
   In other words, the matrices $(H_{uv})\equiv (\langle u|H|v\rangle)$
and $(\Phi_{uv})\equiv(\langle u|\Phi(0)|v \rangle)$ commute and can be 
diagonalized simultaneously.
   Choosing an appropriate basis, one can write
$$\langle u|\Phi(0)|v\rangle=\delta_{uv}v,\quad v\in R,$$
where $v$ denotes the expectation value of $\Phi$ in the state $|v\rangle$.

   In the above example, the ground states $|0,+\rangle$ and $|0,-\rangle$ 
with vacuum expectation values $\pm \Phi_0$ are thus indeed orthogonal and 
satisfy $$\langle 0,+|H|0,-\rangle=\langle 0,-|H|0,+\rangle=0.$$

\section{Spontaneous Breakdown of a Global, Continuous,
Non-Abelian Symmetry}
\label{sec_sbgcnas}
\begin{figure}
\begin{center}
\epsfig{file=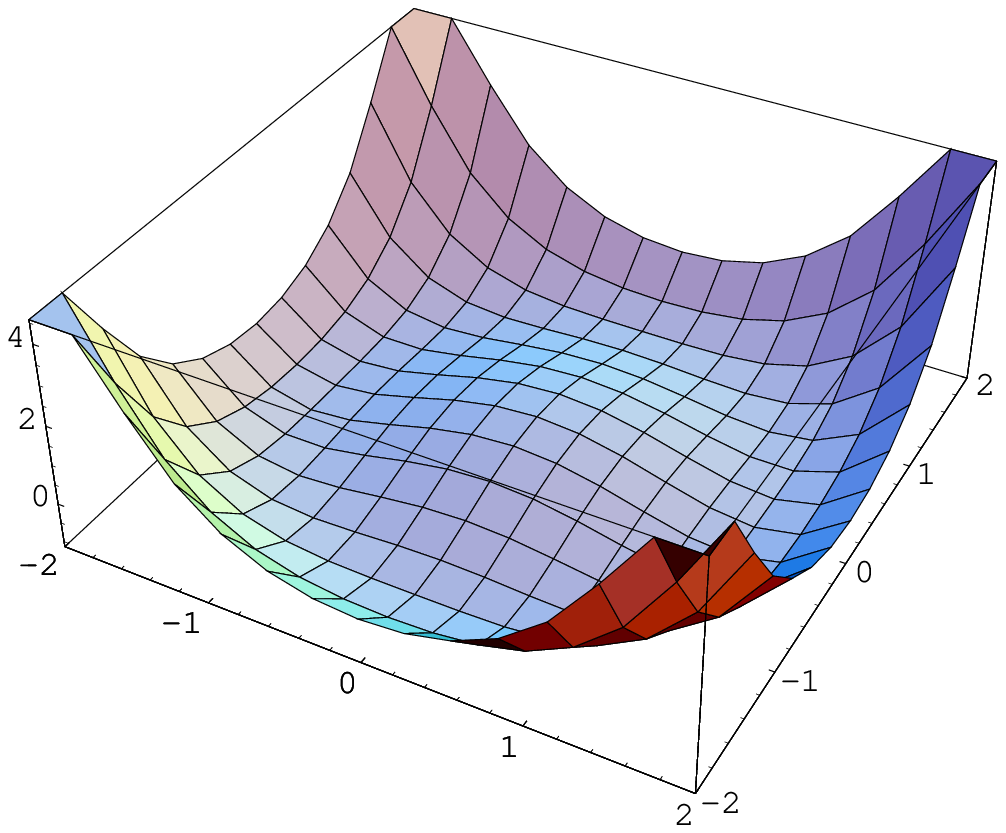,width=6cm}
\caption{\label{3:2:pot2dim}}
Two-dimensional rotationally invariant potential:
${\cal V}(x,y)=-(x^2+y^2)+\frac{(x^2+y^2)^2}{4}$.
\end{center}
\end{figure}

   We now extend the discussion to a system with a continuous, non-Abelian
symmetry such as SO(3).
   To that end, we consider the Lagrangian
\begin{eqnarray}
\label{3:2:lphi}
{\cal L}(\vec{\Phi},\partial_\mu\vec{\Phi})
&=&{\cal L}(\Phi_1,\Phi_2,\Phi_3,\partial_\mu\Phi_1,
\partial_\mu\Phi_2,\partial_\mu\Phi_3)\nonumber\\
&=&\frac{1}{2}\partial_\mu \Phi_i\partial^\mu \Phi_i-\frac{m^2}{2}\Phi_i\Phi_i
-\frac{\lambda}{4}(\Phi_i\Phi_i)^2,
\end{eqnarray}
where $m^2<0$, $\lambda>0$, with Hermitian fields $\Phi_i$.
The Lagrangian of Eq.\ (\ref{3:2:lphi}) is invariant under a global 
``isospin'' rotation,\footnote{Of course, the Lagrangian is invariant under
the full group O(3) which can be decomposed into its two components: the
proper rotations connected to the identity, SO(3), and the 
rotation-reflections. For our purposes it is sufficient to 
discuss SO(3).}
\begin{equation}
\label{3:2:phitrafo}
g\in \mbox{SO(3)}:\,\,\Phi_i\to\Phi_i'=D_{ij}(g)\Phi_j=
(e^{-i\alpha_k T_k})_{ij}\Phi_j.
\end{equation}
   For the $\Phi_i'$ to also be Hermitian, the Hermitian $T_k$ must be purely
imaginary and thus antisymmetric.
   The $iT_k$ provide the basis of a representation of the so(3) Lie algebra
and satisfy the commutation relations $[T_i,T_j]=i\epsilon_{ijk} T_k$.
   We will use the representation with the matrix elements 
given by $t_{jk}^i=-i\epsilon_{ijk}$.
   As in Sec.\ 3.1, we now look for a minimum of the potential which does 
not depend on $x$ and find
\begin{equation}
\label{3:2:phimin}
|\vec{\Phi}_{\rm min}|=\sqrt{\frac{-m^2}{\lambda}}\equiv v,
\quad |\vec{\Phi}|=\sqrt{\Phi_1^2+\Phi_2^2+\Phi_3^2}.
\end{equation}
   Since $\vec{\Phi}_{\rm min}$ can point in any direction in isospin
space we now have a non-countably infinite number of degenerate
vacua.
   In analogy to the discussion of the last section, any infinitesimal 
external perturbation which is not invariant under SO(3) will select a 
particular direction which, by an appropriate orientation of the internal 
coordinate frame, we denote as the 3 direction, 
\begin{equation}
\label{3:2:phimin3}
\vec{\Phi}_{\rm min}=v \hat{e}_3.
\end{equation}
   Clearly, $\vec{\Phi}_{\rm min}$ of Eq.\ (\ref{3:2:phimin3}) is {\em not}
invariant under the full group $G=\mbox{SO(3)}$ since rotations about
the 1 and 2 axis change $\vec{\Phi}_{\rm min}$.\footnote{We say, somewhat
loosely, that $T_1$ and $T_2$ do not annihilate the ground state or,
 equivalently, finite group elements generated by $T_1$ and $T_2$ do not 
leave the ground state invariant. This should become clearer later on.}
   To be specific, if 
$$
\vec{\Phi}_{\rm min}=v\left(\begin{array}{r}0\\0\\1\end{array}\right),
$$
we obtain
\begin{equation}
\label{3:2:t12phimin}
T_1 \vec{\Phi}_{\rm min}=
v\left(\begin{array}{r}0\\-i\\0\end{array}\right),\quad
T_2 \vec{\Phi}_{\rm min}=
v\left(\begin{array}{r}i\\0\\0\end{array}\right),
\quad 
T_3 \vec{\Phi}_{\rm min}=0.
\end{equation}
   Note that the set of transformations which do not leave 
$\vec{\Phi}_{\rm min}$ invariant does {\em not} form a group, because it does 
not contain the identity.
   On the other hand, $\vec{\Phi}_{\rm min}$ is invariant under a 
subgroup $H$ of $G$, namely, the rotations about the 3 axis:
\begin{equation}
\label{3:2:phimintrafoh}
h\in H:\quad \vec{\Phi}'=D(h)\vec{\Phi}=e^{-i\alpha_3 T_3}\vec{\Phi},
\quad
D(h)\vec{\Phi}_{\rm min}=\vec{\Phi}_{\rm min}.
\end{equation}
   In analogy to Eq.\ (\ref{3:1:entw}), we expand $\Phi_3$ with respect
to $v$,
\begin{equation}
\label{3:2:entw}
\Phi_3=v+\eta,
\end{equation}
where $\eta(x)$ is a new field replacing $\Phi_3(x)$,
and obtain the new expression for the potential
\begin{eqnarray}
\label{3:2:ventw}
\tilde{\cal V}
&=&\frac{1}{2}(-2m^2)\eta^2 
+\lambda v\eta (\Phi_1^2+\Phi_2^2+\eta^2)
+\frac{\lambda}{4}(\Phi_1^2+\Phi_2^2+\eta^2)^2-\frac{\lambda}{4}v^4.
\nonumber\\
\end{eqnarray}
   Upon inspection of the terms quadratic in the fields, one finds 
after spontaneous symmetry breaking two massless Goldstone bosons and one 
massive boson:
\begin{eqnarray}
\label{3:2:masses}
m_{\Phi_1}^2=m_{\Phi_2}^2&=&0,\nonumber\\
m_\eta^2&=&-2m^2.
\end{eqnarray}
   The model-independent feature of the above example is given by the
fact that for each of the two generators $T_1$ and $T_2$ which do not
annihilate the ground state one obtains a {\em massless} Goldstone
boson.
   By means of a two-dimensional simplification (see the ``Mexican hat''
potential shown in Fig.\ \ref{3:2:pot2dim}) the mechanism at hand can easily 
be visualized.
    Infinitesimal variations orthogonal to the circle of the minimum
of the potential generate quadratic terms, i.e., ``restoring forces 
linear in the displacement,'' whereas tangential variations experience
restoring forces only of higher orders.

   Now let us generalize the model to the case of an arbitrary compact
Lie group $G$ of order $n_G$ resulting in $n_G$ infinitesimal 
generators.\footnote{The restriction to compact groups 
allows for a complete decomposition into finite-dimensional irreducible 
unitary representations.}
   Once again, we start from a Lagrangian of the form \cite{Goldstone:es}
\begin{equation}
\label{3:2:lallg}
{\cal L}(\vec{\Phi},\partial_\mu\vec{\Phi})
=\frac{1}{2}\partial_\mu \vec{\Phi}\cdot \partial^\mu
\vec{\Phi}- {\cal V}(\vec{\Phi}),
\end{equation}
where $\vec{\Phi}$ is a multiplet of scalar (or pseudoscalar)
Hermitian fields.
   The Lagrangian ${\cal L}$ and thus also ${\cal V}(\vec{\Phi})$ are 
supposed to be globally invariant under $G$, where the infinitesimal 
transformations of the fields are given by
\begin{equation}
\label{3:2:symmtrans}
g\in G:\quad \Phi_i\to\Phi_i+\delta\Phi_i,\quad
\delta \Phi_i=-i\epsilon_a t^a_{ij}\Phi_j.
\end{equation}
   The Hermitian representation matrices $T^a=(t^a_{ij})$ 
are again antisymmetric and purely imaginary.
   We now assume that, by choosing an appropriate form of ${\cal V}$,
the Lagrangian generates a spontaneous symmetry breaking resulting in
a ground state with a vacuum expectation
value $\vec{\Phi}_{\rm min}=\langle\vec{\Phi}\rangle$ 
which is invariant under a continuous subgroup $H$ of $G$.
   We expand 
${\cal V}(\vec{\Phi})$ with respect to  $\vec{\Phi}_{\rm min}$,
$|\vec{\Phi}_{\rm min}|=v$, i.e.,  
$\vec{\Phi}=\vec{\Phi}_{\rm min}+\vec{\chi}$,
\begin{equation}
\label{3:2:venta}
{\cal V}(\vec{\Phi})={\cal V}(\vec{\Phi}_{\rm min})
+\underbrace{\frac{\partial {\cal V}(\vec{\Phi}_{\rm min})}{\partial \Phi_i}}_{
\mbox{$0$}}\chi_i
+\frac{1}{2}\underbrace{\frac{\partial^2 {\cal V}(\vec{\Phi}_{\rm 
min})}{\partial
\Phi_i\partial \Phi_j}}_{\mbox{$m^2_{ij}$}}\chi_i\chi_j+\cdots.
\end{equation}
  The matrix $M^2=(m^2_{ij})$ must be symmetric and, since one is expanding 
around a minimum,  positive semidefinite, i.e.,
\begin{equation}
\label{3:2:m2}
\sum_{i,j}m^2_{ij}x_i x_j\ge 0\quad \forall \quad \vec{x}.
\end{equation}
   In that case, all eigenvalues of $M^2$ are nonnegative.
   Making use of the invariance of  ${\cal V}$ under the symmetry group
$G$,
\begin{eqnarray}
\label{3:2:vinv}
{\cal V}(\vec{\Phi}_{\rm min})&=&{\cal V}(D(g)\vec{\Phi}_{\rm min})
={\cal V}(\vec{\Phi}_{\rm min}+\delta\vec{\Phi}_{\rm min})\nonumber\\
&\stackrel{\mbox{(\ref{3:2:venta})}}{=}&
{\cal V}(\vec{\Phi}_{\rm min})+\frac{1}{2}m^2_{ij}\delta\Phi_{\rm min,i}
\delta{\Phi}_{\rm min,j}+\cdots,
\end{eqnarray}
one obtains, by comparing coefficients,
\begin{equation}
\label{3:2:kv}
m^2_{ij}\delta\Phi_{\rm min,i}\delta\Phi_{\rm min,j}=0.
\end{equation}
   Differentiating Eq.\ (\ref{3:2:kv}) with respect to 
$\delta\Phi_{\rm min,k}$ and using $m^2_{ij}=m^2_{ji}$
results in the matrix equation
\begin{equation}
\label{3:2:rel1}
M^2\delta\vec{\Phi}_{\rm min}=\vec{0}.
\end{equation}
   Inserting the variations of Eq.\ (\ref{3:2:symmtrans}) for arbitrary
$\epsilon_a$, 
$\delta\vec{\Phi}_{\rm min}=-i\epsilon_a T^a\vec{\Phi}_{\rm min}$,
we conclude 
\begin{equation}
\label{3:2:result}
M^2T^a \vec{\Phi}_{\rm min}=\vec{0}.
\end{equation}
   The solutions of Eq.\ (\ref{3:2:result}) can be classified into
two categories:
\begin{enumerate}
\item $T^a$, $a=1,\cdots, n_H$, is a representation of an element of
the Lie algebra belonging to the subgroup $H$ of $G$, leaving
the selected ground state invariant. 
   In that case one has
$$ T^a \vec{\Phi}_{\rm min}=\vec{0}, \quad a=1,\cdots,n_H,$$
such that Eq.\ (\ref{3:2:result}) is automatically satisfied without
any knowledge of $M^2$.

\item $T^a$, $a=n_H+1,\cdots, n_G,$ is {\em not} a representation of an 
element of the Lie algebra belonging to the subgroup $H$. In that case
$T^a\vec{\Phi}_{\rm min}\neq\vec{0}$, and $T^a\vec{\Phi}_{\rm min}$ is
an eigenvector of $M^2$ with eigenvalue 0.
   To each such eigenvector corresponds a massless Goldstone boson.
   In particular, the different $T^a\vec{\Phi}_{\rm min}\neq \vec{0}$ are
linearly independent, resulting in $n_G-n_H$ independent
Goldstone bosons.
   (If they were not linearly independent, there would exist a nontrivial
linear combination
$$\vec{0}=\sum_{a=n_H+1}^{n_G}c_a (T^a\vec{\Phi}_{\rm min})=
\underbrace{\left(\sum_{a=n_H+1}^{n_G}c_a T^a\right)}_{\mbox{$:=T$}}
\vec{\Phi}_{\rm min},
$$
such that $T$ is an element of the Lie algebra of $H$ in contradiction 
to our assumption.)
\end{enumerate}
   Let us check these results by reconsidering the example of 
Eq.\ (\ref{3:2:lphi}). 
   In that case $n_G=3$ and $n_H=1$, generating 2 Goldstone bosons
[see Eq.\ (\ref{3:2:masses})].

   We conclude this section with two remarks. First, the number of
Goldstone bosons is determined by the structure of the symmetry groups.
   Let $G$ denote the symmetry group of the Lagrangian, with $n_G$ generators
and $H$ the subgroup with $n_H$ generators 
which leaves the ground state after spontaneous symmetry
breaking invariant.
   For each generator which does not annihilate the vacuum one obtains
a massless Goldstone boson, i.e., the total number of
Goldstone bosons equals $n_G-n_H$.
   Second, the Lagrangians used in {\em motivating} the phenomenon
of a spontaneous symmetry breakdown are typically constructed in such a 
fashion that the degeneracy of the ground states is built into
the potential at the classical level (the prototype being the ``Mexican hat''
potential of Fig.\ \ref{3:2:pot2dim}).
   As in the above case, it is then argued that an 
{\em elementary} Hermitian field of a multiplet transforming non-trivially
under the symmetry group $G$ acquires a vacuum expectation
value signaling a spontaneous symmetry breakdown.
   However, there also exist theories such as QCD where one cannot infer
from inspection of the Lagrangian whether the theory exhibits spontaneous
symmetry breaking.
   Rather, the criterion for spontaneous symmetry breaking is a non-vanishing
vacuum expectation value of some Hermitian operator, not an elementary field,
which is generated through the dynamics of the underlying theory.
   In particular, we will see that the quantities developing a vacuum
expectation value may also be local Hermitian operators composed of more 
fundamental degrees of freedom of the theory.
   Such a possibility was already emphasized in the derivation of Goldstone's
theorem in Ref.\ \cite{Goldstone:es}.

\section{Goldstone's Theorem}
\label{sec_gt}
   By means of the above example, we motivate another approach to 
Goldstone's theorem without delving into all the subtleties of
a quantum field-theoretical approach \cite{Bernstein}. 
   Given a Hamilton operator with a global symmetry group $G=\mbox{SO(3)}$,
let $\vec{\Phi}(x)=(\Phi_1(x),\Phi_2(x),\Phi_3(x))$ denote a triplet
of local Hermitian operators transforming as a vector
under $G$,
\begin{eqnarray}
\label{3:3:phitrafo}
g\in G:&&\vec{\Phi}(x)\mapsto \vec{\Phi}'(x)=
e^{i\sum_{k=1}^3\alpha_k Q_k}\vec{\Phi}(x)e^{-i\sum_{l=1}^3\alpha_l Q_l}
\nonumber\\
&&=e^{-i\sum_{k=1}^3 \alpha_k T_k}
\vec{\Phi}(x)
\neq \vec{\Phi}(x),
\end{eqnarray}
where the $Q_i$ are the generators of the SO(3) transformations on the Hilbert
space satisfying $[Q_i,Q_j]=i\epsilon_{ijk}Q_k$
and the $T_i=(t^i_{jk})$ are the matrices of the three dimensional 
representation satisfying $t^i_{jk}=-i\epsilon_{ijk}$.
   We assume that one component of the multiplet acquires a non-vanishing
vacuum expectation value:
\begin{equation}
\label{3:3:phi3vac}
\langle0|\Phi_1(x)|0\rangle
=\langle0|\Phi_2(x)|0\rangle=0,\quad \langle0|\Phi_3(x)|0\rangle=v\neq 0.
\end{equation}
  Then the two generators $Q_1$ and $Q_2$ do not annihilate the ground state,
and to each such generator corresponds a massless Goldstone boson.

   In order to prove these two statements 
let us expand Eq.\ (\ref{3:3:phitrafo}) 
to first order in the $\alpha_k$:
$$\vec{\Phi}'=\vec{\Phi}+i\sum_{k=1}^3\alpha_k[Q_k,\vec{\Phi}]
=(1-i\sum_{k=1}^3\alpha_k T_k)\vec{\Phi}
=\vec{\Phi}+\vec{\alpha}\times\vec{\Phi}.
$$
   Comparing the terms linear in the  $\alpha_k$ 
$$i[\alpha_k Q_k,\Phi_l]=\epsilon_{lkm}\alpha_k\Phi_m$$
and noting that all
three $\alpha_k$ can be chosen independently, we obtain
$$i[Q_k,\Phi_l]=-\epsilon_{klm}\Phi_m,$$
   which, of course, simply expresses the fact that the field operators 
$\Phi_i$ transform as a vector.
   Using $\epsilon_{klm}\epsilon_{kln}=2\delta_{mn}$, we find
$$-\frac{i}{2}\epsilon_{kln}[Q_k,\Phi_l]=\delta_{mn}\Phi_m=\Phi_n.$$
   In particular, 
\begin{equation}
\label{3:3:phi3zykl}
\Phi_3=-\frac{i}{2}([Q_1,\Phi_2]-[Q_2,\Phi_1]),
\end{equation}
with cyclic permutations for the other two cases.
   
   In order to prove that $Q_1$ and $Q_2$ do not annihilate the ground
state, let us consider Eq.\ (\ref{3:3:phitrafo})
for $\vec{\alpha}=(0,\pi/2,0)$,
\begin{eqnarray*}
e^{-i\frac{\pi}{2} T_2}\vec{\Phi}&=&
\left(\begin{array}{ccc} \cos(\pi/2)&0&\sin(\pi/2)\\
0&1&0\\
-\sin(\pi/2)&0&\cos(\pi/2)\end{array}\right)
\left(\begin{array}{c}
\Phi_1\\ \Phi_2 \\ \Phi_3\end{array}\right)
=\left(\begin{array}{c}\Phi_3\\ \Phi_2 \\ -\Phi_1\end{array}\right)\\
&=&e^{i\frac{\pi}{2}Q_2}\left(\begin{array}{c}
\Phi_1\\ \Phi_2 \\ \Phi_3\end{array}\right)
e^{-i\frac{\pi}{2}Q_2}.
\end{eqnarray*}
   From the first row we obtain
$$\Phi_3=e^{i\frac{\pi}{2}Q_2}\Phi_1 e^{-i\frac{\pi}{2}Q_2}.$$
   Taking the vacuum expectation value    
$$v=\langle 0| e^{i\frac{\pi}{2}Q_2}\Phi_1 e^{-i\frac{\pi}{2}Q_2}|0\rangle
$$
   and using Eq.\ (\ref{3:3:phi3vac}) clearly $Q_2|0\rangle\neq 0$,
since otherwise the exponential operator could be replaced by unity
and the right-hand side would vanish.
   A similar argument shows $Q_1|0\rangle\neq 0$.

   At this point let us make two remarks.
   The ``states''  $Q_{1(2)}|0\rangle$ cannot be normalized.
In a more rigorous derivation  one makes use of integrals of the form
$$\int d^3 x \langle0|[J^{0,b}(\vec{x},t),\Phi_c(0)]|0\rangle,$$
and first determines the commutator before evaluating the integral
\cite{Bernstein}.
   Some derivations of Goldstone's theorem right away start by
assuming $Q_{1(2)}|0\rangle$ $\neq 0$.
   However, for the discussion of spontaneous symmetry breaking in
the framework of QCD it is advantageous to establish the connection
between the existence of Goldstone bosons and a non-vanishing
expectation value.

   Let us now turn to the existence of Goldstone bosons,
taking the vacuum expectation value of Eq.\ (\ref{3:3:phi3zykl})
$$0\neq v=\langle0|\Phi_3(0)|0\rangle
=
-\frac{i}{2}\langle0|\left([Q_1,\Phi_2(0)]-[Q_2,\Phi_1(0)]\right)|0\rangle
\equiv -\frac{i}{2}(A-B).
$$
   We will first show $A=-B$. To that end we perform a rotation of the
fields as well as the generators by $\pi/2$ about the 3 axis
[see Eq.\ (\ref{3:3:phitrafo}) with $\vec{\alpha}=(0,0,\pi/2)$]:
\begin{eqnarray*}
e^{-i\frac{\pi}{2}T_3}\vec{\Phi}
&=&\left(\begin{array}{r}-\Phi_2\\ \Phi_1\\ \Phi_3\end{array}\right)=
e^{i\frac{\pi}{2}Q_3}
\left(\begin{array}{c}\Phi_1\\ \Phi_2\\ \Phi_3\end{array}\right)
e^{-i\frac{\pi}{2}Q_3},
\end{eqnarray*}
and analogously for the charge operators
$$
\left(\begin{array}{r} -Q_2\\Q_1\\Q_3\end{array}\right)
=e^{i\frac{\pi}{2}Q_3}\left(\begin{array}{r}Q_1\\Q_2\\Q_3\end{array}
\right)e^{-i\frac{\pi}{2}Q_3}.
$$
   We thus obtain
\begin{eqnarray*}
B=\langle0|[Q_2, \Phi_1(0)]|0\rangle&=& 
\langle0|\Big(e^{i\frac{\pi}{2}Q_3}(- Q_1)
\underbrace{e^{-i\frac{\pi}{2}Q_3}e^{i\frac{\pi}{2}Q_3}}_{\mbox{1}}
\Phi_2(0) e^{-i\frac{\pi}{2}Q_3}\\
&&- e^{i\frac{\pi}{2}Q_3}\Phi_2(0)
e^{-i\frac{\pi}{2}Q_3}e^{i\frac{\pi}{2}Q_3}
(-Q_1) e^{-i\frac{\pi}{2}Q_3}\Big)|0\rangle\\
&=&-\langle0|[Q_1,\Phi_2(0)]|0\rangle=-A,
\end{eqnarray*}
where we made use of $Q_3|0\rangle=0$, i.e., the vacuum is invariant
under rotations about the 3 axis.
   In other words, the non-vanishing vacuum expectation value $v$ 
can also be written as
\begin{eqnarray}
\label{3:3:vq1phi2}
0\neq v&=&\langle0|\Phi_3(0)|0\rangle=-i\langle0|[Q_1,\Phi_2(0)]|0\rangle
\nonumber\\
&=&-i\int d^3 x\langle0|[J^1_0(\vec{x},t),\Phi_2(0)]|0\rangle.
\end{eqnarray}
   We insert a complete set of states $1=\sum_n\hspace{-1.4em}\int
\hspace{0.5em} |n\rangle
\langle n|$ into the commutator\footnote{The abbreviation 
$\sum_n\hspace{-1.4em}\int\hspace{0.5em} |n\rangle\langle n|$
includes an integral over the total momentum $\vec{p}$ as well
as all other quantum numbers necessary to fully specify the states.} 
\begin{eqnarray*}
v&=&-i\sum_n\hspace{-1.1em}\int 
\int d^3x
\left(\langle0|J^{1}_0(\vec{x},t)|n\rangle
\langle n|\Phi_2(0)|0\rangle-\langle0|\Phi_2(0)|n\rangle
\langle n|J^{1}_0(\vec{x},t)|0\rangle\right),
\end{eqnarray*}
and make use of translational invariance
\begin{eqnarray*}
&=&-i\sum_n\hspace{-1.1em}\int \int d^3x\left(e^{-iP_n\cdot x}
\langle0|J^{1}_0(0)|n\rangle\langle n|\Phi_2(0)|0\rangle
-\cdots\right)\\
&=&-i\sum_n\hspace{-1.1em}\int (2\pi)^3\delta^3(\vec{P}_n)
\left(e^{-iE_n t}
\langle0|J^{1}_0(0)|n\rangle\langle n|\Phi_2(0)|0\rangle\right.\nonumber\\
&&\left.-e^{iE_n t}\langle 0|\Phi_2(0)|n\rangle
\langle n|J^{1}_0(0)|0\rangle\right).
\end{eqnarray*}
Integration with respect to the momentum of the inserted intermediate
states yields an expression of the form
$$
=-i(2\pi)^3 \sum_n'
 \left(e^{-iE_n t}\cdots
-e^{iE_n t}\cdots\right),
$$
   where the prime indicates that only states with $\vec{P}=0$
need to be considered.
   Due to the Hermiticity of the symmetry current operators 
$J^{\mu,a}$ as well as the $\Phi_l$, we have
$$c_n:=\langle 0|J^{1}_0(0)|n\rangle\langle n|\Phi_2(0)|0\rangle
=\langle n|J^{1}_0(0)|0\rangle^\ast \langle0|\Phi_2(0)|n\rangle^\ast,$$
such that 
\begin{equation}
\label{3:3:vresult}
v=-i(2\pi)^3\sum_n'
 \left(c_n e^{-iE_n t}-c_n^\ast e^{iE_n t}\right).
\end{equation}
   From Eq.\ (\ref{3:3:vresult}) we draw the following conclusions.
\begin{enumerate}
\item Due to our assumption of a non-vanishing vacuum expectation value $v$,
there must exist states $|n\rangle$ for which both 
$\langle0|J^{0}_{1(2)}(0)|n\rangle$ and $\langle n|\Phi_{1(2)}(0)|0\rangle$
do not vanish.
   The vacuum itself cannot contribute to Eq.\ (\ref{3:3:vresult})
because $\langle0|\Phi_{1(2)}(0)|0\rangle=0$.
\item States with $E_n>0$ contribute ($\varphi_n$ is the phase of $c_n$)
\begin{eqnarray*}
\frac{1}{i}\left(c_n e^{-iE_n t}-c_n^\ast e^{iE_n t}\right)
&=&\frac{1}{i}|c_n|\left(e^{i\varphi_n}e^{-iE_n t}
-e^{-i\varphi_n}e^{iE_n t}\right)\\
&=&2|c_n|\sin(\varphi_n-E_n t)
\end{eqnarray*}
to the sum.
   However, $v$ is time-independent and therefore the sum over states
with $(E_n>0,\vec{0})$ must vanish.
\item The right-hand side of Eq.\ (\ref{3:3:vresult}) must therefore 
contain the contribution from states with zero energy as well as
zero momentum thus zero mass.
   These zero-mass states are the Goldstone bosons.
\end{enumerate}
 
\section{Explicit Symmetry Breaking: A First Look}
\label{sec_esbfl} 
  Finally, let us illustrate the consequences of adding to our Lagrangian
of Eq.\ (\ref{3:2:lphi}) a small perturbation which {\em explicitly} breaks
the symmetry.
   To that end, we modify the potential of Eq.\ (\ref{3:2:lphi}) 
by adding a term $a\Phi_3$,
\begin{equation}
\label{3:4:pot}
{\cal V}(\Phi_1,\Phi_2,\Phi_3)=       
\frac{m^2}{2}\Phi_i\Phi_i
+\frac{\lambda}{4}(\Phi_i\Phi_i)^2 + a\Phi_3,
\end{equation}
where $m^2<0$, $\lambda>0$, and $a>0$, with Hermitian fields $\Phi_i$.
   Clearly, the potential no longer has the original O(3) symmetry but 
is only invariant under O(2).
   The conditions for the new minimum, obtained from 
$\vec{\nabla}_\Phi {\cal V}=0$, read
\begin{displaymath}
\Phi_1=\Phi_2=0,\quad \lambda \Phi_3^3+m^2\Phi_3+a=0.
\end{displaymath}
   Let us solve the cubic equation for $\Phi_3$ using the perturbative ansatz
\begin{equation}
\label{3:4:phi3ansatz}
\langle \Phi_3 \rangle=\Phi^{(0)}_3+a\Phi^{(1)}_3+{\cal O}(a^2),
\end{equation}
from which we obtain
\begin{displaymath}
\Phi^{(0)}_3=\pm \sqrt{-\frac{m^2}{\lambda}},\quad
\Phi^{(1)}_3=\frac{1}{2m^2}.
\end{displaymath}
   Of course, $\Phi^{(0)}_3$ corresponds to our result without explicit 
perturbation. 
   The condition for a {\em minimum} [see Eq.\ (\ref{3:2:m2})] excludes 
$\Phi^{(0)}_3=+\sqrt{-\frac{m^2}{\lambda}}$. 
   Expanding the potential with $\Phi_3=\langle\Phi_3\rangle +\eta$ we obtain,
after a short calculation, for the masses
\begin{eqnarray}
\label{3:4:masses}
m_{\Phi_1}^2=m_{\Phi_2}^2&=&a \sqrt{\frac{\lambda}{-m^2}},\nonumber\\
m_\eta^2&=&-2m^2+3 a\sqrt{\frac{\lambda}{-m^2}}.
\end{eqnarray}
   The important feature here is that the original Goldstone bosons 
of Eq.\ (\ref{3:2:masses}) are now massive.
   The squared masses are proportional to the symmetry breaking parameter $a$.
   Calculating {\em quantum} corrections to observables in terms of 
Goldstone-boson loop diagrams will generate corrections which are non-analytic
in the symmetry breaking parameter such as $a\ln(a)$ \cite{Li:1971vr}.
   Such so-called chiral logarithms originate from the mass terms in the 
Goldstone boson propagators entering the calculation of loop integrals.
   We will come back to this point in Chapter 4 when we discuss the masses 
of the pseudoscalar octet in terms of the quark masses which, in QCD, represent
the analogue to the parameter $a$ in the above example.

\chapter{Chiral Perturbation Theory for Mesons}
\label{chap_cptm}

   Chiral perturbation theory provides a systematic method for discussing
the consequences of the global flavor symmetries of QCD at low energies
by means of an {\em effective field theory}.
   The effective Lagrangian is expressed in terms of those hadronic 
degrees of freedom which, at low energies, show up as observable asymptotic 
states.
   At very low energies these are just the members of the pseudoscalar octet 
($\pi,K,\eta$)
which are regarded as the Goldstone bosons of the {\em spontaneous} breaking 
of the chiral $\mbox{SU(3)}_L\times\mbox{SU}(3)_R$ symmetry down to 
$\mbox{SU}(3)_V$.
   The non-vanishing masses of the light pseudoscalars in the ``real'' world
are related to the explicit symmetry breaking in QCD due to the light quark 
masses. 

   We will first consider the indications for a spontaneous breakdown of chiral
symmetry in QCD and then, in quite general terms, discuss the transformation 
properties of Goldstone bosons under the symmetry groups of the Lagrangian
and the ground state, respectively.
   This will lead us to the concept of a nonlinear realization of a symmetry.
   After introducing the lowest-order effective Lagrangian relevant to
the spontaneous breakdown from $\mbox{SU(3)}_L\times\mbox{SU}(3)_R$ to 
$\mbox{SU}(3)_V$, we will illustrate how Weinberg's power counting scheme 
allows for a systematic classification of Feynman diagrams in the so-called 
momentum expansion. 
   We will then outline the principles entering the construction of
the effective Lagrangian and discuss how, at lowest order, the
results of current algebra are reproduced.
   After presenting the Lagrangian of Gasser and Leutwyler and the
Wess-Zumino-Witten action we will discuss some applications at chiral order
${\cal O}(p^4)$.
   We will conclude the presentation of the mesonic sector with referring
to some selected examples at ${\cal O}(p^6)$.

\section{Spontaneous Symmetry Breaking in QCD}  
\label{sec_ssbqcd}
   While the toy model of Sec.\ \ref{sec_sbgcnas} by construction led to
a spontaneous symmetry breaking, it is not fully understood
theoretically why QCD should exhibit this phenomenon
\cite{Jaffe:2000}.
   We will first motivate why experimental input, the hadron spectrum 
of the ``real'' world, indicates that spontaneous symmetry breaking
happens in QCD.
   Secondly, we will show that a non-vanishing singlet scalar quark 
condensate is a sufficient condition for a spontaneous symmetry breaking in 
QCD.

\subsection{The Hadron Spectrum}
\label{subsec_hs}
   We saw in Sec.\ \ref{sec_agsl} that the QCD Lagrangian possesses a 
$\mbox{SU(3)}_L\times\mbox{SU(3)}_R\times \mbox{U(1)}_V$
symmetry in the chiral limit in which the light quark masses vanish.
   From symmetry considerations involving the Hamiltonian $H^0_{\rm QCD}$
only, one would naively expect that hadrons organize themselves into 
approximately degenerate multiplets fitting the dimensionalities 
of irreducible representations of the group 
$\mbox{SU(3)}_L\times\mbox{SU(3)}_R\times\mbox{U(1)}_V$.
   The $\mbox{U(1)}_V$ symmetry results in 
baryon number conservation\footnote{See Ref.\ \cite{Groom:in} for empirical
limits on nucleon decay as well as baryon-number violating
$Z$ and $\tau$ decays.} 
and leads to a classification of hadrons into mesons ($B=0$) and
baryons ($B=1$).
   The linear combinations $Q^a_V=Q^a_R+Q^a_L$ and $Q^a_A=Q^a_R-Q^a_L$
of the left- and right-handed charge operators commute with 
$H^0_{\rm QCD}$, have opposite parity, and thus for any state of positive 
parity one would expect the existence of a degenerate state of negative 
parity (parity doubling) which can be seen as follows.
   Let $|i,+\rangle$ denote an eigenstate of $H^0_{\rm QCD}$ with eigenvalue 
$E_i$, 
$$H^0_{\rm QCD}|i,+\rangle=E_i|i,+\rangle,$$ 
having positive parity,
$$P|i,+\rangle=+ |i,+\rangle,$$
such as, e.g., a member of the ground state baryon octet (in the chiral limit).
   Defining $|\phi\rangle= Q_A^a|i,+\rangle$, because of 
$[H^0_{\rm QCD},Q_A^a]=0$, we have
\begin{displaymath}
H^0_{\rm QCD}|\phi\rangle
=H^0_{\rm QCD} Q_A^a|i,+\rangle
= Q_A^a H^0_{\rm QCD}|i,+\rangle
= E_i Q_A^a|i,+\rangle
= E_i |\phi\rangle,
\end{displaymath}
i.e, the new state $|\phi\rangle$ is also an eigenstate of
$H^0_{\rm QCD}$ with the same eigenvalue $E_i$ but of
opposite parity:
$$P|\phi\rangle= PQ_A^a P^{-1} P|i,+\rangle=-Q_A^a(+|i,+\rangle)
=-|\phi\rangle.
$$
   The state $|\phi\rangle$ can be expanded in terms of the members of the 
multiplet with negative parity,
\begin{displaymath}
|\phi\rangle=Q^a_A|i,+\rangle=-t^a_{ij}|j,-\rangle.
\end{displaymath}
   However, the low-energy spectrum of baryons does not contain a degenerate
baryon octet of negative parity.
   Naturally the question arises whether the above chain of arguments is 
incomplete. 
   Indeed, we have tacitly assumed that the ground state of QCD is annihilated 
by $Q^a_A$.

   Let $a^\dagger_i$ symbolically denote an operator which creates quanta
with the quantum numbers of the state $|i,+\rangle$, whereas 
$b_j^\dagger$ creates degenerate quanta of opposite parity. 
   Let us assume the states $|i,+\rangle$ and $|j,-\rangle$ to be members of 
a basis of an irreducible representation of 
$\mbox{SU(3)}_L\times\mbox{SU(3)}_R$.
   In analogy to Eq.\ (\ref{2:3:qphi}), we assume that under 
$\mbox{SU(3)}_L\times\mbox{SU(3)}_R$ the creation operators are related
by
$$
[Q^a_A,a^\dagger_i]= -t^a_{ij} b_j^\dagger. 
$$
   The usual chain of arguments then works as 
\begin{eqnarray}
\label{4:1:pardoub}
Q^a_A|i,+\rangle&=&Q^a_A a^\dagger_i|0\rangle
=\Big([Q^a_A,a^\dagger_i]+a_i^\dagger \underbrace{Q_A^a}_{
\mbox{$\hookrightarrow 0$}}\Big)|0\rangle
= -t^a_{ij} b_j^\dagger |0\rangle.
\end{eqnarray}
   However, if the ground state is {\em not} annihilated by $Q_A^a$, the
reasoning of Eq.\ (\ref{4:1:pardoub}) does no longer apply.

   Two empirical facts about the hadron spectrum suggest that a spontaneous
symmetry breaking happens in the chiral limit of QCD.
   First, $\mbox{SU(3)}$ instead of $\mbox{SU(3)}_L\times\mbox{SU(3)}_R$
is approximately realized as a symmetry of the hadrons.
   Second, the octet of the pseudoscalar mesons is special in
the sense that the masses of its members are small in comparison with
the corresponding $1^-$ vector mesons.
   They are candidates for the Goldstone bosons of a spontaneous
symmetry breaking.

   In order to understand the origin of the SU(3) symmetry let us consider
the vector charges $Q^a_V=Q_R^a+Q^a_L$ [see Eq.\ (\ref{2:3:v})].
   They satisfy the commutation relations of an SU(3) Lie algebra
[see Eqs.\ (\ref{2:2:crqll}) - (\ref{2:3:crqlr})],
\begin{equation}
\label{4:1:su3v}
[Q_R^a+Q_L^a,Q_R^b+Q_L^b]=[Q_R^a,Q_R^b]+[Q_L^a,Q_L^b]
=if_{abc} Q_R^c+if_{abc}Q_L^c=if_{abc}Q^c_V.
\end{equation}
  In Ref.\ \cite{Vafa:tf} it was shown that, in the chiral limit, the
ground state is necessarily invariant under 
$\mbox{SU(3)}_V\times\mbox{U(1)}_V$, 
i.e., the eight vector charges $Q^a_V$ as well 
as the baryon number operator\footnote{Recall that each quark
is assigned a baryon number 1/3.} $Q_V/3$ annihilate the ground state,
\begin{equation}
Q^a_V|0\rangle =Q_V|0\rangle =0.
\end{equation}
   If the vacuum is invariant under $\mbox{SU(3)}_V\times\mbox{U(1)}_V$,
then so is the Hamiltonian \cite{Coleman:1966} (but not vice versa).
   Moreover, the invariance of the ground state {\em and} the Hamiltonian
implies that the physical states of the spectrum of $H^0_{\rm QCD}$ 
can be organized according to irreducible representations of 
$\mbox{SU(3)}_V\times\mbox{U(1)}_V$.   
   The index $V$ (for vector) indicates that the generators result from
integrals of the zeroth component of vector current operators and
thus transform with a positive sign under parity.

   Let us now turn to the linear combinations $Q^a_A=Q^a_R-Q^a_L$ 
satisfying the commutation relations 
[see Eqs.\ (\ref{2:2:crqll}) - (\ref{2:3:crqlr})]
\begin{eqnarray}
\label{4:1:crqaa}
[Q^a_A,Q^b_A]&=&[Q^a_R-Q^a_L,Q^b_R-Q^b_L]
=[Q^a_R,Q^b_R]+[Q^a_L,Q^b_L]\nonumber\\
&=&if_{abc}Q^c_R+if_{abc}Q^c_L=
if_{abc}Q^c_V,\nonumber\\
\label{4:1:crqva}
{[Q_V^a,Q^b_A]}&=&[Q_R^a+Q_L^a,Q_R^b-Q_L^b]=
[Q_R^a,Q^b_R]-[Q^a_L,Q^b_L]\nonumber\\
&=&if_{abc}Q^c_R-if_{abc}Q^c_L=
if_{abc}Q^c_A.
\end{eqnarray}
   Note that these charge operators do {\em not} form a closed algebra,
i.e., the commutator of two axial charge operators is not again an
axial charge operator.
   Since the parity doubling is not observed for the low-lying states,
one assumes that the $Q_A^a$ do {\em not} 
annihilate the ground state, 
\begin{equation}
\label{4:1:qav}
Q^a_A|0\rangle\neq 0,
\end{equation}
i.e., the ground state of QCD is not invariant under ``axial'' transformations.
   According to Goldstone's theorem \cite{Nambu:xd,Nambu:tp,Nambu:fr,%
Goldstone:eq,Goldstone:es},
to each axial generator $Q^a_A$, which does not annihilate the ground state,
corresponds a massless Goldstone boson field $\phi^a(x)$ with spin 0,
whose symmetry properties are tightly connected to the generator
in question.
   The Goldstone bosons have the same transformation behavior under
parity,
\begin{equation}
\label{4:1:parityphi}
\phi^a(\vec{x},t)\stackrel{P}{\mapsto}-\phi^a(-\vec{x},t),
\end{equation}
i.e., they are pseudoscalars, and transform under the subgroup
$H=\mbox{SU(3)}_V$, which leaves the vacuum invariant, as
an octet [see Eq.\ (\ref{4:1:crqva})]:
 \begin{equation}
\label{4:1:transformationphiqv}
[Q^a_V,\phi^b(x)]=if_{abc}\phi^c(x).
\end{equation}
   In the present case, $G=\mbox{SU(3)}_L\times\mbox{SU(3)}_R$ with $n_G=16$
and $H=\mbox{SU(3)}_V$ with $n_H=8$ and we expect eight
Goldstone bosons.

\subsection{The Scalar Quark Condensate $\langle \bar{q}q\rangle$}
\label{subsec_sqc}
   In the following, we will show that a non-vanishing scalar quark condensate
in the chiral limit is a sufficient (but not a necessary) condition for a
spontaneous symmetry breaking in QCD.\footnote{In this Section all physical
quantities such as the ground state, the quark operators etc.\ are considered
in the chiral limit.}
   The subsequent discussion will parallel that of the toy model in 
Sec.\ \ref{sec_gt} after replacement of the elementary fields $\Phi_i$ by
appropriate composite Hermitian operators of QCD. 

   Let us first recall the definition of the nine scalar and pseudoscalar
quark densities:
\begin{eqnarray}
\label{4:1:sqd}    
S_a(y)&=&\bar{q}(y)\lambda_a q(y),
\quad a=0,\cdots,8,\\
\label{4:1:psqd}  
P_a(y)&=&i\bar{q}(y)\gamma_5\lambda_a  q(y),
\quad a=0,\cdots,8.
\end{eqnarray}
   The equal-time commutation relation of two quark operators of the form 
$A_i(x)=q^\dagger(x)\hat{A}_i q(x)$,
where $\hat{A}_i$ symbolically denotes Dirac- and flavor matrices and
a summation over color indices is implied, can compactly be written
as [see Eq.\ (\ref{2:3:fkf})]
\begin{equation}
\label{4:1:comrel}
[A_1(\vec{x},t),A_2(\vec{y},t)]=\delta^3(\vec{x}-\vec{y})
q^\dagger(x)[\hat{A}_1,\hat{A}_2]q(x).
\end{equation}
   With the definition
\begin{displaymath}
Q_V^a(t)=\int d^3 x q^\dagger(\vec{x},t) \frac{\lambda^a}{2} q(\vec{x},t),
\end{displaymath}
   and using   
\begin{eqnarray*}
[\frac{\lambda_a}{2},\gamma_0\lambda_0]&=&0,\\
{[}\frac{\lambda_a}{2},\gamma_0\lambda_b]&=&
\gamma_0 i f_{abc} \lambda_c,
\end{eqnarray*}
we see, after integration of Eq.\ (\ref{4:1:comrel}) 
over $\vec{x}$, that the scalar quark densities of 
Eq.\ (\ref{4:1:sqd}) transform under 
$\mbox{SU(3)}_V$ as a singlet and as an octet, respectively, 
\begin{eqnarray}
\label{4:1:sitr}
[Q^a_V(t),S_0(y)]&=&0,\quad a=1,\cdots,8,\\
\label{4:1:octr}
{[Q^a_V(t),S_b(y)]}&=&i\sum_{c=1}^8f_{abc}S_c(y),
\quad a,b=1,\cdots,8,
\end{eqnarray}
with analogous results for the pseudoscalar quark densities.
   In the $\mbox{SU(3)}_V$ limit and, of course, also in the even more
restrictive chiral limit, the charge operators in Eqs.\ (\ref{4:1:sitr})
and (\ref{4:1:octr}) are actually time independent.\footnote{
The commutation relations also remain valid for {\em equal} times if the 
symmetry is
explicitly broken.}
   Using the relation
\begin{equation}
\sum_{a,b=1}^8 f_{abc}f_{abd}=3\delta_{cd}
\end{equation}
for the structure constants of SU(3), we re-express the octet components of 
the scalar quark densities as
\begin{equation}
\label{4:1:soktett}
S_a(y)=-\frac{i}{3}\sum_{b,c=1}^8f_{abc}[Q_V^b(t),S_c(y)],
\end{equation}
which represents the 
analogue of Eq.\ (\ref{3:3:phi3zykl}) in the discussion of Goldstone's 
theorem. 

   In the chiral limit the ground state is necessarily invariant under 
$\mbox{SU(3)}_V$ \cite{Vafa:tf}, i.e., $Q_V^a|0\rangle=0$, and we 
obtain from Eq.\ (\ref{4:1:soktett})
\begin{equation}
\label{4:1:saun}
\langle 0|S_a(y)|0\rangle
=\langle 0|S_a(0)|0\rangle
\equiv\langle S_a\rangle =0,\quad a=1,\cdots,8,
\end{equation}
where we made use of translational invariance of the ground state.
   In other words, the octet components of the scalar quark condensate
{\em must} vanish in the chiral limit.  
   From Eq.\ (\ref{4:1:saun}), we obtain for $a=3$  
$$\langle\bar{u}u\rangle-\langle\bar{d}d\rangle=0,$$
i.e.\ $\langle\bar{u}u\rangle=\langle\bar{d}d\rangle$
and for $a=8$
$$\langle\bar{u}u\rangle+\langle\bar{d}d\rangle
-2\langle\bar{s}s\rangle=0,
$$
i.e.\ $\langle\bar{u}u\rangle=\langle\bar{d}d\rangle=
\langle\bar{s}s\rangle.
$

   Because of Eq.\ (\ref{4:1:sitr}) a similar argument cannot be used
for the singlet condensate, and if we assume a non-vanishing singlet
scalar quark condensate in the chiral
limit, we thus find using Eq.\ (\ref{4:1:saun})
\begin{equation}
\label{4:1:cqc}
0\neq \langle \bar{q}q\rangle
=\langle\bar{u}u+\bar{d}d+\bar{s}s\rangle
=3\langle\bar{u}u\rangle =
3\langle\bar{d}d\rangle
=3\langle \bar{s}s\rangle.
\end{equation} 
   Finally, we make use of (no summation implied!) 
$$ (i)^2 [\gamma_5 \frac{\lambda_a}{2},\gamma_0\gamma_5\lambda_a]
=\lambda^2_a\gamma_0$$
in combination with
\begin{eqnarray*}
\lambda_1^2=\lambda_2^2=\lambda_3^2&=&
\left(
\begin{array}{rrr}
1&0&0\\
0&1&0\\
0&0&0
\end{array}
\right),\\
\lambda_4^2=\lambda_5^2&=&
\left(
\begin{array}{rrr}
1&0&0\\
0&0&0\\
0&0&1
\end{array}
\right),\\
\lambda_6^2=\lambda_7^2&=&
\left(
\begin{array}{rrr}
0&0&0\\
0&1&0\\
0&0&1
\end{array}
\right),\\
\lambda_8^2&=&
\frac{1}{3}
\left(
\begin{array}{rrr}
1&0&0\\
0&1&0\\
0&0&4
\end{array}
\right)
\end{eqnarray*}
to obtain
\begin{equation}
\label{4:1:crqapsqd}
i[Q_a^A(t), P_a(y)]
=
\left \{\begin{array}{cl} 
\bar{u}u+\bar{d}d, & a=1,2,3\\
\bar{u}u+\bar{s}s, & a=4,5\\
\bar{d}d+\bar{s}s, & a=6,7\\
\frac{1}{3}(\bar{u}u+\bar{d}d+4\bar{s}s), & a=8
\end{array}
\right.
\end{equation}
where we have suppressed the $y$ dependence on the right-hand side.
   We evaluate Eq.\ (\ref{4:1:crqapsqd}) for a ground state which is
invariant under $\mbox{SU(3)}_V$, assuming a non-vanishing singlet scalar 
quark condensate,
\begin{equation}
\label{4:1:crqc}
\langle 0|i[Q_a^A(t),P_a(y)]|0\rangle
=\frac{2}{3}\langle\bar{q}q\rangle,\quad a=1,\cdots,8,
\end{equation}
   where, because of translational invariance, the right-hand side
is independent of $y$.
   Inserting a complete set of states into the commutator of 
Eq.~(\ref{4:1:crqc}) yields, in complete analogy to Sec.\ \ref{sec_gt}
[see the discussion following Eq.\ (\ref{3:3:vq1phi2})] that 
both the pseudoscalar density $P_a(y)$ as well as the axial
charge operators $Q^a_A$ must have a non-vanishing matrix element
between the vacuum and massless one particle states $|\phi^b\rangle$.
   In particular, because of Lorentz covariance, the
matrix element of the axial-vector current operator between the
vacuum and these massless states,
appropriately normalized, 
 can be written as
\begin{equation}
\label{4:1:acc}
\langle 0|A^a_\mu(0)|\phi^b(p)\rangle=ip_\mu F_0 \delta^{ab},
\end{equation}
where $F_0\approx 93$ MeV denotes the ``decay'' constant of
the Goldstone bosons in the chiral limit.
   Assuming $Q_A^a|0\rangle\neq 0$, a non-zero value of $F_0$ is a necessary 
and sufficient criterion for spontaneous chiral symmetry breaking.    
   On the other hand, because of Eq.\ (\ref{4:1:crqc}) 
a non-vanishing scalar quark condensate $\langle \bar{q}
q\rangle$ is a sufficient (but not a necessary) condition for 
a spontaneous symmetry breakdown in QCD.

   Table \ref{table:4:1:comparison} contains a summary of the patterns of
spontaneous symmetry breaking as discussed in Sec.\ \ref{sec_gt},
 the generalization
of Sec.\ \ref{sec_sbgcnas} to the so-called O($N$) linear sigma model, and QCD.
\begin{table}
\begin{center}
\begin{tabular}{|c|c|c|c|}
\hline
&Sec.\ 3.3&O($N$) linear  &QCD\\
&&sigma model&\\
\hline
Symmetry group $G$ of&O(3)&O($N$)&$\mbox{SU(3)}_L\times\mbox{SU(3)}_R$\\
the Lagrangian density&&&\\
\hline
Number of &3&$N(N-1)/2$&16\\
generators $n_G$&&&\\
\hline
Symmetry group $H$& O(2)&O($N-1$)&SU(3)$_V$\\
of the ground state&&&\\
\hline
Number of &1&$(N-1)(N-2)/2$&8\\
generators $n_H$&&&\\
\hline
Number of &2&$N-1$&8\\
Goldstone bosons&&&\\
$n_G-n_H$&&&\\
\hline
Multiplet of 
&$(\Phi_1(x),\Phi_2(x))$
&$(\Phi_1(x),\cdots,\Phi_{N-1}(x))$&
$i\bar{q}(x)\gamma_5\lambda_a q(x)$\\
Goldstone boson fields&&&\\
\hline
Vacuum expectation & $v=\langle\Phi_3\rangle$&$v=\langle\Phi_N\rangle$&
$v=\langle\bar{q}q\rangle$\\
value&&&\\
\hline
\end{tabular}
\end{center}
\caption{\label{table:4:1:comparison}Comparison of spontaneous symmetry
breaking.}
\end{table}

\section{Transformation Properties of the Goldstone Bosons}
\label{sec_tpgb}
   The purpose of this section is to discuss the transformation properties 
of the field variables describing the Goldstone bosons
\cite{Weinberg:de,Coleman:sm,Callan:sn,Balachandran:zj,Leutwyler:1991mz}.
   We will need the concept of a {\em nonlinear realization} of a group in
addition to a {\em representation} of a group which one usually encounters in
Physics.
   We will first discuss a few general group-theoretical properties before 
specializing to QCD.   

\subsection{General Considerations}
\label{subsec_gc}
   Let us consider a physical system with a Hamilton operator $\hat{H}$ which 
is invariant under a compact Lie group $G$.
   Furthermore we assume the ground state of the system to be invariant under 
only a subgroup $H$ of $G$, giving rise to $n=n_G-n_H$ Goldstone bosons.
   Each of these Goldstone bosons will be described by an independent 
field $\phi_i$ which is a continuous real function on Minkowski space 
$M^4$.\footnote{Depending on the equations of motion, we will
require more restrictive properties of the functions $\phi_i$.
}
   We collect these fields in an $n$-component vector $\Phi$ and define the 
vector space 
\begin{equation}
\label{4:2:m1}
M_1\equiv\{\Phi:M^4\to R^n|\phi_i:M^4\to R\,\,\mbox{continuous}\}.
\end{equation}
   Our aim is to find a mapping $\varphi$ which uniquely associates with each 
pair $(g,\Phi)\in G\times M_1$ an element $\varphi(g,\Phi)\in M_1$ with
the following properties:
\begin{eqnarray}
\label{4:2:condmap1}
&&\varphi(e,\Phi)=\Phi\,\,\forall\,\,\Phi\in M_1,\, e\,\,
\mbox{identity of}\,\, G,\\
\label{4:2:condmap2}
&&\varphi(g_1,\varphi(g_2,\Phi))=\varphi(g_1 g_2,\Phi)\,\,\forall\,\,
g_1,g_2\in G,\,\forall\,\Phi\in M_1.
\end{eqnarray}
   Such a mapping defines an {\em operation} of the group $G$ on $M_1$.
   The second condition is the so-called group-homomorphism property
\cite{Balachandran:ab,O'Raifeartaigh:vq,Jones:ti}.
   The mapping will, in general, {\em not} define a {\em representation} of 
the group $G$, because we do not require the mapping to be linear, i.e.,
$\varphi(g,\lambda \Phi)\neq \lambda\varphi(g,\Phi)$.

   Let $\Phi=0$ denote the ``origin'' of $M_1$ \cite{Leutwyler:1991mz} which,
in a theory containing Goldstone bosons only, loosely speaking corresponds
to the ground state configuration.
   Since the ground state is supposed to be invariant under the subgroup
$H$ we require the mapping $\varphi$ to be such that all elements 
$h\in H$ map the origin onto itself. 
   In this context the subgroup $H$ is also known as the little group of 
$\Phi=0$.
   Given that such a mapping indeed exists, we need to verify for 
infinite groups that (see Chap.\ 2.4 of \cite{Jones:ti}): 
\begin{enumerate}
\item $H$ is not empty, because the identity $e$ maps the origin onto itself.
\item If $h_1$ and $h_2$ are elements satisfying $\varphi(h_1,0)=
\varphi(h_2,0)=0$, so does $\varphi(h_1 h_2,0)=
\varphi(h_1,\varphi(h_2,0))=\varphi(h_1,0)=0$,
i.e., because of the homomorphism property also the product $h_1 h_2\in H$. 
\item For $h\in H$ we have
$$\varphi(h^{-1},0)=\varphi(h^{-1},\varphi(h,0))
=\varphi(h^{-1}h,0)=\varphi(e,0).
$$
i.e., $h^{-1}\in H$.
\end{enumerate}

   Following Ref.\ \cite{Leutwyler:1991mz} we will establish a connection
between the Goldstone boson fields and the set of all left cosets 
$\{gH|g\in G\}$ which is also referred to as the quotient $G/H$.
   For a subgroup $H$ of $G$ the set $gH=\{gh|h\in H\}$ defines the left 
coset of $g$ (with an analogous definition for the right coset) which is one
element of $G/H$.\footnote{
An {invariant} subgroup has the additional property that the left and right
cosets coincide for each $g$ which allows for a definition of the
factor group $G/H$ in terms of the complex product.
However, here we do not need this property.}
   For our purposes we need the property 
that cosets either completely overlap or are
completely disjoint (see, e.g.,  \cite{Jones:ti}), i.e, the quotient is a 
set whose elements themselves are sets of group elements, and these sets
are completely disjoint.
  
   Let us first show that for all elements of a given coset,
$\varphi$ maps the origin onto the same vector in $R^n$:
$$\varphi(gh,0)=\varphi(g,\varphi(h,0))
=\varphi(g,0)\,\,\forall\,\, g\in G\,
\mbox{and}\, h\in H.$$
   Secondly, the mapping is injective with respect to the cosets, 
which can be proven as follows. 
Consider two elements $g$ and $g'$ of $G$ where $g'\not\in g H$. 
   We need to show $\varphi(g,0)\neq \varphi(g',0)$.
   Let us assume $\varphi(g,0)=\varphi(g',0)$:
$$0=\varphi(e,0)=\varphi(g^{-1}g,0)
=\varphi(g^{-1},\varphi(g,0))
=\varphi(g^{-1},\varphi(g',0))=\varphi(g^{-1}g',0).$$
   However, this implies $g^{-1}g'\in H$ or $g'\in gH $
in contradiction to the assumption.
   Thus $\varphi(g,0)=\varphi(g',0)$ cannot be true. 
   In other words, the mapping can be inverted on the image of 
$\varphi(g,0)$.
   The conclusion is that there exists an {\em isomorphic mapping} 
between the quotient $G/H$ and the Goldstone boson 
fields.\footnote{Of course, the Goldstone boson fields are not
constant vectors in $R^n$ but functions on Minkowski space 
[see Eq.\ (\ref{4:2:m1})].
   This is accomplished by allowing the cosets $gH$ to also
depend on $x$.}

   Now let us discuss the transformation behavior of the Goldstone boson
fields under an arbitrary $g\in G$ in terms of the 
isomorphism established above.
   To each $\Phi$ corresponds a  coset $\tilde{g}H$ with appropriate 
$\tilde{g}$.
   Let $f=\tilde{g}h\in \tilde{g}H$ denote a representative of this
coset such that  
$$\Phi=\varphi(f,0)=\varphi(\tilde{g}h,0).$$
   Now apply the mapping $\varphi(g)$ to $\Phi$:
$$\varphi(g,\Phi)=\varphi(g,\varphi(\tilde{g}h,0))
=\varphi(g\tilde{g}h,0)=\varphi(f',0)=\Phi',\quad
f'\in g(\tilde{g}H).
$$
   In other words, in order to obtain the transformed $\Phi'$ from a given 
$\Phi$ we simply need to multiply the left coset $\tilde{g}H$ representing
$\Phi$ by $g$  in order to obtain the new left coset representing $\Phi'$.
   This procedure uniquely determines the transformation behavior of
the Goldstone bosons up to an appropriate choice of variables 
parameterizing the elements of the quotient $G/H$. 

\subsection{Application to QCD}
\label{subsec_aqcd}

   Now let us apply the above general considerations to the specific case 
relevant to QCD and consider the group 
$G=\mbox{SU($N$)}\times\mbox{SU($N$)}=\{(L,R)|
L\in \mbox{SU($N$)}, R\in
\mbox{SU($N$)}\}$ and $H=\{(V,V)|V\in \mbox{SU($N$)}\}$ which is
isomorphic to $\mbox{SU($N$)}$.
Let $\tilde{g}=(\tilde{L},\tilde{R})\in G$. 
   We may uniquely characterize the left coset of $\tilde{g}$,
$\tilde{g}H=\{(\tilde{L}V,\tilde{R}V)|V\in \mbox{SU($N$)}\}$, 
through the SU($N$) matrix $U=\tilde{R}\tilde{L}^\dagger$ 
\cite{Balachandran:zj}, 
$$
(\tilde{L}V,\tilde{R}V)=(\tilde{L}V,\tilde{R}\tilde{L}^\dagger\tilde{L}V)
=(1,\tilde{R}\tilde{L}^\dagger)\underbrace{(\tilde{L}V,\tilde{L}V)}_{
\mbox{$\in H$}},\quad
\mbox{i.e.} \quad \tilde{g}H=(1,\tilde{R}\tilde{L}^\dagger)H,
$$
   if we follow the convention that we choose the representative of the
coset such that the unit matrix stands in its first argument.
   According to the above derivation, $U$ is isomorphic to a $\Phi$.
   The transformation behavior of $U$ under $g=(L,R)\in G$ is obtained by 
multiplication in the left coset:
$$g \tilde{g}H=(L, R\tilde{R}\tilde{L}^\dagger)H=
(1,R\tilde{R}\tilde{L}^\dagger L^\dagger)(L,L)H=
(1,R(\tilde{R}
\tilde{L}^\dagger)L^\dagger)H,$$
i.e.\
\begin{equation}
\label{4:2:utrafo}
U=\tilde{R}\tilde{L}^\dagger \mapsto U'=R(\tilde{R}\tilde{L}^\dagger)L^\dagger
=RUL^\dagger.
\end{equation}
   As mentioned above, we finally need to introduce an $x$ dependence so
that 
\begin{equation}
\label{4:2:utrfafo}
U(x)\mapsto R U(x) L^\dagger.
\end{equation}

    Let us now restrict ourselves to the physically relevant cases of 
$N=2$ and $N=3$ and define
$$
M_1\equiv\left\{
\begin{array}{l}
\{\Phi: M^4\to  R^3|\phi_i: M^4\to
R\,\,\mbox{continuous}\}\,\,\mbox{for $N=2$,}\\
\{\Phi: M^4\to R^8|\phi_i: M^4 \to
R\,\,\mbox{continuous}\}\,\,\mbox{for $N=3$}.
\end{array}
\right.
$$
   Furthermore let $\tilde{\cal H}(N)$ denote the set of all
Hermitian and traceless $N\times N$ matrices,
\begin{displaymath}
\tilde{\cal H}(N)\equiv\{A\in 
\mbox{gl}(N, C)|A^\dagger=A\wedge \mbox{Tr}(A)=0\},
\end{displaymath}
   which under addition of matrices defines a real vector space.
   We define a second set $M_2\equiv\{\phi:M^4\to\tilde{\cal H}(N)|\phi\,\,
\mbox{continuous}\}$, where the entries are continuous functions.
   For $N=2$ the elements of $M_1$ and $M_2$ are related to each
other according to   
\begin{eqnarray*}
\phi(x)&=&\sum_{i=1}^3\tau_i\phi_i(x)
=\left(\begin{array}{cc} \phi_3 & \phi_1-i\phi_2\\
\phi_1+i\phi_2&-\phi_3
\end{array}\right)
\equiv
\left(\begin{array}{cc}
\pi^0&\sqrt{2}\pi^+\\
\sqrt{2}\pi^-&-\pi^0
\end{array}\right),\\
\end{eqnarray*}
where the $\tau_i$ are the usual Pauli matrices and $\phi_i(x)=\frac{1}{2}
\mbox{Tr}[\tau_i \phi(x)]$.
   Analogously for $N=3$,
\begin{eqnarray*}
\phi(x)=\sum_{a=1}^8 \lambda_a \phi_a(x)
&=&\left(\begin{array}{ccc}
\phi_3+ \frac{1}{\sqrt{3}}\phi_8&\phi_1-i\phi_2&\phi_4-i\phi_5\\
\phi_1+i\phi_2& -\phi_3+ \frac{1}{\sqrt{3}}\phi_8&\phi_6-i\phi_7\\
\phi_4+i\phi_5&\phi_6+i\phi_7&-\frac{2}{\sqrt{3}}\phi_8
\end{array}\right)\\
&\equiv& 
\left(\begin{array}{ccc}
\pi^0+\frac{1}{\sqrt{3}}\eta &\sqrt{2}\pi^+&\sqrt{2}K^+\\
\sqrt{2}\pi^-&-\pi^0+\frac{1}{\sqrt{3}}\eta&\sqrt{2}K^0\\
\sqrt{2}K^- &\sqrt{2}\bar{K}^0&-\frac{2}{\sqrt{3}}\eta
\end{array}\right),\\
\end{eqnarray*}
with the Gell-Mann matrices $\lambda_a$ and
$\phi_a(x)=\frac{1}{2}\mbox{Tr}[\lambda_a \phi(x)]$.
   Again, $M_2$ forms a real vector space.
   Let us finally define
\begin{displaymath}
M_3\equiv\left\{U:M^4\to \mbox{SU}(N)|U(x)
=\exp\left(i\frac{\phi(x)}{F_0}\right),
\phi\in M_2\right\}.
\end{displaymath}
   At this point it is important to note that $M_3$ does not define
a vector space because the sum of two SU($N$) matrices is not
an SU($N$) matrix.

   We are now in the position to discuss the so-called nonlinear
realization of $\mbox{SU($N$)}\times\mbox{SU($N$)}$ on $M_3$.
   The homomorphism
\begin{displaymath}
\varphi: G\times M_3 \to M_3\quad\mbox{with}\quad 
\varphi[(L,R),U](x)\equiv R U (x)L^\dagger,
\end{displaymath}
defines an operation of $G$ on $M_3$, because
\begin{enumerate}
\item  $RUL^\dagger\in M_3$, since $U\in M_3$ and 
$R, L^\dagger\in \mbox{SU}(N)$. 
\item $\varphi[(1_{N\times N},1_{N\times N}),U](x)=1_{N\times N}U(x)
 1_{N\times N}=U(x).$
\item Let $g_i=(L_i,R_i)\in G$ and thus $g_1 g_2=(L_1 L_2,R_1 R_2)\in G$.
\begin{eqnarray*}
\varphi[g_1,\varphi[g_2,U]](x)
&=&\varphi[g_1,(R_2 U L_2^\dagger)](x)=R_1 R_2 U(x) L_2^\dagger 
L_1^\dagger,\\
\varphi[g_1g_2,U](x)&=&R_1 R_2 U(x)(L_1 L_2)^\dagger= R_1 R_2 U(x) L_2^\dagger
L_1^\dagger.
\end{eqnarray*}
\end{enumerate}
   The mapping $\varphi$ is called a nonlinear realization, because $M_3$ is 
{\em not} a vector space.
 
   The origin $\phi(x)=0$, i.e.\ $U_0=1$, denotes the ground state of
the system.
   Under transformations of the subgroup $H=\{(V,V)|V\in \mbox{SU($N$)}\}$ 
corresponding to rotating both left- and right-handed quark fields in 
QCD by the same $V$, the ground state remains invariant,
$$\varphi[g=(V,V),U_0]=VU_0 V^\dagger=V V^\dagger= 1=U_0.$$
   On the other hand, under ``axial transformations,'' i.e.\ rotating
the left-handed quarks by $A$ and the right-handed quarks by $A^\dagger$,
the ground state does {\em not} remain invariant,
$$\varphi[g=(A,A^\dagger),U_0]=A^\dagger U_0 A^\dagger=A^\dagger A^\dagger
\neq U_0,$$
   which, of course, is consistent with the assumed spontaneous symmetry
breakdown.

   Let us finally discuss the transformation behavior of $\phi(x)$ under
the subgroup $H=\{(V,V)|V\in \mbox{SU($N$)}\}$. 
   Expanding 
$$U=1+i\frac{\phi}{F_0}-\frac{\phi^2}{2F_0^2}+\cdots,$$
we immediately see that the realization restricted to the subgroup $H$,
\begin{equation}
\label{4:2:uhtrafo}
1+i\frac{\phi}{F_0}-\frac{\phi^2}{2F_0^2}+\cdots\mapsto
V(1+i\frac{\phi}{F_0}-\frac{\phi^2}{2F_0^2}+\cdots)V^\dagger
=1+i\frac{V \phi V^\dagger}{F_0}-\frac{V\phi V^\dagger V\phi V^\dagger}{2F_0^2}
+\cdots,
\end{equation}
defines a linear representation on 
$M_2 \ni \phi\mapsto V\phi V^\dagger \in M_2$, because
\begin{eqnarray*}
&&
(V\phi V^\dagger)^\dagger= V\phi V^\dagger,\quad
\mbox{Tr}(V\phi V^\dagger)=\mbox{Tr}(\phi)=0,\\
&&V_1 (V_2 \phi V_2^\dagger) V_1^\dagger=(V_1 V_2) \phi (V_1 V_2)^\dagger.
\end{eqnarray*}
   Let us consider the SU(3) case and parameterize
$$V=\exp\left(-i\Theta^V_a\frac{\lambda_a}{2}\right),$$
from which we obtain, by comparing both sides of Eq.\ (\ref{4:2:uhtrafo}),
\begin{equation}
\label{4:2:phihtrafo}
\phi=\lambda_b\phi_b\stackrel{\mbox{$h\in$ SU(3)$_V$}}{\mapsto}
V\phi V^\dagger=\phi-i\Theta^V_a
\underbrace{[\frac{\lambda_a}{2},\phi_b\lambda_b]}_{
\mbox{$\phi_b if_{abc}\lambda_c$}}
+\cdots
=\phi+f_{abc}\Theta^V_a\phi_b\lambda_c +\cdots.
\end{equation}
   However, this corresponds exactly to the adjoint representation,
i.e., in SU(3) the fields $\phi_a$ transforms as an octet which is
also consistent with the transformation behavior we discussed in Eq.\
(\ref{4:1:transformationphiqv}):
\begin{eqnarray}
\label{4:2:phivergf}
e^{i\Theta^V_a Q^a_V}\lambda_b\phi_b e^{-i\Theta^V_a Q^a_V}
&=&\lambda_b\phi_b +i\Theta^V_a\lambda_b\underbrace{[Q^a_V,\phi_b]}_{
\mbox{$if_{abc}\phi_c$}}+\cdots\nonumber\\
&=&\phi+f_{abc}\Theta^V_a\phi_b\lambda_c +\cdots.
\end{eqnarray}

   For group elements of $G$ of the form $(A,A^\dagger)$ one may
proceed in a completely analogous fashion.
   However, one finds that the fields $\phi_a$ do {\em not} have
a simple transformation behavior under these group elements.
   In other words, the commutation relations of the fields with
the {\em axial} charges are complicated nonlinear functions of
the fields \cite{Weinberg:de}.

\section{The Lowest-Order Effective Lagrangian}
\label{sec_loel}
   Our goal is the construction of the most general theory describing the
dynamics of the Goldstone bosons associated with the spontaneous 
symmetry breakdown in QCD.
   In the chiral limit, we want the effective Lagrangian to be invariant under 
 $\mbox{SU(3)}_L\times\mbox{SU(3)}_R\times\mbox{U(1)}_V$.
   It should contain exactly eight pseudoscalar degrees of freedom transforming
as an octet under the subgroup $H=\mbox{SU(3)}_V$. 
   Moreover, taking account of spontaneous symmetry breaking, the ground 
state should only be invariant under $\mbox{SU(3)}_V\times\mbox{U(1)}_V$.

   Following the discussion of Sec.\ \ref{subsec_aqcd} we collect the 
dynamical variables in the SU(3) matrix $U(x)$,
\begin{eqnarray}
\label{4:3:upar}
U(x)&=&\exp\left(i\frac{\phi(x)}{F_0}\right),\nonumber\\
\phi(x)&=&\sum_{a=1}^8 \lambda_a \phi_a(x)\equiv
\left(\begin{array}{ccc}
\pi^0+\frac{1}{\sqrt{3}}\eta &\sqrt{2}\pi^+&\sqrt{2}K^+\\
\sqrt{2}\pi^-&-\pi^0+\frac{1}{\sqrt{3}}\eta&\sqrt{2}K^0\\
\sqrt{2}K^- &\sqrt{2}\bar{K}^0&-\frac{2}{\sqrt{3}}\eta
\end{array}\right).
\end{eqnarray}
   The most general, chirally invariant, effective Lagrangian density with the 
minimal number of derivatives reads
\begin{equation}
\label{4:3:l2}
{\cal L}_{\rm eff}
=\frac{F^2_0}{4}\mbox{Tr}\left(\partial_\mu U \partial^\mu U^\dagger
\right),
\end{equation}
   where $F_0\approx 93$ MeV is a free parameter which later on will be 
related to the pion decay $\pi^+\to\mu^+\nu_\mu$ (see Sec.\
\ref{subsec_pdpmn}).

   First of all, the Lagrangian is invariant under the {\em global} 
$\mbox{SU(3)}_L\times\mbox{SU(3)}_R$ transformations of 
Eq.\ (\ref{4:2:utrafo}):
\begin{eqnarray*}
U&\mapsto& R U L^\dagger,\\
\partial_\mu U&\mapsto&\partial_\mu(R U L^\dagger)=
\underbrace{\partial_\mu R}_{\mbox{0}}UL^\dagger+R\partial_\mu U L^\dagger
+RU\underbrace{\partial_\mu L^\dagger}_{\mbox{0}}=R\partial_\mu U L^\dagger,\\
U^\dagger&\mapsto& L U^\dagger R^\dagger,\\
\partial_\mu U^\dagger&\mapsto&L\partial_\mu U^\dagger R^\dagger,
\end{eqnarray*}
because
\begin{displaymath}
{\cal L}_{\rm eff}
\mapsto\frac{F^2_0}{4}\mbox{Tr}\Big(R\partial_\mu U 
\underbrace{L^\dagger L}_{\mbox{1}}\partial^\mu U^\dagger R^\dagger\Big)
=\frac{F^2_0}{4}\mbox{Tr}\Big(\underbrace{R^\dagger R}_{\mbox{1}}
\partial_\mu U
\partial^\mu U^\dagger\Big)
={\cal L}_{\rm eff},
\end{displaymath}
   where we made use of the trace property $\mbox{Tr}(AB)=\mbox{Tr}(BA)$.
   The global $\mbox{U(1)}_V$ invariance is trivially satisfied, because
the Goldstone bosons have baryon number zero, thus transforming
as $\phi\mapsto\phi$ under $\mbox{U(1)}_V$ which also implies $U \mapsto U$.

   The substitution $\phi_a(\vec{x},t)\mapsto -\phi_a(\vec{x},t)$ or,
equivalently, $U(\vec{x},t)\mapsto U^\dagger(\vec{x},t)$ provides a simple 
method of testing, whether an expression is of so-called even or odd 
{\em intrinsic} parity,\footnote{
Since the Goldstone bosons are pseudoscalars, a true parity transformation
is given by $\phi_a(\vec{x},t)\mapsto -\phi_a(-\vec{x},t)$ or,
equivalently, $U(\vec{x},t)\mapsto U^\dagger(-\vec{x},t)$.}
i.e., even or odd in the number of Goldstone boson fields.
   For example, it is easy to show, using the trace property, that
the Lagrangian of Eq.\ (\ref{4:3:l2}) is even. 

   The purpose of the multiplicative constant $F^2_0/4$ in Eq.\ (\ref{4:3:l2})
is to generate the standard form of the kinetic term 
$\frac{1}{2}\partial_\mu \phi_a\partial^\mu \phi_a$, which can be seen by 
expanding the exponential 
$U=1+i\phi/F_0+\cdots$, $\partial_\mu U=i\partial_\mu\phi/F_0
+\cdots$, resulting in 
\begin{eqnarray*}
{\cal L}_{\rm eff}&=&
\frac{F^2_0}{4}\mbox{Tr}\left[\frac{i\partial_\mu\phi}{F_0}
\left(-\frac{i\partial^\mu\phi}{F_0}\right)\right]+\cdots
=\frac{1}{4}\mbox{Tr}(\lambda_a\partial_\mu\phi_a\lambda_b\partial^\mu
\phi_b)+\cdots\\
&=&\frac{1}{4}\partial_\mu\phi_a\partial^\mu
\phi_b\mbox{Tr}(\lambda_a\lambda_b)+\cdots
=\frac{1}{2}\partial_\mu\phi_a\partial^\mu\phi_a +{\cal L}_{\rm int},
\end{eqnarray*}
where we made use of $\mbox{Tr}(\lambda_a \lambda_b)=2\delta_{ab}$.
   In particular, since there are no other terms containing only two fields
(${\cal L}_{\rm int}$ starts with interaction terms
containing at least four Goldstone bosons)
the eight fields $\phi_a$ describe eight independent {\em massless} 
particles.\footnote{At this stage, this is only a tree-level argument.
We will see in Sec.\ \ref{subsec_mgb} that the Goldstone bosons remain
massless in the chiral limit even when loop corrections have been
included.}

   A term of the type $\mbox{Tr}[(\partial_\mu\partial^\mu U) U^\dagger]$ may 
be re-expressed as\footnote{In the present case 
$\mbox{Tr}(\partial^\mu U U^\dagger)=0.$}
$$\mbox{Tr}[(\partial_\mu\partial^\mu U) U^\dagger]
=\partial_\mu[\mbox{Tr}(\partial^\mu U U^\dagger)]
-\mbox{Tr}(\partial^\mu U \partial_\mu U^\dagger),$$
i.e., up to a total derivative it is proportional to the Lagrangian
of Eq.\ (\ref{4:3:l2}).
   However, in the present context, total derivatives do not have a dynamical 
significance, i.e.\ they leave the equations of motion unchanged and can thus 
be dropped.
   The product of two invariant traces is excluded at lowest order,
because $\mbox{Tr}(\partial_\mu U U^\dagger)=0$.
   Let us prove the general SU($N$) case by considering an 
SU($N$)-valued field 
$$U=\exp\left(i \frac{\Lambda_a\phi_a(x)}{F_0}\right),$$
with $N^2-1$ Hermitian, traceless matrices $\Lambda_a$ and real fields
$\phi_a(x)$.
   Defining $\Phi=\Lambda_a\phi_a/F_0$, we expand the exponential
$$U=1+i\Phi+\frac{1}{2}(i\Phi)^2+\frac{1}{3!}(i\Phi)^3+\cdots$$
and consider the derivative\footnote{$\Phi$ and $\partial_\mu\Phi$ are
matrices which, in general, do not commute.} 
$$\partial_\mu U=i\partial_\mu \Phi+\frac{1}{2}(i\partial_\mu \Phi i\Phi
+i\Phi i\partial_\mu \Phi)+\frac{1}{3!}[i\partial_\mu\Phi
(i\Phi)^2+i\Phi i\partial_\mu\Phi i\Phi+(i\Phi)^2i\partial_\mu\Phi]+
\cdots.$$
    We then find
\begin{eqnarray}
\label{4:3:trpmuuud}
\mbox{Tr}(\partial_\mu U U^\dagger)
&=&\mbox{Tr}[i\partial_\mu\Phi U^\dagger +\frac{1}{2}(i\partial_\mu \Phi
i\Phi+i\Phi i\partial_\mu\Phi)U^\dagger +\cdots]\nonumber\\
&=&
\mbox{Tr}[i\partial_\mu \Phi U^\dagger +i\partial_\mu \Phi i\Phi U^\dagger
+\frac{1}{2}i\partial_\mu\Phi (i\Phi)^2 U^\dagger +\cdots]\nonumber\\
&=&\mbox{Tr}(i\partial_\mu \Phi 
\underbrace{U U^\dagger}_{\mbox{1}})=\mbox{Tr}(i\partial_\mu\Phi)=
i\partial_\mu \phi_a\underbrace{\mbox{Tr}(\Lambda_a)}_{\mbox{0}}=0,
\end{eqnarray}
   where we made use of $[\Phi, U^\dagger]=0$.

   Let us turn to the vector and axial-vector currents associated with
the global $\mbox{SU(3)}_L\times\mbox{SU(3)}_R$ symmetry of the effective
Lagrangian of Eq.\ (\ref{4:3:l2}).
   To that end, we parameterize 
\begin{eqnarray}
\label{4:3:l}
L&=&\exp\left(-i\Theta^L_a\frac{\lambda_a}{2}\right),\\
\label{4:3:r}
R&=&\exp\left(-i\Theta^R_a\frac{\lambda_a}{2}\right).
\end{eqnarray}
   In order to construct $J^{\mu,a}_L$, set $\Theta^R_a=0$ and choose
$\Theta^L_a=\Theta^L_a(x)$ (see Sec.\ \ref{subsec_nt}).
   Then, to first order in $\Theta^L_a$,
\begin{eqnarray}
\label{dul}
U&\mapsto& U'=R U L^\dagger=U\left(1+i\Theta^L_a\frac{\lambda_a}{2}\right),
\nonumber\\
U^\dagger&\mapsto&U'^\dagger=
\left(1-i\Theta^L_a\frac{\lambda_a}{2}\right)U^\dagger,
\nonumber\\
\partial_\mu U&\mapsto&\partial_\mu U'
=\partial_\mu U \left(1+i\Theta^L_a\frac{\lambda_a}{2}\right)
+U i\partial_\mu\Theta_a^L\frac{\lambda_a}{2},
\nonumber\\
\partial_\mu U^\dagger&\mapsto&\partial_\mu U'^\dagger
=\left(1-i\Theta_a^L\frac{\lambda_a}{2}\right)\partial_\mu U^\dagger
-i\partial_\mu \Theta_a^L\frac{\lambda_a}{2} U^\dagger,
\end{eqnarray}
from which we obtain for $\delta \cal L_{\rm eff}$:
\begin{eqnarray}
\label{4:3:dll}
\delta{\cal L}_{\rm eff}&=&\frac{F^2_0}{4}\mbox{Tr}\left[
U i\partial_\mu \Theta_a^L\frac{\lambda_a}{2}\partial^\mu U^\dagger
+\partial_\mu U \left(-i\partial^\mu \Theta_a^L\frac{\lambda_a}{2}U^\dagger
\right)\right]
\nonumber\\
&=& \frac{F^2_0}{4}i\partial_\mu \Theta^L_a\mbox{Tr}\left[
\frac{\lambda_a}{2}(\partial^\mu U^\dagger U-U^\dagger\partial^\mu U)
\right]\nonumber\\
&=&\frac{F^2_0}{4}i\partial_\mu \Theta^L_a\mbox{Tr}\left(\lambda_a
\partial^\mu U^\dagger U\right).
\end{eqnarray}
   (In the last step we made use of 
$$
\partial^\mu U^\dagger U=-U^\dagger \partial^\mu U,
$$
   which follows from differentiating $U^\dagger U=1$.)
   We thus obtain for the left currents 
\begin{equation}
\label{4:3:jl}
J^{\mu,a}_L=\frac{\partial \delta {\cal L}_{\rm eff}}{
\partial \partial_\mu \Theta_a^L}
=i\frac{F^2_0}{4}\mbox{Tr}\left(\lambda_a \partial^\mu U^\dagger U\right),
\end{equation}
   and, completely analogously, choosing $\Theta_a^L=0$ and 
$\Theta_a^R=\Theta_a^R(x)$, 
\begin{equation}
\label{4:3:jr}
J^{\mu,a}_R=\frac{\partial \delta {\cal L}_{\rm eff}
}{\partial \partial_\mu \Theta^R_a}
=-i\frac{F^2_0}{4}\mbox{Tr}\left(\lambda_a U \partial^\mu U^\dagger\right)
\end{equation}
   for the right currents.
   Combining Eqs.\ (\ref{4:3:jl}) and (\ref{4:3:jr}) the vector and 
axial-vector currents read
\begin{eqnarray}
\label{4:3:jv}
J^{\mu,a}_V&=&J^{\mu,a}_R+J^{\mu,a}_L=-i\frac{F^2_0}{4}
\mbox{Tr}\left(\lambda_a[U,\partial^\mu U^\dagger]\right),\\
\label{4:3:ja}
J^{\mu,a}_A&=&J^{\mu,a}_R-J^{\mu,a}_L=-i\frac{F^2_0}{4}
\mbox{Tr}\left(\lambda_a\{U,\partial^\mu U^\dagger\}\right).
\end{eqnarray}
   Furthermore, because of the symmetry of ${\cal L}_{\rm eff}$ under 
$\mbox{SU(3)}_L\times\mbox{SU(3)}_R$, both vector and axial-vector
currents are conserved.
   The vector current densities $J^{\mu,a}_V$ of Eq.\ (\ref{4:3:jv})
contain only terms with an even number of Goldstone bosons,
\begin{eqnarray*}
J^{\mu,a}_V&\stackrel{\mbox{$\phi\mapsto-\phi$}}{\mapsto}&
-i\frac{F^2_0}{4}\mbox{Tr}[\lambda_a(U^\dagger\partial^\mu 
U-\partial^\mu U U^\dagger)]\\
&=&-i\frac{F^2_0}{4}\mbox{Tr}[\lambda_a(-\partial^\mu U^\dagger U
+U \partial^\mu U^\dagger)]
=J^{\mu,a}_V.
\end{eqnarray*}
   On the other hand, the expression for the axial-vector currents is
{\em odd} in the number of Goldstone bosons,
\begin{eqnarray*}
J^{\mu,a}_A&\stackrel{\mbox{$\phi\mapsto-\phi$}}{\mapsto}&
-i\frac{F^2_0}{4}\mbox{Tr}[\lambda_a(U^\dagger\partial^\mu 
U+\partial^\mu U U^\dagger)]\\
&=&i\frac{F^2_0}{4}\mbox{Tr}[\lambda_a(\partial^\mu U^\dagger U
+U \partial^\mu U^\dagger)]
=-J^{\mu,a}_A.
\end{eqnarray*}
   To find the leading term let us expand Eq.\ (\ref{4:3:ja}) in the fields,
\begin{displaymath}
J^{\mu,a}_A=-i\frac{F^2_0}{4}\mbox{Tr}\left(\lambda_a\left\{1+\cdots,
-i\frac{\lambda_b\partial^\mu \phi_b}{F_0}+\cdots\right\}\right)=
-F_0\partial^\mu\phi_a+\cdots
\end{displaymath}
   from which we conclude that the axial-vector current has a non-vanishing 
matrix element when evaluated between the vacuum and a one-Goldstone boson 
state [see  Eq.\ (\ref{4:1:acc})]:
\begin{eqnarray*}
\langle 0|J^{\mu,a}_A(x)|\phi^b(p)\rangle
&=&\langle 0|-F_0\partial^\mu\phi_a(x)|\phi^b(p)\rangle\\
&=&-F_0\partial^\mu \exp(-ip\cdot x)\delta^{ab}
=ip^\mu F_0\exp(-ip\cdot x)\delta^{ab}.
\end{eqnarray*}
   In Sec.\ \ref{subsec_pdpmn} $F_0$ will be related to the
pion-decay constant entering $\pi^+\to\mu^+\nu_\mu$.

    So far we have assumed a perfect
$\mbox{SU(3)}_L\times\mbox{SU(3)}_R$ symmetry.
   However, in Sec.\ \ref{sec_esbfl} we saw, by means of a simple example,
how an explicit symmetry breaking may lead to finite masses of the 
Goldstone bosons.
   As has been discussed in Sec.\ \ref{subsec_csbdqm},
the quark mass term of QCD results
in such an explicit symmetry breaking,
\begin{equation}
\label{4:3:qmt}
{\cal L}_M=-\bar{q}_RM q_L-\bar{q}_L M^\dagger q_R,\quad 
M=\left(\begin{array}{ccc}m_u&0&0\\0&m_d&0\\0&0&m_s\end{array}\right).
\end{equation}
   In order to incorporate the consequences of Eq.\ (\ref{4:3:qmt})
into the effective-Lagrangian framework, one makes use of 
the following argument \cite{Georgi}:
   Although $M$ is in reality just a constant matrix and does not
transform along with the quark fields, ${\cal L}_M$ of Eq.\ (\ref{4:3:qmt}) 
{\em would be} invariant {\em if} $M$ transformed as
\begin{equation}
\label{4:3:mgtrafo}
M\mapsto R M L^\dagger.
\end{equation}
   One then constructs the most general Lagrangian ${\cal L}(U,M)$ which is 
invariant under Eqs.\ (\ref{4:2:utrfafo}) and
(\ref{4:3:mgtrafo}) and expands this function in powers of $M$.
   At lowest order in $M$ one obtains 
\begin{equation}
\label{4:3:lqm}
{\cal L}_{\rm s.b.}=\frac{F^2_0 B_0}{2}\mbox{Tr}(MU^\dagger+UM^\dagger),
\end{equation}
   where the subscript s.b.\ refers to symmetry breaking.
   In order to interpret the new parameter $B_0$ let us consider the
energy density of the ground state ($U=U_0=1$),
\begin{equation}
\label{4:3:Heff}
\langle{\cal H}_{\rm eff}\rangle=-F_0^2 B_0(m_u+m_d+m_s),
\end{equation} 
   and compare its derivative with respect to (any of) the light quark masses
$m_q$ with the corresponding quantity in QCD,
\begin{displaymath}
\left.\frac{\partial \langle 0|{\cal H}_{\rm QCD}|0\rangle}{\partial m_q}
\right|_{m_u=m_d=m_s=0}=\frac{1}{3}\langle 0|\bar{q}{q}|0\rangle_0
=\frac{1}{3}\langle \bar{q}q\rangle,
\end{displaymath}
   where $\langle\bar{q}{q}\rangle$ is the chiral quark condensate of 
Eq.\ (\ref{4:1:cqc}).
   Within the framework of the lowest-order effective
Lagrangian, the constant $B_0$ is thus related to the chiral quark condensate 
as 
\begin{equation}
\label{4:3:b0}
3 F^2_0B_0=-\langle\bar{q}q\rangle.
\end{equation}

   Let us add a few remarks.
\begin{enumerate}
\item A term $\mbox{Tr}(M)$ by itself is not invariant.
\item The combination $\mbox{Tr}(MU^\dagger-U M^\dagger)$ has the wrong 
behavior under parity $\phi(\vec{x},t)\mapsto-\phi(-\vec{x},t)$, because
\begin{eqnarray*}
\mbox{Tr}[M U^\dagger(\vec{x},t)-U(\vec{x},t)M^\dagger]
&\stackrel{P}{\mapsto}
&\mbox{Tr}[M U(-\vec{x},t)-U^\dagger(-\vec{x},t)M^\dagger]
\\
&\stackrel{M=M^\dagger}{=}&-\mbox{Tr}[M U^\dagger(-\vec{x},t)-
U(-\vec{x},t) M^\dagger].
\end{eqnarray*}
\item Because $M=M^\dagger$, ${\cal L}_{\rm s.b.}$ contains only terms even
in $\phi$.
\end{enumerate}
   In order to determine the masses of the Goldstone bosons, we identify the 
terms of second order in the fields in ${\cal L}_{\rm s.b.}$,
\begin{equation}
\label{4:3:lmzo}
{\cal L}_{\rm s.b}=-\frac{B_0}{2}\mbox{Tr}(\phi^2M) +\cdots.
\end{equation}
   Using Eq.\ (\ref{4:3:upar}) we find
\begin{eqnarray*}
\mbox{Tr}(\phi^2M)
&=&2(m_u+m_d)\pi^+\pi^-
+2(m_u+m_s)K^+ K^-
+2(m_d+m_s)K^0\bar{K}^0\\
&&+(m_u+m_d)\pi^0\pi^0 +\frac{2}{\sqrt{3}}(m_u-m_d)\pi^0\eta
+\frac{m_u+m_d+4m_s}{3}\eta^2.
\end{eqnarray*}
   For the sake of simplicity we consider the isospin-symmetric limit
$m_u=m_d=m$ so that the $\pi^0\eta$ term vanishes and there is no
$\pi^0$-$\eta$ mixing.
   We then obtain for the masses of the Goldstone bosons, to lowest order in 
the quark masses,
\begin{eqnarray}
\label{4:3:mpi2}
M^2_\pi&=&2 B_0 m,\\
\label{4:3:mk2}
M^2_K&=&B_0(m+m_s),\\
\label{4:3:meta2}
M^2_\eta&=&\frac{2}{3} B_0\left(m+2m_s\right).
\end{eqnarray}   
   These results, in combination with Eq.\ (\ref{4:3:b0}),
$B_0=-\langle\bar{q}q\rangle/(3 F_0^2)$, correspond relations
obtained in Ref.\ \cite{Gell-Mann:rz} and are referred to as
the Gell-Mann, Oakes, and Renner relations.
   Furthermore, the masses of Eqs.\ (\ref{4:3:mpi2}) - (\ref{4:3:meta2}) 
satisfy the Gell-Mann-Okubo relation
\begin{equation}
\label{4:3:gmof}
4M^2_K=4B_0(m+m_s)=2B_0(m+2m_s)+2B_0m=3M^2_\eta+M^2_\pi
\end{equation}
   independent of the value of $B_0$.
   Without additional input regarding the numerical value of $B_0$,
Eqs.\ (\ref{4:3:mpi2}) - (\ref{4:3:meta2}) do not allow for an extraction of 
the  absolute values of the quark masses $m$ and $m_s$, because rescaling 
$B_0\to \lambda B_0$ in combination with $m_q\to m_q/\lambda$ leaves the 
relations invariant.
   For the ratio of the quark masses one obtains, using the empirical
values of the pseudoscalar octet,
\begin{eqnarray}
\frac{M^2_K}{M^2_\pi}=\frac{m+m_s}{2m}&\Rightarrow&\frac{m_s}{m}=25.9,
\nonumber\\
\frac{M^2_\eta}{M^2_\pi}=\frac{2m_s+m}{3m}&\Rightarrow&\frac{m_s}{m}=24.3.
\end{eqnarray}

   Let us conclude this section with the following remark. 
   We saw in Sec.\ \ref{subsec_sqc} that a non-vanishing quark condensate in 
the chiral
limit is a sufficient but not a necessary condition for a spontaneous 
chiral symmetry breaking.
   The effective Lagrangian term of Eq.\ (\ref{4:3:lqm}) not only results in 
a shift of the vacuum energy but also in finite Goldstone boson 
masses.\footnote{Later on we will also see that the $\pi\pi$ scattering 
amplitude is effected by ${\cal L}_{\rm s.b.}$.}
   These are related via the parameter $B_0$ and we recall that it was
a symmetry argument which excluded a term $\mbox{Tr}(M)$ which,
at leading order in $M$, would decouple the vacuum energy shift
from the Goldstone boson masses.
   The scenario underlying ${\cal L}_{\rm s.b.}$ of Eq.\ (\ref{4:3:lqm}) 
is similar to that of a Heisenberg ferromagnet \cite{Ashcroft,Leutwyler:1991mz}
which exhibits a spontaneous magnetization $\langle \vec{M}\rangle$, breaking 
the O(3) symmetry of the Heisenberg Hamiltonian down to O(2).
   In the present case the analogue of the order parameter 
$\langle \vec{M}\rangle$ is the quark condensate $\langle \bar{q} q\rangle$. 
   In the case of the ferromagnet, the interaction with an external magnetic 
field is given by $-\langle \vec{M}\rangle\cdot \vec{H}$, which corresponds 
to Eq.\ (\ref{4:3:Heff}), with the quark masses playing the role of the 
external field $\vec{H}$.
   However, in principle, it is also possible that $B_0$ vanishes or is 
rather small.
   In such a case  the quadratic masses of the Goldstone bosons might
be dominated by terms which are nonlinear in the quark masses, i.e., by 
higher-order terms in the expansion of ${\cal L}(U,M)$.
   Such a scenario is the origin of the so-called generalized chiral
perturbation theory \cite{Knecht:1995tr,Knecht:1995ai,Stern:1997}.
   The analogue would be an antiferromagnet which shows a spontaneous
symmetry breaking but with $\langle \vec{M}\rangle=0$.
   
    The analysis of recent data on $K^+\to \pi^+\pi^-e^+\nu_e$ 
\cite{Pislak:2001bf} in terms of the isoscalar $s$-wave scattering length
$a_0^0$ \cite{Colangelo:2001sp} supports the conjecture that the quark 
condensate is indeed the leading order parameter of the spontaneously broken
chiral symmetry.
   For a recent discussion on the relation between the quark condensate 
and $s$-wave $\pi\pi$ scattering the interested reader is referred 
to Ref.\ \cite{Leutwyler:2001zc}.

\section{Effective Lagrangians and Weinberg's Power Counting Scheme}
\label{sec_elwpcs}
   An essential prerequisite for the construction of effective field theories
is a ``theorem'' of Weinberg stating that a perturbative
description in terms of the most general effective Lagrangian containing all
possible terms compatible with assumed symmetry principles yields the most
general $S$ matrix consistent with the fundamental principles of 
quantum field theory and the assumed symmetry principles
\cite{Weinberg:1978kz}. 
   The corresponding effective Lagrangian will contain an infinite number of 
terms with an infinite number of free parameters.
   Turning Weinberg's theorem into a practical tool requires two
steps: one needs some scheme to organize the effective Lagrangian and 
a systematic method of assessing the importance of diagrams generated by 
the interaction terms of this Lagrangian when calculating a physical matrix 
element.

    In the framework of mesonic chiral perturbation theory, the most general
chiral Lagrangian describing the dynamics of the Goldstone bosons is organized
as a string of terms with an increasing number of derivatives and quark mass 
terms,
\begin{equation}
\label{4:4:ll2l4}
{\cal L}_{\rm eff}={\cal L}_2 + {\cal L}_4 + {\cal L}_6 +\cdots,
\end{equation}
   where the subscripts refer to the order in the momentum and
quark mass expansion.
   The index 2, for example, denotes either two derivatives or one quark 
mass term.
   In the context of Feynman rules, derivatives generate four-momenta, 
whereas the convention of counting quark mass terms as being of the same order
as two derivatives originates from Eqs.\ (\ref{4:3:mpi2}) - (\ref{4:3:meta2})
in conjunction with the on-shell condition $p^2=M^2$.
   In an analogous fashion, ${\cal L}_4$ and ${\cal L}_6$ denote more
complicated terms of so-called chiral orders ${\cal O}(p^4)$ and 
${\cal O}(p^6)$ with corresponding numbers of derivatives and quark mass terms.
   With such a counting scheme, the chiral orders in the mesonic sector are 
always even [${\cal O}(p^{2n})$] because Lorentz indices of derivatives always
have to be contracted with either the metric tensor $g^{\mu\nu}$ or the
Levi-Civita tensor $\epsilon^{\mu\nu\rho\sigma}$ to generate scalars,
and the quark mass terms are counted as ${\cal O}(p^2)$.

   Weinberg's power counting scheme \cite{Weinberg:1978kz} analyzes
the behavior of a given diagram under a linear rescaling of all the 
{\em external} momenta, $p_i\mapsto t p_i$, and a quadratic rescaling of the 
light quark masses, $m_q\mapsto t^2 m_q$, which, in terms of the Goldstone 
boson masses, corresponds to $M^2\mapsto t^2 M^2$.
   The chiral dimension $D$ of a given diagram with amplitude
${\cal M}(p_i,m_q)$ is defined by
\begin{equation}
\label{4:4:mr1}
{\cal M}(tp_i, t^2 m_q)=t^D {\cal M}(p_i,m_q),
\end{equation}
and thus
\begin{equation}
\label{4:4:mr2}
D=2+\sum_{n=1}^\infty2(n-1)N_{2n} +2N_L,
\end{equation}
where $N_{2n}$ denotes the number of vertices originating from 
${\cal L}_{2n}$, and $N_L$ is the number of independent loops.
   Clearly, for small enough momenta and masses diagrams with small $D$, such 
as $D=2$ or $D=4$, should dominate.
   Of course, the rescaling of Eq.\ (\ref{4:4:mr1}) must be viewed as 
a mathematical tool.
   While external three-momenta can, to a certain extent, be made arbitrarily
small, the rescaling of the quark masses is a theoretical instrument only.
   Note that loop diagrams are always suppressed due to the term $2N_L$ in 
Eq.\ (\ref{4:4:mr2}).    
   It may happen, though, that the leading-order tree diagrams vanish and
therefore that the lowest-order contribution to a certain process is
a one-loop diagram.
   An example is the reaction 
$\gamma\gamma\to\pi^0\pi^0$ \cite{Bijnens:1987dc}.

   In order to prove Eq.\ (\ref{4:4:mr2}) we start from the usual Feynman
rules for evaluating an $S$-matrix element (see, e.g., Appendix A-4 of 
Ref.\ \cite{Itzykson:rh}).
   Each internal meson line contributes a factor
\begin{eqnarray}
\label{4:4:intlines}
\int\frac{d^4k}{(2\pi)^4} \frac{i}{k^2-M^2+i\epsilon}
&\stackrel{\mbox{$(M^2\mapsto t^2 M^2)$}}{\mapsto}&
t^{-2}\int\frac{d^4k}{(2\pi)^4} \frac{i}{k^2/t^2-M^2+i\epsilon}\nonumber\\
&\stackrel{\mbox{$(k=tl)$}}{=}&
t^2 \int\frac{d^4l}{(2\pi)^4} \frac{i}{l^2-M^2+i\epsilon}.
\end{eqnarray}
   For each vertex, originating from ${\cal L}_{2n}$, we obtain symbolically
a factor $p^{2n}$ together with a four-momentum conserving delta function 
resulting in $t^{2n}$ for the vertex factor and $t^{-4}$ for the delta 
function.
   At this point one has to take into account the fact that, although 
Eq.\ (\ref{4:4:mr1}) refers to a rescaling of {\em external} momenta,
a substitution $k=tl$ for internal momenta as in Eq.\ (\ref{4:4:intlines}) 
acts in exactly the same way as a rescaling of external momenta:
\begin{eqnarray*}
\delta^4(p+k)&\stackrel{\mbox{$p\mapsto tp, k=tl$}}{\mapsto}t^{-4}
\delta^4(p+l),\\
p^{2n-m}k^m&\stackrel{\mbox{$p\mapsto tp, k=tl$}}{\mapsto}t^{2n}p^{2n-m}l^m,
\end{eqnarray*}
where $p$ and $k$ denote external and internal momenta, respectively.

   So far we have discussed the rules for determining the power $D_S$
referring to the $S$-matrix element which is related to the invariant amplitude
through a four-momentum conserving delta function, 
$$S\sim \delta^4(P_f-P_i){\cal M}.$$
   The delta function contains external momenta only, and thus re-scales
under $p_i\mapsto tp_i$ as $t^{-4}$, so
$$t^{D_S}=t^{-4}t^D.$$
   We thus find as an intermediate result 
\begin{equation}
\label{4:4:d}
D=4+2N_I+\sum_{n=1}^\infty N_{2n}(2n-4),
\end{equation}  
   where $N_I$ denotes the number of internal lines.
   The number of independent loops, total number of vertices, and number of
internal lines are related by\footnote{Note that the number of independent
momenta is {\em not} the number of faces or closed circuits that may be
drawn on the internal lines of a diagram. 
This may, for example, be seen using a diagram with the topology of 
a tetrahedron which has four faces but 
$N_L=6-(4-1)=3$ (see, e.g., Chap.\ 6-2 of Ref.\ 
\cite{Itzykson:rh}).} 
$$N_L=N_I-(N_V-1),$$
   because each of the $N_V$ vertices generates a delta function.
   After extracting one overall delta function this yields $N_V-1$ 
conditions for the internal momenta.
   Using $N_V=\sum_{n} N_{2n}$ we finally obtain from Eq.\ (\ref{4:4:d}) 
$$
D=4+2(N_L+N_V-1)+\sum_{n=1}^\infty N_{2n}(2n-4)
=2+2 N_L +\sum_{n=1}^\infty N_{2n}(2n-2).\quad
$$

\begin{figure}
\begin{center}
\epsfig{file=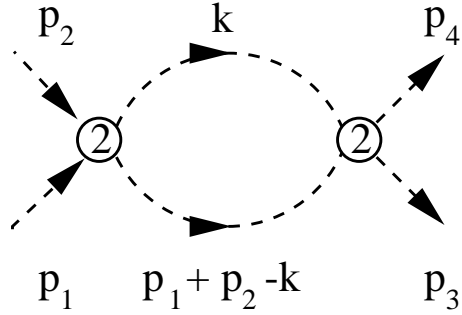,width=6cm}
\caption{\label{4:4:figexamplerescaling} Typical one-loop diagram contributing
to the scattering of two particles.}
\end{center}
\end{figure}

   By means of a simple example we will illustrate how the mechanism
of rescaling actually works.
   To that end we consider as a toy model of an effective field theory
the self interaction of a scalar field,
\begin{equation}
\label{4:4:l2bsp}
{\cal L}_2=g \Phi^2\partial_\mu\Phi\partial^\mu\Phi,
\end{equation}
   where the coupling constant $g$ has the dimension of 
$\mbox{energy}^{-2}$.\footnote{Recall that the dimensions of a Lagrangian 
density and a field $\Phi$ are energy$^4$ and energy,
respectively.}
\begin{figure}
\begin{center}
\epsfig{file=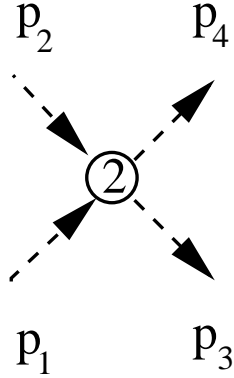,width=3cm}
\caption{\label{4:4:figexamplerescaling1} Tree-level diagram corresponding
to Eq.\ (\ref{4:4:frl2bsp}).}
\end{center}
\end{figure}
   The Feynman rules give the amplitude corresponding to the simple
tree diagram of Fig.\ \ref{4:4:figexamplerescaling1} for the scattering
of two particles,
\begin{eqnarray}
\label{4:4:frl2bsp}
{\cal M}(p_1,p_2,;p_3,p_4)
&=&4ig \left[(p_1+p_2)\cdot(p_3+p_4)-p_1\cdot p_2-p_3\cdot p_4\right]
\nonumber\\
&\stackrel{\mbox{$p_i\mapsto t p_i$}}{\mapsto}& t^2 {\cal M}(p_1,p_2;p_3,p_4).
\end{eqnarray} 
   As expected, the behavior under rescaling is in agreement with 
Eq.\ (\ref{4:4:mr2}) for $N_L=0$, $N_2=1$, and $N_{2n}=0$ for all
remaining $n$.
   Now let us consider a typical loop diagram of Fig.\ 
\ref{4:4:figexamplerescaling} contributing to the same process,
where the 2 in the interaction blob indicates the ${\cal L}_2$ term
in the Lagrangian containing two derivatives.   
   Applying the usual Feynman rules, with the vertex of 
Eq.\ (\ref{4:4:frl2bsp}), we obtain
\begin{eqnarray}
\label{4:4:mbsp2}
{\cal M}&=&\frac{1}{2}
\int \frac{d^4k}{(2\pi)^4}\nonumber\\
&&\times4ig\left[
(p_1+p_2-k+k)\cdot(p_3+p_4)-(p_1+p_2-k)\cdot k-p_3\cdot p_4\right]\nonumber\\
&&\times\frac{i}{k^2-M^2+i\epsilon}\nonumber\\
&&\times\frac{i}{(p_1+p_2-k)^2-M^2+i\epsilon}\nonumber\\
&&\times
4ig\left[(p_1+p_2)\cdot(p_3+p_4-k+k)-p_1\cdot p_2-(p_1+p_2-k)\cdot k\right]
\nonumber\\
&=& 8g^2\int \frac{d^4k}{(2\pi)^4}
\left[(p_1+p_2)\cdot(p_3+p_4)-(p_1+p_2-k)\cdot k-p_3\cdot p_4\right]
\nonumber\\
&&\times\frac{1}{k^2-M^2+i\epsilon}
\frac{1}{(p_1+p_2-k)^2-M^2+i\epsilon}\nonumber\\
&&\times
\left[(p_1+p_2)\cdot(p_3+p_4)-p_1\cdot p_2-(p_1+p_2-k)\cdot k\right]
\nonumber\\
&\stackrel{\stackrel{p_i\mapsto tp_i}{M^2\mapsto t^2 M^2}}{\mapsto}&
8 g^2 \int \frac{d^4k}{(2\pi)^4}\left[
(p_1+p_2)\cdot(p_3+p_4)-(p_1+p_2-\frac{k}{t})\cdot \frac{k}{t}- 
p_3\cdot p_4\right]
\nonumber\\
&&\times \frac{1}{\frac{k^2}{t^2}-M^2+i\epsilon}
\frac{1}{(p_1+p_2-\frac{k}{t})^2-M^2+i\epsilon}\nonumber\\
&&\times
\left[(p_1+p_2)\cdot(p_3+p_4)-p_1\cdot p_2-(p_1+p_2-\frac{k}{t})
\cdot \frac{k}{t}\right]\nonumber\\
&\stackrel{\mbox{$tl=k$}}{=}
&8 g^2 \int \frac{t^4d^4l}{(2\pi)^4}\left[
(p_1+p_2)\cdot(p_3+p_4)-(p_1+p_2-l)\cdot l-p_3\cdot p_4\right]
\nonumber\\
&&\times \frac{1}{l^2-M^2+i\epsilon}
\frac{1}{(p_1+p_2-l)^2-M^2+i\epsilon}\nonumber\\
&&\times\left[(p_1+p_2)\cdot(p_3+p_4)-p_1\cdot p_2-(p_1+p_2-l)\cdot l\right]
\nonumber\\
&=&t^4 {\cal M},
\end{eqnarray} 
   This agrees with the value $D=4$ given by Eq.\ (\ref{4:4:mr2}) for 
$N_L=1$ and $N_{2}=2$.

   For the sake of completeness, let us comment on the symmetry factor 
1/2 in Eq.\ (\ref{4:4:mbsp2}). 
   When deriving the Feynman rule of Eq.\ (\ref{4:4:frl2bsp}),
we took account of $4!=24$ distinct combinations of contracting 
four field operators with four external lines.
   The ``product'' of two such vertices thus contains $24\times 24$ 
combinations. 
   However, from each vertex two lines have to be selected as 
internal lines and there exist 6 possibilities to choose one pair out of 4 
field operators to form internal lines.
   For the two remaining operators one has two possibilities of
contracting them with external lines.
   Finally, the respective pairs of internal lines of the first
and second vertices may be contracted in two ways with each other,
leaving us with $12\times 12\times 2=
(24\times 24)/2$ combinations.
   
   In the discussion of the loop integral we did not address the question 
of convergence.
   This needs to be addressed since applying the substitution $tl=k$ in 
Eq.\ (\ref{4:4:mbsp2}) is well-defined only for convergent integrals.
   Later on we will regularize the integrals by use of the method of
dimensional regularization, introducing a renormalization scale $\mu$ which
also has to be rescaled linearly.
   However, at a given chiral order, the sum of all diagrams will, by 
construction, not depend on the renormalization scale.

   Finally, the proof of Weinberg's theorem 
\cite{Leutwyler:1993iq,D'Hoker:1994ti} for chiral perturbation theory is 
rather technical and lengthy and beyond the scope of this review.
   In Ref.\ \cite{Leutwyler:1993iq} it was shown that global symmetry 
constraints alone do not suffice to fully determine the low-energy structure
of the effective Lagrangian.
   In fact, a determination of the (low-energy) Green functions of 
QCD off the mass shell, i.e., for momenta which do not correspond 
to the mass-shell conditions for Goldstone bosons, one needs to study
the Ward identities, and therefore the symmetries have to be
extended to the local level.
   One thus considers a {\em locally} invariant, effective
Lagrangian although the symmetries of the underlying theory originate
in a global symmetry.
   If the Ward identities contain anomalies, they show up as a modification
of the generating functional, which can explicitly be incorporated through
the Wess-Zumino-Witten construction \cite{Wess:yu,Witten:tw}.

\section{Construction of the Effective Lagrangian}
\label{sec_cel}
   In Sec.\ \ref{sec_loel} we have derived the lowest-order effective 
Lagrangian for a {\em global} $\mbox{SU(3)}_L\times\mbox{SU(3)}_R$ symmetry.
   On the other hand, the Ward identities originating in the global 
$\mbox{SU(3)}_L\times\mbox{SU(3)}_R$ symmetry of QCD are obtained from a 
{\em locally} invariant generating functional involving a coupling to 
external fields (see Sec.\ \ref{subsec_qcdpefgf} and App.\ \ref{app_gfwi}).
   Our goal is to approximate the ``true'' generating functional 
$Z_{\rm QCD}[v,a,s,p]$ of Eq.\ (\ref{2:4:genfun}) by a sequence 
$Z^{(2)}_{\rm eff}[v,a,s,p] + Z^{(4)}_{\rm eff}[v,a,s,p]
+\cdots$, where the effective generating 
functionals are obtained using the effective field theory.
   Therefore, we need to promote the global symmetry of the effective 
Lagrangian to a local one and introduce a coupling to the {\em same} external 
fields $v$, $a$, $s$, and $p$ as in QCD.

   In the following we will outline the principles entering the construction 
of the effective Lagrangian for a local 
$G=\mbox{SU(3)}_L\times\mbox{SU(3)}_R$ symmetry
(see Refs.\ \cite{Fearing:1994ga,Bijnens:1999sh,Ebertshauser:2001nj}
for details).\footnote{In principle, we could also ``gauge'' the
U(1)$_V$ symmetry. However, this is primarily of relevance to the SU(2)
sector in order to fully incorporate the coupling to the electromagnetic
field [see Eq.\ (\ref{2:4:rlasu2})]. 
Since in SU(3), the quark-charge matrix is traceless, this important
case is included in our considerations. For further discussions,
see Ref.\ \cite{Ebertshauser:2001nj}.} 
   The matrix $U$ transforms as $U \mapsto U'=V_R U V_L^{\dagger}$, 
where $V_L(x)$ and $V_R(x)$ are independent space-time-dependent SU(3) 
matrices.
   As in the case of gauge theories, we need external fields 
$l_\mu^a(x)$ and $r_\mu^a(x)$ 
[see Eqs.\ (\ref{2:4:mch}), (\ref{2:4:vrlarl}), and 
(\ref{2:4:sg}) and Table \ref{4:5:table_trafprop}]
 corresponding to the parameters $\Theta^L_a(x)$
and $\Theta^R_a(x)$ of $V_L(x)$ and $V_R(x)$, respectively.
   For any object $A$ transforming as $V_R A V_L^\dagger$ such as, e.g., $U$ 
we define the covariant derivative $D_\mu A$ as
\begin{eqnarray}
\label{4:5:kaa}
D_\mu A&\equiv&\partial_\mu A -i r_\mu A+iA l_\mu\nonumber\\
&\mapsto&\partial_\mu(V_R A V_L^\dagger)-i(V_R r_\mu V_R^\dagger+i
V_R\partial_\mu V_R^\dagger)V_R A V_L^\dagger\nonumber\\
&&+iV_R A V_L^\dagger (V_L l_\mu V_L^\dagger +i V_L \partial_\mu V_L^\dagger)
\nonumber\\
&=&\partial_\mu V_R A V_L^\dagger+ V_R\partial_\mu A V_L^\dagger
+V_R A \partial_\mu V_L^\dagger-i V_R r_\mu A V_L^\dagger
-\partial_\mu V_R A V_L^\dagger\nonumber\\
&& +i V_R A l_\mu V_L^\dagger
-V_R A \partial_\mu V_L^\dagger\nonumber\\
&=& V_R(\partial_\mu A -i r_\mu A+iA l_\mu)V_L^\dagger= 
V_R (D_\mu A) V_L^\dagger, 
\end{eqnarray}
   where we made use of 
$V_R\partial_\mu V_R^\dagger=-\partial_\mu V_R V_R^\dagger$.
   Again, the defining property for the covariant derivative is that it 
should transform in the same way as the object it acts on.\footnote{Under
certain circumstances it is advantageous to introduce for each object with
a well-defined transformation behavior a separate covariant derivative.
   One may then use a product rule similar to the one
of ordinary differentiation [see Eqs.\ (18) and (19) of Ref.\
\cite{Fearing:1994ga}].} 
   Since the effective Lagrangian will ultimately contain arbitrarily high
powers of derivatives we also need the field strength tensors 
$f^L_{\mu\nu}$ and $f^R_{\mu\nu}$ corresponding to the gauge fields,
\begin{eqnarray}
\label{4:5:fr}
f_{\mu\nu}^R&\equiv&\partial_\mu r_\nu-\partial_\nu r_\mu-i{[r_\mu,r_\nu]},\\
\label{4:5:fl}
f_{\mu\nu}^L&\equiv&\partial_\mu l_\nu-\partial_\nu l_\mu-i{[l_\mu,l_\nu]}.
\end{eqnarray}
   The field strength tensors are traceless,
\begin{equation}
\label{4:5:trflfr}
\mbox{Tr}(f^L_{\mu\nu})=\mbox{Tr}(f^R_{\mu\nu})=0,
\end{equation}
because $\mbox{Tr}(l_\mu)=\mbox{Tr}(r_\mu)=0$ and the trace of any
commutator vanishes.
   Finally, following the convention of Gasser and Leutwyler we introduce
the linear combination $\chi\equiv 2B_0(s+ip)$ with the scalar and pseudoscalar
external fields of Eq.\ (\ref{2:4:mch}),  where $B_0$ is defined in 
Eq.\ (\ref{4:3:b0}).
   Table \ref{4:5:table_trafprop} contains the transformation 
properties of all building blocks under the group ($G$), 
charge conjugation ($C$), and parity ($P$).

\begin{table}
\caption[test]{\label{4:5:table_trafprop}
Transformation properties under the group ($G$), charge conjugation ($C$),
and parity ($P$).  
    The expressions for adjoint matrices are trivially obtained 
by taking the Hermitian conjugate of each entry. 
   In the parity transformed 
expression it is understood that the argument is $(-\vec{x},t)$ and that
partial derivatives $\partial_{\mu}$ act with respect to $x$ and not with 
respect to the argument of the corresponding function.}
\begin{center}
\begin{tabular}{|c|c|c|c|}
\hline
element&$G$&$C$&$P$\\
\hline
$U$&$V_R U V_L^\dagger$&$U^T$&$U^\dagger$\\
\hline
$D_{\lambda_1}\cdots D_{\lambda_n}U$&
$V_R D_{\lambda_1}\cdots D_{\lambda_n}U V_L^\dagger$&
$(D_{\lambda_1}\cdots D_{\lambda_n}U)^T$&
$(D^{\lambda_1}\cdots D^{\lambda_n}U)^\dagger$\\
\hline
$\chi$&$V_R \chi V_L^\dagger$&$\chi^T$&$\chi^\dagger$\\
\hline
$D_{\lambda_1}\cdots D_{\lambda_n}\chi$&
$V_R D_{\lambda_1}\cdots D_{\lambda_n}\chi V_L^\dagger$&
$(D_{\lambda_1}\cdots D_{\lambda_n}\chi)^T$&
$(D^{\lambda_1}\cdots D^{\lambda_n}\chi)^\dagger$\\
\hline
$r_\mu$&$V_R r_\mu V_R^\dagger+iV_R\partial_\mu 
V^\dagger_R$&$-l_\mu^T$&$l^\mu$\\
\hline
$l_\mu$&$V_L l_\mu V_L^\dagger+iV_L\partial_\mu 
V^\dagger_L$&$-r_\mu^T$&$r^\mu$\\
\hline
$f^R_{\mu\nu}$&$V_R f^R_{\mu\nu}V_R^\dagger$&
$-(f_{\mu\nu}^L)^T$&$f_L^{\mu\nu}$\\
\hline
$f^L_{\mu\nu}$&$V_L f^L_{\mu\nu}V_L^\dagger$&
$-(f_{\mu\nu}^R)^T$&$f_R^{\mu\nu}$\\
\hline
\end{tabular}
\end{center}
\end{table}

  In the chiral counting scheme of chiral perturbation theory the elements
are counted as:
\begin{equation}
\label{4:5:powercounting}
U =  {\cal O}(p^0),\, D_{\mu} U  = {\cal  O}(p),\, r_{\mu},l_{\mu}  
= {\cal O}(p),\,
f^{L/R}_{\mu\nu}  =  {\cal O}(p^2),\, \chi  =  {\cal O}(p^2).
\end{equation}
   The external fields $r_{\mu}$ and $l_{\mu}$ count as ${\cal O}(p)$ 
to match $\partial_\mu A$, and $\chi$ is of ${\cal O}(p^2)$ because
of Eqs.\ (\ref{4:3:mpi2}) - (\ref{4:3:meta2}).
   Any additional covariant derivative counts as ${\cal O}(p)$.

   The construction of the effective Lagrangian in terms of the building blocks
of Eq.\ (\ref{4:5:powercounting}) proceeds as follows.\footnote{There is
a certain freedom in the choice of the elementary building blocks.
   For example, by a suitable multiplication with $U$ or $U^\dagger$ any
building block can be made to transform as $V_R \cdots V_R^\dagger$ without
changing its chiral order \cite{Fearing:1994ga}.
   The present approach most naturally leads to the Lagrangian of Gasser
and Leutwyler \cite{Gasser:1984gg}.}
   Given objects $A,B,\dots$, all of which transform as
\mbox{$A'=V_R A V_L^{\dagger},$} \mbox{$B'= V_R B V_L^{\dagger},\,\dots,$}
one can form invariants by taking the trace of products of the
type $A B^{\dagger}$:
\begin{eqnarray*}
\mbox{Tr}(A B^\dagger)&\mapsto& \mbox{Tr}[V_R A V_L^\dagger
(V_R B V_L^\dagger)^\dagger]=\mbox{Tr}(V_R A V_L^\dagger V_L B^\dagger 
V_R^\dagger)=\mbox{Tr}(A B^\dagger V_R^\dagger V_R)\\
&=&\mbox{Tr}(A B^\dagger).
\end{eqnarray*}
The generalization to more terms is obvious and, of course, the product of
invariant traces is invariant:
\begin{equation}
\label{4:5:ketten}
\mbox{Tr}(AB^\dagger C D^\dagger),\quad
\mbox{Tr}(A B^\dagger)\mbox{Tr}(C D^\dagger),\quad \cdots.
\end{equation}
   The complete list of elements up to and including order ${\cal O}(p^2)$ 
transforming as $V_R\cdots V_L^\dagger$ reads
\begin{equation}
\label{4:5:lis}
U, D_\mu U, D_\mu D_\nu U,\chi, U f_{\mu\nu}^L, f^R_{\mu\nu}U.
\end{equation} 
   For the invariants up to ${\cal O}(p^2)$ we then obtain
\begin{eqnarray}
\label{4:5:invariants}
{\cal O}(p^0)&:& \mbox{Tr}(UU^\dagger)=3,\nonumber\\
{\cal O}(p)&:& \mbox{Tr}(D_\mu U U^\dagger)
\stackrel{\ast}{=}-\mbox{Tr}[U (D_\mu U)^\dagger]
\stackrel{\ast}{=}0,\nonumber
\\
{\cal O}(p^2)&:& \mbox{Tr}(D_\mu D_\nu U U^\dagger)
\stackrel{\ast\ast}{=}-\mbox{Tr}[D_\nu U (D_\mu U)^\dagger]
\stackrel{\ast\ast}{=}\mbox{Tr}[U(D_\nu D_\mu U)^\dagger],\nonumber\\
&&\mbox{Tr}(\chi U^\dagger),\nonumber\\
&&\mbox{Tr}(U \chi^\dagger),\nonumber\\
&&\mbox{Tr}(U f^L_{\mu\nu}U^\dagger)=\mbox{Tr}(f^L_{\mu\nu})=0,\nonumber\\
&&\mbox{Tr}(f^R_{\mu\nu})=0.
\end{eqnarray}
   In $\ast$ we made use of two important properties of the covariant
derivative $D_\mu U$:
\begin{eqnarray}
\label{4:5:kauprop1}
D_\mu U U^\dagger &=& -U (D_\mu U)^\dagger,\\
\label{4:5:kauprop2}
\mbox{Tr}(D_\mu U U^\dagger)&=&0.
\end{eqnarray}
   The first relation results from the unitarity of $U$ in combination
with the definition of the covariant derivative, Eq.\ (\ref{4:5:kaa}).
   Equation (\ref{4:5:kauprop2}) is shown using 
$\mbox{Tr}(r_\mu)=\mbox{Tr}(l_\mu)=0$ 
together with Eq.\ (\ref{4:3:trpmuuud}), 
$\mbox{Tr}(\partial_\mu U U^\dagger)=0$:
\begin{eqnarray*}
\mbox{Tr}(D_\mu U U^\dagger)&=&\mbox{Tr}(\partial_\mu U U^\dagger
-ir_\mu U U^\dagger
+iUl_\mu U^\dagger)=0.
\end{eqnarray*}
   The relations $\ast\ast$ can either be verified by explicit calculation
or, more elegantly, using the product rule of Ref.\ \cite{Fearing:1994ga}
for the covariant derivatives.
 
   Finally, we impose Lorentz invariance, i.e., Lorentz indices
have to be contracted, resulting in three candidate terms:
\begin{eqnarray}
\label{4:5:lis1}
&&\mbox{Tr}[D_\mu U (D^\mu U)^\dagger],\\
\label{4:5:lis2}
&&\mbox{Tr}(\chi U^\dagger\pm U \chi^\dagger).
\end{eqnarray}
   The term in Eq.\ (\ref{4:5:lis2}) with the minus sign is excluded because
it has the wrong sign under parity (see Table \ref{4:5:table_trafprop}), 
and we end up with the most general, {\em locally} invariant, 
effective Lagrangian at lowest chiral order,\footnote{At ${\cal O}(p^2)$ 
invariance under $C$ does not
provide any additional constraints.}
\begin{equation}
\label{4:5:l2}
{\cal L}_2=\frac{F_0^2}{4}\mbox{Tr}[D_\mu U (D^\mu U)^\dagger]
+\frac{F^2_0}{4}\mbox{Tr}(\chi U^\dagger + U\chi^\dagger).
\end{equation}
   Note that ${\cal L}_2$ contains two free parameters:
the pion-decay constant $F_0$ and $B_0$ of Eq.\ (\ref{4:3:b0})
(hidden in the definition of $\chi$).

   Let us finally derive the equations of motion associated with
the lowest-order Lagrangian. 
   These are important because they can be used to eliminate so-called
equation-of-motion terms in the construction of the higher-order
Lagrangians 
\cite{Georgi:1991ch,Leutwyler:1991mz,Grosse-Knetter:1993td,Arzt:gz,%
Scherer:1994wi} by applying field transformations
\cite{Chisholm,Kamefuchi:sb}.
   To that end we need to consider an infinitesimal change of the
SU(3) matrix $U(x)$.
   Since the set of SU(3) matrices forms a group, to each pair of 
elements $U$ and $U'$ corresponds a unique element $\tilde{U}$, 
connecting the two via $U'=\tilde{U}U$.
   Let us parameterize $\tilde{U}$ by means of the Gell-Mann matrices,
\begin{equation}
\label{4:5:utilde}
\tilde{U}=\exp(i\Delta),\quad \Delta=\sum_{a=1}^8 \lambda_a \Delta_a,
\quad \Delta_a\in R,
\end{equation}
and consider small variations of the SU(3) matrix as
\begin{equation}
\label{4:5:deltau}
U'(x)=U(x)+\delta U(x)=\left(1+i\sum_{a=1}^8 \Delta_a(x)\lambda_a\right) U(x),
\end{equation}
where the $\Delta_a(x)$ are now real functions.
   With such an ansatz, the matrix $U'$ satisfies both conditions
\begin{equation}
\label{4:5:upcond}
U' U'^\dagger= 1, \quad \mbox{det}(U')=1,
\end{equation}
up to and including the terms linear in $\Delta_a(x)$.\footnote{
Some derivations in the literature neglect the second condition
of Eq.\ (\ref{4:5:upcond}) and thus obtain the wrong equations of motion.}
   Given the fields at $t_1$ and $t_2$, the dynamics is determined by the 
principle of stationary action.   
   We obtain for the variation of the action
\begin{eqnarray}
\label{4:5:deltas}
\delta S&=&\frac{F^2_0}{4}\int_{t_1}^{t_2}dt\int d^3 x\, \mbox{Tr}\left[
D_\mu \delta U(D^\mu U)^\dagger +D_\mu U (D^\mu \delta U)^\dagger
+\chi \delta U^\dagger +\delta U \chi^\dagger\right]\nonumber\\
&=&\frac{F^2_0}{4}\int_{t_1}^{t_2}dt\int d^3 x\,\mbox{Tr}\left[
-\delta U (D_\mu D^\mu U)^\dagger
-D_\mu D^\mu U \delta U^\dagger
+\chi \delta U^\dagger + \delta U \chi^\dagger \right]\nonumber\\
&=&i\frac{F^2_0}{4}\int_{t_1}^{t_2}dt \int  d^3 x \sum_{a=1}^8
\Delta_a(x)\nonumber\\
&&
\times \mbox{Tr}\left\{\lambda_a[
D_\mu D^\mu U U^\dagger- U(D_\mu D^\mu U)^\dagger -\chi U^\dagger
+ U\chi^\dagger]\right\}.
\end{eqnarray}
   In the second equation we made use of the standard boundary conditions
$\Delta_a(t_1,\vec{x})=\Delta_a(t_2,\vec{x})=0$, the divergence theorem,
and the definition of the covariant derivative of Eq.\ (\ref{4:5:kaa}).
   The third equality results from 
$\delta U^\dagger=-U^\dagger\delta U U^\dagger$ and the invariance
of the trace with respect to cyclic permutations.
   The functions $\Delta_a(x)$ may be chosen arbitrarily, and we obtain
eight Euler-Lagrange equations
\begin{equation}
\label{4:5:eom8}
\mbox{Tr}\left\{\lambda_a[D^2U U^\dagger - U (D^2 U)^\dagger
-\chi U^\dagger + U\chi^\dagger]\right\}=0, \quad a=1,\cdots, 8.
\end{equation}
Since any $3\times 3$ matrix $A$ can be written as
\begin{equation}
\label{4:5:apar}
A=a_0 1_{3\times 3}+\sum_{i=1}^8 a_i \lambda_i,\quad 
a_0=\frac{1}{3}\mbox{Tr}(A),
\quad a_i=\frac{1}{2}\mbox{Tr}(\lambda_i A),
\end{equation}
the eight equations of motion of Eq.\ (\ref{4:5:eom8}) may compactly
be written in matrix form\footnote{
Applying Eq.\ (\ref{4:5:invariants}) one finds 
$\mbox{Tr}[D^2UU^\dagger-U(D^2 U)^\dagger]=0$.}
\begin{equation}
\label{4:5:eom}
{\cal O}_{\rm EOM}^{(2)}(U)\equiv D^2 U U^\dagger - U (D^2 U)^\dagger 
-\chi U^\dagger + U \chi^\dagger
+\frac{1}{3}\mbox{Tr}(\chi U^\dagger- U \chi^\dagger)=0.
\end{equation}
   The additional term involving the trace is included to 
guarantee that the component proportional to the identity matrix
vanishes identically and thus one does not erroneously generate
a ninth equation of motion.

\section{Applications at Lowest Order}
\label{sec_alo}

   Let us consider two simple 
examples at lowest order $D=2$.
   According to Eq.\ (\ref{4:4:mr2}) we only need to consider tree-level
diagrams with vertices of ${\cal L}_2$.

\subsection{Pion Decay $\pi^+\to \mu^+\nu_\mu$}   
\label{subsec_pdpmn}
\begin{figure}
\epsfig{file=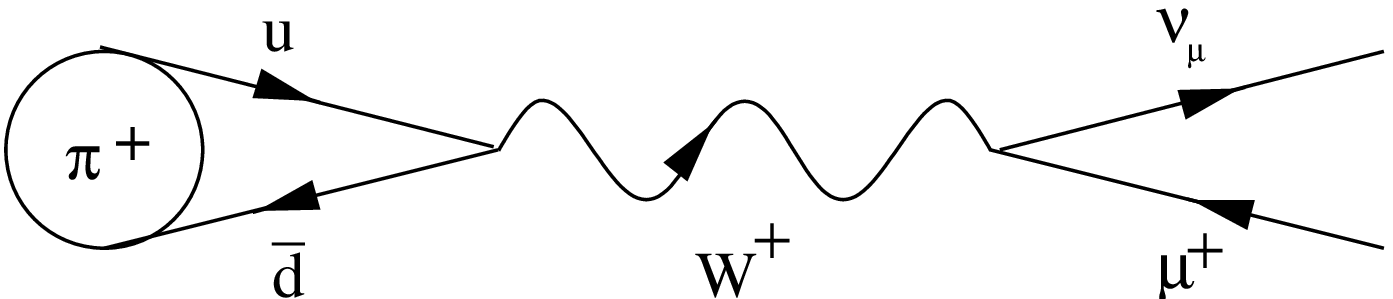,width=12cm}
\caption{\label{4:6:piondecay}
Pion decay $\pi^+\to \mu^+\nu_\mu$.}
\end{figure}

   Our first example deals with the weak decay $\pi^+\to \mu^+\nu_\mu$ which 
will allow us to relate the free parameter $F_0$ of ${\cal L}_2$ 
to the pion-decay constant.
   At the level of the degrees of freedom of the Standard Model,
pion decay is described by the annihilation of a $u$ quark and a $\bar{d}$
antiquark, forming the $\pi^+$, into a $W^+$ boson, propagation of the 
intermediate $W^+$, and creation of the leptons $\mu^+$ and 
$\nu_\mu$ in the final state (see Fig.\ \ref{4:6:piondecay}).
    The coupling of the $W$ bosons to the leptons is given by 
\begin{equation}
\label{4:6:lwl}
{\cal L}=-\frac{g}{2\sqrt{2}}\left[{\cal W}^+_\mu\bar{\nu}_\mu
\gamma^\mu(1-\gamma_5)\mu+{\cal W}^-_\mu\bar{\mu}\gamma^\mu
(1-\gamma_5)\nu_\mu\right],
\end{equation}
    whereas their interaction with the quarks forming the Goldstone
bosons is effectively taken into account by inserting Eq.\ (\ref{2:4:rlw}) 
into the Lagrangian of Eq.\ (\ref{4:5:l2}).
   Let us consider the first term of Eq.\ (\ref{4:5:l2}) and set $r_\mu=0$
with, at this point, still arbitrary $l_\mu$. 
   Using $D_\mu U=\partial_\mu U+iUl_\mu$ we find 
\begin{eqnarray*}
\frac{F^2_0}{4}\mbox{Tr}[D_\mu U (D^\mu U)^\dagger]
&=&\frac{F^2_0}{4}\mbox{Tr}[(\partial_\mu U +iU l_\mu)
(\partial^\mu U^\dagger -i l^\mu U^\dagger)]\\
&=&\cdots +i\frac{F^2_0}{4}\mbox{Tr}(Ul_\mu \partial^\mu U^\dagger 
-l^\mu\underbrace{U^\dagger\partial_\mu U}_{\mbox{$-\partial_\mu U^\dagger U$}}
)+\cdots\\
&=&i\frac{F^2_0}{2}\mbox{Tr}(l_\mu \partial^\mu U^\dagger U)+\cdots,
\end{eqnarray*}
   where only the term linear in $l_\mu$ is shown.
   If we parameterize
$$l_\mu =\sum_{a=1}^8\frac{\lambda_a}{2}l^a_\mu,$$
the interaction term linear in $l_\mu$ reads 
\begin{equation}
\label{4:6:lwli}
{\cal L}_{\rm int}=\sum_{a=1}^8l_\mu^a\left[i \frac{F_0^2}{4}\mbox{Tr}(
\lambda_a \partial^\mu U^\dagger U)\right]=
\sum_{a=1}^8l^a_\mu J^{\mu,a}_L,
\end{equation}
   where we made use of Eq.\ (\ref{4:3:jl}) defining $J^{\mu,a}_L$.
   Again, we expand $J^{\mu,a}_L$ by using Eq.\ (\ref{4:3:upar}) to first 
order in $\phi$,
\begin{eqnarray}
\label{4:6:jlent}
J^{\mu,a}_L&=&
\frac{F_0}{2}\partial^\mu \phi^a+O(\phi^2),
\end{eqnarray}
from which we obtain the matrix element 
\begin{equation}
\label{4:6:lpi}
\langle 0|J^{\mu,a}_L(0)|\phi^b(p)\rangle
=\frac{F_0}{2}\langle 0|\partial^\mu \phi^a(0)|\phi^b(p)\rangle
=-ip^\mu \frac{F_0}{2}\delta^{ab}.
\end{equation}
   Inserting $l_\mu$ of Eq.\ (\ref{2:4:rlw}), we find for the interaction
term of a single Goldstone boson with a $W$
$$
{\cal L}_{W\phi}=\frac{F_0}{2}\mbox{Tr}(l_\mu \partial^\mu \phi)
= -\frac{g}{\sqrt{2}}\frac{F_0}{2}\mbox{Tr}
[({\cal W}_\mu^+T_+ + {\cal W}^-_\mu T_-)\partial^\mu\phi].
$$
   Thus, we need to calculate\footnote{Recall that the entries $V_{ud}$ 
and $V_{us}$ of the Cabibbo-Kobayashi-Maskawa matrix are real.}
\begin{eqnarray*}
\lefteqn{\mbox{Tr}(T_+\partial^\mu\phi)}\\
&=&
\mbox{Tr}\left[
\left(\begin{array}{ccc}0&V_{ud}&V_{us}\\0&0&0\\0&0&0\end{array}
\right)
\partial^\mu
\left(\begin{array}{ccc}
\pi^0+\frac{1}{\sqrt{3}}\eta &\sqrt{2}\pi^+&\sqrt{2}K^+\\
\sqrt{2}\pi^-&-\pi^0+\frac{1}{\sqrt{3}}\eta&\sqrt{2}K^0\\
\sqrt{2}K^- &\sqrt{2}\bar{K}^0&-\frac{2}{\sqrt{3}}\eta
\end{array}\right)\right]\\
&=&
V_{ud}\sqrt{2}\partial^\mu\pi^-+V_{us}\sqrt{2}\partial^\mu K^-,\\
\lefteqn{\mbox{Tr}(T_-\partial^\mu\phi)}\\
&=&\mbox{Tr}\left[
\left(\begin{array}{ccc}0&0&0\\
V_{ud}&0&0\\
V_{us}&0&0
\end{array}
\right)
\partial^\mu
\left(\begin{array}{ccc}
\pi^0+\frac{1}{\sqrt{3}}\eta &\sqrt{2}\pi^+&\sqrt{2}K^+\\
\sqrt{2}\pi^-&-\pi^0+\frac{1}{\sqrt{3}}\eta&\sqrt{2}K^0\\
\sqrt{2}K^- &\sqrt{2}\bar{K}^0&-\frac{2}{\sqrt{3}}\eta
\end{array}\right)\right]\\
&=&V_{ud} \sqrt{2}\partial^\mu\pi^++V_{us}\sqrt{2}\partial^\mu K^+.
\end{eqnarray*}
   We then obtain for the interaction term 
\begin{equation}
\label{4:6:lwphi}
{\cal L}_{W\phi}=
-g \frac{F_0}{2}
[{\cal W}_\mu^+(V_{ud}\partial^\mu\pi^-+V_{us}\partial^\mu K^-)
+{\cal W}_\mu^-(V_{ud}\partial^\mu \pi^++V_{us}\partial^\mu K^+)].
\end{equation} 
   In combination with the Feynman propagator for W bosons,
\begin{equation}
\label{4:6:wprop}
\frac{-g_{\mu\nu}+\frac{k_\mu k_\nu}{M^2_W}}{k^2-M^2_W}
=\frac{g_{\mu\nu}}{M^2_W}+O(\frac{kk}{M^4_W}),
\end{equation}
the Feynman rule for the invariant amplitude for the weak pion decay
reads 
\begin{eqnarray}
\label{4:6:mpionzerfall}
{\cal M}&=&i\left[-\frac{g}{2\sqrt{2}}\bar{u}_{\nu_\mu}
\gamma^\nu (1-\gamma_5)v_{\mu^+}\right] \frac{ig_{\nu\mu}}{M^2_W}
i\left[-g\frac{F_0}{2}V_{ud}
(-ip^\mu)\right]\nonumber\\
&=&-G_F V_{ud} F_0 \bar{u}_{\nu_\mu} p\hspace{-.4em}/(1-\gamma_5)
v_{\mu^+},
\end{eqnarray}
where  $p$ denotes the four-momentum of the pion and 
$$G_F=\frac{g^2}{4\sqrt{2}M^2_W}=
1.16639(1)\times
10^{-5}\,\mbox{GeV}^{-2}
$$ 
is the Fermi constant.
   The evaluation of the decay rate is a standard textbook exercise
and we only quote the final result\footnote{See 
Chap.\ 10.14 of Ref.\ \cite{Bjorken_1964}
with the substitution $a/\sqrt{2}\to V_{ud} F_0$ in Eq.\ (10.140).}
\begin{equation}
\label{4:6:zr}
\frac{1}{\tau}=\frac{G^2_F |V_{ud}|^2}{4\pi} F^2_0 M_\pi m_\mu^2
\left(1-\frac{m_\mu^2}{M_\pi^2}\right)^2.
\end{equation}
   The constant $F_0$ is referred to as the pion-decay constant in the
chiral limit.\footnote{Of course, in the chiral limit, the pion is
massless and, in such a world, the massive leptons would decay into Goldstone 
bosons, e.g., $e^-\to\pi^-\nu_e$.
   However, at ${\cal O}(p^2)$, the symmetry breaking term of 
Eq.\ (\ref{4:3:lqm}) gives rise to Goldstone-boson masses, whereas the 
decay constant is not modified at ${\cal O}(p^2)$.}
   It measures the strength of the matrix element of the axial-vector
current operator between a one-Goldstone-boson state and the vacuum
[see Eq.\ (\ref{4:1:acc})].
   Since the interaction of the $W$ boson with the quarks is of the type 
$l_\mu^a L^{\mu,a}=l_\mu^a(V^{\mu,a}-A^{\mu,a})/2$ [see Eq.\ (\ref{2:4:rlw})]
and the vector current operator does not contribute to the
matrix element between a single pion and the vacuum, pion decay
is completely determined by the axial-vector current.
   The degeneracy of a single constant $F_0$ in Eq.\ (\ref{4:1:acc}) is 
lifted at ${\cal O}(p^4)$ \cite{Gasser:1984gg} once SU(3) symmetry breaking
is taken into account.
   The empirical numbers for $F_\pi$ and $F_K$ are $92.4$ MeV and
$113$ MeV, respectively.\footnote{In the analysis of Ref.\ \cite{Groom:in}
$f_\pi=\sqrt{2} F_\pi$ is used.}

\subsection{Pion-Pion Scattering} 
\label{subsec_pps}
   Our second example deals with the prototype of a Goldstone boson reaction:
$\pi\pi$ scattering.
   For the sake of simplicity we will restrict ourselves to the 
SU(2)$\times$SU(2) version of Eq.\ (\ref{4:5:l2}). 
   We will contrast two different methods of calculating the scattering
amplitude: the ``direct'' calculation in terms of the Goldstone boson
fields of the effective Lagrangian versus the calculation of the QCD Green 
function in combination with the LSZ reduction formalism.
   Loosely speaking, the ``direct'' calculation is somewhat more along the
spirit of Weinberg's original paper \cite{Weinberg:1978kz}:  one 
considers the most general Lagrangian satisfying the general symmetry 
constraints and calculates $S$-matrix elements with that Lagrangian.
   The second method will allow one to also consider QCD Green functions
``off shell,'' i.e., for arbitrary squared invariant momenta.
   We will discuss under which circumstances the two methods
are equivalent and also work out the more general scope of the Green 
function approach.

   For the ``direct'' calculation 
we set to zero all external fields except for the quark mass term,
$\chi=2 B_0 \mbox{diag}(m_q,m_q)=M_\pi^2 1_{2\times 2}$
[see Eq.\ (\ref{4:3:mpi2})], 
\begin{equation}
\label{4:6:l21}
{\cal L}_2=\frac{F^2_0}{4}\mbox{Tr}(\partial_\mu U\partial^\mu U^\dagger)
+\frac{F_0^2 M_\pi^2}{4}\mbox{Tr}(U^\dagger+U).
\end{equation}
   In our general discussion of the transformation behavior of Goldstone 
bosons at the end of Sec.\ \ref{subsec_gc} we argued that we still have a 
choice how to represent the variables parameterizing the elements of 
the set of cosets $G/H$. 
   In the present case these are elements of SU(2) and we will illustrate 
this freedom by making use of two different parameterizations of the 
matrix $U$ \cite{Fearing:1999fw},\footnote{The first parameterization 
is popular, because the pion field appears only linearly in the term
proportional to the Pauli matrices, leading to a substantial simplification 
when deriving Feynman rules. 
   It is specific to SU(2) because, in contrast to the general case of
SU($N$), in SU(2) the totally symmetric $d$ symbols
vanish [see Eq.\ (\ref{2:1:dabc})].
   On the other hand, the exponential parameterization can be used for 
any $N$.}
\begin{eqnarray}
\label{4:6:u1}
U(x)&=&\frac{1}{F_0}\left[\sigma(x)+i\vec{\tau}\cdot\vec{\pi}(x)
\right],\quad \sigma(x)=\sqrt{F^2_0-\vec{\pi}\,^2(x)},
\\
\label{4:6:u2}
U(x)&=&\exp\left[i\frac{\vec{\tau}\cdot\vec{\phi}(x)}{F_0}\right],
\end{eqnarray}
   where in both cases the three Hermitian fields $\pi_i$ and $\phi_i$
describe pion fields transforming as isovectors under SU(2)$_V$.
   The fields in the two parameterizations are non-linearly related,
\begin{equation}
\label{4:6:ft}
\frac{\vec{\pi}}{F_0}=\hat{\phi}\sin\left(\frac{|\vec\phi|}{F_0}\right)
=\frac{\vec{\phi}}{F_0}\left(1-\frac{1}{6}\frac{\vec{\phi}\,^2}{F^2_0}
+\cdots\right).
\end{equation}
   This can be interpreted in terms of a change of variables which
leaves the free-field part of the Lagrangian unchanged 
\cite{Chisholm,Kamefuchi:sb}.
   As a consequence of the equivalence theorem of field theory 
\cite{Chisholm,Kamefuchi:sb} the result for a physical observable should not 
depend on the choice of variables.

   The substitution $U\leftrightarrow U^\dagger$ corresponding, respectively, 
to $\vec{\pi}\mapsto -\vec{\pi}$ and $\vec{\phi}\mapsto -\vec{\phi}$ 
tells us that ${\cal L}_2$ generates only interaction terms containing
an even number of pion fields.
   Since there exists no vertex involving 3 Goldstone bosons, 
$\pi\pi$ scattering must be described by a contact interaction at 
${\cal O}(p^2)$.

   By inserting the expressions for $U$ of Eqs.\ (\ref{4:6:u1}) 
and (\ref{4:6:u2}) into Eq.\ (\ref{4:6:l21}) and collecting 
those terms containing four pion fields we obtain the interaction
Lagrangians
\begin{eqnarray}
\label{4:6:l24pi}
{\cal L}_2^{4\pi}&=&\frac{1}{2F^2_0}\partial_\mu \vec{\pi}\cdot\vec{\pi}
\partial^\mu \vec{\pi}\cdot\vec{\pi}
-\frac{M_\pi^2}{8 F^2_0}(\vec{\pi}\,^2)^2,\\
\label{4:6:l24phi}
{\cal L}_2^{4\phi}&=&\frac{1}{6F^2_0}(\partial_\mu \vec{\phi}\cdot\vec{\phi}
\partial^\mu \vec{\phi}\cdot\vec{\phi}-\vec{\phi}\,^2 \partial_\mu\vec{\phi}
\cdot \partial^\mu\vec{\phi})
+\frac{M_\pi^2}{24 F^2_0}(\vec{\phi}\,^2)^2.
\end{eqnarray}
   Observe that the two interaction Lagrangians depend differently on the
respective pion fields.
   The corresponding Feynman rules 
are obtained in the usual fashion by considering all possible ways of 
contracting pion fields of $i{\cal L}_{\rm int}$ with initial and final pion
lines, with the derivatives $\partial_\mu$ generating $-i p_\mu$ ($i p_\mu$) 
for an initial (final) line.
   For Cartesian isospin indices $a,b,c,d$ the Feynman rules for
the scattering process $\pi^a(p_a)+\pi^b(p_b)\to\pi^c(p_c)+\pi^d(p_d)$
as obtained from Eqs.\ (\ref{4:6:l24pi}) and (\ref{4:6:l24phi}) read, 
respectively,
\begin{eqnarray}
\label{4:6:mpipi1}
{\cal M}_2^{4\pi}&=&i\left[\delta^{ab}\delta^{cd}\frac{s-M^2_\pi}{F^2_0}
+\delta^{ac}\delta^{bd}\frac{t-M^2_\pi}{F^2_0}
+\delta^{ad}\delta^{bc}\frac{u-M^2_\pi}{F^2_0}\right],\\
\label{4:6:mpipi2}
{\cal M}_2^{4\phi}&=&i\left[\delta^{ab}\delta^{cd}\frac{s-M^2_\pi}{F^2_0}
+\delta^{ac}\delta^{bd}\frac{t-M^2_\pi}{F^2_0}
+\delta^{ad}\delta^{bc}\frac{u-M^2_\pi}{F^2_0}\right]\nonumber\\
&&-\frac{i}{3F_0^2}
\left(\delta^{ab}\delta^{cd}+\delta^{ac}\delta^{bd}+\delta^{ad}\delta^{bc}
\right)
\left(\Lambda_a+\Lambda_b+\Lambda_c+\Lambda_d\right),
\end{eqnarray}
where we introduced $\Lambda_k=p_k^2-M^2_\pi$ and the usual  
Mandelstam variables 
$$s=(p_a+p_b)^2=(p_c+p_d)^2,
$$
$$
t=(p_a-p_c)^2=(p_d-p_b)^2,
$$
$$
u=(p_a-p_d)^2=(p_c-p_b)^2,
$$
which are related by $s+t+u=p_a^2+p_b^2+p_c^2+p_d^2$.
  If the initial and final pions are all on the mass shell, 
i.e., $\Lambda_k=0$, 
the scattering amplitudes are the same, in agreement with
the equivalence theorem
\cite{Chisholm,Kamefuchi:sb}.\footnote{For a general
proof of the
equivalence of $S$-matrix elements evaluated at tree level (phenomenological
approximation), see Sec.\ 2 of Ref.\ \cite{Coleman:sm}.}
   The on-shell result also agrees with the current-algebra prediction
for low-energy $\pi\pi$ scattering \cite{Weinberg:1966kf}.
   We will come back to $\pi\pi$ scattering in Sec.\ 
\ref{subsec_eppsop6}  when we also discuss
corrections of higher order \cite{Bijnens:1995yn}.   
   On the other hand, if one of the momenta of the external lines is off mass
shell, the amplitudes of Eqs.\ (\ref{4:6:mpipi1}) and (\ref{4:6:mpipi2})
differ.
   In other words, a ``direct'' calculation gives a unique result
independent of the parameterization of $U$ only
for the on-shell matrix element.

   The second method, developed by Gasser and Leutwyler \cite{Gasser:1983yg},
deals with the Green functions of QCD and their interrelations as expressed in
the Ward identities.
   In particular, these Green functions can, in principle, be calculated
for any value of squared momenta even though ChPT is set up 
only for a low-energy description.
   For the discussion of $\pi\pi$ scattering one considers the four-point
function \cite{Gasser:1983yg}
\begin{equation}
\label{4:6:fpfpppp}
G_{PPPP}^{abcd}(x_a,x_b,x_c,x_d)\equiv
\langle 0|T[P_a(x_a)P_b(x_b)P_c(x_c)P_d(x_d)]|
0\rangle
\end{equation}
with the pseudoscalar quark densities of 
Eq.\ (\ref{4:1:psqd}).

   In order so see that Eq.\ (\ref{4:6:fpfpppp}) can indeed be related to 
$\pi\pi$ scattering, let us first investigate the matrix element of the 
pseudoscalar density evaluated between a single-pion state and the 
vacuum, which is defined in terms of the coupling constant 
$G_\pi$ \cite{Gasser:1983yg}:
\begin{equation}
\label{4:6:pipiv}
\langle 0|P_i(0)|\pi_j(q)\rangle =\delta_{ij}G_\pi.
\end{equation}
\begin{figure}
   \begin{center}
   \mbox{\epsfig{file=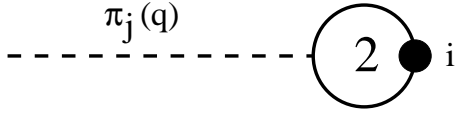,width=6.cm}}
   \end{center}
   \caption{\label{4:6:figppion} Coupling of an external pseudoscalar
field $p_i$ (denoted by ``$\bullet $ i'')  
to a pion $\pi_j$ at ${\cal O}(p^2)$.}
\end{figure}
   At ${\cal O}(p^2)$ we determine the coupling of an external
pseudoscalar source $p$ to the Goldstone bosons by inserting $\chi=2 B_0 i p$
into the Lagrangian of Eq.\ (\ref{4:5:l2})
(see Fig.\ \ref{4:6:figppion}),
\begin{equation}
\label{4:6:l2ext2}
{\cal L}_{\rm ext}
=i\frac{F_0^2B_0}{2}\mbox{Tr}(pU^\dagger-Up)=
\left\{\begin{array}{l}
2B_0F_0p_i \pi_i,\\
2B_0F_0p_i\phi_i[1-\vec{\phi}\,^2/(6F_0^2)+\cdots],
\end{array}\right.
\end{equation}
where the first and second lines refer to the parameterizations of 
Eqs.\ (\ref{4:6:u1}) and (\ref{4:6:u2}), respectively.
   From Eq.\ (\ref{4:6:l2ext2}) we obtain $G_\pi = 2B_0 F_0$ independent of 
the parameterization used which, since the pion is on-shell, is a consequence 
of the equivalence theorem \cite{Chisholm,Kamefuchi:sb}.
   As a consistency check, let us verify the PCAC relation of 
Eq.\ (\ref{2:4:divasc}) (without an external electromagnetic field)  
evaluated between a single-pion state and the vacuum.
   For the axial-vector current matrix element, we found at ${\cal O}(p^2)$
\begin{equation}
\label{4:6:axialcurrentpion}
\langle 0|A^\mu_i(x)|\pi_j(q)\rangle = i q^\mu F_0 
e^{-iq\cdot x}\delta_{ij}.
\end{equation}
   Taking the divergence we obtain
\begin{eqnarray*}
\langle 0|\partial_\mu A^\mu_i(x)|\pi_j(q)\rangle &=&
 i  q^\mu F_0 \partial_\mu e^{-iq\cdot x}\delta_{ij}
=M^2_\pi F_0 e^{-iq\cdot x}\delta_{ij}
= 2m_q B_0F_0 e^{-iq\cdot x}\delta_{ij},
\end{eqnarray*}
   where we made use of Eq.\ (\ref{4:3:mpi2}) for the pion mass.
   Multiplying Eq.\ (\ref{4:6:pipiv}) by $m_q$ and using $G_\pi = 2B_0 F_0$
we explicitly verify the PCAC relation.

   Every field $\Phi_i(x)$, which satisfies the relation 
\begin{equation}
\label{4:6:intpolfield}
\langle 0| \Phi_i(x)|\pi_j(q)\rangle = \delta_{ij} e^{-iq \cdot x},
\end{equation}
can serve as a so-called {\em interpolating} pion field \cite{Borchers:1960}
in the LSZ reduction formulas \cite{Lehmann:1955rq,Itzykson:rh}.
   For the case of $\pi^a(p_a)+\pi^b(p_b)\to\pi^c(p_c)+\pi^d(p_d)$ the 
reduction formula relates the $S$-matrix element to the Green function 
of the interpolating field as 
\begin{eqnarray*}
S_{fi}&=& i^4 \int d^4 x_a \cdots d^4 x_d \,e^{-i p_a \cdot x_a}
\cdots e^{i p_d \cdot x_d}\\
&&\times (\Box_a+M_\pi^2)\cdots
(\Box_d+M_\pi^2)
\langle 0 |T[\Phi_a(x_a)\Phi_b(x_b)\Phi_c(x_c)\Phi_d(x_d)]|0\rangle.
\end{eqnarray*}
   After partial integrations, the Klein-Gordon operators convert into
inverse free propagators
\begin{eqnarray*}
S_{fi}&=& (-i)^4(p_a^2-M_\pi^2)\cdots(p_d^2-M_\pi^2)\\
&&\times \int d^4 x_a \cdots d^4 x_d \,e^{-i p_a \cdot x_a}
\cdots e^{i p_d \cdot x_d}
\langle 0 |T[\Phi_a(x_a)\Phi_b(x_b)\Phi_c(x_c)\Phi_d(x_d)]|0\rangle.
\end{eqnarray*}
   In the present context, we will use 
\begin{equation}
\label{4:6:pionfield}
  \Phi_i(x) = \frac{P_i(x)}{G_\pi}=\frac{P_i(x)}{2B_0F_0}=
  \frac{m_q P_i(x)}{M_\pi^2F_0},
\end{equation}
   which then relates the $S$-matrix element of $\pi\pi$ scattering to the
QCD Green function involving four pseudoscalar densities
\begin{eqnarray*}
S_{fi}&=& \left(\frac{-i}{G_\pi}\right)^4(p_a^2-M_\pi^2)\cdots(p_d^2-M_\pi^2)\\
&&\times\int d^4 x_a \cdots d^4 x_d \,e^{-i p_a \cdot x_a}
\cdots e^{i p_d \cdot x_d}
G_{PPPP}^{abcd}(x_a,x_b,x_c,x_d).
\end{eqnarray*}
   Using translational invariance, let us define the momentum space
Green function as
\begin{eqnarray}
\label{4:6:msgf}
\lefteqn{
(2\pi)^4 \delta^4(p_a+p_b+p_c+p_d)F^{abcd}_{PPPP}(p_a,p_b,p_c,p_d)=}\nonumber
\\
&&
\int d^4 x_a d^4 x_b d^4 x_c d^4 x_d\,  e^{-i p_a \cdot x_a}
e^{-ip_b\cdot x_b} e^{-i p_c\cdot x_c} e^{-i p_d x_d}
G_{PPPP}^{abcd}(x_a,x_b,x_c,x_d),\nonumber\\
\end{eqnarray} 
   where we define all momenta as incoming. 
   The usual relation between the $S$ matrix and the $T$ matrix,
$S=I+iT$, implies for the $T$-matrix element $\langle f|T|i\rangle=
(2\pi)^4\delta^4(P_f-P_I){\cal T}_{fi}$ and, finally, for 
${\cal M}=i{\cal T}_{fi}$:
\begin{equation}
\label{4:6:calMlsz}
{\cal M}= \frac{1}{G_\pi^4} 
\left[\prod_{k=a,b,c,d}\lim_{p_k^2\to M_\pi^2}
(p_k^2-M_\pi^2)\right] F^{abcd}_{PPPP}(p_a,p_b,-p_c,-p_d).
\end{equation}

   We will now determine the Green function 
$F^{abcd}_{PPPP}(p_a,p_b,-p_c,-p_d)$ using the parameterizations of Eqs.\
(\ref{4:6:u1}) and (\ref{4:6:u2}) for $U$.
\begin{figure}
   \begin{center}
   \mbox{\epsfig{file=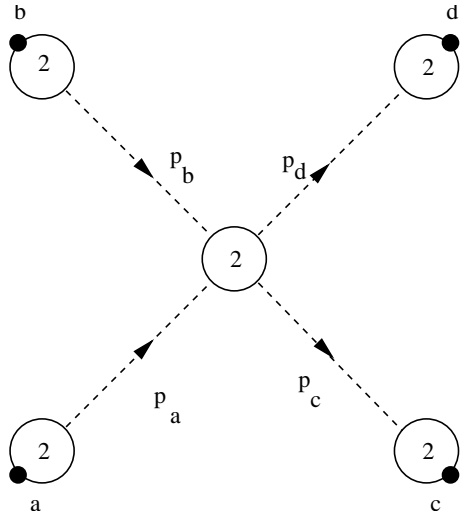,width=6.cm}}
   \end{center}
   \caption{\label{4:6:figpppp1} Four-point Green function 
$F^{abcd}_{PPPP}(p_a,p_b,-p_c,-p_d)$ at ${\cal O}(p^2)$ in the
parameterization of Eq.\ (\ref{4:6:u1}). The $\bullet$ denote the 
pseudoscalar sources which are ``removed'' from the diagram.}
\end{figure}
   In the first parameterization we only obtain a linear coupling between
the external pseudoscalar field and the pion field 
[see Eq.\ (\ref{4:6:l2ext2})] so that only the Feynman diagram of 
Fig.\ \ref{4:6:figpppp1} contributes
\begin{eqnarray}
\label{4:6:fabcd1}
F^{abcd}_{PPPP}(p_a,p_b,-p_c,-p_d)&=&(2B_0 F_0)^4 
\frac{i}{p_a^2-M_\pi^2}\cdots\frac{i}{p_d^2-M_\pi^2} {\cal M}^{4\pi}_2,
\nonumber\\
\end{eqnarray}
where ${\cal M}^{4\pi}_2$ is given in Eq.\ (\ref{4:6:mpipi1}).
   The Green function depends on six independent Lorentz scalars which
can be chosen as the squared invariant momenta $p^2_k$ and the three
Mandelstam variables $s$, $t$, and $u$ satisfying the constraint
$s+t+u=\sum_k p_k^2$.

\begin{figure}
   \begin{center}
   \mbox{\epsfig{file=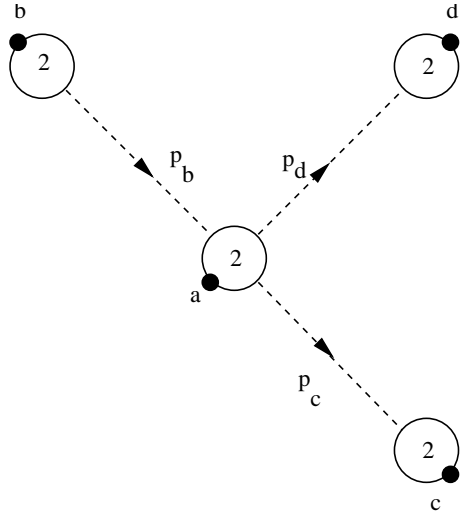,width=6.cm}}
   \end{center}
   \caption{\label{4:6:addfigpppp2} Additional contribution to the four-point
Green function $F^{abcd}_{PPPP}(p_a,p_b,-p_c,-p_d)$ at ${\cal O}(p^2)$ in the
parameterization of Eq.\ (\ref{4:6:u2}). 
   The remaining three permutations are not shown.
The $\bullet$ denote the 
pseudoscalar sources which are ``removed'' from the diagram.}
\end{figure}
   Using the second parameterization we will obtain a contribution 
which is of the same form as Fig.\ \ref{4:6:figpppp1} but with 
${\cal M}^{4\pi}_2$ replaced by ${\cal M}^{4\phi}_2$ of 
Eq.\ (\ref{4:6:mpipi2}).
   Clearly, this is not yet the same result as Eq.\ (\ref{4:6:fabcd1})
because of the terms proportional to $\Lambda_k$ in Eq.\ (\ref{4:6:mpipi2}).
   However, in this parameterization the external pseudoscalar field 
also couples to three pion fields [see Eq.\ (\ref{4:6:l2ext2})],
resulting in four additional diagrams of the type shown 
in Fig.\ \ref{4:6:addfigpppp2}.
   For example, the contribution shown in Fig.\ \ref{4:6:addfigpppp2} reads
\begin{eqnarray}
\label{4:6:deltafabcd2}
\lefteqn{\Delta_a F^{abcd}_{PPPP}(p_a,p_b,-p_c,-p_d)}\nonumber\\
&=&
(2B_0 F_0)^3 
\frac{i}{p_b^2-M_\pi^2}
\frac{i}{p_c^2-M_\pi^2}
\frac{i}{p_d^2-M_\pi^2} 
\left(-\frac{2B_0}{3F_0}\right)
(\delta^{ab}\delta^{cd}
+\delta^{ac}\delta^{bd}+\delta^{ad}\delta^{bc})\nonumber\\
&=&(2B_0 F_0)^4 \frac{i}{p_a^2-M_\pi^2}\cdots\frac{i}{p_d^2-M_\pi^2}
\frac{i\Lambda_a}{3 F_0^2}(\delta^{ab}\delta^{cd}
+\delta^{ac}\delta^{bd}+\delta^{ad}\delta^{bc}),
\end{eqnarray}
where $\Lambda_a=(p_a^2-M_\pi^2)$.
   In combination with the contribution of the remaining three diagrams, 
we find a complete cancelation with those terms proportional to $\Lambda_k$ 
of Fig.\ \ref{4:6:figpppp1} (in the second parameterization)
and the end result is identical with Eq.\ (\ref{4:6:fabcd1}).
   Finally, using $G_\pi=2 B_0F_0$ and inserting the result of 
Eq.\ (\ref{4:6:fabcd1}) into Eq.\ (\ref{4:6:calMlsz}) we obtain
the same scattering amplitude as in the ``direct'' calculation
of Eqs.\  (\ref{4:6:mpipi1}) and (\ref{4:6:mpipi2}) evaluated
for on-shell pions.

   This example serves as an illustration that the method of Gasser and 
Leutwyler generates unique results for the Green functions of QCD
for arbitrary four-momenta.
   There is no ambiguity resulting from the choice of variables used
to parameterize the matrix $U$ in the effective Lagrangian.
   These Green functions can be evaluated for arbitrary (but small)
four-momenta.
   Using the reduction formalism, on-shell matrix elements such as the 
$\pi\pi$ scattering amplitude can be calculated from the QCD Green 
functions.
   The result for the $\pi\pi$ scattering amplitude as derived from 
Eq.\ (\ref{4:6:calMlsz}) agrees with the ``direct'' calculation of the 
on-shell matrix elements of Eqs.\  (\ref{4:6:mpipi1}) and 
(\ref{4:6:mpipi2}). 
   On the other hand, the Feynman rules of Eqs.\ (\ref{4:6:mpipi1}) and 
(\ref{4:6:mpipi2}), when taken {\em off shell}, have to be considered as 
intermediate building blocks only and thus need not be unique.

\section{The Chiral Lagrangian at Order ${\cal O}(p^4)$}
\label{sec_clop4}
   Applying the ideas outlined in Sec.\ \ref{sec_cel} it is possible to 
construct the most general Lagrangian at ${\cal O}(p^4)$.
   Here we only quote the result of Ref.\ \cite{Gasser:1984gg}:
\begin{eqnarray}
\label{4:7:l4gl}
\lefteqn{{\cal L}_4=
L_1 \left\{\mbox{Tr}[D_{\mu}U (D^{\mu}U)^{\dagger}] \right\}^2
+ L_2 \mbox{Tr} \left [D_{\mu}U (D_{\nu}U)^{\dagger}\right]
\mbox{Tr} \left [D^{\mu}U (D^{\nu}U)^{\dagger}\right]}\nonumber\\
& & + L_3 \mbox{Tr}\left[ 
D_{\mu}U (D^{\mu}U)^{\dagger}D_{\nu}U (D^{\nu}U)^{\dagger}
\right ] 
+ L_4 \mbox{Tr} \left [ D_{\mu}U (D^{\mu}U)^{\dagger} \right ]
\mbox{Tr} \left( \chi U^{\dagger}+ U \chi^{\dagger} \right )
\nonumber \\
& & +L_5 \mbox{Tr} \left[ D_{\mu}U (D^{\mu}U)^{\dagger}
(\chi U^{\dagger}+ U \chi^{\dagger})\right]
+ L_6 \left[ \mbox{Tr} \left ( \chi U^{\dagger}+ U \chi^{\dagger} \right )
\right]^2
\nonumber \\
& & + L_7 \left[ \mbox{Tr} \left ( \chi U^{\dagger} - U \chi^{\dagger} \right )
\right]^2
+ L_8 \mbox{Tr} \left ( U \chi^{\dagger} U \chi^{\dagger}
+ \chi U^{\dagger} \chi U^{\dagger} \right )
\nonumber \\
& & -i L_9 \mbox{Tr} \left [ f^R_{\mu\nu} D^{\mu} U (D^{\nu} U)^{\dagger}
+ f^L_{\mu\nu} (D^{\mu} U)^{\dagger} D^{\nu} U \right ]
+ L_{10} \mbox{Tr} \left ( U f^L_{\mu\nu} U^{\dagger} f_R^{\mu\nu} \right )
\nonumber \\
& & + H_1 \mbox{Tr} \left ( f^R_{\mu\nu} f^{\mu\nu}_R +
f^L_{\mu\nu} f^{\mu\nu}_L \right )
+ H_2 \mbox{Tr} \left ( \chi \chi^{\dagger} \right ).
\end{eqnarray}
   The numerical values of the low-energy coupling constants $L_i$ are not
determined by chiral symmetry.
   In analogy to $F_0$ and $B_0$ of ${\cal L}_2$ they are parameters  
containing information on the underlying dynamics and should, 
in principle, be calculable in terms of the (remaining) parameters of
QCD, namely, the heavy-quark masses and the QCD scale $\Lambda_{\rm QCD}$. 
   In practice, they parameterize our inability to solve the
dynamics of QCD in the non-perturbative regime.
   So far they have either been fixed using empirical input 
\cite{Gasser:1983yg,Gasser:1984gg,Bijnens:1994qh} 
or theoretically using QCD-inspired models 
\cite{Ebert:1985kz,Espriu:1989ff,Ebert:1991xd,Bijnens:1992uz},
meson-resonance saturation 
\cite{Ecker:yg,Ecker:1988te,Donoghue:ed,Knecht:2001xc,Leupold:2001vs}, and
lattice QCD \cite{Myint:yw,Golterman:2000mg}.

   From a practical point of view the coefficients are also required for
another purpose.
   When calculating one-loop graphs, using vertices from ${\cal L}_2$ of 
Eq.\ (\ref{4:5:l2}), one generates infinities which,
according to Weinberg's power counting of Eq.\ (\ref{4:4:mr2}),
are of ${\cal O}(p^4)$, i.e., which cannot be absorbed by a renormalization 
of the coefficients $F_0$ and $B_0$.
   In the framework of dimensional regularization (see App.\ \ref{app_drb})
these divergences appear as poles at space-time dimension $n=4$.
   In Refs.\ \cite{Gasser:1983yg,Gasser:1984gg} the poles, together with
the relevant counter terms, were given in closed form.
   To that end, Gasser and Leutwyler made use of the so-called saddle-point 
method which, in the path-integral approach, allows one to identify
the one-loop contribution to the generating functional.
   The action is expanded around the classical solution
and the path integral is performed with respect to the terms quadratic
in the fluctuations about the classical solution.
   The resulting one-loop piece of the generating functional is treated
within the dimensional-regularization procedure and the poles are isolated
by applying the so-called heat-kernel technique.\footnote{Since the whole 
procedure is rather technical, we will restrict ourselves, by means of 
the example to be discussed in Sec.\ \ref{subsec_mgb}, to an explicit 
verification that the renormalization procedure 
indeed leads to finite predictions for physical observables.}
   Except for $L_3$ and $L_7$ the low-energy coupling constants $L_i$ and the
``contact terms''---i.e., pure external field terms---$H_1$ and $H_2$ 
are required in the renormalization of the one-loop graphs
\cite{Gasser:1984gg}.
   Since $H_1$ and $H_2$ contain only external fields, they are of no 
physical relevance \cite{Gasser:1984gg}.

   By construction Eq.\ (\ref{4:7:l4gl}) represents the most general
Lagrangian at ${\cal O}(p^4)$, and it is thus possible to absorb the one-loop 
divergences by an appropriate renormalization of the coefficients
$L_i$ and $H_i$ \cite{Gasser:1984gg}:
\begin{eqnarray}
\label{4:7:li}
L_i&=&L_i^r+\frac{\Gamma_i}{32\pi^2}R, \quad i=1,\cdots,10,\\
\label{4:7:hi}
H_i&=&H^r_i+\frac{\Delta_i}{32\pi^2}R,\quad i=1,2,
\end{eqnarray}
   where $R$ is defined as (see App.\ \ref{app_drb})
\begin{equation}
\label{4:7:R}
R=\frac{2}{n-4}-[\mbox{ln}(4\pi)-\gamma_E+1],
\end{equation}
   with $n$ denoting the number of space-time dimensions and
$\gamma_E=-\Gamma'(1)$ being Euler's constant.
   The constants $\Gamma_i$ and $\Delta_i$ are given in Table
\ref{4:7:tableli}.
   The renormalized coefficients $L_i^r$ depend on the scale $\mu$ 
introduced by dimensional regularization [see Eq.\ (\ref{app:drb:im22})] and
their values at two different scales $\mu_1$ and $\mu_2$ 
are related by 
\begin{equation}
\label{4:7:limu1mu2}
L^r_i(\mu_2)=L^r_i(\mu_1)
+\frac{\Gamma_i}{16\pi^2}\ln\left(\frac{\mu_1}{\mu_2}\right).
\end{equation} 
   We will see that the scale dependence of the coefficients and
the finite part of the loop-diagrams compensate each other in
such a way that physical observables are scale independent.

\begin{table}
\caption[test]{\label{4:7:tableli} 
Renormalized low-energy coupling constants $L_i^r$ in units of $10^{-3}$ 
at the scale $\mu=M_\rho$ \cite{Bijnens:1994qh}. $\Delta_1=-1/8$,
$\Delta_2=5/24$.
}
\begin{center}
\begin{tabular}{|c|r|r|}
\hline
Coefficient &Empirical Value &$\Gamma_i$\\
\hline
$L_1^r$ &    $ 0.4\pm 0.3$  &$\frac{3}{32}$\\
$L_2^r$ &    $ 1.35\pm 0.3$ &$\frac{3}{16}$\\
$L_3^r$ &    $-3.5\pm 1.1$  &$0$\\
$L_4^r$ &    $-0.3\pm 0.5$  &$\frac{1}{8}$\\
$L_5^r$ &    $ 1.4\pm 0.5$  &$\frac{3}{8}$\\
$L_6^r$ &    $-0.2\pm 0.3$  &$\frac{11}{144}$\\
$L_7^r$ &    $-0.4\pm 0.2$  &$0$\\
$L_8^r$ &    $ 0.9\pm 0.3$  &$\frac{5}{48}$\\
$L_9^r$ &    $ 6.9\pm 0.7$  &$\frac{1}{4}$\\
$L_{10}^r$ & $-5.5\pm 0.7$  &$-\frac{1}{4}$\\
\hline
\end{tabular}
\end{center}
\end{table}

   We finally discuss the method of using field transformations
to eliminate redundant terms in the most general effective Lagrangian 
\cite{Georgi:1991ch,Leutwyler:1991mz,Grosse-Knetter:1993td,Arzt:gz,%
Scherer:1994wi}.
   From a ``naive'' point of view the two structures
\begin{equation}
\label{4:7:addstruc}
\mbox{Tr}[D^2 U (D^2 U)^\dagger],\quad
\mbox{Tr}[D^2 U \chi^\dagger +\chi (D^2 U)^\dagger]
\end{equation}
   would qualify as independent terms of order ${\cal O}(p^4)$.
   Loosely speaking, by using the classical equation of motion of
Eq.\ (\ref{4:5:eom}) these terms can be shown to be redundant. 
   We will justify this statement in terms of field transformations.
   To that end let us consider another SU(3) matrix $U'(x)$ which is
related to $U(x)$ by a field transformation of the form
\begin{equation}
\label{4:7:up}
U(x)=\exp[iS(x)] U'(x).
\end{equation}
   Since both $U$ and $U'$ are SU(3) matrices, $S(x)$ must be a Hermitian
traceless $3\times 3$ matrix.
   We demand that $U'(x)$ satisfies the same symmetry properties as $U(x)$
(see Table \ref{4:5:table_trafprop}), 
\begin{equation}
\label{4:7:uptrafo}
U'\stackrel{G}{\mapsto} V_R U' V_L^\dagger,\quad
U'(\vec{x},t)\stackrel{P}{\mapsto} U'^\dagger (-\vec{x},t),\quad
U'\stackrel{C}{\mapsto}U'^T,
\end{equation}
from which we obtain the following conditions for $S$:
\begin{equation}
\label{4:7:strafo}
S\stackrel{G}{\mapsto} V_R S V_R^\dagger,\quad
S(\vec{x},t)\stackrel{P}{\mapsto} -U'^\dagger(-\vec{x},t)S(-\vec{x},t)
U'(-\vec{x},t),\quad
S\stackrel{C}{\mapsto}(U'^\dagger SU')^T.
\end{equation}
   The most general transformation is constructed iteratively in the
momentum and quark-mass expansion,  
\begin{equation}
\label{4:7:trafoit}
U=\exp[iS_2(x)]U^{(1)}(x),\quad  
U^{(1)}(x)=\exp[iS_4(x)]U^{(2)}(x),\quad \cdots, 
\end{equation}
where the matrices $S_{2n}$ are of ${\cal O}(p^{2n})$, satisfy the properties 
of Eq.\ (\ref{4:7:strafo}), and have to be constructed from the same building
 blocks as the effective Lagrangian.
 
   To be explicit, let us derive the most general matrix $S_2(x)$.
   At ${\cal O}(p^2)$, the field strength tensors cannot contribute as 
building blocks because of their antisymmetry under interchange of the 
Lorentz indices.
   Imposing the transformation behavior under the group 
$G=\mbox{SU(3)}_L\times\mbox{SU(3)}_R$, we obtain a list of five terms
\begin{equation}
\label{4:7:bausteines2}
D^2 U' U'^\dagger,\quad 
U' (D^2 U')^\dagger,\quad
D_\mu U' (D^\mu U')^\dagger,\quad 
\chi U'^\dagger,\quad
U'\chi^\dagger.
\end{equation}
   Parity eliminates three combinations and we are left with
\begin{equation}
\label{4:7:bausteines22}
D^2 U' U'^\dagger-U'(D^2 U')^\dagger,\quad
\chi U'^\dagger - U' \chi^\dagger.
\end{equation}
   Demanding Hermiticity and a vanishing trace, we end up with two terms
at ${\cal O}(p^2)$:
\begin{equation}
\label{4:7:s2}
S_2=i\alpha_1[D^2 U' U'^\dagger- U'(D^2 U')^\dagger]
+i\alpha_2[\chi U'^\dagger - U' \chi^\dagger
-\frac{1}{3}\mbox{Tr}(\chi U'^\dagger- U'\chi^\dagger)],
\end{equation}
   where $\alpha_1$ and $\alpha_2$ are real numbers.
   At ${\cal O}(p^2)$, charge conjugation does not provide an additional
constraint.

   What are the consequences of working with $U'(x)$ instead of $U(x)$? 
   In Sec.\ \ref{subsec_pps} we  have already argued, by means of a 
simple example, that the results for the Green functions are independent 
of the parameterizations of $U(x)$ of Eqs.\ (\ref{4:6:u1}) and (\ref{4:6:u2}).
   Expressing $U(x)$ of Eq.\ (\ref{4:7:up}) by using Eq.\ (\ref{4:7:s2}) 
and inserting the result into ${\cal L}_2$ of Eq.\ (\ref{4:5:l2}), we obtain
\begin{equation}
{\cal L}_2(U)={\cal L}_2(U')+\Delta {\cal L}_2(U'),
\end{equation}
where $\Delta {\cal L}_2$, to leading order in $S_2$, is given by
\begin{equation}
\label{4:7:dl2}
\Delta {\cal L}_2(U')=\frac{F^2_0}{4}\mbox{Tr}[iS_2 {\cal O}_{\rm 
EOM}^{(2)}(U')]+ O(S^2_2).
\end{equation}
   The functional form of ${\cal O}_{\rm EOM}^{(2)}$ has been defined in
Eq.\ (\ref{4:5:eom}). Note, however, that we do {\em not} assume  
${\cal O}_{\rm EOM}^{(2)}=0$.
   We have dropped a total derivative, since it does not modify the dynamics.
   Both $S_2$ and ${\cal O}_{\rm EOM}^{(2)}$ are of 
order ${\cal O}(p^2)$ so that $\Delta{\cal L}_2$ is of  order ${\cal O}(p^4)$.
   Of course, higher powers of $S_2$ in Eq.\ (\ref{4:7:dl2}) induce
additional terms of higher orders in the momentum expansion which we
will discuss in a moment.

   Through a suitable choice of the parameters $\alpha_1$ and 
$\alpha_2$ it is possible to eliminate two structures at order
${\cal O}(p^4)$, i.e., one generates a new Lagrangian with a different
functional form which, however, according to the equivalence theorem
leads to the same observables \cite{Chisholm,Kamefuchi:sb}.
   Such a procedure is commonly referred to as using the classical
equation of motion to eliminate terms.
   For example, it is straightforward but tedious to re-express the
two structures of Eq.\ (\ref{4:7:addstruc}) through the terms of
Gasser and Leutwyler, Eq.\ (\ref{4:7:l4gl}), and the following
two terms
\begin{equation}
\label{4:7:addstruc2}
c_1 \mbox{Tr}\left([D^2 U U^\dagger- U (D^2 U)^\dagger]
{\cal O}_{\rm EOM}^{(2)}\right)
+ c_2 \mbox{Tr}
\left((\chi U^\dagger - U\chi^\dagger){\cal O}_{\rm EOM}^{(2)}\right).
\end{equation}
   Choosing $\alpha_1=4 c_1/F_0^2$ and $\alpha_2=4 c_2/F_0^2$ in
Eq.\ (\ref{4:7:s2}), the two terms of Eq.\ (\ref{4:7:addstruc2})
and the modification $\Delta {\cal L}_2$ of Eq.\ (\ref{4:7:dl2})
precisely cancel and one is left with the canonical form of
Gasser and Leutwyler.

   A field redefinition, of course, also leads to modifications of 
the functional form of the effective Lagrangians of higher orders.
   However, for $S_2$ such terms are at least of order ${\cal O}(p^6)$
as are the higher-order terms in Eq.\ (\ref{4:7:dl2}).
   Thus one proceeds iteratively \cite{Scherer:1994wi}.
   Using $S_2$ one generates the simplest form of ${\cal L}_4$.
   Next one constructs $S_4$, inserts it again into ${\cal L}_2$ to
simplify ${\cal L}_6$, etc.

   From a point of view of {\em constructing} the simplest Lagrangian at a 
given order it is sufficient to identify those terms proportional to the 
classical, i.e.\ lowest-order, equation of motion and drop them right from 
the beginning using the argument that, by choosing appropriate generators, 
they can be transformed away.
   A completely different situation arises if one tries to express the 
effective Lagrangian obtained within the framework of a specific model
in the canonical form.
   In such a case it is necessary to explicitly perform the iteration process 
consistently to a given order and, in particular, take into account
the modification of the higher-order coefficients due to the transformation.
    An explicit example is given in Appendix AII of Ref.\ \cite{Belkov:1994qg}.

\section{The Effective Wess-Zumino-Witten Action}
\label{sec_ewzwa}
   The Lagrangians ${\cal L}_2$ and ${\cal L}_4$ discussed so far exhibit
a larger symmetry than the ``real world.''
   For example, if we consider the case of ``pure'' QCD, i.e., no external
fields except for $\chi=2B_0 M$ with the quark mass matrix $M$ of 
Eq.\ (\ref{4:3:qmt}), the two Lagrangians are invariant under the substitution 
$\phi(x)\mapsto -\phi(x)$.
   As discussed in Sec.\ \ref{sec_loel} they contain interaction terms
with an even number of Goldstone bosons only, i.e., they are of even
intrinsic parity, and it would not be
possible to describe the reaction 
$K^+K^-\to\pi^+\pi^-\pi^0$.\footnote{The $\phi$ meson can decay
into both $K^+K^-$ and $\pi^+\pi^-\pi^0$.}
   Analogously, the process $\pi^0\to\gamma\gamma$ cannot be described
by ${\cal L}_2$ and ${\cal L}_4$ in the presence of external electromagnetic
fields.

   These observations lead us to a discussion of the effective 
Wess-Zumino-Witten action \cite{Wess:yu,Witten:tw}.
   Whereas normal Ward identities are related to the {\em invariance} of the
generating functional under local transformations of the 
external fields,
the anomalous Ward identities 
\cite{Adler:1969gk,Adler:1969er,Bardeen:1969md,Bell:1969ts,Adler:1970},
which were first obtained in the framework of renormalized perturbation
theory, 
give a particular form to the {\em variation} of
the generating functional \cite{Wess:yu,Gasser:1983yg}.
   Wess and Zumino derived consistency or integrability 
relations which are satisfied by the anomalous Ward identities
and then explicitly constructed a functional involving the pseudoscalar octet 
which satisfies the anomalous Ward identities \cite{Wess:yu}.
   In particular, Wess and Zumino emphasized that their interaction 
Lagrangians cannot be obtained as part of a chiral invariant Lagrangian.
 
   In the construction of Witten \cite{Witten:tw} the simplest term possible
which breaks the symmetry of having only an even number of Goldstone bosons
at the Lagrangian level is added to the equation of motion of 
Eq.\ (\ref{4:5:eom}) for the case of massless Goldstone bosons without any 
external fields,\footnote{
In order to conform with our previous convention of Eq.\ (\ref{4:2:utrafo}),
we need to substitute $U_{\rm W}\to U^\dagger$. Furthermore $F_\pi$ of
Ref.\ \cite{Witten:tw} corresponds to $2F_0$. Finally,
\begin{displaymath}
\partial^2 U U^\dagger- U\partial^2 U^\dagger=2\partial_\mu(\partial^\mu U
U^\dagger).
\end{displaymath}
}
\begin{equation}
\label{4:8:eomadd}
\partial_\mu\left(\frac{F_0^2}{2}U\partial^\mu U^\dagger\right)
+\lambda \epsilon^{\mu\nu\rho\sigma} U\partial_\mu U^\dagger
U\partial_\nu U^\dagger U\partial_\rho U^\dagger
U\partial_\sigma U^\dagger=0,
\end{equation}
where $\lambda$ is a (purely imaginary) constant.
   Substituting $U\leftrightarrow U^\dagger$ in Eq.\ (\ref{4:8:eomadd})
and subsequently multiplying from the left by $U$ and from the right
by $U^\dagger$, we verify that the two terms transform with opposite 
relative signs.
   Recall that a term which is even (odd) in the Lagrangian leads to
a term which, in the equation of motion, is odd (even).

   However, the action functional corresponding to the new term cannot be 
written as the four-dimensional integral of a Lagrangian expressed in terms 
of $U$ and its derivatives. 
   Rather, one has to extend the range of definition of the fields to a 
hypothetical fifth dimension,
\begin{equation}
\label{4:8:ualpha}
U(y)=\exp\left(i\alpha\frac{\phi(x)}{F_0}\right),
\quad y^i=(x^\mu,\alpha),\,\,i=0,\cdots, 4,\,\, 0\leq\alpha\leq 1, 
\end{equation}
   where Minkowski space is defined as the surface of the five-dimensional
space for $\alpha =1$.
   Let us first quote the result of the effective Wess-Zumino-Witten action
in the absence of external fields (denoted by a superscript 0):
\begin{eqnarray}
\label{4:8:swzw1}
S_{\rm ano}^0&=&n S_{\rm WZW}^0,\\
\label{4:8:swzw2}
S_{\rm WZW}^0&=&-\frac{i}{240\pi^2}\int_0^1 d \alpha \int d^4 x 
\epsilon^{ijklm} \mbox{Tr}\left(
{\cal U}^L_i
{\cal U}^L_j
{\cal U}^L_k
{\cal U}^L_l
{\cal U}^L_m
\right),
\end{eqnarray}
   where the indices $i,\cdots,m$ run from 0  to 4,
$y_4=y^4=\alpha$, $\epsilon_{ijklm}$ is the completely antisymmetric
tensor with $\epsilon_{01234}=-\epsilon^{01234}=1$,
and  ${\cal U}^L_i=U^\dagger \partial U/\partial y^i$.
   By calculating the variation of the action functional as in Eq.\
(\ref{4:5:deltas}) we find that the constant $\lambda$ of 
Eq.\ (\ref{4:8:eomadd}) and $n$ of Eq.\ (\ref{4:8:swzw1}) are related
by $\lambda=in/(48\pi^2)$.
   Using topological arguments Witten showed that the constant $n$ 
appearing in Eq.\ (\ref{4:8:swzw1}) must be an integer. 
   Below, $n$ will be identified with the number of colors $N_c$. 
   Expanding the SU(3) matrix $U(y)$ in terms of the Goldstone
boson fields, $U(y)=1+i\alpha \phi(x)/F_0+O(\phi^2)$, one obtains
an infinite series of terms, each involving an odd number of
Goldstone bosons, i.e., the WZW action $S_{\rm WZW}^0$
is of odd intrinsic parity.
   For each individual term the $\alpha$ integration can be performed 
explicitly resulting in an ordinary action in terms of a four-dimensional
integral of a local Lagrangian.
   For example, the term with the smallest number of Goldstone bosons
reads  
\begin{eqnarray}
\label{4:8:swzw5phi}
S_{\rm WZW}^{5\phi}&=&\frac{1}{240\pi^2F^5_0}\int_0^1 d\alpha\int d^4 x 
\epsilon^{ijklm}\mbox{Tr}[\partial_i(\alpha\phi)
\partial_j(\alpha\phi)
\partial_k(\alpha\phi)
\partial_l(\alpha\phi)
\partial_m(\alpha\phi)]\nonumber\\
&=&\frac{1}{240\pi^2F^5_0}\int_0^1 d\alpha \int d^4 x
\epsilon^{ijklm}\partial_i\mbox{Tr}[\alpha\phi
\partial_j(\alpha\phi)
\partial_k(\alpha\phi)
\partial_l(\alpha\phi)
\partial_m(\alpha\phi)]\nonumber\\
&=&\frac{1}{240\pi^2 F^5_0}\int d^4 x
\epsilon^{\mu\nu\rho\sigma}\mbox{Tr}(\phi\partial_\mu\phi\partial_\nu\phi
\partial_\rho\phi\partial_\sigma\phi).
\end{eqnarray}
   In the last step we made use of the fact that exactly one index can 
take the value 4. 
   The term involving $i=4$ has been integrated with respect to $\alpha$
whereas the other four possibilities cancel each other because the
$\epsilon$ tensor
in four dimensions is antisymmetric under a cyclic permutation
of the indices whereas the trace is symmetric under a cyclic permutation.
   In particular, the WZW action without external fields involves at
least five Goldstone bosons \cite{Wess:yu}.

   The connection to the number of colors $N_c$ is established by 
introducing a coupling to electromagnetism \cite{Wess:yu,Witten:tw}. 
   In the presence of external fields there will be an additional
term in the anomalous action,
\begin{equation}
\label{4:8:sanofull}
S_{\rm ano}=
S_{\rm ano}^0+
S_{\rm ano}^{\rm ext}=n(S_{\rm WZW}^0+S_{\rm WZW}^{\rm ext}),
\end{equation}
given by \cite{Chou:1983qy,Pak:1984bn,Manes:1984gk,Bijnens:xi}
\begin{equation}
\label{4:8:deltaswzw}
S_{\rm WZW}^{\rm ext}= -\frac{i}{48\pi^2}\int d^4 x\, 
\epsilon^{\mu\nu\rho\sigma}\mbox{Tr}(Z_{\mu\nu\rho\sigma})
\end{equation}
with
\begin{eqnarray}
\label{4:8:zmunialphabeta}
Z_{\mu\nu\rho\sigma}&=&
\frac{1}{2}\ U l_\mu U^\dagger r_\nu U l_\rho U^\dagger r_\sigma\nonumber\\
&&+U l_\mu l_\nu l_\rho U^\dagger r_\sigma
 - U^\dagger r_\mu r_\nu r_\rho U l_\sigma\nonumber\\
&&+i  U\partial_\mu l_\nu l_\rho U^\dagger r_\sigma
 - i U^\dagger\partial_\mu r_\nu r_\rho U l_\sigma\nonumber\\
&&+i \partial_\mu r_\nu U l_\rho U^\dagger r_\sigma
 - i \partial_\mu l_\nu U^\dagger r_\rho U l_\sigma\nonumber\\
&&-i {\cal U}_\mu^L l_\nu U^\dagger r_\rho U l_\sigma
 + i {\cal U}_\mu^R r_\nu U l_\rho U^\dagger r_\sigma\nonumber\\
&&-i {\cal U}_\mu^L l_\nu l_\rho l_\sigma
 + i {\cal U}_\mu^R r_\nu r_\rho r_\sigma\nonumber\\
&&+\frac{1}{2}{\cal U}_\mu^L U^\dagger\partial_\nu r_\rho U l_\sigma
 -\frac{1}{2}{\cal U}_\mu^R U\partial_\nu l_\rho U^\dagger r_\sigma\nonumber\\
&&+\frac{1}{2}{\cal U}_\mu^L U^\dagger r_\nu U\partial_\rho l_\sigma
-\frac{1}{2}{\cal U}_\mu^R U l_\nu U^\dagger\partial_\rho r_\sigma\nonumber\\
&&-{\cal U}_\mu^L{\cal U}_\nu^L U^\dagger r_\rho U l_\sigma
 + {\cal U}_\mu^R{\cal U}_\nu^R U l_\rho U^\dagger r_\sigma\nonumber\\
&&+{\cal U}_\mu^L l_\nu\partial_\rho l_\sigma
 - {\cal U}_\mu^R r_\nu\partial_\rho r_\sigma\nonumber\\
&&+{\cal U}_\mu^L\partial_\nu  l_\rho l_\sigma
 - {\cal U}_\mu^R\partial_\nu  r_\rho r_\sigma\nonumber\\
&&+\frac{1}{2}\ {\cal U}_\mu^L l_\nu {\cal U}_\rho^L l_\sigma
 - \frac{1}{2}\ {\cal U}_\mu^R r_\nu {\cal U}_\rho^R r_\sigma\nonumber\\
&&-i {\cal U}_\mu^L{\cal U}_\nu^L{\cal U}_\rho^L l_\sigma
 + i {\cal U}_\mu^R{\cal U}_\nu^R{\cal U}_\rho^R r_\sigma,
\end{eqnarray}
where we defined the abbreviations ${\cal U}_\mu^L=U^\dagger\partial_\mu U$
and ${\cal U}_\mu^R=U\partial_\mu U^\dagger$.
   In the commonly used expression \cite{Bijnens:xi}, 
we performed the replacement
\begin{eqnarray*}
\lefteqn{{\cal U}_\mu^L U^\dagger\partial_\nu r_\rho U l_\sigma-
{\cal U}_\mu^R U\partial_\nu l_\rho U^\dagger r_\sigma\to}\\
&& 
\frac{1}{2}{\cal U}_\mu^L U^\dagger\partial_\nu r_\rho U l_\sigma
 -\frac{1}{2}{\cal U}_\mu^R U\partial_\nu l_\rho U^\dagger r_\sigma
+\frac{1}{2}{\cal U}_\mu^L U^\dagger r_\nu U\partial_\rho l_\sigma
-\frac{1}{2}{\cal U}_\mu^R U l_\nu U^\dagger\partial_\rho r_\sigma,
\end{eqnarray*}
in order to generate a manifestly $C$ invariant and Hermitian action.
   Without this replacement charge-conjugation invariance and Hermiticity 
are satisfied up to a total derivative only.
   As a special case, let us consider a coupling to external electromagnetic
fields by inserting 
\begin{displaymath}
r_\mu=l_\mu=-e Q {\cal A}_\mu,
\end{displaymath}
where $Q$ is the quark charge matrix [see Eq.\ (\ref{2:4:rla})].
   The terms involving three and four electromagnetic four-potentials
vanish upon contraction with the totally antisymmetric tensor 
$\epsilon^{\mu\nu\rho\sigma}$, because their contributions to 
$Z_{\mu\nu\rho\sigma}$ are symmetric in at least two indices, and
we obtain
\begin{eqnarray}
\label{4:8:lanoelm}
n{\cal L}^{\rm ext}_{\rm WZW}&=&-e n {\cal A}_\mu J^\mu
+i \frac{n e^2 }{48\pi^2}\epsilon^{\mu\nu\rho\sigma}
\partial_\nu {\cal A}_\rho{\cal A}_\sigma
\mbox{Tr}[2Q^2(U\partial_\mu U^\dagger - U^\dagger \partial_\mu U)
\nonumber\\
&&
- Q U^\dagger Q \partial_\mu U 
+Q U Q \partial_\mu U^\dagger].
\end{eqnarray}
   We note that the current
\begin{equation}
\label{4:8:jmu}
J^\mu=\frac{\epsilon^{\mu\nu\rho\sigma}}{48\pi^2}
\mbox{Tr}(Q\partial_\nu U U^\dagger \partial_\rho U U^\dagger
\partial_\sigma U U^\dagger
+Q U^\dagger \partial_\nu U U^\dagger \partial_\rho U U^\dagger
\partial_\sigma U),\quad \epsilon_{0123}=1,
\end{equation}
by itself is not gauge invariant and the additional terms of 
Eq.\ (\ref{4:8:lanoelm}) are required to obtain a gauge-invariant action.

   The identification of the constant $n$ with the number of colors $N_c$
\cite{Witten:tw} results from finding in Eq.\ (\ref{4:8:lanoelm}) 
the interaction Lagrangian which is relevant to the decay
$\pi^0\to\gamma\gamma$.
   Since $U=1+i\mbox{diag}(\pi^0,-\pi^0,0)/F_0+\cdots$, 
Eq.\ (\ref{4:8:lanoelm}) contains a piece
\begin{equation}
\label{4:8:lpi0gg}
{\cal L}_{\pi^0\gamma\gamma}=-\frac{n e^2}{96\pi^2}\epsilon^{\mu\nu
\rho\sigma}{\cal F}_{\mu\nu}{\cal F}_{\rho\sigma}\frac{\pi^0}{F_0},
\end{equation}
where we made use of a partial integration to shift the derivative from
the pion field onto the electromagnetic four-potential.
   The corresponding invariant amplitude reads
\begin{equation}
\label{4:8:mpi0gg}
{\cal M}=i\frac{n e^2}{12\pi^2 F_0}\epsilon^{\mu\nu\rho\sigma}
q_{1\mu}\epsilon_{1\nu}^\ast q_{2\rho}\epsilon_{2\sigma}^\ast,
\end{equation}
   which agrees with a direct calculation of the anomaly term in terms of $u$ 
and $d$ quarks with $n=N_c$ colors (see, e.g.,  Ref.\ \cite{Veltman:wz}).
   After summation over the final photon polarizations and
integration over phase space, the decay rate reads
\begin{equation}
\label{}
\Gamma_{\pi^0\to\gamma\gamma}=
\frac{\alpha^2 M^3_{\pi^0} n^2}{576 \pi^3 F^2_0}
=7.6\,\mbox{eV}\times\left(\frac{n}{3}\right)^2,
\end{equation}
   which is in good agreement with the experimental value
$(7.7\pm 0.6)$ eV for $n=N_c=3$ \cite{Groom:in}.

\section{Applications at Order ${\cal O}(p^4)$}
\label{sec_aop4}
\subsection{Masses of the Goldstone Bosons}
\label{subsec_mgb}
   A discussion of the masses at ${\cal O}(p^4)$ will allow us to illustrate
various properties typical of chiral perturbation theory:
\begin{enumerate}
\item The relation between the bare low-energy coupling constants $L_i$ and 
the renormalized coefficients $L_i^r$ in Eq.~(\ref{4:7:li}) is such that
the divergences of one-loop diagrams are canceled.
\item Similarly, the scale dependence of the coefficients $L^r_i(\mu)$ on the
one hand and of the finite contributions of the one-loop diagrams on
the other hand leads to scale-independent predictions for physical observables.
\item A perturbation expansion in the explicit symmetry breaking
with respect to a symmetry that is realized in the Nambu-Goldstone mode
generates corrections which are non-analytic in the symmetry breaking 
parameter \cite{Li:1971vr}, here the quark masses.
\end{enumerate} 
   Let us consider ${\cal L}_2 + {\cal L}_4$ for QCD with finite quark
masses but in the absence of external fields.
   We restrict ourselves to the limit of isospin symmetry, i.e.,
$m_u=m_d=m$.
   In order to determine the masses we calculate the self energies
$\Sigma(p^2)$ of the Goldstone bosons.

   The propagator of a (pseudo-) scalar field is defined as the
Fourier transform of the two-point Green function:
\begin{equation}
\label{4:8:propdef}
i\Delta(p)=\int d^4 x e^{-ip\cdot x}\langle
0|T\left[\Phi_0(x)\Phi_0(0)\right]|0\rangle,
\end{equation}
   where the index 0  refers to the fact that we still deal with the
bare unrenormalized field---not to be confused with a free field without 
interaction.
   At lowest order ($D=2$) the propagator simply reads
\begin{equation}
\label{4:8:prop}
i\Delta(p)=\frac{i}{p^2-M^2_0+i0^+},
\end{equation}
where the lowest-order masses $M_0$ are given in 
Eqs.\ (\ref{4:3:mpi2}) - (\ref{4:3:meta2}):
\begin{eqnarray*}
M^2_{\pi,0}&=&2 B_0 m,\\
M^2_{K,0}&=&B_0(m+m_s),\\
M^2_{\eta,0}&=&\frac{2}{3} B_0\left(m+2m_s\right).
\end{eqnarray*} 
   The loop diagrams with ${\cal L}_2$ and the contact diagrams with
${\cal L}_4$ result in so-called proper self-energy insertions
$-i\Sigma(p^2)$, which may be summed using a geometric series
(see Fig.\ \ref{4:8:fullprop}):
\begin{figure}
\begin{center}
\epsfig{file=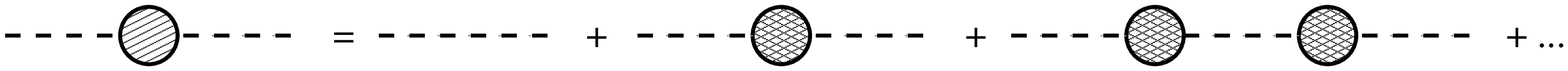,width=12cm}
\caption{\label{4:8:fullprop}
Unrenormalized propagator as the sum of irreducible self-energy diagrams.
Hatched and cross-hatched ``vertices'' denote one-particle-reducible
and one-particle-irreducible contributions, respectively.}
\end{center}
\end{figure}
\begin{eqnarray}
\label{4:8:prop1}
i\Delta(p)&=&\frac{i}{p^2-M^2_0+i0^+}+\frac{i}{p^2-M^2_0+i0^+}
[-i\Sigma(p^2)]\frac{i}{p^2-M^2_0+i0^+}+\cdots\nonumber\\
&=&\frac{i}{p^2-M^2_0-\Sigma(p^2)+i0^+}.
\end{eqnarray}
   Note that $-i\Sigma(p^2)$ consists of one-particle-irreducible
diagrams only, i.e., diagrams which do not fall apart into
two separate pieces when cutting an arbitrary internal line.
   The physical mass, including the interaction, is defined as the position
of the pole of Eq.\ (\ref{4:8:prop1}),
\begin{equation}
\label{4:8:mdef}
M^2-M^2_0-\Sigma(M^2)\stackrel{!}{=}0.
\end{equation}
   Let us assume that $\Sigma(p^2)$ can be expanded in a series around
$p^2=\lambda^2$,
\begin{equation}
\label{4:8:sigmaexp}
\Sigma(p^2)=\Sigma(\lambda^2)
+(p^2-\lambda^2)\Sigma'(\lambda^2)+\tilde{\Sigma}(p^2),
\end{equation}
where the remainder $\tilde{\Sigma}(p^2)$ depends on the choice
of $\lambda^2$ and satisfies 
$\tilde{\Sigma}(\lambda^2)=\tilde{\Sigma}'(\lambda^2)=0$.
   We then obtain for the propagator
\begin{equation}
\label{4:8:prop2}
i\Delta(p)=\frac{i}{p^2-M^2_0
-\Sigma(\lambda^2)-(p^2-\lambda^2)\Sigma'(\lambda^2)
-\tilde{\Sigma}(p^2)+i0^+}.
\end{equation}
   Taking $\lambda^2=M^2$ in Eq.\ (\ref{4:8:prop2})
and applying the condition of Eq.\ 
(\ref{4:8:mdef}), the propagator may be written as   
$$
i\Delta(p)=\frac{i}{(p^2-M^2)[1-\Sigma'(M^2)]-\tilde{\Sigma}(p^2)+i0^+}
=\frac{iZ_\Phi}{p^2-M^2-Z_\Phi\tilde{\Sigma}(p^2)+i0^+},
$$
   where we have introduced the wave function renormalization constant
$$Z_\Phi=\frac{1}{1-\Sigma'(M^2)}.
$$
  Introducing renormalized fields as $\Phi_R=\Phi_0/\sqrt{Z_\Phi}$,
the renormalized propagator is given by  
\begin{eqnarray*}
i\Delta_R(p)&=&\int d^4 x e^{-ip\cdot x}
\langle 0|T[\Phi_R(x)\Phi_R(0)]|0\rangle\nonumber\\
&=&\frac{i}{p^2-M^2-Z_\Phi\tilde{\Sigma}(p^2)+i0^+}.
\end{eqnarray*}
   In particular, since $\tilde{\Sigma}(M^2)
=\tilde{\Sigma}'(M^2)=0$,
in the vicinity of the pole, the renormalized propagator
behaves as a free propagator with physical mass $M^2$.

\begin{figure}
\begin{center}
\epsfig{file=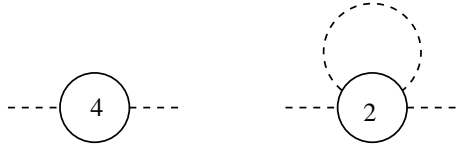,width=6cm}
\caption{\label{4:8:selfenergy}
Self-energy diagrams at ${\cal O}(p^4)$.
   Vertices derived from ${\cal L}_{2n}$ are denoted by $2n$ in the
interaction blobs.}
\end{center}
\end{figure}

   Let us now turn to the calculation within the framework of ChPT 
(see, e.g., Ref.\ \cite{Rudy:1994qb}).
   Since ${\cal L}_2$ and ${\cal L}_4$ without external fields generate
vertices with an even number of Goldstone bosons only, 
the candidate terms at $D=4$ contributing to the self energy 
are those shown in Fig.\ \ref{4:8:selfenergy}.
   For our particular application with exactly two external meson lines,
the relevant interaction Lagrangians can be written as \cite{Rudy:1994qb}
\begin{equation}
{\cal L}_{\rm int}={\cal L}_2^{4\phi}+{\cal L}_4^{2\phi},
\end{equation}
   where ${\cal L}_2^{4\phi}$ is given by
\begin{equation}
\label{4:8:l24phi}
{\cal L}^{4\phi}_2=\frac{1}{24 F^2_0}\left\{\mbox{Tr}(
[\phi,\partial_\mu \phi]\phi \partial^\mu \phi)
+B_0\mbox{Tr}(M\phi^4)\right\}.
\end{equation}
   The terms of ${\cal L}_4$ proportional to  $L_9$, $L_{10}$, $H_1$, and 
$H_2$ do not contribute, because they either contain field-strength tensors
or external fields only.
   Since $\partial_\mu U=O(\phi)$, the $L_1$, $L_2$, and $L_3$ terms of
Eq.\ (\ref{4:7:l4gl}) are $O(\phi^4)$ and need not be considered.
   The only candidates are the $L_4$ - $L_8$ terms, of which we consider
the $L_4$ term as an explicit example,\footnote{For pedagogical reasons, 
we make use of the physical fields. 
From a technical point of view, it is often advantageous
to work with the Cartesian fields and, at the end of the calculation, 
express physical processes in terms of the Cartesian components.}
\begin{eqnarray*}
\lefteqn{L_4\mbox{Tr}(\partial_\mu U \partial^\mu U^\dagger)
\mbox{Tr}(\chi U^\dagger
+U \chi^\dagger)=}\\
&&L_4 \frac{2}{F_0^2}[\partial_\mu \eta \partial^\mu \eta
+\partial_\mu \pi^0 \partial^\mu \pi^0
+2\partial_\mu \pi^+\partial^\mu\pi^-+2\partial_\mu K^+\partial^\mu K^-\\
&&+2\partial_\mu K^0\partial^\mu \bar{K}^0+O(\phi^4)]
[4B_0(2m+m_s)+O(\phi^2)].
\end{eqnarray*} 
   The remaining terms are treated analogously and we obtain for
${\cal L}_4^{2\phi}$ 
\begin{eqnarray}
\label{4:8:l42phi}
{\cal L}_4^{2\phi}&=&
\frac{1}{2}\left(a_\eta\partial_\mu\eta\partial^\mu\eta-b_\eta \eta^2\right)
\nonumber\\
&&+\frac{1}{2}\left(a_\pi\partial_\mu\pi^0\partial^\mu\pi^0-b_\pi\pi^0\pi^0
\right)\nonumber\\
&&+a_\pi\partial_\mu\pi^+\partial^\mu\pi^--b_\pi\pi^+\pi^-\nonumber\\
&&+a_K\partial_\mu K^+\partial^\mu K^- - b_K K^+K^-\nonumber\\
&&+a_K\partial_\mu K^0\partial^\mu\bar{K}^0-b_K K^0\bar{K}^0,
\end{eqnarray}
where the constants $a_\phi$ and $ b_\phi$ are given by
\begin{eqnarray}
\label{4:8:ab}
a_\eta&=&\frac{16B_0}{F^2_0}\left[(2m+m_s)L_4+\frac{1}{3}(m+2m_s)L_5\right],
\nonumber\\
b_\eta&=&\frac{64 B^2_0}{3F^2_0}\left[(2m+m_s)(m+2m_s)L_6+2(m-m_s)^2L_7
+(m^2+2m_s^2)L_8\right],\nonumber\\
a_\pi&=&\frac{16 B_0}{F^2_0}\left[(2m+m_s)L_4+mL_5\right],\nonumber\\
b_\pi&=&\frac{64 B^2_0}{F^2_0}\left[(2m+m_s)mL_6+m^2L_8\right],\nonumber\\
a_K&=&\frac{16 B_0}{F^2_0}\left[(2m+m_s)L_4+\frac{1}{2}(m+m_s)L_5\right],
\nonumber\\
b_K&=&\frac{32B^2_0}{F^2_0}\left[(2m+m_s)(m+m_s)L_6+\frac{1}{2}(m+m_s)^2L_8
\right].
\end{eqnarray}
   At ${\cal O}(p^4)$ the self energies are of the form
\begin{equation}
\label{4:8:sigmaphi}
\Sigma_\phi(p^2)=A_\phi+B_\phi p^2,
\end{equation}   
   where the constants $A_\phi$ and $B_\phi$ receive a tree-level contribution
from ${\cal L}_4$ and a one-loop contribution with a vertex from
${\cal L}_2$ (see Fig.\ \ref{4:8:selfenergy}).
   For the tree-level contribution of ${\cal L}_4$ this is easily seen, because
the Lagrangians of Eq.\ (\ref{4:8:l42phi}) contain either exactly
two derivatives of the fields or no derivatives at all.
   For example, the contact contribution for the $\eta$ reads
$$-i\Sigma_\eta^{\rm contact}(p^2)=i 2\left[\frac{1}{2}a_\eta (ip_\mu)(-ip^\mu)
-\frac{1}{2}b_\eta\right]=i(a_\eta p^2-b_\eta),$$
where, as usual, $\partial_\mu \phi$ generates $-ip_\mu$ and $ip_\mu$ 
for initial and final lines, respectively, and the factor two takes
account of two combinations of contracting the fields with external lines.

   For the one-loop contribution the argument is as follows.
   The Lagrangian ${\cal L}_2^{4\phi}$ contains either two derivatives
or no derivatives at all which, symbolically, can be written as
$\phi\phi\partial\phi\partial\phi$ and $\phi^4$, respectively.
   The first term results in $M^2$ or $p^2$, depending on whether the
$\phi$ or the $\partial \phi$ are contracted with the external fields.
   The ``mixed'' situation vanishes upon integration.
   The second term, $\phi^4$, does not generate a momentum dependence.

\begin{figure}
\begin{center}
\epsfig{file=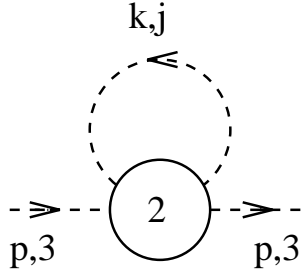,width=4cm}
\caption{\label{4:8:pi0seloop} Contribution of the pion loops to
the $\pi^0$ self energy.}
\end{center}
\end{figure}

   As a specific example, we evaluate the pion-loop contribution to the
$\pi^0$ self energy (see Fig.\ \ref{4:8:pi0seloop}) by applying the
Feynman rule of Eq.\ (\ref{4:6:mpipi2}) 
for $a=c=3$, $p_a=p_c=p$, $b=d=j$, and $p_b=p_d=k$:\footnote{Note that we work
in SU(3) and thus with the exponential parameterization of $U$.}
\begin{eqnarray}
\label{4:8:diag}
&&
\frac{1}{2}\int \frac{d^4k}{(2\pi)^4}i
\left[
\underbrace{\delta^{3j}\delta^{3j}}_{\mbox{$1$}}
\frac{(p+k)^2-M_{\pi,0}^2}{F_0^2}
+\underbrace{\delta^{33}\delta^{jj}}_{\mbox{$3$}}
\frac{-M_{\pi,0}^2}{F_0^2}
+\underbrace{\delta^{3j}\delta^{j3}}_{\mbox{$1$}}
\frac{(p-k)^2-M_{\pi,0}^2}{F_0^2}\right.\nonumber\\
&&\left.-\frac{1}{3 F_0^2}
\underbrace{(\delta^{3j}\delta^{3j}+\delta^{33}\delta^{jj}
+\delta^{3j}\delta^{j3})}_{\mbox{5}}(2p^2+2k^2-4M_{\pi,0}^2)\right]
\frac{i}{k^2-M_{\pi,0}^2+i0^+}\nonumber\\
&&=\frac{1}{2}\int \frac{d^4k}{(2\pi)^4}\frac{i}{3F_0^2}
[-4p^2-4k^2+5 M^2_{\pi,0}]
\frac{i}{k^2-M_{\pi,0}^2+i0^+},
\end{eqnarray}
   where $1/2$ is a symmetry factor, as explained in Sec.\ \ref{sec_elwpcs}.
   The integral of Eq.\ (\ref{4:8:diag}) diverges and we thus consider
its extension to $n$ dimensions in order to make use of the 
dimensional-regularization technique described in App.\ \ref{app_drb}.
   In addition to the loop-integral of Eq.\ (\ref{app:drb:im22}),
\begin{eqnarray}
\label{4:8:im22}
I(M^2,\mu^2,n)&=&\mu^{4-n}\int\frac{d^nk}{(2\pi)^n}\frac{i}{k^2-M^2+i0^+}
\nonumber\\
&=&\frac{M^2}{16\pi^2}\left[
R+\ln\left(\frac{M^2}{\mu^2}\right)\right]+O(n-4),
\end{eqnarray}
where $R$ is given in Eq.\ (\ref{4:7:R}), 
we need
\begin{displaymath}
\mu^{4-n}i\int \frac{d^n k}{(2\pi)^n}\frac{k^2}{k^2-M^2+i0^+}=
\mu^{4-n}i\int \frac{d^n k}{(2\pi)^n}\frac{k^2-M^2+M^2}{k^2-M^2+i0^+},
\end{displaymath}
   where we have added $0=-M^2+M^2$ in the numerator.
We make use of 
$$\mu^{4-n}i\int \frac{d^n k}{(2\pi)^n}=0$$
in dimensional regularization 
(see the discussion at the end of Appendix \ref{subsec_jpin})
and obtain
$$
\mu^{4-n}i\int \frac{d^n k}{(2\pi)^n}\frac{k^2}{k^2-M^2+i0^+}=
M^2 I(M^2,\mu^2,n),
$$ 
with $I(M^2,\mu^2,n)$ of Eq.\ (\ref{4:8:im22}). 
   The pion-loop contribution to the $\pi^0$ self energy is thus
$$\frac{i}{6 F^2_0}(-4p^2+M^2_{\pi,0})I(M^2_{\pi,0},\mu^2,n),$$
which is indeed of the type discussed in Eq.\ (\ref{4:8:sigmaphi})
and diverges as $n\to 4$.

   After analyzing all loop contributions and combining them with the contact 
contributions of Eqs.\ (\ref{4:8:ab}), the constants 
$A_\phi$ and $B_\phi$ of Eq.\ (\ref{4:8:sigmaphi}) are given by
\begin{eqnarray}
\label{4:8:AB}
A_\pi&=&\frac{M^2_\pi}{F^2_0}\Bigg\{
\underbrace{-\frac{1}{6}I(M^2_\pi)
-\frac{1}{6}I(M^2_\eta)-\frac{1}{3}I(M^2_K)}_{\mbox{one-loop contribution}}
\nonumber\\
&&\underbrace{+32[(2m+m_s)B_0L_6+mB_0L_8]}_{\mbox{contact contribution}}
\Bigg\},\nonumber\\
B_\pi&=&\frac{2}{3}\frac{I(M^2_\pi)}{F^2_0}+\frac{1}{3}
\frac{I(M^2_K)}{F^2_0}-\frac{16B_0}{F^2_0}\left[
(2m+m_s)L_4+mL_5\right],\nonumber\\
A_K&=&\frac{M^2_K}{F^2_0}\Bigg\{\frac{1}{12}I(M^2_\eta)
-\frac{1}{4}I(M^2_\pi)-\frac{1}{2}I(M^2_K)\nonumber\\
&&+32\left[(2m+m_s)B_0L_6+\frac{1}{2}(m+m_s)B_0L_8\right]\Bigg\},\nonumber\\
B_K&=&\frac{1}{4}\frac{I(M^2_\eta)}{F^2_0}
+\frac{1}{4}\frac{I(M^2_\pi)}{F^2_0}
+\frac{1}{2}\frac{I(M^2_K)}{F^2_0}\nonumber\\
&&-16 \frac{B_0}{F^2_0}\left[(2m+m_s)L_4+\frac{1}{2}(m+m_s)L_5\right],
\nonumber\\
A_\eta&=&\frac{M^2_\eta}{F^2_0}\left[-\frac{2}{3}I(M^2_\eta)\right]
+\frac{M^2_\pi}{F^2_0}\left[\frac{1}{6}I(M^2_\eta)-\frac{1}{2}I(M^2_\pi)
+\frac{1}{3}I(M^2_K)\right]\nonumber\\
&&+\frac{M^2_\eta}{F^2_0}[16M^2_\eta L_8+32(2m+m_s)B_0L_6]\nonumber\\
&&+\frac{128}{9}\frac{B^2_0(m-m_s)^2}{F^2_0}(3L_7+L_8),\nonumber\\
B_\eta&=&\frac{I(M^2_K)}{F^2_0}-\frac{16}{F^2_0}(2m+m_s)B_0L_4
-8\frac{M^2_\eta}{F^2_0}L_5,
\end{eqnarray}
   where, for simplicity, we have suppressed the dependence on the
scale $\mu$ and the number of dimensions $n$ in
the integrals $I(M^2,\mu^2,n)$ [see Eq.\ (\ref{4:8:im22})].
   Furthermore, the squared masses appearing in the loop integrals of
Eq.\ (\ref{4:8:AB}) are given by the predictions of lowest order, 
Eqs.\ (\ref{4:3:mpi2}) - (\ref{4:3:meta2}). 
   Finally, the integrals $I$ as well as the bare coefficients $L_i$ 
(with the exception of $L_7$) have $1/(n-4)$ poles and finite pieces.
   In particular, the coefficients $A_\phi$ and $B_\phi$ are {\em not}
finite as $n\to 4$.

   The masses at ${\cal O}(p^4)$ are determined by solving the
general equation
\begin{equation}
\label{4:8:mse}
M^2=M_0^2+\Sigma(M^2)
\end{equation}
with the predictions of Eq.\ (\ref{4:8:sigmaphi}) for the self energies,
$$
M^2=M_0^2+A+BM^2,
$$
where the lowest-order terms, $M^2_0$, are given in Eqs.\   
(\ref{4:3:mpi2}) - (\ref{4:3:meta2}).
   We then obtain
\begin{displaymath}
M^2=\frac{M_0^2+A}{1-B}=M_0^2(1+B)+A+
{\cal O}(p^6), 
\end{displaymath}
   because $A={\cal O}(p^4)$ and $\{B, M_0^2\}={\cal O}(p^2)$.
   Expressing the bare coefficients $L_i$ in Eq.\ (\ref{4:8:AB}) in terms of 
the renormalized coefficients by using Eq.\ (\ref{4:7:li}),
the results for the masses of the Goldstone bosons at ${\cal O}(p^4)$
read
\begin{eqnarray}
\label{4:8:mpi24}
M^2_{\pi,4}&=&M^2_{\pi,2}\Bigg\{1+\frac{M^2_{\pi,2}}{32\pi^2F^2_0}
\ln\left(\frac{M^2_{\pi,2}}{\mu^2}\right)-\frac{M^2_{\eta,2}}{96\pi^2F^2_0}
\ln\left(\frac{M^2_{\eta,2}}{\mu^2}\right)\nonumber\\
&&+\frac{16}{F^2_0}\left[(2m+m_s)B_0(2L^r_6-L^r_4)
+mB_0(2L^r_8-L^r_5)\right]\Bigg\},\\
\label{4:8:mk24}
M^2_{K,4}&=&M^2_{K,2}\Bigg\{1+\frac{M^2_{\eta,2}}{48\pi^2F^2_0}
\ln\left(\frac{M^2_{\eta,2}}{\mu^2}\right)\nonumber\\
&&+\frac{16}{F^2_0}\left[(2m+m_s)B_0(2L^r_6-L^r_4)
+\frac{1}{2}(m+m_s)B_0(2L^r_8-L^r_5)\right]\Bigg\},\nonumber\\
&&\\
\label{4:8:meta24}
M^2_{\eta,4}&=&M^2_{\eta,2}\left[1+\frac{M^2_{K,2}}{16\pi^2F^2_0}
\ln\left(\frac{M^2_{K,2}}{\mu^2}\right)
-\frac{M^2_{\eta,2}}{24\pi^2F^2_0}\ln\left(\frac{M^2_{\eta,2}}{\mu^2}\right)
\right.\nonumber\\
&&\left.+\frac{16}{F^2_0}(2m+m_s)B_0(2L^r_6-L^r_4)
+8\frac{M^2_{\eta,2}}{F^2_0}(2L^r_8-L^r_5)\right]\nonumber\\
&&+M^2_{\pi,2}\left[\frac{M^2_{\eta,2}}{96\pi^2F^2_0}
\ln\left(\frac{M^2_{\eta,2}}{\mu^2}\right)
-\frac{M^2_{\pi,2}}{32\pi^2F^2_0}
\ln\left(\frac{M^2_{\pi,2}}{\mu^2}\right)\right.\nonumber\\
&&\left.
+\frac{M^2_{K,2}}{48\pi^2F^2_0}
\ln\left(\frac{M^2_{K,2}}{\mu^2}\right)\right]\nonumber\\
&&+\frac{128}{9}\frac{B^2_0(m-m_s)^2}{F^2_0}
(3L^r_7+L^r_8).
\end{eqnarray}
   In Eqs.\ (\ref{4:8:mpi24}) - (\ref{4:8:meta24}) we have included the
subscripts 2 and 4 in order to indicate from which chiral order the predictions
result.
   First of all, we note that the expressions for the masses are finite.
   The bare coefficients $L_i$ of the Lagrangian of Gasser and Leutwyler
must be infinite in order to cancel the infinities resulting from the 
divergent loop integrals.
   Furthermore, at ${\cal O}(p^4)$ the masses of the Goldstone bosons 
vanish, if the quark masses are sent to zero.
   This is, of course, what we had expected from QCD in the chiral limit
but it is comforting to see that the self interaction in ${\cal L}_2$ 
(in the absence of quark masses) does not generate Goldstone
boson masses at higher order.
   At ${\cal O}(p^4)$, the squared Goldstone boson masses contain terms
which are analytic in the quark masses, namely, of the form $m^2_q$ 
multiplied by the renormalized low-energy coupling constants $L_i^r$.
   However, there are also non-analytic terms  of the 
type $m^2_q \ln(m_q)$---so-called  chiral logarithms---which do not involve 
new parameters.
   Such a behavior is an illustration of the mechanism found by Li and
Pagels \cite{Li:1971vr}, who noticed that a perturbation theory around
a symmetry which is realized in the Nambu-Goldstone mode results in both 
analytic as well as non-analytic expressions in the perturbation.
   Finally, the scale dependence of the renormalized coefficients 
$L_i^r$ of Eq.\ (\ref{4:7:limu1mu2}) is by construction such that it cancels 
the scale dependence of the chiral logarithms.
   Thus, physical observables do not depend on the scale $\mu$.
   Let us verify this statement by differentiating Eqs.\ (\ref{4:8:mpi24}) -
(\ref{4:8:meta24}) with respect to $\mu$.
   Using Eq.\ (\ref{4:7:limu1mu2}), 
$$L_i^r(\mu)=L_i^r(\mu')+\frac{\Gamma_i}{16\pi^2}\ln\left(\frac{\mu'}{\mu}
\right), 
$$
we obtain
$$   
\frac{d L_i^r(\mu)}{d\mu}=-\frac{\Gamma_i}{16\pi^2\mu}
$$
and, analogously, for the chiral logarithms
$$\frac{d}{d\mu}\ln\left(\frac{M^2}{\mu^2}\right)=
2\frac{d}{d\mu}\left[\ln(M)-\ln(\mu)\right]=-\frac{2}{\mu}.
$$
   As a specific example, let us differentiate the expression for the
pion mass
\begin{eqnarray*}
\frac{d M^2_{\pi,4}}{d\mu}&=&\frac{M^2_{\pi,2}}{16\pi^2\mu F_0^2}\Bigg\{
\frac{M^2_{\pi,2}}{2}(-2)-\frac{M^2_{\eta,2}}{6}(-2)\\
&&+16[(2m+m_s)B_0(-2\Gamma_6+\Gamma_4)+mB_0(-2\Gamma_8+\Gamma_5)]\Bigg\}\\
&=&\frac{M^2_{\pi,2}}{16\pi^2\mu F_0^2}\Bigg\{-2B_0m +\frac{2}{9}(m+2m_s)B_0\\
&&+16\Bigg[(2m+m_s)B_0
\underbrace{\left(-2\frac{11}{144}+\frac{1}{8}\right)}_{\mbox{$-\frac{1}{36}$}}
+mB_0\underbrace{\left(-2\frac{5}{48}+\frac{3}{8}\right)
}_{\mbox{$\frac{1}{6}$}}
\Bigg]\Bigg\}\\
&=&\frac{M^2_{\pi,2}}{16\pi^2\mu F_0^2}\Bigg\{
B_0m
\left(-2+\frac{2}{9}-\frac{8}{9}+\frac{8}{3}\right)
+B_0m_s\left(\frac{4}{9}-\frac{16}{36}\right)\Bigg\}\\
&=&0,
\end{eqnarray*}
where we made use of the $\Gamma_i$ of Table \ref{4:7:tableli}.

\subsection{The Electromagnetic Form Factor of the Pion}
\label{subsec_emffp}
   As a second application at ${\cal O}(p^4)$, we discuss
the electromagnetic (or vector) form factor of the pion in 
SU(2)$\times$SU(2) chiral perturbation theory.  
   We will work with two commonly used versions of the ${\cal O}(p^4)$
$\mbox{SU(2)}\times\mbox{SU(2)}$ mesonic Lagrangian
\cite{Gasser:1983yg,Gasser:1987rb} which are related
by a field transformation (see App.\ \ref{app_sec_glvgss}).
   Furthermore, we will perform the calculation with the two parameterizations
for $U$ of Eqs.\ (\ref{4:6:u1}) and (\ref{4:6:u2}).
   We will thus be able to extend the observations 
of Sec.\ \ref{subsec_pps} regarding the invariance of physical results
under a change of variables to the one-loop level.
    
   According to Eq.\ (\ref{2:4:rlasu2}), in the two-flavor sector the coupling
to the electromagnetic field ${\cal A}_\mu$ contains both isoscalar and
isovector terms:
\begin{displaymath}
{\cal L}_{\rm ext}=\bar{q}\gamma^\mu\left(\frac{1}{3}v_\mu^{(s)}+v_\mu
\right)q
=-e{\cal A}_\mu\bar{q}\gamma^\mu\left(\frac{1}{6}+\frac{\tau_3}{2}\right)q
=-e {\cal A}_\mu J^\mu,
\end{displaymath}
i.e.
\begin{eqnarray}
\label{4:9rlasu2}
v_\mu&=&r_\mu=l_\mu=-e\frac{\tau_3}{2}{\cal A}_\mu,\\
\label{4:9vsasu2}
v_\mu^{(s)}&=&-\frac{e}{2}{\cal A}_\mu.
\end{eqnarray}
   When evaluating the electromagnetic current operator 
\begin{displaymath}
J^\mu=\frac{1}{6}\bar{q}\gamma^\mu q+\bar{q}\gamma^\mu\frac{\tau_3}{2} q
\end{displaymath} 
between $|\pi^i(p)\rangle$ and $\langle \pi^j(p')|$, the isoscalar first
term does not contribute,\footnote{\label{fnic}
The matrix element $\langle \pi^j(p')|\bar{q}\gamma^\mu q|\pi^i(p)\rangle$
must be of the form $\delta^{ij}(p'+p)^\mu f(q^2)$ 
which results in $(p+p')^\mu f(q^2)$ for the neutral pion ($i=j=3$). 
   On the other hand, under charge conjugation 
$\bar{q}\gamma^\mu q\mapsto -\bar{q}\gamma^\mu q$ and $|\pi^0\rangle\mapsto
|\pi^0\rangle$, and thus $f(q^2)=-f(q^2)=0$.}
and the matrix element of the electromagnetic current operator
must be of the form\footnote{A second structure proportional to $q^\mu$ 
vanishes for on-mass-shell pions because of current conservation.}
\begin{equation}
\langle \pi^j(p')|J^\mu(0)|\pi^i(p)\rangle=i\epsilon_{3ij}(p'+p)^\mu F(q^2),
\quad q=p'-p.
\end{equation}
   In other words, we only need to consider Eq.\ (\ref{4:9rlasu2}) 
which corresponds to a coupling to the third component of the isovector 
current operator.

\begin{figure}[htb]
\begin{center}
\epsfig{file=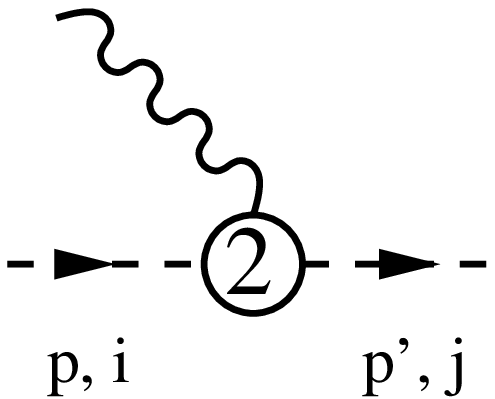,width=4cm}
\caption{\label{4:9:pionffl2} Tree-level diagram at ${\cal O}(p^2)$.}
\end{center}
\end{figure}

   In the calculations that follow we make use of the parameterization
of Eq.\ (\ref{4:6:u1}) for the SU(2) matrix $U(x)$:
\begin{equation}
\label{4:9:sqrt}
U(x)=\frac{1}{F_0}\left[\sigma(x)+i\vec{\tau}\cdot\vec{\pi}(x)
\right],\quad \sigma(x)=\sqrt{F^2_0-\vec{\pi}\,^2(x)},
\end{equation}
but we will comment along the way on the features, which would
differ when using the parameterization of Eq.\ (\ref{4:6:u2}).  
   (The equivalence theorem guarantees that physical observables
do not depend on the specific choice of parameterization of $U$
\cite{Chisholm,Kamefuchi:sb}.)
   The covariant derivative of $U$ with the external fields of 
Eq.\ (\ref{4:9rlasu2}) reads
[see Eq.\ (\ref{4:5:kaa})]
\begin{displaymath}
D_\mu U=\partial_\mu U+\frac{i}{2}e{\cal A}_\mu[\tau_3,U]
\end{displaymath}
and generates, when inserted into the lowest-order Lagrangian of 
Eq.\ (\ref{4:5:l2}), the interaction term 
\begin{equation}
\label{4:9:l2gpp}
{\cal L}_2^{\gamma \pi\pi}=-e\epsilon_{3ij}\pi_i\partial^\mu \pi_j 
{\cal A}_\mu.
\end{equation}
   At ${\cal O}(p^2)$, therefore, the interaction is that of point-like pions
with form factor $F(q^2)=1$, resulting in the Feynman amplitude (see Fig.\
\ref{4:9:pionffl2})
\begin{equation}
\label{4:9:m2gpp}
e\epsilon_{3ij}\epsilon\cdot(p'+p),
\end{equation}
   where $\epsilon$ denotes the polarization vector of the external real
or virtual photon.\footnote{For example, in electron scattering reactions
often the polarization vector $\epsilon_\mu=e\bar{u}(k_f)\gamma_\mu
u(k_i)/q^2$ is used, with four-momentum transfer $q=k_i-k_f$.}
   In particular, using the parameterization of Eq.\ (\ref{4:9:sqrt}),
all interaction terms containing one electromagnetic field and $2n$ pions
vanish for $n\geq 2$.
   This is not the case for the exponential parameterization of 
Eq.\ (\ref{4:6:u2}) which generates the more complicated interaction
Lagrangian
\begin{equation}
\label{4:9:l2exp}
-e\epsilon_{3ij}\phi_i\partial^\mu \phi_j{\cal A}_\mu
\underbrace{\frac{F_0^2}{|\vec\phi|^2}
\sin^2\left(\frac{|\vec{\phi}|}{F_0}\right)}_{
\mbox{$1-\frac{1}{3}\frac{|\vec{\phi}|^2}{F_0^2}+\cdots$}}
\end{equation}
   which is the same as Eq.\ (\ref{4:9:l2gpp}) only at lowest order in the 
fields.

\begin{figure}[htb]
\begin{center}
\epsfig{file=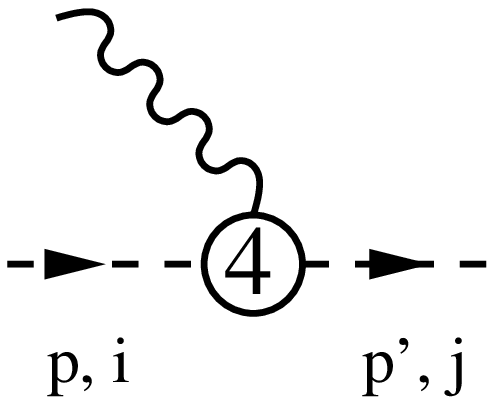,width=4cm}
\caption{\label{4:9:pionffl4} Tree-level diagram at ${\cal O}(p^4)$.}
\end{center}
\end{figure}

\begin{figure}[htb]
\begin{center}
\epsfig{file=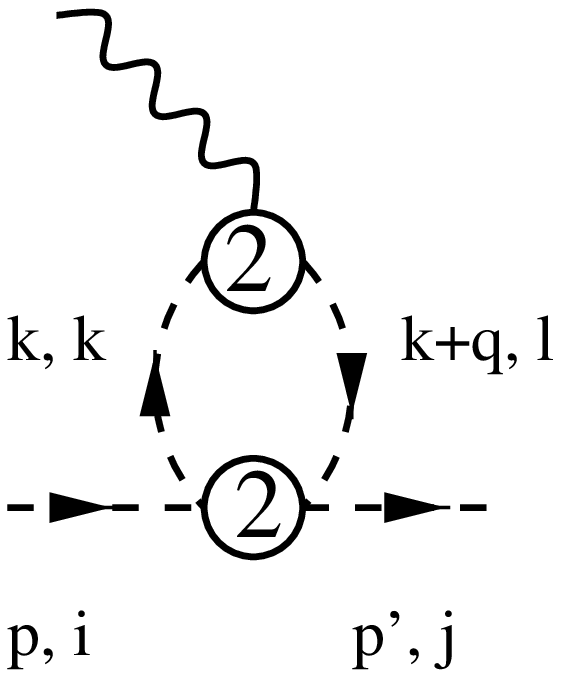,width=4cm}
\caption{\label{4:9:pionffloop1} One-loop diagram at ${\cal O}(p^4)$.}
\end{center}
\end{figure}

\begin{figure}[htb]
\begin{center}
\epsfig{file=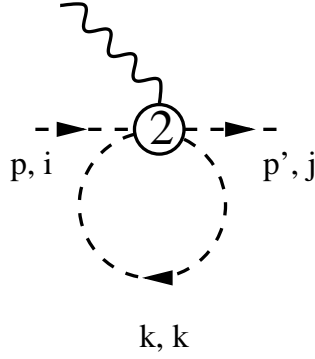,width=4cm}
\caption{\label{4:9:pionffloop2} One-loop diagram at ${\cal O}(p^4)$
contributing in the parameterization of Eq.\ (\ref{4:6:u2})
only.}
\end{center}
\end{figure}

   At ${\cal O}(p^4)$ we need to consider a contact term of ${\cal L}_4$ 
(Fig.\ \ref{4:9:pionffl4}) and one-loop diagrams with vertices from 
${\cal L}_2$ (Figs.\ \ref{4:9:pionffloop1} and \ref{4:9:pionffloop2}).

   We will first work with the Lagrangian of Gasser and Leutwyler
\cite{Gasser:1983yg},
Eq.\ (\ref{app:glvgss:l4gl}) of Appendix \ref{app_sec_glvgss},\footnote{The
low-energy coupling constants of the SU(2)$\times$SU(2) Lagrangian are
denoted by $l_i$ in distinction to the $L_i$ of the
SU(3)$\times$SU(3) Lagrangian.}
\begin{equation}
\label{4:9:l4gl}
{\cal L}^{\rm GL}_4=\cdots 
+i\frac{l_6}{2}\mbox{Tr}[ f^R_{\mu\nu} D^{\mu} U (D^{\nu} U)^{\dagger}
+ f^L_{\mu\nu} (D^{\mu} U)^{\dagger} D^{\nu} U]+\cdots,
\end{equation}
which produces the contact interaction
\begin{equation}
\label{4:9:l4gpp}
{\cal L}_4^{\gamma \pi\pi}=e \frac{l_6}{F_0^2} 
\epsilon_{3ij}\partial^\mu\pi_i\partial^\nu 
\pi_j {\cal F}_{\mu\nu},
\end{equation}
resulting in the Feynman amplitude
\begin{equation}
\label{4:9:m4gpp}
\frac{e l_6 \epsilon_{3ij}}{F_0^2}
\left[-q^2 \epsilon\cdot(p'+p)+\epsilon\cdot(p'-p)(p'^2-p^2)\right]
\end{equation}
   which vanishes for real photons, $q^2=q\cdot\epsilon=0$.
   The second term vanishes if both pions are on the mass
shell, $p^2=p'^2=M_\pi^2$, but can be of relevance if the vertex
is used as an intermediate building block in a calculation such
as virtual Compton scattering off the pion \cite{Unkmeir:1999md}. 
   The Feynman amplitude resulting from Eq.\ (\ref{4:9:l4gl}) is the same
for both parameterizations.

   When using the ${\cal L}_4$ Lagrangian of Ref.\ \cite{Gasser:1987rb},
Eq.\ (\ref{app:glvgss:l4gss}), one obtains an additional contact
interaction,
\begin{equation}
\label{4:9:addl4gpp}
-2 e \frac{l_4 M^2_{\pi,2}}{F_0^2} 
\epsilon_{3ij}\pi_i\partial^\mu \pi_j 
{\cal A}_\mu,
\end{equation}
where $M^2_{\pi,2}=2 B_0 m$.
   For both parameterizations of $U$ we obtain the additional term
\begin{equation}
\label{4:9:addlmgpp}
 \frac{2 l_4 M^2_{\pi,2}}{F_0^2} 
e \epsilon_{3ij}\epsilon\cdot(p'+p).
\end{equation}

   Let us now turn to the one-loop diagram of Fig.\ \ref{4:9:pionffloop1}.
   The corresponding Feynman amplitude in the parameterization of 
Eq.\ (\ref{4:9:sqrt}) using 
the pion-pion vertex of Eq.\ (\ref{4:6:mpipi1}) reads  
\begin{eqnarray}
\label{4:9:loopc1}
&&\frac{1}{2} \frac{ie\epsilon_{3ij}}{F_0^2}
\int \frac{d^4 k}{(2\pi)^4} 
\frac{(2\epsilon\cdot k+\epsilon\cdot q)[(2p+q)\cdot(2k+q)]}{
[k^2-M^2_{\pi,2}+i0^+][(k+q)^2-M^2_{\pi,2}+i0^+]},
\end{eqnarray}
where the $1/2$ is a symmetry factor.
   The integral diverges and its extension to $n$ dimensions is given by
\begin{eqnarray}
\label{4:9:loopc1a}
&&\frac{1}{2} \frac{e\epsilon_{3ij}}{F_0^2}
\left\{\epsilon\cdot(p'+p)4 q^2 B_{21}(q^2,M_{\pi,2}^2)\right.\nonumber\\
&&\left.
+\epsilon\cdot q (p'+p)\cdot q 
[4 B_{20}(q^2,M_{\pi,2}^2)
+4 B_1(q^2,M_{\pi,2}^2)
+B_0(q^2,M_{\pi,2}^2)]\right\},\nonumber
\end{eqnarray}
   where the functions $B_0$, $B_1$, $B_{20}$, and $B_{21}$
are defined in Eqs.\ (\ref{app:ipipib}), (\ref{app:B1}),
and (\ref{app:B20B21}) of Appendix \ref{app_sec_olims}.
   Inserting the results of Eqs.\ (\ref{app:q2B21}) and
(\ref{app:B20}) the one-loop contribution of Fig.\ \ref{4:9:pionffloop1}
finally reads
\begin{eqnarray}
\label{4:9:loopc1b}
&&e\epsilon_{3ij}\left\{\epsilon\cdot(p'+p) 
\frac{M_{\pi,2}^2}{16\pi^2F_0^2}
\left[R+\ln\left(\frac{M_{\pi,2}^2}{\mu^2}\right)\right]\right.\nonumber\\
&&
-\frac{1}{96\pi^2 F_0^2}[q^2 \epsilon\cdot(p'+p)-\epsilon\cdot q (p'^2-p^2)]
\nonumber\\
&&\left.\times
\left[R+\ln\left(\frac{M_{\pi,2}^2}{\mu^2}\right)
+\frac{1}{3}+\left(1-\frac{M^2_{\pi,2}}{q^2}\right)
J^{(0)}\left(\frac{q^2}{M_{\pi,2}^2}\right)\right]\right\}+O(n-4).\nonumber\\
\end{eqnarray}
   The (infinite) contribution of the first term will be precisely canceled
by taking the pion wave function renormalization into account.
   The second structure is separately gauge invariant and also contains
an infinite piece which will be canceled by a corresponding infinite piece
of the bare coefficient $l_6$ [see Eq.\ (\ref{4:9:m4gpp})].
   Finally, a calculation of the one-loop diagram of 
Fig.\ \ref{4:9:pionffloop1} with the exponential parameterization of 
Eq.\ (\ref{4:6:u2}) and the pion-pion vertex of Eq.\ (\ref{4:6:mpipi2}) 
yields exactly the same result as Eq.\ (\ref{4:9:loopc1b}).

   Using the exponential parameterization
there is a $4\pi\gamma$ vertex at ${\cal O}(p^4)$ [see Eq.\ (\ref{4:9:l2exp})]
resulting in the additional loop diagram of Fig.\ \ref{4:9:pionffloop2}.
   The corresponding contribution in dimensional regularization,
\begin{equation}
\label{4:9:loopc2a}
-\frac{5}{3} e\epsilon_{3ij} \epsilon\cdot(p'+p)
\frac{M^2_{\pi,2}}{16\pi^2 F_0^2}\left[R+\ln\left(\frac{M^2_{\pi,2}}{\mu^2}
\right)
\right]+O(n-4),
\end{equation}
   also generates an infinite contribution to the vertex at zero
four-momentum transfer.

   The renormalized vertex is obtained by adding the bare contributions
and multiplying the result by a factor $\sqrt{Z_\pi}$ for each external
pion line.
   The wave function renormalization constant $Z_\pi$ is not an observable
and depends on both the parameterization for $U$ and the ${\cal O}(p^4)$
Lagrangian. 
   The corresponding results are summarized in Table \ref{app:dp:tab:abz} of
Appendix \ref{app_sec_dp}.
   We add the bare contributions which were obtained using the 
parameterization of Eq.\ (\ref{4:9:sqrt}) and the ${\cal O}(p^4)$ Lagrangian 
of Eq.\ (\ref{4:9:l4gpp}),
Eqs.\ (\ref{4:9:m2gpp}), (\ref{4:9:m4gpp}), and (\ref{4:9:loopc1b}),
and multiply the result by the appropriate
wave function renormalization constant
[see entry ``GL, Eq.\ (\ref{app:dp:sqrt})'' of Table \ref{app:dp:tab:abz}],
\begin{eqnarray}
\label{4:9:mr}
\lefteqn{{\cal M}_R=e\epsilon_{3ij}\left(
\epsilon\cdot(p'+p)\left\{1+\frac{M_{\pi,2}^2}{16\pi^2F_0^2}
\left[R+\ln\left(\frac{M_{\pi,2}^2}{\mu^2}\right)\right]\right\}\right.
}\nonumber\\
&&-\frac{q^2 \epsilon\cdot(p'+p)-\epsilon\cdot q (p'^2-p^2)}{F_0^2}
\nonumber\\
&&\left.\times
\left\{l_6
+\frac{1}{96\pi^2}
\left[R+\ln\left(\frac{M_{\pi,2}^2}{\mu^2}\right)
+\frac{1}{3}+\left(1-\frac{M^2_{\pi,2}}{q^2}\right)
J^{(0)}\left(\frac{q^2}{M_{\pi,2}^2}\right)\right]\right\}\right)
\nonumber\\
&&\times 
\left\{1-\frac{M_{\pi,2}^2}{16\pi^2F_0^2}
\left[R+\ln\left(\frac{M_{\pi,2}^2}{\mu^2}\right)\right]\right\}
+O(n-4).
\end{eqnarray}
   The factor $\epsilon\cdot(p'+p)$ counts as ${\cal O}(p^2)$, because
the external electromagnetic field, represented by the polarization
vector $\epsilon$, counts as ${\cal O}(p)$ 
[see Eq.\ (\ref{4:5:powercounting})].
It is multiplied by 
\begin{eqnarray*}
1+I(M^2_{\pi,2})&=&1+\frac{M^2_{\pi,2}}{16\pi^2}\left[
R+\ln\left(\frac{M^2_{\pi,2}}{\mu^2}\right)\right]+O(n-4)\\
&=&1+{\cal O}(p^2).
\end{eqnarray*}
   Since the wave function renormalization constant is 
$Z_\pi=1-I(M^2_{\pi,2})=1+{\cal O}(p^2)$ (see Appendix \ref{app_sec_dp}), 
it is only the tree-level contribution derived
from ${\cal L}_2$ which gets modified. 
   The product $[1+I(M^2_{\pi,2})][1-I(M^2_{\pi,2})]=1+{\cal O}(p^4)$
is such that the renormalized vertex is properly normalized to the charge
at ${\cal O}(p^4)$.
   The factor $[q^2 \epsilon\cdot(p'+p)-\epsilon\cdot q (p'^2-p^2)]$ is
${\cal O}(p^4)$ and thus the apparent infinity $R$ cannot be canceled
through the wave function renormalization.
   Here it is the connection between the bare parameter $l_6$ and the
renormalized parameter $l_6^r(\mu)$, $l_6^r=l_6+R/(96\pi^2)$, 
which cancels the divergence [see Eq.\ (9.6) of Ref.\ \cite{Gasser:1983yg}].
   Moreover, the explicit dependence on the renormalization scale $\mu$ 
cancels with a corresponding scale dependence of the parameter $l_6^r$.
   Finally, using the exponential parameterization or the Lagrangian of
Ref.\ \cite{Gasser:1987rb} results in the same expression as
Eq.\ (\ref{4:9:mr}).
   The additional contributions from Eqs.\ (\ref{4:9:addl4gpp}) and/or
(\ref{4:9:loopc2a}) to the unrenormalized vertex are precisely canceled
by modified wave function renormalization constants, resulting in
the same renormalized vertex.

   On the mass shell $p^2=p'^2=M_\pi^2$, and we obtain for the electromagnetic
form factor \cite{Gasser:1983yg}
\begin{equation}
\label{4:9:emffpi}
F(q^2)=1-l_6^r\frac{q^2}{F^2_\pi}-\frac{1}{6}\frac{q^2}{(4\pi F_\pi)^2}
\left[\ln\left(\frac{M^2_\pi}{\mu^2}\right)
+\frac{1}{3}
+\left(1-4\frac{M^2_\pi}{q^2}\right)J^{(0)}\left(\frac{q^2}{M^2_\pi}
\right)\right],
\end{equation}
   where we replaced the ${\cal O}(p^2)$ quantities $F_0$ and
$M^2_{\pi,2}$ by their physical values,
the error introduced being of order ${\cal O}(p^6)$.
   Given a spherically symmetric charge distribution $e Z \rho(r)$ normalized
so that $\int d^3 x \rho(r)=1$, the form factor $F(|\vec{q}\,|)$ in a
nonrelativistic framework is given by
\begin{displaymath}
F(|\vec{q}\,|)=\int d^3 x e^{i\vec{q}\cdot\vec{x}}\rho(r)
=4\pi \int_0^\infty dr r^2 j_0(|\vec{q}\,|r) \rho(r) 
=1-\frac{1}{6}|\vec{q}\,|^2\langle r^2\rangle+\cdots, 
\end{displaymath}
where $\langle r^2\rangle$ denotes the mean square radius.\footnote{For
neutral particles such as the neutron or the $K^0$  one has
$e\int d^3 x \rho(r)=0$.} 
   In analogy, the Lorentz-invariant form factor of Eq.\ (\ref{4:9:emffpi})
is expanded for small $q^2$ as\footnote{Breit-frame kinematics, i.e.
$q^2=-\vec{q}\,^2$, comes closest to the nonrelativistic 
situation.}
\begin{equation}
F(q^2)=1+\frac{q^2}{6}\langle r^2\rangle +\cdots,
\end{equation}
and the charge radius of the pion is {\em defined} as
\begin{equation}
\langle r^2\rangle_\pi=6 \left.\frac{d F(q^2)}{dq^2}\right|_{q^2=0}
=
-\frac{6}{F_\pi^2}\left\{l_6^r(\mu)+\frac{1}{96\pi^2}
\left[1+\ln\left(\frac{M^2_\pi}{\mu^2}\right)\right]\right\},
\end{equation}
where we made use of $J^{(0)}(x)=-x/6+O(x^2)$.
   Following Ref.\ \cite{Gasser:1983yg}, we introduce a scale-independent
quantity (see Appendix \ref{app_sec_glvgss})
\begin{displaymath}
\bar{l}_6=-96\pi^2 l_6^r(\mu)-\ln\left(\frac{M^2_\pi}{\mu^2}\right)
\end{displaymath}
   which can be determined using the empirical information on the
charge radius of the pion:
$\bar{l}_6=16\pi^2 F_\pi^2\langle r^2\rangle_\pi+1$.
   In a two-loop calculation of the vector form factor \cite{Bijnens:1998fm},
higher-order terms in the chiral expansion terms were also
taken into account and a fit to several experimental data sets was performed 
with the result $\bar{l}_6=16.0\pm 0.5\pm 0.7$,
where the last error is of theoretical origin.
   Once the value of the parameter $\bar{l}_6$ has been determined 
it can be used to predict other processes such as, e.g., virtual
Compton scattering off the pion \cite{Unkmeir:1999md,Unkmeir:2001gw}.
          
   The results for the electromagnetic form factors of the charged pion,
and the charged and neutral kaons in SU(3)$\times$SU(3)
chiral perturbation theory at ${\cal O}(p^4)$ can be found in 
Refs.\ \cite{Gasser:1984ux,Rudy:1994qb}.
   The calculation is very similar to the SU(2)$\times$SU(2) case and
the mean square radii of the charged pions and kaons are dominated 
by the low-energy parameter $L^r_9$, whereas the one-loop diagrams
generate a small contribution only:\footnote{The numerical values
in Ref.\ \cite{Gasser:1984ux} were obtained with 
$M_\pi= 135$ MeV, $M_K=495$ MeV, 
$F_0\approx F_\pi=93.3$ MeV, $\mu=M_\rho
=770$ MeV, and $L^r_9(M_\rho)=(6.9\pm 0.7)\cdot 10^{-3}$.}
\begin{eqnarray}
\label{4:9:r2pipkp}
\langle r^2\rangle_{\pi^+}&=&12 \frac{L^r_9}{F^2_0}
-\frac{1}{32 \pi^2 F^2_0}\left[3 + 2 \ln\left(\frac{M^2_{\pi,2}}{\mu^2}\right)
+\ln\left(\frac{M^2_{K,2}}{\mu^2}\right)\right]\nonumber\\
&=&(0.37\pm 0.04+0.07)\,\mbox{fm}^2=(0.44\pm 0.04)\,\mbox{fm}^2,\nonumber\\
\langle r^2\rangle_{K^+}&=&12 \frac{L^r_9}{F^2_0}
-\frac{1}{32 \pi^2 F^2_0}\left[3 + 2 \ln\left(\frac{M^2_{K,2}}{\mu^2}\right)
+\ln\left(\frac{M^2_{\pi,2}}{\mu^2}\right)\right]\nonumber\\
&=&(0.37 \pm 0.04 + 0.03)\,\mbox{fm}^2 = (0.40 \pm 0.04)\,\mbox{fm}^2.
\end{eqnarray}
   In Ref.\ \cite{Gasser:1984ux} the empirical value 
$\langle r^2\rangle_\pi=(0.439\pm 0.030)\,\mbox{fm}^2$  of \cite{Dally:1982zk}
was used to fix $L^r_9$.\footnote{A more recent value is given by
$\langle r^2\rangle_\pi=(0.439\pm 0.008)\,\mbox{fm}^2$
\cite{Amendolia:1986wj}. 
   Also (model-dependent) results have been obtained from 
pion-electroproduction
experiments \cite{Liesenfeld:1999mv,Volmer:2001ek}.
}
   The result for the mean square radius of the charged kaon is then a 
prediction which has to be compared with the empirical values
$\langle r^2 \rangle_{K^-}=(0.28\pm 0.05)\,\mbox{fm}^2$ 
of \cite{Dally:1980dj} and 
$\langle r^2 \rangle_{K^-}=(0.34\pm 0.05)\,\mbox{fm}^2$ 
of \cite{Amendolia:1986ui}.
   In Ref.\ \cite{Bijnens:2002hp} the empirical data on the charged
pion and kaon form factors were analyzed at two-loop order and
the low-energy constant including the $p^6$ terms was determined 
as $L_9^r(\mu=770\,\mbox{MeV}) =(5.93\pm 0.43)\times 10^{-3}$.
   
   At ${\cal O}(p^4)$ the form factor of the $K^0$ receives one-loop 
contributions only and thus is predicted in terms of 
the pion-decay constant and the Goldstone boson masses.
   The mean square radius is given by
\begin{equation}
\langle r^2\rangle_{K^0}=-\frac{1}{32\pi^2F^2_0}\ln\left(\frac{M^2_K}{M^2_\pi}\right)
=-0.037\, \mbox{fm}^2
\end{equation}
which has to be compared with the empirical value 
$\langle r^2\rangle_{K^0}=
(-0.054\pm 0.026)\, \mbox{fm}^2$ \cite{Molzon:1978py}.
   For a two-loop analysis of the neutral-kaon form factor, see
Refs.\ \cite{Post:1997dk,Post:2000gk}.

   Since the neutral pion and the eta are their own antiparticles, their
electromagnetic vertices vanish because of charge conjugation symmetry
as noted in footnote \ref{fnic} for the case of the $\pi^0$.

\section{Chiral Perturbation Theory at ${\cal O}(p^6)$}
\label{sec_cptop6}

   Mesonic chiral perturbation theory at order ${\cal O}(p^4)$ has led
to a host of successful applications and may be considered as a full-grown
and mature area of low-energy particle physics.
   In this section we will briefly touch upon its extension to ${\cal O}(p^6)$ 
\cite{Issler:1990nj,Akhoury:1990px,Fearing:1994ga,Bijnens:1999sh,%
Ebertshaeuser:2001,Ebertshauser:2001nj,Bijnens:2001bb}, 
which naturally divides into the even- and odd-intrinsic-parity sectors.
   Calculations in the even-intrinsic-parity sector start at ${\cal O}(p^2)$
and two-loop calculations at ${\cal O}(p^6)$ are thus of 
next-to-next-to-leading order (NNLO).
   NNLO calculations at ${\cal O}(p^6)$ have been performed for
$\gamma\gamma\to\pi^0\pi^0$ \cite{Bellucci:1994eb}, 
vector two-point functions 
\cite{Golowich:1995kd,Maltman:1995jg,Durr:1999dp,Amoros:1999dp}
and axial-vector two-point functions \cite{Amoros:1999dp},
$\pi\pi$ scattering \cite{Bijnens:1995yn}, 
$\gamma\gamma\to\pi^+\pi^-$ \cite{Burgi:1996mm},
$\tau\to \pi\pi \nu_\tau$ \cite{Colangelo:1996hs},
$\pi\to l \nu \gamma$ \cite{Bijnens:1996wm},
Sirlin's combination of SU(3) form factors \cite{Post:1997dk},
scalar and electromagnetic form factors of the pion \cite{Bijnens:1998fm},
the $K\to\pi\pi l\nu$ ($K_{l4}$) form factors \cite{Amoros:2000qq},
the electromagnetic form factor of the $K^0$ \cite{Post:2000gk},
the $K\to\pi l\nu$ ($K_{l3}$) form factors \cite{Post:2001si},
and the electromagnetic form factors of pions and kaons in SU(3)$\times$SU(3)
ChPT \cite{Bijnens:2002hp}.
   Further applications deal with more technical aspects such as
the evaluation of specific two-loop integrals 
\cite{Post:1996gg,Gasser:1998qt}
and the renormalization of the even-intrinsic-parity Lagrangian at
${\cal O}(p^6)$ \cite{Bijnens:1999hw}.

   The odd-intrinsic-parity sector starts at ${\cal O}(p^4)$ with the 
anomalous WZW action, as discussed in Sec.\ \ref{sec_ewzwa}.
   In this sector next-to-leading-order (NLO), i.e.\ one-loop calculations, 
are of ${\cal O}(p^6)$.
   It has been known for some time that quantum corrections to
the WZW classical action do not renormalize the coefficient of
the WZW term 
\cite{Donoghue:ct,Issler:1990nj,Bijnens:1989jb,Akhoury:1990px,%
Ebertshaeuser:2001,Bijnens:2001bb}.
   The counter terms needed to renormalize the one-loop singularities
at ${\cal O}(p^6)$ are of a conventional chirally invariant structure.
   The inclusion of the photon as a dynamical degree of freedom in
the odd-intrinsic-parity sector has been discussed in Ref.\ 
\cite{Ananthanarayan:2002kj}.
   For an overview of applications in the odd-intrinsic-parity sector,
we refer to Ref.\ \cite{Bijnens:xi}.
   A two-loop calculation at ${\cal O}(p^8)$ for $\gamma\pi\to\pi\pi$ was
performed in Ref.\ \cite{Hannah:2001ee}.

   Here, we will mainly be concerned with some aspects of the 
construction of the most general mesonic chiral Lagrangian at ${\cal O}(p^6)$
and discuss as an application a two-loop calculation 
for the $s$-wave $\pi\pi$ scattering lengths \cite{Bijnens:1995yn}.

\subsection{The Mesonic Chiral Lagrangian at Order ${\cal O}(p^6)$}
\label{subsec_mclop6}
   The rapid increase in the number of free parameters when going from 
${\cal L}_2$ to ${\cal L}_4$ naturally leads to the expectation of a very 
large number of chirally invariant structures at ${\cal O}(p^6)$.
   One of the problems with the construction of effective Lagrangians at higher
orders is that it is far too easy to think of terms satisfying the necessary
criteria of Lorentz invariance and invariance under the discrete symmetries
as well as chiral transformations.
   To our knowledge there is neither a general formula, 
even at ${\cal O}(p^4)$, for determining the number of independent structures 
to expect nor an algorithm to decide whether a set of given structures is
independent or not.
   Experience has shown that for almost any sector of higher-order 
effective chiral Lagrangians the number of terms found to be independent 
has gone down with time (see Refs.\ 
\cite{Issler:1990nj,Akhoury:1990px,Fearing:1994ga,Ebertshauser:2001nj,%
Bijnens:2001bb} for the odd-intrinsic-parity sector,
\cite{Fearing:1994ga,Bijnens:1999sh} for the even-intrinsic-parity sector,
and \cite{Ecker:1995rk,Fettes:2000gb}
for the heavy-baryon $\pi N$ Lagrangian, respectively).
   For that reason, it is important to define a strategy for obtaining
all of the independent terms without generating a lot of extraneous terms
which have to be eliminated by hand.
   In the following, we will outline the main principles entering the
construction of the ${\cal O}(p^6)$ Lagrangian and refer the reader to 
Refs.\ \cite{Fearing:1994ga,Bijnens:1999sh,Ebertshauser:2001nj,Bijnens:2001bb}
for more details.

   The effective Lagrangian is constructed from the elements $U$, $U^\dagger$, 
$\chi$, $\chi^\dagger$, and the field strength tensors $f^L_{\mu\nu}$
and $f^R_{\mu\nu}$ (see Sec.\ \ref{sec_cel}, in particular, 
Table \ref{4:5:table_trafprop}).
   The external fields $l_\mu$ and $r_\mu$ only appear in the field strength 
tensors or the covariant derivatives which we define as
\begin{eqnarray}
\label{4:10:covder1}
A\stackrel{G}{\mapsto}V_RAV_L^\dagger:
&\quad&D_\mu A\equiv\partial_\mu A-ir_\mu A+iA l_\mu,\quad\mbox{e.g.,}
\,\, U,\chi,\nonumber\\
B\stackrel{G}{\mapsto}V_LBV_R^\dagger:
&\quad&D_\mu B\equiv\partial_\mu B+iBr_\mu -il_\mu B,
\quad\mbox{e.g.,}\,\, U^\dagger,\chi^\dagger,\nonumber\\
C\stackrel{G}{\mapsto}V_RCV_R^\dagger:
&\quad&D_\mu C\equiv\partial_\mu C-ir_\mu C+iCr_\mu,
\quad\mbox{e.g.,}\,\, f^R_{\mu\nu},\nonumber\\
D\stackrel{G}{\mapsto}V_LDV_L^\dagger:
&\quad&D_\mu D\equiv\partial_\mu D-il_\mu D+iDl_\mu,
\quad\mbox{e.g.,}\,\, f^L_{\mu\nu},\nonumber\\
E\stackrel{G}{\mapsto}E:
&\quad&D_\mu E\equiv\partial_\mu E,\quad
\mbox{e.g.,}\,\,\mbox{Tr}(\chi\chi^\dagger).
\end{eqnarray}
   In other words, the covariant derivative knows about the 
transformation property under $G=\mbox{SU(3)}_L\times
\mbox{SU(3)}_R$ of the object it acts on and adjusts itself accordingly.
   With such a convention a product rule analogous to that for ordinary 
derivatives holds. 
   Given the product $Z=XY$ where $X,Y,Z$ have, according to 
Eq.\ (\ref{4:10:covder1}), well-defined but not necessarily the same 
transformation behavior, the product rule 
\begin{equation}
\label{4:10:pr}
D_\mu Z = D_\mu (XY)=(D_\mu X) Y+X( D_\mu Y)
\end{equation} 
applies, which can be easily verified using the definitions of 
Eq.\ (\ref{4:10:covder1}).
   This product rule is valuable as an intermediate step in a number of the 
derivations of various relations.

   In order to avoid unnecessary and tedious repetitions during the process 
of construction, one would like to perform as many manipulations as possible 
on a formal level without explicit reference to the specific building blocks. 
   It is thus convenient to handle the external field terms $\chi$,
$f^R_{\mu\nu}$, and $f^L_{\mu\nu}$ in the same way.
   To that end we define
\begin{equation}
\label{4:10:ghdefinition}
G^{\mu\nu}\equiv f_R^{\mu\nu}U+Uf_L^{\mu\nu},\quad
H^{\mu\nu}\equiv f_R^{\mu\nu}U-Uf_L^{\mu\nu},
\end{equation}
and introduce $\chi^{\mu\nu}$ as a common abbreviation for any of the
building blocks $\chi$ ($\equiv \chi^{\mu\mu}$), 
$G^{\mu\nu}$ and $H^{\mu\nu}$ ($\mu\neq\nu$).
   With these definitions, we have only two basic building blocks
$U$, $\chi^{\mu\nu}$, covariant derivatives acting on them and
the respective adjoints.
   Due to the product rule of Eq.\ (\ref{4:10:pr}) it is not necessary to 
consider derivatives acting on products of these basic terms.
   All building blocks then transform as $U$ (or $U^\dagger$). 
   In terms of the momentum expansion, $U$ is of order 1,
$\chi^{\mu\nu}$ of order $p^2$ and each covariant derivative $D_\mu$ of
order $p$. 
 
  Up to this point we have treated a building block and its adjoint on 
a different footing which we will now remedy by defining the Hermitian 
and anti-Hermitian combinations
\begin{equation}
\label{4:10:apm}
(A)_\pm=u^\dagger A u^\dagger \pm u A^\dagger u,
\end{equation}
   where $A$ is taken as  $\chi^{\mu\nu}$ or $D_\mu U$, or as 
some number of covariant derivatives acting on $\chi^{\mu\nu}$ or
$D_\mu U$.\footnote{In Ref.\ \cite{Fearing:1994ga} the building blocks
$[A]_\pm\equiv\frac{1}{2}(A U^\dagger\pm U A^\dagger)$ transforming as
$V_R\cdots V_R^\dagger$ were used. The notation of Eq.\ (\ref{4:10:apm})
has some advantages when implementing the total-derivative procedure to 
be discussed below.
Moreover, it is more closely related to the conventions used in 
the baryonic sector (see Sec.\ \ref{sec_tpf}).}
   Here $u$ is defined as the square root of $U$, i.e., $u^2\equiv U$.
   In order to discuss the transformation behavior of Eq.\ 
(\ref{4:10:apm}), we define the SU(3)-valued function $K(V_L,V_R,U)$,
referred to as the compensator field \cite{Ecker:1995gg},
through \cite{Gasser:1987rb}
\begin{equation}
\label{4:10:kdef}
u(x)\mapsto u'(x)=\sqrt{V_R U V_L^\dagger}\equiv V_R u K^{-1}(V_L,V_R,U),
\end{equation}
from which one obtains
\begin{equation}
\label{4:10:k}
K(V_L,V_R,U)=u'^{-1} V_R u=\sqrt{V_R U V_L^\dagger}^{-1}V_R\sqrt{U}.
\end{equation}
   From a group-theoretical point of view, $K$ defines a nonlinear realization
of SU(3)$\times$SU(3) \cite{Gasser:1987rb}, 
because\footnote{$K$ does not define an operation of SU(3)$\times$SU(3) on  
SU(3), because $K(1,1,U)=1\neq U\,\forall\, U$ (see Sec.\ \ref{subsec_gc}).}
\begin{eqnarray}
\label{4:10:kreal}
\lefteqn{K(V_{L1}, V_{R1}, V_{R2} U V_{L2}^\dagger)
K (V_{L2},V_{R2},U)}\nonumber\\
&=&
\sqrt{V_{R1}(V_{R2}U V_{L2}^\dagger) V_{L1}^\dagger}^{-1}
V_{R1} \sqrt{V_{R2}U V_{L2}^\dagger} 
\sqrt{V_{R2}U V_{L2}^\dagger}^{-1} V_{R2}\sqrt{U}\nonumber\\
&=&\sqrt{V_{R1} V_{R2} U (V_{L1} V_{L2})^\dagger}^{-1} V_{R1} V_{R2} \sqrt{U}
\nonumber\\
&=&K((V_{L1} V_{L2}), (V_{R1}V_{R2}),U).
\end{eqnarray}
   It is important to note that the first $K$ has the transformed
$U'=V_{R2} U V_{L2}^\dagger$ as its argument. 
   With these definitions, the building blocks $(A)_\pm$ transform as
\begin{equation}
\label{4:10:apmtrans}
(A)_\pm\mapsto K (A)_\pm K^\dagger.
\end{equation}
   The corresponding covariant derivative is defined as\footnote{From an 
aesthetical point of view it would have been more satisfactory to introduce
the covariant derivative as 
$\nabla_\mu(A)_\pm\equiv\partial_\mu(A)_\pm-i[\Gamma_\mu,(A)_\pm]$
to generate a closer formal correspondence to Eqs.\ (\ref{4:10:covder1}).
   However, we follow the standard convention used in the literature.}
\begin{equation}
\label{4:10:covder}
\nabla_\mu(A)_\pm\equiv\partial_\mu(A)_\pm+[\Gamma_\mu,(A)_\pm],
\end{equation}
where $\Gamma_\mu$ is the so-called connection \cite{Ecker:1995gg}, 
and is given by
\begin{equation}
\label{4:10:chiralconnection}
\Gamma_\mu=\frac{1}{2}\left[u^\dagger,\partial_\mu u\right]
-\frac{i}{2}u^\dagger r_\mu u-\frac{i}{2}u l_\mu u^\dagger.
\end{equation}
   As usual, the covariant derivative transforms in the same way
as the object it acts on,
\begin{displaymath}
\nabla_\mu(A)_\pm\stackrel{G}{\mapsto} K \nabla_\mu(A)_\pm K^\dagger.
\end{displaymath}

   Invariants under $\mbox{SU(3)}_L\times\mbox{SU(3)}_R$
are constructed by forming products of objects, each 
transforming as $K\cdots K^\dagger$, and then taking the trace.
   For example, consider the trace
\begin{equation}
\label{4:10:twoobjects}
\mbox{Tr}[(A_1)_\pm \cdots (A_n)_\pm]
\stackrel{G}{\mapsto}
\mbox{Tr}[K (A_1)_\pm K^\dagger \cdots K (A_n)_\pm K^\dagger]
=\mbox{Tr}[(A_1)_\pm \cdots (A_n)_\pm],
\end{equation}
where we made use of $K^\dagger K=1$ and the invariance of the trace under
cyclic permutations.
   Obviously, products of such traces are also invariant.
   Basically, the construction of the most general Lagrangian then
proceeds by forming products of elements $(A)_\pm$, where $A$ is either 
$\chi^{\mu\nu}$ or $D^\mu U$ or covariant derivatives of these objects,
taking appropriate traces, and forming Lorentz scalars by contracting
the Lorentz indices with the metric tensor $g^{\mu\nu}$ or the
totally antisymmetric tensor $\epsilon^{\mu\nu\rho\sigma}$.
     
   After having chosen the set of building blocks and their transformation
behavior, we define a strategy concerning the order of constructing
invariants at ${\cal O}(p^6)$.
   As shown in Refs.\ \cite{Fearing:1994ga,Ebertshauser:2001nj}, by 
applying the product rule, it is sufficient to restrict oneself to 
$(D^m U)_-$, $(D^n G)_+$, $(D^n H)_+$, and $(D^n \chi)_\pm$
($m, n$ integer with $m>0, n\ge 0$),
because the other combinations $(D^m U)_+$, $(D^n G)_-$, and $(D^n H)_-$
can either be expressed in terms of these or vanish.
   One then immediately finds that all possible terms at 
${\cal{O}}{(p^6)}$ can either include no, one, two, or three 
$(D^n \chi_{\mu\nu})_\pm$ blocks which naturally defines four distinct 
levels to be considered:
\begin{enumerate}
\item terms with six $D_\mu$'s;
\item terms with four $D_\mu$'s and one $\chi^{\mu\nu}$;
\item terms with two $D_\mu$'s and two $\chi^{\mu\nu}$'s;
\item terms with three $\chi^{\mu\nu}$'s.
\end{enumerate}
  We always try to get rid of terms as high in the hierarchy (with the
most $D_\mu$'s) as possible.
   In particular, with this strategy one ensures that the number of terms is
minimal also for the special case in which all external fields are set equal 
to zero.\footnote{In that sense, the final 
$\mbox{SU}(N_f)_L\times \mbox{SU}(N_f)_R$ set given in 
Ref.\ \cite{Bijnens:1999sh} for the even-intrinsic-parity sector
is not minimal when setting all external fields to 
zero. In that case the structures $Y_1$ and $Y_4$ are not independent and can 
be eliminated. However, if one replaces these two terms by the terms (120) 
and (139) of Ref.~\cite{Fearing:1994ga}, then the entire set will remain 
independent whether or not there are external fields and both structures, 
(120) and (139), vanish explicitly when the fields are set to zero.}
   The motivation for such an approach is that at each level there exist
relations which allow one to eliminate structures, as long as one
keeps {\em all} terms at the lower levels of the hierarchy.
   To be specific, when considering multiple covariant derivatives,
just one general (i.e., non-contracted) index combination is actually 
independent in the sense of the hierarchy defined above.
   For double derivatives this statement reads
\begin{equation}
\label{4:10:dda}
(D_\mu D_\nu A)_\pm=(D_\nu D_\mu A)_\pm+\frac{i}{4} 
[ (A)_\pm , (G_{\mu\nu})_+ ] - \frac{i}{4} \{ (A)_\mp , (H_{\mu\nu})_+ \},
\end{equation}
i.e., it is sufficient to only keep the block $(D_\mu D_\nu A)_\pm$
as long as one considers all terms lower in the hierarchy.

   The construction then proceeds as follows.
   First of all, write down all conceivable Lorentz-invariant structures 
satisfying $P$ and $C$ invariance, Hermiticity, and chiral order $p^6$ 
in terms of the basic building blocks defined above.
   Then collect as many relations as possible among these structures and
use these to eliminate structures.
   The relations can follow from any of the following mechanisms:
\begin{enumerate}
\item[(1)] partial integration; 
\item[(2)] equation-of-motion argument; 
\item[(3)] epsilon relations;
\item[(4)] Bianchi identities;
\item[(5)] trace relations.
\end{enumerate} 
   Let us illustrate the meaning of each of the above items by selected
examples.

   The partial-integration or total-derivative argument refers to the
fact that a total derivative in the Lagrangian density does 
not change the equation of motion.
   One thus generates relations of the following type
\begin{eqnarray}
\lefteqn{
\underbrace{\partial_\mu\mbox{Tr}[(A_1)_\pm
\cdots (A_m)_\pm]}_{\mbox{tot.~der.}}
+\underbrace{\mbox{Tr}\{[\Gamma_\mu,(A_1)_\pm\cdots (A_m)_\pm ]\}}_{\mbox{0}}}
\nonumber\\
&=&\mbox{Tr}\{\nabla_\mu 
[ (A_1)_\pm\cdots (A_m)_\pm ]\}\nonumber\\
&=&\mbox{Tr}[\nabla_\mu (A_1)_\pm\cdots (A_m)_\pm]+\cdots
+\mbox{Tr}[(A_1)_\pm\cdots \nabla_\mu (A_m)_\pm],
\end{eqnarray}
   where we made use of Eq.\ (\ref{4:10:covder}) and the product rule.
   This derivative shifting procedure is also valid for multiple traces.
   At this stage we note the advantage of working with the basic building 
blocks of Eq.\ (\ref{4:10:apm}) in comparison with those of Ref.\ 
\cite{Fearing:1994ga} due to the relatively simple connection between the 
covariant derivative $\nabla_\mu$ outside the block brackets and the covariant 
derivative $D_\mu$ inside when $(A)_\pm$ are used:
\begin{eqnarray}
\label{4:10:nabla}
\nabla_\mu (A)_\pm&=&(D_\mu A)_\pm - \frac{1}{4}\{(D_\mu U)_-,(A)_\mp\}.
\end{eqnarray}
   From a technical point of view Eq.\ (\ref{4:10:nabla}) is important because 
it helps avoid extremely tedious algebraic manipulations one had to 
perform in the old framework of Ref.\ \cite{Fearing:1994ga}.
   The combination of shifting derivatives back and forth and interchanging 
indices of multiple derivatives is referred to as index exchange.
   In Ref.\ \cite{Fearing:1994ga} not all total-derivative terms were 
properly identified which has led to subsequent reductions in the number of
terms in both the even-intrinsic-parity sector \cite{Bijnens:1999sh}
and the odd-intrinsic-parity sector \cite{Ebertshauser:2001nj,Bijnens:2001bb}.

   The equation-of-motion argument makes use of the invariance of physical
observables under field transformations, as discussed in 
Sec.\ \ref{sec_clop4}.
   The aim is to collect as many terms as possible containing
a factor $(D^2 U)_-$.
   Such terms can be supplemented by corresponding $(\chi)_-$ terms lower
in the hierarchy to generate an equation-of-motion term which can
be eliminated by a field redefinition.

   The epsilon relations refer to the odd-intrinsic-parity sector with
the basic idea being as follows. 
   Consider a structure with six Lorentz indices transforming under
parity as a Lorentz pseudotensor, i.e.,  
$Q_{\lambda\mu\nu\rho\sigma\tau}(\vec{x},t)
\mapsto -Q^{\lambda\mu\nu\rho\sigma\tau}(-\vec{x},t)$.
   In order to form a Lorentz scalar, one needs to contract two indices
pairwise and the remaining four with the totally antisymmetric tensor
$\epsilon^{\alpha\beta\gamma\delta}$ in four dimensions.
   Suppose $Q_{\lambda\mu\nu\rho\sigma\tau}(\vec{x},t)$ is neither symmetric
nor antisymmetric under the exchange of any pair of indices.
  Naively one would then expect $5+4+\cdots+1=15$
independent contractions.
   However, such a counting does not take the totally 
antisymmetric nature of the epsilon tensor into account \cite{Akhoury:1990px},
from which one obtains, for the above case of no symmetry in the 
indices, five additional conditions \cite{Akhoury:1990px,Fearing:1994ga}.
  These additional identities have not been considered in the pioneering
construction of Ref.\ \cite{Issler:1990nj}.

   In general, $Q_{\lambda\mu\nu\rho\sigma\tau}(\vec{x},t)$ has some
symmetry in its indices, and not all five epsilon relations are independent.
   For example, using the transformation behavior of 
Table \ref{4:5:table_trafprop} it is easy to verify that
\begin{displaymath}
\mbox{Tr}\{(G_{\lambda\mu})_+[(G_{\nu\rho})_+(H_{\sigma\tau})_+
+(H_{\sigma\tau})_+(G_{\nu\rho})_+]\}
\end{displaymath}
is an example for a pseudotensor which is invariant under $G$ (and Hermitian).
   Its symmetries are given by
\begin{displaymath}
Q_{\lambda\mu\nu\rho\sigma\tau}=-Q_{\mu\lambda\nu\rho\sigma\tau}
=-Q_{\lambda\mu\rho\nu\sigma\tau}=-Q_{\lambda\mu\nu\rho\tau\sigma}
=Q_{\nu\rho\lambda\mu\sigma\tau},
\end{displaymath}
from which one would naively end up with a single combination 
\begin{displaymath}
\mbox{Tr}\{(G_{\mu\nu})_+[(G_{\lambda\alpha})_+({H^\lambda}_\beta)_+
+({H^\lambda}_\beta)_+(G_{\lambda\alpha})_+]\}\epsilon^{\mu\nu\alpha\beta}
\end{displaymath}
which, however, vanishes due to the epsilon relation.

   The Bianchi identities refer to certain relations among covariant 
derivatives of field-strength tensors.
   Starting from the Jacobi identity
\begin{equation}
\label{4:10:jacobi}
[A,[B,C]]+[B,[C,A]]+[C,[A,B]]=0,
\end{equation}
   we consider the linear combination 
\begin{eqnarray}
D_\mu f_{\nu\rho}^R+D_\nu f^R_{\rho\mu}+ D_\rho f^R_{\mu\nu}
&\equiv&\sum_{\mbox{\scriptsize c.p.}\{\mu,\nu,\rho\}}D_\mu f^R_{\nu\rho}
=\sum_{\mbox{\scriptsize c.p.}\{\mu,\nu,\rho\}}
(\partial_\mu f^R_{\nu\rho}-i[r_\mu,f^R_{\nu\rho}])\nonumber\\
&=&\sum_{\mbox{\scriptsize c.p.}\{\mu,\nu,\rho\}}
\bigg(\partial_\mu\partial_\nu r_\rho-\partial_\mu\partial_\rho r_\nu
-i[\partial_\mu r_\nu, r_\rho]\nonumber\\
&-&i[r_\nu,\partial_\mu r_\rho]
-i[r_\mu,\partial_\nu r_\rho-\partial_\rho r_\nu]
-[r_\mu,[r_\nu, r_\rho]]\bigg)\nonumber\\
&=&0,
\label{sumcp}
\end{eqnarray}
where use of the Schwarz theorem, $\partial_\mu\partial_\nu \cdots=
\partial_\nu\partial_\mu\cdots$, relabeling of indices, and the Jacobi 
identity, Eq.\ (\ref{4:10:jacobi}), has been made. 
Observe that the cyclic permutations of the indices $\mu,\nu,$ and $\rho$ has 
been denoted by $\mbox{c.p.}\{\mu,\nu,\rho\}$.
   The same arguments hold for the independent field strength tensor 
$f^L_{\mu\nu}$, and we can summarize the constraints as
\begin{equation}
\label{4:10:bi}
\sum_{\mbox{\scriptsize c.p.}\{\mu,\nu,\rho\}}D_\mu f^{L/R}_{\nu\rho}=0,
\end{equation}
which, because of their similarity to an analogous equation for the
Riemann-Christoffel curvature
 tensor in general relativity, are referred to as the Bianchi 
identities (see, e.g., Refs.~\cite{Ryder:wq,Weinberg:kr}). 
   Equation (\ref{4:10:bi}) does not require that $f^{R/L}_{\mu\nu}$ satisfy
any equations of motion. 
   In terms of the building blocks $(D_\mu U)_-$, $(G_{\mu\nu})_+$, and 
$(H_{\mu\nu})_+$ the Bianchi identities read
\begin{eqnarray}
\label{4:10:bianchig}
\sum_{\mbox{\scriptsize c.p.}\{\mu,\nu,\rho\}}(D_\mu G_{\nu\rho})_+&=&
-\frac{1}{4}\sum_{\mbox{\scriptsize c.p.}\{\mu,\nu,\rho\}}
[(D_\mu U)_-,(H_{\nu\rho})_+],\\
\label{4:10:bianchih}
\sum_{\mbox{\scriptsize c.p.}\{\mu,\nu,\rho\}}(D_\mu H_{\nu\rho})_+&=&
-\frac{1}{4}\sum_{\mbox{\scriptsize c.p.}\{\mu,\nu,\rho\}}
[(D_\mu U)_-,(G_{\nu\rho})_+].
\end{eqnarray}   
   Given the definition of Eq.\ (\ref{4:10:ghdefinition}), the Bianchi
identities can be used to generate relations among the building blocks
in terms of structures kept in lower orders of the hierarchy defined above.
   In Ref.\ \cite{Fearing:1994ga} each term of the sum on the left-hand side 
of Eqs.\ (\ref{4:10:bianchig}) and (\ref{4:10:bianchih}) was treated as an 
independent element so that the final list of supposedly independent structures
contained redundant elements.

   Finally, the trace relations refer to the fact that the construction
of invariants uses traces and products of traces.
   One is thus particularly interested in finding any relations among
those traces. 
   We know from the Cayley-Hamilton theorem that any $n\times n$ matrix $A$ 
is a solution of its associated characteristical polynomial $\chi_A$. 
   For $n=2$ this statement reads
\begin{eqnarray}
\label{4:10:CalHam}
0=\chi_A (A) & = & A^2 - \mbox{Tr}(A) A + \det (A) 1_{2\times 2}\nonumber\\
&=& A^2 - \mbox{Tr}(A) A 
+ \frac{1}{2} \{ [\mbox{Tr}(A)]^2 - \mbox{Tr}(A^2)\} 1_{2\times 2}.
\end{eqnarray}
   Setting $A=A_1+A_2$ in (\ref{4:10:CalHam}) and making use of 
$\chi_{A_1} (A_1) = 0 =\chi_{A_2} (A_2) $ one ends up with the matrix equation
\begin{eqnarray}
\label{4:10:F2}
0=F_2(A_1,A_2)&\equiv&\{A_1,A_2\} 
- \mbox{Tr}(A_1) A_2 - \mbox{Tr}(A_2) A_1 \nonumber\\
&&+
\mbox{Tr}(A_1)\mbox{Tr}(A_2) 1_{2\times 2} - \mbox{Tr}(A_1 A_2) 1_{2\times 2}
\end{eqnarray}
which is the central piece of information needed to derive the trace 
relations in the SU(2)$\times$SU(2) sector. 
     The analogous $n=3$ equation is slightly more complex 
\begin{eqnarray}
\label{4:10:F3}
0&=&F_3(A_1,A_2,A_3)\nonumber\\
&\equiv& A_1\{A_2,A_3\} + A_2\{A_3,A_1\} + A_3\{A_1,A_2\}\nonumber\\
&&-\mbox{Tr}(A_1)\{A_2,A_3\} - \mbox{Tr}(A_2)\{A_3,A_1\} 
- \mbox{Tr}(A_3)\{A_1,A_2\}\nonumber\\
&&+\mbox{Tr}(A_1)\mbox{Tr}(A_2)A_3 
  +\mbox{Tr}(A_2)\mbox{Tr}(A_3)A_1 
  +\mbox{Tr}(A_3)\mbox{Tr}(A_1)A_2\nonumber\\
&&-\mbox{Tr}(A_1 A_2)A_3 
  -\mbox{Tr}(A_3 A_1)A_2
  -\mbox{Tr}(A_2 A_3)A_1\nonumber\\
&&-\mbox{Tr}(A_1 A_2 A_3) 1_{3\times 3} 
  -\mbox{Tr}(A_1 A_3 A_2) 1_{3\times 3}\nonumber\\ 
&&+\mbox{Tr}(A_1 A_2)\mbox{Tr}(A_3)1_{3\times 3} 
  +\mbox{Tr}(A_3 A_1)\mbox{Tr}(A_2)1_{3\times 3} 
  +\mbox{Tr}(A_2 A_3)\mbox{Tr}(A_1)1_{3\times 3}\nonumber\\
&&-\mbox{Tr}(A_1)\mbox{Tr}(A_2)\mbox{Tr}(A_3) 1_{3\times 3}.
\end{eqnarray}
   We can now derive trace relations by simply multiplying Eq.\ 
(\ref{4:10:F2}) or Eq.\ (\ref{4:10:F3}) 
with another arbitrary matrix of the same dimension and 
finally taking the trace of the whole construction, i.e., 
\begin{eqnarray}
\label{4:10:Tracesu2}
0&=&\mbox{Tr}[F_2(A_1,A_2)A_3],\\
\label{4:10:tracesu3}
0&=&\mbox{Tr}[F_3(A_1,A_2,A_3)A_4].
\end{eqnarray}
   Note that $A_i$ may be any $n\times n$ matrix, even a string of our basic 
building blocks. 
     For example, Eq.\ (\ref{app:glvgss:tr2by2}) of Appendix 
\ref{app_sec_glvgss}, is identical to Eq.\ (\ref{4:10:Tracesu2}).

  In principle, the ideas developed above apply to the general 
$\mbox{SU}(N_f)_L\times\mbox{SU}(N_f)_R$ case and only at the end it
is necessary to specify the number of flavors $N_f$.
   The reduction to the cases $N_f=2$ and $N_f=3$ is
achieved in terms of the trace relations summarized in Eqs.\ 
(\ref{4:10:Tracesu2}) and (\ref{4:10:tracesu3}).
   Although we have never come across a trace relation that could not be 
obtained in the manner explained above, we are not aware of a general proof 
showing that any kind of trace relation must be related to the Caley-Hamilton 
theorem.

  In the even-intrinsic-parity sector the Lagrangian has 112 in principle
measurable + 3 contact terms for the general 
$\mbox{SU}(N_f)_L\times\mbox{SU}(N_f)_R$ case,
90 + 4 for the $\mbox{SU}(3)_L\times\mbox{SU}(3)_R$ case,
and 53 + 4 for the $\mbox{SU}(2)_L\times\mbox{SU}(2)_R$ case
\cite{Bijnens:1999sh}.
   The contact terms refer to structures which can be expressed in terms
of only external fields such as the $H_1$ and $H_2$ terms of the 
${\cal L}_4$ Lagrangian of Eq.\ (\ref{4:7:l4gl}).
   The reduction in the number of terms in comparison with the 111  
$\mbox{SU}(3)_L\times\mbox{SU}(3)_R$ structures of Ref.\ \cite{Fearing:1994ga}
is due to a more complete application of the partial-integration relations,
the use of additional trace relations, 
and the use of four relations due to the Bianchi identities which were not 
taken into account in Ref.\ \cite{Fearing:1994ga}.
   The odd-intrinsic-parity sector was reconsidered in Refs.\ 
\cite{Ebertshauser:2001nj,Bijnens:2001bb}.
   Both analyses found 24 $\mbox{SU}(N_f)_L\times\mbox{SU}(N_f)_R$,
23 $\mbox{SU}(3)_L\times\mbox{SU}(3)_R$, and 5
$\mbox{SU}(2)_L\times\mbox{SU}(2)_R$ terms.
   Moreover, 8 additional terms due to the extension of the chiral
group to $\mbox{SU}(N_f)_L\times\mbox{SU}(N_f)_R\times U(1)_V$ were
found, which are of some relevance when considering the electromagnetic
interaction for the two-flavor case.
   In comparison to Ref.\ \cite{Fearing:1994ga}, the new analysis 
of Ref.\ \cite{Ebertshauser:2001nj} could eliminate two structures
via partial integration, 6 via Bianchi identities and one by a trace relation.

   It is unlikely that the coefficients of all the terms of ${\cal L}_6$
will be determined from experiment. 
   However, usually a much smaller subset actually contributes to most
simple processes, and it is possible to get information on some of
the corresponding coefficients.

\subsection{Elastic Pion-Pion Scattering at ${\cal O}(p^6)$}
\label{subsec_eppsop6}     
   Elastic pion-pion scattering represents a nice example of the
success of mesonic chiral perturbation theory.
   A complete analytical calculation at two-loop order was performed in
Ref.\ \cite{Bijnens:1995yn}.

   Let us consider the $T$-matrix element of the scattering process
$\pi^a(p_a)+\pi^b(p_b)\to\pi^c(p_c)+\pi^d(p_d)$,
\begin{equation}
\label{4:10:tpipi}
T^{ab;cd}(p_a,p_b;p_c,p_d)=\delta^{ab}\delta^{cd}A(s,t,u)
                  +\delta^{ac}\delta^{bd}A(t,s,u)
                  +\delta^{ad}\delta^{bc}A(u,t,s),
\end{equation}
where $s=(p_a+p_b)^2$, $t=(p_a-p_c)^2$, and $u=(p_a-p_d)^2$ denote the 
usual Mandelstam variables, the indices $a,\cdots,d$ refer to the Cartesian 
isospin components, and the function $A$ satisfies 
$A(s,t,u)=A(s,u,t)$ \cite{Weinberg:1966kf}.
   Since the pions form an isospin triplet, the two isovectors of both
the initial and final states may be coupled to $I=0,1,2$.
   For $m_u=m_d=m$ the strong interactions are invariant under isospin
transformations, implying that scattering matrix elements can be decomposed 
as 
\begin{equation}
\label{4:10:tdec}
\langle I',I_3'|T|I,I_3\rangle=T^I \delta_{II'}\delta_{I_3 I_3'}.
\end{equation}
    For the case of $\pi\pi$ scattering the three isospin amplitudes
are given in terms of the invariant amplitude $A$ of Eq.\ (\ref{4:10:tpipi})
by \cite{Gasser:1983yg}
\begin{eqnarray}
\label{4:10:isk}
T^{I=0}&=&3A(s,t,u)+A(t,u,s)+A(u,s,t),\nonumber\\
T^{I=1}&=&A(t,u,s)-A(u,s,t),\nonumber\\
T^{I=2}&=&A(t,u,s)+A(u,s,t).
\end{eqnarray}
   For example, the physical $\pi^+\pi^+$ scattering process is described by 
$T^{I=2}$.
   Other physical processes are obtained using the appropriate Clebsch-Gordan
coefficients.
   Evaluating the $T$ matrices at threshold, one obtains the $s$-wave
$\pi\pi$-scattering lengths\footnote{The definition differs by a factor of
$(-M_\pi)$ \cite{Gasser:1983yg} from the conventional definition of 
scattering lengths in the effective range expansion (see, e.g., 
Ref.\ \cite{Preston:1962}).}
\begin{equation}
\label{$:10:swsl}
T^{I=0}|_{\rm thr}=32\pi a^0_0,\quad T^{I=2}|_{\rm thr}=32\pi a^2_0,
\end{equation}
   where the subscript $0$ refers to $s$ wave and the superscript to 
the isospin.
   ($T^{I=1}|_{\rm thr}$ vanishes because of Bose symmetry). 
   The current-algebra prediction of Ref.\ \cite{Weinberg:1966kf} is identical
with the lowest-order result obtained from Eqs.\ (\ref{4:6:mpipi1}) or
(\ref{4:6:mpipi2}),
\begin{equation}
\label{4:10:a00a02lo}
a_0^0=\frac{7 M_\pi^2}{32 \pi F_\pi^2}=0.156,\quad
a_0^2=-\frac{M_\pi^2}{16 \pi F_\pi^2}=-0.045,
\end{equation} 
   where we replaced $F_0$ by $F_\pi$ and made use of the numerical values 
$F_\pi=93.2$ MeV and $M_\pi=139.57$ MeV of Ref.\ \cite{Bijnens:1995yn}.
   In order to obtain the results of Eq.\ (\ref{4:10:a00a02lo}), use has
been made of $s_{\rm thr}=4 M_\pi^2$ and $t_{\rm thr}=u_{\rm thr}=0$.

   The predictions for the $s$-wave scattering lengths at ${\cal O}(p^6)$ 
read \cite{Bijnens:1995yn}
\begin{eqnarray*}
a_0^0&=& 
\overbrace{0.156}^{\mbox{${\cal O}(p^2)$}} 
+\overbrace{\underbrace{0.039}_{\mbox{L}} 
+\underbrace{0.005}_{\mbox{anal.}}}^{\mbox{${\cal O}(p^4)$: +28\%}}
+\overbrace{\underbrace{0.013}_{\mbox{$k_i$}}
+\underbrace{0.003}_{\mbox{L}}
+\underbrace{0.001}_{\mbox{anal.}}}^{\mbox{${\cal O}(p^6)$: +8.5\%}}=
\overbrace{0.217}^{\mbox{total}},\\
a_0^0-a_0^2&=& 
\overbrace{0.201}^{\mbox{${\cal O}(p^2)$}}
+\overbrace{\underbrace{0.036}_{\mbox{L}}
+\underbrace{0.006}_{\mbox{anal.}}}^{\mbox{${\cal O}(p^4)$: +21\%}}
+\overbrace{\underbrace{0.012}_{\mbox{$k_i$}}
+\underbrace{0.003}_{\mbox{L}}
+\underbrace{0.001}_{\mbox{anal.}}}^{\mbox{${\cal O}(p^6)$: +6.6\%}}=
\overbrace{0.258}^{\mbox{total}}.
\end{eqnarray*}
   The corrections at ${\cal O}(p^4)$ consist of a dominant part from the
chiral logarithms (L) of the one-loop diagrams and a less important
analytical contribution (anal.) resulting from the one-loop diagrams as well
as the tree graphs of ${\cal L}_4$.
   The total corrections at ${\cal O}(p^4)$ amount to 28\% and 21\% of the 
${\cal O}(p^2)$ predictions, respectively.
   At ${\cal O}(p^6)$ one obtains two-loop corrections, one-loop corrections,
and ${\cal L}_6$ tree-level contributions.
   Once again, the loop corrections ($k_i$, involving double chiral
logarithms, and L) are more important than the analytical contributions.
   The influence of ${\cal L}_6$ was estimated via scalar- and vector-meson
exchange and found to be very small.

   The empirical results for the $\pi\pi$ $s$-wave scattering lengths
are, so far, obtained from the $K_{e4}$ decay $K^+\to\pi^+\pi^-e^+\nu_e$
and the $\pi^\pm p\to \pi^\pm \pi^+ n$ reactions.
   In the former case, the connection with low-energy $\pi\pi$ scattering 
stems from a partial-wave analysis of the form factors relevant for the 
$K_{e4}$ decay in terms of $\pi\pi$ angular momentum eigenstates.
   In the low-energy regime the phases of these form factors are related by 
(a generalization of) Watson's theorem \cite{Watson:1954uc} to the 
corresponding phases of $I=0$ $s$-wave and $I=1$ $p$-wave elastic 
scattering \cite{Colangelo:2001sp}.
   Using a dispersion-theory approach in terms of the Roy equations 
\cite{Roy:1971tc,Ananthanarayan:2000ht}, the most recent analysis of 
$K^+\to\pi^+\pi^-e^+\nu_e$ \cite{Pislak:2001bf} has obtained
\begin{equation}
\label{4:10:swslexpold}
a_0^0=0.228\pm 0.012 \pm 0.003.
\end{equation}
   This result has to be compared with older determinations 
\cite{Rosselet:1976pu,Froggatt:hu,Nagels:xh}
\begin{equation}
\label{swslexp}
a_0^0=0.26\pm 0.05,\quad a_0^2=-0.028 \pm 0.012,
\end{equation}
   and the more recent one from $\pi^\pm p\to \pi^\pm \pi^+ n$
\cite{Kermani:gq}
\begin{equation}
\label{4:10:swslexptriumf}   
a_0^0=0.204\pm 0.014\, (stat.) \pm 0.008\, (syst.),
\end{equation}
   which makes use of an extrapolation to the pion pole to extract the
$\pi\pi$  amplitude.

   In particular, when analyzing the data of Ref.\ \cite{Pislak:2001bf}
in combination with the Roy equations, an upper limit $|\bar{l}_3|\leq 16$
was obtained in Ref.\ \cite{Colangelo:2001sp} for the 
scale-independent low-energy coupling constant which is related to
$l_3$ of the SU(2)$\times$SU(2) Lagrangian of
Gasser and Leutwyler (see Appendix \ref{app_sec_glvgss}).
   The great interest generated by this result is to be understood in
the context of the pion mass at ${\cal O}(p^4)$ [see Eq.\ (\ref{app:dp:Mpi2})
of App.\ \ref{app_sec_dp}],
\begin{equation}
\label{4:10:Mpi2}
M_\pi^2=M^2-\frac{\bar{l}_3}{32\pi^2 F^2_0} M^4+ {\cal O}(M^6),
\end{equation}
   where $M^2=(m_u+m_d) B_0$. Recall that the constant $B_0$ is related
to the scalar quark condensate in the chiral limit [see Eq.\ (\ref{4:3:b0})]
and that a nonvanishing quark condensate is a sufficient criterion
for spontaneous chiral symmetry breakdown in QCD (see Sec.\ \ref{subsec_sqc}).
   If the expansion of $M_\pi^2$ in powers of the quark masses is dominated
by the linear term in Eq.\ (\ref{4:10:Mpi2}), the result is often referred 
to as the Gell-Mann-Oakes-Renner relation \cite{Gell-Mann:rz}.
   If the terms of order $m^2$ were comparable or even larger than the linear
terms, a different power counting or bookkeeping
in ChPT would be required \cite{Knecht:1995tr,Knecht:1995ai,Stern:1997}.
   The estimate $|\bar{l}_3|\leq 16$ implies that the 
Gell-Mann-Oakes-Renner relation \cite{Gell-Mann:rz} is indeed a decent starting
point, because the contribution of the second term of Eq.\ (\ref{4:10:Mpi2})
to the pion mass is approximately given by
\begin{displaymath}
-\frac{\bar{l}_3 M_\pi^2}{64\pi^2 F_\pi^2} M_\pi
=-0.054 M_\pi\,\,\mbox{for}\,\, \bar{l}_3=16,
\end{displaymath}
i.e., more than 94 \% of the pion mass must stem from the quark
condensate \cite{Colangelo:2001sp}.

\chapter{Chiral Perturbation Theory for Baryons}
\label{chap_cptb}

   So far we have considered the purely mesonic sector involving 
the interaction of Goldstone bosons with each other and with 
the external fields. 
   However, ChPT can be extended to also describe the dynamics of baryons
at low energies. 
   Here we will concentrate on matrix elements with a single baryon in the 
initial and final states.
   With such matrix elements we can, e.g,  describe static properties such as 
masses or magnetic moments, form factors, or, finally, more complicated
processes, such as pion-nucleon scattering, Compton scattering, pion 
photoproduction etc.
   Technically speaking, we are interested in the baryon-to-baryon 
transition amplitude in the presence of external fields
(as opposed to the vacuum-to-vacuum transition amplitude of 
Sec.\ \ref{subsec_qcdpefgf}) 
\cite{Gasser:1987rb,Krause:xc},
\begin{equation}
\label{4:btbta}
{\cal F}(\vec{p}\,',\vec{p};v,a,s,p)=\langle {\vec{p}\,'};{\rm out}|
{\vec{p}\,};{\rm in}\rangle^{\rm c}_{v,a,s,p},
\quad 
\vec{p}\neq\vec{p}\,',
\end{equation}
determined by the Lagrangian of Eq.\ (\ref{2:4:lqcds}),
\begin{equation}
\label{5:lqcds}
{\cal L}={\cal L}^0_{\rm QCD}+{\cal L}_{\rm ext}
={\cal L}^0_{\rm QCD}+\bar{q}\gamma_\mu (v^\mu +\frac{1}{3}v^\mu_{(s)}
+\gamma_5 a^\mu )q
-\bar{q}(s-i\gamma_5 p)q.
\end{equation}
   In Eq.\ (\ref{4:btbta}) $|\vec{p};{\rm in}\rangle$ and    
$|\vec{p}\,';{\rm out}\rangle$ denote asymptotic one-baryon in- and 
out-states, i.e., states which in the remote past and distant future behave as 
free one-particle states of momentum $\vec{p}$ and $\vec{p}\,'$, respectively.
   The functional ${\cal F}$ consists of connected diagrams only (superscript
c). 
   For example, the matrix elements of the vector and axial-vector currents
between one-baryon states are given by \cite{Krause:xc}
\begin{eqnarray}
\label{5:vc}
\langle \vec{p}\,'|V^{\mu,a}(x)|\vec{p}\,\rangle
&=&\left.
\frac{\delta}{i\delta v^a_\mu(x)}
{\cal F}(\vec{p}\,',\vec{p};v,a,s,p)\right|_{v=0,a=0,s=M,p=0},\\
\label{5:avc}
\langle \vec{p}\,'|A^{\mu,a}(x)|\vec{p}\,\rangle
&=&\left.
\frac{\delta}{i\delta a^a_\mu(x)}
{\cal F}(\vec{p}\,',\vec{p};v,a,s,p)\right|_{v=0,a=0,s=M,p=0},
\end{eqnarray}
where $M=\mbox{diag}(m_u,m_d,m_s)$ denotes the quark-mass matrix
and
\begin{displaymath}
V^{\mu,a}(x)=\bar{q}(x)\gamma^\mu\frac{\lambda^a}{2} q(x),\quad
A^{\mu,a}(x)=\bar{q}(x)\gamma^\mu \gamma_5
\frac{\lambda^a}{2} q(x).
\end{displaymath}   
   As in the mesonic sector the method of calculating the Green functions
associated with the functional of Eq.\ (\ref{4:btbta}) consists of
an effective Lagrangian-approach in combination with an appropriate 
power counting.
   Specific matrix elements will be calculated applying the Gell-Mann
and Low formula of perturbation theory \cite{Gell-Mann:1951rw}. 
   The group-theoretical foundations of constructing phenomenological 
Lagrangians in the presence of spontaneous symmetry breaking have been
developed in Refs.\ \cite{Weinberg:de,Coleman:sm,Callan:sn}.
   The fields entering the Lagrangian are assumed to transform under
irreducible representations of the subgroup $H$ which leaves the vacuum
invariant whereas the symmetry group $G$ of the Hamiltonian or Lagrangian
is nonlinearly realized (for the transformation behavior of the Goldstone
bosons, see Sec.\ \ref{sec_tpgb}).

\section{Transformation Properties of the Fields}
\label{sec_tpf}
   Our aim is a description of the interaction of baryons with the Goldstone
bosons as well as the external fields at low energies.
   To that end we need to specify the transformation properties of the 
dynamical fields entering the Lagrangian.
   Our discussion follows Refs.\ \cite{Georgi,Gasser:1987rb}. 

   To be specific, we consider the octet of the $\frac{1}{2}^+$ baryons (see
Fig.\ \ref{sec:tf:tab:octetbaryons}).
   With each member of the octet we associate a complex, four-component Dirac 
field which we arrange in a traceless $3\times 3$ matrix $B$,
\begin{figure}[htb]
\caption{\label{sec:tf:tab:octetbaryons} The baryon octet in an $(I_3,S)$
diagram. We have included the masses in MeV as well as the 
quark content.} 
\vspace{2em}
\unitlength1cm
\begin{center}
\begin{picture}(10,6)
\thicklines
\put(0,0){\vector(1,0){10}}
\put(10,-1){$I_3$}
\put(0,0){\vector(0,1){6}}
\put(-1,6){$S$}
\put(3,1){\circle*{0.2}}
\put(2.0,0.5){$n(940)(udd)$}
\put(7,1){\circle*{0.2}}
\put(6.0,0.5){$p(938)(uud)$}
\put(1,3){\circle*{0.2}}
\put(0,3.5){$\Sigma^-(1197)(dds)$}
\put(5,3){\circle*{0.2}}
\put(4.0,3.5){$\Sigma^0(1193)(uds)$}
\put(4.0,2.5){$\Lambda(1116)(uds)$}
\put(9,3){\circle*{0.2}}
\put(8.0,3.5){$\Sigma^+(1189)(uus)$}
\put(3,5){\circle*{0.2}}
\put(2.0,4.5){$\Xi^-(1321)(dss)$}
\put(7,5){\circle*{0.2}}
\put(6.0,4.5){$\Xi^0(1315)(uss)$}
\put(0,1){\circle*{0.1}}
\put(-0.5,1){0}
\put(0,3){\circle*{0.1}}
\put(-0.5,3){-1}
\put(0,5){\circle*{0.1}}
\put(-0.5,5){-2}
\put(1,0){\circle*{0.1}}
\put(1,-0.5){-1}
\put(3,0){\circle*{0.1}}
\put(3,-0.5){-1/2}
\put(5,0){\circle*{0.1}}
\put(5,-0.5){0}
\put(7,0){\circle*{0.1}}
\put(7,-0.5){1/2}
\put(9,0){\circle*{0.1}}
\put(9,-0.5){1}
\end{picture}
\vspace{2em}
\end{center}
\end{figure}
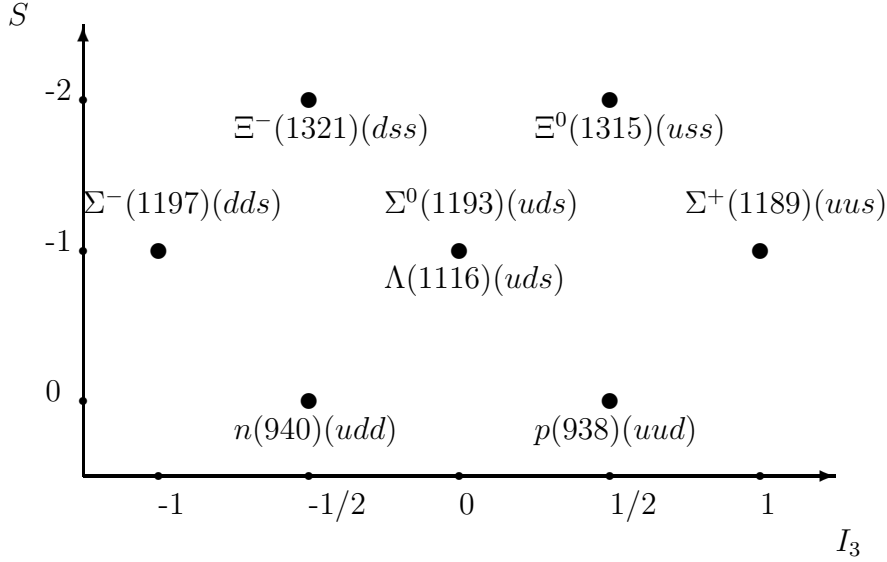
\begin{equation}
\label{5:1:su3oktett}
B=\sum_{a=1}^8 \lambda_a B_a=
\left(\begin{array}{ccc}
\frac{1}{\sqrt{2}}\Sigma^0+\frac{1}{\sqrt{6}}\Lambda&\Sigma^+&p\\
\Sigma^-&-\frac{1}{\sqrt{2}}\Sigma^0+\frac{1}{\sqrt{6}}\Lambda&n\\
\Xi^-&\Xi^0&-\frac{2}{\sqrt{6}}\Lambda
\end{array}\right),
\end{equation}
   where we have suppressed the dependence on $x$.
   For later use, we have to keep in mind that each entry of 
Eq.\ (\ref{5:1:su3oktett}) is a Dirac field, but for the purpose of 
discussing the transformation properties under global flavor SU(3) 
this can be ignored, because these transformations
act on each of the four components in the same way.
   In contrast to the mesonic case of Eq.\ (\ref{4:3:upar}),
where we collected the fields of the Goldstone octet in a Hermitian
traceless matrix $\phi$, the $B_a$ of the spin $1/2$-case are not
real (Hermitian), i.e., $B\neq B^\dagger$.
   Now let us define the set
\begin{equation}
\label{5:1:setm}
M\equiv\{B(x)|B(x)\, \mbox{complex, traceless $3\times 3$ matrix}\}
\end{equation}   
which under the addition of matrices is a complex vector space.
   The following homomorphism is a representation of the abstract
group $H=\mbox{SU}(3)_V$ on the vector space $M$ 
[see also Eq.\ (\ref{4:2:uhtrafo})]:
\begin{eqnarray}
\label{5:1:su3hom}
&&\varphi: H\to \varphi(H),\quad V\mapsto \varphi(V)\quad \mbox{where}\quad 
\varphi(V): M\to M,\nonumber\\
&&B(x)\mapsto B'(x)=\varphi(V)B(x)\equiv V B(x) V^\dagger.
\end{eqnarray}
   First of all, $B'(x)$ is again an element of $M$, because $\mbox{Tr}[B'(x)]
=\mbox{Tr}[VB(x)V^\dagger]=\mbox{Tr}[B(x)]=0$. 
   Equation (\ref{5:1:su3hom}) satisfies the homomorphism property
\begin{eqnarray*}
\varphi(V_1)\varphi(V_2) B(x)&=&\varphi(V_1)V_2B(x) V_2^\dagger
=V_1 V_2 B(x) V_2^\dagger V_1^\dagger=(V_1 V_2)B(x) (V_1 V_2)^\dagger\\
&=&\varphi(V_1V_2) B(x)
\end{eqnarray*}
and is indeed a {\em representation} of SU(3), because
\begin{eqnarray*}
\varphi(V)[\lambda_1B_1(x)+\lambda_2 B_2(x)]&=&
V[\lambda_1B_1(x)+\lambda_2 B_2(x)]V^\dagger\\
&=&\lambda_1 VB_1(x)V^\dagger+\lambda_2VB_2(x)V^\dagger\\
&=&\lambda_1\varphi(V)B_1(x)
+\lambda_2\varphi(V)B_2(x).
\end{eqnarray*} 
   Equation (\ref{5:1:su3hom}) is just the familiar statement that $B$ 
transforms as an octet under (the adjoint representation of) 
SU(3)$_V$.\footnote{Technically speaking the adjoint representation 
is faithful (one-to-one) modulo the center $Z$ of SU(3) which is defined
as the set of all elements commuting with all elements of SU(3) and
is given by 
$Z=\{1_{3\times 3}, \exp(2\pi i/3)1_{3\times 3}, \exp (4\pi i/3) 1_{3\times 3}
\}$ \cite{O'Raifeartaigh:vq}.}

   Let us now turn to various representations and realizations of the 
group $\mbox{SU(3)}_L\times\mbox{SU(3)}_R$.
   We consider two explicit examples and refer the interested reader
to Ref.\ \cite{Georgi} for more details.
   In analogy to the discussion of the quark fields in QCD, we may introduce
left- and right-handed components of the baryon fields 
[see Eq.\ (\ref{2:3:qlr})]:
\begin{equation}
\label{5:1:blr}
B_1=P_L B_1+P_R B_1=B_L + B_R.
\end{equation}
   We define the set $M_1\equiv\{(B_L(x),B_R(x))\}$ which under the
addition of matrices is a complex vector space.
   The following homomorphism is a representation of the abstract group
$G=\mbox{SU}(3)_L\times \mbox{SU(3)}_R$ on $M_1$:
\begin{eqnarray}
\label{5:1:su3lrhom}
(B_L,B_R)\mapsto (B'_L,B'_R)%=\varphi(L,R)(B_L,B_R)
\equiv (L B_L L^\dagger,R B_R R^\dagger),
\end{eqnarray}
   where we have suppressed the $x$ dependence. 
   The proof proceeds in complete analogy to that of Eq.\ (\ref{5:1:su3hom}).
   
   As a second example, consider the set $M_2\equiv\{B_2(x)\}$ with the
homomorphism 
\begin{equation}
\label{5:1:su3lrhomb2}
B_2\mapsto B_2'\equiv L B_2 L^\dagger,
\end{equation}
i.e.\ the transformation behavior is independent of $R$.
   The mapping defines a representation of the group 
$\mbox{SU(3)}_L\times\mbox{SU(3)}_R$, although the transformation 
behavior is drastically different from the first example.
   However, the important feature which both mappings have in common
is that under the subgroup $H=\{(V,V)|V\in \mbox{SU($3$)}\}$ of $G$
both fields $B_i$ transform as an octet:
\begin{eqnarray*}
B_1=B_L+B_R&\stackrel{H}{\mapsto}&VB_LV^\dagger+VB_R V^\dagger=VB_1V^\dagger,\\
B_2&\stackrel{H}{\mapsto}&VB_2V^\dagger.
\end{eqnarray*}

   We will now show how in a theory also containing Goldstone bosons
the various realizations may be connected to each other using field 
redefinitions.
   The procedure is actually very similar to Sec.\ \ref{subsec_mclop6},
where we discussed how, by an appropriate multiplication with $U$ or 
$U^\dagger$, all building blocks of the mesonic effective Lagrangian
could be made to transform in the same way.
   Here we consider the second example, with the fields $B_2$ of 
Eq.\ (\ref{5:1:su3lrhomb2}) and $U$ of
Eq.\ (\ref{4:3:upar}) transforming as
\begin{displaymath}
B_2\mapsto LB_2L^\dagger,\quad U\mapsto RUL^\dagger,
\end{displaymath}
and define new baryon fields by
\begin{displaymath}
\tilde{B}\equiv UB_2,
\end{displaymath}
   so that the new pair $(\tilde B,U)$ transforms as
$$\tilde{B}\mapsto RUL^\dagger L B L^\dagger=R\tilde{B}L^\dagger,\quad
U\mapsto RU L^\dagger.$$
   Note in particular that $\tilde{B}$ still transforms as an octet under
the subgroup $H=\mbox{SU(3)}_V$.

   Given that physical observable are invariant under field transformations
we may choose a description of baryons that is maximally convenient for
the construction of the effective Lagrangian \cite{Georgi} and which is
commonly used in chiral perturbation theory.
   We start with $G=\mbox{SU(2)}_L\times\mbox{SU(2)}_R$ and consider 
the case of $G=\mbox{SU(3)}_L\times\mbox{SU(3)}_R$ later. 
   Let 
\begin{equation}
\label{5:1:Psi}
\Psi=\left(\begin{array}{c}p\\n\end{array}\right)
\end{equation}
denote the nucleon field with two four-component Dirac fields for the proton 
and the neutron and $U$ the SU(2) matrix containing the pion fields.
   We have already seen in Sec.\ \ref{subsec_aqcd} that the mapping
$U\mapsto RUL^\dagger$ defines a nonlinear realization of $G$.
   We denote the square root of $U$ by $u$, $u^2(x)=U(x)$, and define
the SU(2)-valued function $K(L,R,U)$ by [see Eqs.\ (\ref{4:10:kdef}) and
(\ref{4:10:k})]
\begin{equation}
\label{5:1:kdef}
u(x)\mapsto u'(x)=\sqrt{RUL^\dagger}\equiv RuK^{-1}(L,R,U),
\end{equation}
i.e.
\begin{displaymath}
K(L,R,U)=u'^{-1}Ru=\sqrt{RUL^\dagger}^{-1}R \sqrt{U}.
\end{displaymath}
   The following homomorphism defines an operation of $G$ on the set 
$\{(U,\Psi)\}$ in terms of a nonlinear realization:
\begin{equation}
\label{5:1:su2real}
\varphi(g):\left(\begin{array}{c}U\\ \Psi\end{array}\right)\mapsto
\left(\begin{array}{c}U'\\ \Psi'\end{array}\right)
=\left(\begin{array}{c}RUL^\dagger\\K(L,R,U)\Psi\end{array}\right),
\end{equation}
because the identity leaves $(U,\Psi)$ invariant and 
[see Sec.\ \ref{subsec_aqcd} and Eq.\ (\ref{4:10:kreal})] 
\begin{eqnarray*}
\varphi(g_1)\varphi(g_2)
\left(\begin{array}{c}U\\ \Psi\end{array}\right)&=&\varphi(g_1)
\left(\begin{array}{c}R_2UL_2^\dagger\\K(L_2,R_2,U)\Psi\end{array}\right)\\
&=&\left(\begin{array}{c}R_1R_2UL_2^\dagger L_1^\dagger\\
K(L_1,R_1,R_2UL_2^\dagger)K(L_2,R_2,U)\Psi
\end{array}\right)\\
&=&\left(\begin{array}{c}R_1 R_2U(L_1L_2)^\dagger\\
K(L_1L_2,R_1R_2,U)\Psi\end{array}\right)\\
&=&\varphi(g_1g_2)\left(\begin{array}{c}U\\ \Psi\end{array}\right).
\end{eqnarray*}
   Note that for a general group element $g=(L,R)$ the transformation behavior
of $\Psi$ depends on $U$.
   For the special case of an isospin transformation, $R=L=V$, one obtains
$u'=VuV^\dagger$, because
$$U'=u'^2=VuV^\dagger VuV^\dagger=Vu^2V^\dagger=VUV^\dagger.$$
   Comparing with Eq.\ (\ref{5:1:kdef}) yields 
$K^{-1}(V,V,U)=V^\dagger$ or $K(V,V,U)=V$,
i.e., $\Psi$ transforms linearly as an isospin doublet 
under the isospin subgroup $\mbox{SU(2)}_V$ of 
$\mbox{SU(2)}_L\times\mbox{SU(2)}_R$.
   A general feature here is that the transformation behavior under
the subgroup which leaves the ground state invariant is independent of $U$.
   Moreover, as already discussed in Sec.\ \ref{subsec_aqcd}, the Goldstone
bosons $\phi$ transform according to the adjoint representation of 
SU(2)$_V$, i.e., as an isospin triplet.

   For the case $G=\mbox{SU(3)}_L\times \mbox{SU(3)}_R$ one uses the
nonlinear realization
\begin{equation}
\label{5:1:su3real}
\varphi(g):\left(\begin{array}{c}U\\ B\end{array}\right)\mapsto
\left(\begin{array}{c}U'\\ B'\end{array}\right)
=\left(\begin{array}{c}RUL^\dagger\\K(L,R,U)B K^\dagger(L,R,U)
\end{array}\right),
\end{equation}
where $K$ is defined completely analogously to Eq.\ (\ref{5:1:kdef}) after
inserting the corresponding SU(3) matrices.

\section{Lowest-Order Effective Baryonic Lagrangian}
\label{sec_loebl}

   Given the dynamical fields of Eqs.\ (\ref{5:1:su2real}) and 
(\ref{5:1:su3real}) and their transformation properties, we will now discuss
the most general effective baryonic Lagrangian at lowest order. 
   As in the vacuum sector, chiral symmetry provides constraints among the 
single-baryon Green functions contained in the functional of 
Eq.\ (\ref{4:btbta}).
   Analogous to the mesonic sector, these Ward identities will be satisfied 
if the Green functions are calculated from the most general effective 
Lagrangian coupled to external fields with a {\em local} invariance under the 
chiral group (see Appendix \ref{app_gfwi}).

   Let us start with the construction of the $\pi N$ effective Lagrangian
${\cal L}^{(1)}_{\pi N}$ which we demand to have a {\em local}
$\mbox{SU}(2)_L\times\mbox{SU(2)}_R\times\mbox{U(1)}_V$ symmetry.
   The transformation behavior of the external fields is given in 
Eq.\ (\ref{2:4:sg}), whereas the nucleon doublet and $U$ transform as
\begin{equation}
\label{5:2:psitrans}
\left(\begin{array}{c}U(x)\\ \Psi(x)\end{array}\right)\mapsto
\left(\begin{array}{c}V_R(x)U(x)V_L^\dagger(x)\\
\exp[-i\Theta(x)]K[V_L(x),V_R(x),U(x)]\Psi(x)\end{array}\right).
\end{equation}
   The local character of the transformation implies that we need to 
introduce a covariant derivative $D_\mu \Psi$ with the usual property that 
it transforms in the same way as $\Psi$
[compare with Eq.\ (\ref{2:2:cdt}) for the case of QED]:
\begin{equation}
\label{5:2:kovder}
D_\mu \Psi(x)\mapsto [D_\mu \Psi(x)]'\stackrel{!}{=}
\exp[-i\Theta(x)]K[V_L(x),V_R(x),U(x)]D_\mu\Psi(x).
\end{equation}
   Since $K$ not only depends on $V_L$ and $V_R$ but also on $U$,
we may expect the covariant derivative to contain $u$ and $u^\dagger$
and their derivatives.
   In fact, the connection of Eq.\ (\ref{4:10:chiralconnection}) 
(recall $\partial_\mu u u^\dagger=-u\partial_\mu u^\dagger$),
\begin{equation}
\label{5:2:gamma}
\Gamma_\mu=\frac{1}{2}\left[u^\dagger(\partial_\mu-ir_\mu)u
+u(\partial_\mu-il_\mu)u^\dagger\right],
\end{equation}
is also an integral part of the covariant derivative of the nucleon doublet:
\begin{equation}
\label{5:2:kovderpsi}
D_\mu\Psi=(\partial_\mu+\Gamma_\mu-iv_\mu^{(s)})\Psi.
\end{equation}
   What needs to be shown is
\begin{equation}
\label{5:2tbs}
D'_\mu\Psi'=[\partial_\mu+\Gamma_\mu'-i(v_\mu^{(s)}-\partial_\mu\Theta)]
\exp(-i\Theta)K\Psi
=\exp(-i\Theta)K(\partial_\mu+\Gamma_\mu-iv_\mu^{(s)})\Psi.
\end{equation}
To that end, we make use of the product rule, 
$$
\partial_\mu[\exp(-i\Theta)K\Psi]=-i\partial_\mu\Theta \exp(-i\Theta)K\Psi
+\exp(-i\Theta)\partial_\mu K\Psi
+\exp(-i\Theta) K \partial_\mu \Psi,
$$
in Eq.\ (\ref{5:2tbs}) and multiply by $\exp(i\Theta)$, 
reducing it to
$$
\partial_\mu K=K\Gamma_\mu-\Gamma_\mu'K.
$$
   Using Eq.\ (\ref{5:1:kdef}),
\begin{eqnarray*}
K&=&u'^\dagger V_R u=\underbrace{u'u'^\dagger}_{\mbox{1}}u'^\dagger V_R u
=u' U'^\dagger V_R u= u' V_L 
\underbrace{U^\dagger}_{\mbox{$u^\dagger u^\dagger$}} 
\underbrace{V_R^\dagger V_R}_{\mbox{1}} u
=u'V_Lu^\dagger,
\end{eqnarray*}
we find 
\begin{eqnarray*}
2(K\Gamma_\mu-\Gamma_\mu'K)&=&
K\left[u^\dagger(\partial_\mu-ir_\mu)u\right]
-\left[u'^\dagger(\partial_\mu-iV_R r_\mu V_R^\dagger
+V_R\partial_\mu V_R^\dagger)u'\right]K\\
&&+(R\to L,u\leftrightarrow u^\dagger,u'\leftrightarrow u'^\dagger)\\
&=&u'^\dagger V_R(\partial_\mu u-ir_\mu u)
-u'^\dagger\partial_\mu u' 
\underbrace{K}_{\mbox{$u'^\dagger V_R u$}}\\
&&
+iu'^\dagger V_R r_\mu\underbrace{V_R^\dagger u'K}_{\mbox{$u$}}
-u'^\dagger V_R\partial_\mu V^\dagger_R\underbrace{u'K}_{\mbox{$V_R u$}}\\
&&+(R\to L,u\leftrightarrow u^\dagger,u'\leftrightarrow u'^\dagger)\\
&=&u'^\dagger V_R\partial_\mu u
-iu'^\dagger V_R r_\mu u
-\underbrace{u'^\dagger\partial_\mu u'u'^\dagger}_{
\mbox{$-\partial_\mu u'^\dagger$}} V_R u\\
&&+iu'^\dagger V_R r_\mu u
-u'^\dagger\underbrace{V_R\partial_\mu V_R^\dagger
V_R}_{\mbox{$-\partial_\mu V_R$}}u\\
&&+(R\to L,u\leftrightarrow u^\dagger,u'\leftrightarrow u'^\dagger)\\
&=&u'^\dagger V_R\partial_\mu u +\partial_\mu u'^\dagger V_R u +
u'^\dagger \partial_\mu V_R u\\
&&+(R\to L,u\leftrightarrow u^\dagger,u'\leftrightarrow u'^\dagger)\\
&=&\partial_\mu(u'^\dagger V_Ru+u'V_Lu^\dagger)=2\partial_\mu K,
\end{eqnarray*}
i.e., the covariant derivative defined in Eq.\ (\ref{5:2:kovderpsi}) indeed
satisfies the condition of Eq.\ (\ref{5:2:kovder}).
   At ${\cal O}(p)$ there exists another Hermitian building block,
the so-called vielbein \cite{Ecker:1995gg},\footnote{The relation with the
notation of Sec.\ \ref{subsec_mclop6} is given by
$(D_\mu U)_-=-2iu_\mu$ \cite{Ebertshauser:2001nj}.}
\begin{equation}
\label{5:2:chvi}
u_\mu\equiv i\left[u^\dagger(\partial_\mu-i r_\mu)u-u(\partial_\mu-i
l_\mu)u^\dagger\right],
\end{equation}
which under parity transforms as an axial vector:
\begin{displaymath}
u_\mu\stackrel{P}{\mapsto} i\left[u(\partial^\mu-il^\mu)u^\dagger
-u^\dagger(\partial^\mu-ir^\mu)u\right]=-u^\mu.
\end{displaymath}
   The transformation behavior under 
$\mbox{SU(2)}_L\times\mbox{SU(2)}_R\times \mbox{U}(1)_V$ is given by
\begin{displaymath}
u_\mu \mapsto Ku_\mu K^\dagger,
\end{displaymath}
which is shown using [see Eq.\ (\ref{5:1:kdef})]
\begin{displaymath}
u'=V_R u K^\dagger=K u V_L^\dagger
\end{displaymath}
and the corresponding adjoints.
   We obtain
\begin{eqnarray*}
u_\mu&\mapsto&i[u'^\dagger(\partial_\mu-iV_R r_\mu V_R^\dagger
+V_R\partial_\mu V_R^\dagger)u'\\
&&-u'(\partial_\mu-iV_L l_\mu V_L^\dagger +V_L\partial_\mu V_L^\dagger)
u'^\dagger]\\
&=&i[Ku^\dagger V_R^\dagger(\partial_\mu-iV_R r_\mu V_R^\dagger
+V_R\partial_\mu V_R^\dagger)V_R u K^\dagger\\
&&-Ku V_L^\dagger(\partial_\mu-iV_L l_\mu V_L^\dagger 
+V_L\partial_\mu V_L^\dagger) V_L u^\dagger K^\dagger]\\
&=&i[Ku^\dagger V_R^\dagger\partial_\mu V_Ru K^\dagger
+Ku^\dagger\partial_\mu u K^\dagger
+K\partial_\mu K^\dagger\\
&&-iK u^\dagger r_\mu u K^\dagger
+K u^\dagger\underbrace{\partial_\mu V_R^\dagger V_R}_{\mbox{$-V_R^\dagger
\partial_\mu V_R$}} u K^\dagger\\
&&-Ku V_L^\dagger\partial_\mu V_L u^\dagger K^\dagger
-Ku\partial_\mu u^\dagger K^\dagger
-K\partial_\mu K^\dagger\\
&&+iKul_\mu u^\dagger K^\dagger
-Ku\underbrace{\partial_\mu V_L^\dagger V_L}_{\mbox{$
-V_L^\dagger \partial_\mu V_L$}}u^\dagger K^\dagger]\\
&=&iK[u^\dagger(\partial_\mu -ir_\mu)u
-u(\partial_\mu -i l_\mu)u^\dagger]K^\dagger\\
&=&Ku_\mu K^\dagger.
\end{eqnarray*}

   The most general effective $\pi N$ Lagrangian describing processes with a 
single nucleon in the initial and final states is then of the type
$\bar{\Psi} \widehat{O} \Psi$, where $\widehat{O}$ is an operator acting
in Dirac and flavor space, transforming under
$\mbox{SU(2)}_L\times\mbox{SU(2)}_R\times \mbox{U}(1)_V$
as $K\widehat{O}K^\dagger$.
   As in the mesonic sector, the Lagrangian must be a Hermitian Lorentz scalar
which is even under the discrete symmetries $C$, $P$, and $T$.   

   The most general such Lagrangian with the smallest number of derivatives
is given by \cite{Gasser:1987rb}\footnote{The power counting will be 
discussed below.}
\begin{equation}
\label{5:2:l1pin}
{\cal L}^{(1)}_{\pi N}= \bar{\Psi}\left(iD\hspace{-.6em}/ 
-\stackrel{\circ}{m}_N
+\frac{\stackrel{\circ}{g}_A}{2}\gamma^\mu \gamma_5 u_\mu\right)\Psi.
\end{equation}
   It contains two parameters not determined by chiral symmetry:
the nucleon mass $\stackrel{\circ}{m}_N$ and the axial-vector coupling
constant $\stackrel{\circ}{g}_A$, both taken in the chiral limit
(denoted by $\circ$).
   The overall normalization of the Lagrangian is chosen such that in the 
case of no external fields and no pion fields it reduces to that 
of a free nucleon of mass $\stackrel{\circ}{m}_N$.

   Since the nucleon mass $m_N$ does not vanish in the chiral limit, the
zeroth component $\partial^0$ of the partial derivative acting on the nucleon
field does not produce a ``small'' quantity.
   We thus have to address the new features of chiral power counting in the 
baryonic sector.
   The counting of the external fields as well as of covariant derivatives 
acting on the mesonic fields remains the same as in mesonic chiral
perturbation theory [see Eq.\ (\ref{4:5:powercounting}) of Sec.\ 
\ref{sec_cel}].
   On the other hand, the counting of bilinears $\bar{\Psi}\Gamma\Psi$ is
probably easiest understood by investigating the matrix elements of 
positive-energy plane-wave solutions to the free Dirac equation in the 
Dirac representation:
\begin{equation}
\psi^{(+)}(\vec{x},t)=\exp(-ip_N\cdot x) \sqrt{E_N+m_N}\left(
\begin{array}{c}
\chi\\
\frac{\vec{\sigma}\cdot\vec{p}_N}{E_N+m_N}\chi
\end{array}
\right),
\end{equation}
where $\chi$ denotes a two-component Pauli 
spinor and $p_N^\mu=(E_N,\vec{p}_N)$ with $E_N=\sqrt{\vec{p}\,_N^2+m_N^2}$.
   In the low-energy limit, i.e.~for nonrelativistic kinematics, the lower 
(small) component is suppressed as $|\vec{p}_N|/m_N$ in comparison with the
upper (large) component.
   For the analysis of the bilinears it is convenient
to divide the 16 Dirac matrices into even and odd ones, ${\cal E}=
\{1, \gamma_0,\gamma_5 \gamma_i,\sigma_{ij}\}$ and 
${\cal O}=\{\gamma_5,\gamma_5 \gamma_0,\gamma_i,\sigma_{i0}\}$
\cite{Foldy:1949wa,Fearing:ii}, respectively,
where odd matrices couple large and small components but not large with large,
whereas even matrices do not.
   Finally, $i\partial^\mu$ acting on the nucleon solution produces $p_N^\mu$
which we write symbolically as $p_N^\mu=(m_N,\vec{0})+(E_N-m_N,\vec{p}_N\,)$ 
where we count the second term as ${\cal O}(p)$, i.e., as a small quantity.
   We are now in the position to summarize the chiral counting scheme for 
the (new) elements of baryon chiral perturbation theory \cite{Krause:xc}:
\begin{eqnarray}
\label{5:2:powercounting}
&&\Psi,\bar{\Psi} =  {\cal O}(p^0),\, D_{\mu} \Psi = {\cal  O}(p^0),\, 
(iD\hspace{-.6em}/ -\stackrel{\circ}{m}_N)\Psi={\cal O}(p),\nonumber\\
&&1,\gamma_\mu,\gamma_5\gamma_\mu,\sigma_{\mu\nu}={\cal O}(p^0),\,
\gamma_5 ={\cal O}(p),
\end{eqnarray}
   where the order given is the minimal one.
   For example, $\gamma_\mu$ has both an ${\cal O}(p^0)$ piece, $\gamma_0$,
as well as an ${\cal O}(p)$ piece, $\gamma_i$. 
   A rigorous nonrelativistic reduction may be achieved in the framework
of the Foldy-Wouthuysen method \cite{Foldy:1949wa} or the heavy-baryon
approach \cite{Jenkins:1990jv,Bernard:1992qa} which will be discussed later
(for a pedagogical introduction see Ref.\ \cite{Holstein:1997}).

   The construction of the $\mbox{SU(3)}_L\times\mbox{SU(3)}_R$ Lagrangian
proceeds similarly except for the fact that the baryon fields are contained
in the $3\times 3$ matrix of Eq.\ (\ref{5:1:su3oktett})
transforming as $K B K^\dagger$.
   As in the mesonic sector, the building blocks are written as products
transforming as $K\cdots K^\dagger$ with a trace taken at the end.
   The lowest-order Lagrangian reads \cite{Georgi,Krause:xc} 
\begin{equation}
\label{5:2:l1su3}
{\cal L}^{(1)}_{MB}=\mbox{Tr}\left[\bar{B}\left(iD\hspace{-.7em}/\hspace{.2em}
-M_0\right)B\right]
-\frac{D}{2}\mbox{Tr}\left(\bar{B}\gamma^\mu\gamma_5\{u_\mu,B\}\right)
-\frac{F}{2}\mbox{Tr}\left(\bar{B}\gamma^\mu\gamma_5[u_\mu,B]\right),
\end{equation}
where $M_0$ denotes the mass of the baryon octet in the chiral limit.
  The covariant derivative of $B$ is defined as
\begin{equation}
\label{5:2:kovderb}
D_\mu B=\partial_\mu B +[\Gamma_\mu,B],
\end{equation}
with $\Gamma_\mu$ of Eq.\ (\ref{5:2:gamma}) [for $\mbox{SU(3)}_L\times\mbox{
SU(3)}_R$]. 
   The constants $D$ and $F$ may be determined by fitting the semi-leptonic
decays $B\to B'+e^-+\bar{\nu}_e$ at tree level \cite{Borasoy:1998pe}:
\begin{equation}
\label{5:2:df}
D=0.80,\quad
F=0.50.
\end{equation}

\section{Applications at Tree Level}
\label{sec_aatl}
\subsection{Goldberger-Treiman Relation and the 
Axial-Vector Current Matrix Element}
\label{subsec_gtravcme}

   We have seen in Sec.\ \ref{subsec_csbdqm} that the quark masses in
QCD give rise to a non-vanishing divergence of the axial-vector current 
operator [see Eq.\ (\ref{2:3:dsva})].
   Here we will discuss the implications for the matrix elements of the 
pseudoscalar density and of the axial-vector current evaluated between 
single-nucleon states in terms of the lowest-order 
Lagrangians of Eqs.\ (\ref{4:5:l2}) and (\ref{5:2:l1pin}).
   In particular, we will see that the Ward identity
\begin{equation}
\label{5:3:axwi}
\langle N(p')|\partial_\mu A^\mu_i(0)|N(p)\rangle=
\langle N(p')|m_q P_i(0)|N(p)\rangle,
\end{equation}
where $m_q=m_u=m_d$, is satisfied.
   
   The nucleon matrix element of the pseudoscalar density 
can be parameterized as 
\begin{equation}
\label{5:3:def_gt}
   m_q\langle N(p')| P_i (0) | N(p) \rangle =
         \frac{M_\pi^2 F_\pi}{M_\pi^2 - t}
         G_{\pi N}(t)i\bar{u}(p') \gamma_5 \tau_i u(p),
\end{equation}
   where $t=(p'-p)^2$.
   Equation (\ref{5:3:def_gt}) {\em defines} the form factor $G_{\pi N}(t)$ in
terms of the QCD operator $m_q P_i(x)$.
   As we have seen in the discussion of $\pi\pi$ scattering of 
Sec.\ \ref{subsec_pps}, the operator $m_q P_i(x)/(M_\pi^2 F_\pi)$ serves as 
an interpolating pion field [see Eq.\ (\ref{4:6:pionfield})], 
and thus $G_{\pi N}(t)$ is also referred to as the pion-nucleon
form factor (for this specific choice of the interpolating pion field).
   The pion-nucleon coupling constant $g_{\pi N}$ is defined through
$G_{\pi N}(t)$ evaluated at $t=M_\pi^2$.

\begin{figure}[htp]
\begin{center}
\caption{\label{5:3:fig:pionnucleonformfactor}
Lowest-order contribution to the single-nucleon matrix element of the
pseudoscalar density. Mesonic and baryonic vertices are denoted by
a circle and a box, respectively, with the numbers 2 and 1 referring
to the chiral order of ${\cal L}_2$ and ${\cal L}^{(1)}_{\pi N}$. }
\vspace{1em}
\epsfig{file=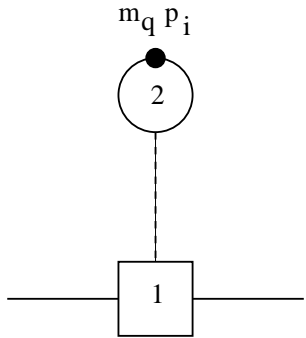,width=4cm}
\end{center}
\end{figure}

   The Lagrangian ${\cal L}_{\pi N}^{(1)}$ of Eq.\ (\ref{5:2:l1pin}) 
does not generate a direct coupling of an external pseudoscalar field
$p_i(x)$ to the nucleon, i.e.,
it does not contain any terms involving $\chi$ or $\chi^\dagger$.
   At lowest order in the chiral expansion, the matrix element of the 
pseudoscalar density is therefore given in terms of the diagram of 
Fig.\ \ref{5:3:fig:pionnucleonformfactor}, i.e., the 
pseudoscalar source produces a pion which propagates and is then absorbed by
the nucleon.
   The coupling of a pseudoscalar field to the pion in the
framework of ${\cal L}_2$ has already been discussed in 
Eq.\ (\ref{4:6:l2ext2}),
\begin{equation}
\label{5:3:l2ext2}
{\cal L}_{\rm ext}
=i\frac{F_0^2B_0}{2}\mbox{Tr}(pU^\dagger-Up)=2B_0F_0p_i\phi_i+\cdots.
\end{equation}
    When working with the nonlinear realization of Eq.\ (\ref{5:1:su2real}) 
it is convenient to use the exponential parameterization of 
Eq.\ (\ref{4:6:u2}),
\begin{displaymath}
U(x)=\exp\left[i\frac{\vec{\tau}\cdot\vec{\phi}(x)}{F_0}\right],
\end{displaymath}
   because in that case the square root is simply given by
\begin{displaymath}
u(x)=\exp\left[i\frac{\vec{\tau}\cdot\vec{\phi}(x)}{2F_0}\right].
\end{displaymath}
   According to Fig.\ \ref{5:3:fig:pionnucleonformfactor}, we need to 
identify the interaction term of a nucleon with a single pion.
   In the absence of external fields the vielbein of Eq.\ (\ref{5:2:chvi}) is 
odd in the pion fields,
\begin{equation}
\label{5:3:umupin}
u_\mu=i\left[u^\dagger\partial_\mu u-u\partial_\mu u^\dagger
\right]\stackrel{\phi^a\mapsto -\phi^a}{\mapsto}
i\left[u\partial_\mu u^\dagger -u^\dagger \partial_\mu u\right]
=-u_\mu.
\end{equation}
   Expanding $u$ and $u^\dagger$ as
\begin{equation}
\label{5:3:uentwi}
u=1+i\frac{\vec{\tau}\cdot\vec{\phi}}{2 F_0}+O(\phi^2),\quad
u^\dagger=1-i\frac{\vec{\tau}\cdot\vec{\phi}}{2 F_0}+O(\phi^2),
\end{equation}
we obtain 
\begin{eqnarray}
\label{5:3:umuentwi}
u_\mu
&=&-\frac{\vec{\tau}\cdot\partial_\mu\vec{\phi}}{F_0}+O(\phi^3),
\end{eqnarray}
which, when inserted into ${\cal L}^{(1)}_{\pi N}$ of Eq.\ (\ref{5:2:l1pin}),
generates the following interaction Lagrangian:
\begin{equation}
\label{5:3:lpin1}
{\cal L}_{\rm int}=-\frac{1}{2}\frac{\stackrel{\circ}{g}_A}{F_0}
\bar{\Psi}\gamma^\mu\gamma_5
\underbrace{\vec{\tau}\cdot\partial_\mu\vec{\phi}}_{\mbox{$
\tau^b\partial_\mu\phi^b$}}\Psi.
\end{equation}
   (Note that the sign is opposite to the conventionally used pseudovector 
pion-nucleon coupling.\footnote{In fact, also the definition of the 
pion-nucleon form factor of Eq.\ (\ref{5:3:def_gt}) contains a sign
opposite to the standard convention so that, in the end, the 
Goldberger-Treiman relation emerges with the conventional sign.})
   The Feynman rule for the vertex of an incoming pion with four-momentum 
$q$ and Cartesian isospin index $a$ is given by
\begin{equation}
\label{5:3:pionnucleonvertex}
i\left(-\frac{1}{2}\frac{\stackrel{\circ}{g}_A}{F_0}\right)
\gamma^\mu\gamma_5\tau^b \delta^{ba}(-i q_\mu)=
-\frac{1}{2}\frac{\stackrel{\circ}{g}_A}{F_0} q\hspace{-.45em}/
\gamma_5 \tau^a.
\end{equation}
   On the other hand, the connection of Eq.\ (\ref{5:2:gamma}) with the 
external fields set to zero is even in the pion fields,
\begin{equation}
\label{5:3:gammapin}
\Gamma_\mu=\frac{1}{2}\left[u^\dagger\partial_\mu u+u\partial_\mu u^\dagger
\right]\stackrel{\phi^a\mapsto -\phi^a}{\mapsto}
\frac{1}{2}\left[u\partial_\mu u^\dagger +u^\dagger \partial_\mu u\right]
=\Gamma_\mu,
\end{equation}
i.e., it does not contribute to the single-pion vertex.

   We now put the individual pieces together and obtain for 
the diagram of Fig.\ \ref{5:3:fig:pionnucleonformfactor} 
\begin{eqnarray*}
\lefteqn{m_q 2 B_0 F_0 \frac{i}{t-M_\pi^2}\bar{u}(p')\left(
-\frac{1}{2}\frac{\stackrel{\circ}{g}_A}{F_0} q\hspace{-.45em}/
\gamma_5 \tau_i\right)u(p)}\\
&=&M_\pi^2 F_0  \frac{\stackrel{\circ}{m}_N \stackrel{\circ}{g}_A}{F_0}
\frac{1}{M_\pi^2-t}\bar{u}(p')\gamma_5 i \tau_i u(p),
\end{eqnarray*}
where we used $M_\pi^2=2B_0 m_q$, and the Dirac equation to show
$\bar{u}q\hspace{-.5em}/ \gamma_5 u=2 \stackrel{\circ}{m}_N\bar{u}\gamma_5 u$.
   At ${\cal O}(p^2)$ $F_\pi=F_0$ so that, by comparison with 
Eq.\ (\ref{5:3:def_gt}), we can read off the lowest-order result
\begin{equation}
\label{5:3:gt1}
G_{\pi N}(t)= \frac{\stackrel{\circ}{m}_N}{F_0}
{\stackrel{\circ}{g}}_A,
\end{equation}
   i.e., at this order the form factor does not depend on $t$.
   In general, the pion-nucleon coupling constant is defined at $t=M_\pi^2$
which, in the present case, simply yields
\begin{equation}
\label{5:3:gpinn}
g_{\pi N}=G_{\pi N}(M_\pi^2)=
\frac{\stackrel{\circ}{m}_N}{F_0}
{\stackrel{\circ}{g}}_A.
\end{equation}
   Equation (\ref{5:3:gpinn}) represents the famous Goldberger-Treiman 
relation \cite{Goldberger:1958tr,Goldberger:1958vp,Nambu:xd} which establishes
a connection between quantities entering weak processes, $F_\pi$ and $g_A$
(to be discussed below), and a typical strong-interaction quantity, namely
the pion-nucleon coupling constant $g_{\pi N}$.
   The numerical violation of the Goldberger-Treiman relation, as expressed
in the so-called Goldberger-Treiman discrepancy
\begin{equation}
\label{5:3:gtd}
\Delta_{\pi N}\equiv1-\frac{g_A m_N}{g_{\pi N}F_\pi},
\end{equation}
is at the percent level,\footnote{
Using $m_N=938.3$ MeV, $g_A=1.267$, $F_\pi=92.4$ MeV, 
and $g_{\pi N}=13.21$  \cite{Schroder:rc},
one obtains $\Delta_{\pi N}=2.6$ \%.}
although one has to keep in mind that 
{\em all four} physical quantities move from their chiral-limit 
values $\stackrel{\circ}{g}_A$ etc.\ to the empirical ones
$g_A$ etc.

   Using Lorentz covariance and isospin symmetry, the matrix element of the 
axial-vector current between initial and final nucleon 
states---excluding second-class currents \cite{Weinberg:1958ut}---can
be parameterized as\footnote{The terminology ``first and second classes'' 
refers to the transformation property of strangeness-conserving 
semi-leptonic weak interactions under ${\cal G}$ conjugation 
\cite{Weinberg:1958ut} which is
the product of charge symmetry and charge conjugation
${\cal G}={\cal C}\exp(i\pi I_2)$.
   A second-class contribution would show up in terms
of a third form factor $G_T$ contributing as
\begin{displaymath}
G_T(t) \bar{u}(p') i\frac{\sigma^{\mu\nu} q_\nu}{2 m_N} \gamma_5 
\frac{\tau_i}{2} u(p).
\end{displaymath}
   Assuming a perfect ${\cal G}$-conjugation symmetry, 
the form factor $G_T$ vanishes.}
\begin{equation}
\label{5:3:axial_current}
   \left<N(p')| A_i^\mu (0) | N(p) \right> =
          \bar{u}(p') \left[ \gamma^\mu G_A(t) + \frac{(p'-p)^\mu}{2m_N}
          G_P(t) \right] \gamma_5 \frac{\tau_i}{2} u(p),
\end{equation}
where $t = (p'-p)^2$, and $G_A(t)$ and $G_P(t)$ are the 
axial and induced pseudoscalar form factors, respectively.

   At lowest order, an external axial-vector field $a_\mu^i$ couples directly 
to the nucleon as 
\begin{equation}
\label{5:3:lnucleonavfc}
{\cal L}_{\rm ext}={\stackrel{\circ}{g}}_A\bar{\Psi}\gamma^\mu \gamma_5
\frac{\tau_i}{2}\Psi a_\mu^i+\cdots,
\end{equation}
which is obtained from ${\cal L}^{(1)}_{\pi N}$ through 
$u_\mu=(r_\mu-l_\mu)+\cdots = 2 a_\mu+\cdots$.
   The coupling to the pions is obtained from ${\cal L}_2$ with 
$r_\mu=-l_\mu=a_\mu$,
\begin{equation}
\label{5:3:lpionavfc}
{\cal L}_{\rm ext}=-F_0 \partial^\mu \phi_i a_\mu^i+\cdots,
\end{equation}
which gives rise to a diagram similar to Fig.\ 
\ref{5:3:fig:pionnucleonformfactor}, with  $m_q p_i$ replaced by $a^\mu_i$.

   The matrix element is thus given by
\begin{eqnarray*}
\bar{u}(p')\left\{{\stackrel{\circ}{g}}_A \gamma^\mu\gamma_5 \frac{\tau_i}{2}
+\left[-\frac{1}{2}\frac{{\stackrel{\circ}{g}}_A}{F_0}
(p'\hspace{-.70em}/\hspace{.45em}-p\hspace{-.45em}/)\gamma_5 \tau_i\right]
\frac{i}{q^2-M^2_\pi} (-iF_0 q^\mu)\right\}u(p),
\end{eqnarray*}
   from which we obtain, by applying the Dirac equation, 
\begin{eqnarray}
\label{5:3:ga}
   G_A(t) &=& {\stackrel{\circ}{g}}_A,\\
\label{5:3:gp}
   G_P(t) &=& - \frac{ 4 \stackrel{\circ}{m}^2_N {\stackrel{\circ}{g}}_A}{
              t-M_\pi^2}.
\end{eqnarray}
   At this order the axial form factor does not yet show a $t$ dependence.
   The axial-vector coupling constant is defined as $G_A(0)$ which is
simply given by ${\stackrel{\circ}{g}}_A$.
   We have thus identified the second new parameter of ${\cal L}^{(1)}_{\pi N}$
besides the nucleon mass $\stackrel{\circ}{m}_N$.
   The induced pseudoscalar form factor is determined by the pion exchange
which is the simplest version of the so-called pion-pole dominance.
   The $1/(t-M_\pi^2)$ behavior of $G_P$ is not in conflict with 
the book-keeping of a calculation
at chiral order ${\cal O}(p)$, because, according to Eq.\ 
(\ref{4:5:powercounting}), the external axial-vector field $a_\mu$ counts
as ${\cal O}(p)$, and the definition of the matrix element contains a
momentum $(p'-p)^\mu$ and the Dirac matrix $\gamma_5$ [see Eq.\ 
(\ref{5:2:powercounting})] so that the combined order of all elements 
is indeed ${\cal O}(p)$.

   It is straightforward to verify that the form factors of 
Eqs.\ (\ref{5:3:gt1}), (\ref{5:3:ga}), and (\ref{5:3:gp}) satisfy the relation 
\begin{equation}
\label{ff_relation}
   2m_N G_A(t) + \frac{t}{2m_N} G_P(t) =
       2\frac{M_\pi^2 F_\pi}{M_\pi^2 - t} G_{\pi N}(t),
\end{equation}
which is required by the Ward identity of Eq.\ (\ref{5:3:axwi}) with
the parameterizations of Eqs.\ (\ref{5:3:def_gt}) and 
(\ref{5:3:axial_current}) for the matrix elements.
   In other words, only two of the three form factors $G_A$, $G_P$, and 
$G_{\pi N}$ are independent.
   Note that this relation is not restricted
to small values of $t$ but holds for any $t$.

   In the chiral limit, Eq.\ (\ref{5:3:axwi}) implies
\begin{equation}
\label{5:3:gagpchirallimit}
2 \stackrel{\circ}{m}_N\, \stackrel{\circ}{G}_A(t)+\frac{t}{2 
\stackrel{\circ}{m}_N} \stackrel{\circ}{G}_P(t)=0,
\end{equation}
   which also follows from Eq.\ (\ref{5:3:ga}) and Eq.\ (\ref{5:3:gp}) for
$M_\pi^2=0$.
   Equation (\ref{5:3:gagpchirallimit}) for $\stackrel{\circ}{G}_A(0)\neq 0$
requires that in the chiral limit
the induced pseudoscalar form factor has a pole in the limit 
$t\to 0$.
   The interpretation of this pole is, of course, given in terms of the
exchange of a massless Goldstone pion.
   To understand this in more detail consider the most general contribution 
of the pion exchange to the axial-vector current matrix element:
\begin{eqnarray*}
\langle N(p')|A^\mu_i(0)|N(p)\rangle_\pi
&=& -\frac{2 F_{\pi}(t) g_{\pi N}(t)}{t-M_\pi^2-\tilde{\Sigma}(t)}
\bar{u}(p')q^\mu\gamma_5 \frac{\tau_i}{2}u(p),
\end{eqnarray*}
where $\tilde{\Sigma}(M^2_\pi)=\tilde{\Sigma}'(M^2_\pi)=0$ 
for the renormalized propagator [see Eq.\ (\ref{4:8:sigmaexp})].
   The functions $F_\pi(t)$ and $g_{\pi N}(t)$ denote the most general
parameterizations for the pion-decay vertex and the pion-nucleon vertex
(note that we have {\em not} specified the interpolating pion field).
   For general $t$ their values depend on the interpolating field,
but for $t=M_\pi^2$ they are identical with the pion-decay constant
$F_\pi$ and the pion-nucleon coupling constant $g_{\pi N}$, respectively.
   In the chiral limit, $M_\pi^2\to 0$, we obtain
\begin{displaymath}
 - \frac{2 F_0(t) \stackrel{\circ}{g}_{\pi N}(t)}{
t-\stackrel{\circ}{\tilde{\Sigma}}(t)}2\stackrel{\circ}{m}_N
\bar{u}(p')\frac{q^\mu}{2\stackrel{\circ}{m}_N}\gamma_5 \frac{\tau_i}{2}u(p),
\end{displaymath}
where $\stackrel{\circ}{\tilde{\Sigma}}(0)=
\stackrel{\circ}{\tilde{\Sigma}'}(0)=0$.
   In other words, the most general contribution of a massless pion to
the induced pseudoscalar form factor in the chiral limit is given by
\begin{displaymath}
\stackrel{\circ}{G}_{P,\pi}(t)=
-\frac{4\stackrel{\circ}{m}_N F_0(t)\stackrel{\circ}{g}_{\pi N}(t)}{
t-\stackrel{\circ}{\tilde{\Sigma}}(t)}.
\end{displaymath}
   We divide the pseudoscalar form factor into the pion contribution and
the rest.
   Making use of Eq.\ (\ref{5:3:gagpchirallimit}), we consider the limit 
\begin{eqnarray*}
\lim_{t\to 0}\, t[\stackrel{\circ}{G}_{P,\pi}(t)
+\stackrel{\circ}{G}_{P,R}(t)]
&=&
-4 \stackrel{\circ}{m}_N F_0 \stackrel{\circ}{g}_{\pi N}\\
&\stackrel{!}{=}&-4 \stackrel{\circ}{m}_N^2 \stackrel{\circ}{G}_A(0)
\end{eqnarray*}
   from which we obtain the Goldberger-Treiman relation 
\begin{displaymath}
\frac{\stackrel{\circ}{g}_A}{F_0}=
\frac{\stackrel{\circ}{g}_{\pi N}}{\stackrel{\circ}{m}_N}.
\end{displaymath}
   Of course, we have assumed that there is no other massless particle
in the theory which could produce a pole in the residual part 
$\stackrel{\circ}{G}_{P,R}(t)$ as $t\to 0$.

\subsection{Pion-Nucleon Scattering at Tree Level}
\label{subsec_apnstl}
   As another example, we will consider pion-nucleon scattering and show
how the effective Lagrangian of Eq.\ (\ref{5:2:l1pin}) reproduces the
Weinberg-Tomozawa predictions for the $s$-wave scattering lengths
\cite{Weinberg:1966kf,Tomozawa}.
   We will contrast the results with those of a tree-level calculation
within pseudoscalar (PS) and pseudovector (PV) pion-nucleon couplings.

   Before calculating the $\pi N$ scattering amplitude within ChPT
we introduce a general parameterization of the invariant amplitude 
${\cal M}=iT$ for the process $\pi^a(q)+N(p)\to\pi^b(q')+N(p')$ 
\cite{Chew:1957,Brown:1971pn}:\footnote{One also finds the parameterization
\cite{Becher:2001hv}
\begin{displaymath}
T=\bar{u}(p')\left(D-\frac{1}{4m_N}
[q\,'\hspace{-.9em}/\hspace{.2em},q\hspace{-.45em}/\hspace{.1em}]B\right)u(p)
\end{displaymath}
with $D=A+\nu B$,
where, for simplicity, we have omitted the isospin indices.}
\begin{eqnarray}
\label{5:3:mpinpar}
T^{ab}(p,q;p',q')&=&\bar{u}(p')\Bigg\{
\underbrace{\frac{1}{2}\{\tau^b,\tau^a\}}_{
\mbox{$\delta^{ab}$}}A^+(\nu,\nu_B)
+\underbrace{\frac{1}{2}[\tau^b,\tau^a]}_{
\mbox{$-i\epsilon_{abc}\tau^c$}}
A^-(\nu,\nu_B)\nonumber\\
&&+\frac{1}{2}(q\hspace{-.45em}/+q'\hspace{-.7em}/\hspace{.2em})\left[
\delta^{ab}B^+(\nu,\nu_B)
-i\epsilon_{abc}\tau^c B^-(\nu,\nu_B)\right]\Bigg\}
u(p),\nonumber\\
\end{eqnarray}
with the two independent scalar kinematical variables
\begin{eqnarray}
\label{5:3:nu}
\nu&=&\frac{s-u}{4m_N}=\frac{(p+p')\cdot q}{2m_N}
=\frac{(p+p')\cdot q'}{2m_N},\\
\label{5:3:nub}
\nu_B&=&-\frac{q\cdot q'}{2 m_N}=\frac{t-2M_\pi^2}{4 m_N},
\end{eqnarray}
where $s=(p+q)^2$, $t=(p'-p)^2$, and $u=(p'-q)^2$ are the usual
Mandelstam variables satisfying $s+t+u=2 m_N^2+ 2M_\pi^2$.
   From pion-crossing symmetry $T^{ab}(p,q;p',q')=
T^{ba}(p,-q';p',-q)$ we obtain for the crossing behavior of
the amplitudes
\begin{eqnarray*}
&&A^+(-\nu,\nu_B)=A^+(\nu,\nu_B),\quad
A^-(-\nu,\nu_B)=-A^-(\nu,\nu_B),\\
&&B^+(-\nu,\nu_B)=-B^+(\nu,\nu_B),\quad
B^-(-\nu,\nu_B)=B^-(\nu,\nu_B).
\end{eqnarray*}
   As in $\pi\pi$ scattering one often also finds the isospin 
decomposition as in Eq.\ (\ref{4:10:tdec}),
\begin{displaymath}
\langle I',{I'}\hspace{-.3em}_3|T|I,I_3\rangle=T^I \delta_{II'}
\delta_{I_3 {I'}\hspace{-.1em}_3},
\end{displaymath}
   where the relation between the two sets is given by \cite{Ericson:gk}
\begin{eqnarray}
\label{5:3:trel}
T^{\frac{1}{2}}&=&T^++2 T^-,\nonumber\\
T^{\frac{3}{2}}&=&T^+- T^-.
\end{eqnarray}

   Let us turn to the tree-level approximation to the $\pi N$ scattering
amplitude as obtained from ${\cal L}^{(1)}_{\pi N}$ of Eq.\ (\ref{5:2:l1pin}).
   In order to derive  the relevant interaction Lagrangians from 
Eq.\ (\ref{5:2:l1pin}), we reconsider the connection of Eq.\ 
(\ref{5:2:gamma}) with the
external fields set to zero and obtain
\begin{eqnarray}
\label{5:3:gammaentwi}
\Gamma_\mu&=&
\frac{i}{4F^2_0}\vec{\tau}\cdot\vec{\phi}\times\partial_\mu\vec{\phi}
+O(\phi^4).
\end{eqnarray}
   The linear pion-nucleon interaction term was already derived in Eq.\
(\ref{5:3:lpin1}) so that we end up with the following
interaction Lagrangian:
\begin{equation}
\label{5:3:lpin}
{\cal L}_{\rm int}=-\frac{1}{2}\frac{\stackrel{\circ}{g}_A}{F_0}
\bar{\Psi}\gamma^\mu\gamma_5
\tau^b\partial_\mu\phi^b\Psi
-\frac{1}{4F^2_0}\bar{\Psi}\gamma^\mu\underbrace{\vec{\tau}\cdot\vec{\phi}
\times\partial_\mu\vec{\phi}}_{\mbox{$\epsilon_{cde}\tau^c
\phi^d\partial_\mu\phi^e$}}\Psi.
\end{equation}
   The first term is the pseudovector pion-nucleon coupling and the second
the contact interaction with two factors of the pion field interacting
with the nucleon at a single point.
   The Feynman rules for the vertices derived from  Eq.\ (\ref{5:3:lpin}) read
\begin{itemize}
\item for an incoming pion with four-momentum $q$ and Cartesian isospin
index $a$: 
\begin{equation}
\label{5:3fr1}
-\frac{1}{2}\frac{\stackrel{\circ}{g}_A}{F_0} q\hspace{-.45em}/
\gamma_5 \tau^a,
\end{equation}
\item for an incoming pion with $q,a$ and an outgoing pion with $q',b$:
\begin{equation}
\label{5:3:fr2}
i\left(-\frac{1}{4F_0^2}\right)\gamma^\mu\epsilon_{cde}\tau^c\left(
\delta^{da}\delta^{eb}iq'_\mu+\delta^{db}\delta^{ea}(-iq)_\mu\right)
=\frac{q\hspace{-.45em}/ +q'\hspace{-.7em}/}{4 F^2_0}\epsilon_{abc}\tau^c.
\end{equation}
\end{itemize}
   The latter gives the contact contribution to ${\cal M}$,
\begin{equation}
\label{5:3:cont}
{\cal M}_{\rm cont}=\bar{u}(p')
\frac{q\hspace{-.45em}/+q'\hspace{-.7em}/\hspace{.2em}}{4 F^2_0}
\underbrace{\epsilon_{abc}\tau^c}_{\mbox{$i\frac{1}{2}[\tau^b,\tau^a]$}}
u(p)
=i \frac{1}{2 F^2_0}\bar{u}(p')\frac{1}{2}[\tau^b,\tau^a]
\frac{1}{2}(q\hspace{-.45em}/+q'\hspace{-.7em}/\hspace{.2em})u(p).
\end{equation}
   We emphasize that such a term is not present in a conventional calculation
with either a pseudoscalar or a pseudovector pion-nucleon interaction.
   For the $s$- and $u$-channel nucleon-pole diagrams the pseudovector
vertex appears twice and we obtain
\begin{eqnarray}
\label{5:3:sukanal}
{\cal M}_{s+u}&=&i\frac{\stackrel{\circ}{g}_A^2}{4 F_0^2}\bar{u}(p')
\tau^b\tau^a(-q'\hspace{-.7em}/\hspace{.2em})\gamma_5 \frac{1}{
p'\hspace{-.7em}/\hspace{.2em}+q'\hspace{-.7em}/\hspace{.2em}
-\stackrel{\circ}{m}_N}
q\hspace{-.45em}/\gamma_5u(p)\nonumber\\
&&+i\frac{\stackrel{\circ}{g}_A^2}{4 F_0^2}\bar{u}(p')
\tau^a\tau^b q\hspace{-.45em}/\gamma_5 
\frac{1}{p'\hspace{-.7em}/\hspace{.2em}-q\hspace{-.45em}/\hspace{.2em}
-\stackrel{\circ}{m}_N}
(-q'\hspace{-.7em}/\hspace{.2em})\gamma_5 u(p).
\end{eqnarray}
   The $s$- and $u$-channel contributions are related to each other
through pion crossing $a\leftrightarrow b$ and $q\leftrightarrow -q'$.
   In what follows we explicitly calculate only the $s$ channel and
make use of pion-crossing symmetry at the end to obtain the $u$-channel
result.
   Moreover, we perform the manipulations such that the result
of pseudoscalar coupling may also be read off.
   Using the Dirac equation, we rewrite 
$$q\hspace{-.45em}/\gamma_5 u(p)=(p'\hspace{-.7em}/\hspace{.2em}
+q'\hspace{-.7em}/\hspace{.2em}
-\stackrel{\circ}{m}_N
+\stackrel{\circ}{m}_N\!-p\hspace{-.45em}/)\gamma_5
u(p)=
(p'\hspace{-.7em}/\hspace{.2em}+q'\hspace{-.7em}/\hspace{.2em}
-\stackrel{\circ}{m}_N)\gamma_5
u(p)
+2\stackrel{\circ}{m}_N\gamma_5 u(p)
$$
and obtain
\begin{eqnarray*}
{\cal M}_s&=&i\frac{\stackrel{\circ}{g}_A^2}{4 F_0^2}\bar{u}(p')\tau^b\tau^a
(-q'\hspace{-.7em}/\hspace{.2em})\gamma_5 
\frac{1}{p'\hspace{-.7em}/\hspace{.2em}
+q'\hspace{-.7em}/\hspace{.2em}-\stackrel{\circ}{m}_N}
\left[(p'\hspace{-.7em}/\hspace{.2em}+q'\hspace{-.7em}/\hspace{.2em}
-\stackrel{\circ}{m}_N)
+2\stackrel{\circ}{m}_N\right]\gamma_5 u(p)\nonumber\\
&\stackrel{\gamma_5^2=1}{=}&
i\frac{\stackrel{\circ}{g}_A^2}{4 F_0^2}\bar{u}(p')\tau^b\tau^a
\left[(-q'\hspace{-.7em}/\hspace{.2em})
+(-q'\hspace{-.7em}/\hspace{.2em})\gamma_5
\frac{1}{p'\hspace{-.7em}/\hspace{.2em}+q'\hspace{-.7em}/\hspace{.2em}
-\stackrel{\circ}{m}_N}
2\stackrel{\circ}{m}_N\gamma_5 \right]u(p).
\end{eqnarray*}
   We repeat the above procedure
$$\bar{u}(p')q'\hspace{-.7em}/\hspace{.2em}\gamma_5
=\bar{u}(p')[-2
\stackrel{\circ}{m}_N\!\gamma_5 -\gamma_5(p\hspace{-.45em}/+q\hspace{-.45em}/
-\stackrel{\circ}{m}_N)],
$$
yielding 
\begin{equation}
\label{5:2ms1}
{\cal M}_s=i\frac{\stackrel{\circ}{g}_A^2}{4 F_0^2}\bar{u}(p')\tau^b\tau^a
[(-q'\hspace{-.7em}/\hspace{.2em})
+\underbrace{4m_N^2\gamma_5
\frac{1}{p'\hspace{-.7em}/\hspace{.2em}+q'\hspace{-.7em}/\hspace{.2em}
-\stackrel{\circ}{m}_N}
\gamma_5}_{\mbox{PS coupling}}
+2\stackrel{\circ}{m}_N]u(p),
\end{equation}
   where, for the identification of the PS-coupling result, one has to
make use of the Goldberger-Treiman relation
\cite{Goldberger:1958tr,Goldberger:1958vp,Nambu:xd} 
(see Sec.\ \ref{subsec_gtravcme})
$$\frac{\stackrel{\circ}{g}_A}{F_0}=\frac{\stackrel{\circ}{g}_{\pi N}
}{\stackrel{\circ}{m}_N},$$
where $\stackrel{\circ}{g}_{\pi N}$ denotes the pion-nucleon coupling
constant in the chiral limit. 
   Using 
$$s-m_N^2=2m_N(\nu-\nu_B),
$$
we find
\begin{eqnarray*}
\bar{u}(p')\gamma_5
\frac{1}{p'\hspace{-.7em}/\hspace{.2em}+q'\hspace{-.7em}/\hspace{.2em}-
\stackrel{\circ}{m}_N}
\gamma_5 u(p)&=&
\bar{u}(p')\gamma_5
\frac{p'\hspace{-.7em}/\hspace{.2em}+q'\hspace{-.7em}/\hspace{.2em}+
\stackrel{\circ}{m}_N}{
(p'+q')^2-\stackrel{\circ}{m}_N^2}\gamma_5 u(p)\nonumber\\
&=&\frac{1}{2\stackrel{\circ}{m}_N(\nu-\nu_B)}\left[-\frac{1}{2}\bar{u}(p')
(q\hspace{-.45em}/+q'\hspace{-.7em}/\hspace{.2em})u(p)\right],
\end{eqnarray*}
where we again made use of the Dirac equation.
   We finally obtain for the $s$-channel contribution
\begin{equation}
\label{5:3:msf}
{\cal M}_{s}=i\frac{\stackrel{\circ}{g}_A^2}{4 F_0^2}
\bar{u}(p')\tau^b\tau^a\left[
2\stackrel{\circ}{m}_N
+\frac{1}{2}(q\hspace{-.45em}/+q'\hspace{-.7em}/\hspace{.2em})\left(
-1-\frac{2\stackrel{\circ}{m}_N}{\nu-\nu_B}\right)\right]u(p).
\end{equation}
   As noted above, the expression for the $u$ channel results from
the substitution $a\leftrightarrow b$ and $q\leftrightarrow -q'$ 
\begin{equation}
\label{5:3:ukanal}
{\cal M}_{u}=i\frac{\stackrel{\circ}{g}_A^2}{4 F_0^2}
\bar{u}(p')\tau^a\tau^b\left[
2\stackrel{\circ}{m}_N
+\frac{1}{2}(q\hspace{-.45em}/+q'\hspace{-.7em}/\hspace{.2em})\left(
1-\frac{2\stackrel{\circ}{m}_N}{\nu+\nu_B}\right)\right]u(p).
\end{equation}
   We combine the $s$- and $u$-channel contributions using
$$\tau^b\tau^a=\frac{1}{2}\{\tau^b,\tau^a\}+\frac{1}{2}[\tau^b,\tau^a],
\quad
\tau^a\tau^b=\frac{1}{2}\{\tau^b,\tau^a\}-\frac{1}{2}[\tau^b,\tau^a],
$$
and
$$
\frac{1}{\nu-\nu_B}\pm\frac{1}{\nu+\nu_B}=
\frac{\left\{\begin{array}{c}2\nu\\ 2\nu_B\end{array}\right\}}{
\nu^2-\nu_B^2}
$$
and summarize the contributions to the functions $A^\pm$ and
$B^\pm$ of Eq.\ (\ref{5:3:mpinpar}) in Table \ref{5:3:tableresults} 
[see also Eq.\ (A.26) of Ref.\ \cite{Gasser:1987rb}].

\begin{table}[htb]
\begin{center}
\caption{\label{5:3:tableresults} Tree-level contributions to the functions
$A^\pm$ and $B^\pm$ of Eq.\ (\ref{5:3:mpinpar}). 
   The second column (PS) denotes the result using pseudoscalar pion-nucleon 
coupling (using the Goldberger-Treiman relation). 
   The sum of the second and third column (PS+$\Delta$PV) represents the 
result of  pseudovector pion-nucleon coupling.
   The contact term is specific to the chiral approach.
The last column, the sum of the second, third, and fourth columns,  
is the lowest-order ChPT result.} 
\vspace{1em}
\begin{tabular}{|l|c|c|c|c|}
\hline
amplitude$\backslash$origin&PS&$\Delta$PV&contact&sum\\
\hline
$A^+$&0&$\frac{\stackrel{\circ}{g}_A^2 \stackrel{\circ}{m}_N}{F^2_0}$&0&
$\frac{\stackrel{\circ}{g}_A^2 \stackrel{\circ}{m}_N}{F^2_0}$\\
\hline
$A^-$&0&0&0&0\\
\hline
$B^+$&$-\frac{\stackrel{\circ}{g}_A^2}{F^2_0}
\frac{\stackrel{\circ}{m}_N\nu}{\nu^2-\nu_B^2}$&0&0&
$-\frac{\stackrel{\circ}{g}_A^2}{F^2_0}
\frac{\stackrel{\circ}{m}_N\nu}{\nu^2-\nu_B^2}$\\
\hline
$B^-$&
$-\frac{\stackrel{\circ}{g}_A^2}{F^2_0}
\frac{\stackrel{\circ}{m}_N\nu_B}{\nu^2-\nu_B^2}$&
$-\frac{\stackrel{\circ}{g}_A^2}{2F^2_0}$&
$\frac{1}{2F^2_0}$&
$\frac{1-\stackrel{\circ}{g}_A^2}{2F^2_0}
-\frac{\stackrel{\circ}{g}_A^2}{F^2_0}
\frac{\stackrel{\circ}{m}_N\nu_B}{\nu^2-\nu_B^2}$
\\
\hline
\end{tabular}
\end{center}
\end{table}

   In order to extract the scattering lengths, let us consider threshold
kinematics
\begin{equation}
\label{5:3:schkin}
p^\mu=p'^\mu=(m_N,0),\quad
q^\mu=q'^\mu=(M_\pi,0),\quad
\nu|_{\rm thr}=M_\pi,\quad
\nu_B|_{\rm thr}=-\frac{M_\pi^2}{2m_N}.
\end{equation}
   Since we only work at lowest-order tree level, we replace 
$\stackrel{\circ}{m}_N\to m_N$, etc.
   Together with\footnote{Recall that we use the normalization 
$\bar{u}u=2 m_N$.}
$$u(p)\to \sqrt{2 m_N}\left(\begin{array}{c}\chi\\ 0\end{array}\right),\quad
\bar{u}(p')\to \sqrt{2 m_N}\left(\chi'^\dagger\,\, 0\right)
$$
we find for the threshold matrix element
\begin{equation}
\label{5:3:mthr}
T|_{\rm thr}=2 m_N 
\chi'^\dagger\left[\delta^{ab}\left(A^++M_\pi B^+
\right)-i\epsilon_{abc}\tau^c\left(A^-+M_\pi B^-\right)\right]_{\rm 
thr}
\chi.
\end{equation}
   Using
$$\left[\nu^2-\nu_B^2\right]_{\rm thr}
=M^2_\pi\left(1-\frac{\mu^2}{4}\right),\quad
\mu=\frac{M_\pi}{m_N}\approx \frac{1}{7}
$$   
we obtain
\begin{eqnarray}
\label{5:3:mthrres}
T|_{\rm thr}&=& 2m_N\chi'^\dagger\Bigg[\delta^{ab}
\underbrace{\Bigg(
\frac{g^2_A m_N}{F_\pi^2}+\underbrace{
M_\pi\left(-\frac{g^2_A}{F_\pi^2}\right)
\frac{m_N}{M_\pi}\frac{1}{1-\frac{\mu^2}{4}}}_{\mbox{PS}}\Bigg)
 }_{\mbox{ChPT = PV}}
\nonumber\\
&&-i\epsilon_{abc}\tau^c M_\pi\underbrace{\Bigg(
\frac{1}{2F_\pi^2}
\underbrace{-\frac{g^2_A}{2F_\pi^2}
\underbrace{-\frac{g^2_A}{F_\pi^2}\left(-\frac{1}{2}\right)
\frac{1}{1-\frac{\mu^2}{4}}}_{\mbox{PS}}}_{\mbox{PV}}
\Bigg)}_{\mbox{ChPT}}\Bigg]\chi,
\end{eqnarray}
  where we have indicated the results for the various coupling schemes.

   Let us discuss the $s$-wave scattering lengths resulting from
Eq.\ (\ref{5:3:mthrres}). 
   Using the above normalization for the Dirac spinors, the
differential cross section in the center-of-mass frame is given
by \cite{Ericson:gk}
\begin{equation}
\label{5:3:dscm}
\frac{d\sigma}{d\Omega}=\frac{|\vec{q}\,'|}{|\vec{q}\,|}
\left(\frac{1}{8\pi \sqrt{s}}\right)^2 |T|^2,
\end{equation}
which, at threshold, reduces to
\begin{equation}
\label{5:3:dsthr}
\left.\frac{d\sigma}{d\Omega}\right|_{\rm thr}
=\left(\frac{1}{8\pi(m_N+M_\pi)}\right)^2
|T|^2\stackrel{!}{=}|a|^2.
\end{equation}
   The $s$-wave scattering lengths are defined as\footnote{
The threshold parameters are defined in terms of a multipole expansion
of the $\pi N$ scattering amplitude \cite{Chew:1957}. 
   The sign convention for the $s$-wave scattering parameters $a_{0+}^{(\pm)}$
is opposite to the convention of the effective range expansion.}
\begin{equation}
\label{5:3:apm}
a^\pm_{0+}=
\frac{1}{8\pi(m_N+M_\pi)}T^{\pm}|_{\rm thr} 
=\frac{1}{4\pi(1+\mu)}\left[A^\pm+M_\pi B^\pm\right]_{\rm thr}.
\end{equation}
   The subscript $0+$ refers to the fact that the $\pi N$ system is in
an orbital $s$ wave ($l=0$) with total angular momentum $1/2=0+1/2$.
   Inserting the results of Table \ref{5:3:tableresults} we obtain\footnote{We
do not expand the fraction $1/(1+\mu)$, because the $\mu$ dependence is not of
dynamical origin.}
\begin{eqnarray}
\label{5:3:aminus}
a^-_{0+}&=&\frac{M_\pi}{8\pi(1+\mu)F_\pi^2}\left(1+\frac{g_A^2\mu^2}{4}
\frac{1}{1-\frac{\mu^2}{4}}\right)
=\frac{M_\pi}{8\pi(1+\mu)F_\pi^2}[1+{\cal O}(p^2)],\nonumber\\
&&\\
\label{5:3:aplus}
a^+_{0+}&=&
-\frac{g_A^2 M_\pi}{16\pi(1+\mu)F_\pi^2}\frac{\mu}{1-\frac{\mu^2}{4}}
= {\cal O}(p^2),
\end{eqnarray}
   where we have also indicated the chiral order.
   Taking the linear combinations $a^\frac{1}{2}=a^+_{0+}+2 a^-_{0+}$
and $a^\frac{3}{2}=a^+_{0+}-a^-_{0+}$ [see Eq.\ (\ref{5:3:trel})], 
we see that the results of Eqs.\
(\ref{5:3:aminus}) and (\ref{5:3:aplus}) indeed satisfy the Weinberg-Tomozawa
relation \cite{Weinberg:1966kf,Tomozawa}:\footnote{The result, in principle,
holds for a general target of isospin $T$ (except for the pion) after 
replacing 3/4 by $T(T+1)$ and $\mu$ by $M_\pi/M_T$.}
\begin{equation}
\label{5:3:weinbergtomozawa}
a^I=-\frac{M_\pi}{8\pi (1+\mu)F_\pi^2}[I(I+1)-\frac{3}{4}-2].
\end{equation}
   As in $\pi\pi$ scattering, the scattering lengths vanish in the chiral
limit reflecting the fact that the interaction of Goldstone bosons 
vanishes in the zero-energy limit.
   The pseudoscalar pion-nucleon interaction produces a scattering length
$a^+_{0+}$ proportional to $m_N$ instead of $\mu M_\pi$ and is clearly in
conflict with the requirements of chiral symmetry.
   Moreover, the scattering length $a^-_{0+}$ of the pseudoscalar coupling
is too large by a factor $g_A^2$ in comparison
with the two-pion contact term of Eq.\ (\ref{5:3:cont}) (sometimes also
referred to as the Weinberg-Tomozawa term) induced by the 
nonlinear realization of chiral symmetry.
   On the other hand, the pseudovector pion-nucleon interaction 
gives a totally wrong result for $a^-_{0+}$, because it misses 
the two-pion contact term of Eq.\ (\ref{5:3:cont}).

   Using the values
\begin{eqnarray}
\label{5:3:par}
&&g_A=1.267,\quad F_\pi=92.4\,\mbox{MeV},\nonumber\\
&&
m_N=m_p=938.3\,\mbox{MeV},\quad 
M_\pi=M_{\pi^+}=139.6\,\mbox{MeV},
\end{eqnarray}
   the numerical results for the scattering lengths are given in 
Table \ref{5:3:tecomp}.
   We have included the full results of Eqs.\ (\ref{5:3:aminus})
and (\ref{5:3:aplus}) and the consistent corresponding prediction
at ${\cal O}(p)$.
   The results of heavy-baryon chiral perturbation theory 
(HBChPT) (see Sec.\ \ref{sec_hbf}) are taken from Ref.\ \cite{Mojzis:1997tu}.
   At ${\cal O}(p^3)$ the calculation involves nine low-energy constants
of the chiral Lagrangian which have been fit to the extrapolated
threshold parameters of the partial wave analysis of Ref.\ \cite{Koch:ay},
the pion-nucleon $\sigma$ term and the Goldberger-Treiman discrepancy.
   Up to and including ${\cal O}(p^4)$ the HBChPT calculation contains 14 free 
parameters \cite{Fettes:2000xg}. 
   In Ref.\ \cite{Fettes:2000xg} the complete one-loop amplitude at
${\cal O}(p^4)$ was fit to the phase shifts provided by three different 
partial wave analyses \cite{Koch:bn} [I], \cite{Matsinos:1997pb} [II],
and SP98 of \cite{SAID} [III].
   Table \ref{5:3:tecomp} includes the results for the $s$-wave scattering
lengths obtained from those fits in combination with the empirical values
of the three analyses.
   Finally, the results of the recently proposed manifestly Lorentz-invariant
form of baryon ChPT [R(elativistic)BChPT] \cite{Becher:1999he} 
(see Sec.\ \ref{sec_mir}) are included up
to ${\cal O}(p^4)$ \cite{Becher:2001hv}.
   The first entries (a) refer to a dispersive representation
of the function $D=A+\nu B$ entering the threshold matrix element
[see Eq.\ (\ref{5:3:apm}) and recall $\nu_{\rm thr}=M_\pi$] 
whereas the second entries (b) involve only the one-loop approximation.
   Whereas for $a_{0+}^-$ there is no difference, the value for
$a_{0+}^+$ differs substantially which has been interpreted as 
the result of an insufficient approximation of the one-loop representation
to allow for an extrapolation from the Cheng-Dashen point 
[$(\nu=0,\nu_B=0)$] to the physical region \cite{Becher:2001hv}.
    
   The empirical results quoted have been taken from 
low-energy partial-wave analyses \cite{Koch:bn,Matsinos:1997pb}
and recent precision X-ray
experiments on pionic hydrogen and deuterium \cite{Schroder:rc}.

\begin{table}[htb]
\begin{center}
\caption{\label{5:3:tecomp} $s$-wave scattering lengths $a_{0+}^\pm$.}
\vspace{1em}
\begin{tabular}{|l|c|c|}
\hline
Scattering length &$a^+_{0+}$ [MeV$^{-1}$]&$a^-_{0+}$ [MeV$^{-1}$]\\
\hline
Tree-level result &$-6.80\times 10^{-5}$ & $+5.71\times 10^{-4}$ \\
\hline
ChPT ${\cal O}(p)$ & $0$ & $+5.66\times 10^{-4}$ \\
\hline
HBChPT ${\cal O}(p^2)$ \cite{Mojzis:1997tu} & $-1.3\times 10^{-4}$
&$+5.5\times 10^{-4}$ \\
\hline
HBChPT ${\cal O}(p^3)$ \cite{Mojzis:1997tu}& $(-7\pm 9)\times 10^{-5}$ & 
$(+6.7\pm1.0)\times 10^{-4}$\\
\hline
HBChPT ${\cal O}(p^4)$  [I] \cite{Fettes:2000xg} & $-6.9\times 10^{-5}$ &
$+6.47\times 10^{-4}$\\
\hline
HBChPT ${\cal O}(p^4)$ [II] \cite{Fettes:2000xg} & $+3.2\times 10^{-5}$ &
$+5.52\times 10^{-4}$\\
\hline
HBChPT ${\cal O}(p^4)$ [III] \cite{Fettes:2000xg} & $+1.9\times 10^{-5}$ &
$+6.21\times 10^{-4}$\\
\hline
RChPT ${\cal O}(p^4)$ (a) \cite{Becher:2001hv} 
&$-6.0\times 10^{-5}$ & $+6.55 \times 10^{-4}$\\
\hline
RChPT ${\cal O}(p^4)$ (b) \cite{Becher:2001hv}
&$-9.4\times 10^{-5}$ & $+6.55 \times 10^{-4}$\\
\hline
PS & $-1.23 \times 10^{-2}$ & $+9.14 \times 10^{-4}$ \\
\hline
PV &$-6.80\times 10^{-5}$ & $+5.06 \times 10^{-6}$\\
\hline
\hline
Empirical values \cite{Koch:bn} & $ (-7\pm 1)\times 10^{-5}$ & 
$(6.6 \pm 0.1) \times
10^{-4}$\\
\hline
Empirical values \cite{Matsinos:1997pb} & $(2.04\pm 1.17) \times 10^{-5}$ & 
$(5.71\pm 0.12) \times 10^{-4}$\\
&& $(5.92\pm 0.11)  \times 10^{-4}$\\
\hline 
Experiment \cite{Schroder:rc} & $(-2.7\pm 3.6)\times 10^{-5}$  &
$(+6.59\pm0.30)\times 10^{-4}$
 \\
\hline
\end{tabular}
\end{center}
\end{table}

\section{Examples of Loop Diagrams}
\label{sec_eld}
   In Sec.\ \ref{sec_elwpcs} we saw that, in the purely mesonic sector, 
contributions of $n$-loop diagrams are at least of order ${\cal O}(p^{2n+2})$,
i.e., they are suppressed by $p^{2n}$ in comparison with tree-level diagrams.
   An important ingredient in deriving this result was the fact that we
treated the squared pion mass as a small quantity of order $p^2$.
   Such an approach is motivated by the observation that the masses of the
Goldstone bosons must vanish in the chiral limit.
   In the framework of ordinary chiral perturbation theory $M_\pi^2\sim m_q$
[see Eq.\ (\ref{4:3:mpi2}) and the discussion at the end of Sec.\ 
\ref{subsec_eppsop6}] which translates into a momentum expansion of observables
at fixed ratio $m_q/p^2$.
   On the other hand, there is no reason to believe that the masses of 
hadrons other than the Goldstone bosons should vanish or become small in the 
chiral limit. 
   In other words, the nucleon mass entering the pion-nucleon Lagrangian
of Eq.\ (\ref{5:2:l1pin}) should---as already anticipated in the discussion
following Eq.\ (\ref{5:2:l1pin})---not be treated as a small quantity of,
say, order ${\cal O}(p)$.

   Naturally the question arises how all this affects the calculation of
loop diagrams and the setup of a consistent power counting scheme.
   We will follow  Ref.\ \cite{Gasser:1987rb} and consider, for illustrative 
purposes, two examples: a one-loop contribution to the
nucleon mass and a loop diagram contributing to $\pi N$ scattering.

\subsection{First Example: One-Loop Correction to the Nucleon Mass}
\label{subsec_feolcnm}
\begin{figure}[htb]
\begin{center} 
\caption{\label{5:4:fig:nsepl}
Example of a pion-loop contribution to the nucleon self energy.}
\vspace{2em}
\epsfig{file=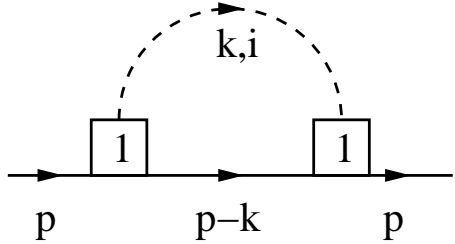,width=6cm}
\end{center}
\end{figure}

   The discussion of the modification of the nucleon mass due to pion
loops is very similar to that of Sec.\ \ref{subsec_mgb} for the masses
of the Goldstone bosons.
   The lowest-order Feynman propagator of the nucleon, corresponding
to the free-field part of ${\cal L}_{\pi N}^{(1)}$ of Eq.\ (\ref{5:2:l1pin}),
\begin{equation}
\label{5:4:sf}
iS_F(p)=\frac{i}{p\hspace{-.45em}/ -\stackrel{\circ}{m}_N+i0^+},
\end{equation}
is modified by the self energy $\Sigma(p)$ (see for example the 
one-loop contribution of Fig.\ \ref{5:4:fig:nsepl}) in a way
analogous to the modification of the meson propagator in 
Eq.\ (\ref{4:8:prop1}), 
\begin{displaymath}
\frac{i}{p\hspace{-.45em}/ -\stackrel{\circ}{m}_N+i0^+}
+
\frac{i}{p\hspace{-.45em}/ -\stackrel{\circ}{m}_N+i0^+}
[-i\Sigma(p)] \frac{i}{p\hspace{-.45em}/ -\stackrel{\circ}{m}_N+i0^+}
+\cdots,
\end{displaymath}
resulting in the full (but still unrenormalized) propagator 
\begin{equation}
\label{5:4:fullurprop}
iS(p)=\frac{i}{p\hspace{-.45em}/ -\stackrel{\circ}{m}_N-\Sigma(p)+i0^+}.
\end{equation}
   In the absence of external fields (but including the quark mass term),
the most general expression for the self energy can be written as
\begin{equation}
\label{5:4:seansatz}
\Sigma(p)=-f(p^2)p\hspace{-.45em}/+g(p^2)  \stackrel{\circ}{m}_N,
\end{equation}
where $f$ and $g$ are as yet undetermined functions of the invariant $p^2$.
   We assume that $f$ and $g$ may be determined in a perturbative 
(momentum or loop) expansion which, symbolically, we denote by some indicator 
$\lambda$, 
\begin{eqnarray}
\label{5:4:fgexp}
f(p^2,\lambda)=f_0(p^2)+\lambda f_1(p^2)+\lambda^2 f_2(p^2)+\cdots,\nonumber\\
g(p^2,\lambda)=g_0(p^2)+\lambda g_1(p^2)+\lambda^2 g_2(p^2)+\cdots.
\end{eqnarray}
   When switching off the interaction, we would like to recover the 
lowest-order result of Eq.\ (\ref{5:4:sf}), i.e.\ $\Sigma\to 0$,
implying $f_0=g_0=0$.
   The mass of the nucleon is defined through the position of the pole of
the full propagator, i.e., for $p\hspace{-.45em}/=m_N$ we require
\begin{displaymath}
m_N-\stackrel{\circ}{m}_N+f(m_N^2)m_N
-g(m_N^2)\stackrel{\circ}{m}_N=0,
\end{displaymath}
from which we obtain
\begin{equation}
\label{5:4:mn}
m_N=\stackrel{\circ}{m}_N\frac{1+g(m_N^2)}{1+f(m_N^2)}.
\end{equation}
   The perturbative result to first order in $\lambda$ reads
\begin{equation}
\label{5:4:mnpert}
m_N=\stackrel{\circ}{m}_N\frac{1+\lambda g_1(\stackrel{\circ}{m}_N^2)+\cdots}{
1+\lambda f_1(\stackrel{\circ}{m}_N^2)+\cdots}
= \stackrel{\circ}{m}_N\left\{1+\lambda\left[g_1(\stackrel{\circ}{m}_N^2)
-f_1(\stackrel{\circ}{m}_N^2)\right]+\cdots\right\}.
\end{equation}
   [Note that the argument $m_N^2$ of the functions $f$ and $g$ also has to 
be expanded in powers of $\lambda$, $m_N^2=\,\,\stackrel{\circ}{m}^2_N
+\,\, O(\lambda)$.]
   The wave function renormalization constant is defined through the residue
at $p\hspace{-.45em}/=m_N$,
\begin{equation}
\label{5:4:zndef}
S(p)\to \frac{Z_N}{p\hspace{-.45em}/ -m_N+i0^+}\,\,
\mbox{for}\,\, p\hspace{-.45em}/\to m_N,
\end{equation}
i.e., the renormalized propagator, defined through $S(p)=Z_N S_R(p)$, 
has a pole at $p\hspace{-.45em}/=m_N$ with residue 1.
   Using $(p^2-m_N^2)^n=(p\hspace{-.45em}/-m_N)^n(p\hspace{-.45em}/+m_N)^n$
and Eq.\ (\ref{5:4:mn}) we find that for $p\hspace{-.45em}/\to m_N$
\begin{eqnarray*}
S(p)&=&\frac{1}{p\hspace{-.45em}/[1+f(p^2)]-\stackrel{\circ}{m}_N[1+g(p^2)]}\\
&=&\left\{\vphantom{\stackrel{\circ}{m}_N}
p\hspace{-.45em}/[1+f(m_N^2)+(p\hspace{-.45em}/-m_N)
(p\hspace{-.45em}/+m_N)f'(m_N^2)+\cdots]\right.\\
&&\left.
-\stackrel{\circ}{m}_N
[1+g(m_N^2)+(p\hspace{-.45em}/-m_N)
(p\hspace{-.45em}/+m_N)g'(m_N^2)+\cdots]\right\}^{-1}
\\
&\to&\frac{1}{(p\hspace{-.45em}/-m_N)[1+f(m_N^2)+2 m_N^2 f'(m_N^2)
-2\stackrel{\circ}{m}_N m_N g'(m_N^2)]},
\end{eqnarray*}
yielding for the wave function renormalization constant
\begin{eqnarray}
\label{5:4:wdrc}
Z_N&=&
\frac{1}{1+f(m_N^2)+2m_N^2 f'(m_N^2)-2\stackrel{\circ}{m}_N m_N g'(m_N^2)}
\nonumber\\
&=&1-\lambda\left\{f_1(\stackrel{\circ}{m}_N^2)
+2\stackrel{\circ}{m}_N^2[f'_1(\stackrel{\circ}{m}_N^2)-
g'_1(\stackrel{\circ}{m}_N^2)]\right\}
+\cdots.
\end{eqnarray}

   With these definitions let us consider the contribution of Fig.\ 
\ref{5:4:fig:nsepl} to the self energy, where, for the sake of simplicity,
we perform the calculation in the chiral limit $M_\pi^2=0$.
   Using the vertex of Eq.\ (\ref{5:3:pionnucleonvertex}) we obtain the
contribution of the self energy
\begin{eqnarray}
\label{5:4:sel1}
-i\stackrel{\circ}{\Sigma}_{\rm loop}(p)&=&
\int \frac{d^4 k}{(2\pi)^4}\left[-\frac{1}{2} \frac{\stackrel{\circ}{g}_A}{F_0}
(-k\hspace{-.45em}/)\gamma_5 \tau_i\right]\frac{i}{k^2+i0^+}\nonumber\\
&&\times \frac{i}{p\hspace{-.45em}/-k\hspace{-.45em}/-
\stackrel{\circ}{m}_N+i0^+}
 \left[-\frac{1}{2} \frac{\stackrel{\circ}{g}_A}{F_0}
k\hspace{-.45em}/\gamma_5 \tau_i\right].
\end{eqnarray}  
    Counting powers we see that the integral has a cubic divergence.
We make use of (normal) dimensional regularization \cite{Jegerlehner:2000dz}, 
where the integrand is first simplified using\footnote{For a recent discussion 
of the problem with $\gamma_5$ in dimensional regularization, see 
Ref.\ \cite{Jegerlehner:2000dz}.
   Since we are neither dealing with matrix elements containing anomalies
nor considering closed fermion loops, we can safely make use of normal
dimensional regularization \cite{Gasser:1987rb,Jegerlehner:2000dz}.}
\begin{equation}
\label{5:4:gammarelations} 
\{\gamma_\mu,\gamma_\nu\}=2 g_{\mu\nu},\quad
\quad {g_\mu}^\mu=n,\quad
\{\gamma_\mu,\gamma_5\}=0, \quad
\gamma_5^2=1.
\end{equation}   
   In the standard fashion, we first insert
\begin{displaymath}
1=\frac{p\hspace{-.45em}/-k\hspace{-.45em}/+ \stackrel{\circ}{m}_N-i0^+}{
p\hspace{-.45em}/-k\hspace{-.45em}/+ \stackrel{\circ}{m}_N-i0^+},
\end{displaymath}
simplify the numerator using Eq.\ (\ref{5:4:gammarelations}),
\begin{displaymath}
k\hspace{-.45em}/\gamma_5 
(p\hspace{-.45em}/-k\hspace{-.45em}/+ \stackrel{\circ}{m}_N)
k\hspace{-.45em}/\gamma_5 
=-(p\hspace{-.45em}/+\stackrel{\circ}{m}_N)k^2
+(p^2-\stackrel{\circ}{m}_N^2)k\hspace{-.45em}/
-[(k-p)^2-\stackrel{\circ}{m}_N^2]k\hspace{-.45em}/,
\end{displaymath}
and obtain, with $\tau_i\tau_i=3$
\begin{eqnarray}
\label{5:4:sigmal}
\stackrel{\circ}{\Sigma}_{\rm loop}(p)&=&
\frac{3 \stackrel{\circ}{g}_A^2}{4 F_0^2}
\left\{-(p\hspace{-.45em}/+\stackrel{\circ}{m}_N)
\mu^{4-n}\int \frac{d^n k}{(2\pi)^n}\frac{i}{(k-p)^2-\stackrel{\circ}{m}_N^2
+i0^+}\right.\nonumber\\
&&+(p^2-\stackrel{\circ}{m}_N^2)\mu^{4-n}\int \frac{d^n k}{(2\pi)^n}
\frac{i k\hspace{-.45em}/}{(k^2+i0^+)[(k-p)^2-\stackrel{\circ}{m}_N^2
+i0^+]}\nonumber\\
&&\left.-\mu^{4-n}\int \frac{d^n k}{(2\pi)^n}
\frac{i  k\hspace{-.45em}/}{k^2+i0^+}\right\}.
\end{eqnarray}
   Indeed, when discussing the contribution to the nucleon mass [see
Eq.\ (\ref{5:4:mnpert})] we only need 
to consider the first integral of Eq.\ (\ref{5:4:sigmal}), because the
second term does not contribute at $p^2=\stackrel{\circ}{m}_N^2$ and
the third term vanishes in dimensional regularization because the
integrand is odd in $k$.
   Using Eqs.\ (\ref{app:ipi}) and (\ref{app:ipib}) of Appendix
\ref{app_subsec_ipi} with the replacement $M_\pi\to\, \stackrel{\circ}{m}_N$ 
we obtain, in the language of Eq.\ (\ref{5:4:fgexp}),
\begin{displaymath}
\lambda f_1(\stackrel{\circ}{m}_N^2)=
\frac{3 \stackrel{\circ}{g}_A^2}{4 F_0^2} I_N(0),\quad
\lambda g_1(\stackrel{\circ}{m}_N^2)=
-\frac{3 \stackrel{\circ}{g}_A^2}{4 F_0^2} I_N(0).
\end{displaymath}
   Applying Eq.\ (\ref{5:4:mnpert}) we find for the nucleon mass including
the one-loop contribution of Fig.\ \ref{5:4:fig:nsepl} [see Eq.\
(4.1) of Ref.\ \cite{Gasser:1987rb}]
\begin{equation}
\label{5:4:mnil}
m_N=\stackrel{\circ}{m}_N\left[1-\frac{3 \stackrel{\circ}{g}_A^2}{2 F_0^2} 
I_N(0)\right],
\end{equation}
where
\begin{eqnarray*}
I_N(0)&=&\frac{\stackrel{\circ}{m}_N^2}{16\pi^2}\left[R+
\ln\left(\frac{\stackrel{\circ}{m}_N^2}{\mu^2}\right)
\right]+O(n-4),\\
R&=&\frac{2}{n-4}-[\ln(4\pi)+\Gamma'(1)+1].
\end{eqnarray*}

   The pion loop of Fig.\ \ref{5:4:fig:nsepl} generates an (infinite)
contribution to the nucleon mass, even in the chiral limit, i.e., the
parameter $\stackrel{\circ}{m}_N$ of ${\cal L}_{\pi N}^{(1)}$ needs
to be renormalized. 
   The same is true for the second parameter $\stackrel{\circ}{g}_A$
\cite{Gasser:1987rb}.
   This situation is completely different from the mesonic sector, where
the two parameters $F_0$ and $B_0$ of the lowest-order 
Lagrangian do not change due to higher-order corrections
in the chiral limit.
   For example, in the SU(2)$\times$SU(2) sector, the pion-decay constant at 
${\cal O}(p^4)$ is given by [see Eq.\ (12.2) of Ref.\ \cite{Gasser:1983yg}]
\begin{equation}
\label{5:4:fpi}
F_\pi=F_0\left[1+\frac{M^2}{16 \pi^2 F_0^2}\bar{l}_4+{\cal O}(M^4)\right],
\end{equation}
where $M^2=2 B_0 m_q$, and the scale-independent low-energy parameter
$\bar{l}_4$ is defined in Eq.\ (\ref{app:barli}).
   Since $F_\pi\to F_0$ in the chiral limit $M^2\to 0$, the pion-decay 
constant in the chiral limit is still given by $F_0$ of ${\cal L}_2$.
   Similarly, in the chiral limit the Goldstone boson masses vanish, 
   not only at ${\cal O}(p^2)$ but also at higher orders,
as we have seen in Eqs.\ (\ref{4:8:mpi24}) - (\ref{4:8:meta24}).

\subsection{Second Example: One-Loop Correction to $\pi N$ Scattering}
   
\begin{figure}[htb]
\begin{center} 
\caption{\label{5:4:schleife}
Example of a loop diagram contributing to pion-nucleon scattering.}
\vspace{1em}
\epsfig{file=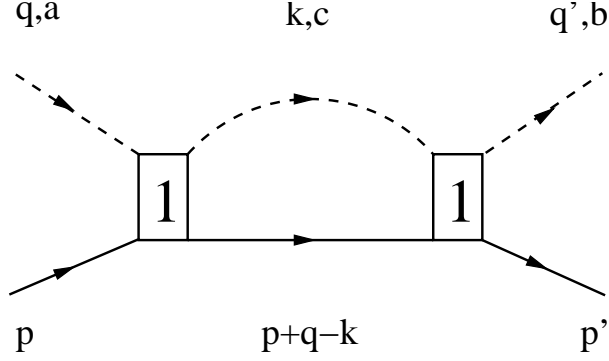,width=8cm}
\end{center}
\end{figure}

   In the previous section we have seen that the parameters of the
lowest-order Lagrangian must be renormalized in the chiral limit.
   As a second example, we will discuss the $\pi N$-scattering loop diagram of 
Fig.\ \ref{5:4:schleife}, which will allow us to draw some further
conclusions regarding the differences between the mesonic and baryonic 
sectors of ChPT.

   Given the Feynman rule of Eq.\ (\ref{5:3:fr2}), the 
contribution of Fig.\ \ref{5:4:schleife} to the invariant amplitude 
reads
\begin{eqnarray}
{\cal M}_{\rm loop}&=&
\int \frac{d^4 k}{(2\pi)^4}
\bar{u}(p')
\frac{k\hspace{-.45em}/+q'\hspace{-.7em}/}{4 F^2_0}
\epsilon_{cbd}\tau^d
\frac{i}{p\hspace{-.45em}/+q\hspace{-.45em}/-k\hspace{-.45em}/-
\stackrel{\circ}{m}_N+i0^+}\nonumber\\
&& \times \frac{i}{k^2-M_\pi^2+i0^+}
\frac{q\hspace{-.45em}/+k\hspace{-.45em}/}{4 F^2_0}
\epsilon_{ace}\tau^e
u(p),\nonumber
\end{eqnarray}
   where, counting powers, we expect the integral to have a cubic divergence.
   The isospin structure is given by
$$
\epsilon_{cbd}\epsilon_{ace}
\tau^d\tau^e=
(\delta_{be}\delta_{da}-\delta_{ba}\delta_{de})\tau^d\tau^e
=\tau^a\tau^b-\delta^{ba}\underbrace{\tau^d\tau^d}_{\mbox{3}}
=-\underbrace{(2\delta^{ab}+\frac{1}{2}[\tau^b,\tau^a])}_{\mbox{
isospin}},
$$
i.e., the diagram contributes to both $\pm$ isospin amplitudes.
   We obtain
\begin{eqnarray*}
{\cal M}_{\rm loop}&=&\mbox{isospin}\,\frac{1}{16F^4_0}
\int\frac{d^4 k}{(2\pi)^4}
\underbrace{\bar{u}(p')
(k\hspace{-.45em}/+q'\hspace{-.7em}/)
(p\hspace{-.45em}/+q\hspace{-.45em}/-k\hspace{-.45em}/+\stackrel{\circ}{m}_N)
(q\hspace{-.45em}/+k\hspace{-.45em}/)
u(p)}_{\mbox{$
\bar{u}(p')\widehat{O}(k)u(p)$}}\\
&&\times\frac{1}{(p+q-k)^2-\stackrel{\circ}{m}_N^2+i0^+}
\frac{1}{k^2-M^2_\pi+i0^+}.
\end{eqnarray*}
   We will outline the evaluation of the integral using dimensional 
regularization.  
   To do this, we first combine the denominators using Feynman's trick, 
Eq.\ (\ref{app:ipipiftrick}) of Appendix \ref{app_subsec_ipipi},
yielding
\begin{displaymath}
\int_0^1 dz \frac{1}{[k^2-2z (p+q)\cdot k 
+z(s-\stackrel{\circ}{m}_N^2)+(z-1)M^2_\pi+i0^+]^2},
\end{displaymath}
where $s=(p+q)^2$.
   Shifting the integration variables as 
$k\to k+z(p+q)$,
the amplitude reads
\begin{eqnarray*}
{\cal M}_{\rm loop}&=&\mbox{isospin}\,\frac{1}{16F^4_0}
\mu^{4-n}\int_0^1 dz\int\frac{d^n k}{(2\pi)^n}
\bar{u}(p')\widehat{O}[k+z(p+q)]u(p)\\
&&\times 
\frac{1}{[k^2+z(s-\stackrel{\circ}{m}_N^2)-z^2s+(z-1)M^2_\pi+i0^+]^2}.
\end{eqnarray*}
   For the final conclusions, it is actually sufficient to consider the
chiral limit, $M_\pi^2=0$, which simplifies the discussion of the loop
integral. 
   We define 
\begin{displaymath}
A(z)\equiv s z^2+(\stackrel{\circ}{m}^2_N-s)z
=z(sz+\stackrel{\circ}{m}^2_N-s)
\end{displaymath}
and will discuss the properties of the function $A$ in more detail below. 
   Note that $A$ is a real, but not necessarily positive, number.
   The numerator of our integral is of the form
\begin{displaymath}
\widehat{O}[k+z(p+q)]=\widehat{O}_0
+\widehat{O}_1^\mu k_\mu
+\widehat{O}_2^{\mu\nu} k_\mu k_\nu
+\widehat{O}_3^{\mu\nu\rho} k_\mu k_\nu k_\rho,
\end{displaymath}
generating integrals of the type
\begin{equation}
\label{5:4:integrals}
\mu^{4-n}\int\frac{d^n k}{(2\pi)^n}
\frac{\{1,k_\mu,k_\mu k_\nu, k_\mu k_\nu k_\rho\}}{(k^2-A+i0^+)^2},
\end{equation}
   where the integrals with an odd power of integration momenta in the 
numerator vanish in dimensional regularization, because of an integration 
over a symmetric interval. (The denominator is even).
   Let us discuss the scalar integral (numerator 1) of Eq.\ 
(\ref{5:4:integrals}).\footnote{It is straightforward to also
determine the second-rank tensor integral of Eq.\ (\ref{5:4:integrals})
using the methods described in Appendices \ref{app_drb} and \ref{app_li}.
   Regarding the analyticity properties we are interested in, one
does not obtain any new information.}
   After a Wick rotation [see Eq.\ (\ref{app:drb:wickrotation})], one
chooses $n$-dimensional spherical coordinates for the Euclidean integral,
and the angular integration is carried out as in 
Eq.\ (\ref{app:drb:winkelintegration}). 
   The remaining one-dimensional integration can be done using Eq.\
(\ref{app:drb:allgint}), and the result is expanded for small 
$\epsilon\equiv4-n$,
\begin{eqnarray}
\label{5:4:integralbsp}
\lefteqn{\mu^{4-n}\int\frac{d^n k}{(2\pi)^n}
\frac{1}{(k^2-A+i0^+)^2}=}\nonumber\\
&&
\frac{i}{(4\pi)^2}
\Bigg[-\frac{2}{n-4}+\ln(4\pi)+\Gamma'(1)
-\ln\left(\frac{A-i0^+}{\mu^2}\right)+O(n-4)\Bigg],
\nonumber\\
\end{eqnarray}
   where $-\Gamma'(1)=\gamma_E=0.5772\cdots$ is Euler's constant.
   The infinity as $n\to 4$ must be canceled by some counter
term of the effective $\pi N$ Lagrangian.
   In order to perform the remaining integration over the Feynman parameter 
$z$, we make use of Eq.\ (\ref{app:drb:lna}),
\begin{equation}
\label{5:4:ln}
\ln(A-i0^+)=\ln(|A|)-i\pi\Theta(-A)\quad\mbox{for}\,A\in R,
\end{equation}
i.e., we need to discuss $A$ as a function of $z\in [0,1]$ (for,
in principle, arbitrary $s$). 
   It is easy to show that $A$ can take negative values in the interval
$0\leq z\leq 1$ only if $s>\,\,\stackrel{\circ}{m}_N^2$ in which case
$A\leq 0$ for $0\leq z\leq 1-\stackrel{\circ}{m}_N^2/s$.
   In combination with Eq.\ (\ref{5:4:ln}) we obtain
\begin{eqnarray*}
\int_0^1 dz\ln\left(\frac{A(z)-i0^+}{\mu^2}\right)&=&
-i\pi\frac{s-\stackrel{\circ}{m}^2_N}{s}\Theta
\left(s-\stackrel{\circ}{m}_N^2\right)\\
&&+
\int_0^1 dz\ln\left(\frac{|sz^2+(\stackrel{\circ}{m}_N^2-s)z|}{\mu^2}\right).
\end{eqnarray*}
   The remaining integral can be evaluated using elementary methods,
and the final expression is
\begin{eqnarray}
\label{5:4:integral}
\int_0^1 dz\ln\left(\frac{A(z)-i0^+}{\mu^2}\right)&=&
-i\pi\frac{s-\stackrel{\circ}{m}_N^2}{s}
\Theta\left(s-\stackrel{\circ}{m}_N^2\right)\nonumber\\
&& +
\ln\left(\frac{\stackrel{\circ}{m}_N^2}{\mu^2}\right)-2
+\frac{s-\stackrel{\circ}{m}_N^2}{s}\ln\left(
\frac{|s-\stackrel{\circ}{m}_N^2|}{\stackrel{\circ}{m}_N^2}\right).\nonumber\\
\end{eqnarray}
   At this point, we refrain from presenting the final expression of 
${\cal M}_{\rm loop}$ in detail, because Eq.\ (\ref{5:4:integral}) 
suffices to point out the difference between one-loop diagrams in the 
mesonic and the baryonic sectors.
   To do this, we expand $s$  for small pion four-momenta in the chiral limit 
about $s_0=\stackrel{\circ}{m}^2_N$: 
\begin{eqnarray}
\frac{s-\stackrel{\circ}{m}^2_N}{\stackrel{\circ}{m}^2_N}
&=&\frac{(p+q)^2-\stackrel{\circ}{m}^2_N}{\stackrel{\circ}{m}^2_N}
=\frac{2p\cdot q}{\stackrel{\circ}{m}^2_N}\equiv\alpha,\nonumber\\
\label{5:4:smmbs}
\frac{s-\stackrel{\circ}{m}_N^2}{s}
&=&\frac{s-\stackrel{\circ}{m}_N^2}{\stackrel{\circ}{m}_N^2
+s-\stackrel{\circ}{m}_N^2}
=\frac{\alpha}{1+\alpha}=\alpha-\alpha^2+\alpha^3+\cdots.
\end{eqnarray}
   [Note that $\alpha$ is a small quantity of chiral order ${\cal O}(p)$.]
   Taking into account that the two extracted Dirac structures (which we have
not displayed) are (at least) of order ${\cal O}(p^2)$ 
[see Eq.\ (4.3) of Ref.\ \cite{Gasser:1987rb}],
one can draw the following conclusions \cite{Gasser:1987rb}:
\begin{itemize}
\item The counter term needed to renormalize the contribution of Fig.\
\ref{5:4:schleife} must contain terms which are of order ${\cal O}(p^2)$
and ${\cal O}(p^3)$.
\item The finite part of the loop diagram has a logarithmic singularity
of the form $p^3\ln(p)$.
\item Expanding the finite part of the diagram in terms of small external
momenta one obtains an infinite series with arbitrary powers of (small)
momenta $p$ [see Eq.\ (\ref{5:4:smmbs})].
\end{itemize}

   In combination with the result of the previous section we see that
a loop calculation with the relativistic Lagrangians ${\cal L}^{(1)}_{\pi N}$
and ${\cal L}_2$ using dimensional regularization leads
to rather different properties in the mesonic and baryonic sectors.
   The example of the nucleon mass shows that loop diagrams may contribute at 
the same order as the tree diagrams which has to be contrasted with the 
mesonic sector where, according to the power counting of Eq.\ (\ref{4:4:mr2}),
loops are always suppressed by a factor $p^{2N_L}$, with $N_L$ denoting the 
number of independent loops.
   In particular, with each new order of the loop expansion one has to expect
that the low-energy coefficients including those of the lowest-order 
Lagrangian ${\cal L}^{(1)}_{\pi N}$ have to be renormalized.
   On the other hand, in the mesonic sector a one-loop calculation in
the even-intrinsic parity sector leads to a renormalization of the
${\cal O}(p^4)$ coefficients (and possibly higher-order coefficients
if vertices of higher order are used), a two-loop calculation to a 
renormalization of the ${\cal O}(p^6)$ and so on. 

   A second difference refers to the orders produced by a loop contribution.
   In the mesonic sector, a one-loop calculation involving vertices of
${\cal L}_2$ produces exclusively an ${\cal O}(p^4)$ contribution. 
   We have seen in the $\pi N$-scattering example above that in the
baryonic sector {\em all} higher orders are generated, even though, in
principle, there is nothing wrong with such a result as long as one can 
organize and predict the leading order of the corresponding contribution 
beforehand.

   In the next section we will discuss the so-called heavy-baryon 
formulation of ChPT \cite{Jenkins:1990jv,Bernard:1992qa}, 
which provides a framework allowing for a power counting 
scheme which is very similar to the mesonic sector.
   One trades the manifestly covariant formulation for the 
systematic power counting.
   Moreover, under certain circumstances, the results obtained in 
HBChPT do not converge in all of the low-energy region. 
   This problem has recently been solved in the framework of the 
so-called infrared regularization \cite{Becher:1999he}
which will be discussed in Sec.\ \ref{sec_mir}.

\section{The Heavy-Baryon Formulation}
\label{sec_hbf}

   We have already seen in Sec.\ \ref{sec_loebl} that the baryonic
sector introduces another energy scale---the nucleon mass---which
does not vanish in the chiral limit.
   Furthermore, the mass of the nucleon has about the same size as the scale
$4\pi F_0$ which appears in the calculation of pion-loop contributions
[see, for example, the discussion of $\pi N$ scattering, where the
tree-level contributions of Table \ref{5:3:tableresults} are $\sim$
$1/F_0^2$, whereas the one-loop diagram of Fig.\ \ref{5:4:schleife} is $\sim$ 
$1/(F_0^2 (4\pi F_0)^2)$].
   The heavy-baryon formulation of ChPT \cite{Jenkins:1990jv,Bernard:1992qa}  
consists in an expansion (of matrix elements) in terms of
$$\frac{p}{4 \pi  F_0}\quad\mbox{and}\quad \frac{p}{\stackrel{\circ}{m}_N},$$
where $p$ represents a  small external momentum.
   Clearly $p$ cannot simply be the four-momentum of the initial and
final nucleons of Eq.\ (\ref{4:btbta}), because the energy components
$E_i$ and $E_f$ are not small.
   Instead, a method has been devised which separates an external
nucleon four-momentum into a large piece of the order of the nucleon mass 
and a small residual component. 
   The approach is similar to the nonrelativistic reduction of
Foldy and Wouthuysen \cite{Foldy:1949wa} which provides a systematic
procedure to block-diagonalize a relativistic Hamiltonian in $1/m$
and produce a decoupling of positive- and negative energy states to
any desired order in $1/m$.
   A criterion for the Foldy-Wouthuysen method to work is that the
potentials in the Dirac Hamiltonian (corresponding to the interaction
with external fields) are small in comparison with the nucleon mass.   
   This may be considered as the analogue of treating external fields 
as small quantities of order ${\cal O}(p)$ ($r_\mu$ and $l_\mu$)
or ${\cal O}(p^2)$ ($f^R_{\mu\nu}$, $f^L_{\mu\nu}$, $\chi$, and $\chi^\dagger$)
in ChPT.

   As in the previous cases we will discuss the lowest-order Lagrangian
in quite some detail.
   For a discussion of the higher-order Lagrangians, the reader
is referred to Refs.\ \cite{Ecker:1995rk,Bernard:1997gq,Fettes:2000gb}.

\subsection{Nonrelativistic Reduction} 
\label{subsec_nr}
   Before discussing the heavy-baryon framework let us start with the
more familiar nonrelativistic limit of the Dirac equation for a charged
particle interacting with an external electromagnetic field.
   Using this example, we will later be able to develop a better
understanding of a peculiarity inherent in the heavy-baryon formulation 
regarding wave function (re)normalization. 
   Our presentation will closely follow Refs.\ \cite{Okubo:54,Das:jx}.
 
   Consider the Dirac equation of a point-particle of charge $q$ and mass $m$
interacting with an electromagnetic four-potential\footnote{In order to 
facilitate the comparison with the Foldy-Wouthuysen result below, we make
use of the ``non-covariant'' form of the Dirac equation.}
\begin{equation}
\label{5:5:dirac1}
i\partial_0 \Psi=[\vec{\alpha}\cdot(\vec{p}-q\vec{A})+\beta m +q A_0]\Psi
\equiv H\Psi,
\end{equation}
where $\alpha_i$ and $\beta$ are the usual Dirac matrices
\begin{displaymath}
\alpha_i=\left(
\begin{array}{cc} 
0_{2\times 2}&\sigma_i\\ \sigma_i&0_{2\times 2}
\end{array}
\right),\quad
\beta=\left(
\begin{array}{rr} 
1_{2\times 2}& 0_{2\times 2}\\ 0_{2\times 2}&-1_{2\times 2}
\end{array}
\right),
\end{displaymath}
and $\vec{p}=\vec{\nabla}/i$ is the momentum operator.
   For simplicity, we consider the interaction with a static external electric 
field, 
\begin{displaymath}
\vec{E}=-\vec{\nabla} A_0,\quad \vec{A}=0.
\end{displaymath}
   Since we want to describe a nonrelativistic particle-like solution, it
is convenient to separate a factor $\exp(-imt)$ from the wave 
function,\footnote{The (second) quantization of the relevant fields will be 
discussed in Sec.\ \ref{subsec_nfs}.}
\begin{displaymath}
\Psi(\vec{x},t)=e^{-imt}\Psi'(\vec{x},t),
\end{displaymath}
so that the Dirac equation (after multiplication with $e^{+imt}$) results in
\begin{equation}
\label{5:5:dirac2}
i\partial_0 \Psi'=[\vec{\alpha}\cdot\vec{p}+(\beta-1)m+q A_0]\Psi'\equiv
H'\Psi'.
\end{equation}
   Note that both $H$ and $H'$ are Hermitian operators.
   In the spirit of the nonrelativistic reduction, we write $\Psi'$ in terms
of a pair of two-component spinors $\Psi_L$ and $\Psi_S$ 
($L$ for large and $S$ for small) 
\begin{equation}
\label{5:5:psiptwocomponent}
\Psi'=
\frac{1}{2}(1+\beta)\Psi'
+\frac{1}{2}(1-\beta)\Psi'
=
\left(
\begin{array}{c}
\Psi_L\\
\Psi_S
\end{array}
\right),
\end{equation}
and obtain, after insertion into Eq.\ (\ref{5:5:dirac2}),
a set of two coupled partial differential equations
\begin{eqnarray}
\label{5:5:dirac3a}
(i\partial_0-q A_0)\Psi_L&=&\vec{\sigma}\cdot\vec{p}\,\Psi_S,\\
\label{5:5:dirac3b}
(i\partial_0-q A_0+2m)\Psi_S&=&\vec{\sigma}\cdot\vec{p}\,\Psi_L.
\end{eqnarray}
   The second equation can formally be solved for $\Psi_S$,
\begin{equation}
\label{5:5:psilsolution}
\Psi_S=(2m+i\partial_0-q A_0)^{-1}\vec{\sigma}\cdot\vec{p}\,\Psi_L
\equiv A \Psi_L,
\end{equation}
where, for later use, we have introduced the abbreviation $A$
for the operator $(2m+i\partial_0-q A_0)^{-1}\vec{\sigma}\cdot\vec{p}$\,.
   We expand Eq.\ (\ref{5:5:psilsolution}) in terms of $1/m$ up to and 
including order $1/m^2$,
\begin{eqnarray}
\label{5:5:apsilexp}
A\Psi_L&=&\frac{\vec{\sigma}\cdot\vec{p}}{2m}\,\Psi_L
-\frac{i\partial_0-q A_0}{2m}
\frac{\vec{\sigma}\cdot\vec{p}}{2m}\,\Psi_L
+O\left(\frac{1}{m^3}\right)\nonumber\\
&=&\frac{\vec{\sigma}\cdot\vec{p}}{2m}
\left(1-\frac{i\partial_0-q A_0}{2m}\right)
\Psi_L-iq\frac{\vec{\sigma}\cdot\vec{E}}{4m^2}\,\Psi_L
+O\left(\frac{1}{m^3}\right)\nonumber\\
&=&\left(\frac{\vec{\sigma}\cdot\vec{p}}{2m}-iq
\frac{\vec{\sigma}\cdot\vec{E}}{4m^2}\right)\Psi_L
+O\left(\frac{1}{m^3}\right),
\end{eqnarray}
where we made use of the commutation relation 
$[A_0,\vec{p}\,]=i(\vec{\nabla}A_0)
=-i\vec{E}$ and 
of $(i\partial_0-q A_0)\Psi_L=O(1/m)\Psi_L$
[see Eqs.\ (\ref{5:5:dirac3a}) and (\ref{5:5:psilsolution})].
   Inserting this result into the right-hand side of Eq.\ (\ref{5:5:dirac3a})
and using $\vec{\sigma}\cdot\vec{A}\,\vec{\sigma}\cdot\vec{B}=\vec{A}\cdot\vec{B}
+i\vec{\sigma}\cdot\vec{A}\times\vec{B}$,
we obtain the Schr\"odinger-type equation
\begin{equation}
\label{5:5:schroedinger}
(i\partial_0-q A_0)\Psi_L=
\left\{\frac{\vec{p}\,^2}{2m}-\frac{q}{4m^2}\left[
(\vec{\nabla}\cdot\vec{E})
+\vec{\sigma}\cdot\vec{E}\times \vec{p}+i\vec{E}\cdot\vec{p}\,\right]
\right\}\Psi_L.
\end{equation}
   As already noted by Okubo \cite{Okubo:54}, 
the last term on the right-hand side of Eq.\ (\ref{5:5:schroedinger}) is 
not Hermitian and, when written as $V=-\vec{d}\cdot\vec{E}$, represents
the interaction of an electric field with a (momentum-dependent) imaginary 
electric dipole moment 
$\vec{d}=iq\vec{p}/(4m^2)$ \cite{Das:jx}.\footnote{The standard textbook
treatment of the nonrelativistic reduction leading to the Pauli equation
considers only terms of $1/m$ and thus does not yet generate
non-Hermitian terms (see, e.g., Refs.\ 
\cite{Bjorken_1964,Itzykson:rh}).}
   As pointed out in Ref.\ \cite{Das:jx}, the non-Hermiticity of the Hamilton 
operator of Eq.\ (\ref{5:5:schroedinger}) is a consequence of the procedure 
for eliminating the small-component spinors.
   The method can be thought of as applying the transformation 
\begin{equation}
\label{5:5:strans}
S=\left(\begin{array}{cc}
1_{2\times 2} & 0_{2\times 2}\\
-A & 1_{2\times 2}
\end{array}
\right)
\end{equation}
to the four-component spinor $\Psi'$ to generate a four-component spinor
consisting exclusively of the upper component $\Psi_L$, and then solving the
corresponding transformed Dirac equation.
   Since $S$ is {\em not} a unitary operator, i.e.,
\begin{displaymath}
\left(\begin{array}{cc}
1_{2\times 2} &-A^\dagger\\
0_{2\times 2} & 1_{2\times 2}
\end{array}\right)
=S^\dagger\neq S^{-1}
=\left(\begin{array}{cc}
1_{2\times 2}&0_{2\times 2}\\
A&1_{2\times 2}
\end{array}
\right),
\end{displaymath}
the norm of the original spinor $\Psi'$ and the transformed spinor $\Psi_L$,
in general, will not be the same
\begin{equation}
\label{5:5:norm}
\int d^3 x \Psi'^\dagger\Psi'=\int d^3 x (\Psi^\dagger_L\Psi_L+
\Psi^\dagger_S\Psi_S)=
\int d^3x \Psi^\dagger_L(1+A^\dagger A)\Psi_L\neq
\int d^3 x \Psi_L^\dagger\Psi_L.
\end{equation}
   Equation (\ref{5:5:norm}) suggests considering a field redefinition of the 
form \cite{Okubo:54,Das:jx}
\begin{equation}
\label{5:5:psird}
\widetilde{\Psi}_L=(1+A^\dagger A)^{\frac{1}{2}}\Psi_L,
\end{equation}
so that the new spinor $\widetilde{\Psi}_L$ has the same norm as $\Psi'$.
   For the specific Hamiltonian of Eq.\ (\ref{5:5:dirac2}) we have
\begin{displaymath}
A=\frac{\vec{\sigma}\cdot\vec{p}}{2m}+O\left(\frac{1}{m^2}\right),
\end{displaymath}
so that we find\footnote{In the framework of plane-wave solutions, 
Eq.\ (\ref{5:5:psilpsiltilde}) already provides a hint that one may have to 
expect ``unconventional normalization factors'' when dealing with 
Feynman rules in the heavy-baryon approach.} 
\begin{equation}
\label{5:5:psilpsiltilde}
\Psi_L=(1+A^\dagger A)^{-\frac{1}{2}}\widetilde{\Psi}_L
=\left[1-\frac{\vec{p}\,^2}{8m^2}+O\left(\frac{1}{m^3}\right)\right]
\widetilde{\Psi}_L.
\end{equation}
   When inserting Eq.\ (\ref{5:5:psilpsiltilde}) into 
Eq.\ (\ref{5:5:schroedinger}), we make use of
\begin{displaymath}
A_0 \vec{p}\,^2 \widetilde{\Psi}_L=
\vec{p}\,^2 (A_0 \widetilde{\Psi}_L)-
(\vec{\nabla}\cdot\vec{E}) \widetilde{\Psi}_L
-2i \vec{E}\cdot\vec{p}\,\widetilde{\Psi}_L
\end{displaymath}
and, as above, $(i\partial_0-q A_0)\widetilde{\Psi}_L=
O(1/m)\widetilde{\Psi}_L$, yielding the Schr\"odinger equation for the
two-component spinor $\widetilde{\Psi}_L$, including relativistic corrections
up to order $1/m^2$,
\begin{equation}
\label{5:5:schroedinger2}
(i\partial_0-q A_0)\widetilde{\Psi}_L=
\left[\frac{\vec{p}\,^2}{2m}
-\frac{q}{4m^2}\vec{\sigma}\cdot\vec{E}\times \vec{p}
-\frac{q}{8m^2}(\vec{\nabla}\cdot\vec{E})
\right]\widetilde{\Psi}_L,
\end{equation}
   where the second term, for a central potential, corresponds to 
the usual spin-orbit interaction and the last term is the 
so-called Darwin term \cite{Bjorken_1964,Itzykson:rh}.
   Note that the Hamiltonian here {\em is} Hermitian, i.e., the 
imaginary dipole moment has disappeared.
   Moreover, because of Eqs.\ (\ref{5:5:norm}) and (\ref{5:5:psird}),
the spinors are normalized conventionally.

   The result of Eq.\ (\ref{5:5:schroedinger2}) is identical with a 
nonrelativistic reduction using the Foldy-Wouthuysen method 
\cite{Foldy:1949wa} which uses a sequence of unitary transformations 
to block-diagonalize a relativistic Hamiltonian of the form
\begin{equation}
\label{5:5:hoe}
H=\beta m+{\cal O}+{\cal E}
\end{equation}
to any desired order in $1/m$. 
   In Eq.\ (\ref{5:5:hoe}) ${\cal O}$ and ${\cal E}$ denote the so-called
odd and even operators of $H$, respectively, where odd operators couple large 
and small components whereas even operators do not.
   In the present case we have
\begin{displaymath}
{\cal O}=\vec{\alpha}\cdot\vec{p},\quad
{\cal E}=q A_0,
\end{displaymath}
and after three successive transformations one obtains the block-diagonal
Hamiltonian
(see, e.g., 
Refs.\ \cite{Bjorken_1964,Itzykson:rh,Fearing:ii})
\begin{eqnarray}
\label{5:5:hfw3}
H_{\rm FW}^{(3)}&=&\beta\left(m+\frac{\vec{p}\,^2}{2m}\right)
+q A_0-\frac{1}{8m^2}
[\vec{\alpha}\cdot\vec{p},[\vec{\alpha}\cdot\vec{p},q A_0]]
+O\left(\frac{1}{m^3}\right)\nonumber\\
&=&\beta\left(m+\frac{\vec{p}\,^2}{2m}\right)+q A_0
-\frac{q}{8m^2}(\vec{\nabla}\cdot\vec{E})
-\frac{q}{4m^2}\vec{\Sigma}\cdot\vec{E}\times \vec{p}\,
+O\left(\frac{1}{m^3}\right),\nonumber\\
\end{eqnarray}
where
\begin{equation}
\label{5:5:Sigma}
\vec{\Sigma}=
\left(
\begin{array}{cc}
\vec{\sigma}&0_{2\times 2}\\
0_{2\times 2}&\vec{\sigma}
\end{array}
\right).
\end{equation}
   Restricting ourselves to the upper left block of Eq.\ (\ref{5:5:hfw3})
and noting that in Eq.\ (\ref{5:5:dirac2}) we have already separated the
time dependence $\exp(-imt)$ from $\Psi$, we find that Eqs.\ 
(\ref{5:5:schroedinger2}) and (\ref{5:5:hfw3}) are indeed identical.
   In Ref.\ \cite{Das:jx}, the equivalence of the two approaches was
explicitly shown to order $1/m^5$.

   We will see that the heavy-baryon approach proceeds along lines very 
similar to the nonrelativistic reduction leading from Eq.\ (\ref{5:5:dirac1}) 
to (\ref{5:5:schroedinger}).
   In analogy to Eqs.\ (\ref{5:5:norm}) and (\ref{5:5:psird}) we thus have to 
be alert to surprises related to the normalization of the relevant wave 
functions.

\subsection{Light and Heavy Components}
\label{subsec_lhc}
   As mentioned above, the idea of the heavy-baryon approach consists
of separating the large nucleon mass from the external four-momenta of the
nucleons in the initial and final states and, in a  sense to be discussed
in Sec.\ \ref{subsec_lol} below, eliminating it from the Lagrangian.

   The starting point is the relativistic Lagrangian of Eq.\ 
(\ref{5:2:l1pin}),
\begin{equation}
\label{5:4:lpin1}
{\cal L}^{(1)}_{\pi N}=\bar{\Psi}\left(iD\hspace{-.7em}/
-m
+\frac{{g}_A}{2}u\hspace{-.5em}/\gamma_5\right)\Psi,
\end{equation}
where the covariant derivative $D_\mu \Psi$ and $u_\mu$ are defined
in Eqs.\ (\ref{5:2:kovderpsi}) and (\ref{5:2:chvi}), respectively.
   The corresponding Euler-Lagrange equation for the nucleon field
reads
\begin{equation}
\label{5:5:eom1}
-\partial_\mu \frac{\partial{\cal L}^{(1)}_{\pi N}}{\partial \partial_\mu 
\bar{\Psi}}
+\frac{\partial{\cal L}^{(1)}_{\pi N}}{\partial \bar{\Psi}}=
\left(iD\hspace{-.7em}/-m
+\frac{{g}_A}{2}u\hspace{-.5em}/\gamma_5\right)
\Psi=0.
\end{equation}
   (For notational convenience we replace $\stackrel{\circ}{m}_N\to m$
and $\stackrel{\circ}{g}_A\to g_A$ in Secs.\ \ref{subsec_lhc} and
\ref{subsec_lol}).
   For a general four-vector $v^\mu$ with the properties $v^2=1$ and
$v^0\geq 1$, we define the projection operators\footnote{It may be worthwhile
to remember that $P_{v\pm}$ do not define orthogonal projectors in the 
mathematical sense, because they do not satisfy $P^\dagger_{v\pm}=
P_{v\pm}$, with the exception of the special case $v^\mu=(1,0,0,0)$
used in Eq.\ (\ref{5:5:psiptwocomponent}).}
\begin{equation}
\label{5:4:ppmdef}
P_{v\pm}\equiv\frac{1\pm v\hspace{-.5em}/}{2},\quad
P_{v+}+P_{v-}=1,\quad
P_{v\pm}^2=P_{v\pm},\quad
P_{v\pm}P_{v\mp}=0,
\end{equation}
and introduce the so-called velocity-dependent fields 
${\cal N}_v$ and ${\cal H}_v$ as
\begin{equation}
\label{5:5:nh}
{\cal N}_v\equiv e^{imv\cdot x}P_{v+}\Psi,
\quad
{\cal H}_v\equiv e^{imv\cdot x}P_{v-}\Psi,
\end{equation}
so that $\Psi$ can be written as
\begin{equation}
\label{5:5:psinh}
\Psi(x)=e^{-imv\cdot x}\left[{\cal N}_v(x)+{\cal H}_v(x)\right].
\end{equation}
   The fields ${\cal N}_v$ and ${\cal H}_v$ satisfy the properties
\begin{equation}
\label{5:5:nhprop}
v\hspace{-.5em}/ {\cal N}_v={\cal N}_v,\quad
v\hspace{-.5em}/ {\cal H}_v=-{\cal H}_v.
\end{equation}
   For a particle with four-momentum $p^\mu=(E,\vec{p}\,)$ the particular 
choice $v^\mu=p^\mu/m$ corresponds to its world velocity which is  
why $v$ is also referred to as a four-velocity.
   The fields ${\cal N}_v$ and ${\cal H}_v$ are often called the
light and heavy components of the field $\Psi$, 
which will become clearer below.

   In order to motivate the ansatz of Eq.\ (\ref{5:5:psinh}) let us consider
a positive-energy plane wave solution to the free Dirac equation with
three-momentum $\vec{p}$:
\begin{eqnarray*}
\psi^{(+)(\alpha)}_{\vec{p}}(\vec{x},t)&=&
u^{(\alpha)}(\vec{p}\,) e^{-ip\cdot x},\\
u^{(\alpha)}(\vec{p}\,)&=& \sqrt{E(\vec{p}\,)+m}
\left(\begin{array}{c}\chi^{(\alpha)} \\ \frac{\vec{\sigma}\cdot
\vec{p}}{E(\vec{p}\,)+m}\chi^{(\alpha)}\end{array}\right),
\end{eqnarray*}
where 
\begin{displaymath}
\chi^{(1)}=\left(\begin{array}{c}1\\0\end{array}\right),\quad
\chi^{(2)}=\left(\begin{array}{c}0\\1\end{array}\right),
\end{displaymath}
are ordinary two-component Pauli spinors, and 
$E(\vec{p}\,)=\sqrt{m^2+\vec{p}\,^2}$.
   We can think of $\psi^{(+)(\alpha)}_{\vec{p}}(\vec{x},t)$ entering
the calculation of, say, an $S$-matrix element through covariant perturbation 
theory in terms of the matrix element of an in-field $\Psi_{\rm in}(x)$
between the vacuum and a single-nucleon state:
\begin{displaymath}
\langle 0| \Psi_{\rm in}(x)|N(\vec{p},\alpha),{\rm in}\rangle= 
u^{(\alpha)}(\vec{p}\,) e^{-ip\cdot x}\chi_N,
\end{displaymath}
where $\chi_N$ denotes the nucleon isospinor.
     For the special case $v^\mu=(1,0,0,0)\equiv v^\mu_1$, i.e.
$$P_{v_1+}=\left(\begin{array}{cc}1_{2\times 2}&0_{2\times 2}\\
0_{2\times 2}&0_{2\times 2}\end{array}\right),\quad
P_{v_1-}=\left(\begin{array}{cc}0_{2\times 2}&0_{2\times 2}\\
0_{2\times 2}&1_{2\times 2}\end{array}\right),$$
the components $N_{v_1}$ and $H_{v_1}$ are, up to
the modified time dependence, equivalent to the large and small
components of the ``one-particle wave function''
\begin{eqnarray}
\label{5:5:nhspwf}
N_{v_1}^{(\alpha)}(x)&=&\sqrt{E(\vec{p}\,)+m}
\left(\begin{array}{c}\chi^{(\alpha)} \\ 0_{2\times 1}\end{array}\right)
e^{-i[E(\vec{p}\,)-m]t+i\vec{p}\cdot\vec{x}},\nonumber\\
H_{v_1}^{(\alpha)}(x)&=&\sqrt{E(\vec{p}\,)+m}
\left(\begin{array}{c}0_{2\times 1}\\ \frac{\vec{\sigma}\cdot
\vec{p}}{E(\vec{p}\,)+m}\chi^{(\alpha)}\end{array}\right)
e^{-i[E(\vec{p}\,)-m]t+i\vec{p}\cdot\vec{x}}.
\end{eqnarray}
   In other words, for this choice of $v$ the light and heavy components of 
the positive-energy solutions are closely related to the large and small 
components of the nonrelativistic reduction discussed in Sec.\ \ref{subsec_nr}.
   Moreover, assuming $|\vec{p}\,|\ll m$, $\exp[-i(E-m)t]$ varies slowly 
with time in comparison with $\exp(-iEt)$ 
of $\psi^{(+)(\alpha)}_{\vec{p}}(\vec{x},t)$,
with the result that a time derivative $i\partial/\partial t$
generates a factor $(E-m)$ which is small in comparison with $m$.

   Another choice is $v^\mu=p^\mu/m\equiv v^\mu_2$, 
in which case $P_{v_2+}$ and $P_{v_2-}$
correspond to the usual projection operators for positive- and 
negative-energy states 
\begin{displaymath}
P_{v_2\pm}=\Lambda_{\pm}(p)
=\frac{\pm p\hspace{-.45em}/\hspace{.2em} + m}{2m}.
\end{displaymath}
   For this case we find
\begin{eqnarray}
\label{5:5:nhspwf2}
N_{v_2}^{(\alpha)}(x)&=&u^{(\alpha)}(\vec{p}\,),\nonumber\\
H_{v_2}^{(\alpha)}(x)&=&0,
\end{eqnarray}
   i.e., the $x$ dependence has completely disappeared in $N_{v_2}$ and, due 
to the projection property $\Lambda_{-}(p) u^{(\alpha)}(\vec{p}\,)=0$, 
$H_{v_2}$ vanishes identically.

   In general, one decomposes the four-momentum $p^\mu$ of a low-energy nucleon
into $m v^\mu$ and a residual momentum $k^\mu$,\footnote{Of course, the
decomposition of Eq.\ (\ref{5:5:pmvk}) alone is not a sufficient criterion
for $v\cdot k \ll m$. Taking, for example, $\vec{p}\perp \vec{v}$ one finds
$v\cdot k=v\cdot p-m= Ev^0-m\gg m$ for large $v^0$.}
\begin{equation}
\label{5:5:pmvk}
p^\mu=m v^\mu+k^\mu,
\end{equation}
so that 
\begin{equation}
\label{5:5:vkcond}
v\cdot k=-\frac{k^2}{2m}\stackrel{v^\mu=(1,0,0,0)}{=}k_0=E-m\ll m.
\end{equation}
   For $v^\mu$ in the vicinity of $(1,0,0,0)$, a partial derivative
$i\partial^\mu$ acting on $e^{-ip\cdot x+imv\cdot x}$ produces 
a small residual momentum $k^\mu$ and, in particular,
$$iv\cdot\partial\mapsto v\cdot k \ll m.$$
   
   The actual choice of $v^\mu$ is, to some extent, a matter of convenience.
   For low-energy processes involving a single nucleon in the initial
and final states, the four-momentum $q^\mu$ transferred in the reaction is
defined as $q=p_f-p_i$, and is considered as a small quantity of chiral order 
${\cal O}(p)$. 
   For $p_i=m v+k_i$ and $p_f=mv+k_f$, where, say, $k_i$ is a small residual
momentum in the sense of Eq.\ (\ref{5:5:vkcond}), also $k_f=k_i+q$ is a small 
four-momentum.
   The implications on a chiral power-counting scheme will be discussed in 
Sec.\ \ref{subsec_pcs} below.

\subsection{Lowest-Order Lagrangian}
\label{subsec_lol}
   In order to proceed with the construction of the lowest-order
heavy-baryon Lagrangian
we insert Eq.\ (\ref{5:5:psinh}) into the EOM of 
Eq.\ (\ref{5:5:eom1}),\footnote{For a derivation in the framework of
the path-integral approach, see Ref.\ \cite{Mannel:1991mc} and Appendix
A of Ref.\ \cite{Bernard:1992qa}.}
\begin{eqnarray*}
\lefteqn{\left(iD\hspace{-.7em}/-m
+\frac{{g}_A}{2}u\hspace{-.5em}/\gamma_5\right)
e^{-imv\cdot x}\left({\cal N}_v+{\cal H}_v\right)=}\\
&&e^{-imv\cdot x}\left(mv\hspace{-.5em}/ 
+iD\hspace{-.7em}/-m
+\frac{{g}_A}{2}u\hspace{-.5em}/\gamma_5\right)
\left({\cal N}_v+{\cal H}_v\right)=0,
\end{eqnarray*}
make use of Eq.\ (\ref{5:5:nhprop}), multiply by $e^{im v\cdot x}$, and
obtain 
\begin{equation}
\label{5:5:eom2}
\left(iD\hspace{-.7em}/+\frac{g_A}{2}u\hspace{-.5em}/\gamma_5\right){\cal N}_v
+\left(iD\hspace{-.7em}/-2m +\frac{g_A}{2}u\hspace{-.5em}/\gamma_5\right)
{\cal H}_v=0.
\end{equation}
   In the next step we would like to separate the $P_{v+}$ and the
$P_{v-}$ part of the EOM of Eq.\ (\ref{5:5:eom2}). 
   To that end we make use of the algebra of the gamma matrices to derive
\begin{eqnarray}
\label{5:5:ps}
P_{v+}A\hspace{-.6em}/\hspace{.2em}P_{v+}&=&v\cdot A\, P_{v+},\nonumber\\
P_{v+}A\hspace{-.6em}/P_{v-}&=&A\hspace{-.6em}/_\bot P_{v-}
=P_{v+}A\hspace{-.6em}/_\bot,
\nonumber\\
P_{v-}A\hspace{-.6em}/\hspace{.2em}P_{v-}&=&-v\cdot A\, P_{v-},\nonumber\\
P_{v-}A\hspace{-.6em}/\hspace{.2em}P_{v+}&=&A\hspace{-.6em}/_\bot P_{v+}=
P_{v-}A\hspace{-.6em}/_\bot,
\nonumber\\
P_{v+}B\hspace{-.6em}/\hspace{.2em}
\gamma_5 P_{v+}&=&B\hspace{-.6em}/_\bot\gamma_5 P_{v+},
\nonumber\\
P_{v+}B\hspace{-.6em}/\hspace{.2em}\gamma_5 P_{v-}&=&v\cdot B\,\gamma_5 P_{v-}
=v\cdot B \,P_{v+}\gamma_5,\nonumber\\
P_{v-}B\hspace{-.6em}/\hspace{.2em}
\gamma_5 P_{v-}&=&B\hspace{-.6em}/_\bot\gamma_5 P_{v-},
\nonumber\\
P_{v-}B\hspace{-.6em}/\hspace{.2em}
\gamma_5 P_{v+}&=&-v\cdot B\gamma_5 P_{v+}
=-v\cdot B P_{v-}\gamma_5,
\end{eqnarray}
where
$$P_{v\pm}=\frac{1\pm v\hspace{-.5em}/}{2}, \quad v^2=1,\quad
A_\bot^\mu= A^\mu- v\cdot A\, v^\mu,\quad
v\cdot A_\bot=0,\quad
A\hspace{-.6em}/_\bot=A_\bot^\mu \gamma_\mu.$$
   As an example, let us explicitly show the first relation of Eq.\
(\ref{5:5:ps})
\begin{eqnarray*}
P_{v+}A\hspace{-.6em}/\hspace{.2em}P_{v+}
&=&\frac{1}{2}(1+v\hspace{-.5em}/)A\hspace{-.6em}/\hspace{.2em}P_{v+}
=\frac{1}{2}(A\hspace{-.6em}/+v\hspace{-.5em}/
\hspace{.2em}A\hspace{-.6em}/\hspace{.2em})P_{v+}
=\frac{1}{2}(A\hspace{-.6em}/-A\hspace{-.6em}/
\hspace{.2em}v\hspace{-.5em}/+2v\cdot A)P_{v+}
\\
&=&(A\hspace{-.6em}/\hspace{.2em}P_{v-}+v\cdot A)P_{v+}
=v\cdot A\, P_{v+}.
\end{eqnarray*}
   The remaining results of Eq.\ (\ref{5:5:ps}) follow analogously.
   Using Eqs.\ (\ref{5:4:ppmdef}) and (\ref{5:5:ps}) we are now in the 
position to project onto the $P_{v+}$ and $P_{v-}$ parts of the EOM of 
Eq.\ (\ref{5:5:eom2}),
\begin{eqnarray}
\label{5:5:eom3a}
\left(iv\cdot D+\frac{g_A}{2}u\hspace{-.5em}/_\bot\gamma_5\right){\cal N}_v
+\left(iD\hspace{-.7em}/_\bot+\frac{g_A}{2}v\cdot u \gamma_5\right)
{\cal H}_v&=&0,\\
\label{5:5:eom3b}
\left(iD\hspace{-.7em}/_\bot-\frac{g_A}{2}v\cdot u \gamma_5\right){\cal N}_v
+\left(-iv\cdot D-2m+\frac{g_A}{2}u\hspace{-.5em}/_\bot\gamma_5\right)
{\cal H}_v&=&0,
\end{eqnarray}
which corresponds to Eqs.\ (\ref{5:5:dirac3a}) and
(\ref{5:5:dirac3b}) of the nonrelativistic reduction of Sec.\ \ref{subsec_nr}.
   We formally solve Eq.\ (\ref{5:5:eom3b}) for ${\cal H}_v$,
\begin{equation}
\label{5:5:solutionhv}
{\cal H}_v=\left(2m+iv\cdot D-\frac{g_A}{2}u\hspace{-.5em}/_\bot
\gamma_5\right)^{-1}
\left(iD\hspace{-.7em}/_\bot-\frac{g_A}{2}v\cdot u \gamma_5\right)
{\cal N}_v,
\end{equation}
which, similar to the discussion of Sec.\ \ref{subsec_nr}, shows that
the heavy component ${\cal H}_v$ is formally
suppressed by powers of $1/m$ relative to 
the light component ${\cal N}_v$.\footnote{In fact, setting all external
fields to zero and dropping the interaction term proportional to 
$g_A$, it is easy to verify that the one-particle wave functions
indeed satisfy the relation implied by Eq.\ (\ref{5:5:solutionhv}).}
    Inserting Eq.\ (\ref{5:5:solutionhv}) into Eq.\ (\ref{5:5:eom3a}),
the EOM for the light component reads
\begin{eqnarray}
\label{5:5:eomnv}
\lefteqn{\left(iv\cdot D+\frac{g_A}{2}u\hspace{-.5em}/_\bot
\gamma_5\right){\cal N}_v
+\left(i D\hspace{-.7em}/_\bot+\frac{g_A}{2}v\cdot u \gamma_5\right)}
\nonumber\\
&&\times
\left(2m+iv\cdot D-\frac{g_A}{2}u\hspace{-.5em}/_\bot
\gamma_5\right)^{-1}
\left(iD\hspace{-.7em}/_\bot-\frac{g_A}{2}v\cdot u \gamma_5\right)
{\cal N}_v=0,
\end{eqnarray}
which represents the analogue of Eq.\ (\ref{5:5:schroedinger}).
   This EOM may be obtained from applying the variational
principle to the effective Lagrangian\footnote{Replacing 
${\cal N}_v\to h^{(+)}$ and ${\cal H}_v\to h^{(-)}$,
omitting all terms containing the chiral vielbein $u_\mu$,
and interpreting the covariant derivative as that of QCD, 
the result of Eq.\ (\ref{5:5:lhneff}) is identical with Eq.\ (7) of
the discussion of heavy quark effective theory in Sec.\ 2 of Ref.\ 
\cite{Balk:1993ev}.}
\begin{eqnarray}
\label{5:5:lhneff}
{\cal L}_{\rm eff}&=&\bar{\cal N}_v\left(iv\cdot D
+\frac{g_A}{2}u\hspace{-.5em}/_\bot\gamma_5\right){\cal N}_v
+
\bar{\cal N}_v \left(i D\hspace{-.7em}/_\bot+\frac{g_A}{2}v\cdot u 
\gamma_5\right)\nonumber\\
&&\times
\left(2m+iv\cdot D -\frac{g_A}{2}u\hspace{-.5em}/_\bot
\gamma_5\right)^{-1}
\left(i D\hspace{-.7em}/_\bot-\frac{g_A}{2}v\cdot u \gamma_5\right)
{\cal N}_v.
\end{eqnarray}
   Note that the second term is suppressed by  $1/m$ relative to the first 
term.
   Equation (\ref{5:5:lhneff}) corresponds to the leading-order result
for Eq.\ (A.10) of Ref.\ \cite{Bernard:1992qa} which was obtained in
the framework of the path-integral approach, but does not yet represent the 
final form commonly used in HBChPT.\footnote{In order to be able to invert 
the operator ${\cal C}$ of Ref.\ \cite{Bernard:1992qa}, strictly
speaking  the projection operators $P_{v-}$ should not be included
in the definition of ${\cal C}$.}
   Having the discussion following Eq.\ (\ref{5:5:strans}) in mind, 
in order for the two Lagrangians of Eqs.\ (\ref{5:4:lpin1}) and 
(\ref{5:5:lhneff}) to describe the same observables, we cannot expect
both fields $\Psi$ and ${\cal N}_v$ to be normalized in the same way.
We will come back to this question in Sec.\ \ref{subsec_nfs}.

   To obtain the heavy-baryon Lagrangian we define the spin matrix $S^\mu_v$ 
as\footnote{For the 
classification of 
the irreducible representations of the Poincar\'e group, one makes use of
the so-called Pauli-Lubanski vector
\begin{displaymath}
W_\mu\equiv \frac{1}{2} \epsilon_{\mu\nu\rho\sigma}J^{\nu\rho} P^\sigma,
\end{displaymath}
where $\epsilon_{\mu\nu\rho\sigma}$ is the completely antisymmetric tensor
in four indices, $\epsilon_{0123}=1$, $J^{\mu\nu}$ denotes the generalized 
angular momentum operator, and $P^\mu$ is the four-momentum operator
(see, e.g, Refs.\ \cite{Itzykson:rh,Jones:ti}). 
   Both $W^2$ and $P^2$ are Lorentz invariant and 
translationally invariant and are thus used as Casimir operators, where the
eigenvalues are denoted by $m^2$ and $-m^2 s(s+1)$, $s=0,1/2,1,\cdots$.
   For the massive spin-1/2 case one obtains
\begin{displaymath}
W_\mu=\frac{1}{4}\epsilon_{\mu\nu\rho\sigma}\sigma^{\nu\rho}P^\sigma.
\end{displaymath}
   Using (in four dimensions) 
\begin{displaymath}
\gamma_5\sigma_{\mu\nu}=-\frac{i}{2}\epsilon_{\mu\nu\rho\sigma}\sigma^{\rho
\sigma},
\end{displaymath}
together with the special choice $\tilde{v}^\mu=P^\mu/m$, one easily
finds that the spin matrix is, for this special case, proportional to the 
Pauli-Lubanski vector, $W^\mu=m S^\mu_{\tilde{v}}$.}
\begin{equation}
\label{5:5:spinop}
S^\mu_v=\frac{i}{2}\gamma_5\sigma^{\mu\nu}v_\nu=-\frac{1}{2}\gamma_5
(\gamma^\mu v\hspace{-.5em}/-v^\mu),\quad 
S_v^{\mu\dagger}=\gamma_0S_v^\mu \gamma_0,
\end{equation}
which, in four dimensions, has the properties
\begin{equation}
\label{5:5:seig}
v\cdot S_v=0,\quad
\{S^\mu_v, S^\nu_v\}=\frac{1}{2}(v^\mu v^\nu-g^{\mu\nu}),\quad
[S^\mu_v,S^\nu_v]=i\epsilon^{\mu\nu\rho\sigma}
v_\rho S_\sigma^v.
\end{equation}
   Using the properties of Eq.\ (\ref{5:5:nhprop}) together with 
Eq.\ (\ref{5:5:seig}), the 16 combinations $\bar{\cal N}_v \Gamma {\cal N}_v$ 
may be written as [see Eqs.\ (9) - (12) of Ref.\ \cite{Jenkins:1990jv}]
\begin{eqnarray}
\label{5:5:nvbilineare}
(\bar{\cal N}_v1_{4\times 4}{\cal N}_v&=&\bar{\cal N}_v1_{4\times 4}
{\cal N}_v,)
\nonumber\\
\bar{\cal N}_v\gamma_5{\cal N}_v&=&0,\nonumber\\
\bar{\cal N}_v\gamma^\mu{\cal N}_v&=&v^\mu \bar{\cal N}_v{\cal N}_v,\nonumber\\
\bar{\cal N}_v\gamma^\mu\gamma_5{\cal N}_v
&=&2 \bar{\cal N}_v S^\mu_v{\cal N}_v,\nonumber\\
\bar{\cal N}_v\sigma^{\mu\nu}{\cal N}_v
&=&2\epsilon^{\mu\nu\rho\sigma}v_\rho \bar{\cal N}_v S_\sigma^v
{\cal N}_v,\nonumber\\
\bar{\cal N}_v\sigma^{\mu\nu}\gamma_5{\cal N}_v
&=&2i (v^\mu \bar{\cal N}_v S^\nu_v{\cal N}_v
-v^\nu\bar{\cal N}_v S^\mu_v{\cal N}_v).
\end{eqnarray}
   For example,
\begin{displaymath}
\bar{\cal N}_v\gamma_5 {\cal N}_v=\bar{\cal N}_v\gamma_5 v\hspace{-.5em}/
\hspace{.2em}
{\cal N}_v=-\bar{\cal N}_v v\hspace{-.5em}/\hspace{.2em}\gamma_5{\cal N}_v
=-\bar{\cal N}\gamma_5{\cal N}_v=0,
\end{displaymath}
where we made use of Eq.\ (\ref{5:5:nhprop}).
   The remaining relations are shown analogously.
   Eqs.\ (\ref{5:5:nvbilineare}) result in a nice simplification of the
Dirac structures in the heavy-baryon approach, because one ultimately only
deals with two groups of $4\times 4$ matrices, the unit matrix and 
$S^\mu_v$, instead of the original six groups on the left-hand side
of Eq.\ (\ref{5:5:nvbilineare}).

   Expanding Eq.\ (\ref{5:5:lhneff}) formally into a series in $1/m$, 
\begin{equation}
\label{5:5:lhneffexp}
{\cal L}_{\rm eff}=\bar{\cal N}_v\left(iv\cdot D
+\frac{g_A}{2}u\hspace{-.5em}/_\bot\gamma_5\right){\cal N}_v
+\sum_{n=1}^\infty \frac{1}{(2m)^n}{\cal L}_{{\rm eff},n},
\end{equation}
and applying Eq.\ (\ref{5:5:nvbilineare}), the
lowest-order term reads
\begin{equation}
\label{5:5:l1pinhb}
\widehat{\cal L}^{(1)}_{\pi N}=
\bar{\cal N}_v\left[iv\cdot D
+g_A S_v\cdot u \right]{\cal N}_v,
\end{equation}
   where we made use of the fourth relation of 
Eq.\ (\ref{5:5:nvbilineare}) and the first relation of 
Eq.\ (\ref{5:5:seig}).
(Recall that the second term of Eq.\ (\ref{5:5:lhneff}) is of order $1/m$.)
   Equation (\ref{5:5:l1pinhb}) represents the lowest-order Lagrangian 
of heavy-baryon chiral perturbation theory (HBChPT), indicated by
the symbol\, $\widehat{}$\,\,.
   When comparing with the relativistic Lagrangian of Eq.\ (\ref{5:4:lpin1}),
we see that the nucleon mass has disappeared from the leading-order
Lagrangian.
   It only shows up in higher orders as powers of $1/m$, as will be discussed
in Sec.\ \ref{subsec_cfo1om}.
   In the power counting scheme $\widehat{\cal L}^{(1)}_{\pi N}$ counts as
${\cal O}(p)$, because the covariant derivative $D_\mu$ and the chiral
vielbein $u_\mu$ both count as ${\cal O}(p)$.

   When calculating loop diagrams with the Lagrangian of Eq.\ 
(\ref{5:5:l1pinhb}) one will encounter divergences which are treated
in the framework of (normal) dimensional regularization
[see Eq.\ (\ref{5:4:gammarelations})].
   Since the definition of the spin matrix $S^\mu_v$ contains $\gamma_5$
and the commutator of two such spin matrices, in four dimensions, 
involves the epsilon tensor, one needs some convention for dealing with 
products of spin matrices when evaluating integrals in $n$ dimensions.
   To be on the safe side, we always reduce the gamma matrices using only 
the rules of Eq.\ (\ref{5:4:gammarelations}).
   Let us consider the following example, which appears in the calculation
of the pion-nucleon form factor at the one-loop level:\footnote{In evaluating
Eq.\ (\ref{5:5:sss}), we made use of 
\begin{displaymath}
\gamma_\mu a\hspace{-.5em}/ \gamma^\mu
=(2-n)a\hspace{-.5em/},\quad
\gamma_\mu a\hspace{-.5em}/\hspace{.2em}
b\hspace{-.5em}/\hspace{.2em} \gamma^\mu=4 a\cdot b
+(n-4)a\hspace{-.5em}/\hspace{.2em}b\hspace{-.5em}/\hspace{.2em},
\end{displaymath}
in $n$ dimensions.}
\begin{eqnarray}
\label{5:5:sss}
S_\mu^v S_\nu^v S_v^\mu&=&-\frac{1}{8}\gamma_5(\gamma_\mu v\hspace{-.5em}/
-v_\mu)\gamma_5(\gamma_\nu v\hspace{-.5em}/-v_\nu)
\gamma_5(\gamma^\mu v\hspace{-.5em}/
-v^\mu)\nonumber\\
&=&-\frac{1}{8}(n-3)\gamma_5(\gamma_\nu v\hspace{-.5em}/-v_\nu)
=\frac{n-3}{4} S_\nu^v, 
\end{eqnarray}
   where we consistently made use of the anticommutation relations
of Eq.\ (\ref{5:4:gammarelations}).
   In contrast, using the anticommutator and commutator of Eq.\ 
(\ref{5:5:seig}), one ends up with $S_\nu^v/4$ which only coincides with 
Eq.\ (\ref{5:5:sss}) for $n=4$.
   However, the factor $(n-3)$ needs to be written as $1+(n-4)$ when it is 
multiplied with a singularity of the form $C/(n-4)$
in order to produce the constant non-divergent term $C$ in the product.  
   Similarly, using the conventions of Eq.\ (\ref{5:4:gammarelations}),
the squared spin operator in $n$ dimensions reads 
\begin{equation}
\label{5:5:s2}
S^2_v=\frac{1-n}{4}.
\end{equation}

\subsection{Normalization of Fields and States}
\label{subsec_nfs}

   So far we have calculated matrix elements of the relativistic Lagrangian
${\cal L}^{(1)}_{\pi N}$ of Eq.\ (\ref{5:2:l1pin}) in the framework of 
covariant perturbation theory 
based on the formula of Gell-Mann and Low \cite{Gell-Mann:1951rw} in 
combination with Wick's theorem \cite{Wick:ee}.
   Let us recall that, for a generic field $\Phi(x)$ described
by the Lagrangian
\begin{displaymath}
{\cal L}={\cal L}_0+{\cal L}_{\rm int},
\end{displaymath}   
the ``magic formula of covariant perturbation theory'' 
\cite{Haag:hx} allows one to calculate the Green functions
\begin{equation}
\label{5:5:tau}
\tau_n(x_1,\cdots,x_n)=\langle \Omega|T[\Phi(x_1)\cdots\Phi(x_n)]|\Omega\rangle
\end{equation}
in terms of the generating functional 
\begin{eqnarray}
\label{5:5:tgf}
{\cal T}[f]&=&\frac{N[f]}{N[0]},\\
\label{5:5:nf}
N[f]&=&\langle\Omega_0|T\exp\left\{i \int d^4 x \left[\Phi^0(x) f(x) + 
{\cal L}^0_{\rm int}(x)\right]\right\}|\Omega_0\rangle.
\end{eqnarray}
   While the Green functions of Eq.\ (\ref{5:5:tau}) involve the interacting
field $\Phi(x)$ and the vacuum $\Omega$ of the corresponding interacting 
theory, the formula of Gell-Mann and Low, in principle, provides an explicit 
expression for the generating functional in terms of the quantities 
$\Phi^0(x)$ and $\Omega_0$ defined in the free theory.
   Note that the Green functions of Eq.\ (\ref{5:5:tau}) are expressed in 
terms of the bare fields and, in the end, still have to be renormalized
(see, e.g, Sec.\ \ref{subsec_mgb}).

   Here we want to address the question of how to establish contact between
matrix elements evaluated perturbatively using the relativistic Lagrangian of 
Eq.\ (\ref{5:2:l1pin}), on the one hand, and the heavy-baryon Lagrangian of 
Eq.\ (\ref{5:5:lhneff}) on the other hand.
   The presentation will make use of the ideas developed in Refs.\ 
\cite{Dugan:1991ak,Balk:1993ev}, where 
this issue was discussed for the case of a heavy-quark effective theory.
   A different route was taken in Refs.\ 
\cite{Ecker:1997dn,Steininger:1998ya,Kambor:1998pi}, where the 
path-integral approach to the generating functional was used to 
define the wave function renormalization constant (including interaction).
   Later on we will explicitly see that the two approaches 
yield identical results at ${\cal O}(p^3)$.

   For later comparison with the heavy-baryon approach, we first need to 
collect a few properties of the free Dirac field operator 
$\Psi^0(x)$ which we decompose in the standard fashion in terms of the
solutions of the free Dirac equation\footnote{For the sake 
of simplicity, we consider one generic fermion field.} 
\begin{equation}
\label{5:5:psifeld}
\Psi^0(x)=\sum_{\alpha=1}^2 \int \frac{d^3 p}{(2\pi)^3 2 E(\vec{p}\,)}
\left[b_\alpha(\vec{p}\,) u^{(\alpha)}(\vec{p}\,)e^{-ip\cdot x}
+d^\dagger_{\alpha}(\vec{p}\,) v^{(\alpha)}(\vec{p}\,) 
e^{ip\cdot x}\right],
\end{equation}
where $p_0=E(\vec{p}\,)=\sqrt{m^2+\vec{p}\,^2}$.
   The Dirac spinors have the following properties
\begin{eqnarray}
\label{5:5:spinorprops}
(p\hspace{-.5em}/-m)u^{(\alpha)}(\vec{p}\,)&=&0,\nonumber\\
(p\hspace{-.5em}/+m)v^{(\alpha)}(\vec{p}\,)&=&0,\nonumber\\
\bar{u}^{(\alpha)}(\vec{p}\,) u^{(\beta)}(\vec{p}\,)&=&
-\bar{v}^{(\alpha)}(\vec{p}\,)v^{(\beta)}(\vec{p}\,)
=2m\delta_{\alpha\beta},\nonumber\\
u^{(\alpha)\dagger}(\vec{p}\,)u^{(\beta)}(\vec{p}\,)
&=&v^{(\alpha)\dagger}(\vec{p}\,)v^{(\beta)}(\vec{p}\,)
=2 E(\vec{p}\,)\delta_{\alpha\beta},\nonumber\\
u^{(\alpha)\dagger}(\vec{p}\,)v^{(\beta)}(-\vec{p}\,)&=&
\bar{u}^{(\alpha)}(\vec{p}\,)v^{(\beta)}(\vec{p}\,)
=0.
\end{eqnarray}
   The creation operators $b_\alpha^\dagger$ and $d_\alpha^\dagger$ 
(annihilation operators $b_\alpha$ and $d_\alpha$) of particles and
antiparticles, respectively, satisfy the 
anticommutation relations 
\begin{eqnarray*}
\{b_\alpha(\vec{p}\,),b_\beta^\dagger(\vec{p}\,')\}&=&(2\pi)^3 2 E(\vec{p}\,)
\delta^3(\vec{p}-\vec{p}\,')\delta_{\alpha\beta},\\
{\{}d_\alpha(\vec{p}\,),d^\dagger_\beta(\vec{p}\,')\}&=&(2\pi)^3 2 E(\vec{p}\,)
\delta^3(\vec{p}-\vec{p}\,')\delta_{\alpha\beta},
\end{eqnarray*}
where all remaining anticommutators such as, e.g., 
${\{}b_\alpha(\vec{p}\,),b_\beta(\vec{p}\,')\}$ vanish.
   With this convention, single-particle states $|\vec{p},\alpha,+\rangle
=b_\alpha^\dagger(\vec{p}\,)|0\rangle$ are normalized as
\begin{equation}
\label{5:5:statenormalization}
\langle \vec{p}\,',\beta,+|\vec{p},\alpha,+\rangle
=(2\pi)^3 2 E(\vec{p}\,) \delta^3(\vec{p}\,'-\vec{p}\,)\delta_{\alpha\beta}.
\end{equation}

   Let us now turn to the heavy-baryon formulation and consider the 
leading-order term of Eq.\ (\ref{5:5:l1pinhb}) which we write as 
\begin{equation}
\label{5:5:widehatl}
\widehat{\cal L}^{(1)}_{\pi N}=
\bar{\cal N}_v iv\cdot\partial {\cal N}_v
+\widehat{\cal L}_{\rm int}\equiv
\widehat{\cal L}_{0}+\widehat{\cal L}_{\rm int}
\end{equation}
   for later use in the formula of Gell-Mann and Low.
   We decompose the solution to the free equation of motion
\begin{equation}
\label{5:5:loeom}
iv\cdot\partial {\cal N}^0_v(x)=0,\quad 
v\hspace{-.5em}/\hspace{.2em}{\cal N}^0_v={\cal N}^0_v,
\end{equation}
as 
\begin{equation}
\label{5:5:n0vdec}
{\cal N}^0_v(x)=\sum_{\alpha=1}^2\int \frac{d^3 k}{(2\pi)^3 2m v_0}
e^{-ik\cdot x}b_{\alpha,v}(\vec{k}) u^{(\alpha)}_v,
\end{equation}
where $k_0$, at leading order, is defined through $v\cdot k=0$ in order
to satisfy Eq.\ (\ref{5:5:loeom}), i.e., 
$$
k_0=\frac{\vec{v}\cdot\vec{k}}{v_0},\quad
k\cdot x=-\vec{k}\cdot\left(\vec{x}-\vec{v}\,\frac{x_0}{v_0}\right).
$$
   The spinors are given by
\begin{equation}
\label{5:5:ualphav0}
u^{(\alpha)}_v=\sqrt{\frac{2m}{v_0}}P_{v+}\left(
\begin{array}{c}
\chi^{(\alpha)}\\
0_{2\times 1}
\end{array}
\right),\quad
\bar{u}^{(\alpha)}_v u^{(\beta)}_v=2m\delta_{\alpha\beta},
\end{equation}
with $\chi^{(\alpha)}$ ordinary two-component Pauli spinors.
   Note that, at lowest order in $1/m$, the spinors do not depend on the 
residual momentum $\vec{k}$.
   Moreover, for an arbitrary choice of $v$, the $u^{(\alpha)}_v$ are 
four-component objects which only for the special case $v^\mu=(1,0,0,0)$ 
effectively reduce to two-component spinors. 
   As will be shown below, the operators $b_{\alpha,v}(\vec{k})$ and 
$b_{\alpha,v}^\dagger(\vec{k})$ destroy and create a nucleon 
(isospin index suppressed) with residual
three-momentum $\vec{k}$.
   They satisfy the anticommutation relations
\begin{equation}
\label{5:5:banticomm}
\{b_{\alpha,v}(\vec{k}), b_{\beta,v}^\dagger(\vec{k}')\}=
2m v_0 (2\pi)^3\delta^3(\vec{k}-\vec{k}\,')\delta_{\alpha\beta},
\end{equation}
where, as usual, the anticommutator of two annihilation or two
creation operators vanishes.
   Accordingly, the single-particle states are normalized as 
\begin{equation}
\label{5:5:statenormalizationhb}
\langle v,\vec{k}\,',\beta|v,\vec{k},\alpha\rangle
=2 m v_0 (2\pi)^3  \delta^3(\vec{k}\,'-\vec{k}\,)\delta_{\alpha\beta}.
\end{equation}
   Note that the normalization of the states
of Eqs.\ (\ref{5:5:statenormalization}) and (\ref{5:5:statenormalizationhb})
coincide only at leading order in $1/m$  (or as $m\to \infty$).
   
  Using Eq.\ (\ref{5:5:banticomm}) it is straightforward to
verify that the (free) theory has been quantized ``canonically''
\cite{Dugan:1991ak}, 
i.e.
\begin{displaymath}
\{\Pi_v^0(\vec{x},t),{\cal N}_v^0(\vec{y},t)\}
=iv_0\{\bar{\cal N}_v^0(\vec{x},t),{\cal N}_v^0(\vec{y},t)\}
=i\delta^3(\vec{x}-\vec{y})P_{v+},
\end{displaymath}
where $\Pi_v^0=\partial\widehat{\cal L}_0/\partial(\partial_0 {\cal N}_v)
=iv_0\bar{\cal N}^0_v$
is the momentum conjugate to ${\cal N}^0_v$ and we
made use of the completeness relation
\begin{equation}
\label{5:5:completeness}
\sum_{\alpha=1}^2 u^{(\alpha)}_v\bar{u}^{(\alpha)}_v=2m P_{v+}.
\end{equation}
   Constructing the energy-momentum tensor corresponding to 
$\widehat{\cal L}_{0}$,
\begin{displaymath}
\Theta^{\mu\nu}_v=\partial^\nu\bar{\cal N}_v
\frac{\partial\widehat{\cal L}_0}{\partial(\partial_\mu\bar{\cal N}_v)}
+\frac{\partial\widehat{\cal L}_0}{\partial(\partial_\mu{\cal N}_v)}
\partial^\nu{\cal N}_v-g^{\mu\nu}\widehat{\cal L}_0=
\bar{\cal N}^0_v iv^\mu\partial^\nu {\cal N}_v,
\end{displaymath}
(we made use of the equation of motion)
we obtain for the four-momentum operator
\begin{equation}
\label{5:5:kmuv}
K^\mu_v=\int d^3 x \bar{\cal N}^0_v iv_0\partial^\mu {\cal N}_v
=\sum_{\alpha=1}^2\int \frac{d^3 k}{(2\pi)^3 2m v_0}
k^\mu b_{\alpha,v}^\dagger(\vec{k}) b_{\alpha,v}(\vec{k}).
\end{equation}
   Using $[ab,c]=a\{b,c\}-\{a,c\}b$ it is then straightforward to verify
the commutation relations
\begin{equation}
\label{5:5:kmucomrel}
[K^\mu_v,b_{\alpha,v}(\vec{k})]=-k^\mu b_{\alpha,v}(\vec{k}),\quad
[K^\mu_v,b_{\alpha,v}^\dagger(\vec{k})]=k^\mu b_{\alpha,v}^\dagger(\vec{k}).
\end{equation}
   Eq.\ (\ref{5:5:kmucomrel}) implies that $b_{\alpha,v}(\vec{k})$
and $b_{\alpha,v}^\dagger(\vec{k})$ destroy and create quanta
with (residual) four-momentum $k^\mu$ and total four-momentum
$p^\mu=mv^\mu+k^\mu$.
   This can be seen by comparing with the four-momentum operator of the
free relativistic theory
\begin{equation}
P^\mu=\int d^3 x \bar{\Psi}^0\gamma^0 i\partial^\mu {\Psi}^0,
\end{equation} 
which, to leading order in $1/m$, is related to Eq.\ (\ref{5:5:kmuv})
by $P^\mu=K^\mu_v+m v^\mu N$, where $N=\int d^3x \Psi^{0\dagger}\Psi^0$ is the 
number operator \cite{Dugan:1991ak}.

   Using the orthogonality relations of Eq.\ (\ref{5:5:ualphav0}) 
we may express 
the creation and annihilation operators in terms of the fields in the standard
way
\begin{eqnarray}
\label{5:5:bbd}
b_{\alpha,v}^\dagger(\vec{k})&=&v_0\int d^3 x\bar{\cal N}_v(x)u_v^{(\alpha)}
e^{-ik\cdot x},\nonumber\\
b_{\alpha,v}(\vec{k})&=&v_0\int d^3 x e^{ik\cdot x}\bar{u}^{(\alpha)}_v
{\cal N}_v(x).
\end{eqnarray}
   Eqs.\ (\ref{5:5:bbd}) are the starting point for the LSZ reduction
\cite{Lehmann:1955rq,Bjorken_1964qf,Itzykson:rh}
in the framework of the heavy-baryon approach.
   We consider the matrix element of Eq.\ (\ref{4:btbta}) 
for the transition in the presence of external fields
$v$, $a$, $s$, and $p$ (we omit spin
and isospin labels)
\begin{eqnarray}
\label{5:5:ffunc}
\lefteqn{{\cal F}(\vec{p}\,',\vec{p};v,a,s,p)=\langle {\vec{p}\,'};{\rm out}|
{\vec{p}\,};{\rm in}\rangle^{\rm c}_{v,a,s,p}}\nonumber\\
&=&\sqrt{\frac{E}{mv_0}}
\langle {\vec{p}\,'};{\rm out}|b_{v,\rm in}^\dagger(\vec{k})|\Omega
\rangle^{\rm c}_{v,a,s,p}\nonumber\\
&=&\sqrt{\frac{E}{mv_0}}
v_0\int d^3 x\langle {\vec{p}\,'};{\rm out}|\bar{\cal N}_{v,\rm in}(x)|
\Omega\rangle^{\rm c}_{v,a,s,p}u_v
e^{-ik\cdot x}\nonumber\\
&=&\sqrt{\frac{E}{mv_0}}
\lim_{t\to-\infty}v_0\int d^3 x\langle {\vec{p}\,'};{\rm out}|
\frac{\bar{\cal N}_v(x)}{\sqrt{Z}}|
\Omega\rangle^{\rm c}_{v,a,s,p}u_v
e^{-ik\cdot x}\nonumber\\
&=&\cdots\nonumber\\
&=&\left(\frac{-i}{\sqrt{Z}}\right)^2
N N'
\int d^4x d^4 y \nonumber\\
&&\times e^{ik'\cdot y}\bar{u}_v iv\cdot \stackrel{\rightarrow}{
\partial_y}\langle\Omega|T[{\cal N}_v(y)\bar{\cal N}_v(x)]|\Omega\rangle^{
\rm c}_{v,a,s,p}(-iv\cdot\stackrel{\leftarrow}{\partial_x})
u_v e^{-ik\cdot x}.\nonumber\\
\end{eqnarray}
   The intermediate steps indicated by $\cdots$ proceed in complete analogy
to the usual reduction formula as described in, e.g., Refs.\ 
\cite{Bjorken_1964qf,Itzykson:rh}.
   In Eq.\ (\ref{5:5:ffunc}), the factors of the type $N=\sqrt{E/(mv_0)}$ are
related to the relative normalization of the states [see Eq.\ 
(\ref{5:5:statenormalization}) vs.~(\ref{5:5:statenormalizationhb})],
whereas $\sqrt{Z}$ refers to the wave function renormalization in
the framework of the heavy-baryon Lagrangian.

   The Green function entering Eq.\ (\ref{5:5:ffunc}) will be calculated
perturbatively using the formula of Gell-Mann and Low
\cite{Gell-Mann:1951rw},
\begin{equation}
\label{5:5:gmlhb}
\langle\Omega|T[{\cal N}_v(y)\bar{\cal N}_v(x)]|\Omega\rangle^{
\rm c}_{v,a,s,p}
=\langle\Omega_0|T[{\cal N}_v^0(y)\bar{\cal N}_v^0(x)
\exp\left(i\int d^4 z \widehat{\cal L}\,^0_{\rm int}(z)\right)
]|\Omega_0\rangle^{
\rm c},
\end{equation}
where, on the right-hand side, $|\Omega_0\rangle$ denotes the vacuum of
the free theory, and the external fields are part of the Lagrangian
$\widehat{\cal L}\,^0_{\rm int}(z)$.\footnote{Strictly speaking we should
also include the mesonic Lagrangian.}

\subsection{Propagator at Lowest Order}
\label{subsec_plo}

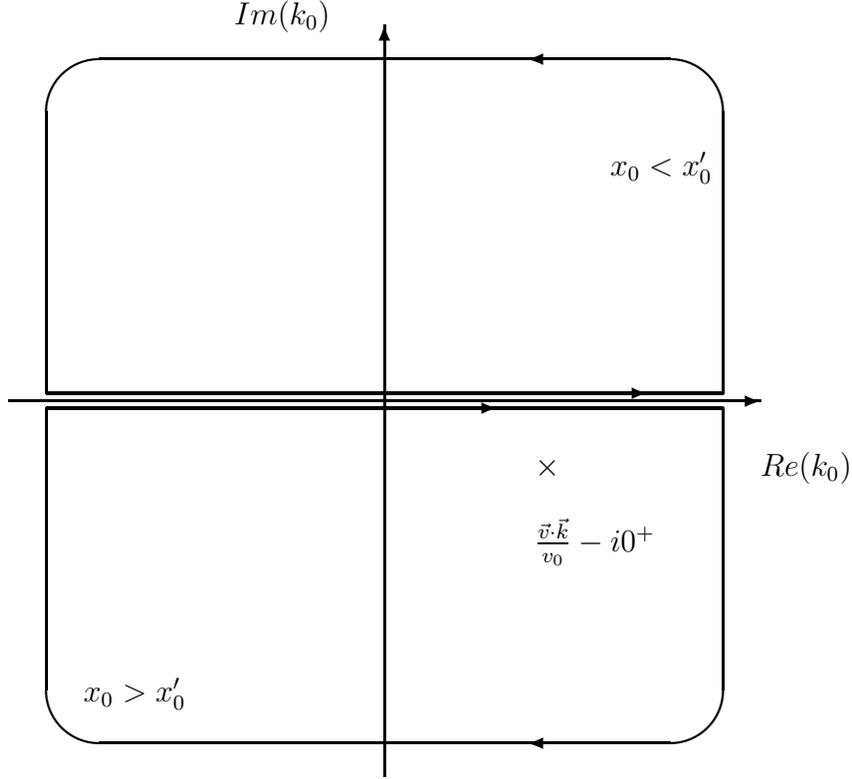
\begin{figure}[htb]
\unitlength1cm
\begin{center}
\caption{\label{5:5:fig:integration} Contour integration in the complex
$k_0$ plane.}
\vspace{1cm}
\begin{picture}(10,10)
\thicklines
\put(0,5){\vector(1,0){10}}
\put(10,4){$Re(k_0)$}
\put(0.5,4.9){\line(1,0){9}}
\put(0.5,5.1){\line(1,0){9}}
\put(5.5,4.9){\vector(1,0){1}}
\put(7.5,5.1){\vector(1,0){1}}
\put(5,0){\vector(0,1){10}}
\put(3,10){$Im(k_0)$}
\put(7,4){$\times$}
\put(7,3){\mbox{$\frac{\vec{v}\cdot\vec{k}}{v_0}-i0^+$}}
\put(5,5.1){\oval(9,8.9)[tl]}
\put(5,5.1){\oval(9,8.9)[tr]}
\put(8,8){$x_0<x_0'$}
\put(5,4.9){\oval(9,8.9)[br]}
\put(5,4.9){\oval(9,8.9)[bl]}
\put(1,1){$x_0>x_0'$}
\put(7,9.55){\vector(-1,0){0.1}}
\put(7,0.45){\vector(-1,0){0.1}}
\end{picture}
\end{center}
\end{figure}

   We will now discuss the propagator of the lowest-order Lagrangian both
on the ``classical level'' as well as in the quantized theory of the last 
section.
   The lowest-order equation of motion corresponding to Eq.\ 
(\ref{5:5:l1pinhb}) reads
\begin{equation}
\label{5:5:l1pinhbeom}
(iv\cdot D+g_A S_v\cdot u){\cal N}_v=0, \quad P_{v+} {\cal N}_v={\cal N}_v,
\end{equation}   
where the second relation implies $P_{v-}{\cal N}_v=0$ 
[see Eq.\ (\ref{5:5:nhprop})].
   We define the propagator corresponding to 
Eq.\ (\ref{5:5:l1pinhbeom}) through
\begin{equation}
\label{5:5:prop}
(iv\cdot D+g_A S_v\cdot u)G_v(x,x')=P_{v+}\delta^4(x-x'),\quad
P_{v-}G_v(x,x')=0. 
\end{equation}
   In order to solve Eq.\ (\ref{5:5:l1pinhbeom}) perturbatively, we
re-write the equation of motion in the standard form as
$$iv\cdot \partial {\cal N}_v(x)=V(x){\cal N}_v(x),$$
where $V$ denotes the interaction term,
and search for the unperturbed 
Green function $G_v^0(x,x')$ satisfying the properties
\begin{eqnarray}
\label{5:5:greenprop1}
i v\cdot \partial G_v^0(x,x')&=&\delta^4(x-x')P_{v+},\\
\label{5:5:greenprop2}
P_{v-} G_v^0(x,x')&=&0,\\
\label{5:5:greenprop3}
G_v^0(x,x')&=&0\,\,\,\mbox{for}\,\,\,x_0'>x_0.
\end{eqnarray}
   In terms of $G^0_v$ the propagator $G_v$ is then given by
\begin{displaymath}
G_v(x,x')=G_v^0(x,x')+\int d^4 y G^0_v(x,y)V(y) G_v(y,x').
\end{displaymath}
   Inserting the standard ansatz in terms of a Fourier decomposition
\begin{equation}
\label{5:5:g0f}
G_v^0(x,x')=\int \frac{d^4 k}{(2\pi)^4}e^{-ik\cdot(x-x')}G_v^0(k)
\end{equation}
into Eq.\ (\ref{5:5:greenprop1}), 
$$
\int \frac{d^4 k}{(2\pi)^4}e^{-ik\cdot(x-x')} v\cdot k G_v^0(k)
=\delta^4(x-x')P_{v+}=
\int \frac{d^4 k}{(2\pi)^4}e^{-ik\cdot(x-x')}P_{v+},
$$
   we obtain by comparing both sides
$$
G_v^0(k)=\frac{P_{v+}}{v\cdot k}\,\,
\mbox{for}\,\, v\cdot k\neq 0.
$$
   The boundary condition of Eq.\ (\ref{5:5:greenprop3}) may be incorporated
by introducing an infinitesimally small imaginary part into the denominator:
\begin{equation}
\label{5:5:gv0k}
G_{v}^0(k)=\frac{P_{v+}}{v\cdot k +i0^+}.
\end{equation}
   That this is indeed the correct choice is easily seen by evaluating
the integral
\begin{displaymath}
\int_{-\infty}^\infty\frac{dk_0}{2\pi}e^{-ik_0(x_0-x'_0)}\frac{1}{
k_0-\frac{\vec{v}\cdot\vec{k}}{v_0}+i0^+}
=-i\Theta(x_0-x_0')
\exp\left[-i\frac{(x_0-x_0')\vec{v}\cdot\vec{k}}{v_0}\right]
\end{displaymath}
as a contour integral in the complex $k_0$ plane 
(see Fig.\ \ref{5:5:fig:integration}).
   For $x_0>x_0'$ the contour is closed in the lower half plane and one makes
use of the residue theorem. 
   On the other hand, for $x_0<x_0'$ the contour is closed in the upper
half plane and, since the contour does not contain a pole, the integral 
vanishes.
   We then obtain
\begin{eqnarray}
\label{5:5:gv0f}
G_v^0(x,x')&=&-i\frac{\Theta(x_0-x_0')}{v_0}
\int \frac{d^3 k}{(2\pi)^3}
\exp\left[i\vec{k}\cdot\left(\vec{x}-\vec{x}\,'-\vec{v}\,\frac{x_0-x_0'}{v_0}
\right)\right]
P_{v+}\nonumber\\
&=&-i\frac{\Theta(x_0-x_0')}{v_0}\delta^3\left(\vec{x}-\vec{x}\,'-\vec{v}\,
\frac{x_0-x_0'}{v_0}\right)P_{v+}.
\end{eqnarray}
   For the special choice $v^\mu=(1,0,0,0)\equiv{\tilde v}^\mu$ the 
propagator reduces to that of a static source 
\begin{displaymath}
G_{\tilde{v}}^0(x,x')=-i\Theta(x_0-x_0')
\delta^3(\vec{x}-\vec{x}\,') 
\left(
\begin{array}{cc}
1_{2\times 2}&0_{2\times 2}\\
0_{2\times 2}&0_{2\times 2}
\end{array}
\right).
\end{displaymath}

   Finally, it is easy to show that a definition of the propagator in terms 
of the field operators ${\cal N}^0_v$ and $\bar{\cal N}^0_v$
\cite{Dugan:1991ak},
\begin{equation}
G^0_v(x,x')=-i\Theta(x_0-x_0')\langle\Omega_0|{\cal N}_v^0(x)
\bar{\cal N}^0_v(x')|\Omega_0\rangle,
\end{equation}
yields the same result as Eq.\ (\ref{5:5:gv0f}).
   To that end, one inserts for each of the two operators a sum according
to Eq.\ (\ref{5:5:n0vdec}), commutes the creation and annihilation operators
using Eq.\ (\ref{5:5:banticomm}), applies the completeness relation
of Eq.\ (\ref{5:5:completeness}), and makes use of $v\cdot k=0$ for the
individual Fourier components.
   Performing the remaining integration over $\vec{k}$ one ends up with
Eq.\ (\ref{5:5:gv0f}), i.e., as expected the two methods yield the same
result.

\subsection{Example: $\pi N$ Scattering at Lowest Order}
   As a simple example, let us return to $\pi N$ scattering, but now
in the framework of the heavy-baryon Lagrangian of Eq.\ (\ref{5:5:l1pinhb}). 
   The four-momenta of the initial and final nucleons are written as
$p=\,\stackrel{\circ}{m}_N v+k$ and $p'=\,\stackrel{\circ}{m}_N v+k'$, 
respectively, with $v\cdot k=0=v\cdot k'$ to leading order in 
$1/\stackrel{\circ}{m}_N$.  
   The relevant interaction Lagrangian is obtained in complete analogy
to Eq.\ (\ref{5:3:lpin}), 
\begin{equation}
\label{5:5:lpinhb}
\widehat{\cal L}^{(1)}_{\rm int}=-\frac{\stackrel{\circ}{g}_A}{F_0}
\bar{\cal N}_v S^\mu_v \vec{\tau}\cdot\partial_\mu\vec{\phi}{\cal N}_v
-\frac{1}{4 F_0^2} v^\mu \bar{\cal N}_v\vec{\tau}\cdot\vec{\phi}\times
\partial_\mu\vec{\phi}{\cal N}_v,
\end{equation}
and the corresponding Feynman rules for the vertices derived from  
Eq.\ (\ref{5:5:lpinhb}) read
\begin{itemize}
\item for a single incoming pion with four-momentum $q$ and Cartesian isospin
index $a$: 
\begin{equation}
\label{5:5fr1}
-\frac{\stackrel{\circ}{g}_A}{F_0} S_v\cdot q \tau^a,
\end{equation}
\item for an incoming pion with $q,a$ and an outgoing pion with $q',b$:
\begin{equation}
\label{5:6:fr2}
\frac{v\cdot(q+q')}{4 F^2_0}\epsilon_{abc}\tau^c.
\end{equation}
\end{itemize}
   As in the case of the relativistic calculation of Sec.\ 
\ref{subsec_apnstl} the latter gives rise to a contact contribution to 
${\cal M}^v$
\begin{eqnarray}
\label{5:5:mconthb}
{\cal M}_{\rm cont}^v&=&
N' N
\bar{u}_v'\frac{v\cdot(q+q')}{4 F^2_0}\epsilon_{abc}\tau^c u_v,
\end{eqnarray}
where the spinors are given in Eq.\ (\ref{5:5:ualphav0}) and $N$ and $N'$
are the normalization factors appearing in the reduction formula of
Eq.\ (\ref{5:5:ffunc}).
   The result for the direct-channel nucleon pole term reads 
\begin{eqnarray}
\label{5:5:mdcconthb}
{\cal M}_{\rm d}^v=-i\frac{\stackrel{\circ}{g}_A^2}{F_0^2}
N'N
\tau^b\tau^a \bar{u}_v'S_v\cdot q' 
\frac{P_{v+}}{v\cdot(k+q)} S_v\cdot{q} u_v,
\end{eqnarray}
   where, at leading order, we can make use of $v\cdot k=0$.
   The crossed channel is obtained from Eq.\ (\ref{5:5:mdcconthb}) by
the replacement $a\leftrightarrow b$ and $q\leftrightarrow -q'$
(pion crossing).

   The evaluation of the total matrix element ${\cal M}^v=
{\cal M}^v_{\rm cont}+{\cal M}^v_{\rm d}+{\cal M}^v_{\rm c}$
is particularly simple for the special choice
$v^\mu=(1,0,0,0)\equiv\tilde{v}^\mu$, for which we have
\begin{displaymath}
P_{\tilde{v}+}=\left(\begin{array}{cc}1_{2\times 2}&0_{2\times 2}\\
0_{2\times 2}&0_{2\times 2}\end{array}\right).
\end{displaymath}
   In that case, the calculation effectively reduces to that of
a two-component
theory as in the Foldy-Wouthuysen transformation, because the $4\times 4$ 
matrices of the vertices are multiplied both from the left and the right by 
$P_{\tilde{v}+}$ originating from either the propagator of 
Eq.\ (\ref{5:5:gv0k}) or the spinors of Eq.\ (\ref{5:5:ualphav0}).
   To be specific, for a $4\times 4$ matrix $\Gamma$ of the type
\begin{displaymath}
\Gamma=
\left(\begin{array}{cc}
A&B\\
C&D
\end{array}
\right),
\end{displaymath}
where each block $A$, $B$, $C$, and $D$ is a $2\times 2$ matrix, one has
\begin{displaymath}
P_{\tilde{v}+}\Gamma P_{\tilde{v}+}=
\left(\begin{array}{cc}
A&0_{2\times 2}\\
0_{2\times 2}&0_{2\times 2}
\end{array}
\right) 
\end{displaymath}
and 
\begin{displaymath}
P_{\tilde{v}+}\Gamma_1P_{\tilde{v}+}\Gamma_2P_{\tilde{v}+}=
\left(\begin{array}{cc}
A_1 A_2&0_{2\times 2}\\
0_{2\times 2}&0_{2\times 2}
\end{array}
\right).
\end{displaymath}
   Moreover, the spin matrix of Eq.\ (\ref{5:5:spinop}) is very simple
for $\tilde{v}$,
\begin{equation}
\label{5:5:spinopvtilde}
S^0_{\tilde{v}}=0,\quad
\vec{S}_{\tilde{v}}=\frac{1}{2}\vec{\Sigma},
\end{equation}
where $\vec{\Sigma}$ has been defined in Eq.\ (\ref{5:5:Sigma}). 
  With this special choice of $v$ the $T$ matrix in the center-of-mass
frame reads
\begin{equation}
\label{5:5:tmatrix}
T=2\stackrel{\circ}{m}_N 
\chi'^\dagger\left[-i\epsilon_{abc}\tau^c\left(\frac{E_\pi}{2F_0^2}
-\frac{\stackrel{\circ}{g}_A^2}{F_0^2}\frac{\vec{q}\cdot\vec{q}\,'}{2E_\pi}
\right)
+\delta^{ab}\frac{\stackrel{\circ}{g}_A^2}{F_0^2}
\left(-\frac{i\vec{\sigma}\cdot\vec{q}\,'\times\vec{q}}{2E_\pi}\right)\right]
\chi.
\end{equation}
   Performing a nonrelativistic reduction of Eq.\ (\ref{5:3:mpinpar}) in
the center-of-mass frame,
\begin{equation}
\label{5:5:tnonrelred}
T=2\stackrel{\circ}{m}_N \chi'^\dagger\left[A+ \left(E_\pi
+\frac{\vec{q}\,^2+\vec{q}\,'\cdot\vec{q}}{2\stackrel{\circ}{m}_N}\right)B
+i\frac{\vec{\sigma}\cdot \vec{q}\,'\times\vec{q}}{2
\stackrel{\circ}{m}_N} B +\cdots\right]
\chi,
\end{equation}
and using
\begin{eqnarray*}
\frac{1}{\nu-\nu_B}+\frac{1}{\nu+\nu_B}&=&\frac{1}{E_\pi}\left[2-
\frac{\vec{q}\,^2+\vec{q}\cdot\vec{q}\,'}{E_\pi
\stackrel{\circ}{m}_N  }+O\left(\frac{1}{\stackrel{\circ}{m}_N^2}\right)
\right],\\
\frac{1}{\nu-\nu_B}-\frac{1}{\nu+\nu_B}&=&\frac{1}{E_\pi}\left[
-\frac{E_\pi}{\stackrel{\circ}{m}_N}+
\frac{\vec{q}\cdot\vec{q}\,'}{E_\pi\stackrel{\circ}{m}_N}
+O\left(\frac{1}{\stackrel{\circ}{m}_N^2}\right)\right]
\end{eqnarray*}
in the expansion of $A$ and $B$ of Table \ref{5:3:tableresults}, one verifies 
that, at leading order in $1/\!\!\stackrel{\circ}{m}_N$, 
the relativistic Lagrangian of 
Eq.\ (\ref{5:2:l1pin}) and the heavy-baryon Lagrangian of Eq.\ 
(\ref{5:5:l1pinhb}) indeed generate the same $\pi N$ scattering amplitude.
   We emphasize that in order to obtain this equivalence of the two approaches
an expansion of Eq.\ (\ref{5:5:tnonrelred}) to 
$1/\!\!\stackrel{\circ}{m}_N$ is mandatory, because
the functions $A^{(+)}$ and $B^{(+)}$ contain terms of leading order 
$\stackrel{\circ}{m}_N$.
   These terms disappear through a cancellation in the final result.\footnote{
The overall factor $2\stackrel{\circ}{m}_N$ in Eq.\ (\ref{5:5:tnonrelred})
is a result of our normalization 
of the spinors [see Eq.\ (\ref{5:3:mthr})].}

\subsection{Corrections at First Order in $1/m$}
\label{subsec_cfo1om}
   So far we have concentrated on the leading-order, $m$-independent, 
heavy-baryon Lagrangian of Eq.\ (\ref{5:5:l1pinhb}).
   In comparison with Eq.\ (\ref{5:2:powercounting}), the chiral counting 
scheme of HBChPT is different, because an ordinary partial derivative acting 
on a heavy-baryon field ${\cal N}_v$ produces a small residual 
{\em four}-momentum
[see also Eq.\ (\ref{4:5:powercounting}) for the mesonic sector]:
\begin{equation}
\label{5:5:hbchppowercounting}
{\cal N}_v,\bar{\cal N}_v =  {\cal O}(p^0),\, 
D_{\mu} {\cal N}_v = {\cal  O}(p),\, 
v_\mu, S^v_\mu, 1_{4\times 4}={\cal O}(p^0).
\end{equation}
   In the heavy-baryon approach four-momenta are considered small if
their components are small in comparison with either the nucleon mass
$m_N$ or the chiral symmetry breaking scale $4\pi F_\pi$, both of
which we denote by a common scale $\Lambda\simeq 1$ 
GeV.\footnote{In reality, the
excitation energy of the $\Delta(1232)$ resonance very often provides
the limit of convergence of the expansion.} 
   It is clear that the Lagrangian of Eq.\ (\ref{5:5:lhneffexp}) also 
generates terms of higher order in $1/m$ and, in analogy to the mesonic 
sector, we also expect additional new chiral structures from the most 
general chiral Lagrangian at higher orders.
   Recall that in the baryonic sector the chiral orders increase in units
of one, because of the additional possibility of forming Lorentz invariants
by contracting (covariant) derivatives with gamma matrices
(see Sec.\ \ref{sec_loebl}). 
   (The relativistic $\pi N$ Lagrangian at ${\cal O}(p^2)$ has (partially)
been given in Ref.\ \cite{Gasser:1987rb}.) 

   Let us first consider the $1/m$ correction resulting from
Eq.\ (\ref{5:5:lhneffexp})
\begin{eqnarray*}
&&\frac{1}{2m}\bar{\cal N}_v
\left(iD\hspace{-.7em}/_\bot+\frac{g_A}{2}v\cdot u \gamma_5\right)
\left(iD\hspace{-.7em}/_\bot-\frac{g_A}{2}v\cdot u \gamma_5\right){\cal N}_v
\nonumber\\
&=&
\frac{1}{2m}\bar{\cal N}_v\Bigg[-D\hspace{-.7em}/_\bot D\hspace{-.7em}/_\bot
-i\frac{g_A}{2}D\hspace{-.7em}/_\bot v\cdot u \gamma_5
+i\frac{g_A}{2}v\cdot u \gamma_5 D\hspace{-.7em}/_\bot
-\frac{g^2_A}{4}(v\cdot u)^2\Bigg]{\cal N}_v.
\end{eqnarray*}
   We make use of Eqs.\ (\ref{5:5:nvbilineare}) to identify the relevant
replacements in the heavy-baryon bilinears:
\begin{eqnarray*}
D\hspace{-.7em}/_\bot\gamma_5&=&
D\hspace{-.7em}/\hspace{.2em}\gamma_5-v\cdot D v\hspace{-.5em}/\hspace{.2em}
\gamma_5
\mapsto
2 D\cdot S_v - 2v\cdot D \underbrace{v\cdot S_v}_{\mbox{0}} 
= 2 D\cdot S_v,\\
\gamma_5 D\hspace{-.7em}/_\bot&\mapsto&-2 D\cdot S_v,\\
D\hspace{-.7em}/_\bot D\hspace{-.7em}/_\bot&=&
(D_\mu-v\cdot D v_\mu)(D_\nu-v\cdot D v_\nu)\hspace{-1em}
\underbrace{\gamma^\mu\gamma^\nu}_{\mbox{$g^{\mu\nu}-i\sigma^{\mu\nu}$}}\\
&=&(D^2-v\cdot D v\cdot D)-i\hspace{-1.8em}\underbrace{\sigma^{\mu\nu}}_{
\mbox{$\mapsto 2 \epsilon^{\mu\nu\rho\sigma}v_\rho S_\sigma^v$}}
\hspace{-1.8em}
(D_\mu-v\cdot D v_\mu)(D_\nu-v\cdot D v_\nu)\\
&\mapsto&D^2-(v\cdot D)^2-i\epsilon^{\mu\nu\rho\sigma}[D_\mu,D_\nu]v_\rho
S_\sigma^v\\
&=& D^2-(v\cdot D)^2 -\frac{i}{4}\epsilon^{\mu\nu\rho\sigma}
[u_\mu,u_\nu]v_\rho S_\sigma^v\nonumber\\
&& -\frac{1}{2}\epsilon^{\mu\nu\rho\sigma}
f^+_{\mu\nu}v_\rho S_\sigma^v
-\epsilon^{\mu\nu\rho\sigma}v^{(s)}_{\mu\nu} v_\rho S_\sigma^v,
\end{eqnarray*}
where the expression for the commutator $[D_\mu,D_\nu]$ of the covariant 
derivative of Eq.\ (\ref{5:2:kovderpsi}) is obtained after 
straightforward algebra.
   The field-strength tensors are defined as
\begin{equation}
\label{fpm}
f^{\pm}_{\mu\nu}
= u f_{\mu\nu}^L u^\dagger \pm u^\dagger f_{\mu\nu}^R u ,
\quad
v^{(s)}_{\mu\nu}=\partial_\mu v^{(s)}_{\nu}-\partial_\nu
v^{(s)}_{\mu},
\end{equation}
where $f_{\mu\nu}^R$ and $f_{\mu\nu}^L$ are given in Eqs.\ 
(\ref{4:5:fr}) and (\ref{4:5:fl}), respectively.

   Collecting all terms, we finally obtain as the contribution of Eq.\  
(\ref{5:5:lhneffexp}) of order $1/m$ (returning to the notation in terms of
expressions in the chiral limit)
\begin{eqnarray}
\label{5:5:l2pinhba}
\widehat{\cal L}^{(2)}_{\pi N,1/m}
&=&\frac{1}{2\stackrel{\circ}{m}_N}
\bar{\cal N}_v\Bigg[(v\cdot D)^2-D^2-i\stackrel{\circ}{g}_A
\{S_v\cdot D,
v\cdot u\}-\frac{\stackrel{\circ}{g}_A^2}{4}(v\cdot u)^2\nonumber\\
&&+\frac{1}{2}\epsilon^{\mu\nu\rho\sigma} v_\rho S_\sigma^v
\left(i u_\mu u_\nu+ f^+_{\mu\nu}+2 v^{(s)}_{\mu\nu}\right)
\Bigg]{\cal N}_v.
\end{eqnarray}
   Applying the counting rules of Eqs.\ (\ref{4:5:powercounting}) and 
(\ref{5:5:hbchppowercounting}), we see that Eq.\ (\ref{5:5:l2pinhba})
is indeed of ${\cal O}(p^2)$, where the suppression relative to 
(\ref{5:5:l1pinhb}) is of the form $p/\stackrel{\circ}{m}_N$.

   At ${\cal O}(p^2)$ the heavy-baryon Lagrangian 
$\widehat{\cal L}^{(2)}_{\pi N}$ contains another contribution which, 
in analogy to $\widehat{\cal L}^{(1)}_{\pi N}$ in Sec.\ \ref{subsec_lol},
may be obtained as the projection of the relativistic Lagrangian 
${\cal L}^{(2)}_{\pi N}$ of \cite{Gasser:1987rb}
onto the light components.
   Here we quote the result in the convention of Ref.\ 
\cite{Bernard:1997gq}(except for the $c_6$ and $c_7$ terms, where,
following Ref.\ \cite{Ecker:1995rk}, we explicitly separate the
traceless and isoscalar terms)\footnote{The 
nomenclature of Refs.\ 
\cite{Gasser:1987rb} and \cite{Bernard:1992qa} differs from the 
(more or less) standard
convention of Eq.\ (\ref{5:5:l2pinhbb}).
   The constants $c_i$ of Ecker and 
Moj\v{z}i\v{s} \cite{Ecker:1995rk}
differ by a factor $1/\stackrel{\circ}{m}_N$ from those of 
Eq.\ (\ref{5:5:l2pinhbb}).}
\begin{eqnarray}
\label{5:5:l2pinhbb}
\widehat{\cal L}^{(2)}_{\pi N, c_i}&=&
\bar{\cal N}_v\Bigg[c_1 \mbox{Tr}(\chi_+)
+c_2 (v\cdot u)^2 +c_3 u\cdot u
+c_4 [S^\mu_v,S^\nu_v]u_\mu u_\nu\nonumber\\
&&
+c_5\left[\chi_+-\frac{1}{2}\mbox{Tr}(\chi_+)\right]
-ic_6[S^\mu_v,S^\nu_v]f^+_{\mu\nu}
-ic_7[S^\mu_v,S^\nu_v]v^{(s)}_{\mu\nu}
\Bigg]{\cal N}_v,\nonumber\\
\end{eqnarray}
where 
\begin{equation}
\label{5:5:chipm}
\chi_{\pm} =  u^\dagger \chi u^\dagger \pm  u\chi^\dagger u.
\end{equation}
   In the parameterization of Eq.\ (\ref{5:5:l2pinhbb}) the constants 
$c_i$ carry the dimension of an inverse mass and should be of
the order of $1/\Lambda$ in order to produce a reasonable convergence 
of the chiral expansion.
   (The details of convergence generally depend on the observable in
question).
    The complete heavy-baryon Lagrangian at ${\cal O}(p^2)$ is then given
by the sum of Eqs.\ (\ref{5:5:l2pinhba}) and (\ref{5:5:l2pinhbb}),
\begin{equation}
\label{5:5:l2pinhb}
\widehat{\cal L}^{(2)}_{\pi N}=
\widehat{\cal L}^{(2)}_{\pi N,1/m}+
\widehat{\cal L}^{(2)}_{\pi N, c_i}.
\end{equation}

   It is worthwhile mentioning that the contribution of 
Eq.\ (\ref{5:5:l2pinhba}) to $\widehat{\cal L}^{(2)}_{\pi N}$ contains 
chirally invariant structures that are {\em not} 
part of Eq.\ (\ref{5:5:l2pinhbb}). 
   Unless such terms can be transformed away by a field transformation (see
below) their coefficients are fixed in terms of the
parameters of the lowest-order Lagrangian. 
   As stressed by Ecker \cite{Ecker:1995gg}, these fixed coefficients
are a consequence of the Lorentz covariance of the whole approach.
   A related issue is the so-called reparameterization invariance,
i.e., if a heavy particle of physical four-momentum $p$ is described by, say,
$p=mv+k$ with $p^2=m^2$ and $v^2=1$ implying $2mv\cdot k+k^2=0$,
physical observables should not change under the replacement 
$(v,k)\to (v+q/m,k-q)$ giving rise to an equivalent parameterization
$p=mv'+k'$, if $q$ satisfies $(v+q/m)^2=1$ \cite{Luke:1992cs}.
   As a result, some coefficients of terms in the effective Lagrangian
which are of different order in the $1/m$ expansion are related.
   For a detailed discussion, the reader is referred to Refs.\ 
\cite{Luke:1992cs,Chen:te,Finkemeier:1997re}.
   
   The seven low-energy constants $c_i$ are determined by comparison
with experimental information. 
   For example, if we consider the interaction with an external electromagnetic
field [see Eq.\ (\ref{2:4:rlasu2})],
\begin{displaymath}   
r_\mu=l_\mu=-e\frac{\tau_3}{2}{\cal A}_\mu,\quad
v_\mu^{(s)}=-\frac{e}{2}{\cal A}_\mu,
\end{displaymath}
we obtain 
\begin{displaymath}
f_{\mu\nu}^+=-e\tau_3 {\cal F}_{\mu\nu}+\cdots,\quad
v_{\mu\nu}^{(s)}=-\frac{e}{2}{\cal F}_{\mu\nu},\quad
{\cal F}_{\mu\nu}=\partial_\mu{\cal A}_\nu-\partial_\nu{\cal A}_\mu,
\end{displaymath} 
so that the interaction with the field-strength tensor is given by
\begin{equation}
\label{5:5:l2int}
\widehat{\cal L}^{(2)}_{\rm int}=
-e\epsilon_{\mu\nu\rho\sigma} {\cal F}^{\mu\nu} v^{\rho}
\bar{\cal N}_v S^\sigma_v\left[\frac{1}{4\stackrel{\circ}{m}_N}+\frac{c_7}{2}
+\tau_3\left(\frac{1}{4\stackrel{\circ}{m}_N}+c_6\right)\right]{\cal N}_v.
\end{equation}
[We made use of Eq.\ (\ref{5:5:seig}).]
   For the special choice $v^\mu=(1,0,0,0)=\tilde{v}^\mu$ we find
[see Eq.\ (\ref{5:5:spinopvtilde})]
\begin{displaymath}
\epsilon_{\mu\nu\rho\sigma}{\cal F}^{\mu\nu} 
\tilde{v}^{\rho}S^\sigma_{\tilde{v}}
=\epsilon_{ijk}{\cal F}^{ij} \frac{1}{2}\Sigma^k
=-\vec{\Sigma}\cdot\vec{B},
\end{displaymath}
and the interaction
term reduces to\footnote{Recall that  ${\cal N}_{\tilde{v}}$ are 
two-component fields.}
\begin{equation}
\label{5:5:lintmm}
\frac{e}{2\stackrel{\circ}{m}_N} \bar{\cal N}_{\tilde{v}}\,
\vec{\sigma}\cdot\vec{B}\,
{\cal N}_{\tilde{v}}\left[\frac{1}{2}
\left(1+2\stackrel{\circ}{m}_N c_7\right)
+\frac{\tau_3}{2}\left(1+4\stackrel{\circ}{m}_N c_6\right)\right],
\end{equation}
which describes the interaction Lagrangian of a magnetic field with the 
magnetic moment of the nucleon.
    We define the isospin decomposition of the magnetic moment (in
units of the nuclear magneton $e/2 m_p$) as
\begin{displaymath}
\mu=\frac{1}{2}\mu^{(s)}+\frac{\tau_3}{2}\mu^{(v)}
=\frac{1}{2}(1+\kappa^{(s)})+\frac{\tau_3}{2}(1+\kappa^{(v)}),
\end{displaymath}
where $\kappa^{(s)}$ and $\kappa^{(v)}$ denote the isoscalar and isovector
{\em anomalous} magnetic moments of the nucleon, respectively,
with empirical values $\kappa^{(s)}=-0.120$ and $\kappa^{(v)}=3.706$.
   A comparison with Eq.\ (\ref{5:5:lintmm}) shows that the constants
$c_6$ and $c_7$ are related to the {\em anomalous} magnetic 
moments of the nucleon in the chiral limit
\begin{displaymath}
\stackrel{\circ}{\kappa}^{(s)}=2\stackrel{\circ}{m}_N c_7,\quad
\stackrel{\circ}{\kappa}^{(v)}=4\stackrel{\circ}{m}_N c_6.
\end{displaymath}
   The results for $\kappa^{(s)}$ and $\kappa^{(v)}$ up to and including
${\cal O}(p^3)$
\cite{Bernard:1995dp,Fearing:1997dp}
\begin{eqnarray*}
\kappa^{(s)}&=&\stackrel{\circ}{\kappa}^{(s)}+\,\,{\cal O}(p^4),\\
\kappa^{(v)}&=&\stackrel{\circ}{\kappa}^{(v)}
-\frac{M_\pi m_N g_A^2}{4\pi F_\pi^2}+{\cal O}(p^4),
\end{eqnarray*}
are used to express the parameters $c_6$ and $c_7$ in terms of physical
quantities.
   Note that the numerical correction 
of $-1.96$ [parameters of Eq.\ (\ref{5:3:par})]
to the isovector anomalous magnetic moment is substantial. 
   Differences by factors of about 1.5 were generally observed for the
determination of the $c_i$ at ${\cal O}(p^2)$ and to one-loop accuracy
${\cal O}(p^3)$ \cite{Bernard:1995dp,Bernard:1997gq}.

    The numerical values of the low-energy constants $c_1,\cdots,c_4$ have
been determined in Ref.\ \cite{Bernard:1997gq} by performing a best fit
to a set of nine pion-nucleon scattering observables at ${\cal O}(p^3)$
which do not contain any new low-energy constants from the ${\cal O}(p^3)$
Lagrangian. 
   Finally, $c_5$ was determined in terms of
the strong contribution to the neutron-proton mass difference.
   The results in units of  GeV$^{-1}$ are given by (see also Ref.\
\cite{Buttiker:1999ap})
\begin{eqnarray}
\label{5:5:ciresults}
&&c_1=-0.93\pm 0.10,\quad
c_2=3.34\pm 0.20,\quad
c_3=-5.29\pm0.25,\nonumber\\
&&c_4=3.63\pm 0.10,\quad
c_5=-0.09\pm 0.01.
\end{eqnarray}
   For a phenomenological interpretation of the low-energy constants in
terms of (meson and $\Delta$) resonance exchanges see 
Ref.\ \cite{Bernard:1997gq}.

   We will see in the next section that the constants $c_i$ are 
{\em not} required to compensate divergences of one-loop integrals.
   Such infinities first appear at ${\cal O}(p^3)$.

   The Lagrangian of Eq.\ (\ref{5:5:l2pinhba}) still contains terms
of the type $v\cdot D$ which appears in the lowest-order equation of 
motion of Eq.\ (\ref{5:5:l1pinhbeom}).
   As discussed in detail for the mesonic sector in Sec.\ \ref{sec_clop4}
and Appendix \ref{app_sec_glvgss}, such terms can be eliminated by appropriate
field redefinitions.
   For example, the field transformation eliminating in 
Eq.\ (\ref{5:5:l2pinhba}) the term
\begin{displaymath}
\frac{1}{2\stackrel{\circ}{m}_N}
\bar{\cal N}_v(v\cdot D)^2{\cal N}_v
\end{displaymath} 
is given by \cite{Ecker:1995rk}
\begin{equation}
\label{5:5:nvft}
{\cal N}_v=\left[1+\frac{iv\cdot D}{4\stackrel{\circ}{m}_N}
-\frac{\stackrel{\circ}{g}_A S_v\cdot u}{4\stackrel{\circ}{m}_N} \right]
\tilde{\cal N}_v.
\end{equation}
   Inserting Eq.\ (\ref{5:5:nvft}) into the lowest-order Lagrangian of 
Eq.\ (\ref{5:5:l1pinhb}) yields
\begin{eqnarray}
\label{5:5:dell1}
&&\bar{\tilde{\cal N}}_v(iv\cdot D+
\stackrel{\circ}{g}_AS_v\cdot u)\tilde{\cal N}_v
-\frac{1}{2\stackrel{\circ}{m}_N}
\bar{\tilde{\cal N}}_v(v\cdot D)^2\tilde{\cal N}_v
-\frac{\stackrel{\circ}{g}_A^2}{2\stackrel{\circ}{m}_N}
\bar{\tilde{\cal N}}S_v\cdot u S_v\cdot u \tilde{\cal N}_v\nonumber\\
&&+\,\,\mbox{total derivative}
+O\left(\frac{1}{\stackrel{\circ}{m}_N^2}\right).
\end{eqnarray}
   The second term cancels the equation-of-motion term, whereas 
rewriting the last term by using Eq.\ (\ref{5:5:seig}),
\begin{displaymath}
S_v\cdot u S_v\cdot u=\frac{1}{4}[(v\cdot u)^2-u\cdot u]+
\frac{1}{2}i\epsilon^{\mu\nu\rho\sigma}v_\rho S^v_\sigma u_\mu u_\nu,
\end{displaymath}
 we find
that some of the coefficients at ${\cal O}(p^2)$
(and at higher orders) are modified.
   As in the case of the SU(2)$\times$SU(2) mesonic Lagrangian 
at ${\cal O}(p^4)$
(see Appendix \ref{app_sec_glvgss}) one finds 
{\em equivalent} parameterizations of $\widehat{\cal L}^{(2)}_{\pi N}$
(and also of the higher-order Lagrangians)
in the baryonic sector.
   For the sake of completeness we quote the result of Ecker and 
Moj\v{z}i\v{s} \cite{Ecker:1995rk},
\begin{eqnarray}
\label{5:5:l2eckermojzis}
\widehat{\cal L}^{(2)}_{\pi N}&=&\bar{\cal N}_v\Bigg\{
-\frac{1}{2\stackrel{\circ}{m}_N}\left(D^2+i\stackrel{\circ}{g}_A
\{S_v\cdot D,v\cdot u\}\right)
+\frac{a_1}{\stackrel{\circ}{m}_N}\mbox{Tr}(u\cdot u)\nonumber\\
&&+\frac{a_2}{\stackrel{\circ}{m}_N}\mbox{Tr}\left[(v\cdot u)^2\right]
+\frac{a_3}{\stackrel{\circ}{m}_N}\mbox{Tr}(\chi_+)
+\frac{a_4}{\stackrel{\circ}{m}_N}
\left[\chi_+-\frac{1}{2}\mbox{Tr}(\chi_+)\right]\nonumber\\
&&+\frac{1}{\stackrel{\circ}{m}_N}\epsilon^{\mu\nu\rho\sigma}v_\rho
S^v_\sigma\left[ia_5 u_\mu u_\nu+a_6 f^+_{\mu\nu}+a_7 v^{(s)}_{\mu\nu}\right]
\Bigg\}{\cal N}_v,
\end{eqnarray}
where the relation to the coefficients $c_i$ of Eq.\ 
(\ref{5:5:l2pinhbb}) is given
by
\begin{eqnarray}
\label{5:5:parrel}
&&a_1=\frac{\stackrel{\circ}{m}_N\! c_3}{2}+\frac{\stackrel{\circ}{g}_A^2}{16},
\quad
a_2=\frac{\stackrel{\circ}{m}_N\! c_2}{2}-\frac{\stackrel{\circ}{g}_A^2}{8},
\quad
a_3=\,\stackrel{\circ}{m}_N\! c_1,\quad
a_4=\,\stackrel{\circ}{m}_N\! c_5,\nonumber\\
&&
a_5=\,\stackrel{\circ}{m}_N\! c_4+\frac{1-\stackrel{\circ}{g}_A^2}{4},
\quad
a_6=\,\stackrel{\circ}{m}_N\! c_6+\frac{1}{4},\quad
a_7=\,\stackrel{\circ}{m}_N\! c_7+\frac{1}{2}.
\quad
\end{eqnarray}
    Of course, the Lagrangians of Eq.\ (\ref{5:5:l2pinhb}) and 
(\ref{5:5:l2eckermojzis}) yield the same results for physical observables,
provided their parameters are related by Eq.\ (\ref{5:5:parrel}).
    However, they will differ for intermediate mathematical
quantities such as vertices or wave function renormalization constants
as observed in Ref.\ \cite{Fearing:1997dp} for the case of
the nucleon wave function renormalization constant.
    We repeat that the coefficients of the first two terms of
Eq.\ (\ref{5:5:l2eckermojzis}) are fixed in terms of $\stackrel{\circ}{m}_N$
and $\stackrel{\circ}{g}_A$, whereas the constants $a_i$ are free
parameters which have to be determined by comparison with experimental
information.

\subsection{The Power Counting Scheme}
\label{subsec_pcs}
   The power counting scheme of HBChPT may be formulated in close
analogy to the mesonic sector (see Sec.\ \ref{sec_elwpcs}).
   On the scale of either the nucleon mass $m_N$ or $4\pi F_\pi$
we consider as small external momenta the four-momenta of pions, 
the four-momenta transferred by external sources, and the residual momenta 
$k^\mu$ of the nucleon appearing in the separation 
$p^\mu=\,\stackrel{\circ}{m}_N\! v^\mu + k^\mu$. 
   For a given Feynman diagram we introduce
\begin{itemize}
\item the number of independent loop momenta $N_L$,
\item the number of internal pion lines $I_M$,
\item the number of pion vertices  $N^M_{2n}$ originating from
${\cal L}_{2n}$,
\item the total number of pion vertices $N_M=\sum_{n=1}^\infty N^M_{2n}$,
\item the number of internal nucleon lines $I_B$,
\item the number of baryonic vertices $N^B_n$ originating from
$\widehat{\cal L}^{(n)}_{\pi N}$,
\item and the total number of baryonic vertices 
$N_B=\sum_{n=1}^\infty N^B_n$.
\end{itemize}
   As in the mesonic sector, the internal momenta appearing in the loop 
integration are not necessarily small.
   However, via the four-momentum conserving delta functions at the vertices 
and a substitution of integration variables, 
the rescaling of the external momenta is transferred to
the internal momenta (see Sec.\ \ref{sec_elwpcs}).
   The chiral dimension $D$ of a given diagram is then given by
\cite{Weinberg:1991um,Ecker:1995gg}
\begin{equation}
\label{5:5:d1}
D=4 N_L - 2 I_M - I_B +\sum_{n=1}^\infty 2n N^M_{2n}
+\sum_{n=1}^\infty n N_n^B.
\end{equation}
   We make use of the topological relation [see, e.g., 
Eq.\ (2.130) of Ref.\ \cite{Cheng:bj}]
\begin{equation}
\label{5:5:NLint}
N_L=I_M +I_B-N_M-N_B+1
\end{equation}
to eliminate $I_M$ from Eq.\ (\ref{5:5:d1})
\begin{equation}
\label{5:5:d2}
D=2 N_L +I_B +2 +\sum_{n=1}^\infty 2(n-1)N^M_{2n}
+\sum_{n=1}^\infty (n-2)N_n^B.
\end{equation}

\begin{figure}[htb]
\begin{center} 
\caption{\label{5:5:fig:twoloopselfenergy}      
Two-loop contribution to the nucleon self energy.}
\vspace{2em}
\epsfig{file=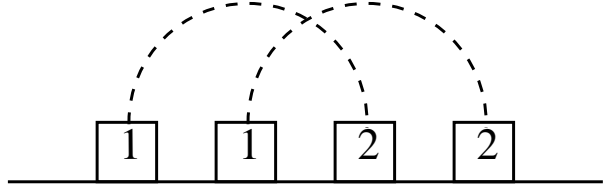,width=8cm}
\end{center}
\end{figure}

   For processes containing exactly one nucleon in the initial and final
states we have\footnote{In the heavy-baryon formulation 
one has no closed fermion loops. In other words, in the single-nucleon
sector exactly one fermion line runs through the diagram connecting
 the initial and final states.}
$N_B=I_B+1$  and we thus obtain
\begin{equation}
\label{5:5:d3}
D=2 N_L+1 +\sum_{n=1}^\infty 2(n-1)N^M_{2n}
+\sum_{n=1}^\infty (n-1)N_n^B.
\end{equation}

   The power counting is very similar to the mesonic sector. 
   We first observe that $D\ge 1$. 
   Moreover, as already mentioned in Sec.\ \ref{subsec_cfo1om}, loops
start contributing at $D=3$. 
   In other words, the low-energy coefficients $c_i$ of 
$\widehat{\cal L}^{(2)}_{\pi N}$ are not needed to
renormalize infinities from one-loop calculations.
   Again, we have a connection between the number of loops and the 
chiral dimension $D$: $N_L\le (D-1)/2$.
   Each additional loop adds two units to the chiral dimension.

   As an example, let us consider the two-loop contribution to the
nucleon self energy of Fig.\ \ref{5:5:fig:twoloopselfenergy}.
   First of all, the number of independent loops is $N_L=2$ in agreement
with Eq.\ (\ref{5:5:NLint}) for $I_M=2$, $I_B=3$, $N_M=0$, and
$N_B=4$.
   The counting of the chiral dimension is most intuitively performed
in the framework of Eq.\ (\ref{5:5:d1}), because it associates with
each building block a unique term which is easy to remember ($+4$ for 
each independent loop, $-2$ for each internal meson propagator, etc). 
   For $N^M_{2n}=0$, $N^B_1=2$, and $N^B_2=2$ 
we obtain $D=8-4-3+0+2+4=7$.

\subsection{Application at ${\cal O}(p^3)$: One-Loop Correction to the
Nucleon Mass }
\label{subsec_aop3olcnm}
   As a simple example, we will return to the modification of the nucleon
mass through higher-order terms in the heavy-baryon approach.
   The calculation will proceed along the lines of Ref.\ \cite{Fearing:1997dp},
where use was made of the Lagrangian of Ecker and 
Moj\v{z}i\v{s} \cite{Ecker:1995rk} [see Eq.\ (\ref{5:5:l2eckermojzis})].
 
   The determination of the physical nucleon mass and the discussion of
the wave function renormalization factor will be very similar to Secs.\
\ref{subsec_mgb} for the masses
of the Goldstone bosons and \ref{subsec_feolcnm} for the nucleon mass
in the relativistic approach.
   Let us denote the four-momentum of the nucleon by 
$p = \, \stackrel{\circ}{m}_N\! v +r$, where, since we are interested in
the propagator, we must allow the four-momentum to be off the mass shell.
   The on-shell case is, of course, given by $p^2=m_N^2$ with $m_N$ denoting
the {\em physical} nucleon mass.
   Let us stress that, due to the interaction, we must expect the physical 
mass to be different from the mass $\stackrel{\circ}{m}_N$ in the chiral
limit. 

   We start from the lowest-order propagator of Eq.\ (\ref{5:5:gv0k}),
\begin{equation}
\label{5:5:prop0}
\frac{P_{v+}}{v\cdot r +i0^+}
=\frac{P_{v+}}{v\cdot p\,-\stackrel{\circ}{m}_N +\,i0^+},
\end{equation}
and first determine its modification in terms of the tree-level contribution 
of Eq.\ (\ref{5:5:l2eckermojzis}) to the self energy.\footnote{In the 
remaining part of this section, we adopt the common practice
of leaving out the projector $P_{v+}$ in the propagator and (possibly) in
vertices with the understanding that all operators act only in the projected
subspace.}
   Neglecting isospin-symmetry breaking effects proportional to $m_u-m_d$,
we obtain at ${\cal O}(p^2)$
\begin{equation}
\label{5:5:sigma2}
\Sigma^{(2)}(p)=-\frac{r^2}{2\stackrel{\circ}{m}_N} 
-\frac{4a_3 M^2}{\stackrel{\circ}{m}_N},
\end{equation}
where $M^2=2B_0 m_q$ denotes the squared pion mass at ${\cal O}(p^2)$. 
   The $r^2$ term comes from the term in $\widehat{\cal L}^{(2)}_{\pi N}$ 
proportional to $-\partial^2/2\stackrel{\circ}{m}_N$ which involves no pions. 
   In the spirit of the reduction formula of Eq.\ (\ref{5:5:ffunc}),
in combination with the formula of Gell-Mann and Low of 
Eq.\ (\ref{5:5:gmlhb}), we choose to include this as part of the interaction 
rather than part of the free Lagrangian and reserve for the free Lagrangian 
the $i v \cdot \partial$ term from $\widehat{\cal L}^{(1)}_{\pi N}$.
   The term involving $a_3$ is a contact term coming ultimately from Eq.\ 
(\ref{5:5:l2pinhbb}), where $a_3=\,\stackrel{\circ}{m}_N\! c_1$.

\begin{figure}[htb]
\begin{center} 
\caption{\label{5:5:fig:hbnseloop}
One-loop contribution to the nucleon self energy at ${\cal O}(p^3)$
in the heavy-baryon approach.
   The diagram (b) vanishes because of its isospin structure.}
\vspace{2em}
\epsfig{file=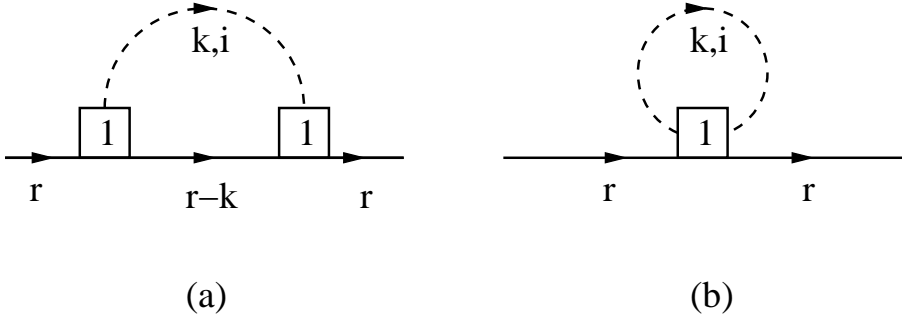,width=12cm}
\end{center}
\end{figure}
   
   The heavy-baryon Lagrangian at ${\cal O}(p^3)$ \cite{Ecker:1995rk} does
not produce a contact contribution to the self energy, because all the
structures contain at least one pion or an external field. 
   Moreover, given the Feynman rule of Eq.\ (\ref{5:6:fr2}), the second
one-loop diagram of Fig.\ \ref{5:5:fig:hbnseloop} (b) vanishes, because
$\epsilon_{iij}\tau_j=0$.
   In other words, the only contribution at ${\cal O}(p^3)$ results from
the one-loop diagram of Fig. \ref{5:5:fig:hbnseloop} (a).
   Using the vertex of Eq.\ (\ref{5:5fr1}) and the propagator of Eq.\
(\ref{5:5:gv0k}) we obtain for the one-loop contribution at ${\cal O}(p^3)$
\begin{eqnarray}
\label{5:5:sigmaloop1}
-i\Sigma^{(3)}_{\rm loop}(p)&=&\int\frac{d^4 k}{(2\pi)^4}\left[
\frac{\stackrel{\circ}{g}_A}{F_0}(-S_v\cdot k)\tau_i\right]
\frac{i}{v\cdot(r-k)+i0^+}\nonumber\\
&&\times \frac{i}{k^2-M^2+i0^+}
\left[\frac{\stackrel{\circ}{g}_A}{F_0}S_v\cdot k\tau_i\right].
\end{eqnarray}
   As in the relativistic case of Eq.\ (\ref{5:4:sel1}), counting powers,
we expect the integral to have a cubic divergence.
   Extending the integral to $n$ dimensions, using $\tau_i\tau_i=3$,
performing the substitution $k\to -k$, and applying Eq.\ (\ref{app:C20C21}) 
of Appendix \ref{subsec_jpin}, we obtain the intermediate result
\begin{displaymath}
\Sigma^{(3)}_{\rm loop}(p)=3 \frac{\stackrel{\circ}{g}_A^2}{F_0^2}
S^\mu_v S^\nu_v[v_\mu v_\nu C_{20}(v\cdot r, M^2)+g_{\mu\nu} C_{21}
(v\cdot r, M^2)].
\end{displaymath}
   Since $S_v\cdot v=0$ and $S^2_v=(1-n)/4$ [see Eq.\ (\ref{5:5:s2})], we
obtain, applying the first equality of Eq.\ (\ref{app:C21}),
\begin{equation}
\Sigma^{(3)}_{\rm loop}(p)=-\frac{3}{4}\frac{\stackrel{\circ}{g}_A^2}{F_0^2}
\left\{[M^2-(v\cdot r)^2]J_{\pi N}(0;
v\cdot r)+v\cdot r I_\pi(0)\right\},
\end{equation}
where the integrals $J_{\pi N}$ and $I_\pi$ are given in Eqs.\
(\ref{app:jpinb}) and (\ref{app:ipi}), respectively.

   Combining with Eq.\ (\ref{5:5:sigma2}) and using Eq.\ (\ref{app:jpinb}) 
we thus obtain for the (unrenormalized)
nucleon self energy at ${\cal O}(p^3)$ \cite{Fearing:1997dp}
\begin{eqnarray}
\label{5:6:sigmaN}
\lefteqn{\Sigma(p) = -\frac{r^2}{2\stackrel{\circ}{m}_N} 
-\frac{4a_3 M^2}{\stackrel{\circ}{m}_N}}\nonumber\\
&&-\frac{3\stackrel{\circ}{g}_A^2}{(4\pi F_0)^2}\left(
\frac{v\cdot r}{4}\left\{[3M^2-2(v\cdot r)^2]
\left[R+\ln\left(\frac{M^2}{\mu^2}\right)\right]
-\frac{1}{2}\left[M^2-(v\cdot r)^2\right]\right\}\right.\nonumber\\
&&\left.+[M^2-(v\cdot r)^2]^\frac{3}{2}\arccos\left(-\frac{v\cdot r}{M}\right)
\right),
\end{eqnarray}
for $(v\cdot r)^2<M^2$.
   Clearly, the self-energy contribution generated by the loop diagram
of Fig.\ \ref{5:5:fig:hbnseloop} (a) contains a divergent piece proportional
to $R$ of Eq.\ (\ref{app:constantR}).

   We have chosen to express the self energy as a function of the 
four-momentum $p$. 
   In the relativistic case of Eq.\ (\ref{5:4:seansatz}) we needed two
scalar functions depending on $p^2$ to parameterize the self energy.
   In contrast to the relativistic case, the heavy-baryon self energy of Eq.\
(\ref{5:6:sigmaN}) is given by one function depending on two scalar variables
for which one can take, say, $r^2$ and $v\cdot r$ or 
\begin{equation}
\label{5:5:etaxi}
\eta\equiv v\cdot p-m_N,\quad
\xi\equiv (p-m_N v)^2.
\end{equation}
   Making use of $r=(m_N-\stackrel{\circ}{m}_N)v+(p-m_N v)$, the two
sets are related by
\begin{eqnarray}
\label{5:5:rvretaxi1}
r^2&=&(m_N-\stackrel{\circ}{m}_N)^2
+2(m_N-\stackrel{\circ}{m}_N)\eta
+\xi,\\
\label{5:5:rvretaxi2}
v\cdot r&=&m_N-\stackrel{\circ}{m}_N+\,\eta.
\end{eqnarray}
   The choice of Eq.\ (\ref{5:5:etaxi}) is convenient for the determination 
of the physical nucleon mass $m_N$ and renormalization constant $Z_N$, 
because, in view
of Eq.\ (\ref{5:5:prop0}), we want the full (but yet unrenormalized) 
propagator to have a pole at $p=m_N v$ which includes both the mass-shell
condition $p^2=m_N^2$ and $v \cdot p =m_N$.
   In the vicinity of the pole at 
$p=m_N v$ the second choice of variables corresponds to terms which
are, respectively, first and second 
order in the (small) distance from the pole.
   Thus in the following discussion we will use both notations
$\Sigma(p)$ and $\Sigma(\eta,\xi)$ for the self energy, where it should
be clear from the context which expression applies.

   In analogy to the mesonic case discussed in Sec.\ \ref{subsec_mgb} the
full heavy-baryon propagator is written as [see Eq.\ (\ref{4:8:prop2})]
\begin{eqnarray}
\label{5:5:fullhbp}
iG_v(p) &=& \frac{i}{v \cdot p\,\, -\stackrel{\circ}{m}_N-\,
\Sigma(p)}\nonumber\\
&=&\frac{i}{v \cdot p\,\, -\stackrel{\circ}{m}_N 
-\,\Sigma(0,0)-\eta \Sigma^\prime(0,0)-\tilde{\Sigma}(\eta,\xi)}\nonumber \\ 
&=& \frac{i}{[1-\Sigma^\prime (0,0)]
\left\{\eta  - \frac{\tilde{\Sigma}(\eta,\xi)}{[1-\Sigma^\prime (0,0)]}
\right\}}\nonumber\\
&=& \frac{i Z_N}{\eta - Z_N 
\tilde{\Sigma}(\eta,\xi)},
\end{eqnarray} 
where
\begin{eqnarray}
\label{5:5:mndef}
m_N&=&\stackrel{\circ}{m}_N+\,\Sigma(0,0),\\
\label{5:5:zndef}
Z_N&=&\frac{1}{1-\Sigma^\prime (0,0)}.
\end{eqnarray}
   In these equations $\Sigma^\prime (0,0)$ denotes the first partial 
derivative of $\Sigma(\eta,\xi)$ with respect to $\eta$ evaluated 
at $(\eta,\xi)=(0,0)$,
\begin{displaymath}
\Sigma^\prime (0,0)=\left.\frac{\partial\Sigma(\eta,\xi)}{\partial\eta}
\right|_{(\eta,\xi)=(0,0)},
\end{displaymath}
and $\tilde{\Sigma}(\eta,\xi)$ is at least of second order in the
distance from the pole.

   For the evaluation of $\Sigma(0,0)$, $\Sigma^\prime (0,0)$, and 
$\tilde{\Sigma}(\eta,\xi)$ we need to expand Eq.\ (\ref{5:6:sigmaN}).
   To the order we are working the $a_3$ term contributes only to 
$\Sigma(0,0)$ whereas the loop piece contributes to all three. 
   In contrast to the mesonic sector at ${\cal O}(p^4)$, $\tilde{\Sigma}$ 
is not zero in this case. 
   Using Eq.\ (\ref{5:5:rvretaxi1}), we obtain for the $r^2$ term 
\begin{equation}
\label{5:5:rsq}
\frac{r^2}{2\stackrel{\circ}{m}_N}
=\frac{(m_N-\stackrel{\circ}{m}_N)^2}{2\stackrel{\circ}{m}_N}
+\frac{(m_N-\stackrel{\circ}{m}_N)}{\stackrel{\circ}{m}_N}
\eta + \frac{\xi}{2\stackrel{\circ}{m}_N}.
\end{equation}
   The first term on the right-hand side contributes to $\Sigma(0,0)$ but is
${\cal O}(1/m_N^3)$, since, as we will see, the difference 
$(m_N-\stackrel{\circ}{m}_N)$ is ${\cal O}(1/m_N)$.
   The second term is ${\cal O}(1/m_N^2)$ and will contribute to 
$\Sigma^\prime (0,0)$. 
   Finally the third term contributes only to $\tilde{\Sigma}$.
 
   Applying Eq.\ (\ref{5:5:mndef}) we obtain for the physical mass 
\begin{equation}
\label{5:5:nucmass1}
m_N  =\, \stackrel{\circ}{m}_N 
- \frac{(m_N-\stackrel{\circ}{m}_N)^2}{2\stackrel{\circ}{m}_N}
-\frac{4a_3 M^2}{\stackrel{\circ}{m}_N}
+\Sigma^{(3)}_{\rm loop}(0,0),
\end{equation}
   which implies $m_N-\stackrel{\circ}{m}_N={\cal O}(1/m_N)$.\footnote{
Strictly speaking we should say 
$m_N-\stackrel{\circ}{m}_N={\cal O}[M^2/\stackrel{\circ}{m}_N,
M^3/(4\pi F_0)^2]$, where the second result originates from the loop 
contribution.}
   We can thus neglect the second term on the right-hand side of Eq.\
(\ref{5:5:nucmass1}).
   The loop contribution is only a function of $v\cdot r$ and
thus a function only of $\eta$, and, neglecting terms of higher 
order in $1/m_N$, we may replace $v\cdot r$ by 0, yielding
\begin{displaymath}
\Sigma^{(3)}_{\rm loop}(0,0)=-\frac{3{\stackrel{\circ}{g}_A^2}M^3}
                {(4\pi F_0)^2}\arccos(0).
\end{displaymath}
   We finally obtain for the physical nucleon mass
\begin{equation}
\label{5:5:nucmass2}
m_N \simeq\, 
\stackrel{\circ}{m}_N \left[1
-\frac{4a_3 M_\pi^2}{m_N^2}-\frac{3\pi{g_A^2}M_\pi^3}
                {2m_N(4\pi F_\pi)^2}\right],
\end{equation}
where, in the expression between the brackets, we have replaced all quantities
in terms of the physical quantities, because the difference is of higher
order in the chiral expansion.
   In the chiral limit, both the counter-term contribution $\sim M^2\sim m_q$ 
and the pion-loop correction $\sim M^3\sim m_q^{3/2}$ disappear.
   In other words, in the heavy-baryon framework
the situation is again as in the mesonic sector, where the parameters
of the lowest-order Lagrangian do not get modified due to higher-order 
corrections in the chiral limit.
   The same is actually true for the second parameter $\stackrel{\circ}{g}_A$
of Eq.\ (\ref{5:5:l1pinhb}) [see, e.g., Eq.\ (50) of Ref.\ 
\cite{Fearing:1997dp}].
   Using the parameters of Eqs.\ (\ref{5:3:par}) and (\ref{5:5:ciresults})
one finds that the counter term and the pion loop generate contributions
to the physical nucleon mass of $0.0733$ and 
$-0.0163$ in units of $\stackrel{\circ}{m}_N$, respectively.

   The wave function renormalization constant $Z_N$ is obtained from
Eq.\ (\ref{5:5:zndef}) as
\begin{equation}
\label{5:5:znresulttemp}
Z_N = \frac{1}{1-\Sigma'(0,0)}
\approx 1+\Sigma'(0,0)
=1-\frac{m_N-\stackrel{\circ}{m}_N}{\stackrel{\circ}{m}_N}
+\Sigma'_{\rm loop}(0,0).
\end{equation}
   To the order we are considering, we have from Eqs.\ (\ref{5:5:nucmass2})
and (\ref{5:6:sigmaN}), respectively,
\begin{eqnarray*}
\frac{m_N-\stackrel{\circ}{m}_N}{\stackrel{\circ}{m}_N}&=&
-\frac{4 a_3 M^2}{\stackrel{\circ}{m}_N^2},\\
\Sigma'_{\rm loop}(0,0)&=&-\frac{9\stackrel{\circ}{g}_A^2M^2}{4(4\pi F_0)^2}
       \left[R+{\rm ln}\left(\frac{M^2}{\mu^2}\right)+\frac{2}{3}\right].
\end{eqnarray*}
   Finally, expressing all quantities in terms of physical quantities,
the wave function renormalization constant $Z_N$ reads
\begin{equation}
\label{5:5:znresult}
Z_N = 1+\frac{4a_3M_\pi^2}{m_N^2}-\frac{9g_A^2M_\pi^2}{4(4\pi{F_\pi})^2}
       \left[R+{\rm ln}\left(\frac{M_\pi^2}{\mu^2}\right)+\frac{2}{3}\right].
\end{equation}

   As in the pion case (see Table \ref{app:dp:tab:abz} of Appendix
\ref{app_sec_dp}) $Z_N$ contains the infinite constant $R$ entering
through dimensional regularization, i.e., $Z_N$ is not a finite quantity.
   However, this is not a problem, because the wave function renormalization 
constant is not a physical observable.
   Moreover, as we have seen explicitly for the pion case,
and as discussed in Ref.\ \cite{Fearing:1997dp} for the  heavy-baryon 
Lagrangian, $Z_N$ will also depend on the specific parameterization of the 
Lagrangian.

   In Ref.\ \cite{Ecker:1997dn} it was shown that the wave function 
renormalization ``constant'' $Z_N$ is in fact a non-trivial
differential operator and should, in momentum space, depend on the
momentum of the initial or final nucleon.
   Here we argue that the findings of Ref.\ \cite{Ecker:1997dn} and
the method used above do not seem to be in conflict with each other.
   To that end, we first note that Ref.\ \cite{Ecker:1997dn} made use
of the external spinor $u_+(\vec{p}\,)=P_{v+}u(\vec{p}\,)$. 
   Using relativistic spinors normalized as in Eq.\ (\ref{5:5:spinorprops})
this corresponds to a normalization of the heavy baryon spinors
to $\bar{u}_+(\vec{p}\,)u_+(\vec{p}\,)=(p\cdot v+ m_N)$.
   To facilitate the comparison, let us consider the special choice
$v^\mu=(1,0,0,0)$. 
   In the framework of the reduction formula of Eq.\ (\ref{5:5:ffunc}),
we work with a factor $N u^{\alpha}_v/\sqrt{Z_N}$ for an external nucleon
in the initial state, where Ecker and Moj\v{z}i\v{s} would have 
$u_+(\vec{p}\,)/\sqrt{Z_N^{\rm EM}}$.
   It is now straightforward to show that the normalization factor $N$
exactly produces the additional term which, in the approach of 
Ref.\ \cite{Ecker:1997dn}, results from the additional term in the
wave function renormalization.
   This explains why the two approaches, at least up to order ${\cal O}(p^3)$,
generate the same result.
   For further discussion on this topic, the reader is referred to
\cite{Fearing:1997dp,Ecker:1997dn,Steininger:1998ya,Kambor:1998pi}.

\section{The Method of Infrared Regularization}
\label{sec_mir}
   In the discussion of the one-loop corrections to the nucleon self energy
and pion-nucleon scattering of Sec.\ \ref{sec_eld}, we saw that the 
relativistic framework for baryons did not naturally provide a simple power 
counting scheme as for mesons.
   One major difference in comparison with the mesonic sector is related
to the fact that the nucleon remains massive in the chiral limit which
also introduces another mass scale into the problem.
   Thus, because of the zeroth component one can no longer argue 
that a derivative acting on the baryon field results in a small four-momentum.
   This problem is avoided in the heavy-baryon 
approach discussed in Sec.\ \ref{sec_hbf}, where, through a field 
redefinition, the mass dependence has been shifted into an (infinite) string 
of vertices which are suppressed by powers of $1/m$.
   Since the derivatives in the heavy-baryon Lagrangian 
produce small residual four-momenta in the low-energy regime, a power
counting scheme analogous to the mesonic sector can be formulated
[see Eqs.\ (\ref{5:5:d1}) and (\ref{5:5:d3})].
   A vast majority of applications of chiral perturbation theory in
the baryonic sector were performed in the framework of the heavy-baryon
approach.
   However, it was realized some time ago that the
heavy-baryon approach, under certain circumstances, may generate
Green functions which do not satisfy the analytic properties resulting
from a (fully) relativistic field theory \cite{Bernard:1996cc}.
 
   Clearly, it would be desirable to have a method which combines the
advantages of the relativistic and the heavy-baryon approaches and,
at the same time, avoids their shortcomings---absence of 
a power counting scheme on the one hand and failure of convergence on the
other hand.
    Such approaches have been proposed and developed by various authors 
\cite{Tang:1996ca,Ellis:1997kc,Becher:1999he,Gegelia:1999qt,%
Lutz:1999yr,Becher:2000mb,Lutz:2001yb}
and here we will briefly outline the ideas of the so-called infrared
regularization \cite{Becher:1999he}.   
   Our presentation will closely follow Refs.\ 
\cite{Becher:1999he,Becher:2000mb} to which we refer the reader for 
technical details.
   Some recent applications of the new approach deal with the
electromagnetic form factors of the nucleon \cite{Kubis:2000zd}
and the baryon octet \cite{Kubis:2000aa},
$\pi N$ scattering \cite{Becher:2001hv}, axial-vector current
matrix elements \cite{Zhu:2000zf}, and
the generalized Gerasimov-Drell-Hearn
sum rule \cite{Bernard:2002bs}.

   In order to understand the problems of the heavy-baryon approach regarding
the analytic behavior of invariant functions let us start with a simple 
example \cite{Becher:2000mb}.
   To that end we consider the $s$ channel of pion-nucleon scattering 
(see Sec.\ \ref{subsec_apnstl}).
   The invariant amplitudes $B^\pm$ of Table \ref{5:3:tableresults} develop 
poles for $\nu=\pm \nu_B$ (the upper and lower signs correspond to 
$s=m_N^2$ and $u=m_N^2$, respectively).
   For example, the singularity due to the nucleon pole in the $s$ channel
is understood in terms of the relativistic propagator
\begin{equation}
\label{5:6:relprop}
\frac{1}{(p+q)^2-m_N^2}=\frac{1}{2p\cdot q+M_\pi^2},
\end{equation}
which, of course, has a pole at $2p\cdot q=-M_\pi^2$ or, equivalently,
$s=m_N^2$.
   (Analogously, a second pole results from the $u$ channel at $u=m_N^2$.)
   We also note 
that the propagator of Eq.\ (\ref{5:6:relprop})
counts as ${\cal O}(p^{-1})$, because it is part of a tree-level diagram
so that the four-momentum $q$ is assumed to be small, i.e., of ${\cal O}(p)$.
   Although both poles are not in the physical region of pion-nucleon 
scattering, analyticity of the invariant amplitudes requires these poles
to be present in the amplitudes.
   Let us compare the situation with a heavy-baryon type of expansion,
where, for simplicity, we choose as the four-velocity $p^\mu=m_N v^\mu$,
\begin{equation}
\label{5:6:relpropv}
\frac{1}{2p\cdot q+M_\pi^2}=
\frac{1}{2m_N}\frac{1}{v\cdot q+ \frac{M_\pi^2}{2m_N}}=
\frac{1}{2 m_N}\frac{1}{v\cdot q}\left(1-\frac{M_\pi^2}{2 m_N v\cdot q}
+\cdots\right).
\end{equation}
   Clearly, to any finite order the heavy-baryon expansion produces poles 
at $v\cdot q=0$ instead of a simple pole at $v\cdot q=-M_\pi^2/(2 m_N)$ and 
will thus not generate the (nucleon) pole structures of the functions $B^\pm$.
   
\begin{figure}[htb]
\begin{center} 
\caption{\label{5:6:fig:triangle}      
Triangle diagram. The symbol $\bullet$ denotes an interaction
which transfers the momentum $q$ to the virtual pion.}
\vspace{2em}
\epsfig{file=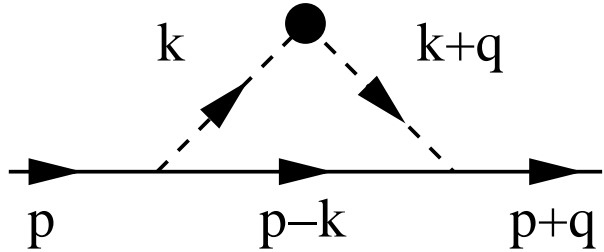,width=8cm}
\end{center}
\end{figure}

   As a second example, we consider the so-called triangle diagram
of Fig.\ \ref{5:6:fig:triangle} which will serve to illustrate the
different analytic properties of invariant functions obtained from
loop diagrams in the relativistic and heavy-baryon approaches.
   A diagram of this type appears in many calculations such as the
scalar or electromagnetic form factors of the nucleon, where $\bullet$
represents an external scalar or electromagnetic field, 
or $\pi N$ or Compton scattering, where $\bullet$ stands for 
two pion or electromagnetic fields.
   In all of these cases a four-momentum $q$ is transferred to the nucleon
and the analytic properties of the Feynman diagram as a function of 
$t\equiv q^2$ are determined by the pole structure of the propagators.

Thus we need to discuss some properties of the integral
\begin{equation}
\label{5:6:gammat}
\gamma(t)\equiv i\int \frac{d^4 k}{(2\pi)^4} \frac{1}{k^2-M^2_\pi+i0^+}
\frac{1}{(k+q)^2-M^2_\pi+i0^+}\frac{1}{(p-k)^2-m^2_N+i0^+},
\end{equation}
where $t=q^2$.
   We assume the initial and final nucleons to be on the mass shell,
$p^2=m^2_N=(p+q)^2$ which implies $2p\cdot q=-t$.
   Counting powers we see that the integral converges.
   The function $\gamma(t)$ is analytic in $t$ except for a cut along the 
positive real axis starting at $t= 4M^2_\pi$ which expresses the fact that two 
on-shell pions can be produced for $t\geq 4M^2_\pi$.
   In the following discussion of the analytic properties of Eq.\ 
(\ref{5:6:gammat}) we will concentrate on the imaginary
part of $\gamma(t)$ which we will derive applying the Cutkosky (or cutting)
rules \cite{Cutkosky:1960sp,LeBellac:cq,Peskin:ev}.
   The rules, as summarized in \cite{Peskin:ev}, read:
   In order to obtain $2i {\rm Im}\{\gamma(t)\}$ first cut through the 
diagram in all possible ways such that the cut propagators
can simultaneously be put on shell. 
   Next, for each cut one need to replace each cut propagator 
$1/(p^2-m^2+i0^+)$ by $-2\pi i\delta(p^2-m^2)$.
   Finally, sum the contributions of all possible cuts.
   In the present case, the two terms where the nucleon propagator is
simultaneously cut with either the first or the second pion propagator
do not contribute.
   The result from cutting the two pion propagators reads
\begin{equation}
\label{5:6:img1t}
2i{\rm Im}\{\gamma(t)\}=(-2\pi i)^2 i \int \frac{d^4 k}{(2\pi)^4}
\frac{\delta(k^2-M_\pi^2)\delta((k+q)^2-M_\pi^2)}{(p-k)^2-m_N^2+i0^+}.
\end{equation}
   In order to evaluate Eq.\ (\ref{5:6:img1t}), we choose a frame where
$q^\mu=(q_0,\vec{0})$ with $q_0=\sqrt{t}>0$, and $p^\mu=(-q_0/2,\vec{p}\,)$.
   Using 
\begin{displaymath}
\delta(k^2-M_\pi^2)\delta((k+q)^2-M_\pi^2)=\delta(k^2-M_\pi^2)
\frac{1}{2\sqrt{t}}\delta\left(k_0+\frac{\sqrt{t}}{2}\right)
\end{displaymath}
we find, as an intermediate result,
\begin{equation}
\label{5:6:img2t}
{\rm Im}\{\gamma(t)\}=-\frac{1}{16\pi^2\sqrt{t}}\int d^3 k
\delta\left(\vec{k}^2
+M_\pi^2-\frac{t}{4}\right)\frac{1}{
-\frac{t}{2}+2\vec{p}\cdot\vec{k}+M_\pi^2+i0^+}.
\end{equation}
   For $t<4 M_\pi^2$, the delta function in Eq.\ (\ref{5:6:img2t}) always
vanishes, showing that the cut starts, as anticipated, at $t=4M_\pi^2$. 
   Applying the mass-shell condition $p^2=m_N^2$, we write
\begin{eqnarray*}
\vec{p}=i\frac{\sqrt{4m_N^2-t}}{2} \hat{e}_z&\mbox{for}&4M_\pi^2\leq t\leq
4m_N^2,\\
\vec{p}=\frac{\sqrt{t-4m_N^2}}{2} \hat{e}_z&\mbox{for}&4m_N^2\leq t.
\end{eqnarray*}
   Performing the integration using spherical coordinates, 
the result for the first case reads
\begin{eqnarray}
\label{5:6:imgammat}
{\rm Im}\{\gamma(t)\}&=&\frac{\sqrt{t-4M_\pi^2}}{16\pi \sqrt{t}}\int_{-1}^1
dz \frac{1}{t-2M_\pi^2-i\sqrt{4m_N^2-t}\sqrt{t-4M_\pi^2}\, z-i0^+}\nonumber\\
&=&\frac{i}{16\pi \sqrt{t}\sqrt{4m_N^2-t}}\ln\left(\frac{1-iy}{1+iy}\right)
\nonumber\\
&=&\frac{1}{8\pi \sqrt{t(4m_N^2-t)}}\arctan(y),
\end{eqnarray}
where
\begin{displaymath}
y=\frac{\sqrt{(t-4M^2_\pi)(4m^2_N-t)}}{t-2M^2_\pi},\quad
4M_\pi^2\leq t\leq 4m_N^2.
\end{displaymath}
   The second case, $t>4m_N^2$, is obtained analogously by the replacement 
$i\sqrt{4m_N^2-t} \to \sqrt{t-4m_N^2}$:
\begin{equation}
\label{5:6:img3ta}
{\rm Im}\{\gamma(t)\}=\frac{1}{16\pi\sqrt{t(t-4m_N^2)}}
\ln\left(\frac{t-2M_\pi^2+\sqrt{t-4m_N^2}\sqrt{t-4M_\pi^2}}{
t-2M_\pi^2-\sqrt{t-4m_N^2}\sqrt{t-4M_\pi^2}}\right).
\end{equation}
   Equations (\ref{5:6:imgammat}) and (\ref{5:6:img3ta}) agree with the results
given in Eq.\ (B.43) of Ref.\ \cite{Gasser:1987rb}.
   In the low-energy region $t\ll m^2_N$, and Eq.\ (\ref{5:6:imgammat})
becomes
\begin{equation}
\label{5:6:imgammatexp}
\mbox{Im}\{\gamma(t)\}\approx\frac{1}{16\pi m_N\sqrt{t}}\arctan(x),\quad
x=\frac{2m_N\sqrt{t-4M^2_\pi}}{t-2M^2_\pi}.
\end{equation}
   Taking the factors resulting from the vertices and the relevant
tensor structures of the loop integral into account, the contribution 
of Fig.\ \ref{5:6:fig:triangle} to the imaginary part of the scalar
form factor of the nucleon reads  \cite{Gasser:1987rb,Becher:1999he}    
\begin{displaymath}
{\rm Im}\{\sigma(t)\}=\frac{3 g^2_A M_\pi^2 m_N}{4 F_\pi^2}(t-2M_\pi^2)
{\rm Im}\{\gamma(t)\},
\end{displaymath}
where $\sigma(t)$ is defined in terms
of the $u$- and $d$-quark scalar densities $\bar{u}u$ and $\bar{d}d$ as
\begin{equation}
\label{5:6:sffdef}
\langle N(p')| m[\bar{u}(0)u(0)+\bar{d}(0)d(0)]|N(p)\rangle=
\bar{u}(p')u(p)\sigma(t),
\end{equation}
where $m=m_u=m_d$ and $t=q^2=(p'-p)^2$.

   We will now investigate two limiting procedures. 
   First, we consider a fixed value $t>4 M_\pi^2$ and let $m_N\to \infty$. 
   In that case $x\gg 1$, and one would use the expansion
\begin{displaymath}
\arctan(x)=\frac{\pi}{2}-\frac{1}{x}+\frac{1}{3 x^3}-\cdots, \quad
x>1.
\end{displaymath}
   Keeping only the leading order term, we find
\begin{equation}
\label{5:6:imsfhbchpt}
{\rm Im}\{\sigma(t)\}=
\frac{3 g_A^2 M_\pi^2}{128 F_\pi^2}\frac{t-2 M_\pi^2}{\sqrt{t}},
\end{equation}
which corresponds exactly to the result of HBChPT at 
${\cal O}(p^3)$ \cite{Bernard:1992qa}.
   This result corresponds to the standard chiral expansion which treats
the quantity $x$ of Eq.\ (\ref{5:6:imgammatexp}) as ${\cal O}(p^{-1})$,
because $m_N={\cal O}(p^0)$ and $t,M_\pi^2={\cal O}(p^2)$.

   However, for a fixed $m_N$ we may also consider a small enough $t$ 
close to the threshold value $t_{\rm thr}=4M^2_\pi$ so that $x<1$. 
   In that case the expansion of the arctan reads
\begin{displaymath}
\arctan(x)=x-\frac{x^3}{3}+\cdots
\end{displaymath}
   yielding   
\begin{equation}
\label{5:6:imsfrchpt}
{\rm Im}\{\sigma(t)\}\approx
\frac{3 g_A^2 M_\pi^2 m_N}{32 \pi F_\pi^2}\frac{\sqrt{t-4 M_\pi^2}}{\sqrt{t}},
\end{equation}
where we have neglected higher powers of $x$.
   The critical value of $t$ corresponding to $x=1$ is given by
\begin{displaymath}
t_{\rm cr}=4M_\pi^2\left[1+\frac{\mu^2}{4}+O(\mu^4)\right],
\end{displaymath}
where $\mu=M_\pi/m_N$.
   Clearly the behavior of Eq.\ (\ref{5:6:imsfrchpt}) is very different
from the chiral expansion of Eq.\ (\ref{5:6:imsfhbchpt}) and, similar
to the discussion of Eq.\ (\ref{5:6:relpropv}), a finite sum of terms
in HBChPT cannot reproduce such a threshold behavior
\cite{Becher:1999he}.
   The rapid variation of the imaginary part can be understood in
terms of the analytic properties of the arctan which, as a function
of the complex variable $z$, is analytic in the entire complex plane
save for cuts along the positive and negative imaginary axis starting at 
$\pm i$. 
   These branch points corresponding to $x=\pm i$ are obtained for
$t=4M_\pi^2(1-\mu^2/4)$ which is just below the physical threshold
$t_{\rm thr}=4M_\pi^2$.  
   For that reason an expansion around $x=0$ corresponding to 
$t=4M_\pi^2$ has a small radius of convergence.
  
   Clearly, the heavy-baryon approach does not produce the correct 
analytic structure as generated by the relativistic loop diagram. 
   Moreover the low-energy behavior of Eq.\ (\ref{5:6:imgammatexp})
cannot be accounted for in the standard chiral analysis because the
argument $x$ is of order ${\cal O}(p^{-1})$.
   What is needed is a method which produces both the relevant analytic
structure and a consistent power counting.

   Here we will illustrate the method of Ref.\ \cite{Becher:1999he} by
means of the nucleon self energy diagram of Fig.\ \ref{5:4:fig:nsepl}.
   For a---at this stage---qualitative discussion of its 
properties we focus on the scalar loop integral
\begin{equation}
\label{5:6:defh}
H(p^2,n)=-i\int \frac{d^n k}{(2\pi)^n}\frac{1}{k^2-M_\pi^2+i0^+}
\frac{1}{k^2-2p\cdot k+(p^2-m_N^2)+i0^+},
\end{equation}
where, as usual, the right-hand side is thought of as a Feynman integral
which has to be analytically continued as a function of the space-time 
dimension $n$. 
   Counting powers, we see that, for $n=4$, the integrand behaves for large
values of the integration variable $k$ as $k^3/k^4$, producing a
logarithmic ultraviolet divergence while, on the other hand, the integral
converges for $n<4$.
   Let us now consider the limit $M_\pi^2\to 0$. 
   In this case, for both $p^2=m_N^2$ and $p^2\neq m_N^2$, the integral
is infrared regular for $n=4$ because, for small momenta, the integrand
behaves as $k^3/k^3$ and $k^3/k^2$, respectively. 
   For $n=3$ the integral is infrared regular for $p^2\neq m_N^2$ but singular
for $p^2=m_N^2$. 
   For any smaller value of $n$ it is infrared singular for arbitrary $p^2$.
   The infrared singularity as $M_\pi^2\to 0$ originates in the region,
where the integration variable $k$ is small, i.e., of the order ${\cal O}(p)$.
   Counting powers of momenta, we (naively) expect  
this part to be of order ${\cal O}(p^{n-3})$.
   On the other hand, for loop momenta of the order of
and larger than the nucleon mass we expect power counting to fail,
   because the momentum of the nucleon propagating in loop integral is
not constrained to be small in contrast to the case of tree-level diagrams
[see Eq.\ (\ref{5:6:relprop})].

   In order to explain these qualitative statements let us discuss the
integral in more detail.
   We first introduce the Feynman parameterization\footnote{In order to
make it easier for the interested reader to follow Ref.\ \cite{Becher:1999he} 
we have used the notation there, omitting a factor $\mu^{n-4}$ and choosing
the opposite overall sign in comparison with previous sections.}
\begin{equation}
\label{5:6:feynmanparh}
H(p^2,n)=-i\int \frac{d^n k}{(2\pi)^n}\int_0^1 dz \frac{1}{[az+b(1-z)]^2},
\end{equation}
with $a=k^2-2k\cdot p+p^2-m_N^2+i0^+$ and $b=k^2-M_\pi^2+i0^+$, perform
the shift $k\to k+pz$, and obtain
\begin{displaymath}
H(p^2,n)=-i
\int_0^1 dz\int\frac{d^n k}{(2\pi)^n}\frac{1}{[k^2-A(z)
+i0^+]^2},
\end{displaymath}
where
\begin{displaymath}
A(z)=z^2 p^2-z(p^2-m_N^2+M_\pi^2)+M_\pi^2.
\end{displaymath}
   We then apply Eq.\ (\ref{app:drb:moregenint}) of Appendix 
\ref{app_drb},
\begin{equation}
\label{5:6:defh1}
H(p^2,n)=
\frac{\Gamma\left(2-\frac{n}{2}\right)}{(4\pi)^{\frac{n}{2}}}
\int_0^1 dz [A(z)-i0^+]^{\frac{n}{2}-2}.
\end{equation}
   The relevant properties can nicely be displayed at the threshold 
$p^2_{\rm thr}=(m_N+M_\pi)^2$, where $A(z)=[z(m_N+M_\pi)-M_\pi]^2$  
is particularly simple.
   The small imaginary part can be dropped in this case, because $A(z)$
is never negative.  
    Splitting the integration interval into $[0,z_0]$ and $[z_0,1]$ with
$z_0=M_\pi/(m_N+M_\pi)$, we have, for $n>3$,
\begin{eqnarray*}
\int_0^1 dz [A(z)]^{\frac{n}{2}-2}&=&\int_0^{z_0}dz [M_\pi-z(m_N+M_\pi)]^{n-4}
\\
&&+\int_{z_0}^1dz [z(m_N+M_\pi)-M_\pi]^{n-4}\\
&=&\frac{1}{(n-3)(m_N+M_\pi)}(M_\pi^{n-3}+m_N^{n-3}),
\end{eqnarray*}
yielding, through analytic continuation, for arbitrary $n$
\begin{equation}
\label{5:6:defhthr}
H((m_N+M_\pi)^2,n)=
\frac{\Gamma\left(2-\frac{n}{2}\right)}{(4\pi)^{\frac{n}{2}}(n-3)}
\left(\frac{M_\pi^{n-3}}{m_N+M_\pi}+\frac{m_N^{n-3}}{m_N+M_\pi}\right).
\end{equation}
   The first term, proportional to $M_\pi^{n-3}$, is defined as the so-called 
infrared singular part $I$ which, as $M_\pi\to 0$, behaves as in the
qualitative discussion above. 
   Since $M_\pi\to 0$ implies $p^2_{\rm thr} \to m_N^2$ this term is
singular for $n\leq 3$.
   The second term, proportional to $m_N^{n-3}$, is defined as the 
infrared regular part $R$ and can be thought of as originating from an 
integration region where $k$ is of order $m_N$ so that the tree-level 
counting rules no longer apply [see Eq.\ (\ref{5:6:relprop})].
   Note that for non-integer $n$ the infrared singular
part contains non-integer powers of $M_\pi$, while an
expansion of the regular part always contains non-negative integer powers 
of $M_\pi$ only.

   Let us now turn to a {\em formal} definition of the infrared singular 
and regular parts \cite{Becher:1999he} which makes use of the Feynman 
parameterization of Eq.\ (\ref{5:6:defh1}).
   Introducing the dimensionless variables
\begin{equation}
\label{5:6:alphaomegadef}
\alpha=\frac{M_\pi}{m_N},\quad \Omega=\frac{p^2-m_N^2-M_\pi^2}{2m_N M_\pi},
\end{equation}
counting as ${\cal O}(p)$ and ${\cal O}(p^0)$ [$p^2-m_N^2={\cal O}(p)$], 
respectively, we rewrite $A(z)$ as
\begin{displaymath}
A(z)=m_N^2[z^2-2\alpha\Omega z(1-z)+\alpha^2(1-z)^2]\equiv m_N^2 C(z),
\end{displaymath}
so that $H$ is now given by 
\begin{equation}
\label{5:6:defh2}
H(p^2,n)=\kappa(n)
\int_0^1 dz [C(z)-i0^+]^{\frac{n}{2}-2},
\end{equation}
where
\begin{equation}
\label{5:6:kappan}
\kappa(n)
=\frac{\Gamma\left(2-\frac{n}{2}\right)}{(4\pi)^{\frac{n}{2}}}m_N^{n-4}.
\end{equation}
   The infrared singularity originates from small values of $z$, where
the function $C(z)$ goes to zero as $M_\pi\to 0$.
   In order to isolate the divergent part one scales the integration variable
$z\equiv \alpha x$ so that the upper limit $z=1$ in Eq.\ (\ref{5:6:defh2})
corresponds to $x=1/\alpha\to
\infty$ as $M_\pi\to 0$.
   An integral $I$ having the same infrared singularity as $H$ is then
defined which is identical to $H$ except that the upper limit is
replaced by $\infty$:
\begin{equation}
\label{5:6:Idef}
I\equiv\kappa(n)\int_0^\infty dz[C(z)-i0^+]^{\frac{n}{2}-2}
=\kappa(n) \alpha^{n-3}\int_0^\infty [D(x)-i0^+]^{\frac{n}{2}-2},
\end{equation}
where 
\begin{displaymath}
D(x)=1-2\Omega x+x^2+2\alpha x(\Omega x-1)+\alpha^2 x^2.
\end{displaymath}
   (The pion mass $M_\pi$ is not sent to zero.)
   Accordingly, the regular part of $H$ is defined as
\begin{equation}
\label{5:6:Rdef}
R\equiv-\kappa (n)\int_1^\infty dz [C(z)-i0^+]^{\frac{n}{2}-2},
\end{equation}
so that 
\begin{equation}
\label{5:6:hir}
H=I+R.
\end{equation}
   Let us verify that the definitions of Eqs.\ (\ref{5:6:Idef}) and
(\ref{5:6:Rdef}) indeed reproduce the behavior of Eq.\ (\ref{5:6:defhthr}).
   To that end we make use of $\Omega_{\rm thr}=1$, yielding
\begin{equation}
\label{5:6:ithr1}
I_{\rm thr}=\kappa(n)\alpha^{n-3}\int_0^\infty dx
\left\{[(1+\alpha)x-1]^2-i0^+
\right\}^{\frac{n}{2}-2},
\end{equation}
which converges for $n<3$. 
   In order to continue the integral to $n>3$, we write 
\cite{Becher:1999he}
\begin{eqnarray*}
\lefteqn{\left\{[(1+\alpha)x-1]^2-i0^+\right\}^{\frac{n}{2}-2}=}\\
&&=\frac{(1+\alpha)x-1}{(1+\alpha)(n-4)}\frac{d}{dx}
\left\{[(1+\alpha)x-1]^2-i0^+\right\}^{\frac{n}{2}-2},
\end{eqnarray*}
and make use of a partial integration 
\begin{eqnarray*}
\lefteqn{\int_0^\infty dx 
\left\{[(1+\alpha)x-1]^2-i0^+\right\}^{\frac{n}{2}-2}
=}\\
&&\left[\frac{(1+\alpha)x-1}{(1+\alpha)(n-4)}
\left\{[(1+\alpha)x-1]^2-i0^+\right\}^{\frac{n}{2}-2}
\right]_0^\infty\\
&&-\frac{1}{n-4}
\int_0^\infty dx \left\{[(1+\alpha)x-1]^2-i0^+\right\}^{\frac{n}{2}-2}.
\end{eqnarray*}
   For $n<3$, the first expression vanishes at the upper limit and,
at the lower limit, yields $1/[(1+\alpha)(n-4)]$.
   Bringing the second expression to the left-hand side, we may then
continue the integral analytically as
\begin{equation}
\label{5:6:intanc}
\int_0^\infty dx\left\{[(1+\alpha)x-1]^2-i0^+
\right\}^{\frac{n}{2}-2}=\frac{1}{(n-3)(1+\alpha)},
\end{equation}
so that we obtain for $I_{\rm thr}$
\begin{equation}
\label{5:6:ithr}
I_{\rm thr}=\kappa(n)\alpha^{n-3}\frac{1}{(n-3)(1+\alpha)}
=
\frac{\Gamma\left(2-\frac{n}{2}\right)}{(4\pi)^{\frac{n}{2}}(n-3)}
\frac{M_\pi^{n-3}}{m_N+M_\pi},
\end{equation}
which agrees with the infrared singular part $I$ of Eq.\ (\ref{5:6:defhthr}).
 
  The threshold value of the regular part of 
Eq.\ (\ref{5:6:Rdef}) is obtained by analytic continuation from $n<3$ 
to $n>3$: 
\begin{eqnarray}
\label{5:6:Rthr}
R_{\rm thr}&=&-\frac{\Gamma\left(2-\frac{n}{2}\right)}{(4\pi)^{\frac{n}{2}}}
\int_1^\infty [z(m_N+M_\pi)-M_\pi]^{n-4}\nonumber\\
&=&
-\frac{\Gamma\left(2-\frac{n}{2}\right)}{(4\pi)^{\frac{n}{2}}}
\frac{1}{(n-3)(m_N+M_\pi)}(\infty^{n-3}-m_N^{n-3})\nonumber\\
&\stackrel{\mbox{$n<3$}}{=}&\frac{\Gamma\left(2-\frac{n}{2}\right)}{(4\pi)^{\frac{n}{2}}(n-3)}
\frac{m_N^{n-3}}{m_N+M_\pi},
\end{eqnarray}
which is indeed the regular part $R$ of Eq.\ (\ref{5:6:defhthr}).

   What distinguishes $I$ from $R$ is that, for non-integer values of
$n$, the chiral expansion of $I$ gives rise to non-integer powers of 
${\cal O}(p)$, whereas the regular part $R$ may be 
expanded in an ordinary Taylor series.
   For the threshold integral, this can nicely be seen by expanding 
$I_{\rm thr}$ and $R_{\rm thr}$ in the pion mass counting as 
${\cal O}(p)$.
   On the other hand, it is the regular part which does not satisfy 
the counting rules valid at tree level.
    The basic idea of the infrared regularization consists of replacing
the general integral $H$ of Eq.\ (\ref{5:6:defh}) 
by its infrared singular part $I$, defined in Eq.\ (\ref{5:6:Idef}),
and dropping the regular part $R$, defined in Eq.\ (\ref{5:6:Rdef}).
   In the low-energy region $H$ and $I$ have the same analytic properties
whereas the contribution of $R$, which is of the type of an infinite
series in the momenta, can be included by adjusting the coefficients of 
the most general effective Lagrangian.  

   As discussed in detail in Ref.\ \cite{Becher:1999he}, the method can 
be generalized to an arbitrary one-loop graph.
   Using techniques similar to those of Appendices \ref{app_subsec_ipipi}
and \ref{subsec_jpin}, it is first argued that tensor integrals involving 
an expression of the type $k^{\mu_1}\cdots k^{\mu_2}$ in the numerator 
may always be reduced to scalar loop integrals of the form
\begin{displaymath}
-i\int\frac{d^n k}{(2\pi)^n}\frac{1}{a_1\cdots a_m}\frac{1}{b_1\cdots b_n},
\end{displaymath}
where $a_i=(q_i+k)^2-M_\pi^2+i0^+$ and $b_i=(p_i-k)^2-m_N^2+i0^+$ are
inverse meson and nucleon propagators, respectively.
   Here, the $q_i$ refer to four-momenta of ${\cal O}(p)$ and the $p_i$ are 
four-momenta which are not far off the nucleon mass shell, i.e.,
$p_i^2=m_N^2+{\cal O}(p)$.
    Using the Feynman parameterization, all pion propagators and all nucleon
propagators are separately combined, and the result is written in such
a way that it is obtained by applying $(m-1)$ and $(n-1)$ partial derivatives
with respect to $M_\pi^2$ and $m_N^2$, respectively, to a master formula.
   A simple illustration is given by
\begin{displaymath}
\frac{1}{a_1 a_2}=\int_0^1 dz \frac{1}{[a_1 z+a_2(1-z)]^2}
=\frac{\partial}{\partial M_\pi^2}\int_0^1 dz \frac{1}{a_1 z+a_2 (1-z)},
\end{displaymath}
where $a_i=(q_i+k)^2-M_\pi^2+i0^+$.
   Of course, the expressions become more complicated for larger numbers
of propagators.
   The relevant property of the above procedure is that the result of 
combining the meson propagators is of the type $1/A$ with 
$A=(k+q)^2-M_\pi^2+i0^+$, where $q$ is a linear combination of the $m$ momenta 
$q_i$, with an analogous expression $1/B$ for the nucleon propagators.
   Finally, the expression
\begin{displaymath}
-i\int\frac{d^n k}{(2\pi)^n}\frac{1}{AB}
\end{displaymath}
may then be treated in complete analogy to $H$ of Eq.\ (\ref{5:6:defh}),
i.e., the denominators are combined as in Eq.\ (\ref{5:6:feynmanparh}),
and the infrared singular and regular pieces are identified by 
writing $\int_0^1 dz\cdots = \int_0^\infty dz \cdots-\int_1^\infty dz \cdots$.

   A crucial question is whether the infrared regularization respects the
constraints of chiral symmetry as expressed through the chiral Ward identities.
   The argument given in Ref.\ \cite{Becher:1999he} that this is indeed the
case is as follows.
   The total nucleon-to-nucleon transition amplitude of Eq.\ (\ref{4:btbta}) is
chirally symmetric, i.e., invariant under a simultaneous local transformation
of the quark fields and the external fields (see Appendix \ref{app_gfwi}
for an illustration).
   In terms of the effective theory, the contribution from all the tree-level
diagrams is chirally symmetric 
so that the loop contribution must also be chirally symmetric.
   Since we work in dimensional regularization this statement holds for
an arbitrary $n$.
   However, as we have seen in the example of Eq.\ (\ref{5:6:defhthr}),
the separation into infrared singular and regular 
parts amounts to distinguishing between contributions of non-integer and 
non-negative integer powers in the momentum expansion.
   Since these powers do not mix for arbitrary $n$, the infrared singular and 
regular parts must be separately chirally symmetric.
   Finally, the regular part can be expanded in powers of either momenta
or quark masses, and thus may as well be absorbed in the (modified) tree-level
contribution.   

\begin{figure}[htb]
\begin{center} 
\caption{\label{5:6:fig:relprophb}      
Expansion of the relativistic propagator (single line)
in terms of heavy-baryon propagators (double line) 
and self-energy insertions (cross).}
\vspace{2em}
\epsfig{file=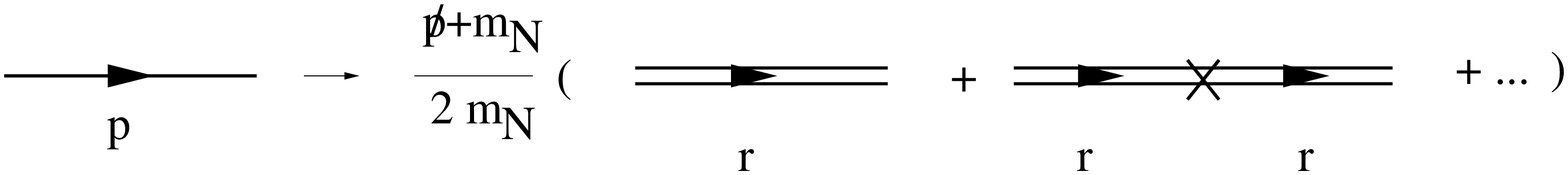,width=12cm}
\end{center}
\end{figure}

   Let us finally establish the connection between the infrared singular part
$I$ and the corresponding result in HBChPT. 
   To that end, we first consider the relativistic propagator by expressing
the (off-shell) four-momentum as $p=m_N v+r$,
\begin{eqnarray}
\label{5:6:relprophb}
\lefteqn{\frac{i}{p\hspace{-.5em}/\hspace{.2em}-m_N+i0^+}=
i\frac{p\hspace{-.5em}/\hspace{.2em}+m_N}{p^2-m_N^2+i0^+}
=i\frac{p\hspace{-.5em}/\hspace{.2em}+m_N}{2m_N v\cdot r+ r^2+i0^+}}
\nonumber\\
&=&i\frac{p\hspace{-.5em}/\hspace{.2em}+m_N}{2m_N v\cdot r+i0^+}
\,\frac{1}{1+\frac{r^2}{2m_N v\cdot r+i0^+}}\nonumber\\
&\mapsto&\frac{p\hspace{-.5em}/\hspace{.2em}+m_N}{2m_N}
\frac{i}{v\cdot r+i0^+}\left[1+\frac{ir^2}{2m_N}\frac{i}{v\cdot r+i0^+}+
\left(\frac{ir^2}{2m_N}\frac{i}{v\cdot r+i0^+}\right)^2+\cdots\right].\nonumber
\\
\end{eqnarray}
   In the last step, we have assumed that $r$ is small enough to allow
for an expansion in terms of a geometric series.
   The result of Eq.\ (\ref{5:6:relprophb}) is displayed in 
Fig.\ \ref{5:6:fig:relprophb} and may be interpreted as an infinite
series in terms of the heavy-baryon propagator $i/(v\cdot r+i0^+)$ and
the self-energy insertion $-i\Sigma=i r^2/2m_N$
which has the form of a non-relativistic kinetic energy.
(Note that the expression still involves the operator 
$(p\hspace{-.5em}/\hspace{.2em}+m_N)/2m_N$.)
   Let us apply Eq.\ (\ref{5:6:relprophb}) to the loop integral $H$ of
Eq.\ (\ref{5:6:defh}) by first expanding the integrand and then
performing the summation. 
   This corresponds to the prescription proposed in Refs.\ 
\cite{Tang:1996ca,Ellis:1997kc} for identifying the low-energy or, in the
nomenclature of Ref.\ \cite{Tang:1996ca}, soft contribution to a
Feynman graph.
   In the case at hand we obtain
\begin{equation}
H(p^2,n)\to\sum_{j=0}^\infty I_j,
\end{equation}
where
\begin{equation}
\label{5:6:ijdef}
I_j=\frac{-i}{2m_N}\int\frac{d^n k}{(2\pi)^n}\frac{1}{k^2-M_\pi^2+i0^+}
\,\frac{1}{v\cdot (r-k)+i0^+}
\left[\frac{-(r-k)^2}{2m_N v\cdot(r-k)+i0^+}\right]^j,
\end{equation}
   which is somewhat easier to handle if we perform the shift $k\to 
k+r$ and then the substitution $k\to -k$,
\begin{equation}
\label{5:6:ijdef2}
I_j=i\frac{(-)^{j+1}}{(2m_N)^{j+1}}\int\frac{d^n k}{(2\pi)^n}
\frac{1}{(k-r)^2-M_\pi^2+i0^+}
\,\frac{(k^2)^j}{(v\cdot k+i0^+)^{j+1}}.
\end{equation}
   As above, we explicitly discuss the threshold $p^2_{\rm thr}=(m_N+M_\pi)^2$
by inserting $r=M_\pi v$ into Eq.\ (\ref{5:6:ijdef2}).
   Defining $X=k^2-2M_\pi v\cdot k+i0^+$ and $Y=v\cdot k+i0^+$,
we have
\begin{equation}
\label{5:6:ijthr}
I_{j,\rm thr}=i \frac{(-)^{j+1}}{(2m_N)^{j+1}}\int\frac{d^n k}{(2\pi)^n}
\frac{(X+2M_\pi Y)^j}{X Y^{j+1}}.
\end{equation}
   The different $I_{j,\rm thr}$ are related by a simple recursion relation
\begin{equation}
\label{5:6:ijp1jthr}
I_{j+1,\rm thr}=-\frac{M_\pi}{m_N} I_{j,\rm thr},\quad j\geq 0,
\end{equation}
implying
\begin{equation}
I_{j,\rm thr}=(-)^j \left(\frac{M_\pi}{m_N}\right)^j I_{0,\rm thr},
\quad j\geq 0.
\end{equation}
   Equation (\ref{5:6:ijp1jthr}) is easily verified:
\begin{eqnarray*}
I_{j+1,\rm thr}&=&i\frac{(-)^{j+2}}{(2m_N)^{j+2}}
\int\frac{d^n k}{(2\pi)^n}
\frac{(X+2M_\pi Y)^j}{X Y^{j+1}}\,
\frac{X+2M_\pi Y}{Y}\\
&=&i\frac{(-)^{j+2}}{(2m_N)^{j+2}}\int \frac{d^n k}{(2\pi)^n}
\frac{(k^2)^j}{(v\cdot k+i0^+)^{j+2}}\\
&&-\frac{M_\pi}{m_N}i\frac{(-1)^{j+1}}{(2m_N)^{j+1}}\int
\frac{d^n k}{(2\pi)^n}\frac{(X+2M_\pi Y)^j}{X Y^{j+1}}\\
&=&-\frac{M_\pi}{m_N}I_{j,\rm thr},
\end{eqnarray*}
where we made use of the fact that the first term in the second
line vanishes in dimensional regularization
[see Eq.\ (\ref{app:dnkk2pvkmq})].
   We then obtain for the series, evaluated at threshold,
\begin{displaymath}
\sum_{j=0}^\infty I_{j,\rm thr}=I_{0,\rm thr}\sum_{j=0}^{\infty}
(-)^j \left(\frac{M_\pi}{m_N}\right)^j
=\frac{m_N}{m_N+M_\pi} I_{0,\rm thr}.
\end{displaymath}
   What remains to be determined is the threshold integral 
\begin{displaymath}
I_{0,{\rm thr}}=\frac{-i}{2m_N}
\int\frac{d^n k}{(2\pi)^n}\frac{1}{k^2-2M_\pi v\cdot k+i0^+}
\frac{1}{v\cdot k+i0^+}.
\end{displaymath}
   Performing a shift $k\to k+M_\pi v$, combining the denominators as in
Eq.\ (\ref{app:feynmantricksphb}), performing another shift $k\to k-yv$,
and making use of Eq.\ (\ref{app:drb:moregenint}), one finds
\begin{displaymath}
I_{0,{\rm thr}}=\frac{1}{m_N}
\frac{\Gamma\left(2-\frac{n}{2}\right)}{(4\pi)^\frac{n}{2}}
\int_0^\infty dy \left\{(y-M_\pi)^2-i0^+\right\}^{\frac{n}{2}-2}.
\end{displaymath}
   Finally, performing a substitution $y=M_\pi x$ and using the analytic
continuation of Eq.\ (\ref{5:6:intanc}) with $\alpha=0$, we obtain
\begin{equation}
\label{5:6:I0thr}
I_{0,{\rm thr}}= \frac{\Gamma\left(2-\frac{n}{2}\right)}{(4\pi)^\frac{n}{2}
(n-3)}\frac{M_\pi^{n-3}}{m_N}.
\end{equation}  
   Inserting Eq.\ (\ref{5:6:I0thr}) into the series, the final 
result reads
\begin{equation}
\label{5:6:series}
\sum_{j=0}^\infty I_{j,\rm thr}
=\frac{\Gamma\left(2-\frac{n}{2}\right)}{(4\pi)^\frac{n}{2}(n-3)}
\frac{M_\pi^{n-3}}{m_N+M_\pi},
\end{equation}
which is the same as $I_{\rm thr}$ of Eq.\ (\ref{5:6:ithr}).
   This example shows that the infrared regularized amplitude is
related to an {\em infinite} sum of heavy-baryon amplitudes with 
self-energy insertions in the heavy-baryon propagator, as depicted 
in Fig.\ \ref{5:6:fig:relprophb}.
   The advantage of the relativistic approach is obvious, because for a
general one-loop amplitude it may be very difficult, if not impossible, to 
obtain a closed expression for the sum of all insertions. 
   To conclude this section, the method of infrared regularization provides a 
fully relativistic framework producing amplitudes having the relevant 
analytic properties and satisfying the chiral power-counting rules.
   At the moment, it is not yet clear, whether it can be generalized
beyond the one-loop level.

\chapter{Summary and Concluding Remarks}
\label{chap_cr}

   As we have discussed in great detail, the chiral 
$\mbox{SU(3)}_L\times\mbox{SU(3)}_R\times\mbox{U(1)}_V$ 
symmetry of QCD in the limit of vanishing $u$-, $d$-, and $s$-quark masses
(Sec.\ \ref{subsec_gsclqs}), 
together with the assumption of its spontaneous breakdown to 
$\mbox{SU(3)}_V\times\mbox{U(1)}_V$ in the ground state (Sec.\ 
\ref{sec_ssbqcd}),
is one of the keys to understanding the phenomenology of
the strong interactions in the low-energy regime.
   The importance of chiral symmetry was realized long before
the formulation of QCD and led to a host of predictions within the
current-algebra and PCAC approaches of the 1960's \cite{Adler:1968}.
   Some of the consequences of an explicit symmetry breaking,
yielding non-analytic terms in the perturbation, were worked out
in the early 1970's, but the development came to a halt \cite{Pagels:se}
because it was not clear how to systematically organize a perturbative 
expansion.
   From the present point of view, the explicit symmetry breaking is due to 
the finite $u$-, $d$-, and $s$-quark masses, leading to divergences of the 
symmetry currents (Sec.\ \ref{subsec_csbdqm}).

   In 1979 Weinberg \cite{Weinberg:1978kz}
laid the foundations for further progress
with his observation that the constraints due to (chiral) symmetry may
perturbatively be analyzed in terms of the most general effective field theory.
   A very important ingredient was the formulation of a consistent 
power-counting scheme (Secs.\ \ref{sec_elwpcs} and \ref{subsec_pcs})
which allowed for a systematic perturbative
analysis in contrast to various commonly used {\em ad hoc} phenomenological 
approaches to the strong interactions at low energies.
   In particular, the inclusion of loop diagrams allowed for a perturbative
restoration of unitarity which would be violated if only tree-level diagrams
were used.
   Subsequently Gasser and Leutwyler \cite{Gasser:1983yg,Gasser:1984gg}
combined the ideas of Weinberg 
with other modern techniques of quantum field theory to analyze the Ward 
identities of QCD Green functions in terms of  a local invariance of the 
generating functional under the chiral group (Sec.\ \ref{sec_gfcwi} and
App.\ \ref{app_gfwi}).
   These papers were the starting point of what is nowadays called 
chiral perturbation theory.
 
   The mesonic sector has generated a host of successful
applications, some of which have reached two-loop accuracy.
   Here, we have concentrated on a few elementary 
observables and processes, namely: masses of the Goldstone bosons
(Secs.\ \ref{sec_loel} and \ref{subsec_mgb}), 
weak and electromagnetic $\pi$ decays (Secs.\ \ref{subsec_pdpmn}
and \ref{sec_ewzwa}), $\pi\pi$ scattering (Secs.\ \ref{subsec_pps}
and \ref{subsec_eppsop6}), and electromagnetic form factors
(Sec.\ \ref{subsec_emffp}).
   Moreover, we have discussed in quite some detail how to construct
the mesonic effective Lagrangian (Secs.\ \ref{sec_tpgb}, \ref{sec_clop4},
and \ref{subsec_mclop6}).

   At first sight, it might appear that the large number of low-energy 
parameters at ${\cal O}(p^6)$ would make any quantitative prediction
at the two-loop level impossible.
   However, there are several reasons why this is not the case.
   To start with, there exist observables which do not depend on {\em any}
new parameters at ${\cal O}(p^6)$, i.e., which can be predicted in terms 
of the ${\cal O}(p^2)$ and ${\cal O}(p^4)$ low-energy constants only.
   An example is given by the correction to Sirlin's theorem discussed
in Ref.\ \cite{Post:1997dk}.
   Clearly, such cases provide a natural testing ground for the convergence
of the approach.
   Secondly, only a limited set of low-energy parameters
contribute to any given process.
   It follows from the nature of the Ward identities that different physical
processes are interrelated due to the underlying symmetries so that 
coefficients which have been fixed using one reaction can be used to
{\em predict} another observable.
   In view of the ordinary implementation of symmetries, such as in the 
Wigner-Eckart theorem, this is not a surprise, because it is well-known
that symmetries imply relations among $S$-matrix elements.
   However, the Ward identities provide {\em additional} constraints among 
Green functions of a different type and allow one to also include
an explicit symmetry breaking (Sec.\ \ref{sec_gfcwi}).
   It is this second case which can systematically be studied in
the framework of ChPT and which provides interesting new
insights into our understanding of both spontaneous and explicit
symmetry breaking within QCD.
   Finally, different methods exist which allow one to estimate the value
of the parameters and thus, in combination with the ChPT result, test our 
physical picture of the strong interactions.

   In this work we have only considered elementary processes 
at an introductory level, not the extensions to and combinations with other 
methods.
   We omitted, for example, the weak interactions 
of kaons which are mediated by the exchange of $W$ bosons {\em between} the 
quark currents \cite{Ecker:1995gg,deRafael:1995zv,Pich:1995bw}.
   We also did not discuss the breaking of isospin symmetry which requires
the inclusion of the electromagnetic interaction in terms of dynamical 
(virtual) photons \cite{Urech:1994hd,Neufeld:1995mu,Ananthanarayan:2002kj}.

   Chiral symmetry also dictates the interaction of the Goldstone bosons with
other hadrons (Secs.\ \ref{sec_tpf} and \ref{sec_loebl}).
   By studying the axial-vector current matrix element 
(Sec.\ \ref{subsec_gtravcme}) and $\pi N$ scattering 
(Sec.\ \ref{subsec_apnstl})
we verified that a tree-level calculation using the lowest-order Lagrangian 
reproduces the Goldberger-Treiman relation and the Weinberg-Tomozawa result
for the $s$-wave scattering lengths, respectively.
   As we have seen, the first systematic study in the pion-nucleon sector 
\cite{Gasser:1987rb}
raised the question of a consistent power counting (Sec.\ \ref{sec_eld}).
   This problem was subsequently overcome in the framework of the 
heavy-baryon approach \cite{Jenkins:1990jv} (Sec.\ \ref{sec_hbf})
and most of the numerous applications in this
sector have been performed in HBChPT \cite{Bernard:1995dp}.

   In the baryonic sector the chiral orders increase in 
units of one, because of the additional possibility of forming Lorentz
invariants by contracting (covariant) derivatives with gamma matrices
(Sec.\ \ref{sec_loebl}).
   As a result, in the SU(2)$\times$SU(2) baryonic sector at the one-loop 
level, up to and including ${\cal O}(p^4)$, one has in total $2+7+23+118=150$
\cite{Fettes:2000gb}
low-energy constants as opposed to the $2+7=9$ free parameters of the
corresponding mesonic sector \cite{Gasser:1983yg}.
   Nevertheless, numerous results have been obtained in the baryonic
sector because, at the same time, a large amount of very precise experimental 
data are available due to the existence of a stable proton target.
   (Neutron data can also be extracted, e.g, from experiments on the deuteron.)
   The availability of new high-precision data in combination with the
techniques of chiral perturbation theory have led to a considerable
improvement of our understanding of the strong interactions at low
energies, in particular since systematic corrections to the old 
current-algebra predictions could be worked out and (successfully) tested.

   We have not discussed the approach of the so-called small scale 
expansion to include the $\Delta(1232)$ resonance as an explicit degree of 
freedom \cite{Hemmert:1996xg}.
   Clearly this is an important issue in the baryonic sector because 
the first nucleon excitation is such a prominent feature of the low-energy
spectrum as seen, e.g., in the total pion-nucleon scattering 
or the total photo absorption cross sections.

   Most recently, the method of infrared regularization
\cite{Becher:1999he} has opened the
possibility of reconciling the relativistic approach with a consistent
chiral power counting scheme (Sec.\ \ref{sec_mir}).
   One may expect that this method will have a large impact insofar as
many of the results obtained within the heavy-baryon framework will have to 
be checked with respect to relativistic corrections.  
   The question regarding the radius of convergence in the baryonic sector
remains a big challenge because, ultimately, a calculation at the two-loop 
level ${\cal O}(p^5)$ is, in general, required to quantitatively assess
higher-order corrections \cite{McGovern:1998tm}.
   In comparison to two-loop calculations in the 
SU(2)$\times$SU(2) mesonic sector such an investigation in the (relativistic)
nucleon sector is even more complicated for two reasons. 
   First, due to the spin of the nucleon, the structure of vertices is richer 
than for spin-0 particles.
   Second, the nucleon mass introduces another mass in the propagators
making the evaluation of the two-loop integrals more difficult than for a
single mass.

   Finally, we would like to mention that a description of the nucleon-nucleon 
interaction within the framework of effective field theory has made tremendous
progress and that a rigorous treatment of nuclei within field theory 
is no longer out of reach
\cite{Weinberg:1991um,Ordonez:1996rz,Kaiser:1997mw,vanKolck:1999mw,%
Epelbaum:2000dj,Beane:2001bc,Finelli:2002na}.

   In conclusion, chiral perturbation theory has added a new and
unprecedented level of systematics to the description of strong-interaction 
processes at low energies and continues to be a very fruitful 
and rich field with promising perspectives.
   If this introductory review encourages students and newcomers to 
chiral perturbation theory to participate in this field of research, 
it has served its purpose.

\section*{Acknowledgments}
   I am greatly indebted to David R.~Harrington for carefully and critically 
reading the whole manuscript (!) and his uncountably numerous suggestions for 
improvement.
   He continuously forced me to explain the meaning of concepts instead of 
hiding behind the chiral jargon.

   I would like to thank my academic teachers Dieter Drechsel, Harold
W.~Fearing, and Justus H.~Koch for sharing their deep insights into theoretical
physics with me.
   Learning from them has been a pleasure!
         
   Numerous discussions with my collaborators and colleagues
\mbox{Thomas} \linebreak
\mbox{Ebertsh\"auser}, \mbox{Thomas} \mbox{Fuchs}, 
\mbox{Thomas} R.~\mbox{Hemmert}, \mbox{Barry} R.~\mbox{Holstein},\linebreak
\mbox{Germar} \mbox{Kn\"ochlein}, \mbox{Anatoly I.} \mbox{L'vov}, 
\mbox{Andreas} \mbox{Metz}, \mbox{Barbara} \mbox{Pasquini},
and \mbox{Christine} \mbox{Unkmeir} are gratefully acknowledged. 

   Special thanks go to Rolf Brockmann for many discussions on
effective field theory, to Jambul Gegelia for his valuable discussions on
the infrared regularization, to Martin Reuter for his kind
help in preparing Appendix A on Green functions and Ward identities,
and to Thomas Walcher for challenging discussions on spontaneous
symmetry breaking.

   Last but not least, I would like to thank Erich Vogt for his incredible
enthusiasm and his continuous encouragement to finish the manuscript.

   I dedicate this work to my family. I hope it was worth it!

\begin{appendix}

\chapter{Green Functions and Ward Identities}
\label{app_gfwi}

   In this appendix we will show how to derive Ward identities for Green 
functions in the framework of canonical quantization on the one hand, 
and quantization via the Feynman path integral on the other hand, 
by means of an explicit example.
   In order to keep the discussion transparent, we will concentrate on 
a simple scalar field theory with a global O(2) or U(1) invariance. 
   To that end, let us consider the Lagrangian
\begin{eqnarray}
\label{app:gfwi:lphi4}
{\cal L}&=&\frac{1}{2}(\partial_\mu \Phi_1\partial^\mu \Phi_1
+\partial_\mu \Phi_2\partial^\mu \Phi_2)
-\frac{m^2}{2} (\Phi_1^2+\Phi_2^2)
-\frac{\lambda}{4}(\Phi_1^2+\Phi_2^2)^2\nonumber\\
&=&
\partial_\mu \Phi^\dagger \partial^\mu \Phi -m^2 \Phi^\dagger\Phi
-\lambda (\Phi^\dagger\Phi)^2,
\end{eqnarray}
where 
\begin{displaymath}
\Phi(x)=\frac{1}{\sqrt{2}}[\Phi_1(x)+i\Phi_2(x)],
\quad
\Phi^\dagger(x)=\frac{1}{\sqrt{2}}[\Phi_1(x)-i\Phi_2(x)],
\end{displaymath}
with real scalar fields $\Phi_1$ and $\Phi_2$.
   Furthermore, we assume $m^2>0$ and $\lambda>0$, so there is
no spontaneous symmetry breaking and the energy is bounded from below.
   Equation (\ref{app:gfwi:lphi4}) is invariant under the global (or rigid)
transformations
\begin{equation}
\label{app:gfwi:inftrans1}
\Phi'_1=\Phi_1-\epsilon \Phi_2,\quad
\Phi'_2=\Phi_2+\epsilon \Phi_1,
\end{equation}
or, equivalently,
\begin{equation}
\label{app:gfwi:inftrans2}
\Phi'=(1+i\epsilon)\Phi,\quad
\Phi'^\dagger=(1-i\epsilon)\Phi^\dagger,
\end{equation}   
where $\epsilon$ is an infinitesimal real parameter.
   Applying the method of Gell-Mann and L{\'e}vy \cite{Gell-Mann:np},
we obtain for a {\em local}
parameter $\epsilon(x)$,
\begin{equation}
\label{app:gfwi:dlphi4}
\delta{\cal L}=\partial_\mu\epsilon(x)(i\partial^\mu\Phi^\dagger \Phi
-i\Phi^\dagger\partial^\mu\Phi),
\end{equation}
from which, via Eqs.\ (\ref{2:3:strom2}) and (\ref{2:3:divergenz}),
we derive for the current corresponding to the global symmetry, 
\begin{eqnarray}
J^{\mu}&=&\frac{\partial \delta\cal L}{\partial \partial_\mu
\epsilon}=(i\partial^\mu\Phi^\dagger \Phi
-i\Phi^\dagger\partial^\mu\Phi),\\
\partial_\mu J^{\mu}&=&\frac{\partial \delta\cal L}{\partial
\epsilon}=0.
\end{eqnarray}
   Recall that the identification of Eq.\ (\ref{2:3:divergenz}) as
the divergence of the current is only true for fields
satisfying the Euler-Lagrange equations of motion.  

   We now extend the analysis to a {\em quantum} field theory.
   In the framework of canonical quantization, we first define 
conjugate momenta,
\begin{equation}
\label{app:gfwi:conjmom}
\Pi_i(x)=\frac{\partial \cal L}{\partial \partial_0\Phi_i},\quad
\Pi(x)=\frac{\partial \cal L}{\partial \partial_0\Phi},\quad
\Pi^\dagger(x)=\frac{\partial \cal L}{\partial \partial_0\Phi^\dagger},
\end{equation}
and interpret the fields and their conjugate momenta
as operators which, in the Heisenberg 
picture, are subject to the equal-time commutation relations
\begin{equation}
[\Phi_i(\vec{x},t),\Pi_j(\vec{y},t)]=i\delta_{ij}\delta^3(\vec{x}-\vec{y}),
\end{equation}
and 
\begin{equation}
\label{app:gfwi:eqtcr}
{[}\Phi(\vec{x},t),\Pi(\vec{y},t)]=
[\Phi^\dagger(\vec{x},t),\Pi^\dagger(\vec{y},t)]
=i\delta^3(\vec{x}-\vec{y}).
\end{equation}
   The remaining equal-time commutation relations, involving fields or momenta
only, vanish.
   For the quantized theory, the current operator then reads
\begin{equation}
J^\mu(x)=:(i\partial^\mu\Phi^\dagger \Phi
-i\Phi^\dagger\partial^\mu\Phi):,
\end{equation}
where $:\quad :$ denotes normal or Wick ordering, i.e., annihilation 
operators appear to the right of creation operators.
   For a conserved current, the charge operator, i.e., the space integral of 
the charge density, is time independent and serves as the generator of 
infinitesimal transformations of the Hilbert space states,
\begin{equation}
Q=\int d^3 x J^0(\vec{x},t).
\end{equation}
   Applying Eq.\ (\ref{app:gfwi:eqtcr}), it is 
straightforward to calculate the equal-time commutation
relations\footnote{The 
transition to normal ordering involves an (infinite) 
constant which does not contribute to the commutator.} 
\begin{eqnarray}
\label{app:gfwi:j0comrel}
[J^0(\vec{x},t),\Phi(\vec{y},t)]&=&\delta^3(\vec{x}-\vec{y})\Phi(\vec{x},t),
\nonumber\\
{[}J^0(\vec{x},t),\Pi(\vec{y},t)]&=&-\delta^3(\vec{x}-\vec{y})
\Pi(\vec{x},t),\nonumber\\
{[}J^0(\vec{x},t),\Phi^\dagger(\vec{y},t)]&=&-\delta^3(\vec{x}-\vec{y})
\Phi^\dagger(\vec{x},t),\nonumber\\
{[}J^0(\vec{x},t),\Pi^\dagger(\vec{y},t)]&=&\delta^3(\vec{x}-\vec{y})
\Pi^\dagger(\vec{x},t).
\end{eqnarray}
   In particular, performing the space integrals in Eqs.\ 
(\ref{app:gfwi:j0comrel}), one obtains
\begin{eqnarray}
\label{app:gfwi:Qcomrel}
[Q,\Phi(x)]&=&\Phi(x),
\nonumber\\
{[}Q,\Pi(x)]&=&-\Pi(x),\nonumber\\
{[}Q,\Phi^\dagger(x)]&=&-\Phi^\dagger(x),\nonumber\\
{[}Q,\Pi^\dagger(x)]&=&\Pi^\dagger(x).
\end{eqnarray}
   In order to illustrate the implications of Eqs.\ (\ref{app:gfwi:Qcomrel}), 
let us take an eigenstate $|\alpha\rangle$ of $Q$ with eigenvalue 
$q_\alpha$ and consider, for example, the action of $\Phi(x)$ on that
state,
\begin{eqnarray*}
Q\left(\Phi(x)|\alpha\rangle\right)=\left([Q,\Phi(x)]+\Phi(x)Q\right)
|\alpha\rangle
=(1+q_\alpha)\left(\Phi(x)|\alpha\rangle\right).
\end{eqnarray*}
   We conclude that the operators $\Phi(x)$ and $\Pi^\dagger(x)$ 
[$\Phi^\dagger(x)$ and $\Pi(x)$] increase (decrease) the Noether charge
of a system by one unit.

   We are now in the position to discuss the consequences of the U(1)
symmetry of Eq.\ (\ref{app:gfwi:lphi4}) for the Green functions of the theory.
   To that end, let us consider as our prototype the Green function
\begin{equation}
\label{app:gfwi:gmuxyz}
G^\mu(x,y,z)=\langle 0|T[\Phi(x) J^\mu(y) \Phi^\dagger(z)]|0\rangle,
\end{equation}
   which describes the transition amplitude for the creation of a quantum
of Noether charge $+1$ at $x$, propagation to $y$, interaction at $y$
via the current operator, propagation to $z$ with annihilation at $z$.
   First of all we observe that under the global infinitesimal transformations
of Eq.\ (\ref{app:gfwi:inftrans2}), 
$J^\mu(x)\mapsto J'^\mu(x)=J^\mu(x)$, or in
other words $[Q,J^\mu(x)]=0$. 
   We thus obtain
\begin{eqnarray}
\label{app:gfwi:gmutrans}
G^\mu(x,y,z)\mapsto G'^\mu(x,y,z)&=&
\langle 0|T[
(1+i\epsilon)\Phi(x) J'^\mu(y) (1-i\epsilon)\Phi^\dagger(z)]|0\rangle
\nonumber\\
&=&\langle 0|T[\Phi(x) J^\mu(y) \Phi^\dagger(z)]|0\rangle\nonumber\\
&=&G^\mu(x,y,z),
\end{eqnarray}
   the Green function remaining invariant under
the U(1) transformation. 
   (In general, the transformation behavior of a Green function depends on the 
irreducible representations under which the fields transform.    
   In particular, for more complicated groups such as SU($N$), standard tensor 
methods of group theory may be applied to reduce the product representations 
into irreducible components 
\cite{Balachandran:ab,O'Raifeartaigh:vq,Jones:ti}.
   We also note that for U(1), the symmetry current is charge neutral, 
i.e.\ invariant, which for more complicated groups, in general, is not the 
case.)   

   Moreover, since $J^\mu(x)$ is the Noether current of the underlying
U(1) there are further restrictions on the Green function beyond
its transformation behavior under the group.
   In order to see this, we consider the divergence of 
Eq.\ (\ref{app:gfwi:gmuxyz})
   and apply the equal-time commutation relations of Eqs.\ 
(\ref{app:gfwi:j0comrel}) to obtain (see Sec.\ \ref{subsec_cgf})
\begin{eqnarray}
\label{app:gfwi:gmuwt}
\partial_\mu^y G^\mu(x,y,z)&=&[\delta^4(x-y)-\delta^4(z-y)]
\langle 0|T[\Phi(x)\Phi^\dagger(z)]|0\rangle,
\end{eqnarray}
   where we made use of $\partial_\mu J^\mu=0$.
   Equation (\ref{app:gfwi:gmuwt}) is the analogue of the Ward identity of QED
[see Eq.\ (\ref{2:4:qedwardidentity})]
\cite{Ward:1950xp,Fradkin:1955jr,Takahashi:xn}.
   In other words, the underlying symmetry not only determines the
transformation behavior of Green functions under the group, but also
relates $n$-point Green functions containing a symmetry current 
to $(n-1)$-point Green functions [see Eq.\ (\ref{2:4:gendmug})].
   In principle, calculations similar to those leading to Eqs.\
(\ref{app:gfwi:gmutrans}) and (\ref{app:gfwi:gmuwt}), can be performed for any 
Green function of the theory.
   However, we will now show that the symmetry constraints can be 
compactly summarized in terms of an invariance property of a
generating functional.

   The generating functional is defined as the vacuum-to-vacuum transition
amplitude in the presence of external fields,
\begin{eqnarray}
\label{app:gfwi:genfunc1}
\lefteqn{
W[j,j^\ast,j_\mu]=\langle 0,+\infty|0,-\infty\rangle_{j,j^\ast,j_\mu}}
\nonumber\\
&=&\exp(iZ[j,j^\ast,j_\mu])\nonumber\\
&=&\langle 0|T\left(\exp\left\{i\int d^4x[j(x)\Phi^\dagger(x)+j^\ast(x) \Phi(x)
+j_\mu(x) J^\mu(x)]\right\}\right)|0\rangle,\nonumber\\
\end{eqnarray} 
   where $\Phi$ and $\Phi^\dagger$ are the field operators and $J^\mu(x)$
is the Noether current.
   Note that the field operators and the conjugate momenta are subject
to the equal-time commutation relations and, in addition, must satisfy 
the Heisenberg equations of motion.
   Via this second condition and implicitly through the ground state, 
the generating functional depends on the dynamics 
of the system which is determined by the Lagrangian of 
Eq.\ (\ref{app:gfwi:lphi4}).
   The Green functions of the theory involving $\Phi$, $\Phi^\dagger$, and
$J^\mu$ are obtained through functional derivatives of 
Eq.\ (\ref{app:gfwi:genfunc1}).
   For example, the Green function of Eq.\ (\ref{app:gfwi:gmuxyz}) is given by
\begin{equation}
\label{app:gfwi:gmuxyz1}
G^\mu(x,y,z)=(-i)^3 \left.\frac{\delta^3 W[j,j^\ast,j_\mu]}{\delta j^\ast(x)
\delta j_\mu(y) \delta j(z)}\right|_{j=0,j^\ast=0,j_\mu=0}.
\end{equation}
   
   In order to discuss the constraints imposed on the generating functional
via the underlying symmetry of the theory, let us consider its path integral
rep\-re\-sentation 
\cite{Zinn-Justin:1989mi,Das:1993gd},\footnote{Up to an irrelevant constant
the measure $[d\Phi_1][d\Phi_2]$ is equivalent to $[d\Phi][d\Phi^\ast]$,
with $\Phi$ and $\Phi^\ast$ considered as independent variables of integration.
}
\begin{eqnarray}
\label{app:gfwi:genfunc2}
W[j,j^\ast,j_\mu]&=&\int [d\Phi_1][d\Phi_2] e^{iS[\Phi,\Phi^\ast,j,j^\ast,
j_\mu]},
\end{eqnarray}
where 
\begin{equation}
\label{app:gfwi:actionsources}
S[\Phi,\Phi^\ast,j,j^\ast,j_\mu]=S[\Phi,\Phi^\ast]+
\int d^4 x [\Phi(x)j^\ast(x)+\Phi^\ast(x) j(x)+J^\mu(x) j_\mu(x)]
\end{equation}
denotes the action corresponding to the Lagrangian of 
Eq.\ (\ref{app:gfwi:lphi4}) 
in combination with a coupling to the external sources. 
    Let us now consider a {\em local} infinitesimal transformation of the
fields [see Eqs.\ (\ref{app:gfwi:inftrans2})] together 
with a {\em simultaneous} 
transformation of the external sources,
\begin{equation}
\label{app:gfwi:inftranssources}
j'(x)=[1+i\epsilon(x)]j(x),\quad
j'^\ast(x)=[1-i\epsilon(x)] j^\ast(x),\quad
j_\mu'(x)=j_\mu(x)-\partial_\mu\epsilon(x).
\end{equation}
   The action of Eq.\ (\ref{app:gfwi:actionsources}) remains invariant under
such a transformation,
\begin{equation}
S[\Phi',\Phi'^\ast,j',j'^\ast,j_\mu']=
S[\Phi,\Phi^\ast,j,j^\ast,j_\mu].
\end{equation}
   We stress that the transformation of the external current $j_\mu$ is
necessary to cancel a term resulting from the kinetic term in
the Lagrangian. 
   We can now verify the invariance of the generating functional as follows,
\begin{eqnarray}
\label{app:gfwi:winvariance}
W[j,j^\ast,j_\mu]&=&\int [d\Phi_1][d\Phi_2] e^{iS[\Phi,\Phi^\ast,j,j^\ast,
j_\mu]}\nonumber\\
&=&\int [d\Phi_1][d\Phi_2] e^{iS[\Phi',\Phi'^\ast,j',j'^\ast,
j'_\mu]}\nonumber\\
&=&\int [d\Phi_1'][d\Phi_2'] \left|\left(\frac{\partial\Phi_i}{\partial
\Phi_j'}\right)\right|
e^{iS[\Phi',\Phi'^\ast,j',j'^\ast,
j'_\mu]}\nonumber\\
&=&\int [d\Phi_1][d\Phi_2] 
e^{iS[\Phi,\Phi^\ast,j',j'^\ast,
j'_\mu]}\nonumber\\
&=&W[j',j'^\ast,j'_\mu].
\end{eqnarray}
   We made use of the fact that the Jacobi determinant is one  
and renamed the integration variables.
   In other words, given the {\em global} U(1) symmetry of the Lagrangian,
Eq.\ (\ref{app:gfwi:lphi4}), the generating functional is invariant under the 
{\em local} transformations of Eq.\ (\ref{app:gfwi:inftranssources}).
   It is this observation which, for the more general case of the
chiral group SU($N$)$\times$SU($N$), was used by Gasser and Leutwyler
as the starting point of chiral perturbation theory.

   We still have to discuss, how this invariance allows us to collect the Ward
identities in a compact formula.
   We start from Eq.\ (\ref{app:gfwi:winvariance}),
\begin{eqnarray*}
0&=&\int[d\Phi_1] [d\Phi_2]\left(e^{iS[\Phi,\Phi^\ast,j',j'^\ast,j'_\mu]}
-e^{iS[\Phi,\Phi^\ast,j,j^\ast,j_\mu]}\right)\\
&=&\int[d\Phi_1] [d\Phi_2]\int d^4x \left\{
\epsilon[\Phi j^\ast-\Phi^\ast j]
-iJ^\mu\partial_\mu\epsilon\right\}
e^{iS[\Phi,\Phi^\ast,j,j^\ast,j_\mu]}.
\end{eqnarray*}
   Observe that 
\begin{displaymath}
\Phi(x) e^{iS[\Phi,\Phi^\ast,j,j^\ast,j_\mu]}
=-i\frac{\delta}{\delta j^\ast(x)} e^{iS[\Phi,\Phi^\ast,j,j^\ast,j_\mu]},
\end{displaymath}
and similarly for the other terms, resulting in
\begin{eqnarray*}
0&=&\int[d\Phi_1] [d\Phi_2]\int d^4 x\left\{
\epsilon(x)\left[-ij^\ast(x)\frac{\delta}{\delta j^\ast(x)}
+ij(x)\frac{\delta}{\delta j(x)}\right]\right.\\
&&\left.
-\partial_\mu\epsilon(x)\frac{\delta}{\delta j_\mu(x)}\right\}
e^{iS[\Phi,\Phi^\ast,j,j^\ast,j_\mu]}.
\end{eqnarray*}
   Finally we interchange the order of integration, make use of
partial integration, and apply the divergence theorem:
\begin{equation}
\label{app:gfwi:0fl}
0=\int d^4 x \epsilon(x)\left[i j(x)\frac{\delta}{\delta j(x)}
-i j^\ast(x)\frac{\delta}{\delta j^\ast(x)}
+\partial_\mu ^x \frac{\delta}{\delta j_\mu(x)}\right]
W[j,j^\ast,j_\mu].
\end{equation}
   Since Eq.\ (\ref{app:gfwi:0fl}) must hold for any $\epsilon(x)$ we obtain 
as the master equation for deriving Ward identities,
\begin{equation}
\label{app:gfwi:mewi}
\left[j(x)\frac{\delta}{\delta j(x)}-
j^\ast(x)\frac{\delta}{\delta j^\ast(x)}
-i\partial_\mu^x \frac{\delta}{\delta j_\mu(x)}\right] W[j,j^\ast,j_\mu]
=0.
\end{equation}
   We note that Eqs.\ (\ref{app:gfwi:winvariance}) and 
(\ref{app:gfwi:mewi}) are equivalent.
   
   As a final illustration let us re-derive the Ward identity of 
Eq.\ (\ref{app:gfwi:gmuwt}) using Eq.\ (\ref{app:gfwi:mewi}). 
   For that purpose we start from Eq.\ (\ref{app:gfwi:gmuxyz1}),
\begin{displaymath}
\partial_\mu^y G^\mu(x,y,z)=
(-i)^3 \partial_\mu^y
\left.\frac{\delta^3 W}{\delta j^\ast(x)\delta j_\mu (y)\delta j(z)},
\right|_{j=0,j^\ast=0,j_\mu=0},
\end{displaymath}
apply Eq.\ (\ref{app:gfwi:mewi}),
\begin{displaymath}
=(-i)^2\left\{
\frac{\delta^2}{\delta j^\ast(x)\delta j(z)}\left[
j^\ast(y)\frac{\delta}{\delta j^\ast(y)}-j(y)\frac{\delta}{\delta j(y)}
\right] W\right\}_{j=0,j^\ast=0,j_\mu=0},
\end{displaymath}
make use of
$\delta j^\ast(y)/\delta j^\ast(x)=
\delta^4(y-x)$ and $\delta j(y)/\delta j(z)=\delta^4(y-z)$
for the functional derivatives,
\begin{displaymath}
=(-i)^2\left\{\delta^4(x-y)\frac{\delta^2 W}{\delta j^\ast(y)
\delta j(z)}
-\delta^4(z-y)\frac{\delta^2 W}{\delta j^\ast(x)
\delta j(y)}\right\}_{j=0,j^\ast=0,j_\mu=0},
\end{displaymath}
and, finally, use the definition of Eq.\ (\ref{app:gfwi:genfunc1}),
\begin{displaymath}
\partial_\mu^y G^\mu(x,y,z)
=[\delta^4(x-y)-\delta^4(z-y)]\langle 0| T[\Phi(x)\Phi^\dagger(z)
]|0\rangle
\end{displaymath}
   which is the same as Eq.\ (\ref{app:gfwi:gmuwt}).
   In principle, any Ward identity can be obtained by taking appropriate 
higher functional derivatives of $W$ and then using Eq.\ (\ref{app:gfwi:mewi}).

\chapter{Dimensional Regularization: Basics}
\label{app_drb}
   For the sake of completeness we provide a simple illustration of
the method of dimensional regularization.
   For a detailed account the interested reader is referred to
Refs.\ \cite{'tHooft:fi,Leibbrandt:1975dj,'tHooft:1978xw,%
Cheng:bj,Collins:xc,Veltman:wz}.

   Let us consider the integral
\begin{equation}
\label{app:drb:int}
I=\int\frac{d^4k}{(2\pi)^4}\frac{i}{k^2-M^2+i0^+}
\end{equation}
which appears in the calculation of the masses of the Goldstone bosons
[see Eq.\ (\ref{4:8:diag})].  
   We introduce 
$$a\equiv\sqrt{\vec{k}^2+M^2}>0$$
so that
\begin{eqnarray*}
k^2-M^2+i0^+
&=&[k_0+(a-i0^+)][k_0-(a-i0^+)],
\end{eqnarray*}
   and define 
$$f(k_0)=\frac{1}{[k_0+(a-i0^+)][k_0-(a-i0^+)]}.$$
   In order to determine $\int_{-\infty}^{\infty} dk_0 f(k_0)$ as part of
the calculation of $I$, we consider
$f$ in the complex $k_0$ plane and make use of Cauchy's theorem 
\begin{equation}
\label{app:drb:cauchy}
\oint_C dz f(z)=0
\end{equation}
for functions which are differentiable in every point inside the closed
contour $C$.
\begin{figure}
\begin{center}
\epsfig{file=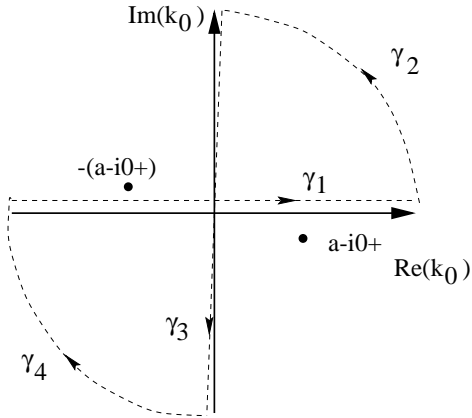,width=6cm}
\caption{\label{app:drb:wickrotation_fig}
Path of integration in the complex $k_0$ plane.}
\end{center}
\end{figure}
   We choose the contour as shown in Fig.\ \ref{app:drb:wickrotation_fig},
$$
0=\sum_{i=1}^4 \int_{\gamma_i} dz f(z),
$$
   and make use of 
$$\int_\gamma f(z)dz=\int_a^b f[\gamma(t)]\gamma'(t)dt
$$
to obtain for the individual integrals 
\begin{eqnarray*}
\int_{\gamma_1} f(z) dz &=& \int_{-\infty}^\infty f(t)dt,\\
\int_{\gamma_2} f(z) dz &=&
\lim_{R\to\infty} \int_{0}^\frac{\pi}{2} f(Re^{it})iRe^{it}dt =0,
\,\,\mbox{since}\,\, 
\lim_{R\to\infty} \underbrace{Rf(Re^{it})}_{\mbox{$\sim \frac{1}{R}$}}=0,\\
\int_{\gamma_3} f(z) dz &=& \int_{\infty}^{-\infty} f(it)idt,\\
\int_{\gamma_4} f(z) dz &=& \lim_{R\to\infty} 
\int_{\frac{3}{2}\pi}^\pi f(Re^{it})iRe^{it}dt =0.
\end{eqnarray*}
   In combination with Eq.\ (\ref{app:drb:cauchy}) we obtain the so-called
Wick rotation
\begin{equation}
\label{app:drb:wickrotation}
\int_{-\infty}^\infty f(t)dt=-i\int_{\infty}^{-\infty}dt f(it)
=i\int_{-\infty}^\infty dt f(it).
\end{equation}
   As an intermediate result the integral of 
Eq.\ (\ref{app:drb:int}) reads
$$I=\frac{1}{(2\pi)^4}i\int_{-\infty}^\infty dk_0\int d^3 k
\frac{i}{(ik_0)^2-\vec{k}^2 -M^2 +i0^+}
=\int \frac{d^4 l}{(2\pi)^4} \frac{1}{l^2+M^2-i0^+},
$$
   where $l^2=l_1^2+l_2^2+l_3^2+l_4^2$ denotes a Euclidian scalar 
product.
   In this {\em special} case, the integrand does not have
a pole and we can thus omit the $-i0^+$ which gave the positions of the
poles in the original integral consistent with the boundary conditions.
   The degree of divergence can be estimated by simply counting the powers of
momenta \cite{Veltman:wz}.
   If the integral behaves asymptotically as $\int d^4 l /l^2$,
$\int d^4 l /l^3$, $\int d^4 l /l^4$
the integral is said to diverge quadratically, linearly, and
logarithmically, respectively.
   Thus, our example $I$ diverges quadratically.
   Various methods have been devised to regularize divergent integrals.
   We will make use of {\em dimensional} regularization,  because it preserves 
algebraic relations between Green functions (Ward identities) if the 
underlying symmetries do not depend on the number of dimensions of space-time. 

   In dimensional regularization, we generalize the integral from 4 to $n$ 
dimensions and introduce polar coordinates
\begin{eqnarray}
\label{app:drb:polkoord}
l_1&=& l\cos(\theta_1),\nonumber\\
l_2&=& l\sin(\theta_1)\cos(\theta_2),\nonumber\\
l_3&=& l\sin(\theta_1)\sin(\theta_2)\cos(\theta_3),\nonumber\\
&\vdots&\nonumber\\
l_{n-1}&=&l\sin(\theta_1)\sin(\theta_2)\cdots\cos(\theta_{n-1}),\nonumber\\
l_{n}&=&l\sin(\theta_1)\sin(\theta_2)\cdots\sin(\theta_{n-1}),
\end{eqnarray}
where $0\leq l$, $\theta_i\in[0,\pi], i=1,\cdots,n-2$, $\theta_{n-1}\in
[0,2\pi]$.
   A general integral is then symbolically of the form 
\begin{equation}
\label{app:drb:volumenelement}
\int d^n l\cdots = \int_0^\infty l^{n-1}dl\int_0^{2\pi}d\theta_{n-1}
\int_0^\pi d\theta_{n-2}\sin(\theta_{n-2})\cdots\int_0^\pi d\theta_1
\sin^{n-2}(\theta_1)\cdots .
\end{equation}
   If the integrand does not depend on the angles, the angular integration
can explicitly be carried out.
   To that end one makes use of
$$
\int_0^\pi \sin^m(\theta) d\theta=\frac{\sqrt{\pi} \Gamma\left(\frac{m+1}{2}
\right)}{\Gamma\left(\frac{m+2}{2}\right)}
$$
which can be shown by induction.
   We then obtain for the angular integration
\begin{eqnarray}
\label{app:drb:winkelintegration}
\int_0^{2\pi}d\theta_{n-1}\cdots \int_0^\pi d\theta_1
\sin^{n-2}(\theta_1)&=&
2\pi\underbrace{\frac{\sqrt{\pi}\Gamma(1)}{\Gamma\left(\frac{3}{2}\right)}
\frac{\sqrt{\pi}\Gamma\left(\frac{3}{2}\right)}{\Gamma(2)}\cdots
\frac{\sqrt{\pi}\Gamma\left(\frac{n-1}{2}\right)}{\Gamma\left(\frac{n}{2}
\right)}}_{\mbox{$(n-2)$ factors}}\nonumber\\
&=&2\frac{\pi^\frac{n}{2}}{\Gamma\left(\frac{n}{2}\right)}.
\end{eqnarray}
   We define the integral for $n$ dimensions ($n$ integer) as 
\begin{equation}
\label{app:drb:im2}
I_n(M^2,\mu^2)=\mu^{4-n}\int\frac{d^nk}{(2\pi)^n}\frac{i}{k^2-M^2+i0^+},
\end{equation}
   where for convenience we have introduced the renormalization scale $\mu$ 
so that the integral has the same dimension for arbitrary $n$.
   (The integral of Eq.\ (\ref{app:drb:im2}) is convergent only for 
$n=1$.)
   After the Wick rotation of Eq.\ (\ref{app:drb:wickrotation}) and
the angular integration of Eq.\ (\ref{app:drb:winkelintegration}) the
integral formally reads
$$
I_n(M^2,\mu^2)=\mu^{4-n}2\frac{\pi^\frac{n}{2}}{\Gamma\left(\frac{n}{2}\right)}
\frac{1}{(2\pi)^n}
\int_0^\infty dl \frac{l^{n-1}}{l^2+M^2}.
$$
  For later use, we investigate the (more general) integral
\begin{equation}
\label{app:drb:mgint}
\int_0^\infty \frac{l^{n-1}dl}{(l^2+M^2)^\alpha}
=\frac{1}{(M^2)^\alpha}\int_0^\infty \frac{l^{n-1}dl}{(\frac{l^2}{M^2}+1
)^\alpha}
=\frac{1}{2}(M^2)^{\frac{n}{2}-\alpha} \int_0^\infty \frac{
t^{\frac{n}{2}-1}dt}{(t+1)^\alpha},
\end{equation}
where we made use of the substitution $t\equiv l^2/M^2$.
   We then make use of the Beta function
\begin{equation}
\label{app:drb:betafunktion}
B(x,y)=\int_0^\infty \frac{t^{x-1}dt}{(1+t)^{x+y}}=\frac{\Gamma(x)\Gamma(y)}{
\Gamma(x+y)},
\end{equation}
where the {\em integral} converges for $x>0$, $y>0$ and diverges if 
$x\leq 0$ or $y\leq 0$.
   For non-positive values of $x$ or $y$ we make use of the analytic 
continuation in terms of the Gamma function to define the Beta 
function and thus the integral of Eq.\ (\ref{app:drb:mgint}).\footnote{
Recall that
$\Gamma(z)$ is single valued and analytic over the entire complex plane, save
for the points $z=-n$, $n=0,1,2,\cdots$, where it possesses simple poles with
residue $(-1)^n/n!$ \cite{Abramowitz}.}
   Putting $x=n/2$, $x+y=\alpha$ and $y=\alpha-n/2$
our (intermediate) integral reads
\begin{equation}
\label{app:drb:allgint}
\int_0^\infty \frac{l^{n-1}dl}{(l^2+M^2)^\alpha}=
\frac{1}{2}(M^2)^{\frac{n}{2}-\alpha}\frac{\Gamma\left(\frac{n}{2}\right)
\Gamma\left(\alpha-\frac{n}{2}\right)}{\Gamma(\alpha)}
\end{equation}
which, for $\alpha=1$, yields for our original integral
\begin{eqnarray}
\label{app:drb:iint}
I_n(M^2,\mu^2)&=&\mu^{4-n}\underbrace{2\frac{\pi^\frac{n}{2}}{\Gamma\left(
\frac{n}{2}\right)}}_{\mbox{angular integration}}
\frac{1}{(2\pi)^n} \frac{1}{2} (M^2)^{\frac{n}{2}-1}\frac{\Gamma
\left(\frac{n}{2}\right)\Gamma\left(1-\frac{n}{2}\right)}{
\underbrace{\Gamma(1)}_{\mbox{1}}}\nonumber\\
&=&\frac{\mu^{4-n}}{(4\pi)^{\frac{n}{2}}} (M^2)^{\frac{n}{2}-1}
\Gamma\left(1-\frac{n}{2}\right).
\end{eqnarray}
   Since $\Gamma(z)$ is an analytic function in the complex plane except
for poles of first order in $0,-1,-2,\cdots$, and
$a^z=\exp[\ln(a)z]$, $a\in R^+$ is an analytic function in $C$,
the right-hand side of Eq.\ (\ref{app:drb:iint}) can be thought of
as a function of a {\em complex} variable $n$ which is 
analytic in $C$ except for poles of first order
for $n=2,4,6,\cdots$.
   Making use of 
$$\mu^{4-n}=(\mu^2)^{2-\frac{n}{2}},\quad
(M^2)^{\frac{n}{2}-1}=M^2 (M^2)^{\frac{n}{2}-2},
\quad
(4\pi)^\frac{n}{2}=(4\pi)^2(4\pi)^{\frac{n}{2}-2},
$$
we define (for complex n)
$$ 
I(M^2,\mu^2,n)=\frac{M^2}{(4\pi)^2}\left(\frac{4\pi\mu^2}{M^2}\right)^{2-
\frac{n}{2}} \Gamma\left(1-\frac{n}{2}\right).
$$
   Of course, for $n\to 4$ the Gamma function has a pole and we want
to investigate how this pole is approached.
   The property $\Gamma(z+1)=z\Gamma(z)$ allows one to rewrite 
\begin{displaymath}
\Gamma\left(1-\frac{n}{2}\right)=
\frac{\Gamma\left(1-\frac{n}{2}+1\right)}{1-\frac{n}{2}}
=\frac{\Gamma\left(2-\frac{n}{2}+1\right)}{\left(1-\frac{n}{2}\right)
\left(2-\frac{n}{2}\right)}=\frac{\Gamma\left(1+\frac{\epsilon}{2}\right)}{
(-1)\left(1-\frac{\epsilon}{2}\right)\frac{\epsilon}{2}},
\end{displaymath}
where we defined $\epsilon\equiv 4-n$. 
   Making use use of $a^x=\exp[\ln(a)x]=1+\ln(a)x+O(x^2)$ we expand
the integral for small $\epsilon$
\begin{eqnarray*}
I(M^2,\mu^2,n)
&=&\frac{M^2}{16\pi^2}\left[1+\frac{\epsilon}{2}\ln\left(
\frac{4\pi\mu^2}{M^2}\right)+O(\epsilon^2)\right]\\
&&\times
\left(-\frac{2}{\epsilon}\right)\left[1+\frac{\epsilon}{2}+O(\epsilon^2)\right]
\left[\underbrace{\Gamma(1)}_{\mbox{1}}
+\frac{\epsilon}{2}\Gamma'(1)+O(\epsilon^2)\right]
\\
&=&\frac{M^2}{16\pi^2}\left[-\frac{2}{\epsilon}\underbrace{-\Gamma'(1)}_{
\mbox{$\gamma_E=0.5772\cdots$}}-1-\ln(4\pi)+\ln\left(\frac{M^2}{\mu^2}\right)
+O(\epsilon)\right],
\end{eqnarray*} 
   where $\gamma_E$ is Euler's constant.
   We finally obtain
  \begin{equation}
\label{app:drb:im22}
I(M^2,\mu^2,n)=\frac{M^2}{16\pi^2}\left[
R+\ln\left(\frac{M^2}{\mu^2}\right)\right]+O(n-4),
\end{equation}
where
\begin{equation}
\label{app:constantR}
R=\frac{2}{n-4}-[\mbox{ln}(4\pi)+\Gamma'(1)+1].
\end{equation}

   Using the same techniques one can easily derive a very useful expression 
for the more general integral
\begin{eqnarray}
\label{app:drb:moregenint}
\lefteqn{\int \frac{d^n k}{(2\pi)^n} \frac{(k^2)^p}{(k^2-M^2+i0^+)^q}=}
\nonumber\\
&&i(-)^{p-q} \frac{1}{(4\pi)^{\frac{n}{2}}}(M^2)^{p+\frac{n}{2}-q}
\frac{\Gamma\left(p+\frac{n}{2}\right)\Gamma\left(q-p-\frac{n}{2}\right)}{
\Gamma\left(\frac{n}{2}\right)\Gamma(q)}.
\end{eqnarray}
   We first assume $M^2>0$, $p=0,1,\cdots$, $q=1,2,\cdots$, and $p<q$.
   The last condition is used in the Wick rotation to guarantee that the
quarter circles at infinity do not contribute to the integral.
   The transition to the Euclidian metric produces the factor $i(-)^{p-q}$.  
   The angular integral in $n$ dimensions is then performed as in 
Eq.\ (\ref{app:drb:winkelintegration}).
   The remaining radial integration is done using Eq.\ (\ref{app:drb:mgint})
with the substitution $n-1\to 2p+n-1$ and $\alpha \to q$.
   The analytic continuation of the right-hand side of Eq.\ 
(\ref{app:drb:moregenint}) is used to also define expressions with 
(integer) $q\leq p$ in dimensional regularization.
 
  In the context of combining propagators by using Feynman's trick
one encounters integrals of the type of Eq.\ 
(\ref{app:drb:moregenint})
with $M^2$ replaced by $A-i0^+$, where $A$ is a real number.
   In this context it is important to consistently deal with the
boundary condition $-i0^+$ \cite{Veltman:wz}.
   For example, let us consider a term of the type
$\ln(A-i0^+)$.
   To that end one expresses a complex number $z$ in its polar form
$z=|z|\exp(i\varphi),$
where the argument $\varphi$ of $z$ is uniquely determined if, in addition,
we demand $-\pi\leq\varphi < \pi$.
   For $A>0$ one simply has $\ln(A-i0^+)=\ln(A)$.
   For $A<0$ the infinitesimal imaginary part indicates that  $-|A|$
is reached in the third quadrant from below the real axis
so that we have to use the $-\pi$.
   We then make use of $\ln(ab)=\ln(a)+\ln(b)$ and obtain
$$
\ln(A-i0^+)=\ln(|A|)+\ln(e^{-i\pi})=\ln(|A|)-i\pi,\quad
A<0.
$$
   Both cases can be summarized in a single expression
\begin{equation}
\label{app:drb:lna}
\ln(A-i0^+)=\ln(|A|)-i\pi\Theta(-A)\quad\mbox{for}\,A\in R.
\end{equation}
   The preceding discussion is of importance for consistently determining
imaginary parts of loop integrals. 

   Let us conclude with the general observation that 
(ultraviolet) divergences of
one-loop integrals in dimensional regularization always show up
as single poles in $\epsilon=4-n$.

\chapter{Loop Integrals}
\label{app_li}
   In Appendix \ref{app_drb} we discussed the basic ideas of the method
of dimensional regularization.   
   Here we outline the calculation of more complicated one-loop integrals
of mesonic as well as heavy-baryon chiral perturbation theory.
   We restrict ourselves to the cases needed to reproduce the examples
discussed in the main text and refer the interested reader to 
Refs.\ \cite{'tHooft:fi,Leibbrandt:1975dj,'tHooft:1978xw,%
Cheng:bj,Collins:xc,Veltman:wz}
for more details.

\section{One-Loop Integrals of the Mesonic Sector}
\label{app_sec_olims}
   In the mesonic sector we will use the following definition and
nomenclature for the scalar loop integrals (i.e., no Lorentz indices)
extended to $n$ dimensions:
\begin{eqnarray}
\label{app:olims:ipidef}
\lefteqn{I_{\pi\cdots\pi}(q_1,\cdots,q_m)\equiv}\nonumber\\
&&i\mu^{4-n} \int \frac{d^n k}{(2\pi)^n}
\frac{1}{(k+q_1)^2-M_\pi^2+i0^+}\cdots \frac{1}{(k+q_m)^2-M_\pi^2+i0^+},
\end{eqnarray}
where we omit an explicit reference to the scale $\mu$ and the
``number of dimensions'' $n$.\footnote{If $m\geq 3$, the integral converges
for $n=4$.}
   In the SU(3)$\times$SU(3) case one also needs loop integrals with different
masses such as, e.g., 
\begin{displaymath}
I_{\pi K}(q_1,q_2)=i\mu^{4-n} \int \frac{d^n k}{(2\pi)^n}
\frac{1}{(k+q_1)^2-M_\pi^2+i0^+}\frac{1}{(k+q_2)^2-M_K^2+i0^+}.
\end{displaymath}

\subsection{$I_\pi$}
\label{app_subsec_ipi}
We define
\begin{equation}
\label{app:ipi}
I_\pi(q)\equiv i\mu^{4-n} \int \frac{d^n k}{(2\pi)^n}
\frac{1}{(k+q)^2-M_\pi^2+i0^+}.
\end{equation}
   Using a shift $k\to k-q$ (in the regularized integral) we obtain
\begin{displaymath}
I_\pi(q)=I_\pi(0).
\end{displaymath}
   However, this is just the basic integral $I$ we discussed in detail
in App.\ \ref{app_drb}:
\begin{equation}
\label{app:ipib}
I_\pi(0)=\frac{M^2_\pi}{16\pi^2}\left[R+\ln\left(\frac{M^2_\pi}{\mu^2}\right)
\right]+O(n-4),
\end{equation}
   where 
\begin{equation}
\label{app:ipiR}
R=\frac{2}{n-4}-[\ln(4\pi)+\Gamma'(1)+1].
\end{equation}
   Later on we will also use the common notation $A_0(M_\pi^2)$ for the 
integral $I_\pi(0)$.

\subsection{$I_{\pi\pi}$}
\label{app_subsec_ipipi}
   In the calculation of the one-loop contribution of 
Fig.\ \ref{4:9:pionffloop1} to the electromagnetic form factor of the pion,
Eq.\ (\ref{4:9:loopc1}), we encounter an integral of the type
\begin{equation}
\label{app:ipipi}
I_{\pi\pi}(q_1,q_2)\equiv i\mu^{4-n} \int \frac{d^n k}{(2\pi)^n}
\frac{1}{(k+q_1)^2-M_\pi^2+i0^+}\frac{1}{(k+q_2)^2-M_\pi^2+i0^+}.
\end{equation}
   Using a shift $k\to k-q_2$ we obtain
\begin{displaymath}
I_{\pi\pi}(q_1,q_2)=I_{\pi\pi}(q_1-q_2,0).
\end{displaymath}
   It is thus sufficient to consider $I_{\pi\pi}(q,0)$. 
   To that end, we first combine the denominators using Feynman's trick:
\begin{equation}
\label{app:ipipiftrick}
\frac{1}{ab}=\int_0^1 dz \frac{1}{[az+b(1-z)]^2},
\end{equation}
with $a=(k+q)^2-M_\pi^2+i0^+$ and $b=k^2-M_\pi^2+i0^+$ to obtain
\begin{displaymath}
I_{\pi\pi}(q,0)=i\mu^{4-n} \int \frac{d^n k}{(2\pi)^n}
\int_0^1 dz \frac{1}{[k^2+2k\cdot q z-(M_\pi^2-zq^2)+i0^+]^2},
\end{displaymath}
and perform the shift $k\to k-zq$, resulting in
\begin{displaymath}
I_{\pi\pi}(q,0)=\int_0^1 dz\left(
i\mu^{4-n} \int \frac{d^n k}{(2\pi)^n}
\frac{1}{\left\{k^2-[M_\pi^2+z(z-1)q^2]+i0^+\right\}^2}\right).
\end{displaymath}
   Performing the Wick rotation, applying the results of 
Eqs.\ (\ref{app:drb:winkelintegration}) and (\ref{app:drb:allgint}),
expanding the result in $(n-4)$, and finally performing the $z$ integration,
we obtain
\begin{equation}
\label{app:ipipib}
I_{\pi\pi}(q,0)=\frac{1}{16\pi^2}\left[
R+\ln\left(\frac{M^2_\pi}{\mu^2}\right)+1+J^{(0)}\left(\frac{q^2}{M^2_\pi}
\right)+O(n-4)\right],
\end{equation}
   where \cite{Unkmeir:1999md}
\begin{eqnarray*}
J^{(0)}(x)
&=&\int_0^1 dz \ln[1+x(z^2-z)-i0^+]\\
&=&
 \left \{ \begin{array}{ll}
-2-\sigma\ln\left(\frac{\sigma-1}{\sigma+1}\right),&x<0,\\
-2+2\sqrt{\frac{4}{x}-1}\,\mbox{arccot}
\left(\sqrt{\frac{4}{x}-1}\right),&0\le x<4,\\
-2-\sigma\ln\left(\frac{1-\sigma}{1+\sigma}\right)-i\pi\sigma,&
4<x,
\end{array} \right.
\end{eqnarray*}
with 
\begin{displaymath}
\sigma(x)=\sqrt{1-\frac{4}{x}},\quad x\notin [0,4].
\end{displaymath}
   Note that Eq.\ (\ref{app:ipipib}) represents a case where the $i0^+$
boundary condition has to be treated consistently, as discussed at the
end of App.\ \ref{app_drb}.
   For later use we introduce the notation
\begin{displaymath}
B_0(q^2,M_\pi^2)=I_{\pi\pi}(q,0),
\end{displaymath}
where the subscript $0$ refers to the scalar character of the integral.

  Next we want to determine the tensor integrals appearing in 
Eq.\ (\ref{4:9:loopc1}) by reducing them to already known integrals.
   The general idea consists of parameterizing the tensor structure in
terms of the metric tensor and products of external four-vectors and  
multiplying the results by invariant functions of Lorentz scalars.
   We first consider
\begin{equation}
\label{app:Ipipik}
i\mu^{4-n} \int \frac{d^n k}{(2\pi)^n}
\frac{k^\mu}{(k+q)^2-M_\pi^2+i0^+}\frac{1}{k^2-M_\pi^2+i0^+},
\end{equation}
which must have the form
\begin{equation}
\label{app:B1}
q^\mu B_1(q^2,M_\pi^2),
\end{equation}
where the subscript $1$ refers to one four-vector $k$ in 
the numerator of the integral.
   We contract Eq.\ (\ref{app:B1}) with $q_\mu$ 
and make use of $q\cdot k=[(k+q)^2 -M_\pi^2-(k^2-M_\pi^2)-q^2]/2$ 
to obtain
\begin{eqnarray*}
q^2 B_1(q^2,M_\pi^2)&=&\frac{1}{2} i\mu^{4-n} \int \frac{d^n k}{(2\pi)^n}
\frac{1}{k^2-M_\pi^2+i0^+}\\
&&-\frac{1}{2} i\mu^{4-n} \int \frac{d^n k}{(2\pi)^n}
\frac{1}{(k+q)^2-M_\pi^2+i0^+}\\
&&-\frac{1}{2}q^2 i\mu^{4-n} \int \frac{d^n k}{(2\pi)^n}
\frac{1}{(k+q)^2-M_\pi^2+i0^+}\frac{1}{k^2-M_\pi^2+i0^+}\\
&=&-\frac{1}{2}q^2 B_0(q^2,M_\pi^2),
\end{eqnarray*}
where we used the argument in Appendix \ref{app_subsec_ipi} to show
that the first two integrals cancel.
   We have thus reduced the determination of Eq.\ (\ref{app:Ipipik}) to an
already known integral:
\begin{equation}
B_1(q^2,M_\pi^2)=-\frac{1}{2}B_0(q^2,M_\pi^2).
\end{equation}
   Finally, we also need
\begin{eqnarray}
\label{app:B20B21}
\lefteqn{q^\mu q^\nu B_{20}(q^2,M_\pi^2)
+g^{\mu\nu}q^2 B_{21}(q^2,M_\pi^2)=}\nonumber\\ 
&&
i\mu^{4-n} \int \frac{d^n k}{(2\pi)^n}
\frac{k^\mu k^\nu}{(k+q)^2-M_\pi^2+i0^+}\frac{1}{k^2-M_\pi^2+i0^+},
\end{eqnarray}
where the first subscript $2$ refers to two four-vectors $k$ in the
numerator of the integral and the second subscripts  $0$ and $1$ refer to the 
number of metric tensors in the parameterization, respectively. 

  Contracting with $g_{\mu\nu}$ and making use of 
$g_{\mu\nu} g^{\mu\nu}=n$ in $n$ dimensions we obtain
\begin{equation}
\label{app:cond1}
q^2 B_{20} + n q^2 B_{21}
=A_0+M_\pi^2 B_0.
\end{equation}
   Similarly, contracting with $q_\mu$ we obtain
\begin{equation}
\label{app:cond2}
q^2 B_{20}
+q^2 B_{21}=
\frac{1}{2} A_0+\frac{q^2}{4} B_0.
\end{equation}
   By subtracting Eq.\ (\ref{app:cond2}) from (\ref{app:cond1}) we find
\begin{eqnarray}
\label{app:q2B21}
q^2 B_{21}&=&\frac{1}{n-1}\left[\frac{1}{2}A_0+(M_\pi^2-\frac{q^2}{4}) 
B_0\right]\nonumber\\
&=&\frac{1}{32\pi^2}\left\{\left(M_\pi^2-\frac{q^2}{6}\right)\left[
R+\ln\left(\frac{M_\pi^2}{\mu^2}\right)\right]\right.\nonumber\\
&&\left.+\frac{2}{3}J^{(0)}\left(\frac{q^2}{M_\pi^2}\right)\left(
M_\pi^2-\frac{q^2}{4}\right)-\frac{q^2}{18}\right\}+O(n-4),
\end{eqnarray}
where we made use of 
\begin{displaymath}
\frac{1}{n-1}=\frac{1}{3+(n-4)}=\frac{1}{3}\left(1-\frac{n-4}{3}+\cdots\right)
\end{displaymath}
and Eqs.\ (\ref{app:ipib}) and (\ref{app:ipipib}).
   From Eq.\ (\ref{app:cond2}) we obtain
\begin{eqnarray}
\label{app:B20}
B_{20}&=&\frac{1}{q^2(n-1)}\left(\frac{n-2}{2}A_0
+n\frac{q^2}{4}B_0-M_\pi^2B_0\right)\nonumber\\
&=&\frac{1}{3q^2}\left[A_0+\frac{6M_\pi^2-q^2}{96\pi^2}+(q^2-M_\pi^2)B_0\right
]+O(n-4)\nonumber\\
&=&\frac{1}{48\pi^2}\left[R+\ln\left(\frac{M^2_\pi}{\mu^2}\right)
+\frac{5}{6}+\left(1-\frac{M_\pi^2}{q^2}\right)J^{(0)}\left(
\frac{q^2}{M_\pi^2}\right)\right]+O(n-4).\nonumber\\
\end{eqnarray}   
   In working out Eqs.\ (\ref{app:q2B21}) or (\ref{app:B20}) it must
be remembered that $R$ contains a term $2/(n-4)$ which, when multiplied
by $(n-4)$, gives a term of order $(n-4)^0$.
   This must be done carefully to obtain, e.g., the $-q^2/18$ term in 
Eq.\ (\ref{app:q2B21}) and the $5/6$ term in Eq.\ (\ref{app:B20}).

\section{One-Loop Integrals of the Heavy-Baryon Sector}
\label{sec_olinhbs}
\subsection{Basic Loop Integral}
\label{subsec_bli} 
  The structure of the one-loop integrals in the heavy-baryon approach is
slightly different from the integrals of the mesonic sector discussed
in Sec.\ \ref{app_sec_olims}.
   Here we will outline the calculation of a basic loop integral which
serves as a starting point for more complicated calculations.
 
   Consider an integral of the type
\begin{equation}
\label{app:int1}
\int \frac{d^4k}{(2\pi)^4}\frac{1}{v\cdot k +\alpha +i0^+}
\frac{1}{k^2-A+i0^+},
\end{equation}
where $\alpha$ and $A$ are arbitrary real numbers and $v$ is the four-velocity
of the heavy-baryon approach.
   Counting powers of the momenta, the integral is linearly divergent.
   An integral of this type appears in the calculation of the nucleon self
energy in Sec.\ \ref{subsec_aop3olcnm}.
   We combine the denominators using the following Feynman trick:
\begin{equation}
\label{app:feynmantricksphb}
\frac{1}{ab}=2\int_0^\infty dy \frac{1}{(2ya+b)^2}.
\end{equation}
    Below, this  choice will allow us most easily to combine the $y$ 
integration
with the ``radial'' integration of the loop momentum after the Wick rotation.
   Inserting $a=v\cdot k +\alpha +i0^+$ and $b=k^2-A+i0^+$, we obtain
\begin{equation}
\label{app:int2}
2\int \frac{d^4k}{(2\pi)^4}\int_0^\infty dy
\frac{1}{(k^2+2yv\cdot k +2 y\alpha-A+i0^+)^2}.
\end{equation}
   We now generalize to $n$ dimensions:
\begin{equation}
\label{app:int3}
2\mu^{4-n}\int_0^\infty dy \int \frac{d^nk}{(2\pi)^n}
\frac{1}{(k^2+2yv\cdot k +2 y\alpha-A+i0^+)^2}
\end{equation}
and perform a shift of integration variables $k\to k-yv$
so that there remain no terms linear in $k$ in the denominator:
\begin{equation}
\label{app:int4}
2\mu^{4-n}\int_0^\infty dy \int \frac{d^n k}{(2\pi)^n}
\frac{1}{(k^2-y^2-A+2y\alpha+i0^+)^2},
\end{equation}
where we made use of $v^2=1$.
   Finally, we shift the integration variable $y\to y+\alpha$ 
in order to eliminate terms linear in $y$ in the denominator:
\begin{equation}
\label{app:int5}
2\mu^{4-n}\int_{-\alpha}^\infty dy \int\frac{d^nk}{(2\pi)^n}\frac{1}
{(k^2-y^2-A+\alpha^2+i0^+)^2}.
\end{equation}
   The $y$ integration is split into $[-\alpha,0]$ and $[0,\infty[$.
   Making use of a Wick rotation and Eqs.\ (\ref{app:drb:winkelintegration}),
(\ref{app:drb:allgint}), and (\ref{app:drb:lna}) 
we obtain for the first integral
\begin{eqnarray}
\label{app:int6a}
\lefteqn{2\mu^{4-n}\int_{-\alpha}^0 dy \int\frac{d^nk}{(2\pi)^n}\frac{1}
{(k^2-y^2-A+\alpha^2+i0^+)^2}}\nonumber\\
&=&-\frac{i}{8\pi^2}\int_{-\alpha}^0 dy\left[R+1+
\ln\left(\frac{|A+y^2-\alpha^2|}{\mu^2}\right)
-i\pi\Theta(\alpha^2-A-y^2)\right]\nonumber\\
&&+O(n-4),
\end{eqnarray}
where 
\begin{displaymath}
R=\frac{2}{n-4}-[\mbox{ln}(4\pi)+\Gamma'(1)+1].
\end{displaymath}
   Performing a Wick rotation and using 
Eq.\ (\ref{app:drb:winkelintegration}),   
the second integral reads 
\begin{eqnarray}
\label{app:int6b}
\lefteqn{2\mu^{4-n}\int_0^\infty dy \int\frac{d^nk}{(2\pi)^n}\frac{1}
{(k^2-y^2-A+\alpha^2+i0^+)^2}}\nonumber\\
&=&\frac{4i\mu^{4-n}}{(4\pi)^\frac{n}{2}\Gamma\left(\frac{n}{2}\right)}
\int_0^\infty dy\int_0^\infty dx \frac{x^{n-1}}{(x^2+y^2+A-\alpha^2-i0^+)^2}.
\end{eqnarray}
   Using polar coordinates $x=r\cos(\varphi)$ and $y=r\sin(\varphi)$ together
with
\begin{displaymath}
\int_0^{\frac{\pi}{2}}\sin^{2\alpha+1}(\varphi)\cos^{2\beta+1}(\varphi)d\varphi
=
\frac{\Gamma(\alpha+1)\Gamma(\beta+1)}{2\Gamma(\alpha+\beta+2)},
\end{displaymath}
we rewrite the second integral as ($\Gamma(1/2)=\sqrt{\pi}$) 
\begin{displaymath}
\frac{2i \mu^{4-n}\sqrt{\pi}}{(4\pi)^\frac{n}{2}
\Gamma\left(\frac{n+1}{2}\right)}
\int_0^\infty dr\frac{r^n}{(r^2+A-\alpha^2-i0^+)^2}.
\end{displaymath}
   Finally, applying again Eq.\ (\ref{app:drb:allgint}) for the radial integral
we obtain for the second integral (in four dimensions)
\begin{equation}
\label{app:int7b}
2\int_0^\infty dy \int\frac{d^4k}{(2\pi)^4}\frac{1}
{(k^2-y^2-A+\alpha^2+i0^+)^2}
=\frac{-i}{8\pi}\sqrt{A-\alpha^2-i0^+},
\end{equation}
where we made use of $\Gamma(2)=1$ and $\Gamma(-1/2)=-2\sqrt{\pi}$.

   The remaining $y$ integration of the first integral is elementary 
and we obtain as the final expression
\begin{eqnarray}
\label{app:intfinal}
\lefteqn{\mu^{4-n}\int \frac{d^n k}{(2\pi)^4}\frac{1}{v\cdot k +\alpha +i0^+}
\frac{1}{k^2-A+i0^+}}\nonumber\\
&=&\frac{-i}{8\pi^2}\Bigg(
\alpha\left[R+\ln\left(\frac{|A|}{\mu^2}\right)-1\right]
\nonumber\\
&&+
\left\{
\begin{array}{ll}
2\sqrt{\alpha^2-A}\,\mbox{arccosh}\left(\frac{\alpha}{\sqrt{A}}\right)
&\\
-2\pi i\sqrt{\alpha^2-A},&A>0\wedge\alpha>\sqrt{A},\\
2\sqrt{A-\alpha^2}\,\arccos\left(-\frac{\alpha}{\sqrt{A}}\right),&
A>\alpha^2,\\
-2\sqrt{\alpha^2-A}\,\mbox{arccosh}\left(-\frac{\alpha}{\sqrt{A}}\right),
&A>0\wedge\alpha<-\sqrt{A},\\
2\sqrt{\alpha^2-A}\,\mbox{arcsinh}\left(\frac{\alpha}{\sqrt{-A}}\right)&\\
-i\pi\left(\alpha+\sqrt{\alpha^2-A}\,\right),&A<0,
\end{array}
\right\}
\nonumber\\
&&+O(n-4)\Bigg),
\end{eqnarray}
with $ R=\frac{2}{n-4}-[\mbox{ln}(4\pi)+\Gamma'(1)+1]$.

\subsection{$J_{\pi N}$}
\label{subsec_jpin}
   In analogy to the mesonic case we define
\begin{equation}
\label{app:jpin}
J_{\pi N}(q;\omega)\equiv i\mu^{4-n} \int \frac{d^n k}{(2\pi)^n}
\frac{1}{(k+q)^2-M_\pi^2+i0^+}\frac{1}{v\cdot k+\omega+i0^+}.
\end{equation}
   Using a shift $k\to k-q$ we obtain
\begin{displaymath}
J_{\pi N}(q;\omega)=J_{\pi N}(0;\omega-v\cdot q).
\end{displaymath}
   It is thus sufficient to consider $J_{\pi N}(0;\omega)$ which, using the
result of Eq.\ (\ref{app:intfinal}), is given 
by\footnote{Our $J_{\pi N}(0;\omega)$ corresponds to
$-\mu^{4-n}J_0(\omega)$ of Ref.\ \cite{Bernard:1995dp}.}
\begin{eqnarray}
\label{app:jpinb}
\lefteqn{J_{\pi N}(0;\omega)}\nonumber\\
&=&
\frac{\omega}{8\pi^2}\left[R+\ln\left(\frac{M^2_\pi}{\mu^2}\right)-1
\right]\nonumber\\
&&+\frac{1}{8\pi^2}\left\{
\begin{array}{ll}
2\sqrt{\omega^2-M_\pi^2}\,
\mbox{arccosh}\left(\frac{\omega}{M_\pi}
\right)-2\pi i\sqrt{\omega^2-M^2_\pi},
&\omega>M_\pi,\\
2\sqrt{M^2_\pi-\omega^2}\,\arccos\left(-\frac{\omega}{M_\pi}
\right),
&\omega^2<M^2_\pi,\\
-2\sqrt{\omega^2-M_\pi^2}\,
\mbox{arccosh}\left(-\frac{\omega}{M_\pi}
\right),
& \omega<-M_\pi,
\end{array}
\right\}\nonumber\\
&&+O(n-4).
\end{eqnarray}

   In the calculation of the nucleon self energy we also need 
tensor integrals which, as in the mesonic case, one may
reduce to already known integrals.
   Let us introduce the notation
\begin{displaymath}
C_0(\omega,M_\pi^2)=J_{\pi N}(0;\omega),
\end{displaymath}
where the subscript $0$ refers to the scalar character of the integral.
   Once again, the general idea in the determination of tensor
integrals consists of 
parameterizing the tensor structure in terms of the metric tensor and 
products of the four-velocity $v$.
   We first consider
\begin{equation}
\label{app:JpiNk}
i\mu^{4-n} \int \frac{d^n k}{(2\pi)^n}
\frac{k^\mu}{k^2-M_\pi^2+i0^+}\frac{1}{v\cdot k+\omega +i0^+},
\end{equation}
which must have the form
\begin{equation}
\label{app:C1}
v^\mu C_1(\omega,M_\pi^2),
\end{equation}
where the subscript $1$ refers to one four-vector $k$ in 
the numerator of the integral.
   We contract Eq.\ (\ref{app:C1}) with $v_\mu$, make use of $v^2=1$,
and add and subtract $\omega$ in the numerator of the integral,
obtaining 
\begin{eqnarray}
\label{app:c1result}
C_1(\omega,M_\pi^2)&=&i\mu^{4-n} \int \frac{d^n k}{(2\pi)^n}
\frac{1}{k^2-M_\pi^2+i0^+}\nonumber\\
&&-\omega
i\mu^{4-n} \int \frac{d^n k}{(2\pi)^n}
\frac{1}{k^2-M_\pi^2+i0^+}\frac{1}{v\cdot k+\omega+i0^+}\nonumber\\
&=&I_\pi(0)-\omega J_{\pi N}(0;\omega),
\end{eqnarray}
where $I_\pi(0)$ is defined in Eq.\ (\ref{app:ipib}).
   We have thus reduced the determination of Eq.\ (\ref{app:JpiNk})
to the already known integrals $I_\pi$ and $J_{\pi N}$.
   As our final example, let us discuss  
\begin{displaymath}
i\mu^{4-n} \int \frac{d^n k}{(2\pi)^n}
\frac{k^\mu k^\nu}{k^2-M_\pi^2+i0^+}\frac{1}{v\cdot k+\omega+i0^+},
\end{displaymath}
which must be of the form
\begin{equation}
\label{app:C20C21}
v^\mu v^\nu C_{20}(\omega,M_\pi^2)
+g^{\mu\nu} C_{21}(\omega,M_\pi^2),
\end{equation}
where the first subscript $2$ refers to two four-vectors $k$ in the
numerator of the integral and the second subscripts  $0$ and $1$ refer to the 
number of metric tensors in the parameterization, respectively. 
  Contracting with $v_\mu$ and adding and subtracting $\omega$
in the numerator,  we obtain
\begin{equation}
\label{app:condt21}
v^\nu C_{20}+v^\nu C_{21}=
-\omega v^{\nu} C_1=-\omega v^\nu[I_\pi(0)
-\omega J_{\pi N}(0;\omega)],
\end{equation}
where we made use of Eq.\ (\ref{app:c1result}) and
\begin{displaymath}
i\mu^{4-n} \int \frac{d^n k}{(2\pi)^n}
\frac{k^\nu}{k^2-M_\pi^2+i0^+}=0.
\end{displaymath}
   Finally, contracting with $g_{\mu\nu}$ and making use of 
$g_{\mu\nu} g^{\mu\nu}=n$ in $n$ dimensions we obtain
\begin{equation}
\label{app:condt22}
C_{20} + n C_{21}
=M_\pi^2 C_0,
\end{equation}
where we made use of 
\begin{equation}
\label{app:intvan}
i\mu^{4-n} \int \frac{d^n k}{(2\pi)^n}\frac{1}{v\cdot k+\omega+i0^+}=0
\end{equation}
in dimensional regularization.   
   In order to verify Eq.\ (\ref{app:intvan}), one writes
\begin{eqnarray*}
i\mu^{4-n} \int \frac{d^n k}{(2\pi)^n}\frac{1}{v\cdot k+\omega+i0^+}
&=&
i\mu^{4-n} \int \frac{d^n k}{(2\pi)^n}
\frac{v\cdot k}{(v\cdot k)^2-(\omega+i0^+)^2}\\
&&
-\omega i\mu^{4-n} \int \frac{d^n k}{(2\pi)^n}
\frac{1}{(v\cdot k)^2-(\omega+i0^+)^2}.
\end{eqnarray*}
   The first term vanishes, because the integrand is odd in $k$.
   For the evaluation of the second term we choose $v^\mu=(1,0,\cdots,0)$.
   Applying the residue theorem, we obtain for the integral
\begin{displaymath}
\int_{-\infty}^\infty dk_0 \frac{1}{k_0^2-(\omega+i0^+)^2}=
\frac{\pi i}{\omega},\quad \omega\neq 0,
\end{displaymath} 
so that we may define for arbitrary $\omega$
\begin{eqnarray*}
i\mu^{4-n} \int \frac{d^n k}{(2\pi)^n}\frac{1}{v\cdot k+\omega+i0^+}
&=&-i\mu^{4-n} \int \frac{d^n k}{(2\pi)^n}\frac{\omega}{
k_0^2-(\omega+i0^+)^2}\\
=\frac{\pi}{(2\pi)^n}\mu^{4-n}\int d^{n-1}k. 
\end{eqnarray*}
   However, the last term vanishes in dimensional regularization.

   From Eqs.\ (\ref{app:condt21}) and (\ref{app:condt22}) we obtain
\begin{eqnarray}
\label{app:C21}
C_{21}&=&\frac{1}{n-1}\left[(M_\pi^2-\omega^2)J_{\pi N}(0;\omega)+\omega
I_\pi(0)\right]\nonumber\\
&=&\frac{1}{3}\left[(M_\pi^2-\omega^2)J_{\pi N}(0;\omega)+\omega
I_\pi(0)\right]
-\frac{\omega}{12\pi^2}\left(\frac{M_\pi^2}{2}-\frac{\omega^2}{3}\right),
\nonumber\\
\end{eqnarray}
and 
\begin{eqnarray}
\label{app:C20}
C_{20}&=&M_\pi^2 J_{\pi N}(0,\omega)-C_{21}\nonumber\\
&=&\frac{1}{3}(2M_\pi^2 +\omega^2)J_{\pi N}(0;\omega)
-\frac{1}{3}\omega I_\pi(0)
+\frac{\omega}{12\pi^2}\left(\frac{M_\pi^2}{2}-\frac{\omega^2}{3}\right),
\nonumber\\
\end{eqnarray}
where, as in the mesonic case, it is important to identify the finite terms
resulting from the product of $O(n-4)$ terms and $R$.

   Finally, when discussing the relation between the infrared regularization
method of Ref.\ \cite{Becher:1999he} and the heavy-baryon approach in
Sec.\ \ref{sec_mir}, we made use of the fact that an integral of the type
\begin{equation}
\label{app:dnkk2pvkmq}
\int d^n k \frac{(k^2)^p}{(v\cdot k+i0^+)^q}
\end{equation}
vanishes in dimensional regularization.
   This can be seen by substituting $k=\lambda k'$ and relabeling $k'=k$
\begin{displaymath}
=\lambda^{n+2p-q}\int d^n k \frac{(k^2)^p}{(v\cdot k+i0^+)^q}.
\end{displaymath}
   Since $\lambda>0$ is arbitrary and, for fixed $p$ and $q$, the result
is to hold for arbitrary $n$, Eq.\ (\ref{app:dnkk2pvkmq})
is set to zero in dimensional regularization.
   We emphasize that the vanishing of Eq.\ (\ref{app:dnkk2pvkmq}) has
the character of a prescription \cite{Itzykson:rh}. 
   The integral does not depend on any scale and its analytic continuation
is ill defined in the sense that there is no dimension $n$ where it is
meaningful. 
  It is ultraviolet divergent for $n+2p-q\geq 0$ and infrared divergent
for $n+2p-q\leq 0$.

\chapter{Different Forms of ${\cal L}_4$ in $\mbox{SU(2)}\times\mbox{SU(2)}$}
\section{GL Versus GSS}
\label{app_sec_glvgss}
   The purpose of this appendix is to explicitly relate the two commonly used
${\cal O}(p^4)$ $\mbox{SU(2)}\times\mbox{SU(2)}$ Lagrangians of Refs.\ 
\cite{Gasser:1983yg} (GL) and \cite{Gasser:1987rb} (GSS).

   In their pioneering work on mesonic $\mbox{SU(2)}\times\mbox{SU(2)}$
chiral perturbation theory \cite{Gasser:1983yg}, Gasser and Leutwyler 
used a notation adopted from the O(4) nonlinear $\sigma$ model, because the two
Lie groups $\mbox{SU(2)}\times\mbox{SU(2)}$ and O(4) are locally isomorphic, 
i.e., their Lie algebras are isomorphic.
   The effective Lagrangian was written in terms of invariant scalar 
products of real four-vectors in contrast to the nowadays standard
trace form.
   The dynamical pion degrees of freedom were expressed in terms of
a four-component real vector field of unit length with 
components $U^A(x)$, $A=0,1,2,3$.
   The connection to the SU(2) matrix $U(x)$ of Sec.\ \ref{subsec_aqcd} 
can be expressed as
\begin{eqnarray}
\label{app:glvgss:uau}
U(x)&=&U^0(x)+i\vec{\tau}\cdot\vec{U}(x),\nonumber\\
U^0(x)&=&\frac{1}{2}\mbox{Tr}[U(x)],\nonumber\\
U^i(x)&=&-\frac{i}{2}\mbox{Tr}[\tau_i U(x)],\quad i=1,2,3,
\end{eqnarray}
with
\begin{displaymath}
\sum_{A=0}^3 [U^A(x)]^2=1,
\end{displaymath}
   so that $U$ is unitary.
   The lowest-order Lagrangian in the trace notation is given by the
$\mbox{SU(2)}\times\mbox{SU(2)}$ version of Eq.\ (\ref{4:5:l2}), 
and the transcription of the ${\cal O}(p^4)$
$\mbox{SU(2)}\times\mbox{SU(2)}$ Lagrangian [see Eq.\ (5.5) of
Ref.\ \cite{Gasser:1983yg}] reads\footnote{Note that Gasser and Leutwyler
called the ${\cal O}(p^2)$ and ${\cal O}(p^4)$ Lagrangians 
${\cal L}_1$ and ${\cal L}_2$, respectively.}
\begin{eqnarray}
\label{app:glvgss:l4gl}
{\cal L}^{\rm GL}_4 &=&
\frac{l_1}{4} \left\{\mbox{Tr}[D_{\mu}U (D^{\mu}U)^{\dagger}] \right\}^2
+\frac{l_2}{4}\mbox{Tr}[D_{\mu}U (D_{\nu}U)^{\dagger}]
\mbox{Tr}[D^{\mu}U (D^{\nu}U)^{\dagger}]
\nonumber \\
&&+\frac{l_3}{16}\left[\mbox{Tr}(\chi U^\dagger+ U\chi^\dagger)\right]^2
+\frac{l_4}{4}\mbox{Tr}[D_\mu U(D^\mu\chi)^\dagger
+D_\mu\chi(D^\mu U)^\dagger]\nonumber\\
&&+l_5\left[\mbox{Tr}(f^R_{\mu\nu}U f^{\mu\nu}_LU^\dagger)
-\frac{1}{2}\mbox{Tr}(f_{\mu\nu}^L f^{\mu\nu}_L
+f_{\mu\nu}^R f^{\mu\nu}_R)\right]\nonumber\\
&&+i\frac{l_6}{2}\mbox{Tr}[ f^R_{\mu\nu} D^{\mu} U (D^{\nu} U)^{\dagger}
+ f^L_{\mu\nu} (D^{\mu} U)^{\dagger} D^{\nu} U]\nonumber\\
&&-\frac{l_7}{16}\left[\mbox{Tr}(\chi U^\dagger-U\chi^\dagger)\right]^2
\nonumber\\
&&+\frac{h_1+h_3}{4}\mbox{Tr}(\chi\chi^\dagger)
+\frac{h_1-h_3}{16}\left\{
\left[\mbox{Tr}(\chi U^\dagger + U\chi^\dagger)\right]^2\right.
\nonumber\\
&&\left.
+\left[\mbox{Tr}(\chi U^\dagger-U\chi^\dagger)\right]^2
-2\mbox{Tr}(\chi U^\dagger\chi U^\dagger + U\chi^\dagger U\chi^\dagger)
\right\}\nonumber\\
&&-2h_2 \mbox{Tr}(f_{\mu\nu}^L f^{\mu\nu}_L
+f_{\mu\nu}^R f^{\mu\nu}_R).
\end{eqnarray}
   When comparing with the SU(3)$\times$SU(3) version of Eq.\ (\ref{4:7:l4gl})
we first note that Eq.\ (\ref{app:glvgss:l4gl}) contains fewer independent
terms which results from the application of the trace relations, as 
discussed in Sec.\ \ref{subsec_mclop6}.
   The bare and the renormalized low-energy constants $l_i$ and $l_i^r$ are
related by
\begin{equation}
\label{app:lilir}
l_i=l_i^r+\gamma_i\frac{R}{32\pi^2},
\end{equation}
where $R=2/(n-4)-[\ln(4\pi)+\Gamma'(1)+1]$ and 
\begin{displaymath}
\gamma_1=\frac{1}{3},\quad \gamma_2=\frac{2}{3},\quad
\gamma_3=-\frac{1}{2},\quad
\gamma_4=2,\quad
\gamma_5=-\frac{1}{6},\quad
\gamma_6=-\frac{1}{3},\quad
\gamma_7=0.
\end{displaymath}
   In the SU(2)$\times$SU(2) sector one often uses the scale-independent
parameters $\bar{l}_i$ which are defined by
\begin{equation}
\label{app:barli}
l_i^r=\frac{\gamma_i}{32\pi^2}\left[\bar{l}_i+\ln\left(\frac{M^2}{\mu^2}\right)
\right],\quad i=1,\cdots,6,
\end{equation}
where $M^2=B_0(m_u+m_d)$.
   Since $\ln(1)=0$, the $\bar{l}_i$ are proportional to the renormalized
coupling constant at the scale $\mu=M$.
   Table \ref{app:tablebarli} contains numerical values for 
the scale-independent low-energy coupling constants $\bar{l}_i$
as obtained in Ref.\ \cite{Gasser:1983yg} together with more recent 
determinations.
\begin{table}[htb]
\caption[test]{\label{app:tablebarli} 
Scale-independent low-energy coupling constants $\bar{l}_i$.}
\begin{center}
\begin{tabular}{|c|c|c|r|}
\hline
&
Value & 
Obtained from &
$\gamma_i$\\
\hline
&
$-2.3\pm 3.7$ \cite{Gasser:1983yg} &
$\pi\pi$ $D$-wave scattering lengths ${\cal O}(p^4)$&
\\
& 
$-1.7\pm 1.0$ \cite{Bijnens:1994ie} &
$\pi\pi$ and $K_{l4}$ &
\\
$\bar{l}_1$ &
$-1.5$ \cite{Bijnens:1997vq} &
$\pi\pi$ $D$-wave scattering lengths ${\cal O}(p^6)$&
$\frac{1}{3}$\\ 
&
$-1.8$ \cite{Colangelo:2001df} & 
$\pi\pi$ scattering ${\cal O}(p^4)$ + Roy equations &
\\
&
$-0.4\pm 0.6$ \cite{Colangelo:2001df} & 
$\pi\pi$ scattering ${\cal O}(p^6)$ + Roy equations &
\\
\hline
&
$6.0 \pm 1.3 $ \cite{Gasser:1983yg} &
$\pi\pi$ $D$-wave scattering lengths ${\cal O}(p^4)$&
\\
& $6.1\pm 0.5$ \cite{Bijnens:1994ie} &
$\pi\pi$ and $K_{l4}$ &
\\
$\bar{l}_2$ &
$4.5$ \cite{Bijnens:1997vq} &
$\pi\pi$ $D$-wave scattering lengths ${\cal O}(p^6)$&
$\frac{2}{3}$
\\ 
& 
$5.4$ \cite{Colangelo:2001df} & 
$\pi\pi$ scattering ${\cal O}(p^4)$ + Roy equations &
\\
&
$4.3\pm 0.1$ \cite{Colangelo:2001df} & 
$\pi\pi$ scattering ${\cal O}(p^6)$ + Roy equations &
\\
\hline
&
$2.9\pm 2.4$ \cite{Gasser:1983yg} & 
SU(3) mass formulae &
\\
$\bar{l}_3$ & $|\bar{l}_3|\leq 16$ \cite{Colangelo:2001sp} &
$K_{l4}$ decay &
$-\frac{1}{2}$
\\
\hline
& 
$4.3\pm 0.9$  \cite{Gasser:1983yg} &
$F_K/F_\pi$ &
\\
$\bar{l}_4$&
$4.4\pm 0.3$ \cite{Bijnens:1998fm} & scalar form factor ${\cal O}(p^6)$&
2\\
&
$4.4\pm 0.2$ \cite{Colangelo:2001df} & 
$\pi\pi$ scattering ${\cal O}(p^6)$ + Roy equations &
\\
\hline
& 
$13.9 \pm 1.3$ \cite{Gasser:1983yg} &
$\pi\to e\nu\gamma$ ${\cal O}(p^4)$&
\\
$\bar{l}_5$  & 
$13.0\pm 0.9$   \cite{Bijnens:1998fm} &
$\pi\to e\nu\gamma$ ${\cal O}(p^6)$ \cite{Bijnens:1996wm} &
$-\frac{1}{6}$
\\
\hline
& 
$16.5\pm 1.1$  \cite{Gasser:1983yg} &
$\langle r^2\rangle_\pi$ ${\cal O}(p^4)$&
\\
$\bar{l}_6$&
$16.0\pm 0.5\pm 0.7$   & 
vector form factor ${\cal O}(p^6)$&
$-\frac{1}{3}$\\
&
\cite{Bijnens:1998fm} &
& 
\\
\hline
$l_7$& 
$O(5\times 10^{-3})$ \cite{Gasser:1983yg} &
$\pi^0$-$\eta$ mixing &
0
\\
\hline
\end{tabular}
\end{center}
\end{table}

   Secondly, the expression proportional to $(h_1-h_3)$ can be rewritten
so that the $U$'s completely drop out, i.e., it contains only external
fields. 
   The trick is to use
\begin{eqnarray}
\label{app:glvgss:trick}
\lefteqn{2\mbox{Tr}(\chi U^\dagger\chi U^\dagger+U\chi^\dagger U\chi^\dagger)=}
\nonumber\\
&&[\mbox{Tr}(\chi U^\dagger+U\chi^\dagger)]^2
+[\mbox{Tr}(\chi U^\dagger- U\chi^\dagger)]^2\nonumber\\
&&+[\mbox{Tr}(\tau_i\chi)]^2
+[\mbox{Tr}(\tau_i\chi^\dagger)]^2
-[\mbox{Tr}(\chi)]^2-[\mbox{Tr}(\chi^\dagger)]^2.
\end{eqnarray}
   The terms proportional to the $l_i$ agree with Eq.\ (4.2) of 
Ref.\ \cite{Bellucci:1994eb} but the $l_4$ term is not yet in the form of 
the SU(3)$\times$SU(3) version of Eq.\ (\ref{4:7:l4gl}).
   
   By means of a total-derivative argument in combination with a field
transformation as discussed in Sec.\ \ref{sec_clop4} we will transform
${\cal L}_4^{\rm GL}$ of Eq.\ (\ref{app:glvgss:l4gl}) into another form
which is often used in the literature.
   To that end let us make use of 
\begin{eqnarray*}
\lefteqn{\mbox{Tr}[D_\mu\chi (D^\mu U)^\dagger+D^\mu U (D_\mu\chi)^\dagger]
=}\\
&&\partial_\mu \mbox{Tr}[\chi (D^\mu U)^\dagger + D^\mu U \chi^\dagger]
-\mbox{Tr}[\chi (D^2 U)^\dagger+D^2 U\chi^\dagger],
\end{eqnarray*} 
rewrite the $D^2 U$ and $(D^2 U)^\dagger$ terms by using
\begin{eqnarray*}
2 D^2 U U^\dagger&=&D^2 U U^\dagger - U (D^2 U)^\dagger
-2D_\mu U (D^\mu U)^\dagger,\\
2 U (D^2 U)^\dagger&=&U (D^2 U)^\dagger -D^2 U U^\dagger-2 D_\mu U 
(D^\mu U)^\dagger,
\end{eqnarray*}
to obtain 
\begin{eqnarray}
\label{app:glvgss:intmed1}
\lefteqn{\mbox{Tr}[D_\mu\chi (D^\mu U)^\dagger+D^\mu U (D_\mu\chi)^\dagger]=}
\nonumber\\
&&\mbox{tot. der.}+
\mbox{Tr}[D_\mu U (D^\mu U)^\dagger(\chi U^\dagger + U\chi^\dagger)]
\nonumber\\
&&-\frac{1}{2}\mbox{Tr}\{(\chi U^\dagger-U\chi^\dagger)
[U (D^2 U)^\dagger- D^2 U U^\dagger]\},
\end{eqnarray}
   where ``tot.\ der.'' refers to a total-derivative term which has no
dynamical significance.
   We make use of a trace relation for arbitrary $2\times 2$
matrices $A_i$ [see Eqs.\ (\ref{4:10:F2}) and (\ref{4:10:Tracesu2})],
\begin{eqnarray}
\label{app:glvgss:tr2by2}
\lefteqn{\mbox{Tr}(A_1 A_2 A_3 + A_1 A_3 A_2)
-\mbox{Tr}(A_1)\mbox{Tr}(A_2 A_3)
-\mbox{Tr}(A_2)\mbox{Tr}(A_3 A_1)}\nonumber\\
&&
-\mbox{Tr}(A_3)\mbox{Tr}(A_1 A_2)
+\mbox{Tr}(A_1)\mbox{Tr}(A_2)\mbox{Tr}(A_3)=0,
\end{eqnarray}
and $\mbox{Tr}(D_\mu U U^\dagger)=0$ [see Eq.\ 
(\ref{4:5:kauprop2})]
to rewrite the first term of (\ref{app:glvgss:intmed1}) as the product
of two trace terms,
$$\mbox{Tr}[D_\mu U (D^\mu U)^\dagger(\chi U^\dagger + U\chi^\dagger)]
=\frac{1}{2}\mbox{Tr}[D_\mu U (D^\mu U)^\dagger]
\mbox{Tr}(\chi U^\dagger + U\chi^\dagger).
$$
   By adding and subtracting appropriate $\chi$ terms to generate an
expression proportional to the lowest-order equation of motion which, 
for SU(2)$\times$SU(2), reads [see Eq.\ (\ref{4:5:eom})]
\begin{equation}
{\cal O}^{(2)}_{\rm EOM}(U)=D^2 U U^\dagger - U(D^2 U)^\dagger 
-\chi U^\dagger + U\chi^\dagger+\frac{1}{2}\mbox{Tr}(\chi U^\dagger
-U\chi^\dagger)=0,
\end{equation}
we re-express the last term of Eq.\ (\ref{app:glvgss:intmed1}) as
\begin{eqnarray}
\label{app:glvgss:intmed2}
\lefteqn{-\frac{1}{2}\mbox{Tr}\{(\chi U^\dagger-U\chi^\dagger)
[U (D^2 U)^\dagger- D^2 U U^\dagger]\}=}\nonumber\\
&&+\frac{1}{2}\mbox{Tr}[(\chi U^\dagger-U\chi^\dagger)
{\cal O}^{(2)}_{\rm EOM}(U)]
\nonumber\\
&&+\frac{1}{2}\mbox{Tr}[(\chi U^\dagger- U\chi^\dagger)
(\chi U^\dagger- U\chi^\dagger)]-\frac{1}{4}[\mbox{Tr} 
(\chi U^\dagger- U\chi^\dagger)]^2.
\end{eqnarray}
   The $l_4$ term can thus be written as
\begin{eqnarray}
\label{app:glvgss:intmed3}
\lefteqn{\mbox{Tr}[D_\mu\chi (D^\mu U)^\dagger+D^\mu U (D_\mu\chi)^\dagger]
=}\nonumber\\ 
&&\mbox{tot. der.}+
\frac{1}{2}\mbox{Tr}[D_\mu U (D^\mu U)^\dagger]
\mbox{Tr}(\chi U^\dagger+ U\chi^\dagger)
+\frac{1}{2}\mbox{Tr}(\chi U^\dagger\chi U^\dagger+U\chi^\dagger U\chi^\dagger)
\nonumber\\
&&-\mbox{Tr}(\chi\chi^\dagger)
-\frac{1}{4}[\mbox{Tr} 
(\chi U^\dagger- U\chi^\dagger)]^2
+\frac{1}{2}\mbox{Tr}[(\chi U^\dagger-U\chi^\dagger){\cal O}^{(2)}_{\rm EOM}],
\end{eqnarray}
   which, except for the total derivative and the equation-of-motion term,
is the same as Eq.\ (5.9) of Gasser, Sainio, and \v{S}varc (GSS) 
\cite{Gasser:1987rb}.
 
   The difference between the Lagrangians of \cite{Gasser:1983yg} and
\cite{Gasser:1987rb} then reads
\begin{eqnarray}
\label{app:glvgss:diffl}
{\cal L}_4^{\rm GL}-{\cal L}^{\rm GSS}_4&=&\frac{l_4}{4}\left\{\mbox{tot. der.}
+\frac{1}{2}\mbox{Tr}\left[(\chi U^\dagger - U\chi^\dagger)
{\cal O}^{(2)}_{\rm EOM}
\right]\right\},
\end{eqnarray}
which agrees with Eq.\ (26) of Ecker and Moj\v{z}i\v{s} \cite{Ecker:1995rk}
once their expressions are rewritten in the above notation.
   Let us also specify the field transformation required to connect
the two Lagrangians. 
   For that purpose we rewrite Eq.\ (\ref{app:glvgss:diffl}) in accord with 
Eq.\ (2.11) of  \cite{Scherer:1994wi},
\begin{eqnarray}
\label{app:glvgss:diffl2}
{\cal L}_4^{\rm GSS}(U)&=&{\cal L}_4^{\rm GL}(U) + \mbox{tot. der.}
-\frac{l_4}{8}\mbox{Tr}[(\chi U^\dagger - U\chi^\dagger)
{\cal O}^{(2)}_{\rm EOM}].
\end{eqnarray}
   According to Eqs.\ (\ref{4:7:dl2}) and (\ref{4:7:addstruc2}) 
we need to insert $\alpha_1=0$ and $\alpha_2=-l_4/(2F_0^2)$ in Eq.\
(\ref{4:7:s2}) in order to relate the two Lagrangians.

   Finally, making use of Eqs.\ (\ref{app:glvgss:intmed3}) and
(\ref{app:glvgss:trick}) and dropping the total derivative as
well as the equation-of-motion term let us 
explicitly write out the GSS Lagrangian:
\begin{eqnarray}
\label{app:glvgss:l4gss}
\lefteqn{{\cal L}^{\rm GSS}_4 =
\frac{l_1}{4} \left\{\mbox{Tr}[D_{\mu}U (D^{\mu}U)^{\dagger}] \right\}^2
+\frac{l_2}{4}\mbox{Tr}[D_{\mu}U (D_{\nu}U)^{\dagger}]
\mbox{Tr}[D^{\mu}U (D^{\nu}U)^{\dagger}]
\nonumber} \\
&&+\frac{l_3+l_4}{16}\left[\mbox{Tr}(\chi U^\dagger+ U\chi^\dagger)\right]^2
+\frac{l_4}{8}\mbox{Tr}[D_\mu U(D^\mu U)^\dagger]\mbox{Tr}(\chi U^\dagger
+U\chi^\dagger)\nonumber\\
&&+l_5\mbox{Tr}(f^R_{\mu\nu}U f^{\mu\nu}_LU^\dagger)
+i\frac{l_6}{2}\mbox{Tr}[ f^R_{\mu\nu} D^{\mu} U (D^{\nu} U)^{\dagger}
+ f^L_{\mu\nu} (D^{\mu} U)^{\dagger} D^{\nu} U]\nonumber\\
&&-\frac{l_7}{16}\left[\mbox{Tr}(\chi U^\dagger-U\chi^\dagger)\right]^2
+\frac{h_1+h_3-l_4}{4}\mbox{Tr}(\chi\chi^\dagger)\nonumber\\
&&+\frac{h_1-h_3-l_4}{16}\left\{
\left[\mbox{Tr}(\chi U^\dagger + U\chi^\dagger)\right]^2
+\left[\mbox{Tr}(\chi U^\dagger-U\chi^\dagger)\right]^2\right.\nonumber\\
&&\left.
-2\mbox{Tr}(\chi U^\dagger\chi U^\dagger + U\chi^\dagger U\chi^\dagger)
\right\}
-\frac{4h_2+l_5}{2}\mbox{Tr}(f_{\mu\nu}^L f^{\mu\nu}_L
+f_{\mu\nu}^R f^{\mu\nu}_R).
\end{eqnarray}

\section{Different Parameterizations}
\label{app_sec_dp}
   In App.\ \ref{app_sec_glvgss} we saw that two versions of the 
${\cal O}(p^4)$ $\mbox{SU(2)}\times\mbox{SU(2)}$ mesonic Lagrangian,
Eqs.\ (\ref{app:glvgss:l4gl}) and (\ref{app:glvgss:l4gss}), 
are used in the literature.
   Since they are related by a field transformation, they must yield
the same results for physical observables \cite{Chisholm,Kamefuchi:sb}.
   Furthermore, in SU(2)$\times$SU(2) two different parameterizations of 
the SU(2) matrix $U(x)$ [see Eqs.\ (\ref{4:6:u1}) and (\ref{4:6:u2})]
are popular,
\begin{eqnarray}
\label{app:dp:exp}
U(x)&=&\exp\left[i\frac{\vec{\tau}\cdot\vec{\phi}(x)}{F_0}\right],\\
\label{app:dp:sqrt}
U(x)&=&\frac{1}{F_0}\left[\sigma(x)+i\vec{\tau}\cdot\vec{\pi}(x)
\right],\quad \sigma(x)=\sqrt{F^2_0-\vec{\pi}\,^2(x)},
\end{eqnarray}   
  where the pion fields of the two parameterizations are non-linearly related
[see Eq.\ (\ref{4:6:ft})].
 
   In this appendix we collect the pion wave function renormalization 
constants entering a calculation at ${\cal O}(p^4)$ depending on which 
Lagrangian and parameterization is used.
   The actual calculation parallels that of Sec.\ \ref{subsec_mgb}
and will not be repeated here.
  The self energies up to ${\cal O}(p^4)$ can be written as 
\begin{equation}
\label{app:dp:selfenergy}
\Sigma(p^2) = A + B p^2.
\end{equation}
   The renormalized mass and the wave function renormalization constant
are, respectively to ${\cal O}(p^4)$ and ${\cal O}(p^2)$, given by
\begin{eqnarray}
\label{app:dp:mass}
M^2_{\pi,4} &=& M^2_{\pi,2}(1+B)+A,\\
\label{app:dp:Z}
Z&=&1+B,
\end{eqnarray} 
   where $M^2_{\pi,2}=2B_0 m$ is the prediction at ${\cal O}(p^2)$.
   The different values for $A$, $B$, and $Z$ are given in Table
\ref{app:dp:tab:abz}.
   Note that the result for the pion mass is, as expected, independent
of the Lagrangian and parameterization used:
\begin{equation}
\label{app:dp:Mpi2}
M_{\pi,4}^2=M^2\left(1+\frac{2}{3} \frac{I}{F_0^2}\right)
-\frac{1}{6}\frac{M^2}{F_0^2}I
+2 l_3 \frac{M^4}{F_0^2}
=M^2-\frac{\bar{l}_3}{32\pi^2 F_0^2} M^4, 
\end{equation}
where $M^2=2 B_0 m$ and 
\begin{displaymath}
l_3=-\frac{1}{64\pi^2}\left[\bar{l}_3+\ln\left(\frac{M^2}{\mu^2}\right)
+R\right]
\end{displaymath}
[see Eqs.\ (\ref{app:lilir}) and (\ref{app:barli})].
   On the other hand, the constants $A$, $B$, and $Z$ are 
auxiliary mathematical quantities and thus depend on both Lagrangian and 
parameterization.

\begin{table}[htb]
\caption{Self-energy coefficients and wave function renormalization
constants for the Lagrangians of Eqs.\ (\ref{app:glvgss:l4gl}) (GL)
and (\ref{app:glvgss:l4gss}) (GSS) 
and the field parameterizations of Eqs.\ (\ref{app:dp:exp})
and (\ref{app:dp:sqrt}).
   $I$ denotes the dimensionally regularized integral of 
Eq.\ (\ref{app:drb:im22}),
$I=I(M^2,\mu^2,n)=\frac{M^2}{16\pi^2}\left[
R+\ln\left(\frac{M^2}{\mu^2}\right)\right]+O(n-4)$,
$R=\frac{2}{n-4}-\left[\ln(4\pi)+\Gamma'(1)+1\right]$, $M^2=2B_0 m$.
}
\label{app:dp:tab:abz}
\begin{tabular}
{|c|c|c|c|}
\hline
&$A$&$B$&$Z$\\
\hline
&&&\\
GL, Eq.\ (\ref{app:dp:exp})&
$-\frac{1}{6}\frac{M^2}{F^2_0} I+2 l_3 \frac{M^4}{F^2_0}$&
$\frac{2}{3}\frac{I}{F^2_0}$&
$1+\frac{2}{3}\frac{I}{F^2_0}$\\
&&&\\
\hline
&&&\\
GL, Eq.\ (\ref{app:dp:sqrt})&
$\frac{3}{2}\frac{M^2}{F^2_0} I+2 l_3 \frac{M^4}{F^2_0}$&
$-\frac{I}{F^2_0}$&
$1-\frac{I}{F^2_0}$\\
&&&\\
\hline
&&&\\
GSS, Eq.\ (\ref{app:dp:exp})&
$-\frac{1}{6}\frac{M^2}{F^2_0} I+2 (l_3+l_4)\frac{M^4}{F^2_0}$&
$\frac{2}{3}\frac{I}{F^2_0}-2l_4\frac{M^2}{F^2_0}$&
$1+\frac{2}{3}\frac{I}{F^2_0}-2l_4\frac{M^2}{F^2_0}$\\
&&&\\
\hline
&&&\\
GSS, Eq.\ (\ref{app:dp:sqrt})&
$\frac{3}{2}\frac{M^2}{F^2_0} I+2 (l_3+l_4)\frac{M^4}{F^2_0}$&
$-\frac{I}{F^2_0}-2l_4\frac{M^2}{F^2_0}$&
$1-\frac{I}{F^2_0}-2l_4\frac{M^2}{F^2_0}$
\\
&&&\\
\hline
\end{tabular}
\end{table}

\end{appendix}

\end{document}